\documentclass[11pt,twoside,a4paper,cmspaper,final,collab]{cms-tdr}
\def\svnVersion{5e66d7d}\def\svnDate{2026/09/30}\def\cmsCernNoTag{CERN-EP-2025-294}\def\cmsCernDate{\today}\def\cmsMessage{Published in the Journal of High Energy Physics as \href{http://dx.doi.org/10.1007/JHEP09(2026)258}{\doi{10.1007/JHEP09(2026)258}.}}
\begin{document}\cmsNoteHeader{SMP-22-003}

\newcommand{\Nsubjettiness}{\ensuremath{N\text{-subjettiness}}\xspace}
\newcommand{\Nsub}[2]{\ensuremath{\tau_{#1}^{(#2)}}\xspace}
\newcommand{\HERWIGvii}{\HERWIG{}\,7\xspace}
\providecommand{\cmsLeft}{left\xspace}
\providecommand{\cmsRight}{right\xspace}

\cmsNoteHeader{SMP-22-003}
\title{Simultaneous measurements of $N$-subjettiness observables in jets from gluons and light-flavour quarks, and in decays of boosted \texorpdfstring{\PW}{W} bosons and top quarks}

\date{\today}

\abstract{
	A simultaneous measurement of 25 substructure observables is presented using large-radius jets with high transverse momentum from proton-proton collisions at $\sqrt{s}=13\TeV$. The measurement is carried out on dijet events and \ttbar events enriched in Lorentz-boosted \PW bosons and top quarks decaying hadronically. The three data samples consist of jets with one, two, or three prongs from the showering and hadronization of a gluon or light-flavour quark, two quarks, or three quarks, respectively. The data correspond to an integrated luminosity of 138\fbinv, recorded by the CMS experiment in 2016--2018. A detailed characterization of the jet substructure is provided using a 6-body basis of $N$-subjettiness observables that overconstrains the phase space of the resolved emissions in the jet. The measurements are unfolded to the level of stable particles, and an estimate of the particle-level correlations between observables is provided, ensuring that the results can be used to systematically assess and refine the modelling of radiation in jets. 	
}

\hypersetup{pdfauthor={CMS Collaboration},pdftitle={Simultaneous measurements of N-subjettiness observables in jets from gluon and light-flavour quarks, and in decays of boosted W bosons and top quarks},
	pdfsubject={CMS},pdfkeywords={CMS, jet, substructure, QCD, basis, subjettiness, unfolding}}

\maketitle 

\section{Introduction}
\label{sec:intro}
Jets are collimated showers of hadrons initiated by the fragmentation of quarks and gluons (partons) produced in particle collisions, and are reconstructed using dedicated clustering algorithms~\cite{Salam:2010nqg} that aim to maintain coherence between the hadron- and parton-level descriptions. 
In the standard model (SM) of particle physics, quantum chromodynamics (QCD) is the theory of the strong force that governs the interactions of these partons. 
The successful operation of the CERN LHC has enabled precise studies of final states containing jets, providing tests of the SM in a previously inaccessible kinematic regime. 

In proton-proton ($\Pp\Pp$) collisions at LHC energies, there is a significant production of massive particles, such as the \PW boson and the top quark, which predominantly decay hadronically. 
When they are sufficiently Lorentz boosted, outgoing partons from their decays are highly collimated and merged into single, large-radius jets. 
This yields the characteristic two- and three-pronged jet substructure topologies of the hadronic decays of boosted \PW bosons and top quarks, respectively. 
Here, prongs are localized clusters of energy within the collection of final-state particles of the jet, corresponding to the fragmentation and hadronization of the coloured partons from the resonance decay. 
The substructure of gluon and light-flavour quark jets is characteristically one-pronged, with diffused radiation around a hard core, but can have contributions from additional energetic emissions. 

At the LHC, jet substructure observables are used to enhance the sensitivity of precision measurements and of searches for physics beyond the SM by characterizing the internal radiation pattern of high transverse-momentum (\pt) jets~\cite{Larkoski2017,Kogler:2018hem,Marzani:2019hun}. 
Unfolded measurements of substructure observables probing the underlying QCD dynamics, with jets from different originating particles, provide an invaluable input for the modelling of parton showering and hadronization in simulations.

One widely used class of infrared- and collinear-safe (IRC-safe) jet substructure observables is $N$-subjettiness \Nsub{N}{\beta}~\cite{Thaler:2010tr,Thaler2012b}, where $N$ denotes the number of specified subjet axes used to characterize the radiation pattern inside a jet.  
Typically, these are used to construct jet discriminants by considering ratios of the form $\tau_{N} / \tau_{N-1}$ for the identification of hadronic multi-prong decays of boosted resonances (jet tagging). While substructure observables with simple analytic forms are powerful discriminants, they are also used as effective inputs to machine-learning (ML) jet classifiers trained on simulations~\cite{Larkoski2017,Mondal:2919392}. 

The challenge of ML-based jet tagging was addressed in Ref.~\cite{Datta:1}, proposing an organizing principle for the information in a jet that leverages sets of $N$-subjettiness observables to minimally and completely resolve the $(3M-4)$-dimensional kinematic phase space of $M$ emissions in a jet. 
This is referred to as a basis for an $M$-body phase space in the following. 
Jet-tagging classifiers trained on these bases are used to identify the relevant $M$-body phase space for a given number of prongs by studying where the discrimination power saturates. 
This motivates the experimental measurement of an adequately expressive basis of substructure observables that captures the corresponding distinguishing features in the hard-collinear and soft-diffuse radiation in the jets, probing both wide-angle radiation and finer substructure features arising from the parton shower.

Several measurements probing the characteristic radiation pattern within jets have been carried out by the ATLAS~\cite{ATLAS:2017pgl,ATLAS:2017zda,ATLAS:2018bvp,ATLAS:2018olo,ATLAS:2018zhf,ATLAS:2019mgf,ATLAS:2019rqw,ATLAS:2019kwg,ATLAS:2019dty,ATLAS:2019dsv,Aad_2020,ATLAS:2021agf,ATLAS:2023jdw,ATLAS:2024dua}, CMS~\cite{CMS:2017qlm,CMS:2018ypj,CMS:2018vzn,CMS:2018fof,CMS:2018jco,CMS:2021iwu,CMS:2024mlf,CMS:2023lpp}, ALICE~\cite{ALICE:2017nij,ALICE:2019ykw,ALICE:2020pga,ALICE:2021njq,ALargeIonColliderExperiment:2021mqf,ALICE:2021aqk,ALICE:2021vrw}, and LHCb~\cite{Aaij_2017,LHCb:2019qoc} Collaborations. 
These include recent measurements of multicount (per jet) observables, such as the density of emissions in the Lund jet plane and energy-energy correlators, probing the modelling of perturbative and nonperturbative QCD dynamics and the scaling behaviour of QCD by direct sensitivity to the time evolution of the jet. 

This analysis similarly aims to provide a detailed picture of the radiation in a jet, relying on a fixed $M$-body jet description, following the prescription of Ref.~\cite{Datta:1}. The measurements are carried out in one-, two-, and three-pronged large-radius jets at high \pt. 
One-pronged jets are obtained from QCD dijet events predominantly composed of gluon and light-flavour quark jets. Events in the muon+jets ($\PGm$+jets) channel of \ttbar production are split into two mutually exclusive samples enriched in two- and three-pronged jets that capture the hadronic decays of boosted \PW bosons and top quarks, respectively. 
The data set from $\Pp\Pp$ collisions at $\sqrt{s}=13\TeV$ recorded by the CMS detector during 2016--2018 is used, corresponding to a total integrated luminosity of approximately 138\fbinv~\cite{CMS-PAS-LUM-17-003, CMS-PAS-LUM-17-004, CMS-PAS-LUM-18-002}. 

Using simulated events in the dijet, \PW boson-, and top quark-enriched event samples, we demonstrate that discrimination power for boosted \PW boson and top quark jets versus QCD jets effectively saturates once the 4- and 5-body phase space of the jets is resolved, respectively, at both the particle and detector levels. 
This motivates the use of a 6-body basis for the measurement in order to constrain any further discriminating information in the higher multi-body phase space of the jets. 
Then, to robustly overconstrain the information in the 6-body kinematic phase space, the minimal and complete formulation of the basis is extended by considering additional $N$-subjettiness observables. 
This yields a collection of 25 $N$-subjettiness observables, and is referred to as the overcomplete (OC) 6-body basis in the following. 

The measured distributions of individual $N$-subjettiness observables in the overcomplete 6-body basis are unfolded simultaneously to correct for detector effects by considering a single, combined distribution of all observables, within a global binning scheme and, correspondingly, a non-diagonal input covariance matrix capturing the correlations between bins of the individual observables. 
This ensures that the statistical correlations among the observables are preserved and propagated through the unfolding procedure to the covariance matrix of the unfolded, combined distribution. 
The unfolded measurements, together with the resulting correlations at the particle level, provide a comprehensive set of inputs for the development of parton shower and hadronization models.

In the following, we define the basis of $N$-subjettiness observables used for the measurement in Section~\ref{sec:basis}. 
We describe the CMS detector in Section~\ref{sec:detector} and pertinent aspects of the event reconstruction in Section~\ref{sec:reconstruction}. 
The data and simulated samples used in this work are detailed in Section~\ref{sec:samples}, and in Section~\ref{sec:selection} we define the particle-level phase spaces, and corresponding detector-level event selections, for the QCD dijet, as well as the boosted \PW boson- and top quark-enriched measurements. 
In Section~\ref{sec:saturation}, we determine in simulation the point of saturation of the discrimination power of jet taggers for \PW boson and top quark jets versus QCD jets. 
The unfolding procedure and sources of systematic uncertainty in the measurement are described in Sections~\ref{sec:unfolding} and~\ref{sec:systematics}, respectively. 
The combined, unfolded results are presented in Section~\ref{sec:results}, along with illustrative examples of individual observables extracted from the simultaneous unfolding in each event category. 
Finally, we summarize the results in Section~\ref{sec:summary}. 
Additional results for pair-wise correlations between the various $N$-subjettiness observables are presented in Appendix~\ref{sec:Correlations}, and particle-level correlations and unfolded results for individual observables, extracted from the simultaneous unfolding, are presented in Appendices~\ref{sec:unfCombinedCorr} and~\ref{sec:resultsPerObs}, respectively. 
Tabulated results for the unfolded data are provided in the HEPData record for this analysis \cite{hepdata}. 
 
\section{Basis of \texorpdfstring{$N$}{N}-subjettiness observables}
\label{sec:basis}
The $N$-subjettiness observables $\tau_N^{(\beta)}$ ~\cite{Thaler:2010tr,Thaler2012b}, inspired by the $N$-jettiness global event shape \cite{Stewart_2010}, provide a measure of the QCD radiation about $N$ subjet axes in a jet and are defined as
\begin{equation}
	\label{eq:nsub}
	\Nsub{N}{\beta} = \frac{1}{d_0} \sum_{i\in \text{Jet}} {\pt}_{i} \min\left\{
	\Delta R_{1i}^\beta,\Delta R_{2i}^\beta,\dotsc,\Delta R_{Ni}^\beta
	\right\}.
\end{equation}
\sloppy{Here, $\beta$ is an angular weighting exponent, $d_0=\sum_{i\in \text{Jet}}{\pt}_{i}R^\beta$ is a normalization factor, $R=0.8$ is the jet radius, ${\pt}_{i}$ is the transverse momentum of particle $i$ in the jet, and $\Delta R_{ki}=\sqrt{\smash[b]{\left(\Delta y_{ki}\right)^2 + \left(\Delta\phi_{ki}\right)^2}}$, for $k=1,2,\dotsc,N$, is the distance in the rapidity-azimuth $y$-$\phi$ plane between particle $i$ and subjet axis $k$ in the jet. 
In the collinear limit, the functional form for the $N$-subjettiness observables presented in Eq.~\eqref{eq:nsub}, can be instructively reformulated as}
\begin{equation}
	\label{eq:nsub2}
	\Nsub{N}{\beta} \propto \sum_{i\in \text{Jet}} {z}_{i} \min\left\{
	\theta_{i1}^\beta,\theta_{i2}^\beta,\dotsc,\theta_{iN}^\beta
	\right\},
\end{equation}
where, for particle $i$ in the jet, ${z}_{i}$ is its relative energy fraction and $\theta_{ik}$ is the angle between its momentum vector and the $k^{\mathrm{th}}$ subjet axis in the jet. 
Individual $N$-subjettiness observables effectively quantify the compatibility of a jet with the hypothesis of having $N$ or fewer subjets. When most of the energy in a jet is well aligned with the $N$ subjet axes, $\tau_N^{(\beta)}$ is small, whereas it is larger when a jet has more than $N$ hard prongs. For a jet with $N$ or fewer constituents, $\tau_N^{(\beta)}=0$. 
By construction, for subjets determined with an IRC-safe algorithm, the observables are collinear-safe for values of the exponent $\beta\geq0$, and infrared-safe since \Nsub{N}{\beta} is linear in the jet constituents' momenta. 

In a Lorentz-invariant system, an $M$-body phase space consists of all possible four-momenta configurations of particles constituting the system. 
Considering the $M$ assumed emissions in a jet as $M$ four-momenta with fixed masses and imposing overall energy and momentum conservation yields a generic dimensionality of $(3M-4)$ for the $M$-body phase space of a jet. 
These correspond to $M-1$ energy fractions $z_i$ of emissions in the jet, and $2M-3$ to the angles $\theta_{ij}$ between them.
Then, the kinematic coordinates of the $M$-body phase space can be minimally and completely determined with a $(3M-4)$-dimensional basis of $N$-subjettiness observables,
\begin{equation}
	\label{eq:Mbodybasis}
	\left\{
	\tau_1^{(0.5)},\tau_1^{(1)},\tau_1^{(2)},\dotsc,\tau_{M-2}^{(0.5)},\tau_{M-2}^{(1)},\tau_{M-2}^{(2)},\tau_{M-1}^{(1)},\tau_{M-1}^{(2)}
	\right\}.
\end{equation}
The basis spans the jet substructure for generic configurations of the momenta of emissions in the jet, for noncollinear emissions with nonzero energy \cite{Datta:1}, that is, the observables define a simultaneous system of equations that can be inverted to uniquely solve for the kinematic phase-space coordinates of the $M$ assumed emissions. Such a basis is thus systematically improvable: following the above organizing principle one can add three further $N$-subjettiness observables to an $M$-body basis to resolve further structure in the jet corresponding to $(M+1)$-body phase space, and, by construction, when $M$ equals the number of particles in the jet, the basis captures all of the IRC-safe information in a jet substructure. 

The individual observables are dominantly sensitive to the information about small-angle radiation in jets and wide-angle features, respectively, for values of $\beta<1$ and $\beta>1$. 
Leveraging various values of the angular weighting $\beta$, for various values of $N$, allows the collection of observables to span the full $M$-body phase space and to capture features from different physics effects in the radiation patterns of the jets. 

To calculate the observables for this measurement, the $N$ subjet axes are determined using the exclusive \kt algorithm \cite{Catani1993a,Ellis1993a} and $E$-scheme recombination \cite{Blazey:2000qt}, as implemented in the \textsc{FastJet} package \cite{Cacciari:2011ma}. 
While recoil-free axis schemes such as winner-take-all (WTA) recombination \cite{Bertolini:2013iqa,Larkoski_2014,Larkoski_2014b} are more robust to contributions from soft radiation in the event, $3M-4$ observables computed, e.g., with the WTA recombination cannot be used to generically span $M$-body kinematic phase space when $M\geq3$~\cite{Datta:1}.

Finally, we extend the aforementioned minimal and complete formulation of the bases, by considering further $N$-subjettiness observables computed with values of the angular exponent, $\beta=0.25, 1.5$, as shown for the 6-body phase space:
\begin{align}
	\label{eq:overcomplete}
	\left\{
	\tau_1^{(0.25)},\tau_1^{(0.5)},\tau_1^{(1)},\tau_1^{(1.5)}, \tau_1^{(2)},\dotsc,\tau_5^{(0.25)},\tau_5^{(0.5)},\tau_5^{(1)},\tau_5^{(1.5)}, \tau_5^{(2)}
	\right\}.
\end{align}
The reason to consider an expanded basis for the measurement is that in experimental data, the information captured by the observables and information of angular relations between resolved emissions therein are smeared out by detector effects, which themselves vary in terms of effects in detector-level simulation and measured data. 
Further, the individual $N$-subjettiness observables with $\beta=0.5, 2$ are found to have limited experimental resolution, compared with the $\beta=1$ case. 
This is particularly true for values of $N\geq3$, where the individual observables are sensitive to $\geq$4-body phase space. 
Thus, the specific values of $\beta=0.25, 1.5$ for observables in the overcomplete 6-body basis are chosen with the intention to provide additional, redundant handles sensitive to information from collinear effects and wide-angle contributions in particular. Then, the overcomplete basis used in the final unfolded measurements is comprised of 25 individual $N$-subjettiness observables.

\section{The CMS detector}
\label{sec:detector}
The CMS apparatus~\cite{CMS:2008xjf,CMS:2023gfb} is a multi-purpose, nearly hermetic detector, designed to trigger on~\cite{CMS:2020cmk,CMS:2016ngn,CMS:2024aqx} and identify electrons, muons, photons, and (charged and neutral) hadrons~\cite{CMS:2020uim,CMS:2018rym,CMS:2014pgm}. 
The central feature of the CMS detector is a superconducting solenoid of 6\unit{m} internal diameter, providing a magnetic field of 3.8\unit{T}. 
Within the solenoid volume are a silicon pixel and strip tracker, a lead tungstate crystal electromagnetic calorimeter (ECAL), and a brass and scintillator hadron calorimeter (HCAL), each composed of a barrel and two endcap sections. 
Forward calorimeters extend the pseudorapidity ($\eta$) coverage provided by the barrel and endcap detectors. 
Muons are reconstructed with gas-ionization detectors embedded in the steel flux-return yoke outside the solenoid. 
The CMS pixel detector was upgraded between data-taking runs in 2016 and 2017, with further layers added in the barrel and endcap regions~\cite{Phase1Pixel} leading to an improvement ~\cite{CMS:2014pgm} in the track resolution~\cite{DP-2020-049,DP-2017-015}. 
A more detailed description of the CMS detector, together with a definition of the coordinate system and relevant kinematic variables, can be found in Refs.~\cite{CMS:2008xjf,CMS:2023gfb}.

Events of interest are selected using a two-tiered trigger system. 
The first level (L1), composed of custom hardware processors, uses information from the calorimeters and muon detectors to select events at a rate of around 100\unit{kHz} within a fixed latency of about 4\mus~\cite{CMS:2020cmk}. 
The second level, referred to as the high-level trigger (HLT), consists of a farm of processors running a version of the full event reconstruction software optimised for fast processing, and reduces the event rate to a few \unit{kHz} before data storage~\cite{CMS:2016ngn,CMS:2024aqx}. 
 
\section{Event reconstruction}
\label{sec:reconstruction}
A particle-flow (PF) algorithm~\cite{CMS:2017yfk} aims to reconstruct and identify each individual particle in an event as a photon, electron, muon, charged hadron or neutral hadron with an optimized combination of information from all CMS subdetectors. 
Muons are identified as tracks in the central tracker consistent with either a track or several hits in the muon system, and associated with calorimeter deposits compatible with the muon hypothesis. 
The energy of muons is obtained from the curvature of the corresponding tracks. 
Photons are identified as ECAL energy clusters not linked to the extrapolation of any charged-particle trajectory to the ECAL, and their energy is determined from the ECAL measurement. 
Electrons are identified as a primary charged-particle track, potentially multiple ECAL energy clusters corresponding to this track extrapolation to the ECAL, and to possible bremsstrahlung photons emitted along the way through the tracker material. 
The energy of electrons is reconstructed from a combination of the track momentum at the main interaction vertex, the corresponding ECAL cluster energy, and the energy sum of all bremsstrahlung photons attached to the track. 
Charged hadrons are identified as charged-particle tracks neither identified as electrons, nor as muons. 
Neutral hadrons are either identified as HCAL energy clusters not linked to any charged-hadron trajectory, or as a combined ECAL and HCAL energy excess with respect to the expected charged-hadron energy deposit. 
The energy of charged hadrons is determined from a combination of the track momentum and the corresponding ECAL and HCAL energies, corrected for the response function of the calorimeters to hadronic showers. 
Finally, the energy of neutral hadrons is obtained from the corresponding corrected ECAL and HCAL energies.

Muons are measured in the pseudorapidity range $\abs{\eta}<2.4$, with detection planes made using three technologies: drift tubes, cathode-strip chambers, and resistive-plate chambers. 
The single-muon trigger efficiency exceeds 90\% over the full $\eta$ range, and the efficiency to reconstruct and identify muons is greater than 96\%~\cite{CMS:2018rym}. 

Hadronic jets are reconstructed from PF candidates using the IRC-safe anti-\kt algorithm~\cite{Cacciari:2008gp}, as implemented in the \FASTJET software package \cite{Cacciari:2011ma}. 
Jet momentum is determined as the vectorial sum of all particle momenta in the jet and is found from simulation to be, on average, within 5 to 10\% of the true momentum over the entire \pt spectrum and detector acceptance.

Two types of anti-\kt jets are utilized in this analysis. 
The first is clustered with the distance parameter for the anti-\kt algorithm $R=0.4$ (AK4 jets). 
Additional $\Pp\Pp$ interactions within the same or nearby bunch crossings (pileup) can contribute extra tracks and calorimetric energy depositions to the jet momentum. 
For the AK4 jets, used only to identify candidate \PQb jets in the $\PGm$+jets \ttbar event selection, charged PF candidates not associated with the primary vertex are removed (charged-hadron subtraction), and the remaining neutral contribution from pileup interactions is corrected on an event-by-event, jet-area basis~\cite{Sirunyan:2020foa,CMS-PAS-JME-14-001,Cacciari:2007fd}. 
The second collection of large-radius anti-\kt jets is clustered with a distance parameter $R=0.8$ (AK8 jets) and is employed for the measurement of the basis of $N$-subjettiness observables in both the dijet and $\PGm$+jets \ttbar event selections. 

To mitigate the effect of pileup on the substructure measurements, the pileup-per-particle identification algorithm (PUPPI)~\cite{Sirunyan:2020foa,Bertolini:2014bba} is used at the detector level, making use of the local shape information, event pileup properties, and tracking information. 
A local variable is defined for each particle, which distinguishes between the collinear and soft diffuse contributions from particles surrounding the one under consideration. 
The local shape for charged pileup is assumed to be the same as neutral pileup, and the median of its distribution is computed on an event-by-event basis. 
The probability for a specific particle to originate from the pileup is then determined by comparing its local shape variable to the median value for charged pileup. 
Based on this information, a weight is computed for each particle to rescale its four-momentum; jet constituents compatible with pileup are down-weighted, and charged PF candidates from pileup vertices receive effectively zero weight to ensure charged pileup does not contribute to the AK8 jet collection. This supersedes the need for jet-based corrections~\cite{Sirunyan:2020foa,CMS-PAS-JME-14-001}.

Jet energy corrections are derived from simulation and applied to both the AK4 and AK8 jet collections to bring the measured response of the jets to that of particle-level jets on average. 
In situ measurements of the momentum balance in dijet, photon+jet, {\PZ}+jet, and multijet events are used to account for any residual differences in the jet energy scale between data and simulation~\cite{CMS:2016lmd}. 
The jet energy resolution amounts typically to 15--20\% at 30\GeV, 10\% at 100\GeV, and 5\% at 1\TeV~\cite{CMS:2016lmd}. 

The primary vertex (PV) is taken to be the vertex corresponding to the hardest scattering in the event, evaluated using tracking information alone, as described in Section 9.4.1 of Ref.~\cite{CMS-TDR-15-02}. 
The missing transverse momentum vector \ptvecmiss is computed as the negative vector sum of the transverse momenta of all the PF candidates in an event, and its magnitude is denoted as \ptmiss~\cite{CMS:2019ctu}. 
The \ptvecmiss is modified to account for corrections to the energy scale of the reconstructed AK4 jets in the event. 
Anomalous high-\ptmiss events can arise from a variety of reconstruction failures, detector malfunctions or noncollision backgrounds, and these are rejected by event filters designed to identify more than 85--90\% of the spurious high-\ptmiss events with a mistagging rate less than 0.1\%~\cite{CMS:2019ctu}.

\section{Data and simulated samples}
\label{sec:samples}
Events of interest in the $\PGm$+jets \ttbar selection rely on a combination of HLT selection algorithms (``paths'') that require the presence of at least one muon with $\pt>50\GeV$ with $\abs{\eta}<2.4$, and without an isolation requirement, which are referred to as nonisolated muon trigger paths. 
The efficiency to select events with a nonisolated muon using these trigger paths is about 90\%~\cite{CMS:2021yvr}. 

The QCD dijet events are selected using several prescaled, and one unprescaled, HLT paths. 
The triggers require at least one AK8 jet in the event to have a \pt larger than the nominal trigger thresholds, in the range of 80 to 450\GeV for 2016, and up to 500\GeV for the 2017 and 2018 data-taking periods.

The prescaled triggers collect a fraction of the collision events passing the leading-jet \pt requirement, and the data collected by them are scaled to match the total luminosity recorded by the CMS detector. 
The efficiency of the AK8 jet triggers was studied for each data-taking period. 
The offline AK8 jet \pt at which each HLT path is at least 99.5\% efficient is identified for all periods, and is considered as the minimum threshold to select events.

The leading AK8 jet HLT paths utilized for taking data in 2016 came online after the first approximately 3\fbinv of data were recorded by the CMS detector. 
Therefore, the measurements in the dijet selection correspond to a slightly lower total integrated luminosity of at most 135\fbinv, compared with the 138\fbinv data set used in the $\PGm$+jets \ttbar selection.

For each data-taking period, the relevant SM processes are simulated using different Monte Carlo (MC) event generators and normalized to the corresponding total integrated luminosity. 
The various event generators simulating the hard scattering are interfaced with parton shower and hadronization programs, followed by a full simulation of the CMS detector based on the \GEANTfour package~\cite{Agostinelli:2002hh, Allison:2006ve}. 
All simulated samples, including both signal and background samples, unless otherwise noted, are showered in \PYTHIA~8.240~\cite{Sjostrand:2014zea} using the CP5 tune~\cite{2020_CMS} with a value of the strong coupling at the \PZ pole mass, $\alpha_S(m_Z^2)=0.118$. 
Samples showered with \HERWIG{}\,7.2.2~\cite{B_hr_2008,Bellm:2015jjp} use the CH3 tune \cite{CMS:2020dqt}, which relies on values of the strong coupling $\alpS(m_Z^2)=0.118$ for the parton shower and $\alpS(m_Z^2)=0.130$ for the modelling of multiple parton interactions (MPIs) and to handle beam remnants. 
In all of the following figures, \PYTHIAviii and \HERWIGvii are abbreviated as P8 and H7, respectively, when used as the parton-shower program in combination with a matrix-element (ME) generator, whereas samples generated and showered with \PYTHIAviii alone are labelled as \PYTHIAviii. 
All simulations include the effects of multiple $\Pp\Pp$ collisions per bunch crossing, and events at the detector level are reweighted to the corresponding pileup conditions in the measured data. 

For the measurement based on dijet events, \MGvATNLO~2.6.5~\cite{Alwall:2014hca} with MLM jet merging~\cite{Alwall_2007,Mangano:2006rw} is used to simulate SM events composed uniquely of jets produced through the strong interaction, referred to as QCD multijet events, at leading order (LO) in bins of the scalar sum of the \pt of all jets in an event \HT. 
The matrix elements for this simulation include up to four hard partons in the final state. 
The nominal sample for the measurements uses \PYTHIAviii for the parton shower and hadronization, while the alternative sample is interfaced with showering in \HERWIGvii. 
The latter is used to estimate the shower and hadronization uncertainties in the unfolded results. 
An additional QCD multijet sample is used, exclusively for the comparison of simulation to data; this is generated and showered with \PYTHIAviii in bins of the transverse momentum exchange $\hat{p}_{\mathrm{T}}$ in the hard scatter. 
No other samples are used because of the negligible contributions from SM background processes in the fiducial phase space defined for the dijet measurement~\cite{CMS:2021iwu}. 

To study jets originating from boosted \PW boson and top quark decays, we use \POWHEG{} v2~\cite{Nason:2004rx,Frixione:2007vw,Alioli:2010xd,Frixione:2007nw, Alioli:2009je,Re:2010bp} at next-to-LO (NLO) to simulate \ttbar events. 
For the nominal simulation used for the measurement, the ME generator is interfaced with \PYTHIAviii to simulate parton showering, hadronization, and MPIs. 
An alternative \ttbar sample, showered with \HERWIGvii, is used to estimate the contribution of shower and hadronization uncertainties to the unfolded results. 
A further \ttbar sample is generated at NLO accuracy with \MGvATNLO using the FxFx~\cite{Frederix_2012} jet merging scheme and showered using \PYTHIAviii; this is used exclusively for studies of the saturation of discrimination power in Section~\ref{sec:saturation} and for data-to-simulation comparisons. 
In the following, we will abbreviate \MGvATNLO as \MADGRAPH{}5 and a\MCATNLO{} in all relevant figures to delineate between samples generated at LO and NLO accuracy, respectively. 
All \ttbar samples are normalized to the next-to-NLO (NNLO) cross section prediction, $\sigma_{\ttbar}=833.9\unit{pb}$~\cite{PhysRevD.110.030001}, for $\Pp\Pp$ collisions at $\sqrt{s} = 13\TeV$ assuming a top quark mass of 172.5\GeV.

Additional SM processes with similar final states are considered as backgrounds for the measurements in the $\PGm$+jets \ttbar selection, including simulations of the single production of top quarks, the associated production of top quarks with a \PW boson, diboson ($\PW\PW$, $\PZ\PZ$, $\PW\PZ$) production, \PW boson production in association with jets, Drell--Yan processes, and QCD multijets enriched in muons in the final state. Contributions from single top quark/anti-quark production are generated at NLO with \MGvATNLO in the four-flavour scheme for the $s$-channel process, while events for the $t$-channel process and for the associated production of a top quark/anti-quark with a \PW boson are simulated using the five-flavour scheme with the \POWHEG generator. 
The simulation of background events from {\PW}+jets and Drell--Yan+jets production is carried out at LO QCD using \MGvATNLO with MLM merging.
The background samples for diboson production are simulated at LO with \PYTHIAviii, as are the muon-enriched QCD multijet samples generated in bins of $\hat{p}_{\mathrm{T}}$.

Variations of the nominal \ttbar simulation are also generated to estimate the systematic uncertainties arising from theoretical or modelling considerations. 
We consider samples with variations of the top quark mass by $\pm1\GeV$ around the nominal value $m_{\PQt} = 172.5\GeV$. 
Uncertainty contributions from the choice of the parton shower matching scale are studied by varying the gluon resummation damping variable used in \POWHEG, $h_{\text{damp}}=(1.379^{+0.92}_{-0.51})m_{\PQt}$, within its uncertainties~\cite{2020_CMS}. 
Alternative colour reconnection (CR) models are utilized to estimate the uncertainty contributions from nonperturbative effects, where the effect of allowing early resonance decays~\cite{CMS:2022awf} is also considered.

\section{Event selection}
\label{sec:selection}

\subsection{Particle-level phase spaces}\label{sec:genLevel}
The collection of all particles with a lifetime longer than $10^{-8}$ seconds defines the particle-level phase space for an event. 
Jets at the particle level are reconstructed from all final-state particles excluding neutrinos. 
Only AK8 jets with $\pt\geq170\GeV$ and $\abs{y}\leq1.7$ are considered while defining the event selections. 

\subsubsection{QCD dijets}
The fiducial region at the particle level for the dijet selection considers events that satisfy the following criteria:
\begin{itemize}
	\label{list:dijet_sel_gen}
	\item at least two AK8 jets with $\pt\geq200$ \GeV and $\abs{y}\leq1.7$ must be present,
	\item the AK8 jet with the highest \pt ($j_1$) must be well-separated in the rapidity-azimuth plane from any other AK8 jet in the event ($j_i$) with $\pt\geq170\GeV$ and $\abs{y}\leq1.7$, $\Delta R(j_1, j_i)\geq1.6$,
    \item the two jets with the largest transverse momentum ($j_1,j_2$) in the event must be well-separated from one another in the azimuthal angle: $\Delta \phi (j_{1}, j_{2})\geq2$.
\end{itemize}
Requiring that the two most energetic AK8 jets in the event have $\pt\geq200\GeV$ and are produced back-to-back in the azimuthal angle, ensures that the fiducial phase space definition selects balanced QCD dijet topologies. 
The more central of these two jets is used for the measurement of the $N$-subjettiness observables. 

\subsubsection{\texorpdfstring{\PW}{W} boson and top quark jets}
The fiducial phase space at the particle level for the measurement of \PW boson and top quark jets is defined by the following criteria. 
Exactly one muon with $\pt\geq55\GeV$ and $\abs{\eta}\leq2.4$ must be present in the event, and its azimuthal separation from the \pt-leading AK8 jet must satisfy $\Delta\phi(\text{AK8~jet},\mu)\geq2$. 
No additional muons or electrons with $\pt>15\GeV$ and $\abs{\eta}\leq2.4$ can be present in the event. 
The leading AK8 jet must have $\pt\geq200\GeV$ and jet invariant mass $m_{\text{jet}}\geq50\GeV$. 
An AK4 jet with $\pt\geq30\GeV$ and $\abs{\eta}\leq2.4$ must be geometrically close to the selected muon with $0.2\leq\Delta R(\mu,\text{AK4~jet})\leq1.6$, isolated from other AK4 jets ($\Delta R(\PQb\text{-candidate, other~AK4})\geq0.4$) and well-separated from the leading AK8 jet with $\Delta R(\text{AK4,AK8})\geq1.2$. 
The \pt-leading AK4 jet satisfying these criteria and ghost-associated with a \PQb hadron~\cite{Cacciari_2008} is considered as the \PQb-tagged jet in the leptonic hemisphere of the event. 
Events in which no \PQb-tagged AK4 jet is found in the leptonic hemisphere are rejected. 
The \ptmiss in the event must be larger than 30\GeV and the \pt of the leptonically decaying \PW boson, reconstructed as the vector sum of \ptvecmiss and the momentum of the selected muon, must exceed 100\GeV.   

The selected AK8 jet is used to define the \PW boson and top quark jet measurement regions. 
The \PW boson-enriched region must satisfy $\pt\geq200\GeV$ and $65\leq m_{\text{jet}}\leq125\GeV$, and the top quark-enriched region $\pt\geq400\GeV$ and $140\leq m_{\text{jet}}\leq300\GeV$. 

\begin{figure}[htb!]
	\centering
	\includegraphics[width=.495\textwidth]{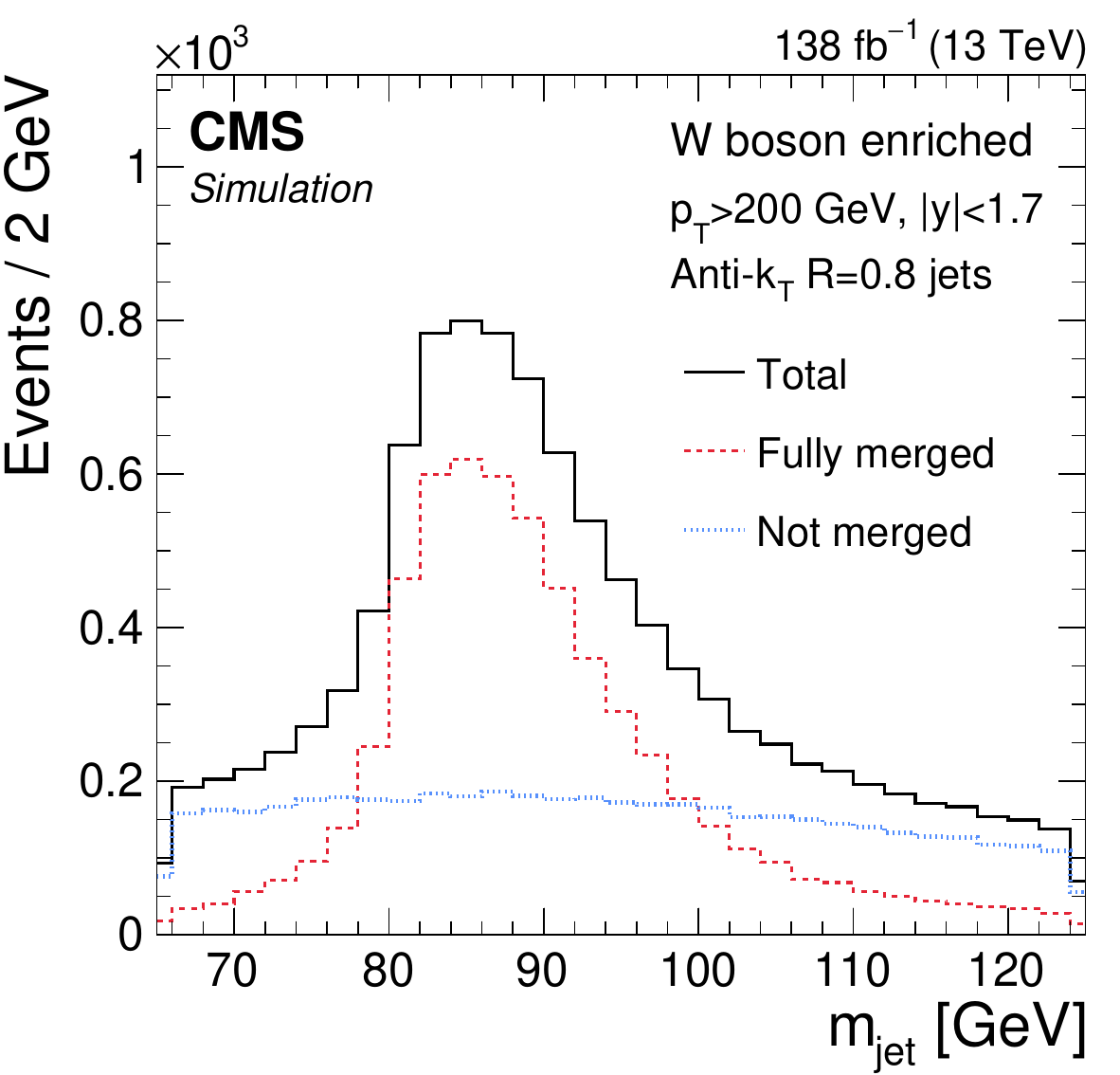}
	\includegraphics[width=.495\textwidth]{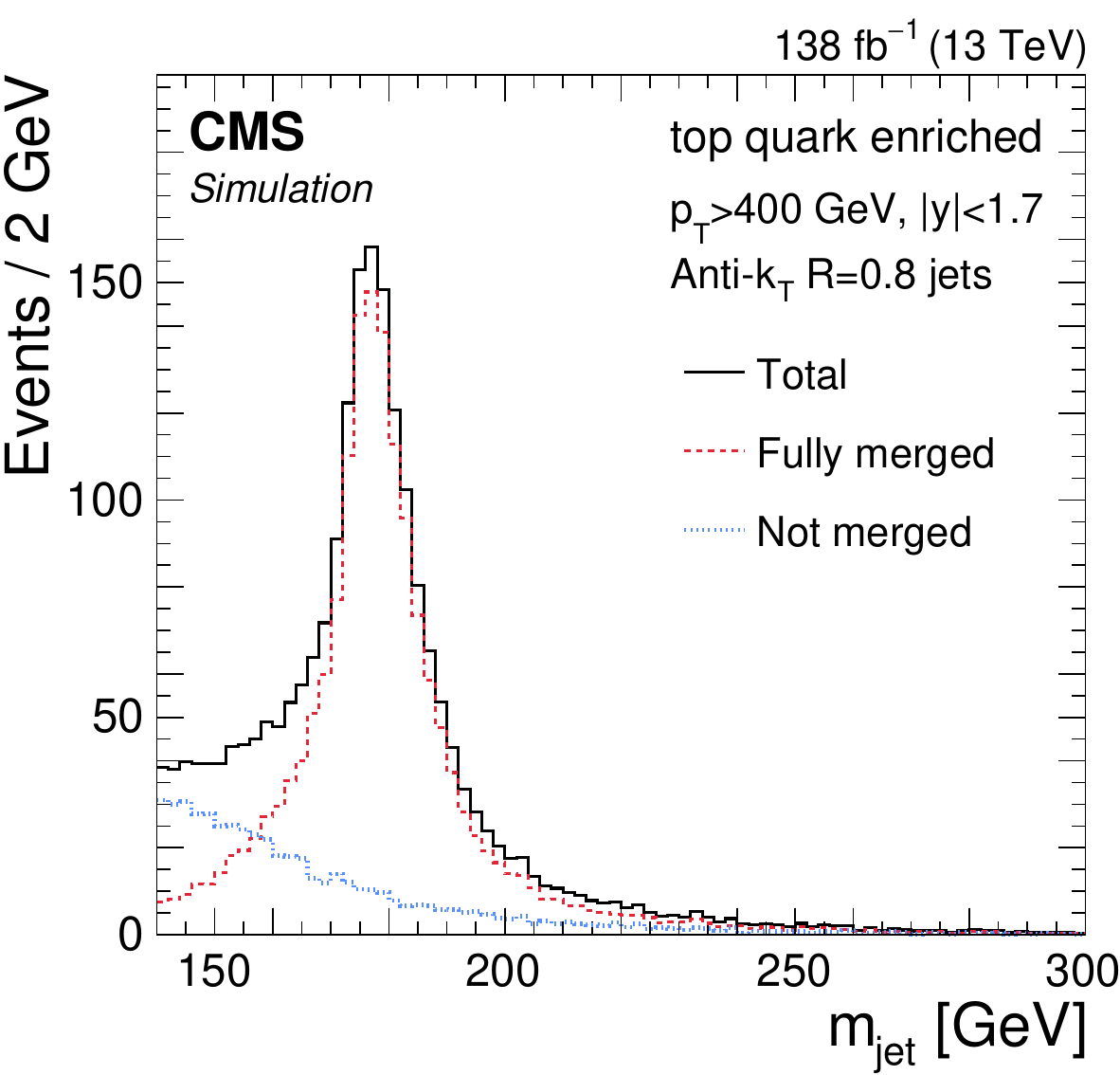}
    \caption{Distributions of the particle-level AK8 jet mass in fiducial regions enriched in hadronic decays of boosted \PW bosons (\cmsLeft) and top quarks (\cmsRight), obtained from events in the muon+jets channel of \ttbar production. The contributions to the total jet mass distribution (black) from fully-merged (red) AK8 jets and not or partially-merged (blue) jets are illustrated in the figures. }
	\label{fig:mass_matchedAndMerged_Wtop}
\end{figure}

The efficiency of the particle-level selections was studied in the nominal \ttbar signal samples. 
In particular, the veto on additional soft leptons reduces the signal yield by about 10\% in the \PW boson-enriched region and 6\% in the top quark-enriched region. 
For events passing the selections, we find average matching efficiencies of 89 and 80\% between the selected particle-level AK8 jet and the generated top quark or the \PW boson from its decay, respectively. 
The efficiencies for selecting fully merged jets, i.e. reconstructing the three partons from the top quark decay and the two partons from the \PW boson decay in the AK8 jet are 76 and 56\% for the top quark- and \PW boson-regions, as shown in Fig.~\ref{fig:mass_matchedAndMerged_Wtop}. For top quark jets, contributions from unmerged jets are largest for $m_{\text{jet}}<160\GeV$, which mostly corresponds to $\pt < 500$--$550\GeV$. 
For $\pt>600\GeV$, more than 80\% of the AK8 jets correspond to fully merged top quark decays. 
For \PW boson jets, the efficiency for selecting fully merged events is limited by the presence of the \PQb quark from the top quark decay and the categorization into \PW boson- and top quark-enriched events. 
The contribution of fully merged \PW boson decays is about 60\% for $\pt<250\GeV$ and drops to about 50\% for higher \pt. 
With increasing \pt it becomes more likely that one of the light-flavour quarks from the \PW boson decay is merged with the \PQb quark from the top quark decay such that the resulting jet mass exceeds 125\GeV. 
In addition, when $\pt>400\GeV$, events are more likely to include a fully merged top quark jet and pass the criteria of the top quark-enriched region.
 \subsection{Detector-level event selection}
\label{sec:recLevel}

The event selection at the detector level is chosen to match the selection of the fiducial phase space at the particle level. 
Most importantly, we consider only AK8 jets with $\pt>170\GeV$ and $\abs{y}<1.7$. 
All AK4 and AK8 jets in the \PW boson- and top quark-enriched selections must satisfy the PF-based tight working point (WP) of the CMS noise-jet identification (jet ID) criteria while AK8 jets in the dijet selection must satisfy the tight-lepton veto jet ID WP \cite{CMS:2017wyc}.

\subsubsection{QCD dijet selection}

Identical to the particle-level selection, two AK8 jets with $\pt>200\GeV$ and an azimuthal separation of $\Delta\phi>2$ must be present at the detector level. 
In addition, the leading AK8 jet must be spatially separated from any other AK8 jet with $\Delta R>1.6$. 
The measurement of the $N$-subjettiness observables is made on the more central of the two jets in each event. 

\begin{figure}[htb!]
	\centering
	\includegraphics[width=.495\textwidth]{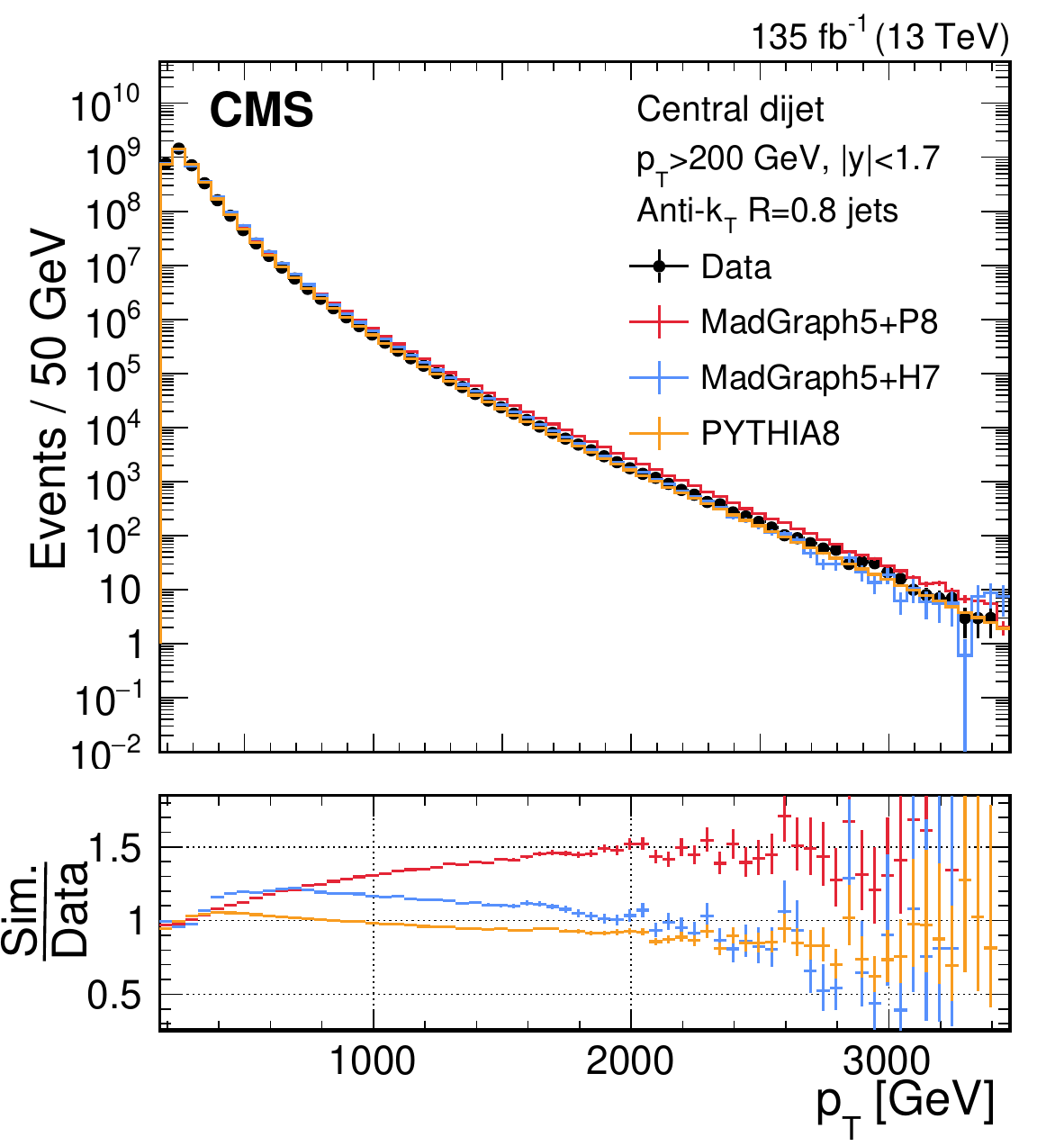}
	\includegraphics[width=.495\textwidth]{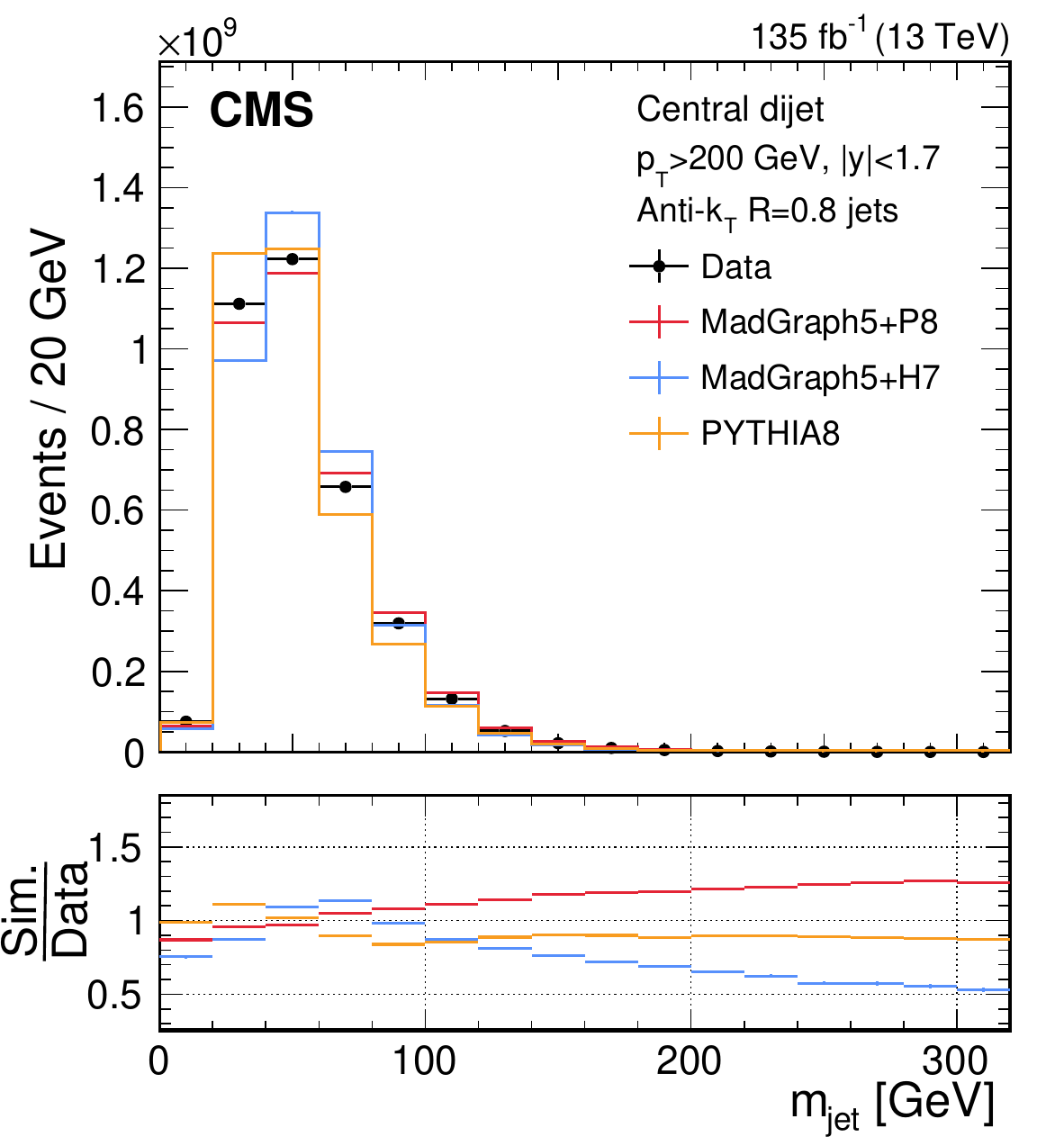}
	\caption{Distributions of the AK8 jet \pt (\cmsLeft) and $m_{\text{jet}}$ (\cmsRight) after the dijet selection, based on the combined 2016--2018 data set. The error bars in the upper panels indicate the statistical uncertainties in the data and simulation. The lower panels of the figures show the ratio of simulation to data with statistical uncertainties following the same colour code as the upper panel. The event yields in the simulated QCD samples are normalized to the yield in data. }
	\label{fig:dijetDefinition}
\end{figure}

\begin{figure}[htb!]
	\centering
	\includegraphics[width=.495\textwidth]{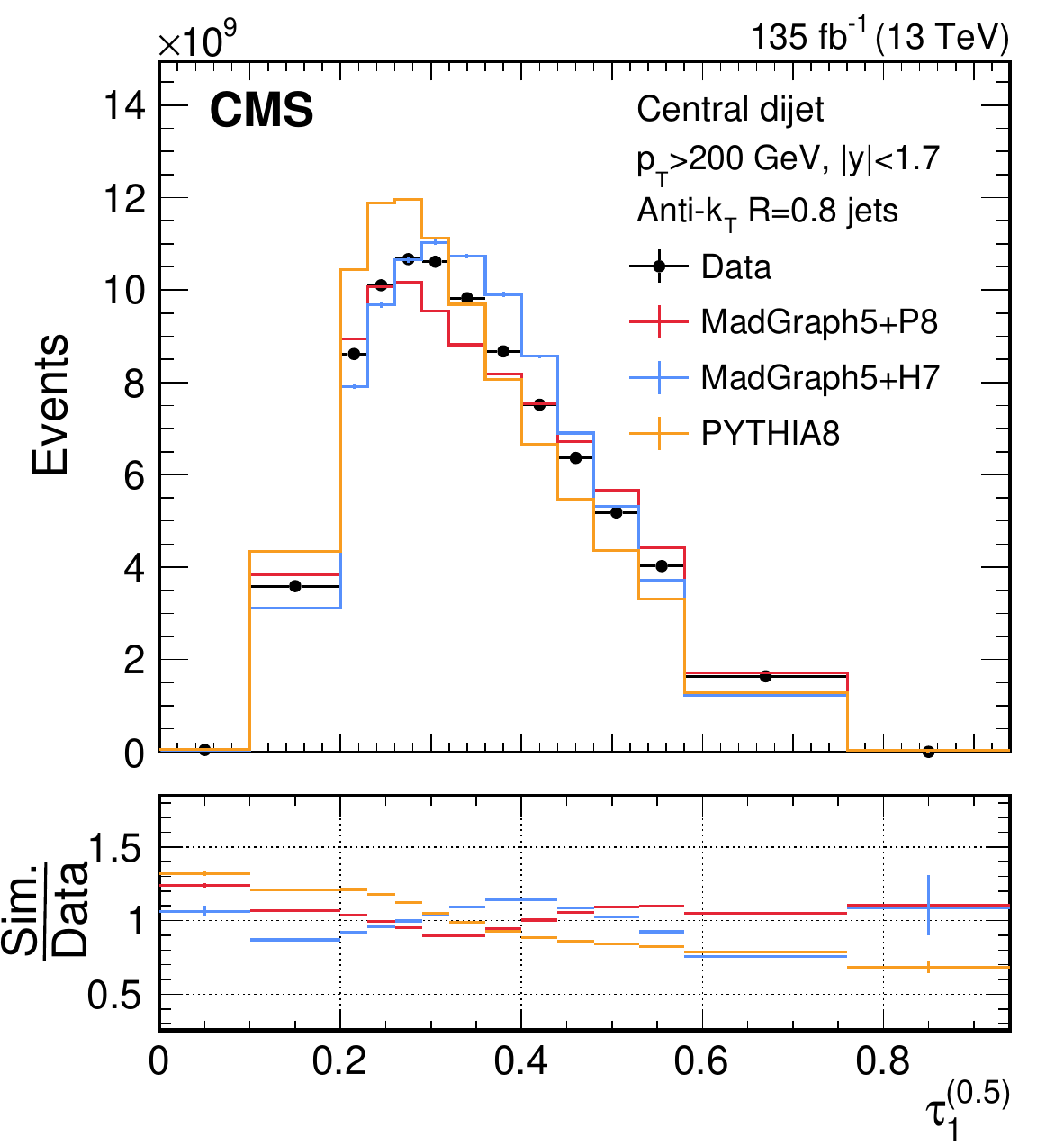}
	\includegraphics[width=.495\textwidth]{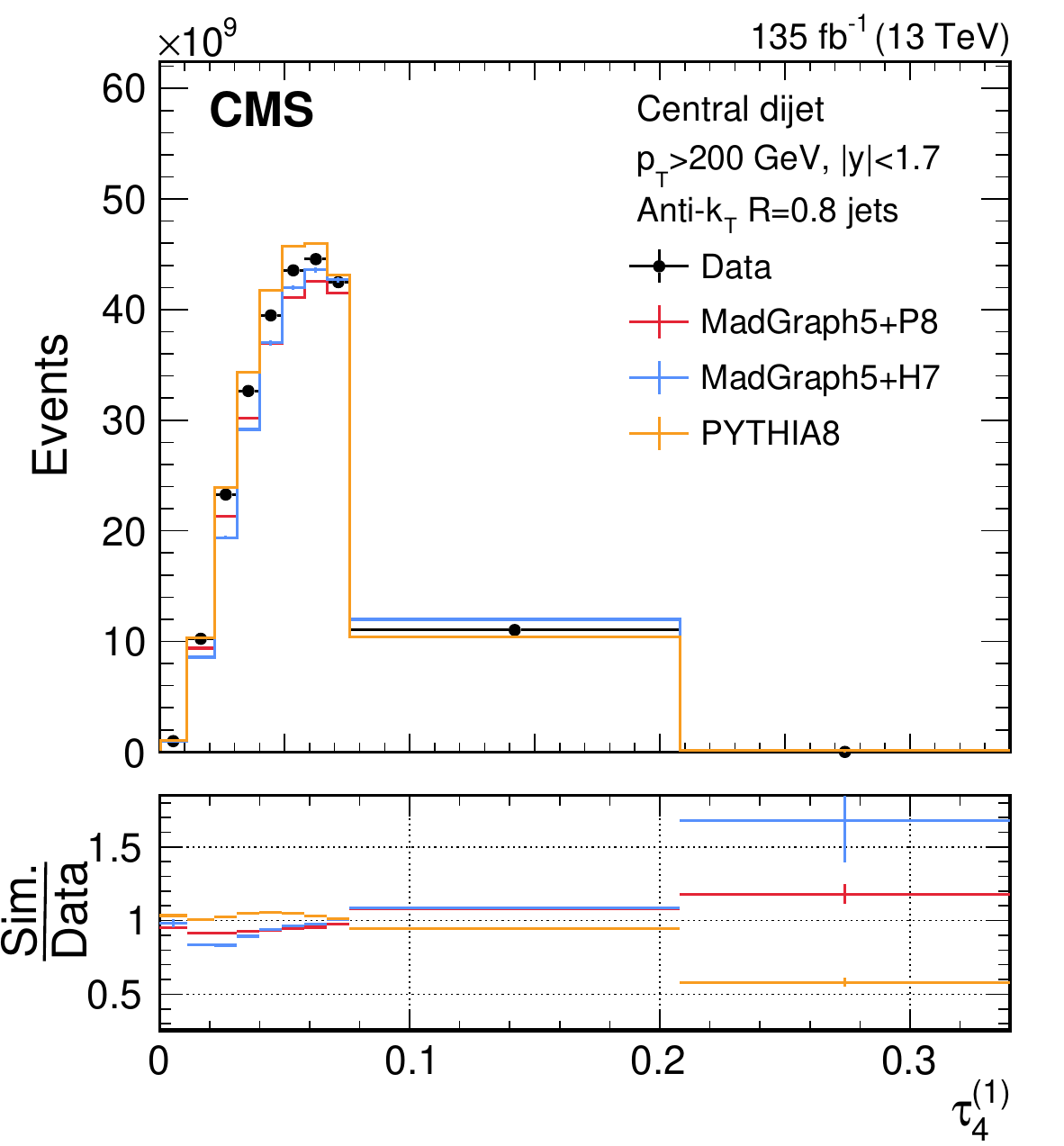}
	\caption{Distributions of the $N$-subjettiness observables \Nsub{1}{0.5} (\cmsLeft) and \Nsub{4}{1} (\cmsRight) at the detector level after the dijet selection, based on the combined 2016--2018 data set. The error bars in the upper panels indicate the statistical uncertainties in the data and simulation. The lower panels of the figures show the ratio of simulation to data with statistical uncertainties following the same colour code as the upper panel. The event yields in the simulated QCD samples are normalized to the yield in data. }
	\label{fig:dijetDataMCtaus}
\end{figure}

The distributions of \pt and $m_{\text{jet}}$ for AK8 jets in data and simulation are shown in Fig.~\ref{fig:dijetDefinition} after the dijet selection. 
The corresponding detector-level distributions of two representative $N$-subjettiness observables, \Nsub{1}{0.5} and \Nsub{4}{1}, for which unfolded results are presented in Section~\ref{sec:results}, are shown in Fig.~\ref{fig:dijetDataMCtaus}. 
The simulated samples are scaled to match the normalization observed in the data for the corresponding data-taking periods. 
It is observed that the disagreements between data and simulation for both distributions follow similar trends across simulations. 
Predictions from the two \MGvATNLO{} samples and the sample generated with \PYTHIAviii tend to envelope the data for the \pt spectrum for the central dijets, whereas for the $m_{\text{jet}}$ distribution, the nominal and alternative \MGvATNLO{} samples show similar levels of disagreement with data, but in opposite directions. 
On average, the events simulated and showered using \PYTHIAviii demonstrate the best agreement with the data. 

\subsubsection{\texorpdfstring{\PW}{W} boson and top quark jet selection}
\label{sec:WtopSel}

\begin{figure}[ht!]
	\centering
	\includegraphics[width=.495\textwidth]{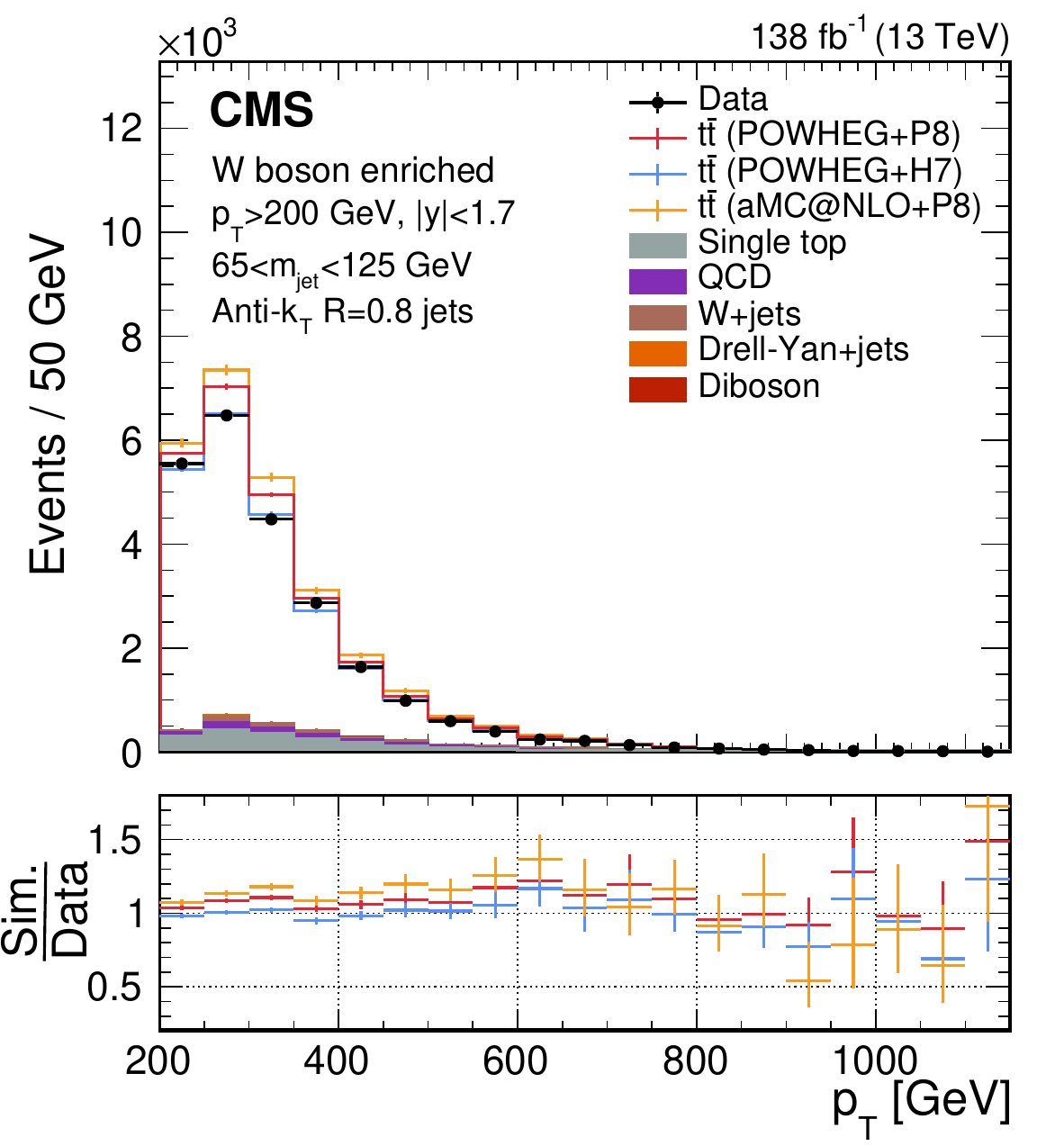}
	\includegraphics[width=.495\textwidth]{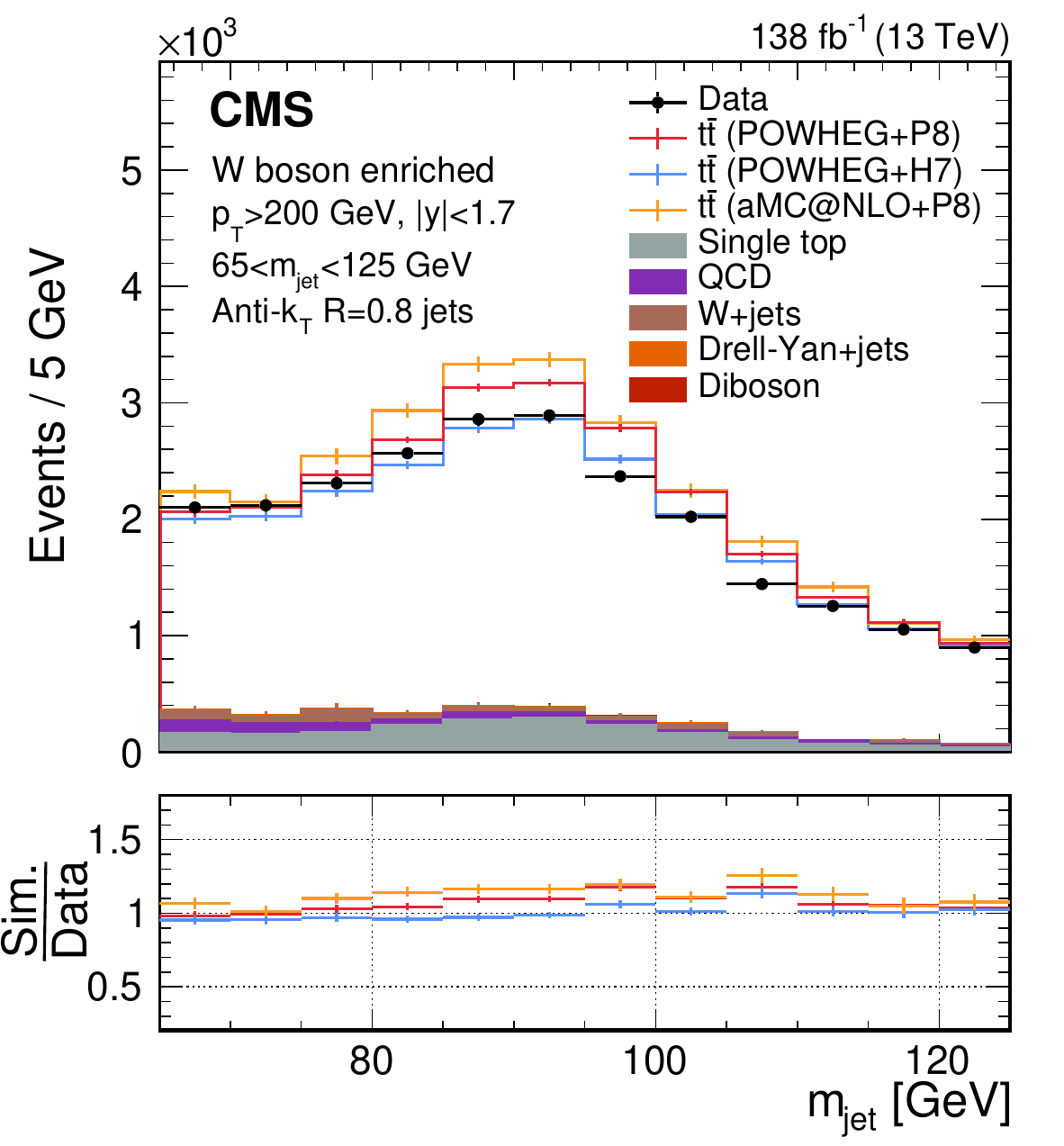}
	\caption{Distribution of the leading AK8 jet \pt (\cmsLeft) and $m_{\text{jet}}$ (\cmsRight) after the boosted \PW boson selection, for the combined 2016--2018 data set. The error bars in the upper panels indicate the statistical uncertainties in the data and simulation. The lower panels of the figures show the ratio of simulation to data with statistical uncertainties following the same colour code as the upper panel. The contributions of \ttbar events in the data, estimated by subtracting contributions from simulated physics background processes, is found to be approximately 85\%. }
	\label{fig:basicWSel1}
\end{figure}

\begin{figure}[ht!]
	\centering
	\includegraphics[width=.495\textwidth]{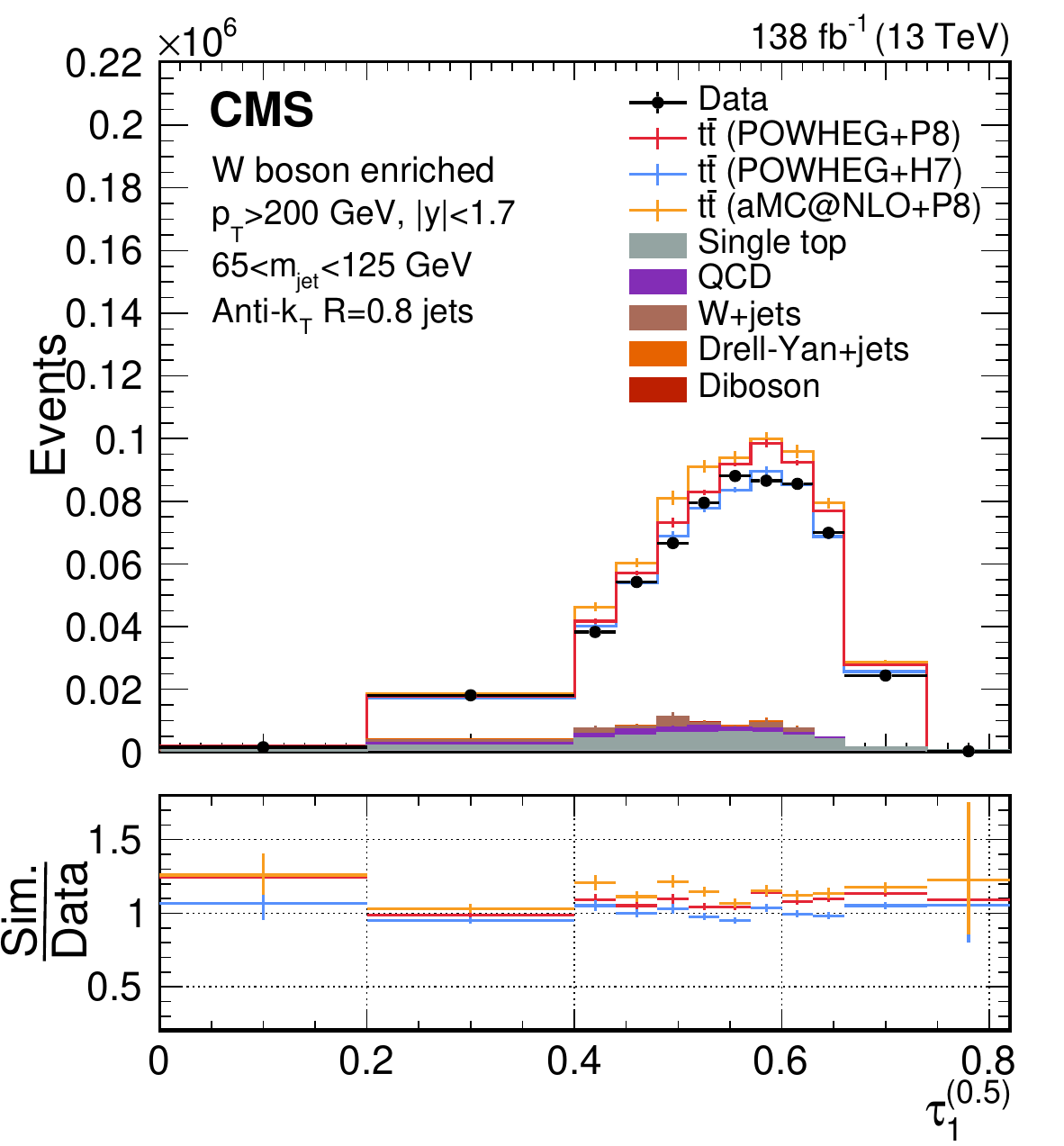}
	\includegraphics[width=.495\textwidth]{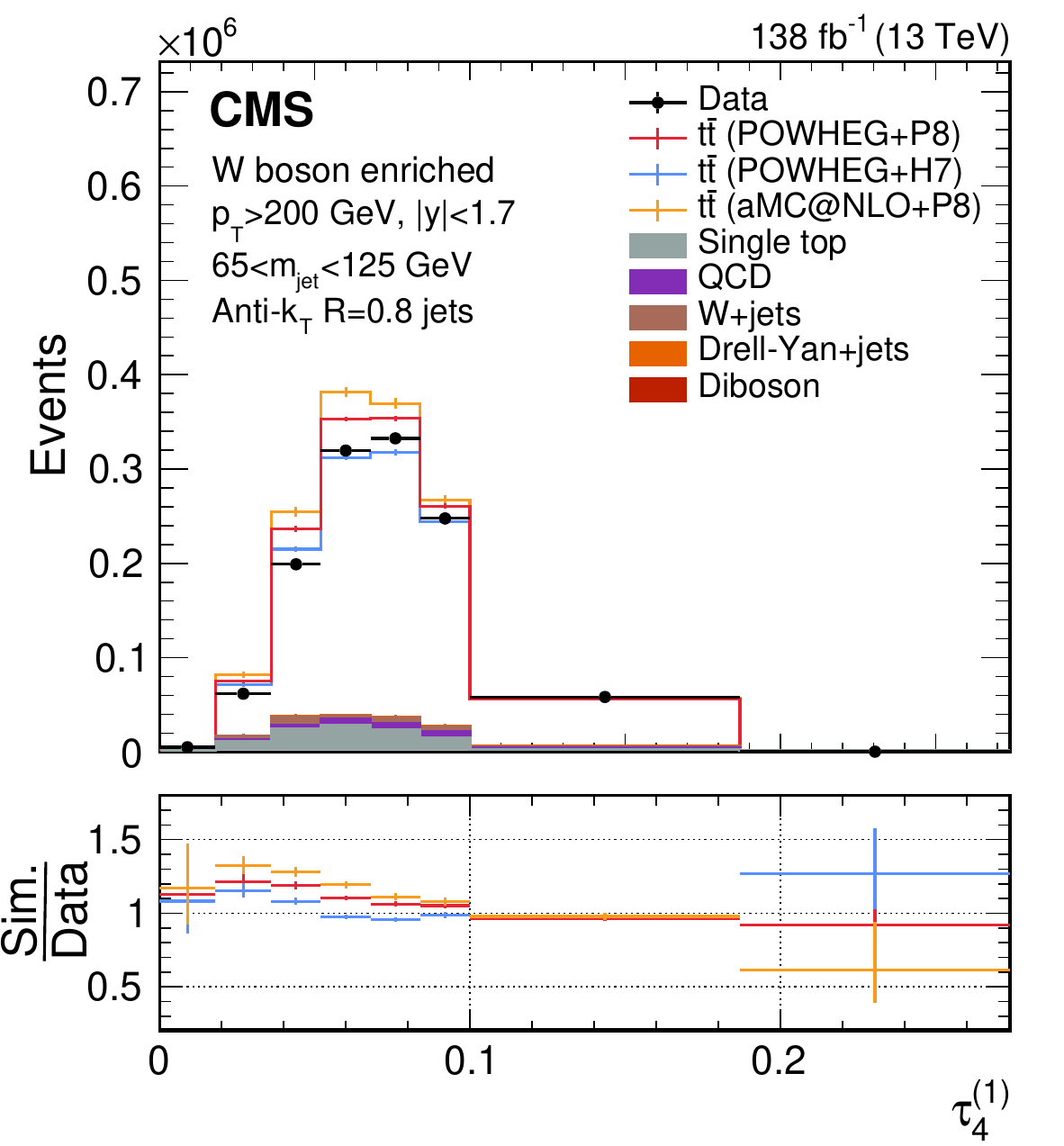}
	\caption{Distributions of the $N$-subjettiness observables \Nsub{1}{0.5} (\cmsLeft) and \Nsub{4}{1} (\cmsRight), measured on the leading AK8 jet at the detector level after the boosted \PW boson selection, for the combined 2016--2018 data set. The error bars in the upper panels indicate the statistical uncertainties in the data and simulation. The lower panels of the figures show the ratio of simulation to data with statistical uncertainties following the same colour code as the upper panel. }
	\label{fig:WDataMCtaus}
\end{figure}

\begin{figure}[ht!]
	\centering
	\includegraphics[width=.495\textwidth]{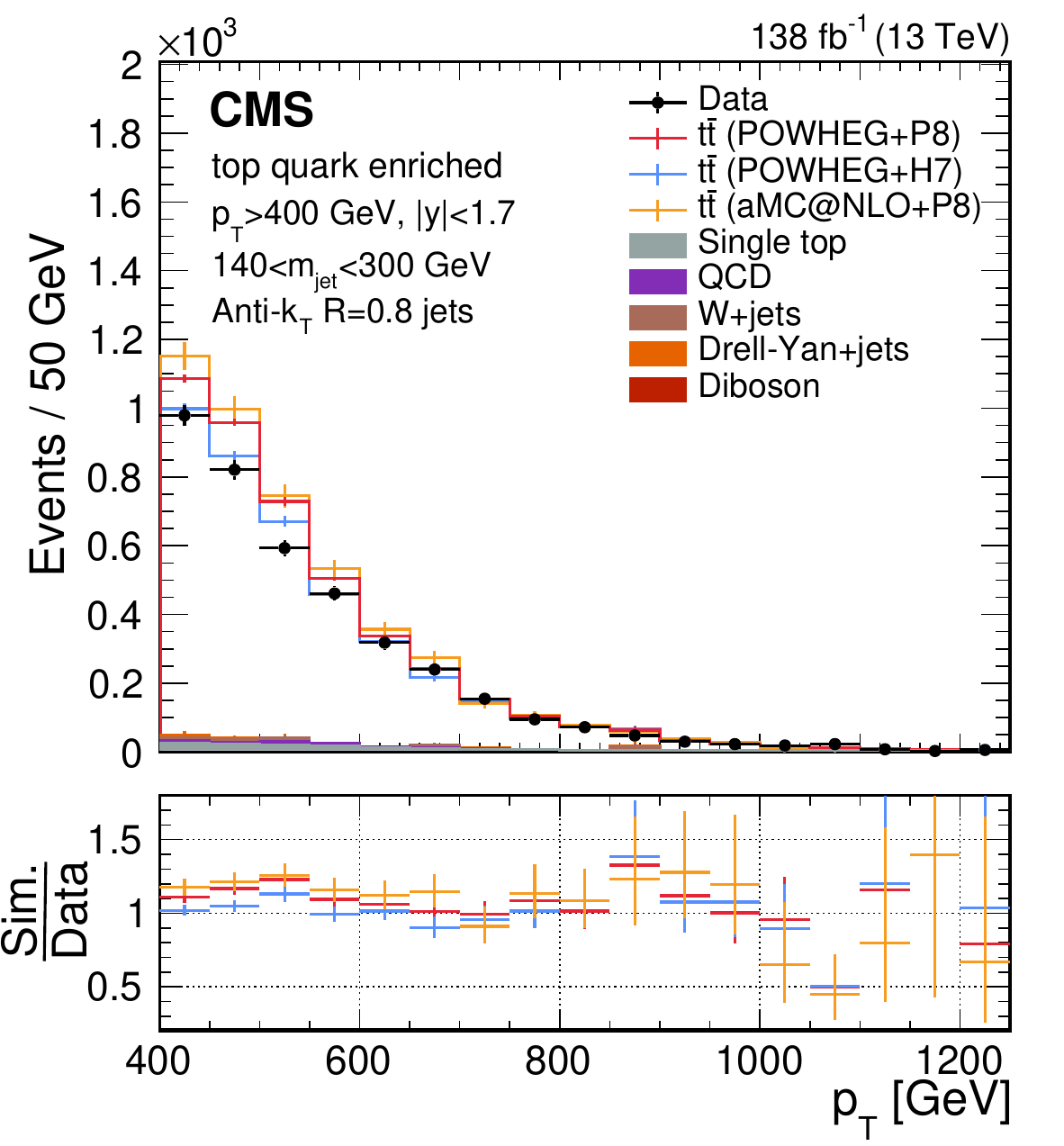}
	\includegraphics[width=.495\textwidth]{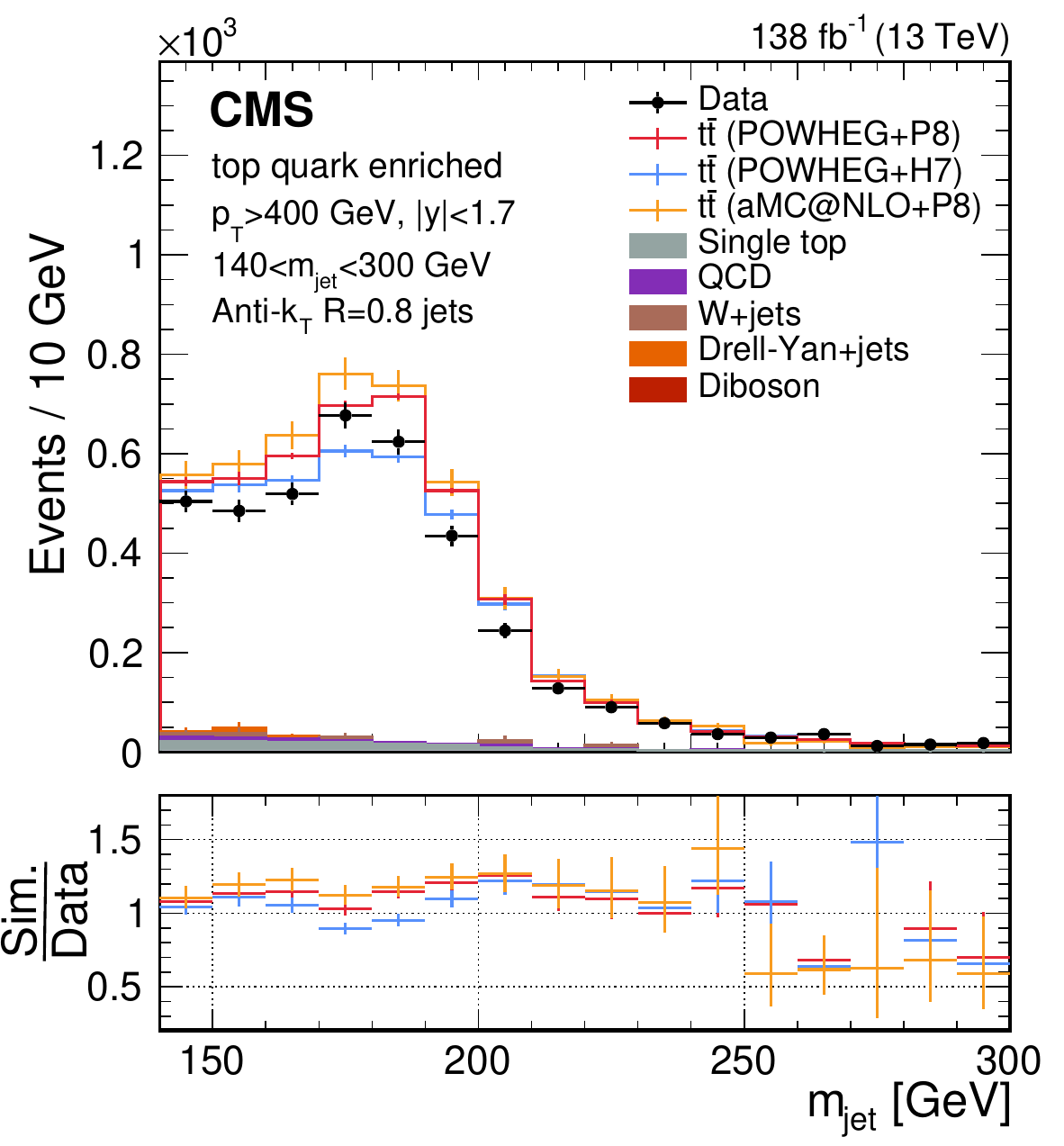}
	\caption{Distribution of the leading AK8 jet \pt (\cmsLeft) and $m_{\text{jet}}$ (\cmsRight) after the boosted top quark selection for the combined 2016--2018 data set. The error bars in the upper panels indicate the statistical uncertainties in the data and simulation. The lower panels of the figures show the ratio of simulation to data with statistical uncertainties following the same colour code as the upper panel. The contributions of \ttbar events in the data, estimated by subtracting contributions from simulated physics background processes, is found to be approximately 94\%. }
	\label{fig:basictopSel1}
\end{figure}

\begin{figure}[ht!]
	\centering
	\includegraphics[width=.495\textwidth]{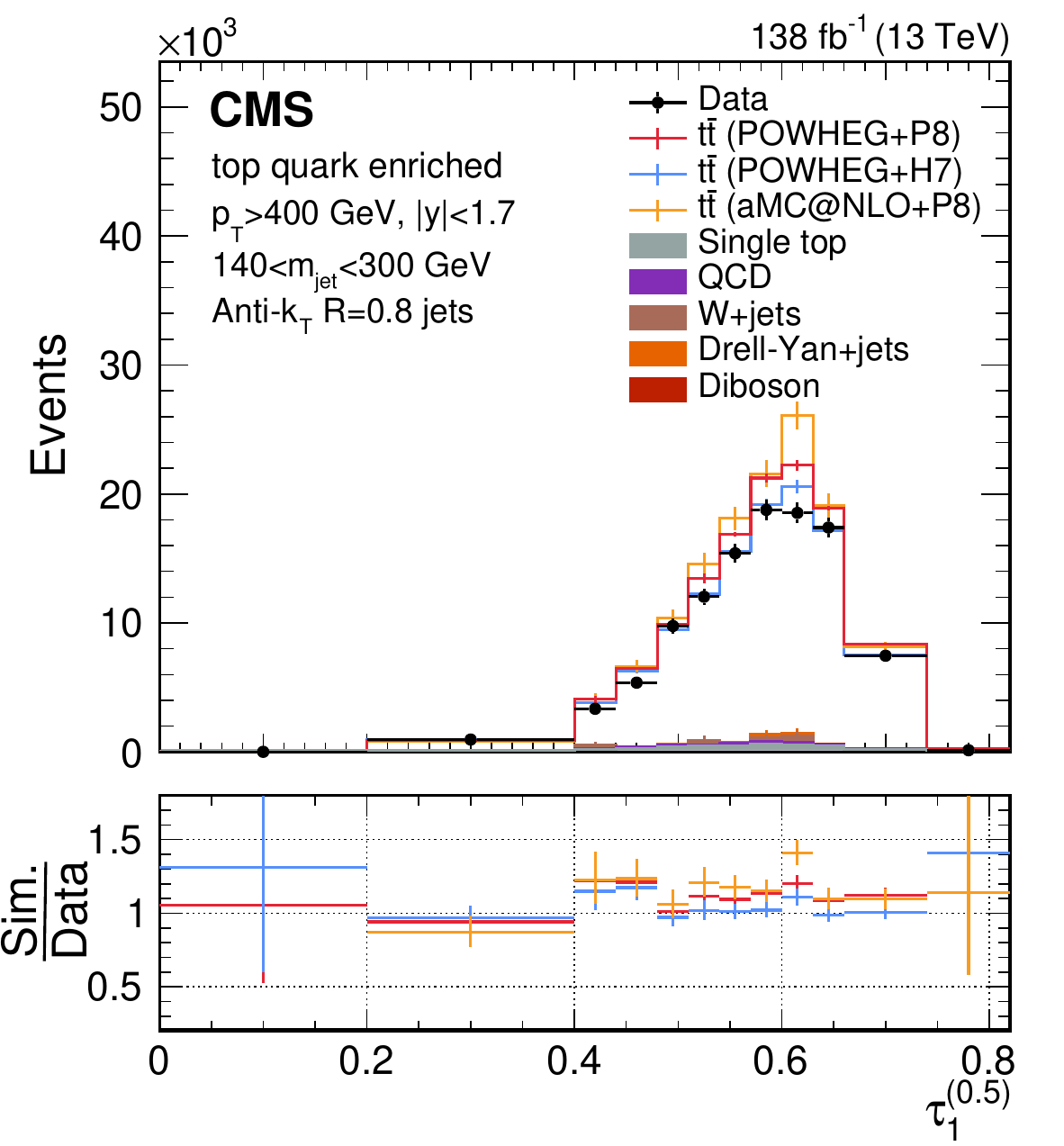}
	\includegraphics[width=.495\textwidth]{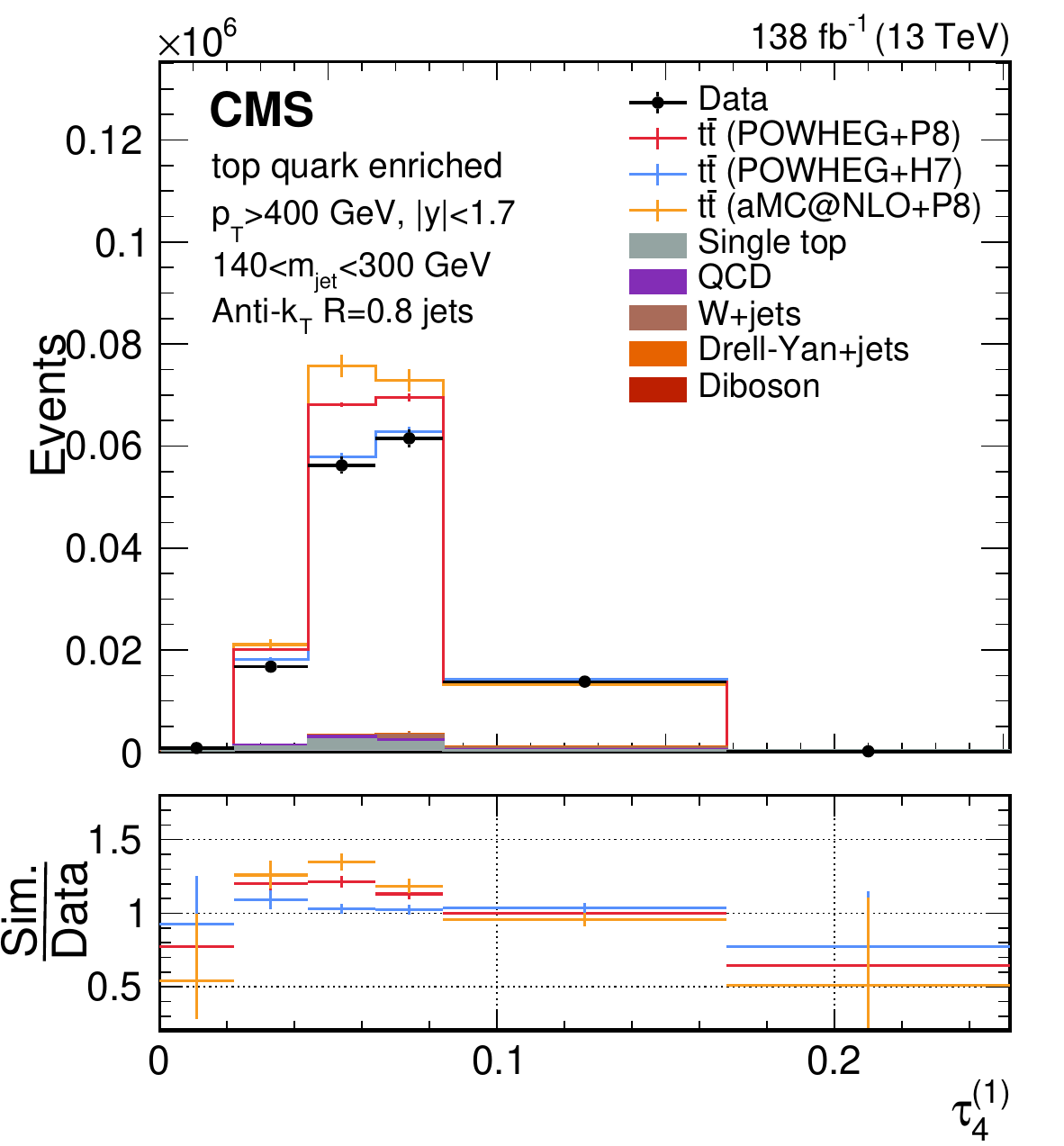}
	\caption{Distributions of the $N$-subjettiness observables \Nsub{1}{0.5} (\cmsLeft) and \Nsub{4}{1} (\cmsRight), measured on the leading AK8 jet at the detector level after the boosted top quark selection, for the combined 2016--2018 data set. The error bars in the upper panels indicate the statistical uncertainties in the data and simulation. The lower panels of the figures show the ratio of simulation to data with statistical uncertainties following the same colour code as the upper panel. }
	\label{fig:topDataMCtaus}
\end{figure}

Events in the boosted \PW boson- and top quark-enriched regions are selected based on the identification of a final state with a single energetic muon, the presence of large \ptmiss in the event and multiple jets. 
Selection criteria are imposed to define the $\PGm$+jets \ttbar decay topology at the detector level, requiring that events are accepted by the combination of nonisolated, single-muon HLT paths discussed in Section~\ref{sec:samples}. 

At the detector level, we require a single muon with $\pt>55\GeV$ and $\abs{\eta}<2.4$. 
It is required to be separated from the \pt-leading AK8 jet with $\pt>200\GeV$ by $\Delta\phi(\text{AK8},\mu)>2$. 
The muon must pass the tight WP for track-based isolation criteria and the global high-\pt muon identification criteria~\cite{CMS:2018rym}. 
These selections have an efficiency of about 95 and 98\%, respectively, over the full detector acceptance for muons with $\pt>53\GeV$~\cite{Sirunyan:2703645}. 
We veto events with additional high-\pt muons, as defined above, or with any additional soft leptons (muons or electrons) with $\pt>15\GeV$ and $\abs{\eta}<2.4$ that pass loose selection criteria. 

Identical to the selection at the particle level, the \pt-leading AK4 jet with $\pt>30\GeV$ and $\abs{\eta}<2.4$ within $0.2<\Delta R(\mu,\text{AK4})<1.6$, must be separated from any other AK4 jet by $\Delta R>0.4$ and from the \pt-leading AK8 jet by $\Delta R>1.2$. 
It is considered as the candidate $\PQb$ jet in the leptonic hemisphere if it passes the tight WP of the \textsc{DeepJet}~\cite{Stoye2018,CMS-DP-2018-058} algorithm which has a misidentification rate of $<$1\%. 

Finally, the presence of \ptmiss exceeding 30\GeV serves as a proxy for the neutrino produced in the leptonic hemisphere of the decay and suppresses QCD multijet backgrounds. 
The leptonically decaying \PW boson, reconstructed from the muon and missing transverse momentum, must have $\pt>100\GeV$.

The comparisons between data and simulation are shown for the leading AK8 jet \pt and $m_{\text{jet}}$ distributions in the boosted \PW boson- and top quark jet-enriched regions in Figs.~\ref{fig:basicWSel1} and~\ref{fig:basictopSel1}, respectively. 
The corresponding detector-level distributions of the representative $N$-subjettiness observables \Nsub{1}{0.5} and \Nsub{4}{1} are shown in Figs.~\ref{fig:WDataMCtaus} and~\ref{fig:topDataMCtaus} for the two selections. 
The simulated detector-level events are additionally reweighted by data-to-simulation scale factors for the muon identification, isolation, and trigger efficiencies, and for the \PQb tagging efficiency. 

We observe that the simulations overestimate the predicted yields relative to the data, consistent with previous measurements of top quarks with high \pt by the CMS and ATLAS Collaborations~\cite{PhysRevD.103.052008,ATLAS:2024dua,ATLAS:2023jdw} compared with NLO predictions. 
For the nominal \ttbar simulation, the ratios of predicted yields in data to simulation for the boosted \PW boson- and top quark jet-enriched regions are respectively about 93\% and 89\%. 
The \POWHEG{}+\HERWIGvii simulation provides the best overall level of agreement between the data and simulated samples in both selections. 
While the different \ttbar signal simulations lead to differing descriptions of the event yield, the shapes of the distributions in the data are well described. 
 
\section{Saturation of discrimination power}
\label{sec:saturation}

In $\Pp\Pp$ collisions, saturation occurs upon the resolution of information from a small number of emissions in a jet, allowing for the extraction of reduced sets of observables that encode all of the IRC-safe, discriminating information within a jet. 
For classifying jets from boosted, hadronically decaying \PZ bosons versus QCD jets, Refs.~\cite{Datta:1, Datta:3} demonstrated that discrimination power saturates at approximately 4-body phase space in particle-level studies. 
Similarly, for distinguishing between boosted Higgs boson and gluon splittings to pairs of \PQb quarks, saturation occurs once $3$-body phase space is resolved \cite{Datta:2,Datta:3}, and at about $5$-body phase space for boosted top quark jets versus QCD jets \cite{nsub_vs_ji}. 
In contrast, distinguishing jets modified by the quark-gluon plasma in heavy-ion collisions from those produced in vacuum requires resolving substantially more of the intrajet radiation: the relevant discriminating information resides not only in hard splittings but also in soft and IRC-unsafe emissions, and is further diluted by the fluctuating underlying event~\cite{Lai:2021ckt}.

We identify the point of saturation of discrimination power for the classification of boosted \PW boson and top quark jets versus QCD jets with the minimal and overcomplete variations of the $N$-subjettiness bases.  
This motivates the number of emissions to be resolved using the basis of substructure observables, in order to ensure the sensitivity of the measurement to all relevant discriminating information in the jet substructure topologies considered in this analysis. 
Deep neural networks (DNNs) are trained using inputs corresponding to the minimal and complete $N$-subjettiness bases for 2- through 6-body phase space, as well as overcomplete 5- and 6-body bases defined as in Eq.~\eqref{eq:overcomplete}. 
The performance of the networks is compared with that of single $N$-subjettiness ratio observables, as a reference for the information captured by standard observables for tagging 2- and 3-prong hadronic decays. 

Jets passing the detector- and particle-level selections in the boosted \PW boson- and top quark-enriched regions in the nominal \ttbar simulation are considered as the signal jets. Then, the background class is constructed from simulated QCD multijet events using the dijet selection. 
The more central of the jets in the QCD dijet topology are categorized as being \PW boson or top quark like, based on whether they lie within the \pt and $m_{\text{jet}}$ windows used to select AK8 jets for the measurements in the \PW boson- and top quark-enriched event selections. 
Balanced data sets are constructed by keeping a subset of the QCD jets, approximately corresponding to the size of the data set in the \PW boson or top quark signal class for detector- and particle-level selections. 
We ensure that the distributions of the AK8 jet \pt and $m_{\text{jet}}$ are not biased by this selection. 

The feed-forward DNNs trained for the classification tasks are composed of several dense hidden layers with leaky ReLU activations~\cite{relu,Maas2013Rectifier} and a single output node with a sigmoid activation function, and are regularized using dropout~\cite{hinton2012}. 
The network depth and layer widths are optimized separately for the particle- and detector-level cases, with one additional hidden layer employed at the particle level.

\begin{figure}[ht!]
	\centering	
	\includegraphics[width=0.495\textwidth]{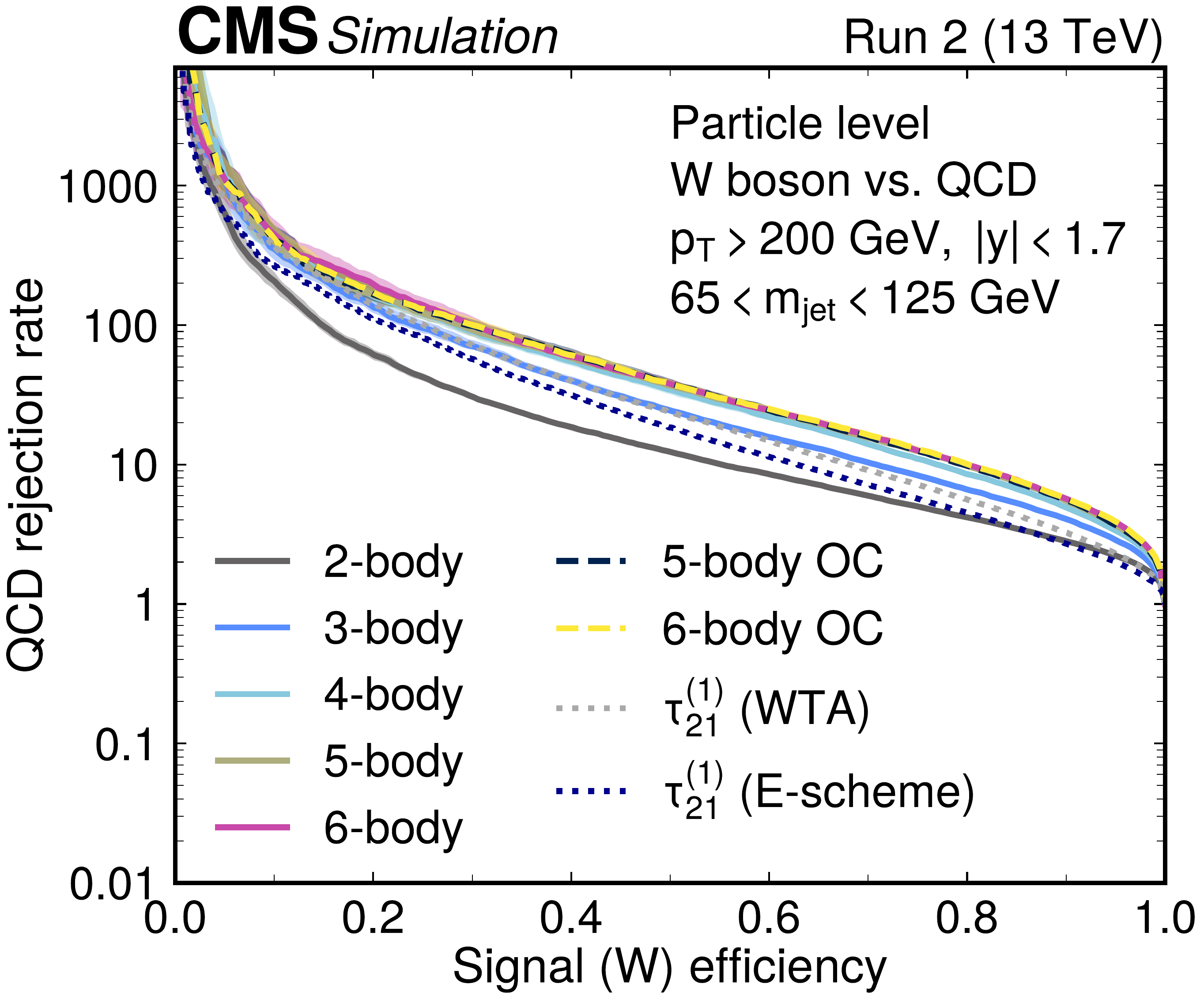}
	\includegraphics[width=0.495\textwidth]{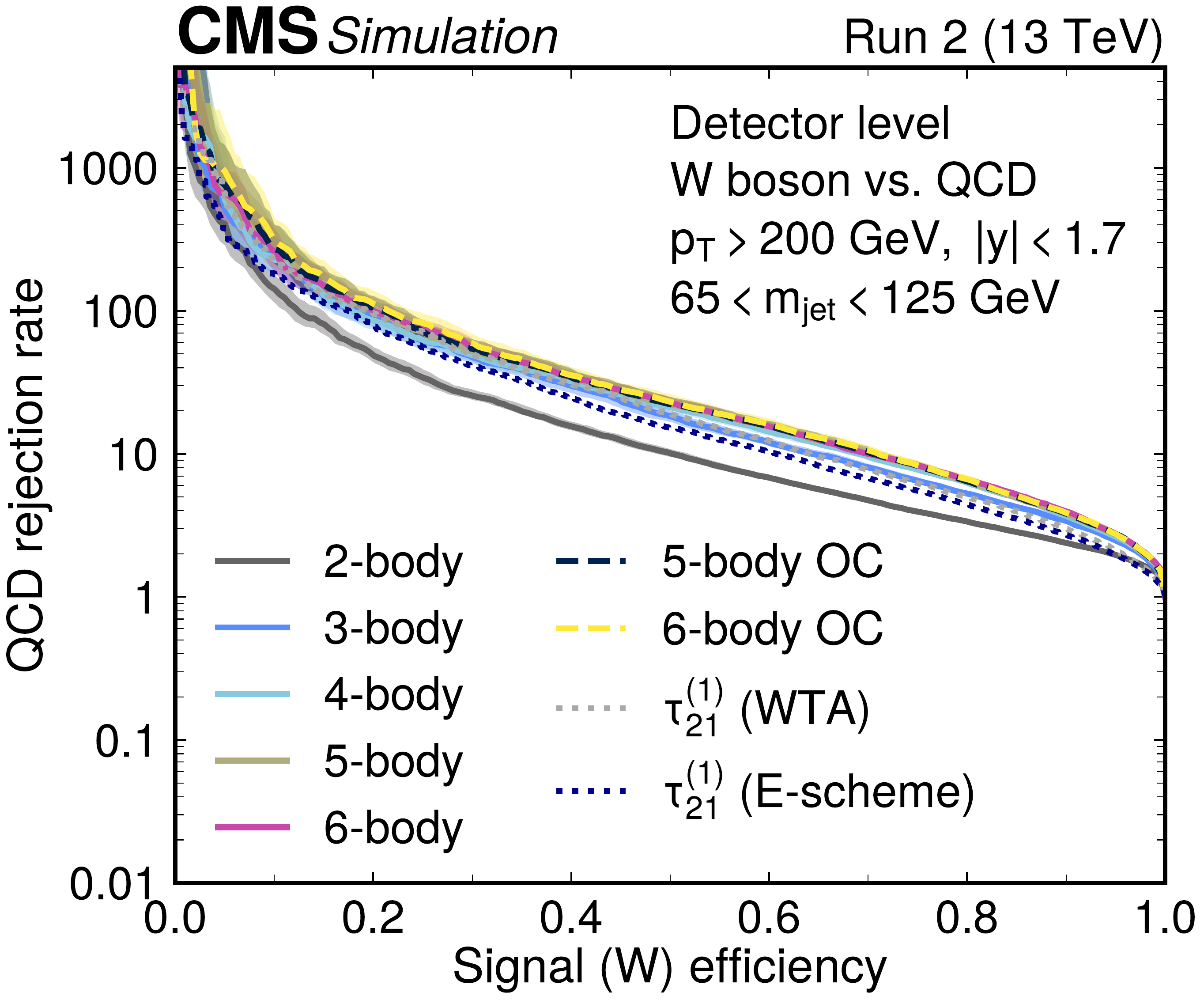}
	
	\caption{Background rejection rate as a function of signal efficiency for boosted \PW boson discrimination using deep neural networks trained on minimal and complete $M$-body bases (solid lines), overcomplete 5-/6-body bases (dashed lines), and $\tau_{2,1}^{(1)}$ (dotted lines) calculated with winner-take-all (WTA) and E-scheme recombination schemes. Shaded bands around each curve show the pointwise 95\% confidence interval on the ROC curves, obtained by a nonparametric bootstrap.   }
	\label{fig:saturationW}
\end{figure}
\begin{figure}[ht]
	\centering	
	\includegraphics[width=0.495\textwidth]{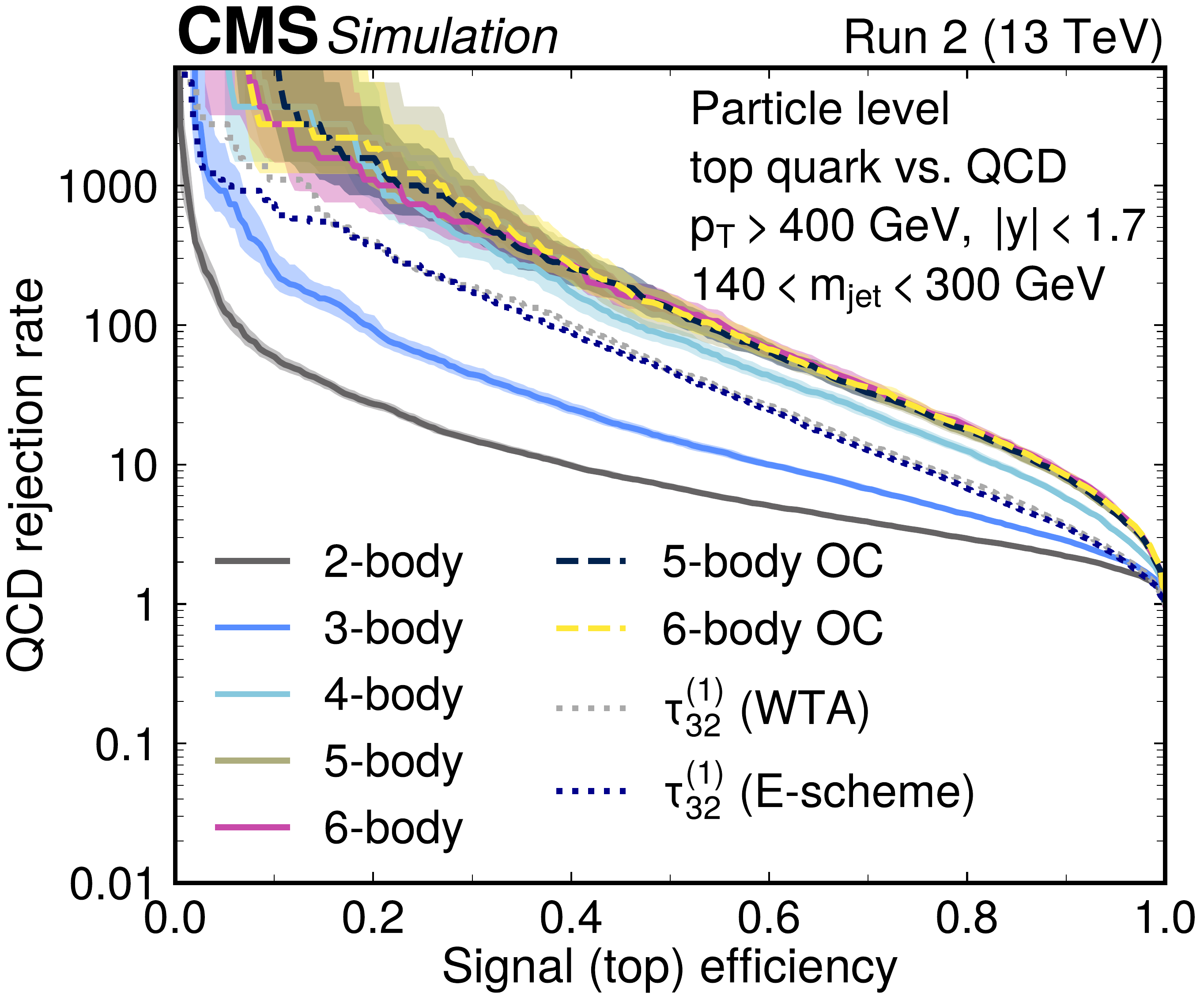} 
	\includegraphics[width=0.495\textwidth]{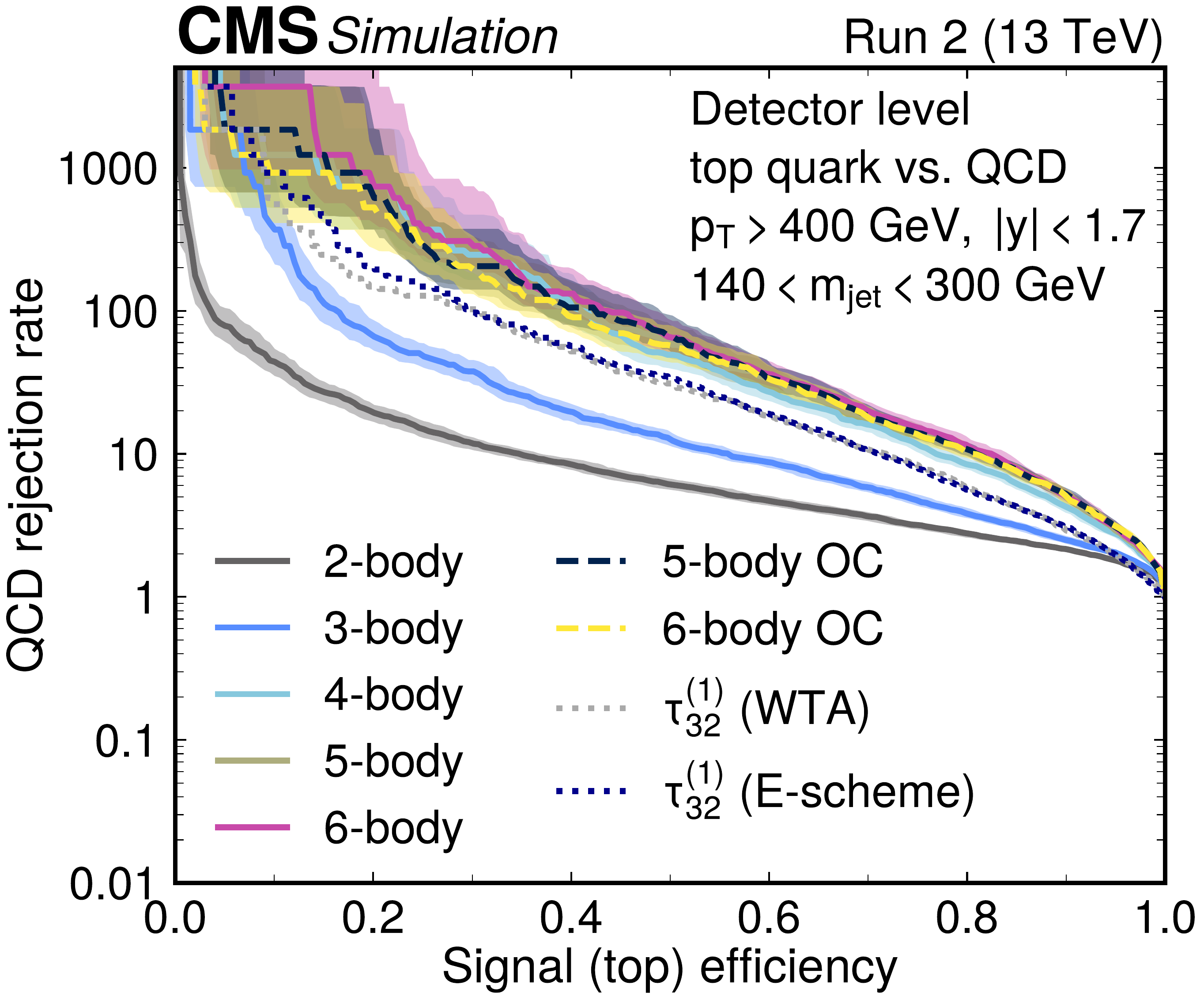}
	
	\caption{Background rejection rate as a function of signal efficiency for boosted top quark discrimination using deep neural networks trained on minimal and complete $M$-body bases (solid lines), overcomplete 5-/6-body bases (dashed lines), and $\tau_{3,2}^{(1)}$ (dotted lines) calculated with winner-take-all (WTA) and E-scheme recombination schemes. Shaded bands around each curve show the pointwise 95\% confidence interval on the ROC curves, obtained by a nonparametric bootstrap.   }
	\label{fig:saturationtop}
\end{figure}

As discussed in Section~\ref{sec:genLevel}, the efficiency of the selection in semileptonic \ttbar events to capture exactly two-pronged decays is limited for the \PW boson-enriched region. Therefore, to ensure that these studies adequately reflect outcomes for tagging purely two- or three-pronged jets, the signal data sets used for particle-level network predictions only consider events with fully merged AK8 jets from the decay of the generated top quark, or its daughter \PW boson. 
In the corresponding events at the detector level, if the AK8 jet is matched within half its radius to the particle-level jet, the events are considered in the signal class for the detector-level data set. 
For the network predictions, a separate set of \ttbar events simulated with \MGvATNLO are used for the signal jets from \PW bosons and top quarks. 
It was checked that similar discrimination power is also demonstrated for fully merged jets from the nominal \ttbar simulation used for the DNN trainings. 

The saturation of discrimination power for the different classification problems is demonstrated by studying the background rejection rate (inverse of the false-positive rate) for the network predictions as a function of signal efficiency (true-positive rate). 
This is illustrated with the receiver operating characteristic (ROC) curves shown in Figs.~\ref{fig:saturationW} and \ref{fig:saturationtop}. 
The nominal ROC curves are shown as central lines, and the shaded regions correspond to the 95\% confidence level bands obtained by recomputing the ROC curves on nonparametric bootstrap resamplings of the simulation. 

At both the particle and detector levels, discrimination power saturates once 4- and 5-body phase space is resolved for classifying jets from \PW bosons and top quarks, respectively. 
The ROC curves at the detector level show a faster saturation of discrimination power compared with the particle level, albeit with a lower maximum, because of the finite detector resolution. 
Studies of the pairwise correlations between the observables of the overcomplete 6-body basis are presented in Appendix~\ref{sec:Correlations} for data and simulations. 
These visualize the information captured by the individual $N$-subjettiness observables at the detector level by relating it to the particle level. 
Similar to the studies of the saturation of discrimination power, only fully-merged particle-level jets and detector-level jets geometrically matched to them are considered for the studies of the pairwise correlations for the boosted \PW boson and top quark jets in simulation.

For both \PW boson and top quark jet versus QCD jet classification at the detector and particle levels, it is observed that networks trained on the overcomplete and minimal versions of the 5-/6-body bases demonstrate the same performance.  
However, to ensure that the unfolded measurements robustly constrain the distinguishing features in the radiation patterns of the jets, and to capture any further uncorrelated information in the jets from beyond the point of saturation of discrimination power, we proceed by utilizing an overcomplete 6-body basis for the measurement.

\section{Simultaneous unfolding}
\label{sec:unfolding}
The measured distributions of \Nsubjettiness observables are unfolded to the level of stable particles with the \textsc{TUnfold} package \cite{Schmitt:2012kp} without regularization. 
The unfolding procedure is formulated in terms of a least-squares minimization problem, 
\begin{equation}
	\chi^2 = \min_x\left[(\mathbf{A}x+b-y)^\mathrm{T}\mathbf{V_{yy}}^{-1}(\mathbf{A}x+b-y)\right],
\end{equation}
where $x$ is the particle-level estimate, $b$ represents contributions from background processes and events in the nominal simulations that pass selection criteria at detector level but not at the particle level, and $\mathbf{A}$ is the probability matrix derived by normalizing the response matrix, encoding bin-to-bin migrations between detector and particle level. 
The matrix $\mathbf{V_{yy}}$ represents the covariance of the measured distribution $y$ and is diagonal for observables that have a single entry per event. 

The response matrix for each $N$-subjettiness observable is derived using the nominal simulation for each event selection. 
The matrix is populated with entries corresponding to the measurement of observables in the selected jet in an event passing both the detector- and particle-level selections, requiring a geometric matching within half the jet radius for jets selected at both levels. 
The distributions of the observables measured on particle-level jets without a matched detector-level counterpart, or for cases where the event does not pass the detector-level selections, are used to compute bin-by-bin corrections for reconstruction inefficiencies. 
The detector-level distributions for observables measured on jets without a matched particle-level counterpart, or where the corresponding event is not selected at the particle level, are subtracted as a background from the input distribution prior to unfolding along with contributions from background processes (in the unfolding of \PW boson and top quark jets) estimated in simulation. 
The former account for about 20\%, 39\%, and 33\% of the simulated detector-level distributions in the central dijet, \PW boson, and top quark selections, respectively, consistent across the nominal and alternative simulations. 
The statistical uncertainties in the detector-level background distributions, and rate uncertainties in individual background processes, are propagated to the input covariance matrix prior to unfolding.

The goal of this measurement is to map out the radiation pattern of a jet, using the overcomplete 6-body basis of $N$-subjettiness observables. 
Thus, all of the observables must be unfolded simultaneously, respecting correlations between bins of their distributions. 
Since the individual $N$-subjettiness observables are single-entry distributions per jet, this requires a strategy similar to that proposed in Ref.~\cite{Collaboration:2888301}.

For the simultaneous unfolding in each signal region, the vectors $x$ and $y$ contain the distributions of all $N$-subjettiness observables at the particle and detector levels, respectively, following a global binning scheme. 
There are twice as many bins in $y$ as there are in $x$, ensuring the stability of the $\chi^2$ minimization. 
The response matrix, used to derive $\mathbf{A}$, is constructed from the individual response matrices. 
It has a block-diagonal structure in the global binning scheme since there are no migrations between the observables. 
The covariance matrix $\mathbf{V_{yy}}$ is populated with multiple entries per event, accounting for statistical correlations between individual observables in its off-diagonal blocks. 
This ensures correlations between distributions are propagated correctly through the unfolding.

The initial particle-level binning schemes, in the physical ranges of the individual $N$-subjettiness observables, account for the detector resolution of the measurements and are chosen based on a study of the purity and stability of the bins. 
Here, purity is defined as the fraction of reconstructed events generated in the same bin, and stability as the fraction of generated events reconstructed in the same bin. 
We note that in the boosted \PW boson and top quark event selections, and to a lesser extent in the QCD dijet selection, there is an unavoidable trade-off between purity and stability in the chosen binning schemes, particularly for \Nsubjettiness observables with $N\geq3$ and/or $\beta>1$. 
In such instances, it is ensured that at least one of the two aforementioned metrics exceeds 50\%, or that both are above 40\%. 
Further, it was checked that the simultaneous unfolding procedure, as posed by the choice of binning schemes, is well-conditioned and that regularization is not required.

We order the observables in the combined distributions following Eq.~\eqref{eq:overcomplete}. 
That is, the ordering proceeds with blocks of five $N$-subjettiness observables, in increasing order of number of exclusive subjets ($N=1 \to N=5$, from left to right) for a set of observables computed with a specific value $\beta$, and these groups of five are ordered in increasing value of $\beta$ ($\beta=0.25\to\beta=2$) from left to right in the combined distributions.

Prior to unfolding the data, it is first validated that the simultaneous unfolding procedure ``closes''.  
The combined detector-level distribution for the overcomplete 6-body basis in the nominal simulations is unfolded with the response matrix constructed from the same set of events after subtracting the combined distributions of misreconstructed detector-level events. 
This agrees exactly with the particle-level predictions. 
In a second step, the physics model dependence in the simultaneous unfolding is studied by taking the ratio of results obtained by unfolding the detector-level distribution in the nominal simulation with a response matrix built from the same events and with a response matrix constructed from an alternative sample for the same event selection. 
Contributions from detector-level jets without a particle-level counterpart are estimated in the nominal simulation and subtracted from the input distribution prior to unfolding; this is done for both the test of the unfolding procedure in simulation, and for unfolding the data. 
It is found that model dependences are typically below 5\% in the bulk of the subdistributions, and in some limited cases rises to 10--15\% in the tails of observables sensitive to $\geq$3 subjets. 

The unfolded distributions for individual observables in the combined distributions are normalized to unit area individually. 
Correspondingly, the output covariance matrices for the unfolding are transformed to the coordinates of the normalized space using a block-diagonal Jacobian, computed such that each block corresponds to the normalizations on a per-observable basis. 
The unfolded results are presented in Section~\ref{sec:results}.

\section{Systematic uncertainties}
\label{sec:systematics}
The individual sources of uncertainty contributing to the total uncertainties in the unfolded distributions are grouped into statistical uncertainties, and systematic uncertainties arising from experimental effects and modelling assumptions.

The statistical uncertainties stem from the finite number of data events entering the input distributions, as well as the finite statistical precision of simulations used to construct the response matrices or estimate background contributions. 
The effect of these is propagated to the input covariance matrix prior to unfolding, along with contributions from rate uncertainties in the background processes considered in the \PW boson and top quark measurements that are subtracted from the input distribution. 
Following previous CMS jet substructure measurements in lepton+jets \ttbar events~\cite{CMS:2022kqg} we assign rate uncertainties of 23\% for single top quark production, 19\% for {\PW}+jets production, and 100\% for the remaining background processes.

Experimental systematic uncertainties include the following sources that are common to the QCD dijet and \PW boson or top quark measurements: jet energy scale (JES) corrections, jet energy resolution (JER) smearing, reweighting of the pileup profile, and a further uncertainty to account for timing shifts in the L1 trigger for a part of data taking~\cite{CMS:2021yvr,CMS:2020cmk}. 
A further energy-scale uncertainty is considered for the jet constituents, where for PF candidates identified as neutral hadrons, photons, and charged particles these amount to variations of their energy by 5, 3, and 1\%, respectively. 
Additionally, in the \PW boson and top quark jet measurements, we consider uncertainties arising from estimated differences in the \PQb tagging efficiencies in data and simulation, as well as shifts in the unclustered energy for reconstructing the \ptvecmiss in an event. Each source is varied up and down within its uncertainty, either by reweighting events or by recomputing event yields to construct alternative response matrices with respect to the nominal simulations. 
The effects of the correlated up/down shifts of systematic sources are propagated through to the output covariance matrix for the unfolding, and the shifts of the unfolded distributions corresponding to each variation of the nominal response matrix enable an estimate of the contribution of a specific source to the unfolded results. 

The impact of modelling uncertainties is estimated by varying various parameters in the parton shower and hadronization or the ME event generator used to produce the nominal simulation for the different event selections. 
The sources of modelling and theoretical uncertainties common to the different event selections include contributions from the choice of the strong coupling $\alpS^{\text{FSR}}(Q^2)$ for initial- and final-state radiation (ISR and FSR) \cite{Mrenna_2016} in parton showering and hadronization in the \PYTHIAviii CP5 tune, and from the choice of parton distribution function (PDF) sets used in simulations. 
Uncertainty contributions from the former consider the variation of the strong coupling $\alpS^{\text{FSR}}(Q^2)$ used to model ISR and FSR, by independently varying the renormalization scale $Q^2$ up and down by a factor of 2.
For the latter, a combined uncertainty for the PDF and \alpS variations is computed following the PDF4LHC~\cite{2016PDF} recommendations, using the NNPDF3.1 set \cite{Ball_2017} for simulations. 
These include 100 MC replicas (which assume a central value for $\alpS(m_Z^2)=0.118$ for the hard scatter) in the PDF set and variations of the central value of $\alpS(m_Z^2)=0.118\pm0.0015$ within its uncertainties.  
Finally, the uncertainty stemming from the choice of the parton shower and hadronization model used in simulations is estimated by unfolding the data with a response matrix constructed from the alternative simulation for the QCD dijet and \ttbar event selection. 
The alternative simulation relies on the same event generator as the nominal (signal) sample for QCD multijet events (\ttbar production) interfaced with showering in \HERWIGvii using the CH3 tune, instead of \PYTHIAviii with the CP5 tune. 

Additional modelling uncertainties are considered for \ttbar event simulations, corresponding to variations of specific parameters for event generation in \POWHEG{}\,v2. 
These are estimated using dedicated simulated samples. 
The MC top quark mass uncertainty contribution is estimated using alternative samples where $m_{\text{top}}$ is varied by $\pm 1$\GeV about the nominal value, and that from the ME-PS matching scale is considered by varying the factor $h_{\mathrm{damp}}$ within its uncertainties. 
The effect of the underlying event (UE) modelling is estimated by varying parameters of the CP5 tune \cite{CMS-PAS-TOP-16-021} used for \PYTHIAviii, and contributions from nonperturbative effects are also included by considering three different CR models. 
The details of the CR1 and CR2 models, as well as one considering the effects of allowing early resonance decays (ERD on), are described in Refs.~\cite{2020_CMS,CMS:2022awf}.

\section{Results and discussion}
\label{sec:results}
We first present the results from the simultaneous unfolding of all observables with comparisons to the particle-level predictions from various simulations. 
Thereafter, we present results for the normalized, unfolded distributions of some individual observables extracted from the combined unfolding, to illustrate key observations from the measurements.
Estimates of the correlations between bins, extracted from the covariance matrices of the total uncertainties in the unfolded results, are presented in Appendix~\ref{sec:unfCombinedCorr}. 

\subsection{Gluon and light-flavour quark jets}
\label{sec:resultsDijetComb}
\begin{figure}[htbp]
	\centering
	\includegraphics[width=0.9\textwidth]{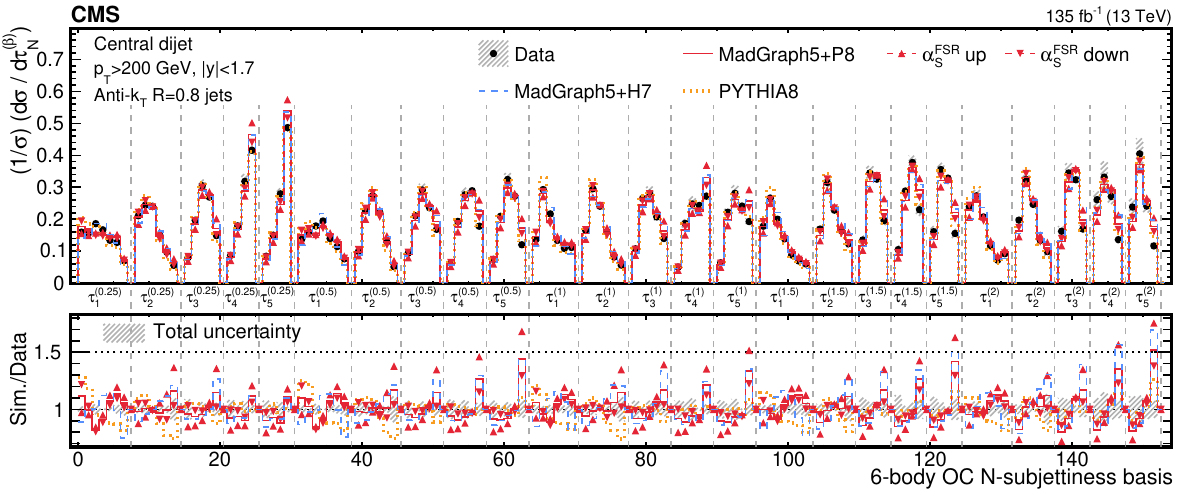}  
	\caption{The unfolded combined distribution of the overcomplete 6-body basis of $N$-subjettiness observables measured with AK8 jets in the QCD dijet selection (upper panel). The unfolded data (black) are compared with the nominal simulation (red), FSR scale variations of the nominal simulation (red, filled triangles), and predictions from the alternative (blue, yellow) simulations, at the particle level. The ratio of the simulated predictions to the unfolded data are shown in the lower panel. The shaded bands (dark grey) for the data markers indicate the total unfolding uncertainties. }
	\label{fig:dataUnfCombined_dijetSel}
\end{figure}

The unfolded measurement of the overcomplete 6-body basis of $N$-subjettiness observables in light quark or gluon-initiated jets in the QCD dijet selection consider 128 particle-level bins. 
The normalized, unfolded results are shown in Fig.~\ref{fig:dataUnfCombined_dijetSel} along with the particle-level distribution of the nominal and alternative simulated samples. 
Also illustrated is the distribution of the nominal sample for variations in $\alpS^{\text{FSR}}$ of 0.122 and 0.115.
The diagonal entries of the total covariance matrix of the normalized distribution are used to assign the per-bin uncertainties in the normalized, combined unfolded distributions and the uncertainty bands shown in the comparisons of unfolded data to predictions for all event selections. 

Across the bulk of the distributions of the observables sensitive to $\geq$4-body phase space, the variation of the nominal simulation with $\alpS^{\text{FSR}}=0.115$ and the standalone \PYTHIAviii sample provide the best overall description of the data. 
For 2-subjettiness observables, sensitive to information in 3-body phase space, both the nominal simulation and its $\alpS^{\text{FSR}}$-down variation describe the data well. 
Nonetheless, residual discrepancies with the unfolded data persist, to varying degrees, for all of the particle-level predictions across the individual $N$-subjettiness observables.

\subsection{Boosted \texorpdfstring{\PW}{W} boson and top quark jets}

\label{sec:resultsWtopComb}

\begin{figure}[htbp]
	\centering
	\includegraphics[width=0.9\textwidth]{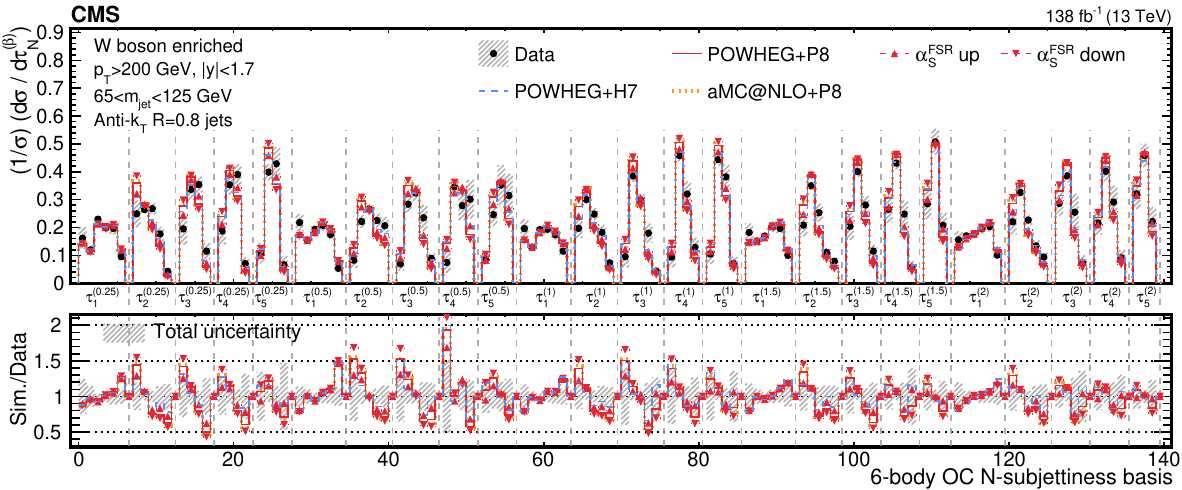}  
	\caption{The unfolded, combined distribution of the overcomplete 6-body basis of $N$-subjettiness observables measured with the selected AK8 jet for $\PGm$+jets \ttbar events enriched in boosted \PW boson jets (upper panel). The unfolded data (black) are compared with the nominal simulation (red), FSR scale variations of the nominal simulation (red, filled triangles), and predictions from the alternative signal (blue, yellow) simulations, at the particle level. The ratio of the simulated predictions to the unfolded data are shown in the lower panel. The shaded bands (dark grey) for the data markers indicate the total unfolding uncertainties.}
	
	\label{fig:dataUnfCombined_WSel}
\end{figure}

\begin{figure}[htbp]
	\centering
	\includegraphics[width=0.9\textwidth]{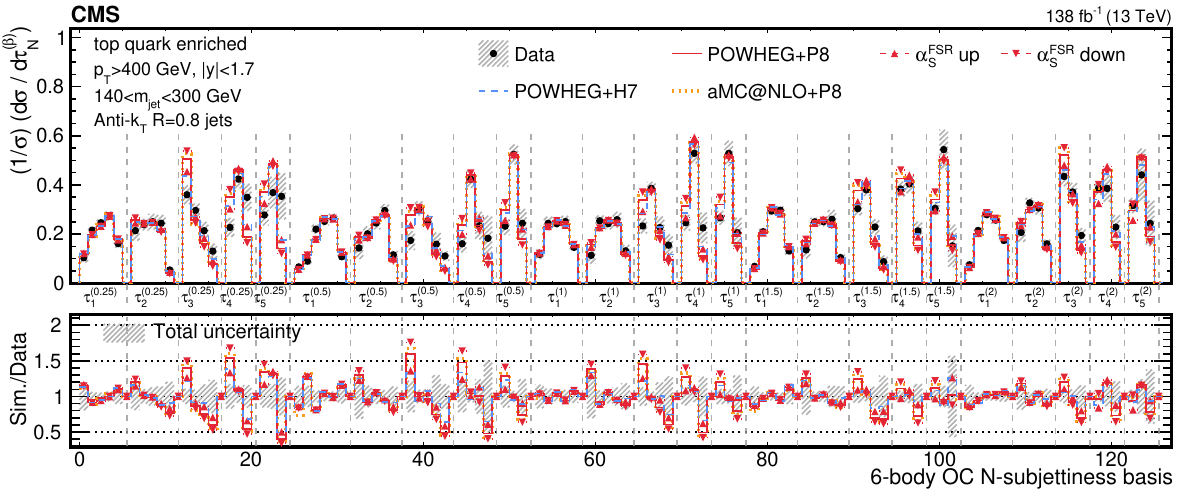}  
	\caption{The unfolded, combined distribution of the overcomplete 6-body basis of $N$-subjettiness observables measured with the selected AK8 jet for $\PGm$+jets \ttbar events enriched in jets from boosted top quark decays (upper panel). The unfolded data (black) are compared with the nominal simulation (red), FSR scale variations of the nominal simulation (red, filled triangles), and predictions from the alternative signal (blue, yellow) simulations, at the particle level. The ratio of the simulated predictions to the unfolded data are shown in the lower panel. The shaded bands (dark grey) for the data markers indicate the total unfolding uncertainties.}
	\label{fig:dataUnfCombined_topSel}
\end{figure}

The unfolded measurements of the $N$-subjettiness observables in $\PGm$+jets \ttbar events are presented in Figs.~\ref{fig:dataUnfCombined_WSel} and~\ref{fig:dataUnfCombined_topSel}. 
The measurements consist of a total of 115 and 101 bins for the \PW boson and top quark jets, respectively. 


Contrary to single-prong jets, we find that the predictions with a larger value of $\alpS^{\text{FSR}}=0.122$ provide better agreement with data across the bulk of all subdistributions. 
This is generally in agreement with previous substructure measurements in boosted \ttbar events by the CMS Collaboration~\cite{CMS:2018ypj,CMS:2022kqg} using simulated samples that rely on the \PYTHIAviii CP5 tune. 
The various simulations are observed to model the 1-subjettiness observables well, as well as the peak of some other subdistributions. 
However, for regions enriched in boosted \PW boson and top quark jets there are typically large disagreements with data in the bulk of all individual $N$-subjettiness observables sensitive to 3-body phase space and beyond.

\subsection{Results for individual observables}

The normalized, unfolded distributions for the individual observables are extracted from the combined distributions. 
Their bin contents and error bars are divided by the bin widths in the physical, particle-level binning scheme chosen for the observables. 
Similarly, the uncertainty contributions from various input sources to the total unfolding uncertainty per bin are obtained for the simultaneous unfolding. 
This is done by extracting the relevant covariance matrices, or shifts with respect to the nominal unfolded results, for the various sources considered in the unfolding, prior to extracting results for individual observables. 

\begin{figure}[!htb]
	\centering
	\includegraphics[width=.495\textwidth]{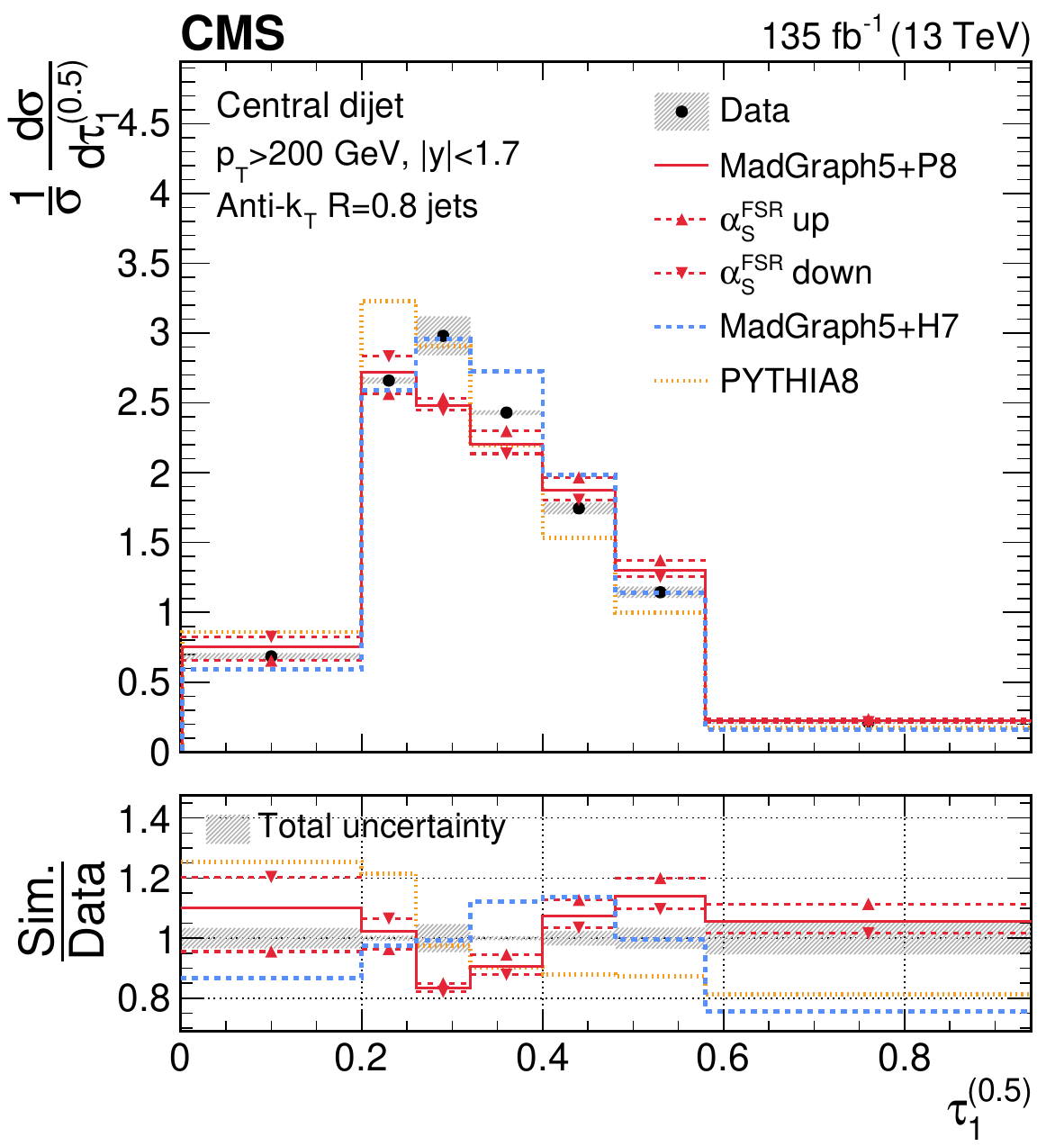}
	\includegraphics[width=.495\textwidth]{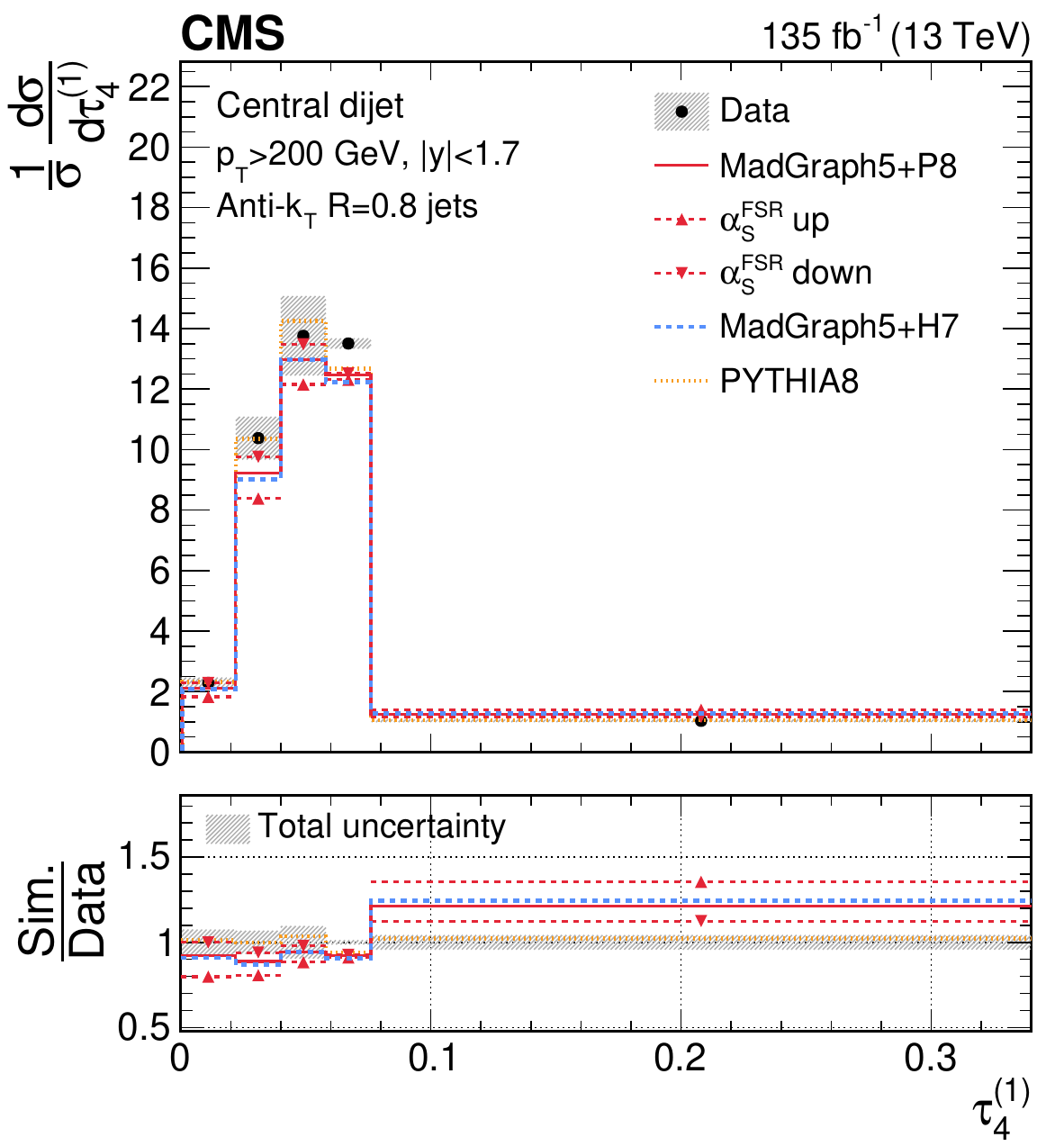}\\
	\caption{Representative unfolded distributions from the simultaneous unfolding are shown for \Nsub{1}{0.5} and \Nsub{4}{1} in the QCD dijet selection. The results are extracted from the normalized simultaneous unfolding, and the bin contents and the error bars are scaled by the corresponding bin widths. In the unfolded results, shown in the upper panel, the data (black) are compared with the nominal simulation (red), FSR scale variations of the nominal simulation (red, filled triangles), and predictions from the alternative signal (blue, yellow) simulations at the particle level. The ratio of the particle-level predictions to the unfolded data are shown in the lower panel. Shaded bands indicate the total uncertainties (dark grey). 
	}
	\label{fig:representativeplotdijet}
\end{figure}

\begin{figure}[!htbp]
	\centering
	\includegraphics[width=.495\textwidth]{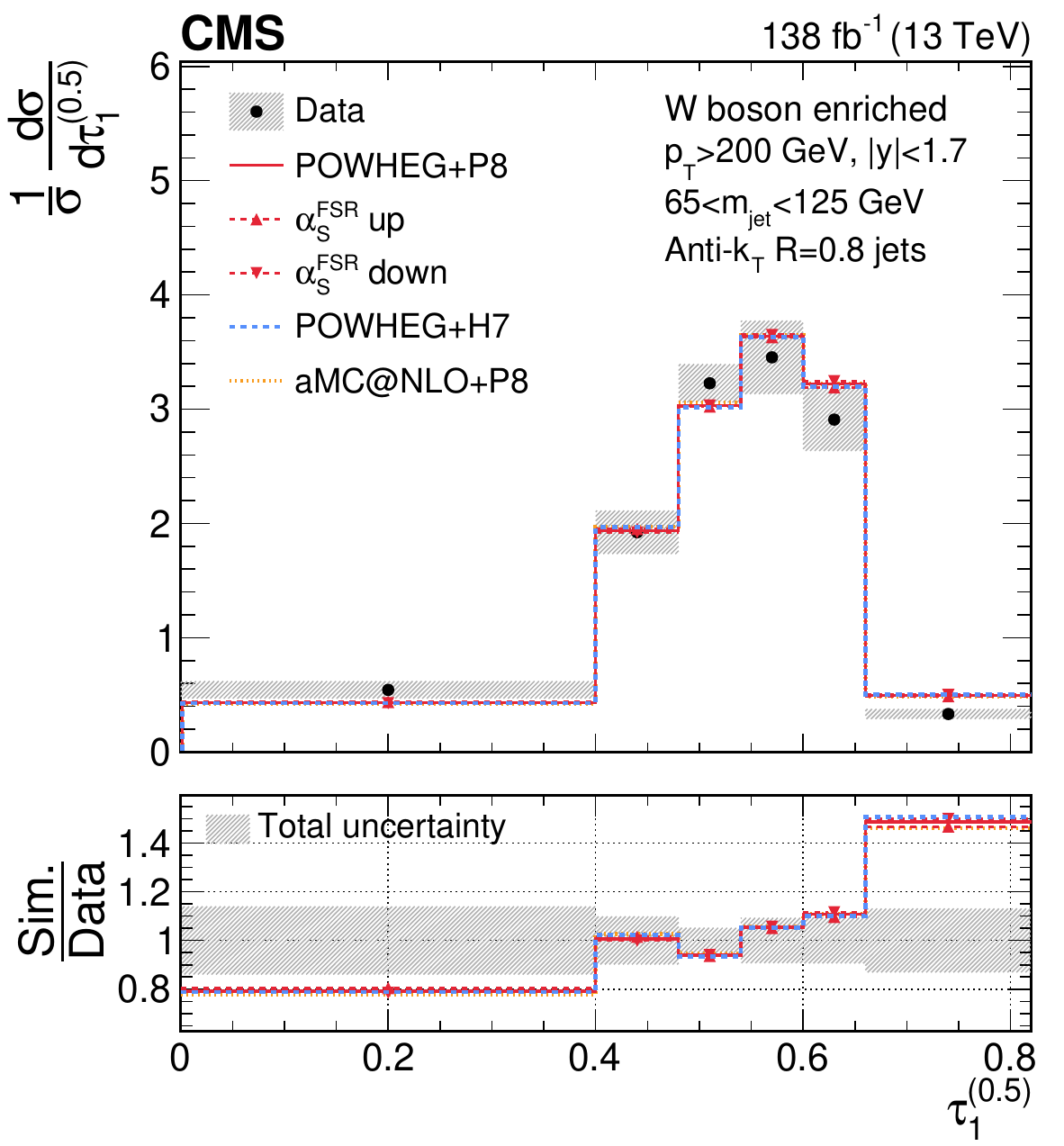} 
	\includegraphics[width=.495\textwidth]{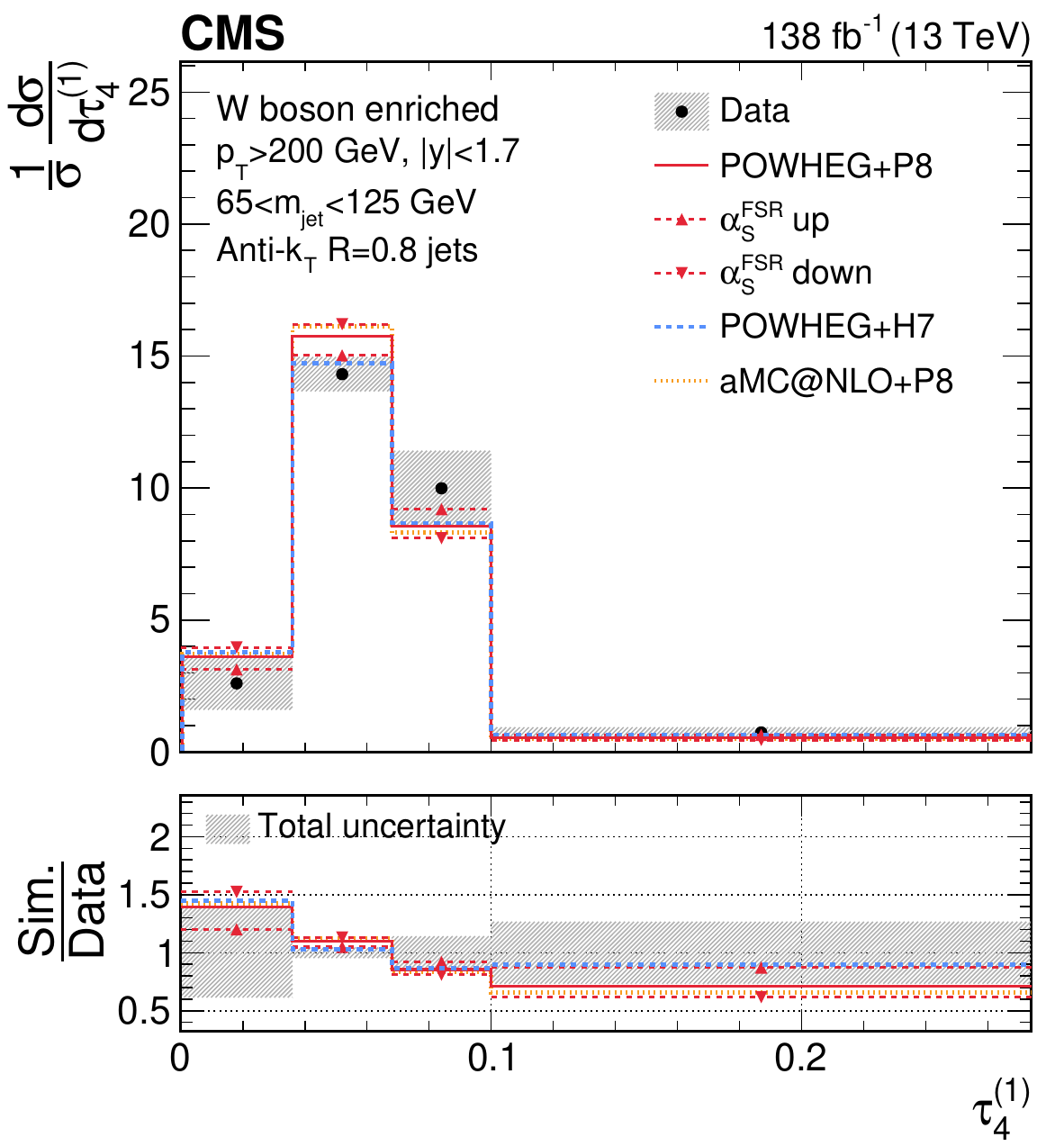} 		
	\caption{Representative unfolded distributions from the simultaneous unfolding are shown for \Nsub{1}{0.5} and \Nsub{4}{1} in the boosted \PW boson-enriched selection. The results are extracted from the normalized simultaneous unfolding, and the bin contents and the error bars are scaled by their corresponding bin widths. More details are provided in the caption of Fig.~\ref{fig:representativeplotdijet}.
	}
	\label{fig:representativeplotW}
\end{figure}

\begin{figure}[!htbp]
	\centering
	\includegraphics[width=.495\textwidth]{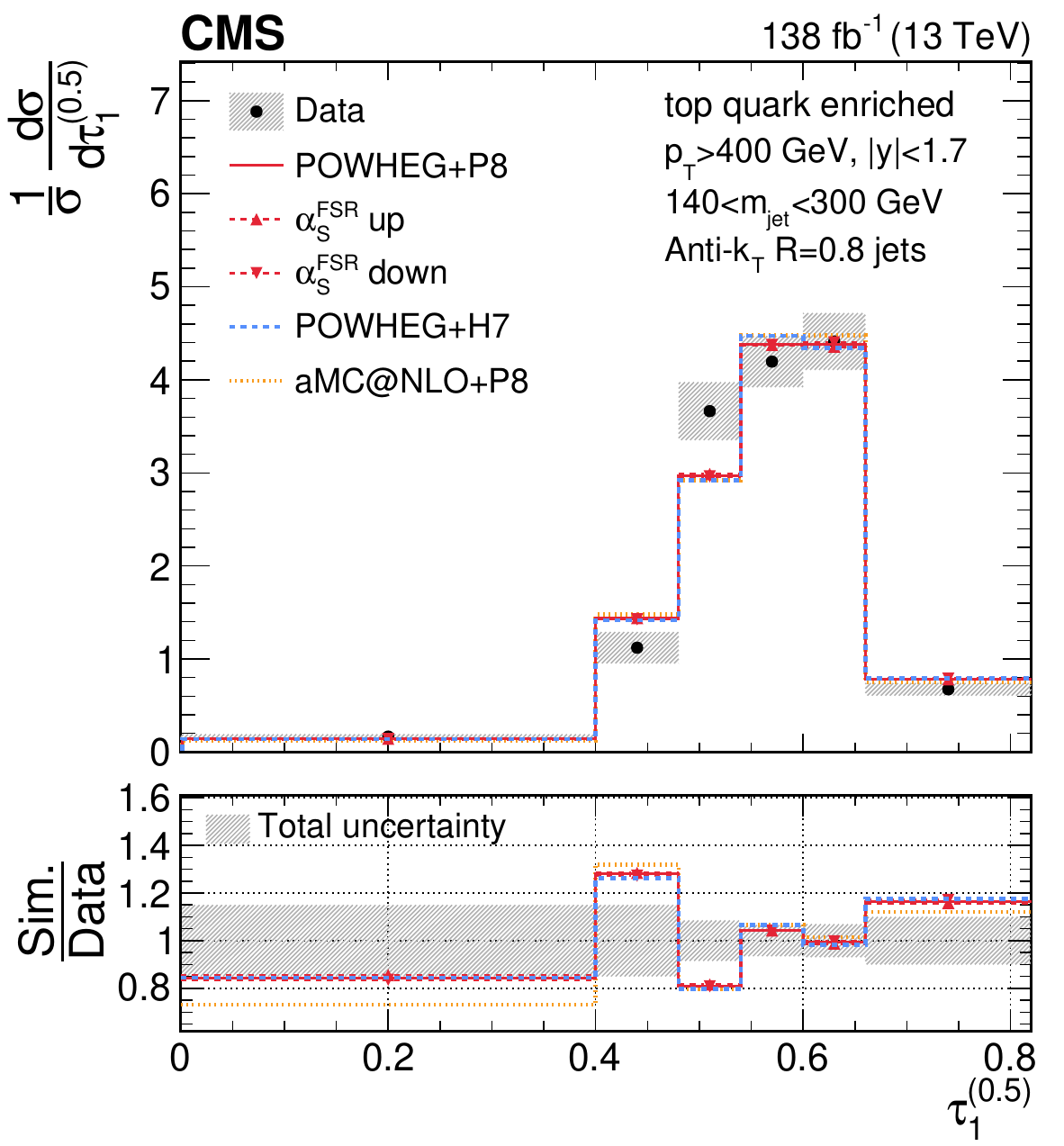} 
	\includegraphics[width=.495\textwidth]{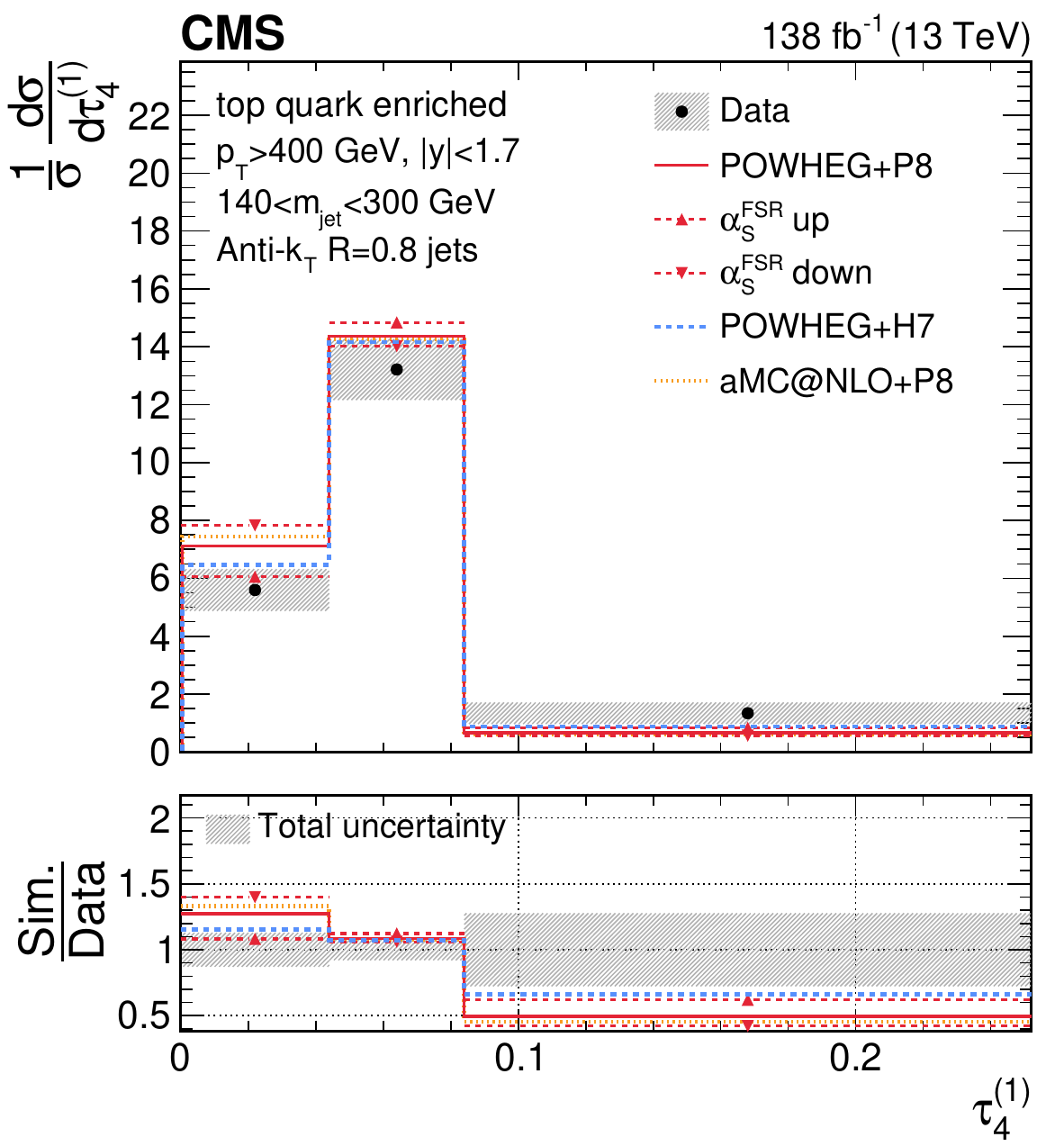} 
	\caption{Representative unfolded distributions of individual observables, \Nsub{1}{0.5} and \Nsub{4}{1}, are shown for measurements in the boosted top quark-enriched selection. The results are extracted from the normalized simultaneous unfolding, and the bin contents and the error bars are scaled by their corresponding bin widths. More details are provided in the caption of Fig.~\ref{fig:representativeplotdijet}.
	}
	\label{fig:representativeplottop}
\end{figure}

We show representative unfolded results for the QCD dijet selection in Fig.~\ref{fig:representativeplotdijet} and for the boosted \PW boson and top quark jet measurements in Figs.~\ref{fig:representativeplotW} and \ref{fig:representativeplottop}, respectively. 
The unfolded results for \Nsub{1}{0.5} and \Nsub{4}{1} presented here are representative of the general trends for the observables sensitive to the 2- and multi-body phase spaces of the jet. 
The particle-level distributions obtained from the nominal simulation, the variations of the nominal simulation obtained by varying the strong coupling for FSR, as well as the predictions from the alternative simulated samples, are shown along with the unfolded results in the upper panels of the figures. 
The ratio of the various particle-level predictions to the unfolded data are presented in the lower panels, where the total unfolding uncertainty, symmetrized about unity, is indicated with a dark-grey hashed region. 

In the QCD dijet selection, the various particle-level predictions generally envelope the data and have disagreements of about 10\% in the bulk of the distributions even for the 1-subjettiness observables. 
Disagreements of the same order are exhibited by the nominal predictions for the rest of the observables as well, particularly for those with $\beta<1$ and $\beta>1$ where the observables are increasingly sensitive to soft/soft-collinear and wide-angle contributions, respectively. 
In addition, while this is inconclusive for the \Nsub{1}{0.5} case shown on the left in Fig.~\ref{fig:representativeplotdijet} and generally other $1$-subjettiness observables, the variation of the nominal simulation corresponding to a smaller value of $\alpS^{\text{FSR}}$ generally shows the best agreement with the data across the bulk of the remaining observables, and for observables sensitive to $\geq$4-body phase space, the standalone \PYTHIAviii simulation provides a comparably good description. 
This is illustrated on the right in Fig.~\ref{fig:representativeplotdijet} for \Nsub{4}{1}. 
For observables sensitive to $\geq$5-body phase space in particular, the \PYTHIAviii-standalone prediction provides the closest description of the data in the highest values of the observables, whereas the \MGvATNLO{}-based predictions systematically tend to lie above the data irrespective of the parton shower used or the value of $\alpS^{\text{FSR}}$.


For the measurements in the boosted \PW boson- and top quark-enriched regions, the various simulations demonstrate similar shape differences to the data for most individual $N$-subjettiness observables relative to the results shown in Figs.~\ref{fig:representativeplotW} and \ref{fig:representativeplottop}. 
While the peaks of the binned distributions are typically in disagreement by only a few percent, going up to 10\% in some cases as observed in the results for \Nsub{1}{0.5} and \Nsub{4}{1} in both selections, the discrepancies in the remainder of the bulk of the distributions are generally larger, varying between 5--20\%. 
The latter is particularly noticeable for observables sensitive to higher ($M\geq3$) $M$-body phase space. 
Across the majority of the individual observables, it is found that the variation of the nominal \ttbar simulation corresponding to a larger value of $\alpS^{\text{FSR}}$ for showering in \PYTHIAviii, and the alternative \ttbar sample showered with \HERWIGvii, demonstrate better agreement with data than the nominal simulated sample. 
In particular, the improved modelling of the jet substructure for higher values of $\alpha_{s}^{\text{FSR}}$ than that used in the CP5 tune is consistent with previous jet substructure measurements in hadronic decays of boosted top quarks~\cite{CMS:2022kqg}.

\begin{figure}[!htb]
	\centering
	\includegraphics[width=.495\textwidth]{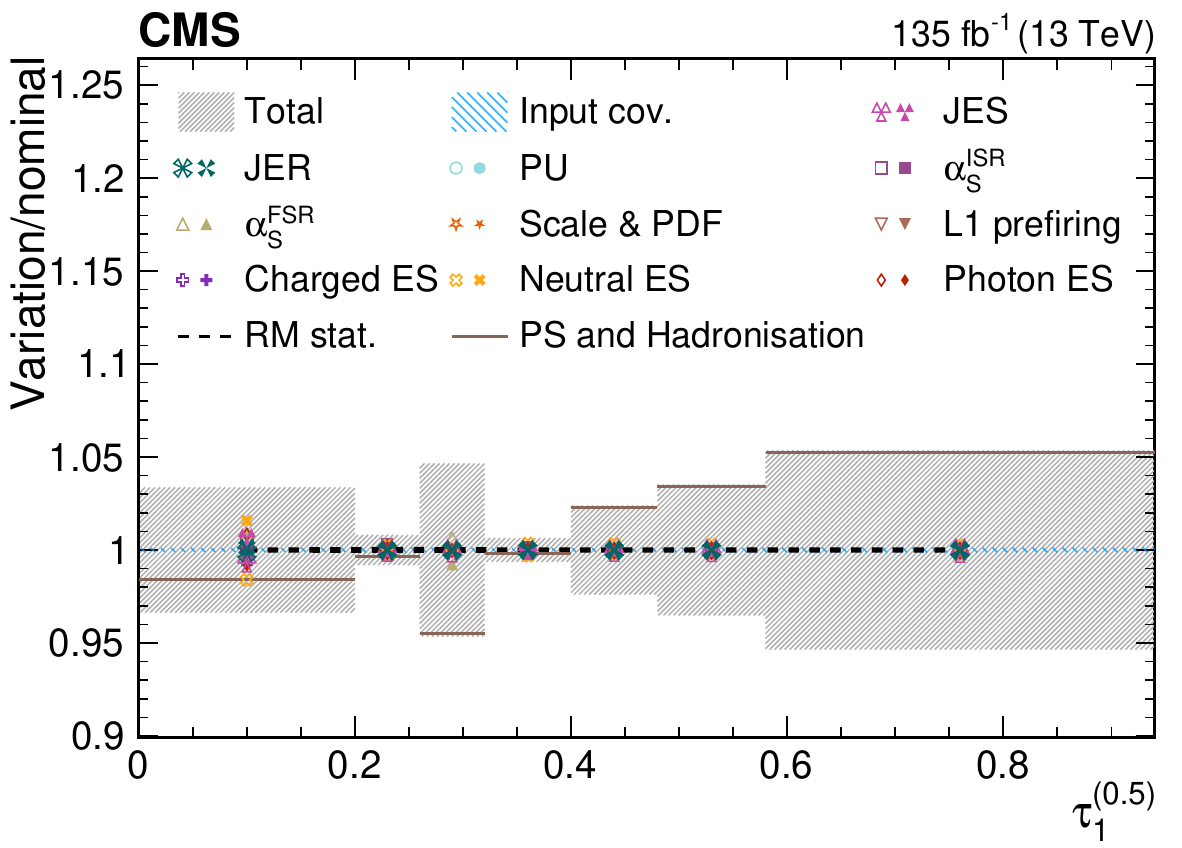} 
	\includegraphics[width=.495\textwidth]{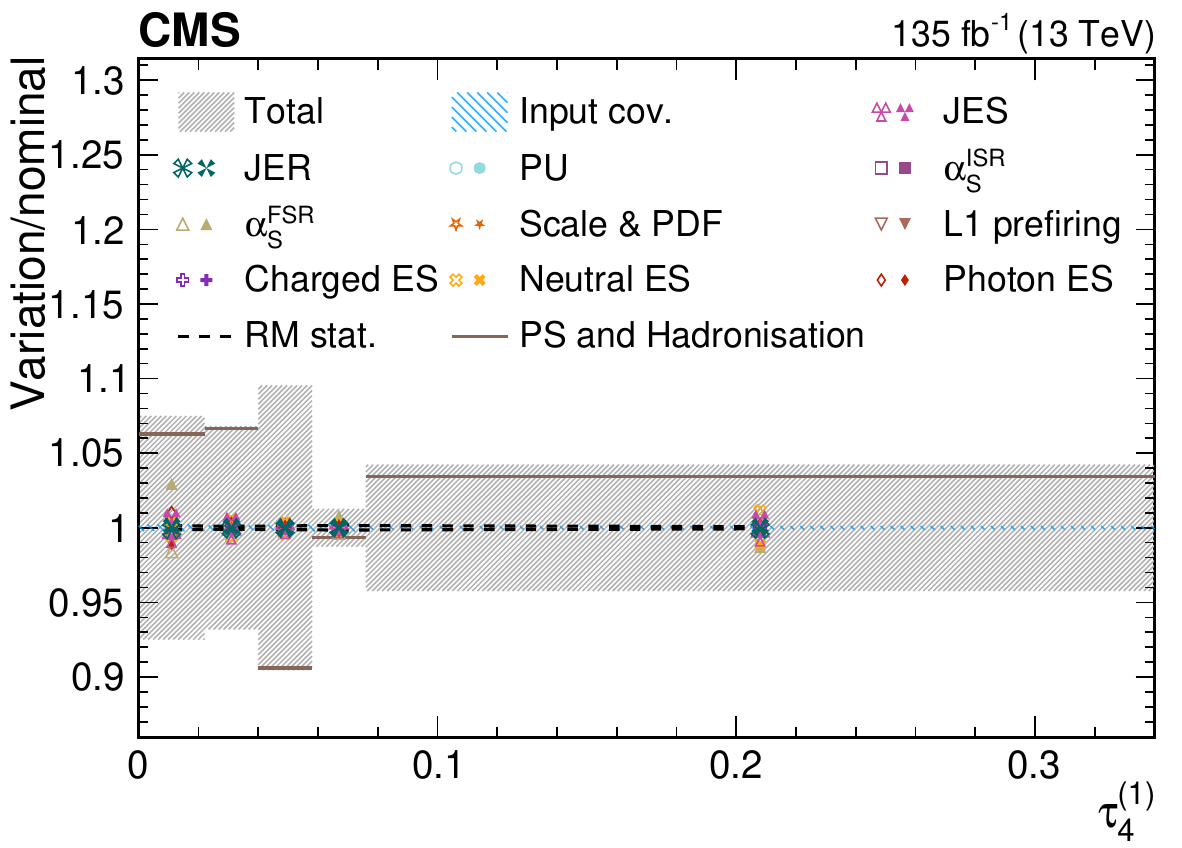} 
	\caption{Uncertainty breakdown estimates for the measurements of \Nsub{1}{0.5} and \Nsub{4}{1} in QCD dijets. These include all sources of experimental and modelling uncertainties that are common between the QCD dijet and \PW boson or top quark measurements. 
		The shaded bands indicate the total (dark grey), and data statistical and background subtraction (blue) uncertainties for the unfolded distribution, the uncertainty from the number of events in simulated samples for the nominal response matrix (RM stat.) is illustrated with a dashed line, and up (down) variations of relevant systematics are shown with filled (open) markers of the same colour and shape. 
		Contributions from the showering and hadronization uncertainty are estimated using \HERWIGvii and are illustrated with a solid line as a one-sided shift.	
	}
	\label{fig:representativeplotdijetUnc}
\end{figure}

\begin{figure}[!htb]
	\centering
	\includegraphics[width=.495\textwidth]{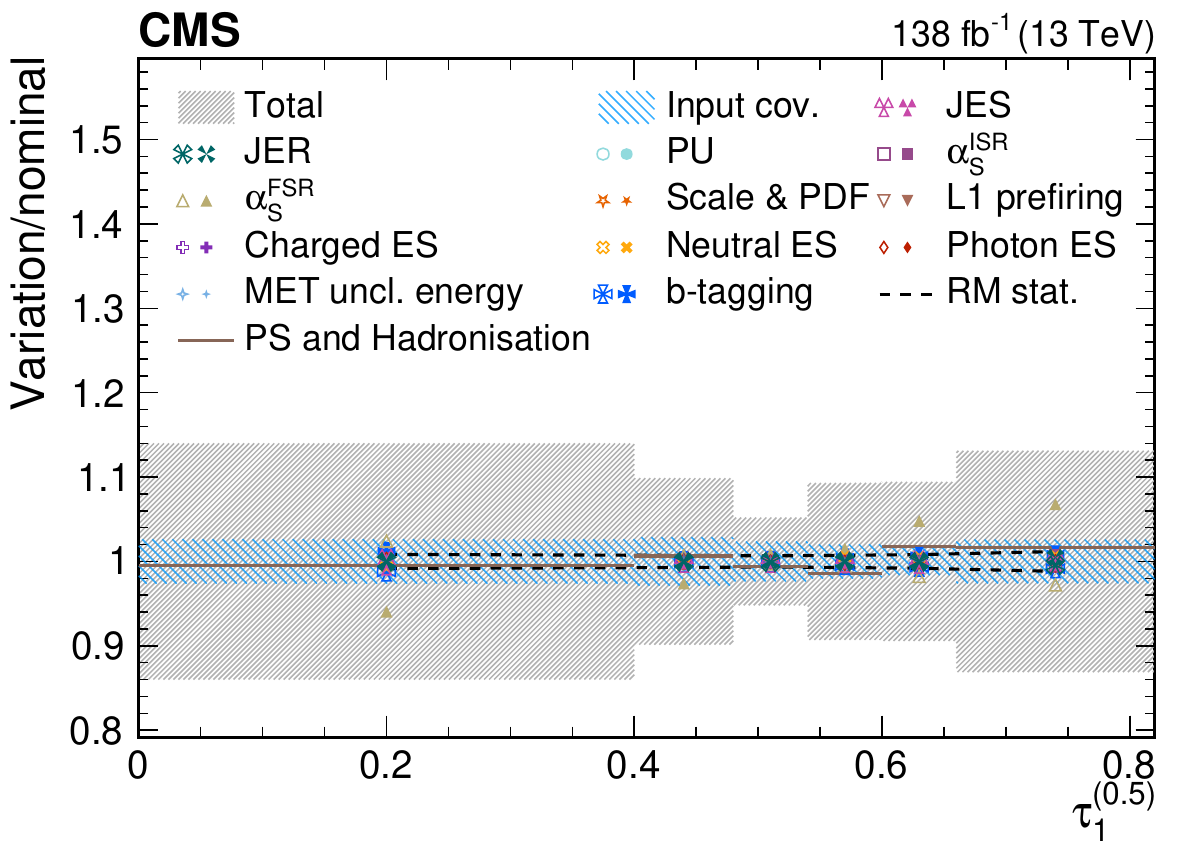} 	
	\includegraphics[width=.495\textwidth]{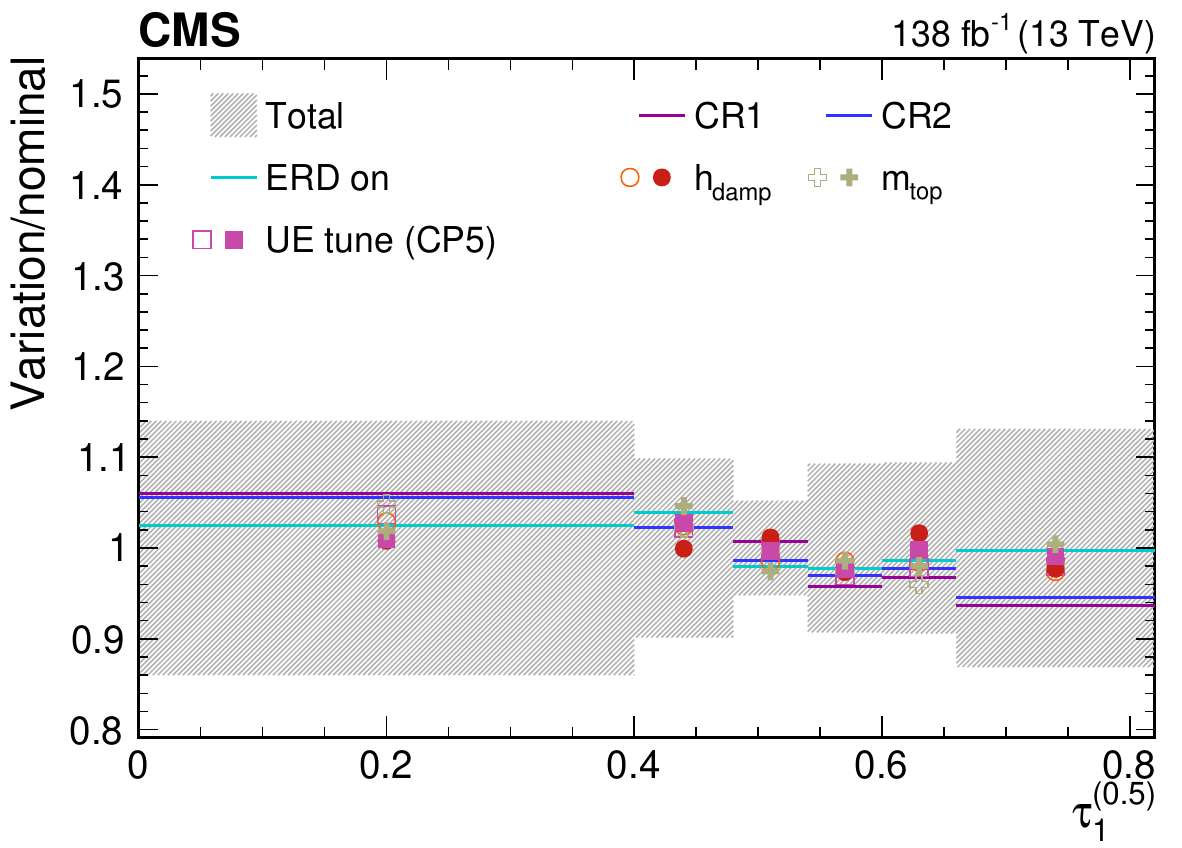}\\
	\includegraphics[width=.495\textwidth]{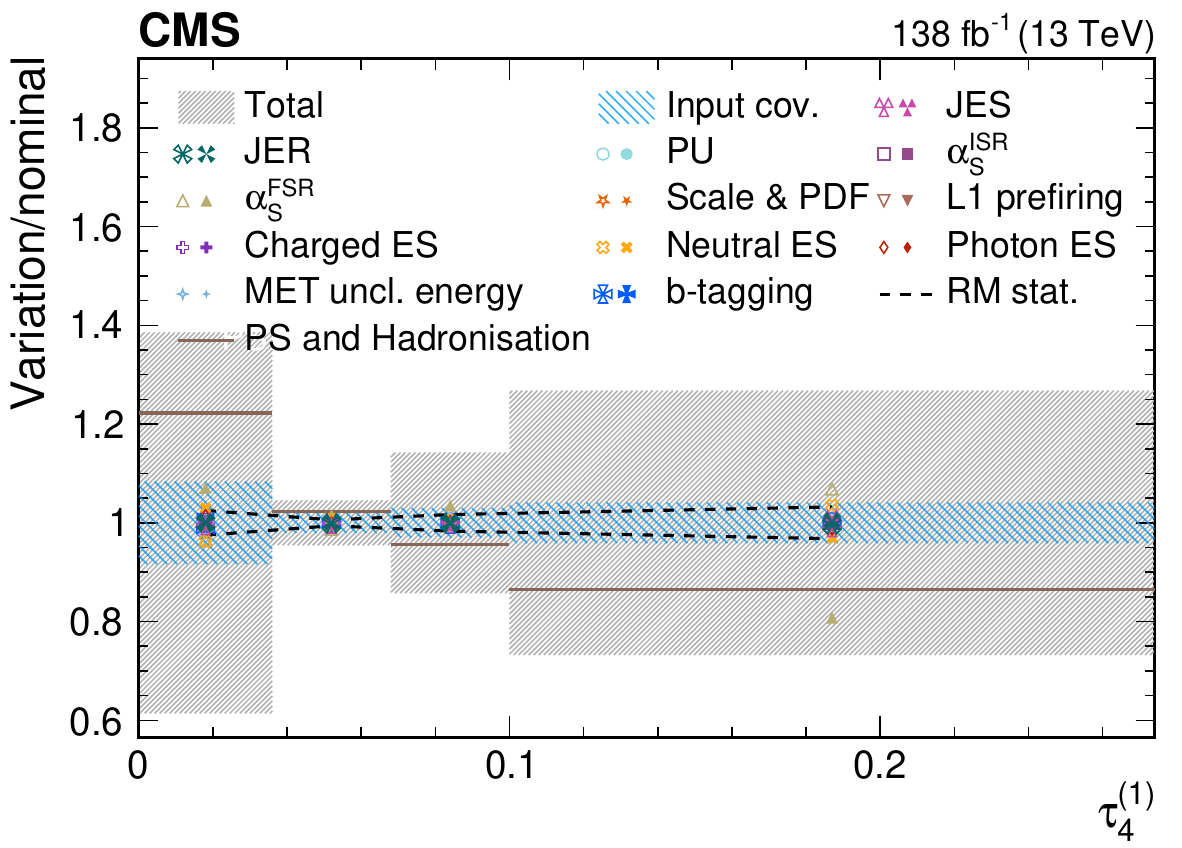} 	
	\includegraphics[width=.495\textwidth]{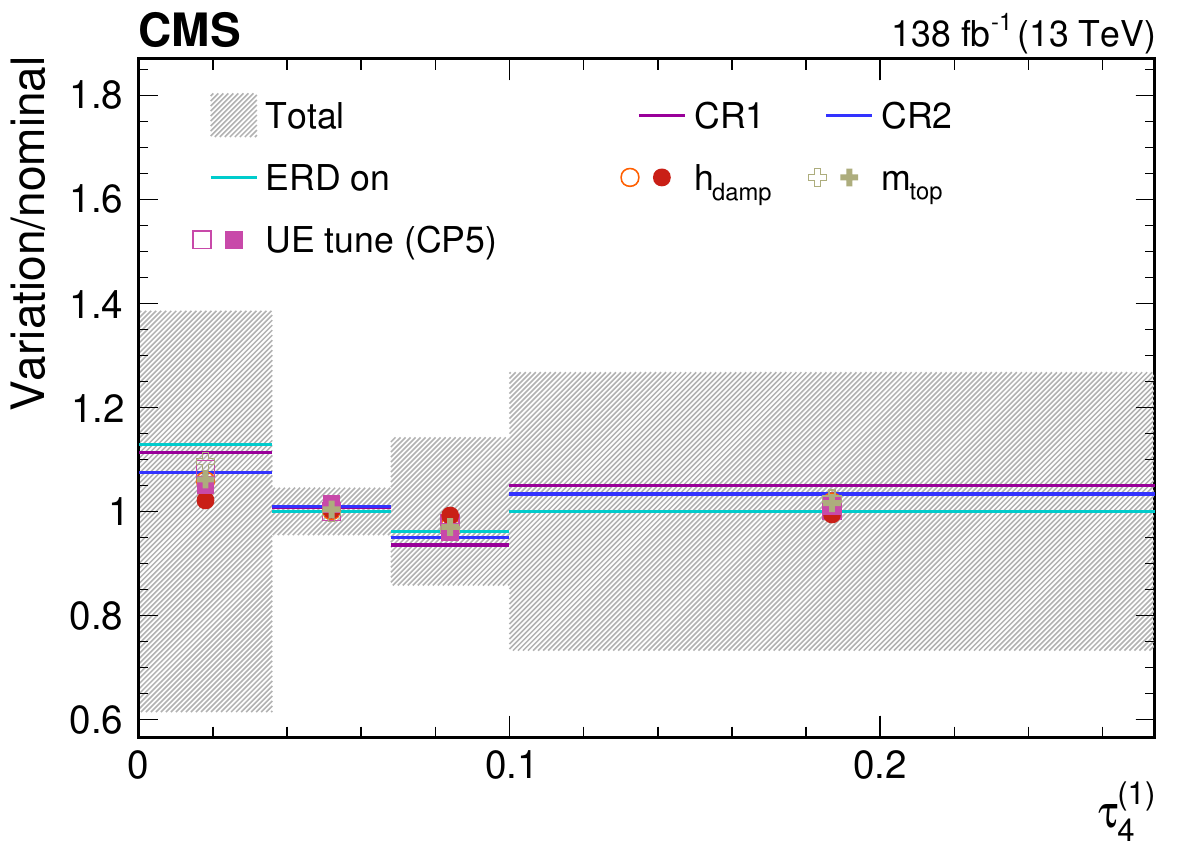} 
	\caption{A representative set of uncertainty breakdown estimates for the unfolded measurement of \Nsub{1}{0.5} and of \Nsub{4}{1} in the boosted \PW boson-enriched selection.
		The breakdowns are split into two separate figures: including all sources of experimental uncertainty, and additional uncertainties that are common between the dijet and \PW boson or top quark measurements (\cmsLeft), and for variations of parameters used to generate events in \POWHEG{}\,v2, or in the parton showering and hadronization in \PYTHIAviii with the CP5 tune, for exclusively the \PW boson and top quark measurements (\cmsRight). 
		The shaded bands indicate the total (dark grey), and data statistical and background subtraction (blue) uncertainties in the unfolded distribution, the uncertainty from the number of events in simulated samples for the nominal response matrix (RM stat.) is illustrated with a dashed line, and up (down) variations of relevant systematics are shown with filled (open) markers of the same colour and shape. 
		Contributions from the showering and hadronization uncertainty estimated using \HERWIGvii, as well as for the various CR models, are illustrated with the solid lines as one-sided shifts.	
	}
	\label{fig:representativeplotWUnc}
\end{figure}

\begin{figure}[!htb]
	\centering
	\includegraphics[width=.495\textwidth]{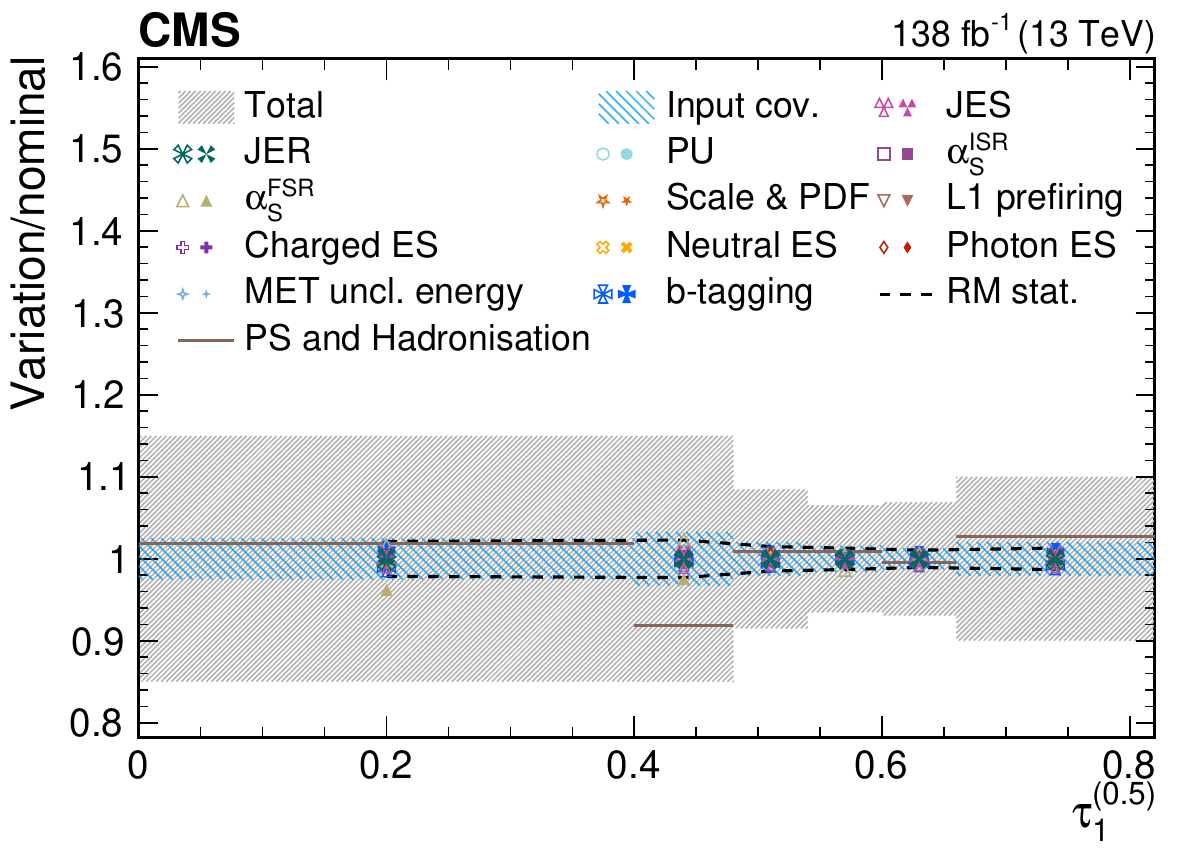} 	
	\includegraphics[width=.495\textwidth]{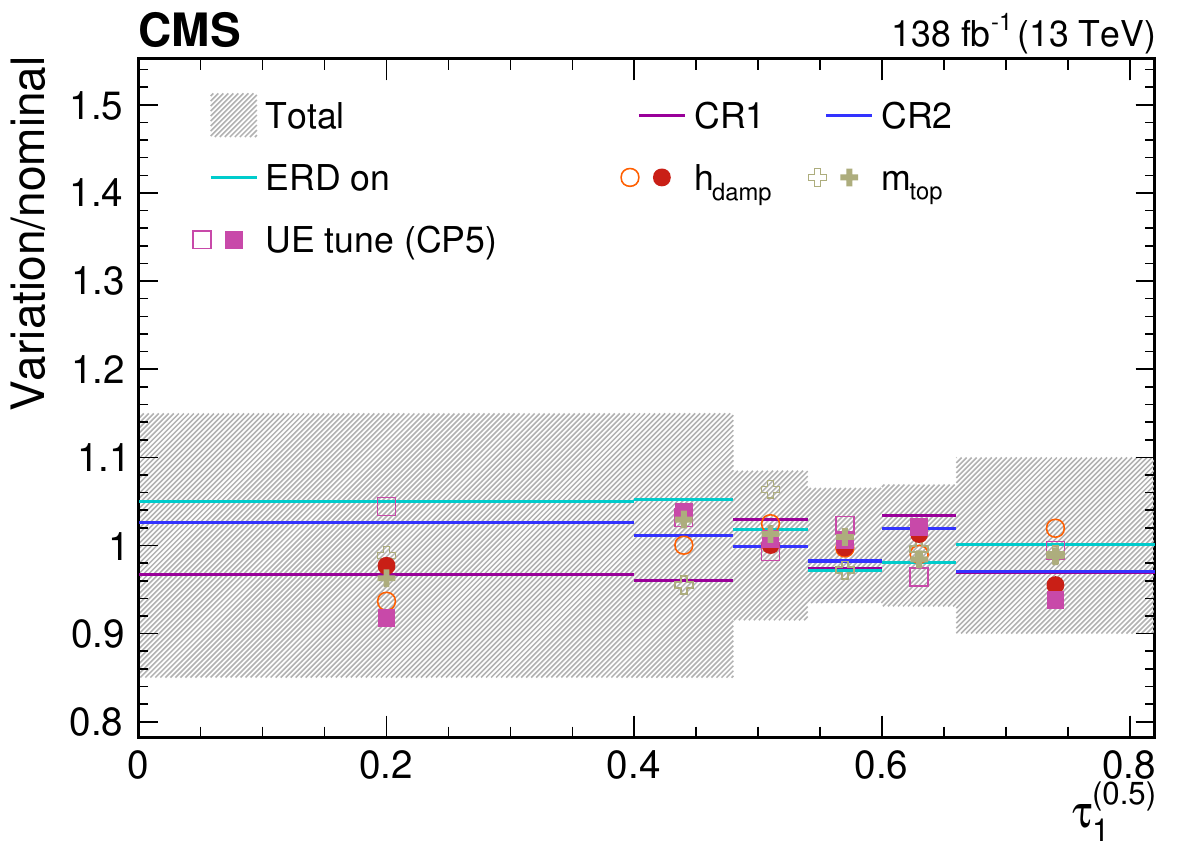}\\
	\includegraphics[width=.495\textwidth]{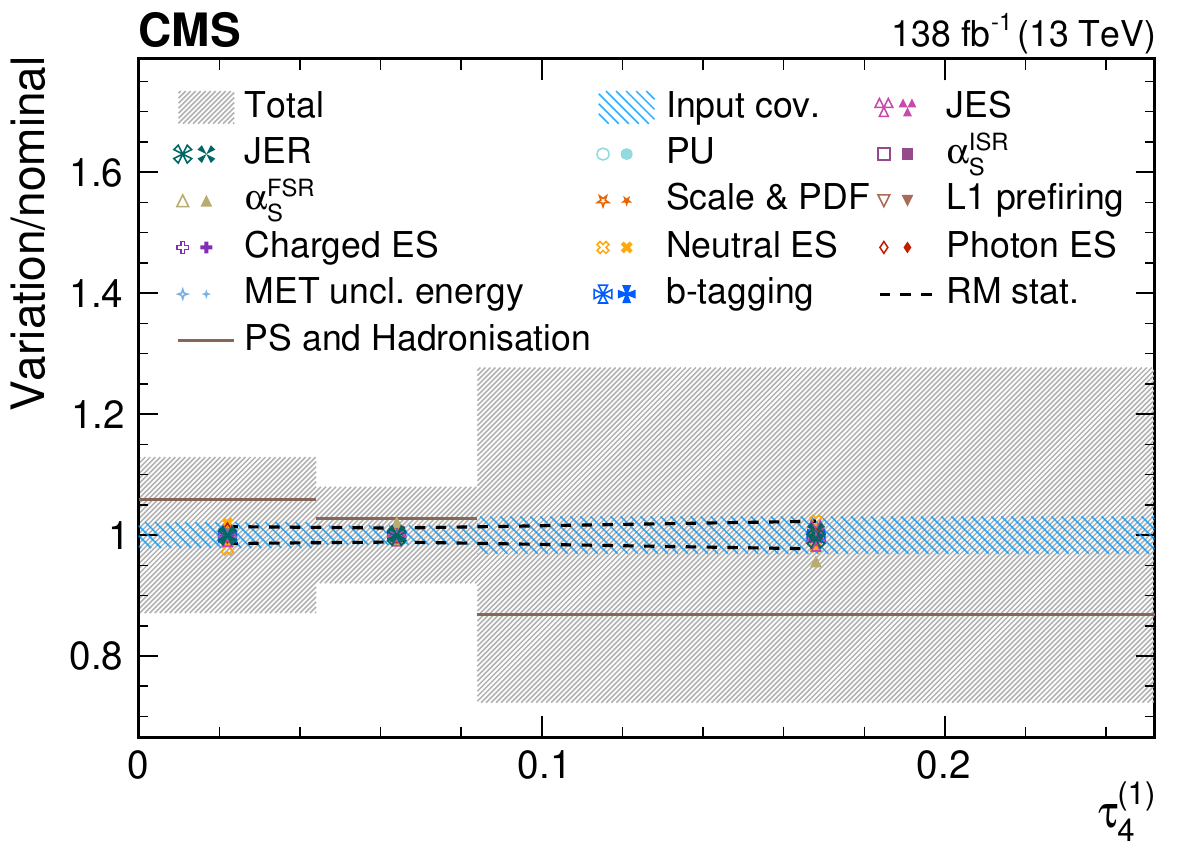} 	
	\includegraphics[width=.495\textwidth]{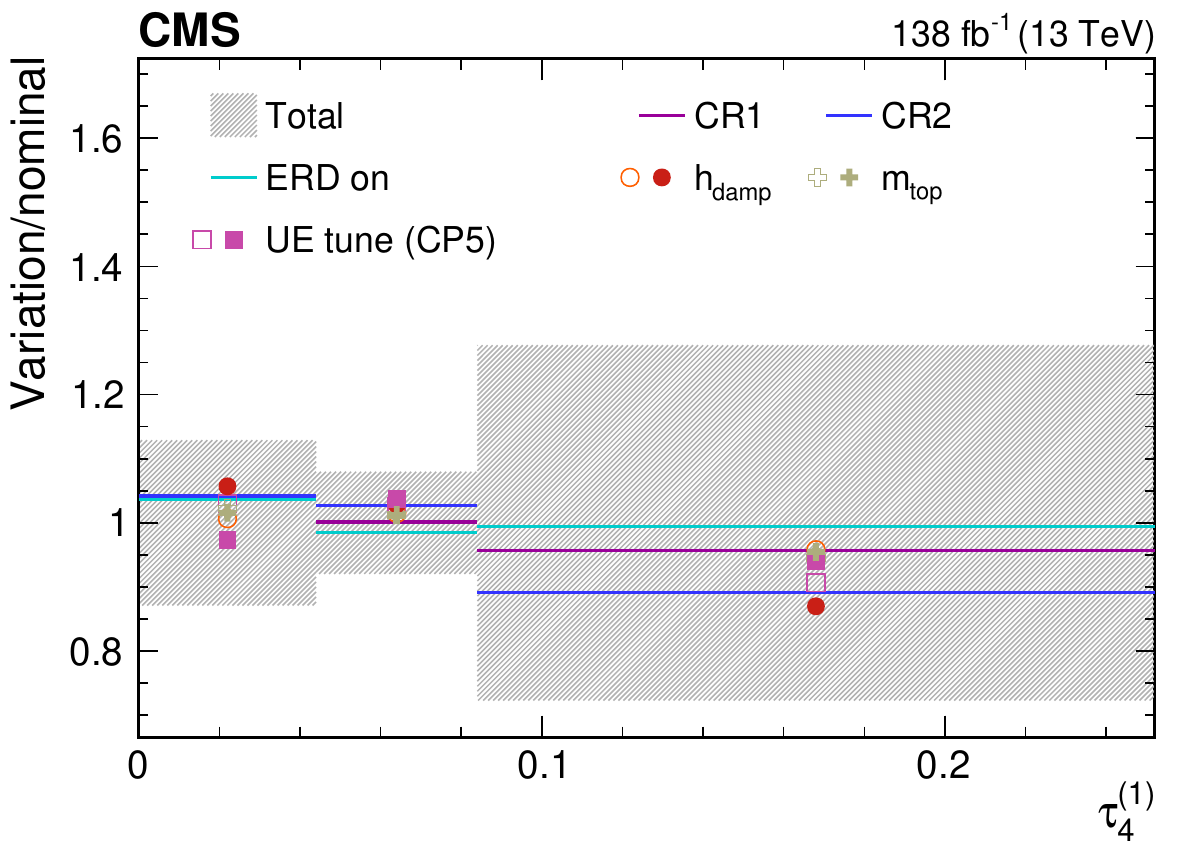} 
	\caption{A representative set of uncertainty breakdown estimates for the unfolded measurements of \Nsub{1}{0.5} and of \Nsub{4}{1} in the boosted top quark-enriched selection.
		The breakdowns are split into two separate figures per the details given in the caption of  Fig.~\ref{fig:representativeplotWUnc}.
	}
	\label{fig:representativeplottopUnc}
\end{figure}

We show the results of the uncertainty breakdowns for \Nsub{1}{0.5} and \Nsub{4}{1} in the QCD dijet selection in Fig.~\ref{fig:representativeplotdijetUnc}, and for the boosted \PW boson and top quark jet measurements in Figs.~\ref{fig:representativeplotWUnc} and \ref{fig:representativeplottopUnc}, respectively. 
For the latter two event selections, the uncertainty breakdowns are split into two figures for the individual observables where, in each case, the figure on the left illustrates contributions from various systematic uncertainties common to the selections. 
However, the figures on the right represent the modelling systematic uncertainties relevant only to the \ttbar event simulation.

The total uncertainty in the unfolded results, represented by the hashed dark-grey band, is obtained from the diagonal entries of the total covariance matrix of the normalized, unfolded distribution. 
This includes contributions from all sources of systematic and statistical uncertainty. 
The contributions from sources propagated to the input covariance prior to the unfolding, namely, the statistical uncertainties in the measured data, uncorrelated uncertainties from the finite statistical precision of the background simulations, and correlated scale uncertainties in selected background processes (for measurements on \PW boson and top quark jets), are combined into a single source. 
This is represented by the blue hashed region in the uncertainty breakdown figures, and is labelled as the `Input cov.' uncertainty. 
The contribution arising from the finite statistical precision of the simulated samples used to estimate the nominal response matrices for the unfoldings is indicated with the dashed black line. 
The effect of uncertainties from various systematic sources are illustrated by showing the ratio of the normalized and systematic-shifted unfolded distributions to the nominal, normalized unfolded distributions for each observable extracted from the simultaneous unfolding.

The dominant source of systematic uncertainty in the representative results for the QCD dijet measurements is the parton shower and hadronization uncertainty and the variation of the strong coupling used in the FSR simulation. 
There are also nonnegligible contributions from the variation of the neutral constituent energy scales, which is typically the largest source of experimental uncertainty in the dijet selection, in some bins in the tails of the distributions. 
All remaining systematic uncertainty sources typically have contributions below the percent-level across the bulk of the distributions. 
These conclusions are qualitatively in agreement with the results for the remaining observables presented in Appendix~\ref{sec:resultsPerObs}. 

Similarly, for the boosted \PW boson- and top quark-enriched regions, the parton shower and hadronization uncertainty has a dominant contribution. 
The other leading contributions arise from the FSR variations, as well as other modelling systematic uncertainties that contribute directly to the radiation pattern of the jets, such as the CR modelling and choice of parameter values for the UE tune and $h_{\text{damp}}$. 
These uncertainties are typically at the order of 5\% and rise to about 10--15\% in some bins. 
While the finite statistical precision of the input data are effectively negligible as a result of the simultaneous unfolding procedure, there are significant contributions from the combination of statistical and rate uncertainties in the background sources; these contributions rise to the level of a few percent. 
The limited statistical power of the nominal simulated sample used to construct the response matrix for the unfolding also contributes to the total uncertainties at the percent level. 
Contributions from most experimental systematic uncertainties are typically subdominant, or negligible, in the bulk of the distributions, but there are percent-level contributions from the neutral constituent energy scale variations and \PQb tagging efficiencies in the tails of some distributions.

The remaining results for the individual unfolded distributions of the $N$-subjettiness observables in the QCD dijet events, and in boosted \PW boson- and top quark-enriched regions in $\PGm$+jets \ttbar events, are presented for the individual observables sensitive to 2, 3, 4, 5, 6-body phase space in Appendix~\ref{sec:resultsPerObs}, along with the corresponding uncertainty breakdowns. 
This enables a systematic assessment of the modelling of a specified $M$-body phase space in simulations. 

We observe that no single simulation can describe all of the unfolded distributions simultaneously. 
Instead, an improvement in the QCD dijet region obtained by a smaller value of $\alpha_{s}^{\text{FSR}}$ leads to a worse description of the data for \PW boson and top quark jets.  
\section{Summary}
\label{sec:summary}
Simultaneous measurements have been presented of $N$-subjettiness observables that form a basis that overconstrains the phase space of up to six emissions in a jet. 
The measurements are performed in various hadronic environments: jets originating from gluons and light-flavour quarks in QCD dijet events, and in selections enriched in hadronic decays of boosted \PW bosons and top quarks. 

The use of a basis of $N$-subjettiness observables enables the analysis to provide a detailed picture of the structure of jets, for a fixed jet description corresponding to the resolved 6-body phase space. 
Multiple handles are provided to robustly overconstrain the sensitivity of the measurements to all the IRC-safe information in the jet substructure that is relevant to distinguish the substructure of light quark- and gluon-initiated jets from jets originating in decays of Lorentz-boosted \PW bosons and top quarks. By simultaneously unfolding all observables, normalized particle-level spectra for the individual observables are provided, along with complete covariance information including correlations between the unfolded distributions. 
These unfolded measurements furnish a comprehensive set of inputs for future tuning and validation of simulations, aiming to refine the modelling of QCD radiation in jets originating from decays of boosted massive electroweak-scale particles and from gluons or light-flavour quarks.

\begin{acknowledgments}
	We congratulate our colleagues in the CERN accelerator departments for the excellent performance of the LHC and thank the technical and administrative staffs at CERN and at other CMS institutes for their contributions to the success of the CMS effort. In addition, we gratefully acknowledge the computing centers and personnel of the Worldwide LHC Computing Grid and other centers for delivering so effectively the computing infrastructure essential to our analyses. Finally, we acknowledge the enduring support for the construction and operation of the LHC, the CMS detector, and the supporting computing infrastructure provided by the following funding agencies: SC (Armenia), BMBWF and FWF (Austria); FNRS and FWO (Belgium); CNPq, CAPES, FAPERJ, FAPERGS, and FAPESP (Brazil); MES and BNSF (Bulgaria); CERN; CAS, MoST, and NSFC (China); MINCIENCIAS (Colombia); MSES and CSF (Croatia); RIF (Cyprus); SENESCYT (Ecuador); ERC PRG and PSG, TARISTU24-TK10 and MoER TK202 (Estonia); Academy of Finland, MEC, and HIP (Finland); CEA and CNRS/IN2P3 (France); SRNSF (Georgia); BMFTR, DFG, and HGF (Germany); GSRI (Greece); MATE and NKFIH (Hungary); DAE and DST (India); IPM (Iran); SFI (Ireland); INFN (Italy); MSIT and NRF (Republic of Korea); MES (Latvia); LMTLT (Lithuania); MOE and UM (Malaysia); BUAP, CINVESTAV, CONACYT, LNS, SEP, and UASLP-FAI (Mexico); MOS (Montenegro); MBIE (New Zealand); PAEC (Pakistan); MSHE, NSC, and NAWA (Poland); FCT (Portugal); MESTD (Serbia); MICIU/AEI and PCTI (Spain); MOSTR (Sri Lanka); Swiss Funding Agencies (Switzerland); MST (Taipei); MHESI (Thailand); TUBITAK and TENMAK (T\"{u}rkiye); NASU (Ukraine); STFC (United Kingdom); DOE and NSF (USA).
	
	\hyphenation{Rachada-pisek} Individuals have received support from the Marie-Curie program and the European Research Council and Horizon 2020 Grant, contract Nos.\ 675440, 724704, 752730, 758316, 765710, 824093, 101115353, 101002207, 101001205, and COST Action CA16108 (European Union); the Leventis Foundation; the Alfred P.\ Sloan Foundation; the Alexander von Humboldt Foundation; the Science Committee, project no. 22rl-037 (Armenia); the Fonds pour la Formation \`a la Recherche dans l'Industrie et dans l'Agriculture (FRIA) and Fonds voor Wetenschappelijk Onderzoek contract No. 1228724N (Belgium); the Beijing Municipal Science \& Technology Commission, No. Z191100007219010, the Fundamental Research Funds for the Central Universities, the Ministry of Science and Technology of China under Grant No. 2023YFA1605804, the Natural Science Foundation of China under Grant No. 12535004, and USTC Research Funds of the Double First-Class Initiative No.\ YD2030002017 (China); the Ministry of Education, Youth and Sports (MEYS) of the Czech Republic; the Shota Rustaveli National Science Foundation, grant FR-22-985 (Georgia); the Deutsche Forschungsgemeinschaft (DFG), among others, under Germany's Excellence Strategy -- EXC 2121 ``Quantum Universe" -- 390833306, and under project number 400140256 - GRK2497; the Hellenic Foundation for Research and Innovation (HFRI), Project Number 2288 (Greece); the Hungarian Academy of Sciences, the New National Excellence Program - \'UNKP, the NKFIH research grants K 131991, K 138136, K 143460, K 143477, K 147557, K 146913, K 146914, K 147048, TKP2021-NKTA-64, and 2025-1.1.5-NEMZ\_KI-2025-00004, and MATE KKP and KKPCs Research Excellence and Flagship Research Groups grants (Hungary); the Council of Science and Industrial Research, India; ICSC -- National Research Center for High Performance Computing, Big Data and Quantum Computing, FAIR -- Future Artificial Intelligence Research, and CUP I53D23001070006 (Mission 4 Component 1), funded by the NextGenerationEU program, the Italian Ministry of University and Research (MUR) under Bando PRIN 2022 -- CUP I53C24002390006, PRIN PRIMULA 2022RBYK7T (Italy); the Latvian Council of Science; the Ministry of Science and Higher Education, project no. 2022/WK/14, and the National Science Center, contracts Opus 2021/41/B/ST2/01369, 2021/43/B/ST2/01552, 2023/49/B/ST2/03273, and the NAWA contract BPN/PPO/2021/1/00011 (Poland); the Funda\c{c}\~ao para a Ci\^encia e a Tecnologia (Portugal); the National Priorities Research Program by Qatar National Research Fund; MICIU/AEI/10.13039/501100011033, ERDF/EU, ``European Union NextGenerationEU/PRTR", projects PID2022-142604OB-C21, PID2022-139519OB-C21, PID2023-147706NB-I00, PID2023-148896NB-I00, PID2023-146983NB-I00, PID2023-147115NB-I00, PID2023-148418NB-C41, PID2023-148418NB-C42, PID2023-148418NB-C43, PID2023-148418NB-C44, PID2024-158190NB-C22, RYC2021-033305-I, RYC2024-048719-I, CNS2023-144781, CNS2024-154769 and Plan de Ciencia, Tecnolog{\'i}a e Innovaci{\'o}n de Asturias, Spain; the Chulalongkorn Academic into Its 2nd Century Project Advancement Project, the National Science, Research and Innovation Fund program IND\_FF\_68\_369\_2300\_097, and the Program Management Unit for Human Resources \& Institutional Development, Research and Innovation, grant B39G680009 (Thailand); the Eric \& Wendy Schmidt Fund for Strategic Innovation through the CERN Next Generation Triggers project under grant agreement number SIF-2023-004; the Kavli Foundation; the Nvidia Corporation; the SuperMicro Corporation; the Welch Foundation, contract C-1845; and the Weston Havens Foundation (USA).
\end{acknowledgments}\section*{Data availability} Release and preservation of data used by the CMS Collaboration as the basis for publications is guided by the  \href{https://doi.org/10.7483/OPENDATA.CMS.1BNU.8V1W}{CMS data preservation, re-use and open access policy}.

\bibliography{auto_generated} \clearpage

\appendix
\section{Pairwise correlations between observables}
\label{sec:Correlations}

The pairwise correlations between the observables constituting the overcomplete 6-body basis are studied. 
The trends in the correlations between the observables for the range of values of $N$ and $\beta$ are systematically different for the substructure of jets from gluons and light-flavour quarks, and those originating from collimated decays of boosted \PW bosons and top quarks. 
This is shown in the following for simulated events at the particle and detector levels, and in the data, where the results at the detector-level reflect the more rapid saturation of discrimination power, as noted in Section~\ref{sec:saturation}. 

For the \PW boson- or top quark-enriched particle-level cases in the \POWHEG{}+P8 simulation, only fully-merged jets are used to compute the particle-level pairwise correlations shown in the following. 
If the corresponding event at the detector level also passes the event selections, and the AK8 jets at the detector and particle levels are matched within $\Delta R<0.4$, the events are used to extract the corresponding correlation coefficients at the detector level. 
The QCD multijet events simulated with \MGvATNLO{}+P8 are used to compute the pairwise correlations for the QCD dijet selection.

The computation of the covariances also does not assume the same weight for all events at the particle level, which are reweighted by their appropriate generator weights. 
The (weighted) covariance between two observables \(a\) and \(b\) is calculated as
\begin{equation}
	\text{C}(a,b) = \frac{\sum w_i (a_i - \bar{a}_w)(b_i - \bar{b}_w)}{\sum w_i},
\end{equation}
where $w_i$ are the event weights, and $\bar{a}_w$, $\bar{b}_w$ are the weighted means of observables $a$ and $b$, respectively. 
Then the weighted Pearson correlation coefficient for a pair of observables is given by
\begin{equation}
	\rho_{ab} = \frac{\text{C}(a,b)}{\sigma_a\ \sigma_b},
\end{equation}
where $\sigma_a$ and $\sigma_b$ are the weighted standard deviations of the observables $a$ and $b$. 

\begin{figure}[!htb]
	\centering	
	\includegraphics[width=1.\textwidth]{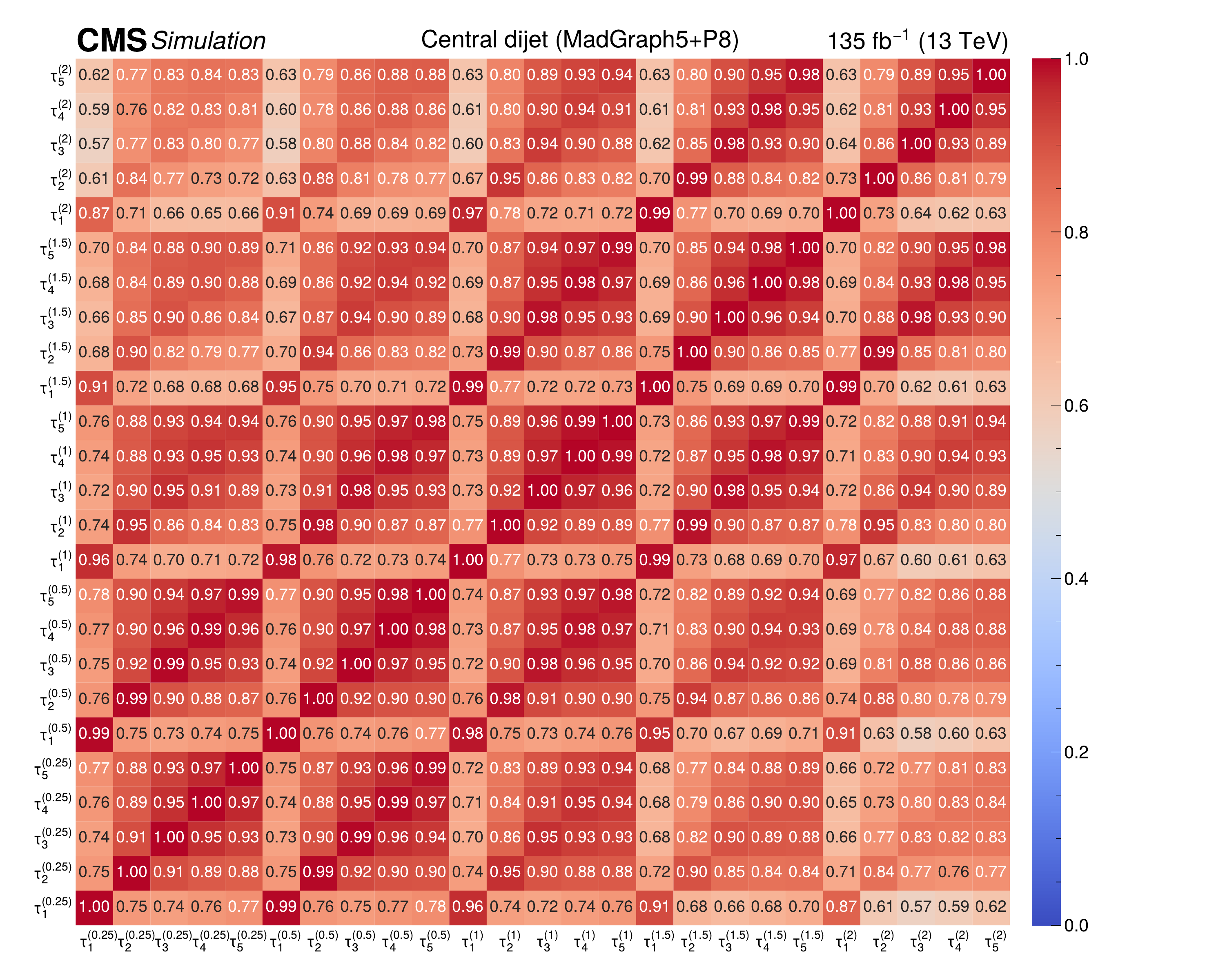}  
	\caption{Pairwise Pearson correlations between $N$-subjettiness observables constituting the overcomplete 6-body basis, in the nominal \MGvATNLO{}+\PYTHIAviii simulation, at the particle level, for the QCD dijet selection. All particle-level events passing selections are considered.}
	\label{fig:correlationDijet_nomMC_gen}
	
\end{figure}

\begin{figure}[ht]
	\centering	
	\includegraphics[width=1.\textwidth]{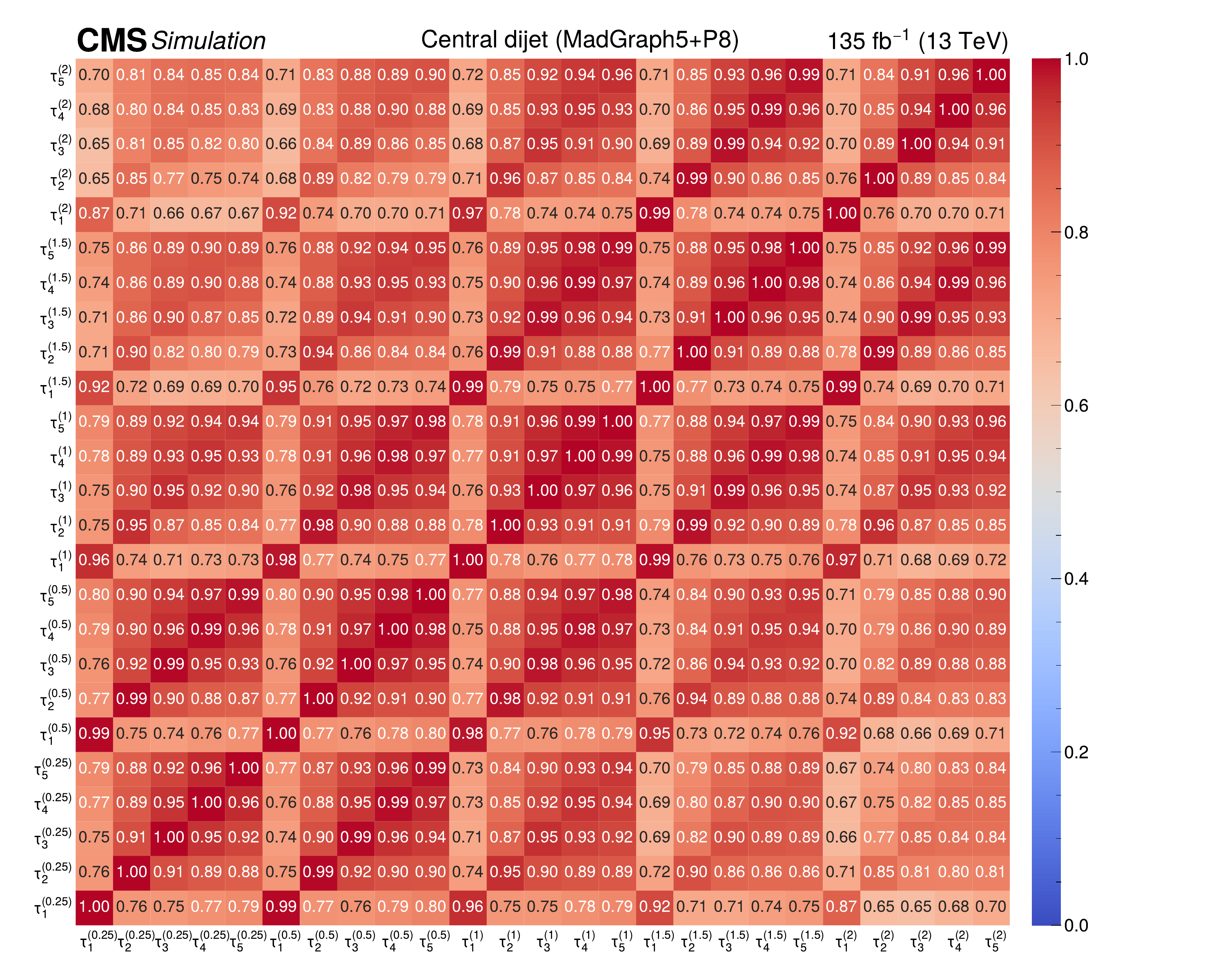}  
	\caption{Pairwise Pearson correlations between $N$-subjettiness observables constituting the overcomplete $6$-body basis, in the nominal \MGvATNLO{}+\PYTHIAviii simulation, at the detector level, for the QCD dijet selection. Only detector-level events with a matched jet in the corresponding particle-level event are considered. }
	\label{fig:correlationDijet_nomMC_truereco}
\end{figure}

\begin{figure}[!htb]
	\centering	
	\includegraphics[width=1.\textwidth]{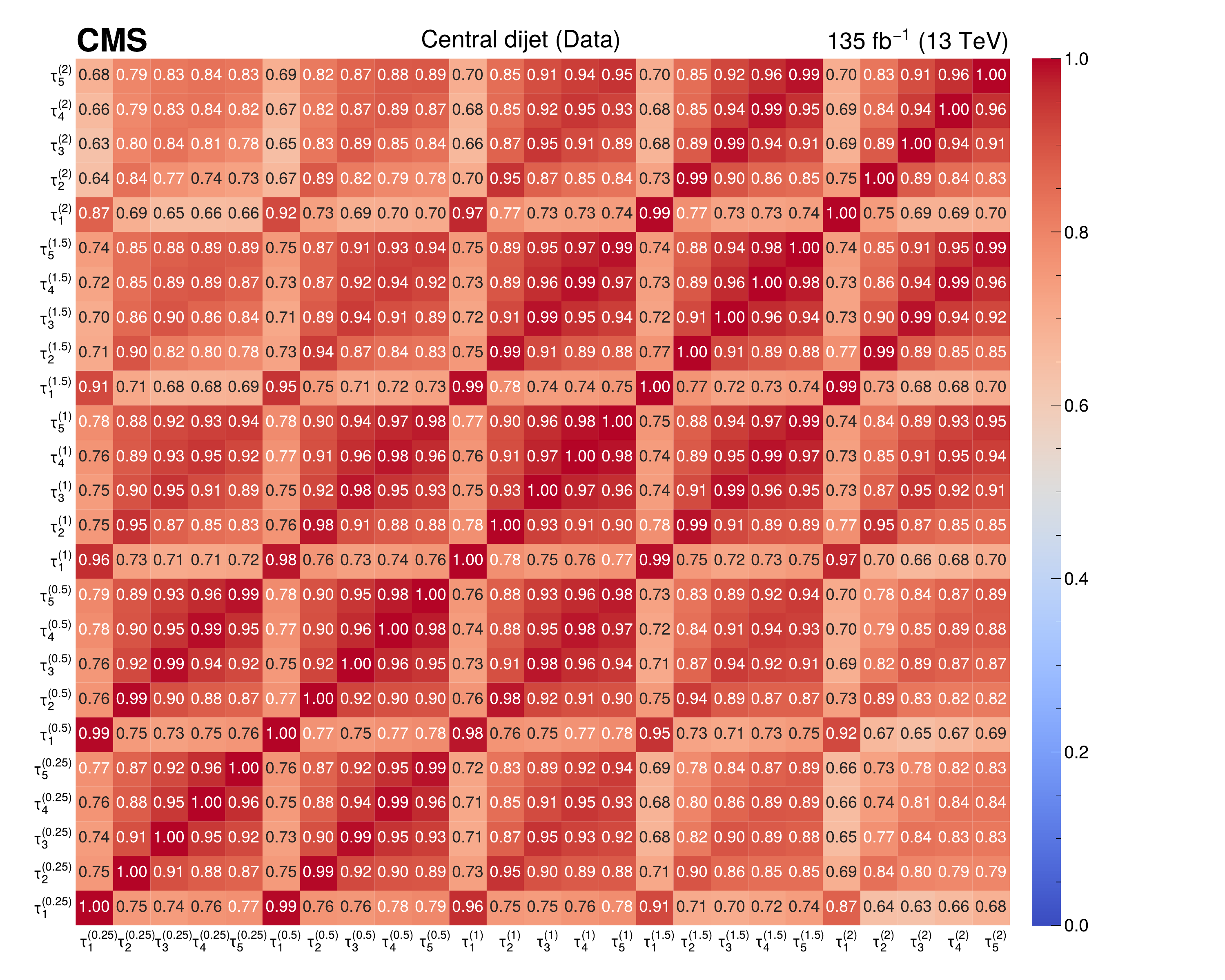}  
	\caption{Pairwise Pearson correlations between $N$-subjettiness observables constituting the overcomplete 6-body basis, using the full Run 2 data set recorded by the CMS detector, for the QCD dijet selection. }
	\label{fig:correlationDijet_data}
\end{figure}

\begin{figure}[!htb]
	\centering	
	\includegraphics[width=1.\textwidth]{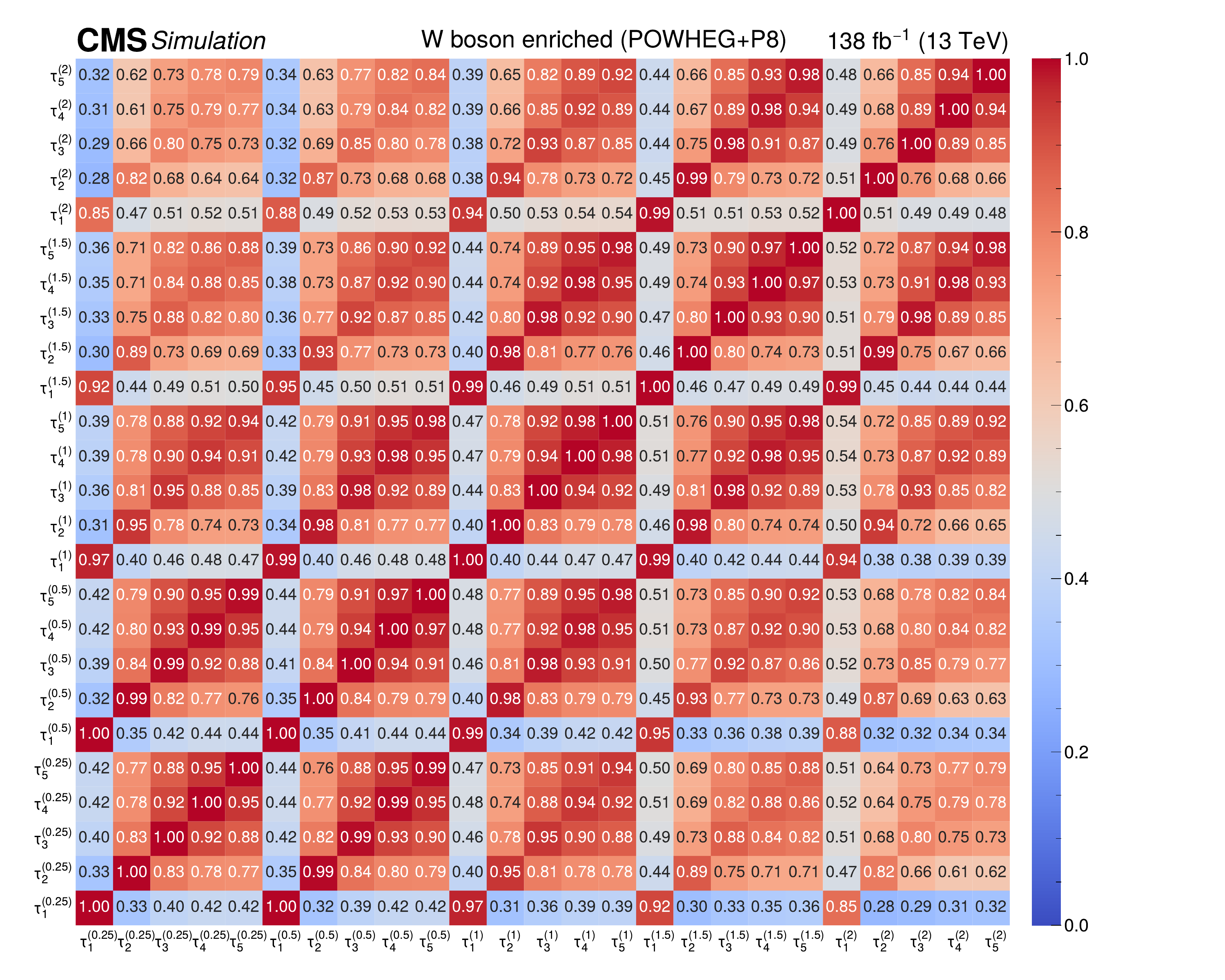}  
	\caption{Pairwise Pearson correlations between $N$-subjettiness observables constituting the overcomplete 6-body basis, in the nominal \POWHEG{}+\PYTHIAviii signal sample, at the particle level, in the boosted \PW boson-enriched region. All particle-level events with fully-merged jets passing the event selections are considered.}
	\label{fig:correlationW_nomMC_gen}
\end{figure}

\begin{figure}[!htb]
	\centering	
	\includegraphics[width=1.\textwidth]{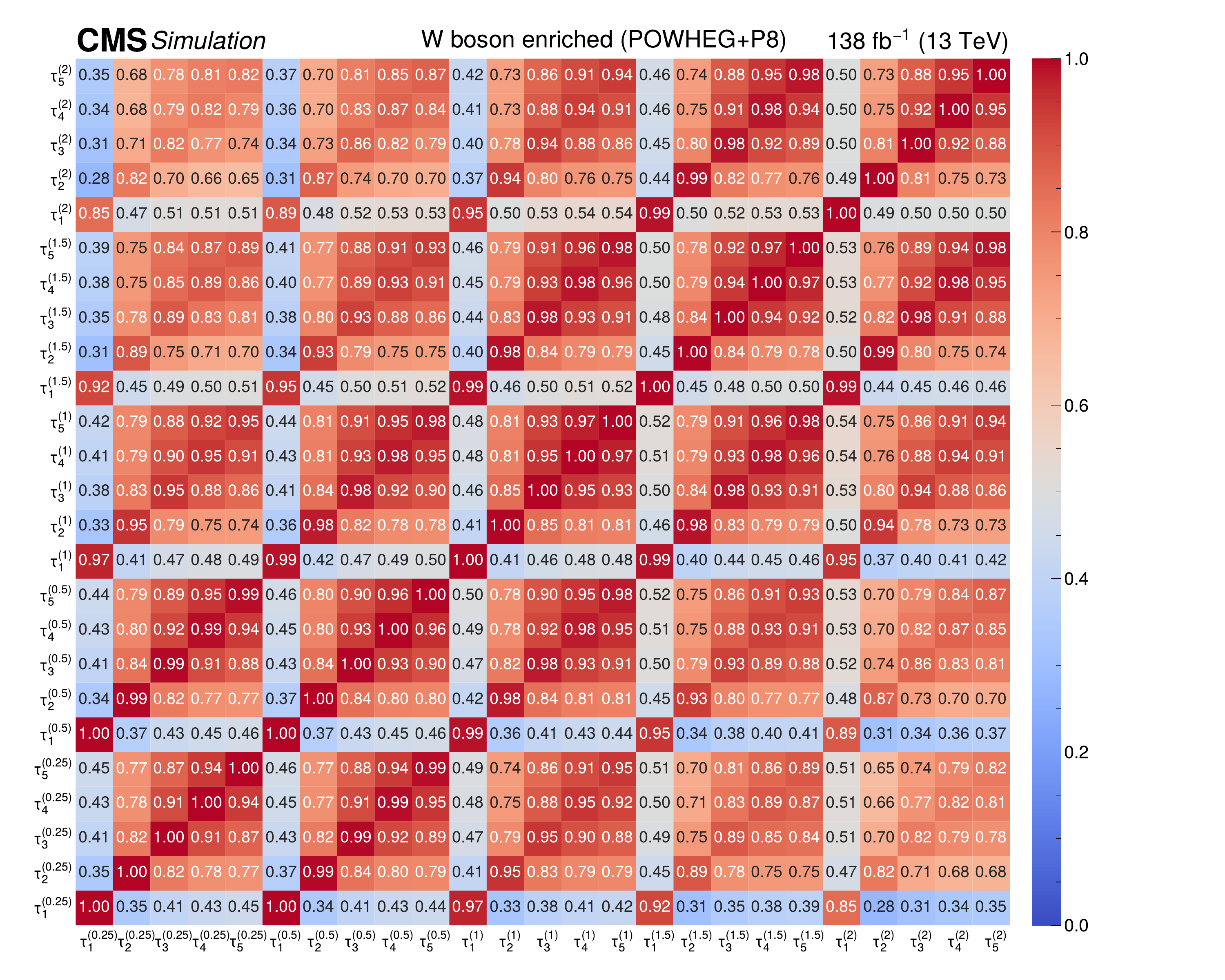}  
	\caption{Pairwise Pearson correlations between $N$-subjettiness observables constituting the overcomplete 6-body basis, in the nominal \POWHEG{}+\PYTHIAviii signal sample, at the detector level, in the boosted \PW boson-enriched region. Only detector-level events with a matched jet in the corresponding fully-merged particle-level event are considered. }
	\label{fig:correlationW_nomMC_truereco}
\end{figure}

\begin{figure}[!htb]
	\centering	
	\includegraphics[width=1.\textwidth]{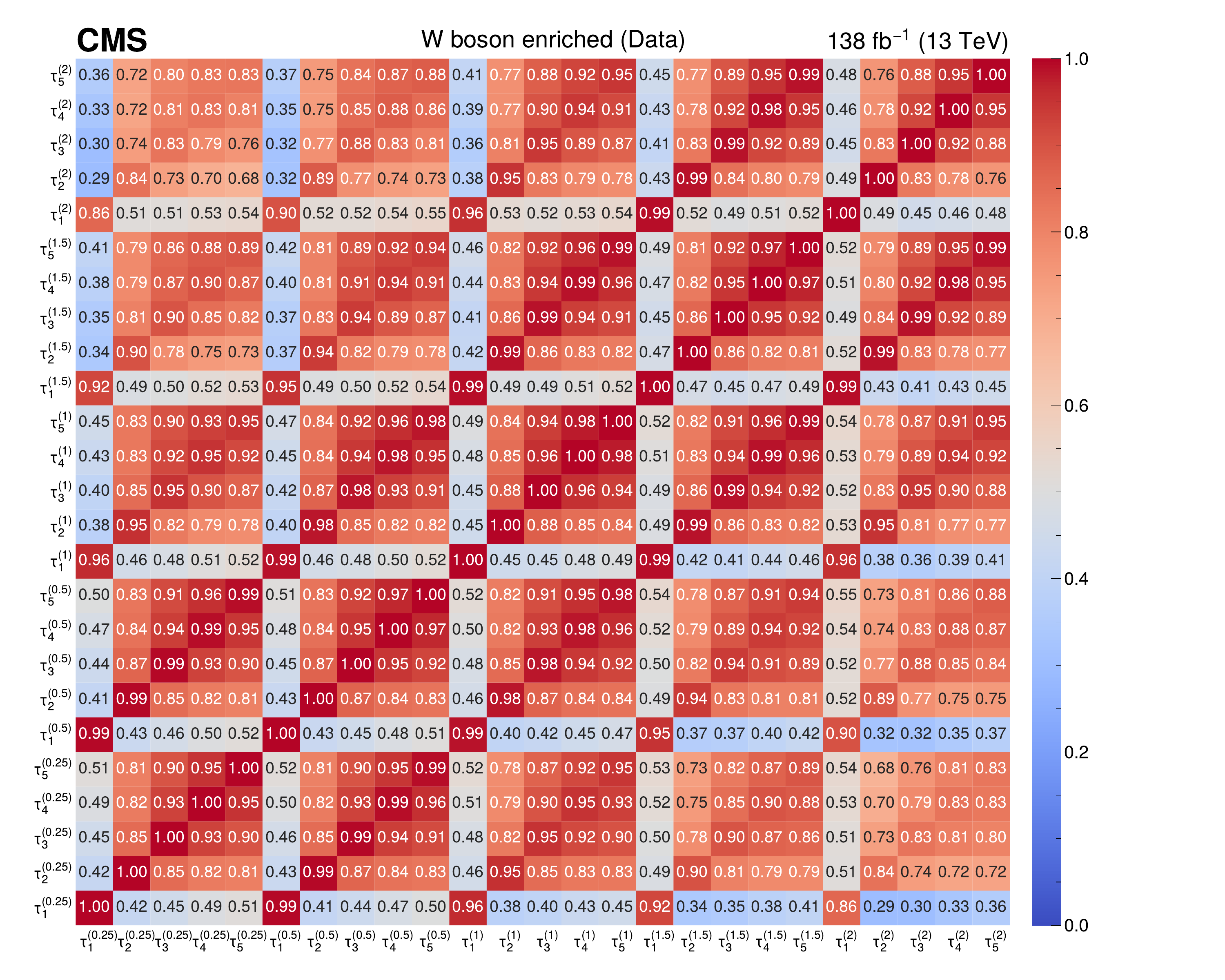}  
	\caption{Pairwise Pearson correlations between $N$-subjettiness observables constituting the overcomplete 6-body basis, using the full Run 2 data set recorded by the CMS detector, in the boosted \PW boson-enriched region. }
	\label{fig:correlationW_data}
\end{figure}

\begin{figure}[!htb]
	\centering	
	\includegraphics[width=1.\textwidth]{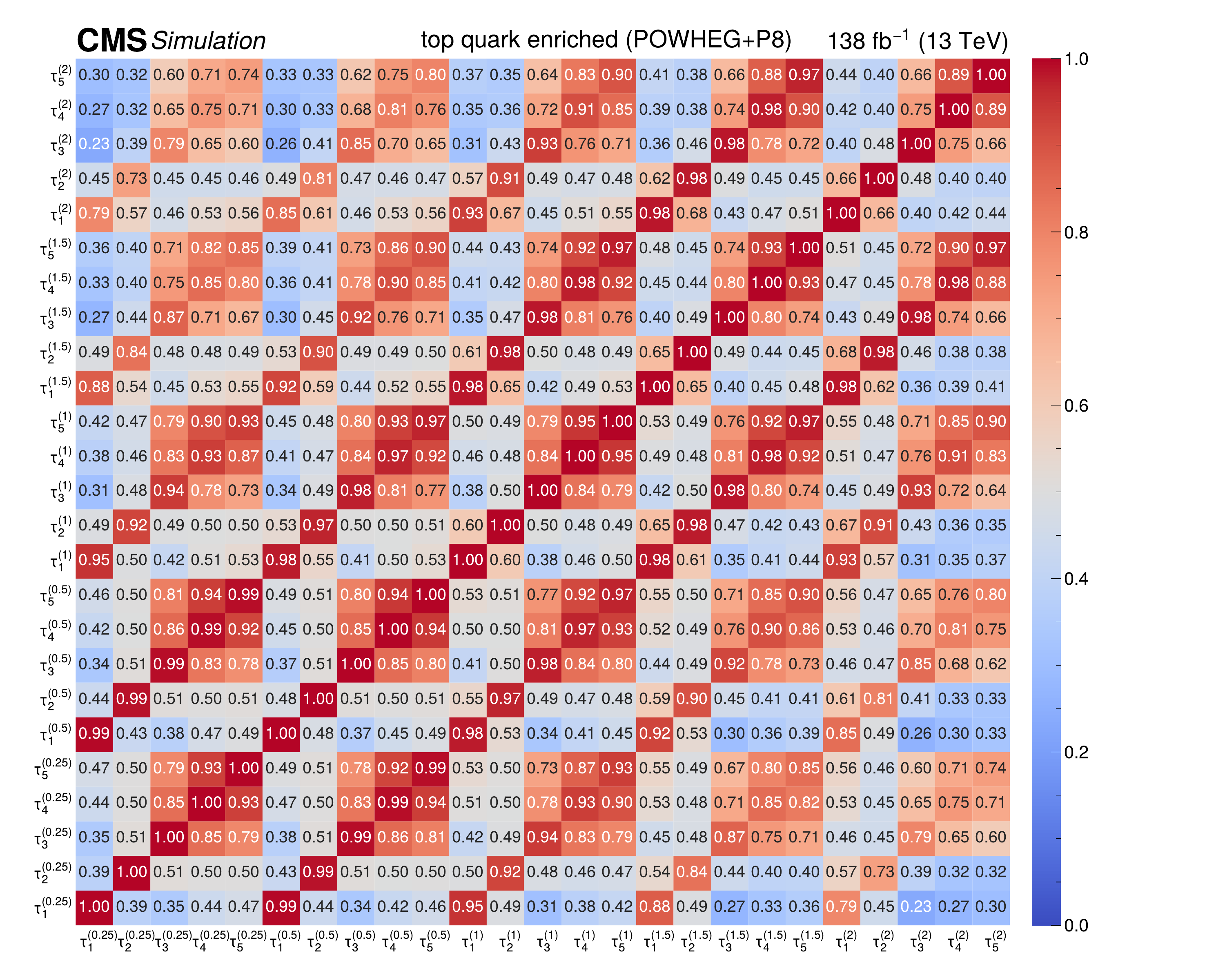}  
	\caption{Pairwise Pearson correlations between $N$-subjettiness observables constituting the overcomplete 6-body basis, in the nominal \POWHEG{}+\PYTHIAviii signal sample, at the particle level, in the boosted top quark-enriched region. All particle-level events with fully-merged jets passing the event selections are considered.}
	\label{fig:correlationtop_nomMC_gen}
\end{figure}

\begin{figure}[!htb]
	\centering	
	\includegraphics[width=1.\textwidth]{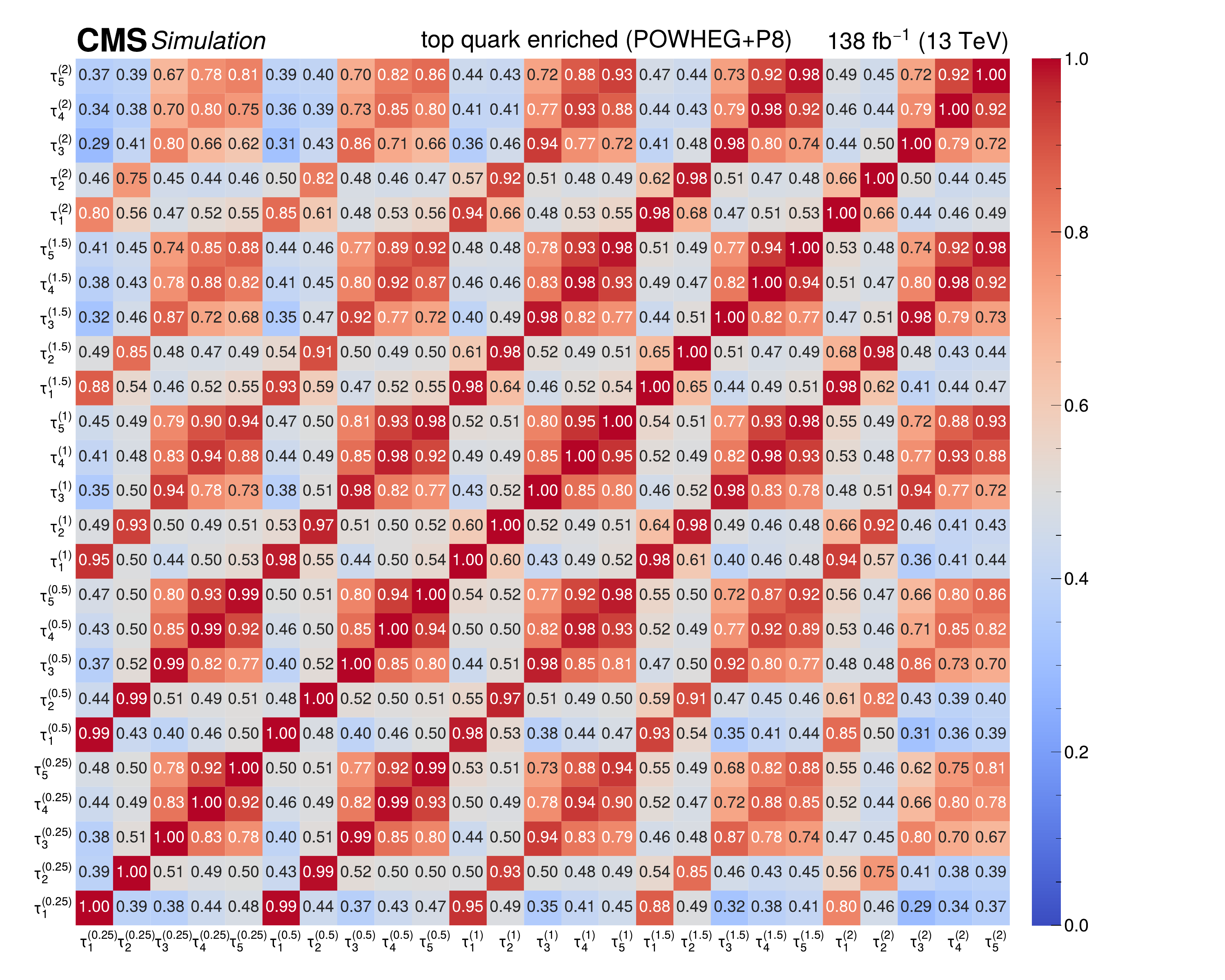}  
	\caption{Pairwise Pearson correlations between $N$-subjettiness observables constituting the overcomplete 6-body basis, in the nominal \POWHEG{}+\PYTHIAviii signal sample, at the detector level, in the boosted top quark-enriched region. Only detector-level events with a matched jet in the corresponding fully-merged particle-level event are considered. }
	\label{fig:correlationtop_nomMC_truereco}
\end{figure}

\begin{figure}[!htb]
	\centering	
	\includegraphics[width=1.\textwidth]{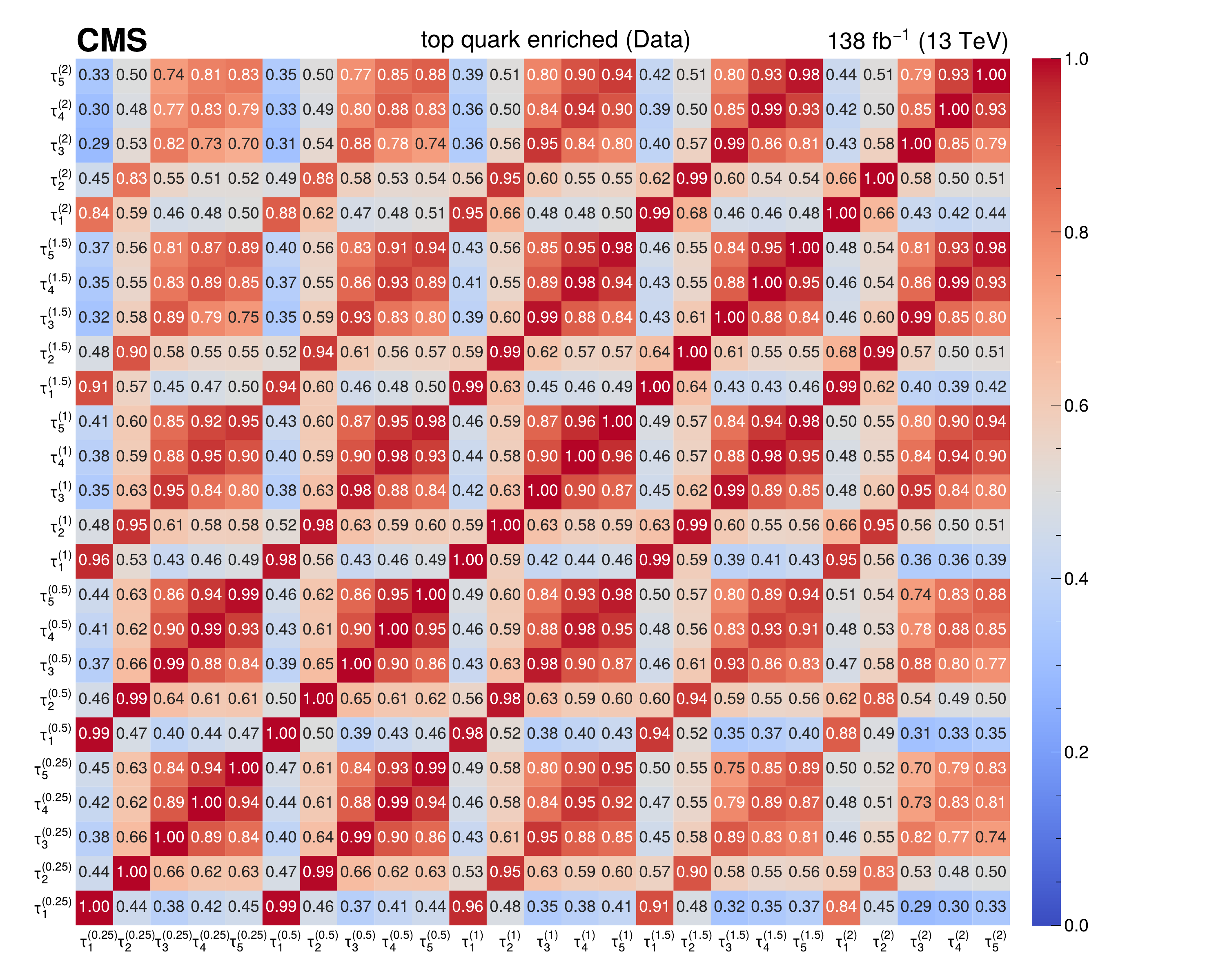}  
	\caption{Pairwise Pearson correlations between $N$-subjettiness observables constituting the overcomplete 6-body basis, using the full Run 2 data set recorded by the CMS detector, in the boosted top quark-enriched region. }
	\label{fig:correlationtop_data}
\end{figure}

The pairwise correlations between the observables are presented for the dijet selection, in simulated samples and data, in Figs.~\ref{fig:correlationDijet_nomMC_gen}--\ref{fig:correlationDijet_data}, and for the boosted \PW boson- and top quark-enriched regions in \ttbar events in Figs.~\ref{fig:correlationW_nomMC_gen}--\ref{fig:correlationtop_data}.

Generally, correlations between most pairs of observables in simulation are nominally higher in the measured data and the simulated detector-level events, in comparison with the simulated particle-level results, although they exhibit similar trends. 
This is qualitatively similar to the conclusions from results shown in Figs.~\ref{fig:saturationtop} and~\ref{fig:saturationW}, which demonstrate that discrimination power saturates relatively quickly at the detector level, and can be understood as a result of detector effects and inefficiencies in object reconstruction. 
In other words, there is less uncorrelated information in a jet that can be resolved at the detector level in comparison to the idealized particle-level picture.

For all event selections, for a specified value of $\beta=0.25,0.5,1,1.5,2$, $(N-1)$-subjettiness observables are increasingly correlated with $N$-subjettiness observables, with increasing value of $N$. 
This can be understood as follows. For $\beta>1$, $\tau_N^{(\beta)}$ only emphasizes contributions from emissions at wider angles to the $k$-th subjet with $k$ in the interval $[1, N]$, while for $\beta<1$ collinear contributions are more important. 
Larger values of $N$ allow for the resolution of information from a higher $M$-body phase space.
In a kinematic regime where QCD emissions are approximately scale invariant, additional emissions beyond the hardest prongs of the jets carry information that is increasingly more correlated with the primary emissions in the jets.

In particular, it is found that, for boosted \PW boson and top quark jets, the $N$-subjettiness observables with $N\geq3$ and $N\geq4$, respectively, are strongly correlated with one another, and for gluon and light-flavour quark jets in the QCD dijet selection this is found to be true for $N\geq2$. 
This trend in the correlations, albeit that individual/pairs of $N$-subjettiness observables only capture a small amount of the information in $(N+1)$-body kinematic phase space, illustrates what is also observed in the ROC curves shown in Figs.~\ref{fig:saturationW} and~\ref{fig:saturationtop}. 
Namely, resolving arbitrarily more structure (emissions) in the hadronic decay topologies of boosted massive particles, beyond the $M$-body phase space at which discrimination power saturates, does not provide meaningful gains in terms of classification performance.

Thus, extending the set of measured observables to be unfolded, beyond the point at which discrimination power effectively saturates, and resolving 6-body kinematic phase space is a conservative approach to mapping out the substructure of jets originating from boosted \PW boson and top quark decays. 
This ensures sensitivity to the limited amount of uncorrelated information in a jet that can be experimentally resolved beyond 4 (5)-body phase space, to effectively distinguish between the \PW boson (top quark) jets and QCD jets. 
Similarly, measuring an overcompletely defined $M$-body basis incorporates redundancies in terms of handles on the angular relations between $M$ resolved emissions.

The high correlations between observables sensitive to 4- and 5-body or 5- and 6-body phase spaces indicate that there is only a minimal amount of uncorrelated information in the jets, relevant to boosted \PW boson (Figs.~\ref{fig:correlationW_nomMC_gen}--\ref{fig:correlationW_data}) and top quark (Figs. \ref{fig:correlationtop_nomMC_gen}--\ref{fig:correlationtop_data}) decays that can be captured by measuring further $N$-subjettiness observables for values beyond $N\geq3$ and 4, respectively. 
This is true even if one considers an improved measurement resolution for the observables with larger $N(\geq3)$, as demonstrated by the qualitatively similar trends in the detector- and particle-level pair-wise correlations computed using simulated samples. 

\clearpage
\newpage

\section{Particle-level correlations for simultaneous unfoldings}
\label{sec:unfCombinedCorr}
Correlations between bins of the unfolded, combined distributions are presented below for measurements in the QCD dijet, and boosted \PW boson and top quark-enriched events. The correlations are extracted from the normalized covariance matrix of the unfolded results, which considers contributions from all statistical and systematic sources of uncertainty. The particle-level correlations and the corresponding covariance matrices are provided in the HEPData record for the analysis \cite{hepdata}. 

\subsection{Gluon and light-flavour quark jets}
\label{sec:unfCombinedCorrQCD}

\begin{figure}[htbp!]
	\centering
	\includegraphics[width=0.8\textwidth]{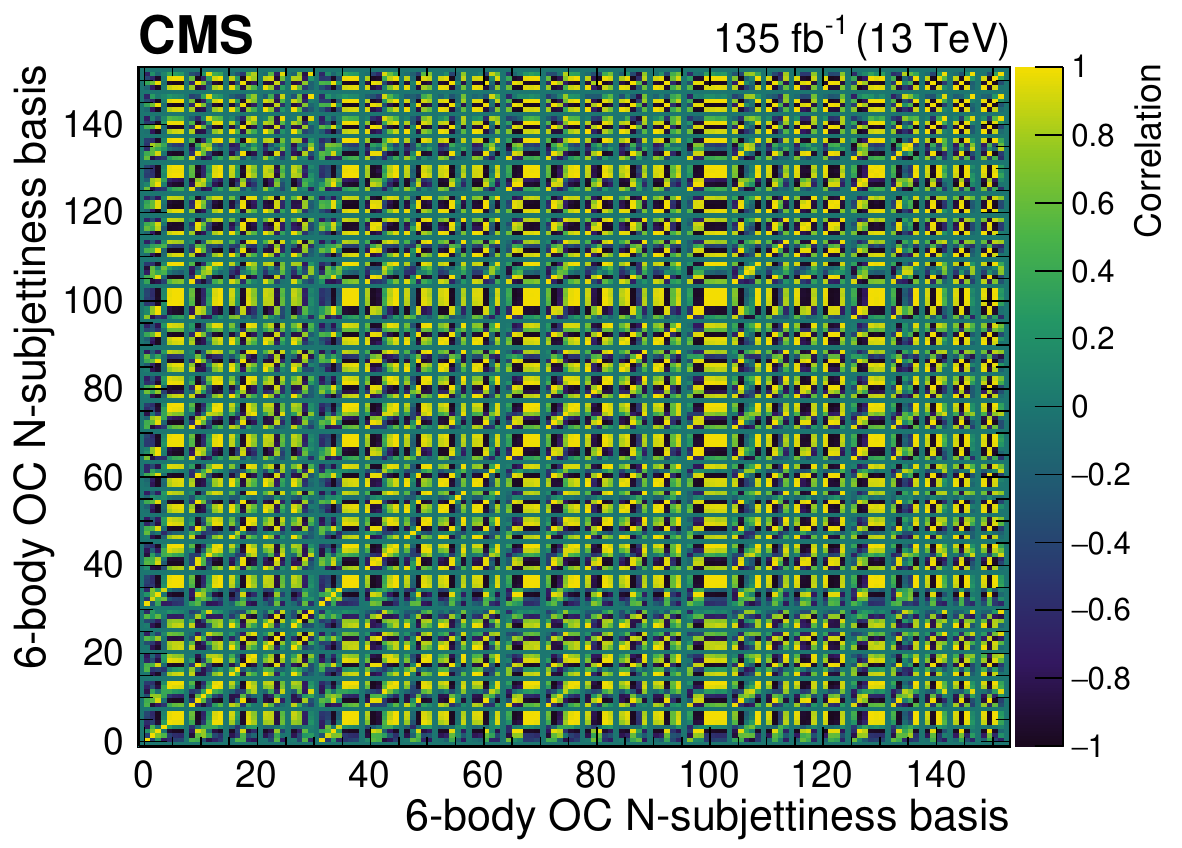}
	\caption{Correlations between bins in the normalized, unfolded data in the QCD dijet selection. The correlations are computed from the total covariance matrix of the normalized, combined unfolded distribution.}
	\label{fig:dijet_unfolded_corr}
\end{figure}
The correlations between the bins for the unfolded results for gluon and light-flavour quark jets are shown in Fig.~\ref{fig:dijet_unfolded_corr}. 
High correlations are observed for different observables sensitive to two or more subjets in the AK8 jet, for all values of the angular weight $\beta$. 
This matches the naive expectations for single-pronged topologies, and results in the pairwise correlation coefficients presented in Appendix~\ref{sec:Correlations}, as a consequence of the approximate scale invariance of QCD radiation at high energies.  

\subsection{Boosted \texorpdfstring{\PW}{W} boson and top quark jets}
\label{sec:unfCombinedCorrWtop}

\begin{figure}[htbp]
	\centering
	\includegraphics[width=0.8\textwidth]{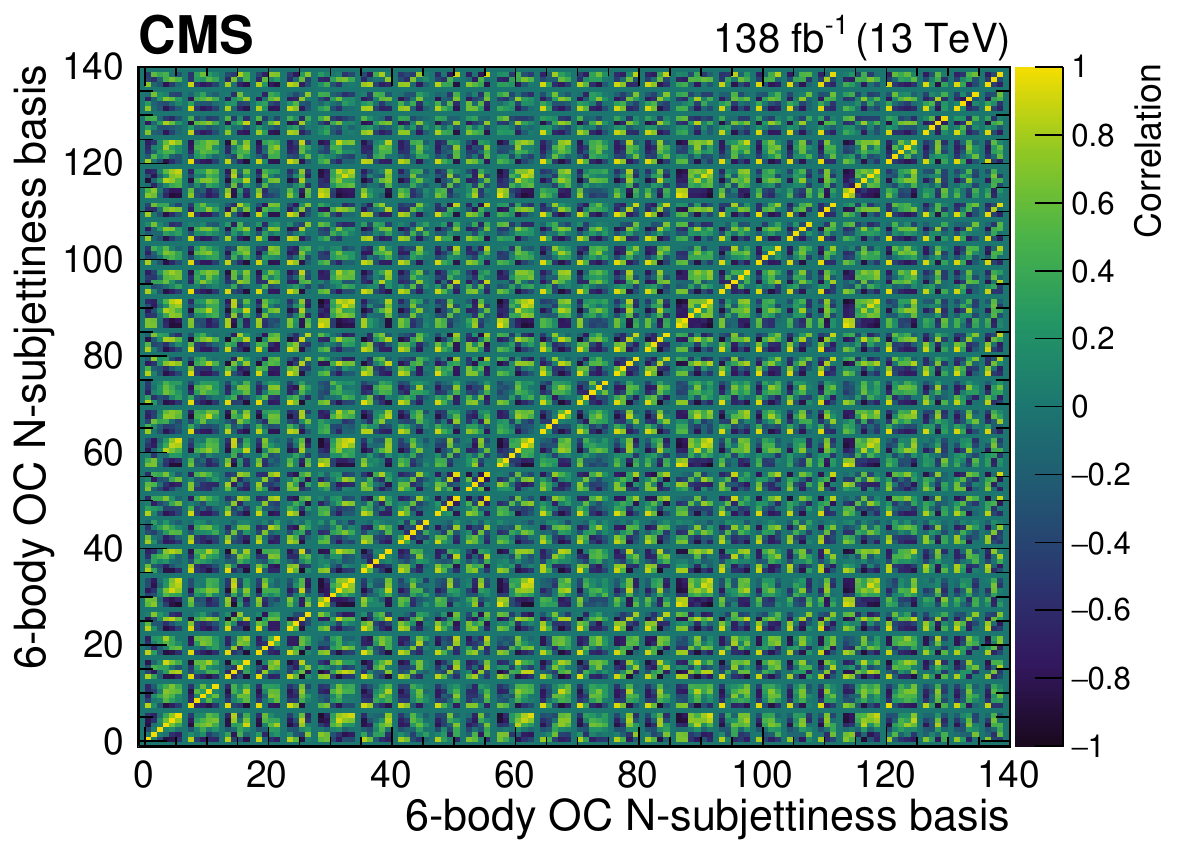}
	\caption{Correlations between the bins of the normalized, unfolded data in the boosted \PW boson-enriched region. The correlations are computed from the total covariance matrix of the normalized, combined unfolded distribution. }
	\label{fig:W_unfolded_corr}
\end{figure}
\begin{figure}[htbp]
	\centering
	\includegraphics[width=0.8\textwidth]{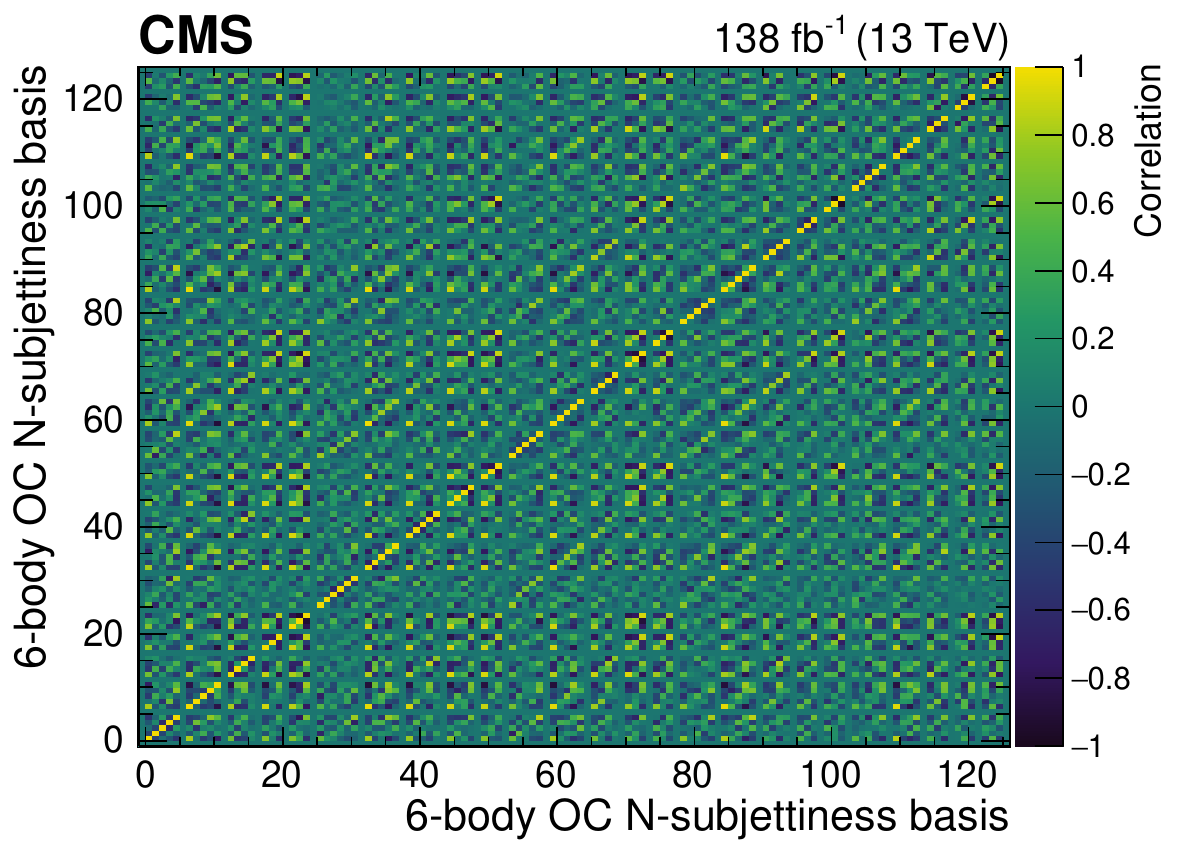}
	\caption{Correlations between the bins of the normalized, unfolded data in the boosted top quark-enriched selection. The correlations are computed from the total covariance matrix of the normalized, combined unfolded distribution.}
	\label{fig:top_unfolded_corr}
\end{figure}

The correlations between the bins of the normalized unfolded results for \PW boson and top quark jets are shown in Figs.~\ref{fig:W_unfolded_corr} and \ref{fig:top_unfolded_corr}, respectively. 
Lower correlations and higher anti-correlations are found between the blocks for specific values of $\beta$ for top quark jets compared to those originating from \PW bosons and single-prong jets. 
In the top quark measurement phase space, where the events contain dominantly three-pronged jets, observables sensitive to 5-/6-body phase space, or more sensitive to contributions at wider angles ($\beta>1$) to the subjet axes, carry relatively more discriminating information with low correlations than in one- or two-prong topologies.

\section{Unfolded results for individual  \texorpdfstring{$N$}{N}-subjettiness observables}
\label{sec:resultsPerObs}

In this section, we present the unfolded distributions of all individual $N$-subjettiness observables in the QCD dijet, and boosted \PW boson- and top quark-enriched regions. 
The distributions of the observables are extracted from the simultaneous unfolding after normalizing the combined distribution and uncertainties on its bins; these results are presented in Sections~\ref{sec:addlmeasurements2body}--\ref{sec:addlmeasurements6body}. The corresponding breakdowns of estimated contributions from various statistical and systematic sources of uncertainty are presented in Section~\ref{sec:dijetUnfUncs} for the dijet selection, and for boosted \PW boson- and top quark-enriched events in Sections~\ref{sec:WUnfUncs} and \ref{sec:topUnfUncs}, respectively.

\subsection{2-body phase space}
\label{sec:addlmeasurements2body}
\begin{figure}[!htb]
	\centering
	\includegraphics[width=.395\textwidth]{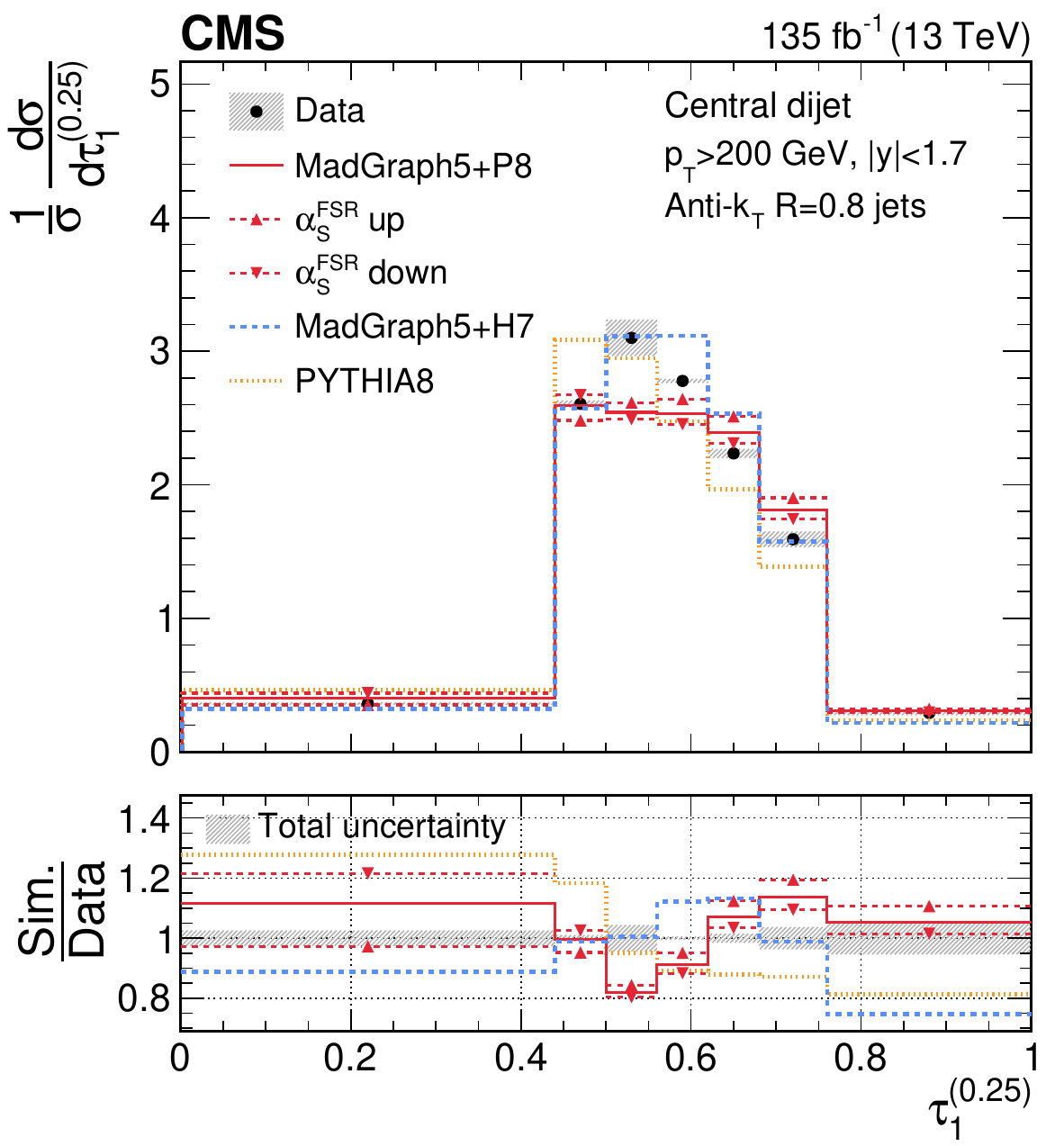} 		
	\includegraphics[width=.395\textwidth]{Figure_013-a.pdf} 
	\includegraphics[width=.395\textwidth]{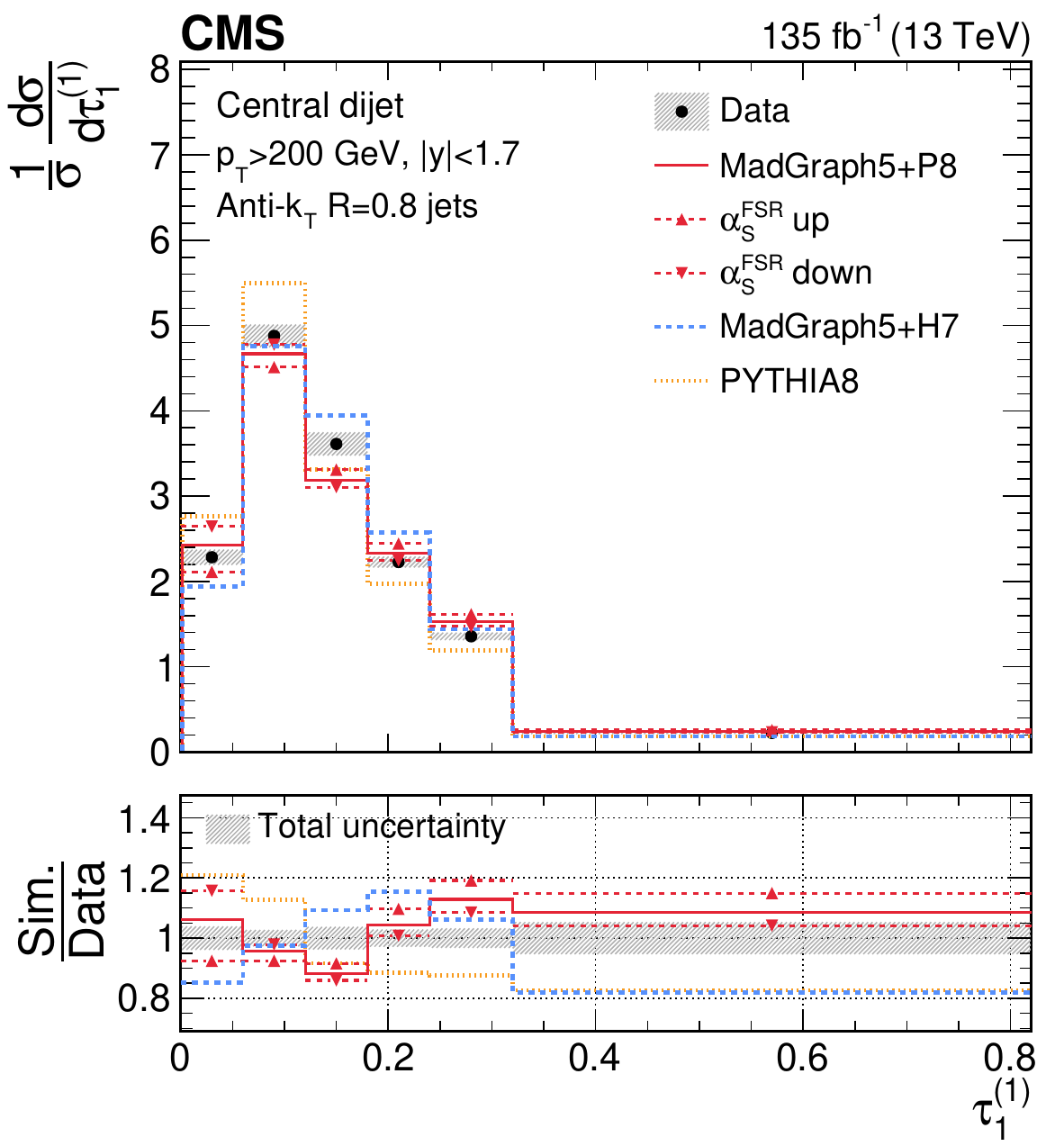} 
	\includegraphics[width=.395\textwidth]{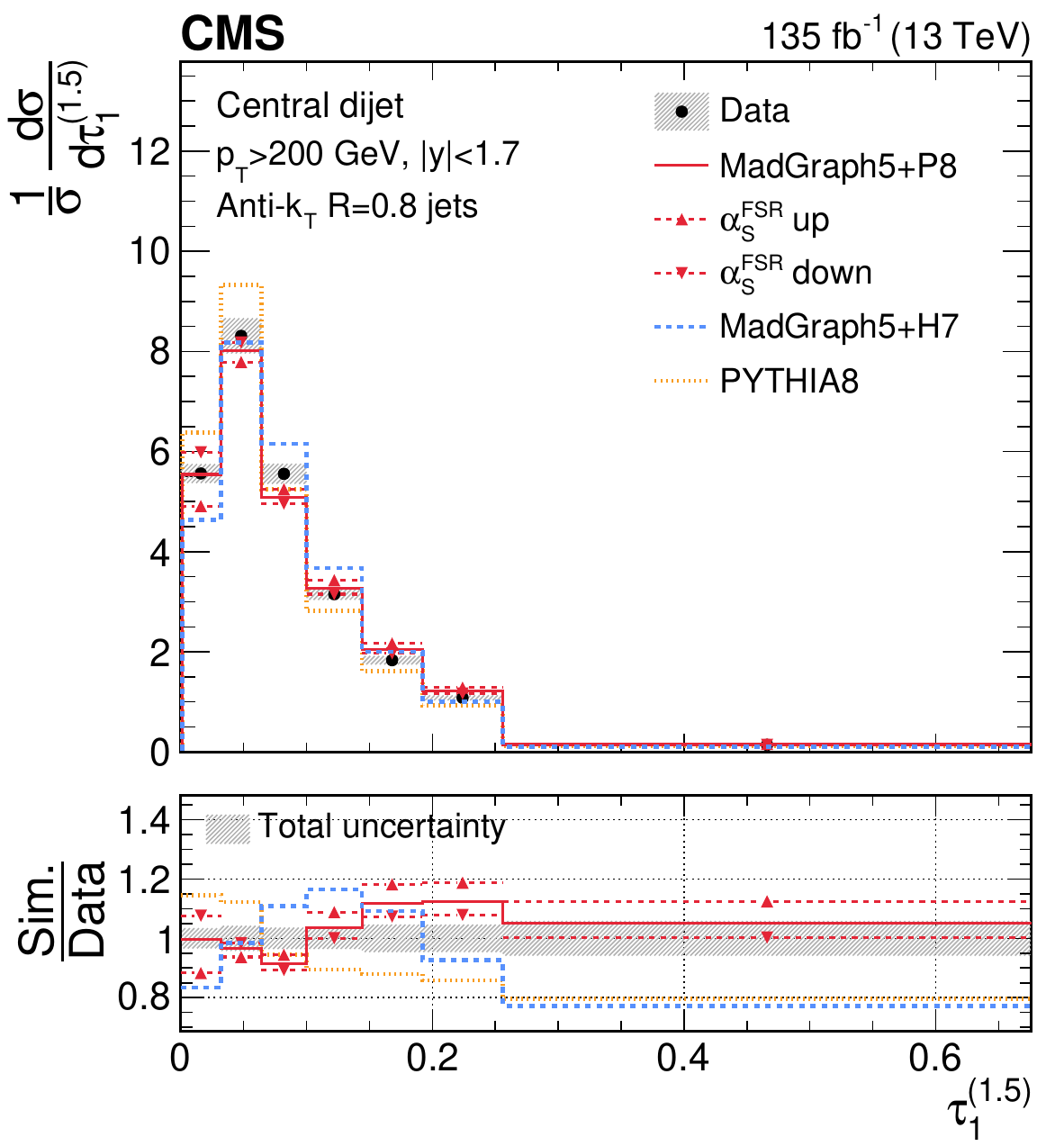} 
	\includegraphics[width=.395\textwidth]{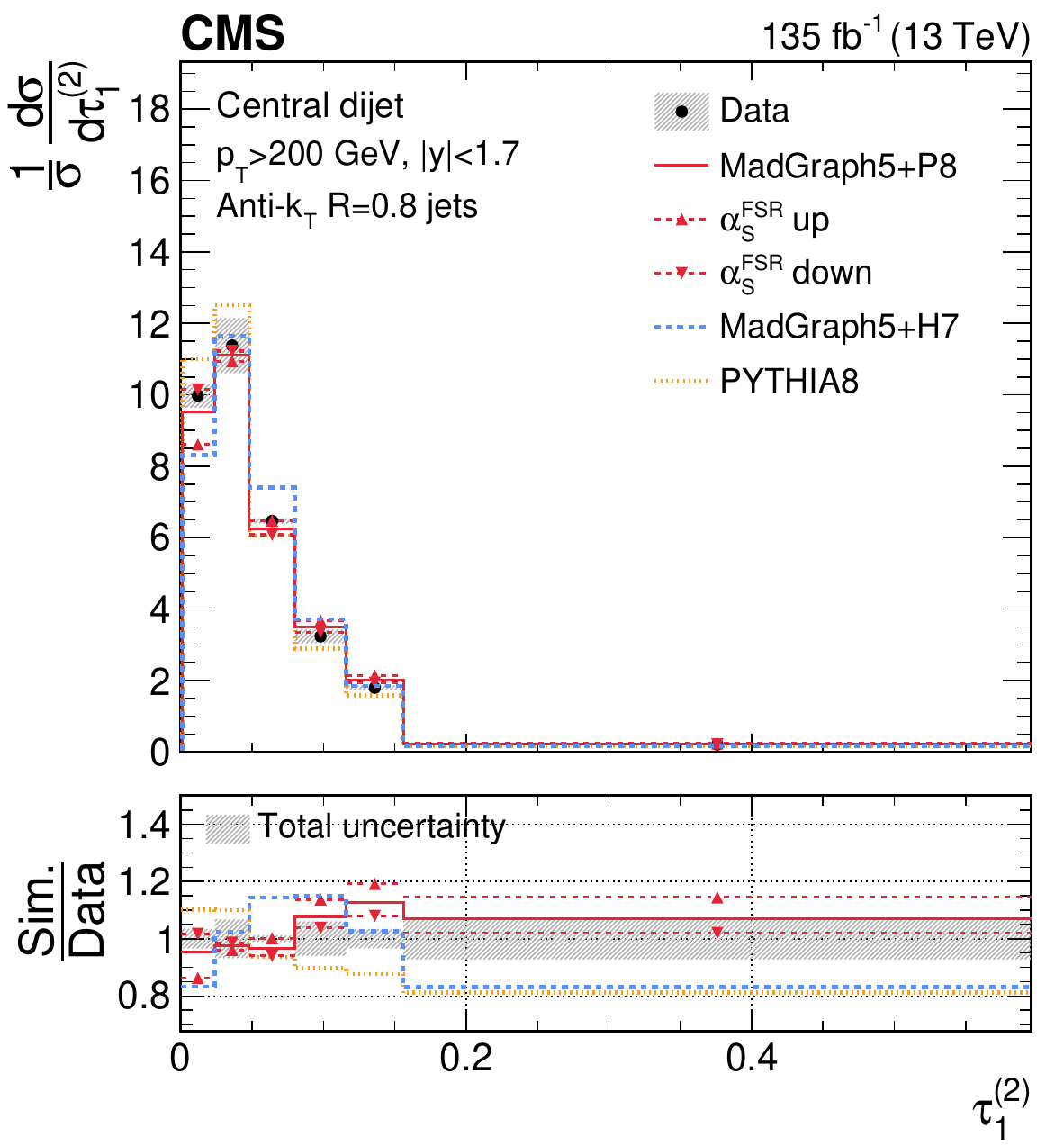} 
	\caption{Unfolded distributions of 1-subjettiness observables, \Nsub{1}{0.25}, \Nsub{1}{0.5}, \Nsub{1}{1}, \Nsub{1}{1.5}, and \Nsub{1}{2}, 
		measured for AK8 jets in the QCD dijet event selection, extracted from the normalized, combined distribution after unfolding; the bin contents and the error bars are scaled by the bin widths for the distributions of the individual observables.  
		For comparisons with particle-level predictions, the error bars in data correspond to the total unfolding uncertainties, 
		and the lower panels present the ratio of particle-level predictions to the unfolded data. 
		The dark grey hashed region illustrates the total uncertainties per bin in the unfolded result.}
	\label{fig:addlresults2bodyDijet}
\end{figure}

\begin{figure}[!htb]
	\centering
	\includegraphics[width=.395\textwidth]{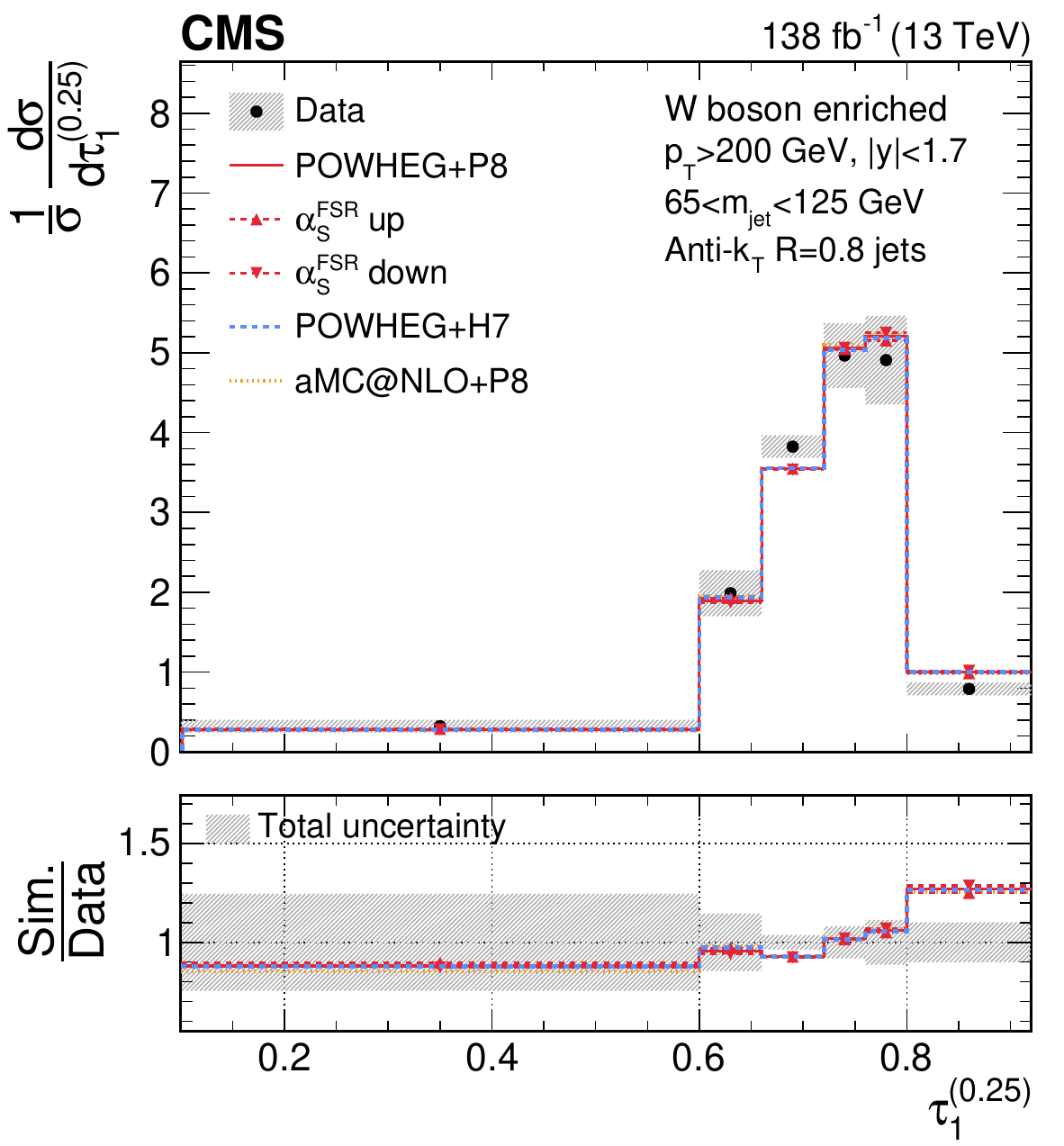} 		
	\includegraphics[width=.395\textwidth]{Figure_014-a.pdf} 
	\includegraphics[width=.395\textwidth]{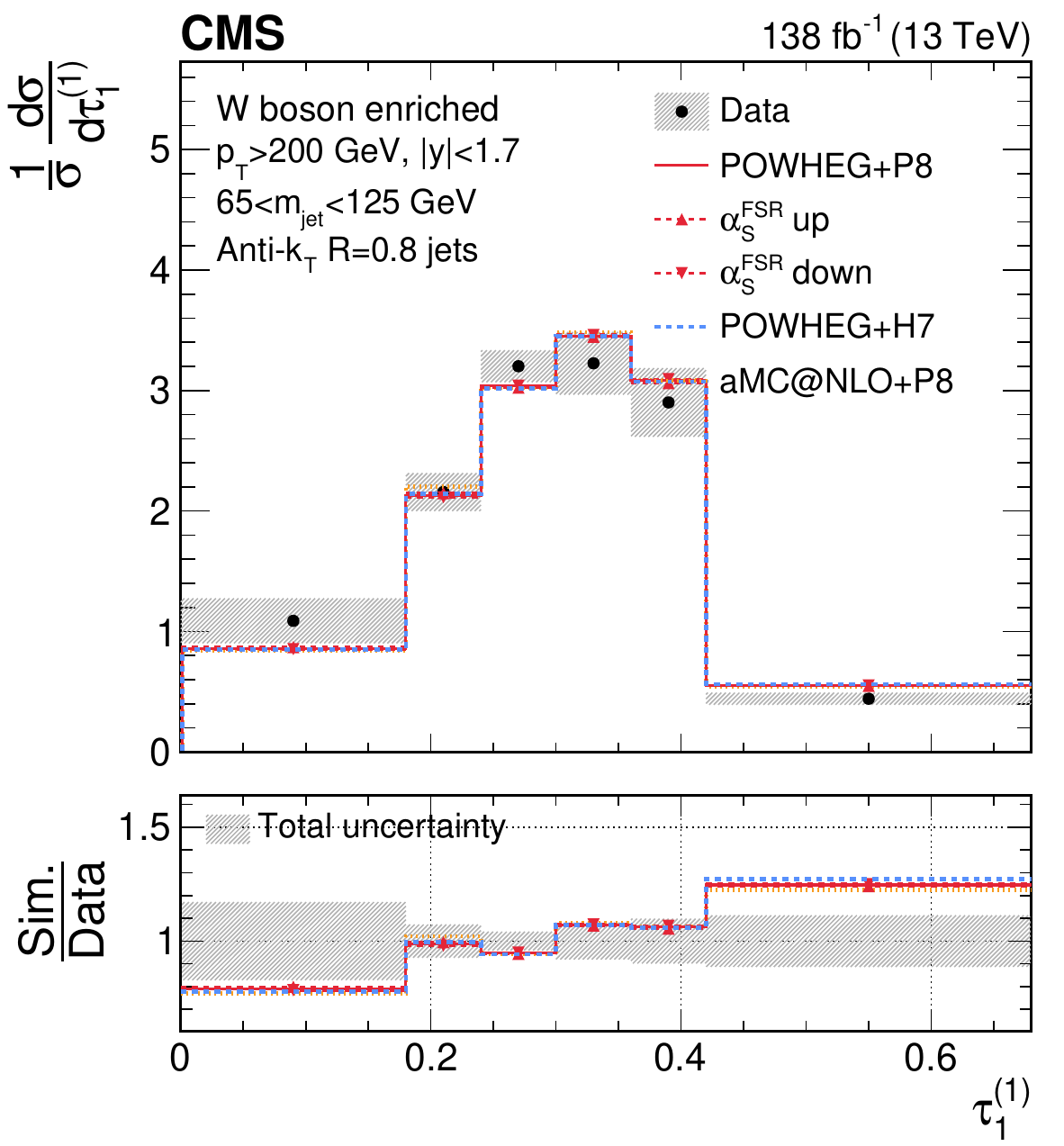} 
	\includegraphics[width=.395\textwidth]{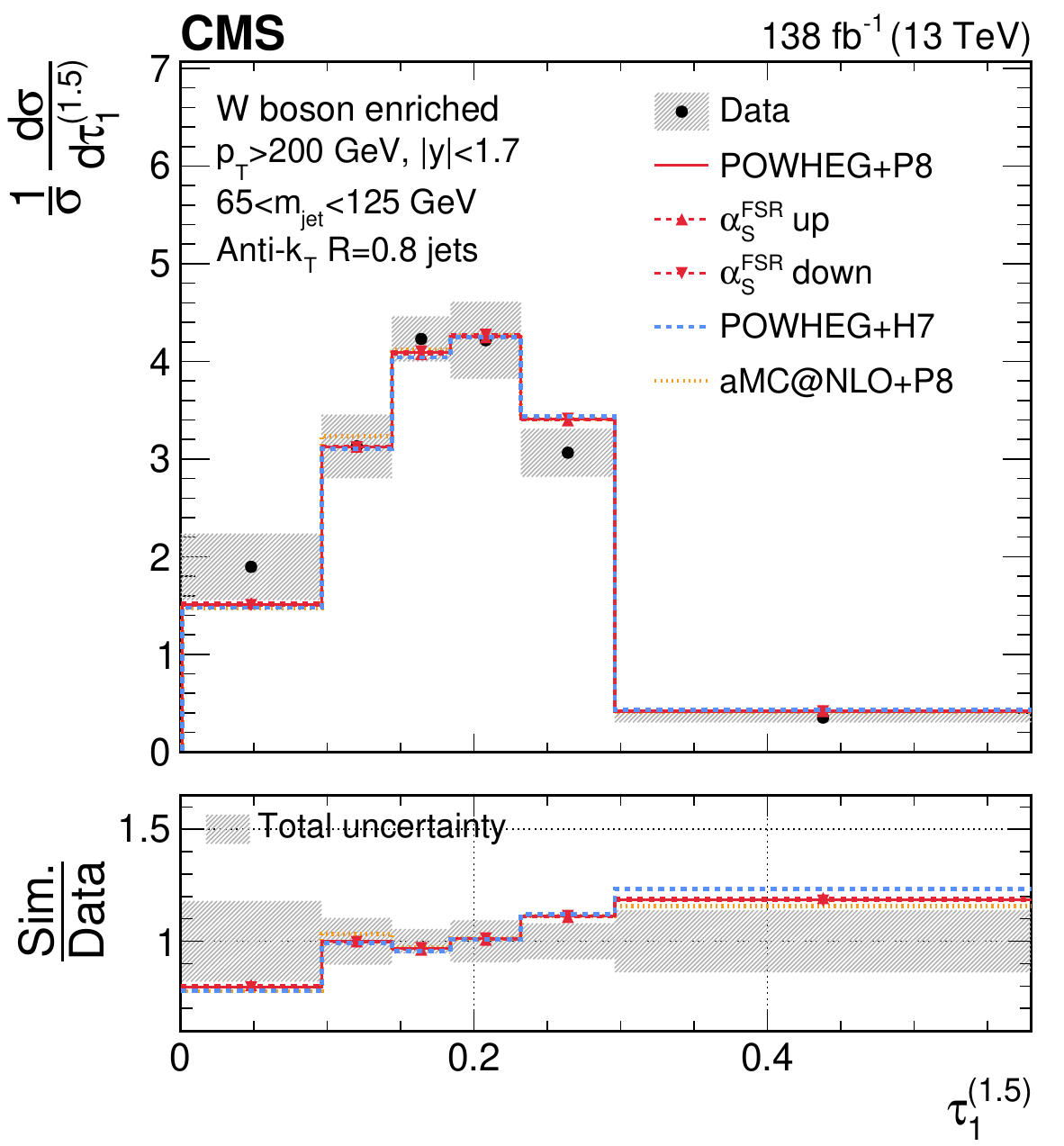} 
	\includegraphics[width=.395\textwidth]{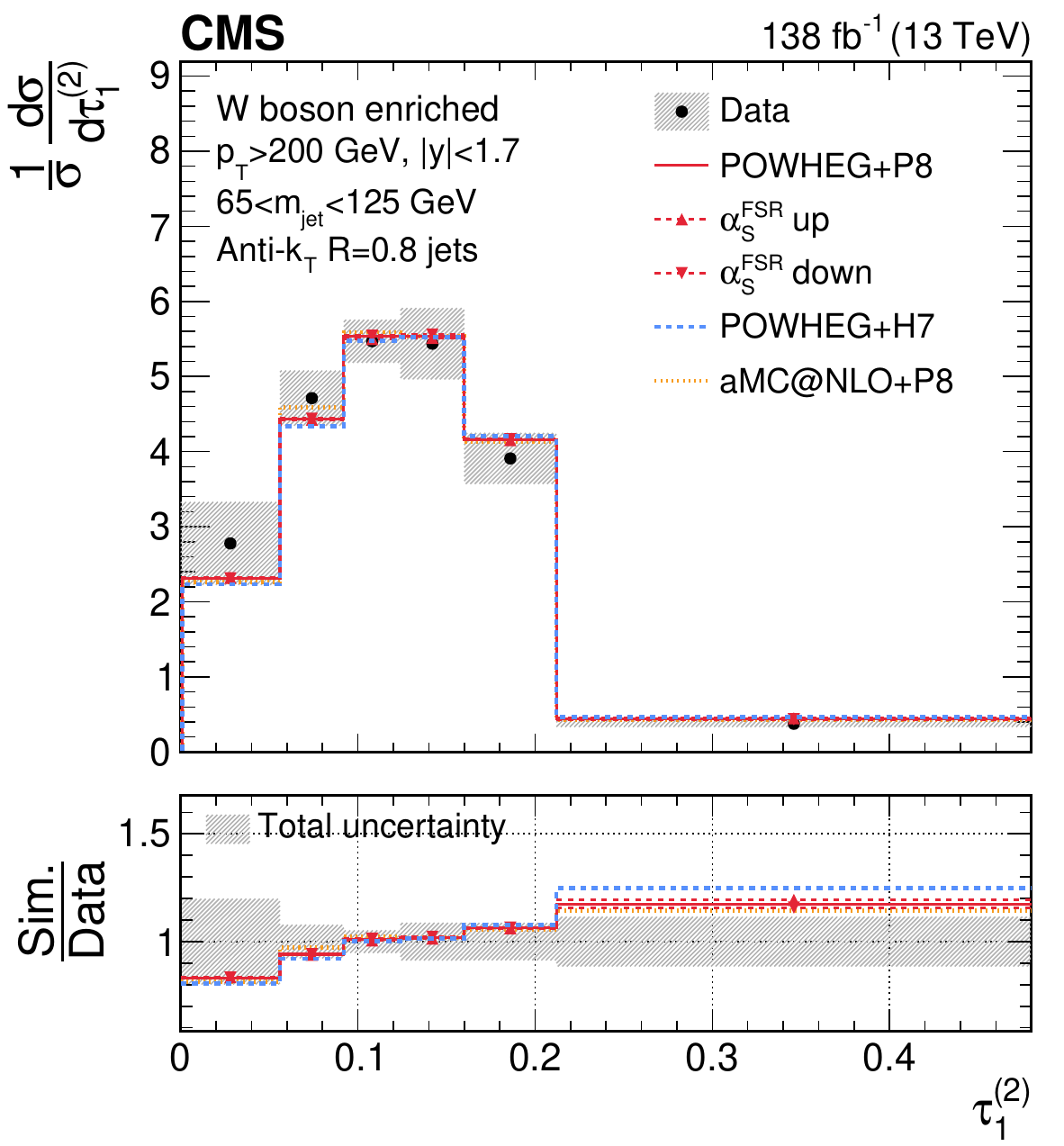} 
	\caption{Unfolded distributions of 1-subjettiness observables, \Nsub{1}{0.25}, \Nsub{1}{0.5}, \Nsub{1}{1}, \Nsub{1}{1.5}, and \Nsub{1}{2}, 
		measured for AK8 jets in boosted \PW boson-enriched events, extracted from the normalized, combined distribution after unfolding; the bin contents and the error bars are scaled by the bin widths for the distributions of the individual observables.  
		For comparisons with particle-level predictions, the error bars in data correspond to the total unfolding uncertainties, 
		and the lower panels present the ratio of particle-level predictions to the unfolded data. 
		The dark grey hashed region illustrates the total uncertainties per bin in the unfolded result.}
	\label{fig:addlresults2bodyW}
\end{figure}

\begin{figure}[!htb]
	\centering
	\includegraphics[width=.395\textwidth]{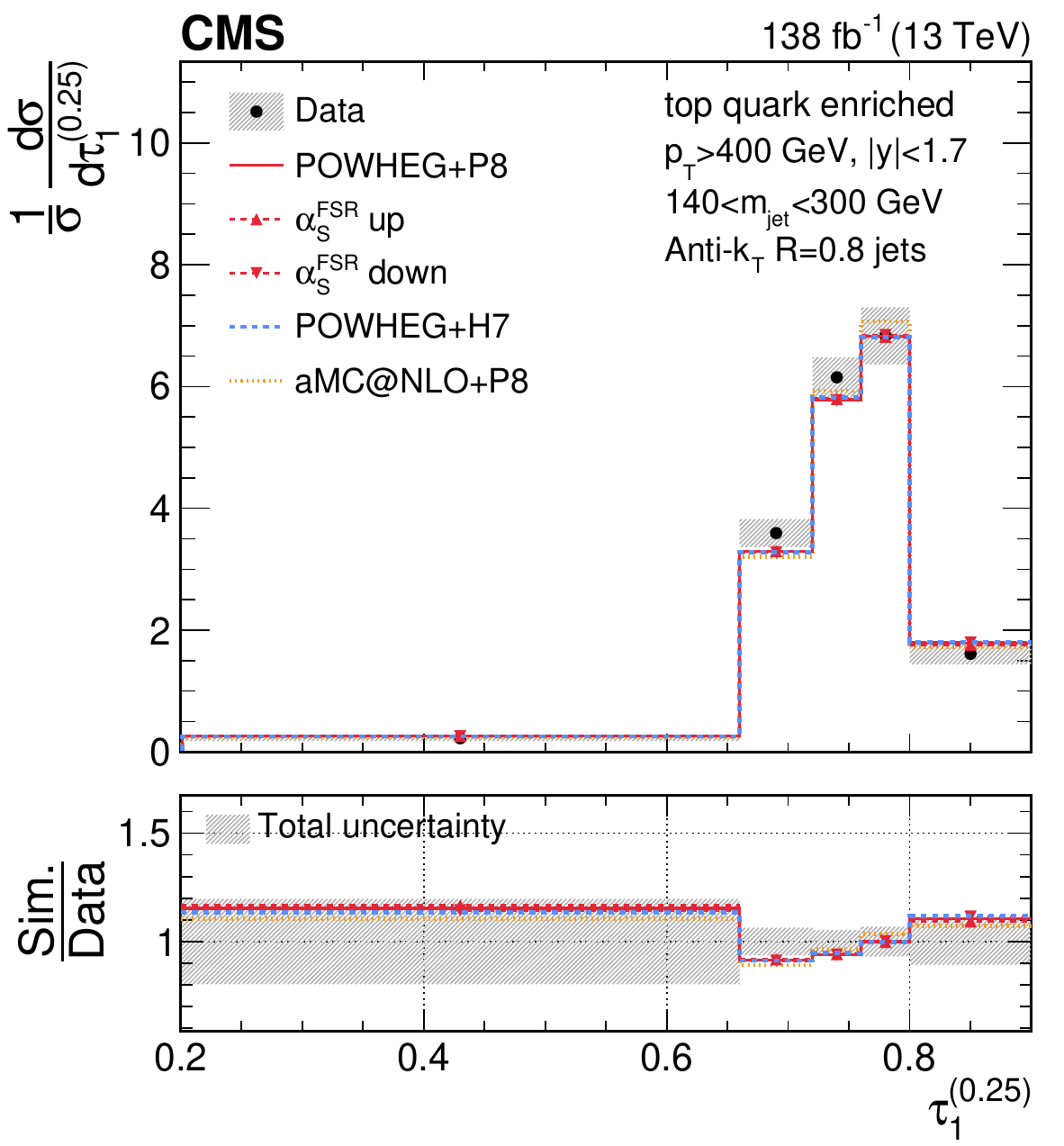} 		
	\includegraphics[width=.395\textwidth]{Figure_015-a.pdf} 
	\includegraphics[width=.395\textwidth]{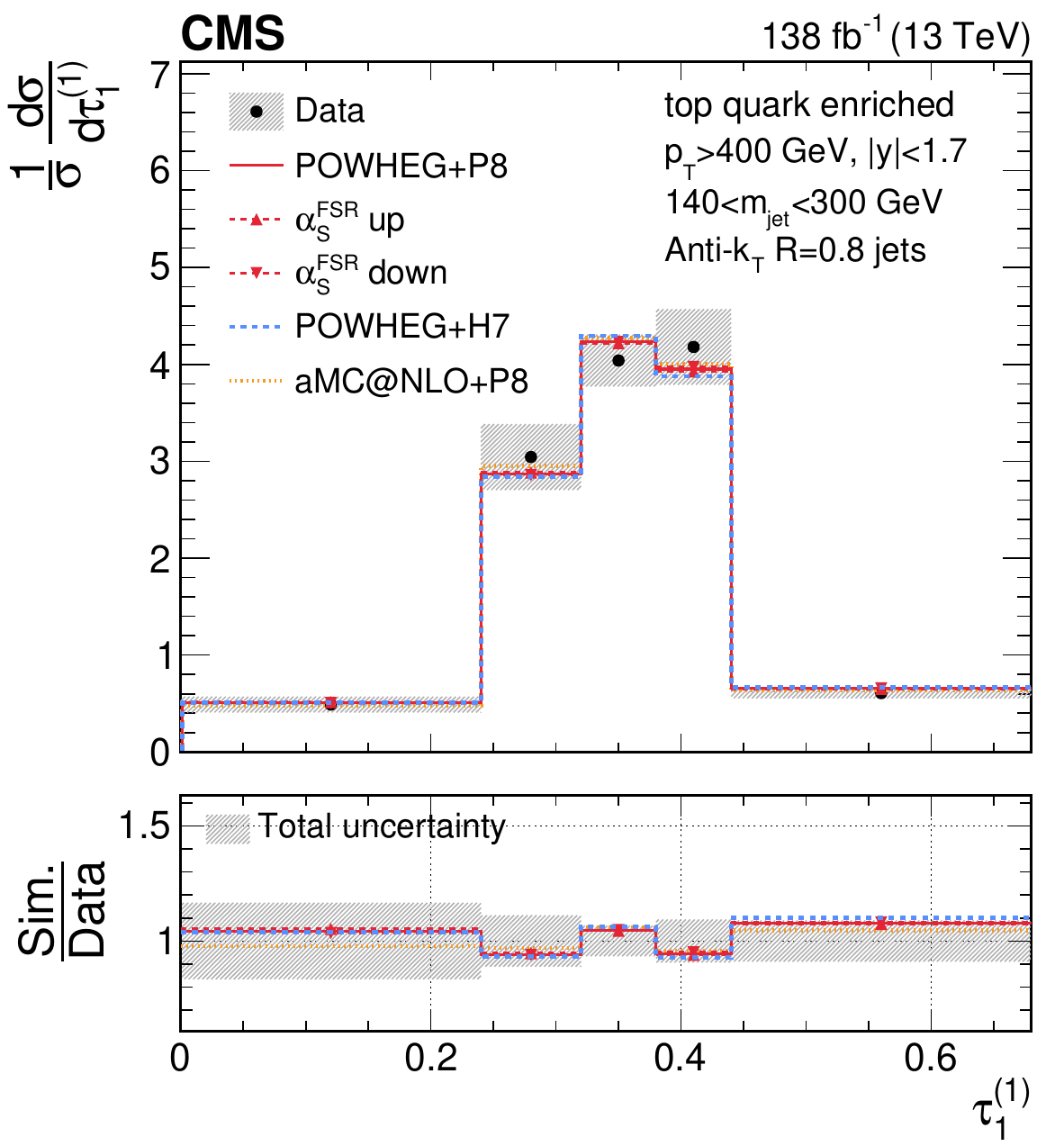} 
	\includegraphics[width=.395\textwidth]{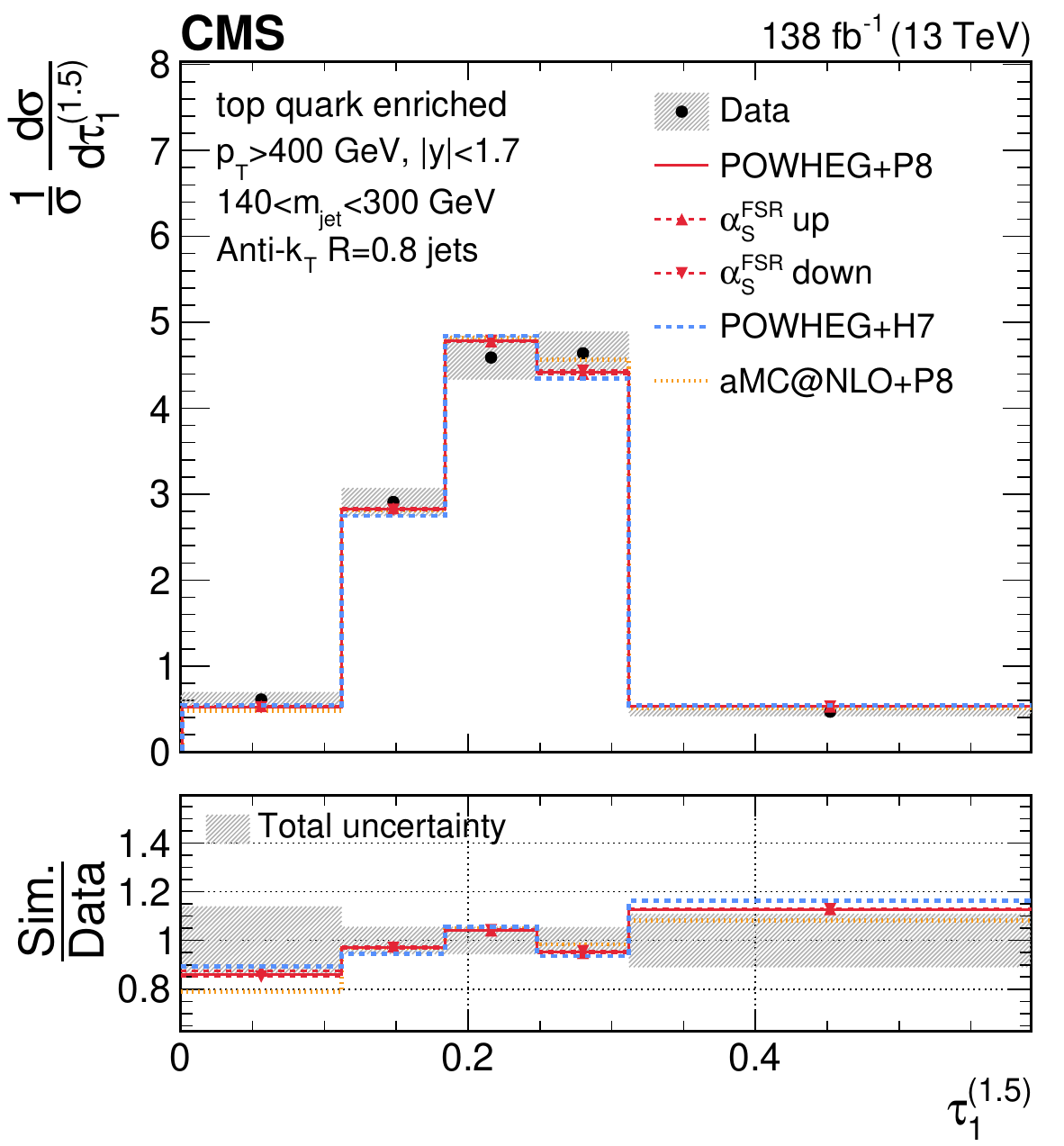} 
	\includegraphics[width=.395\textwidth]{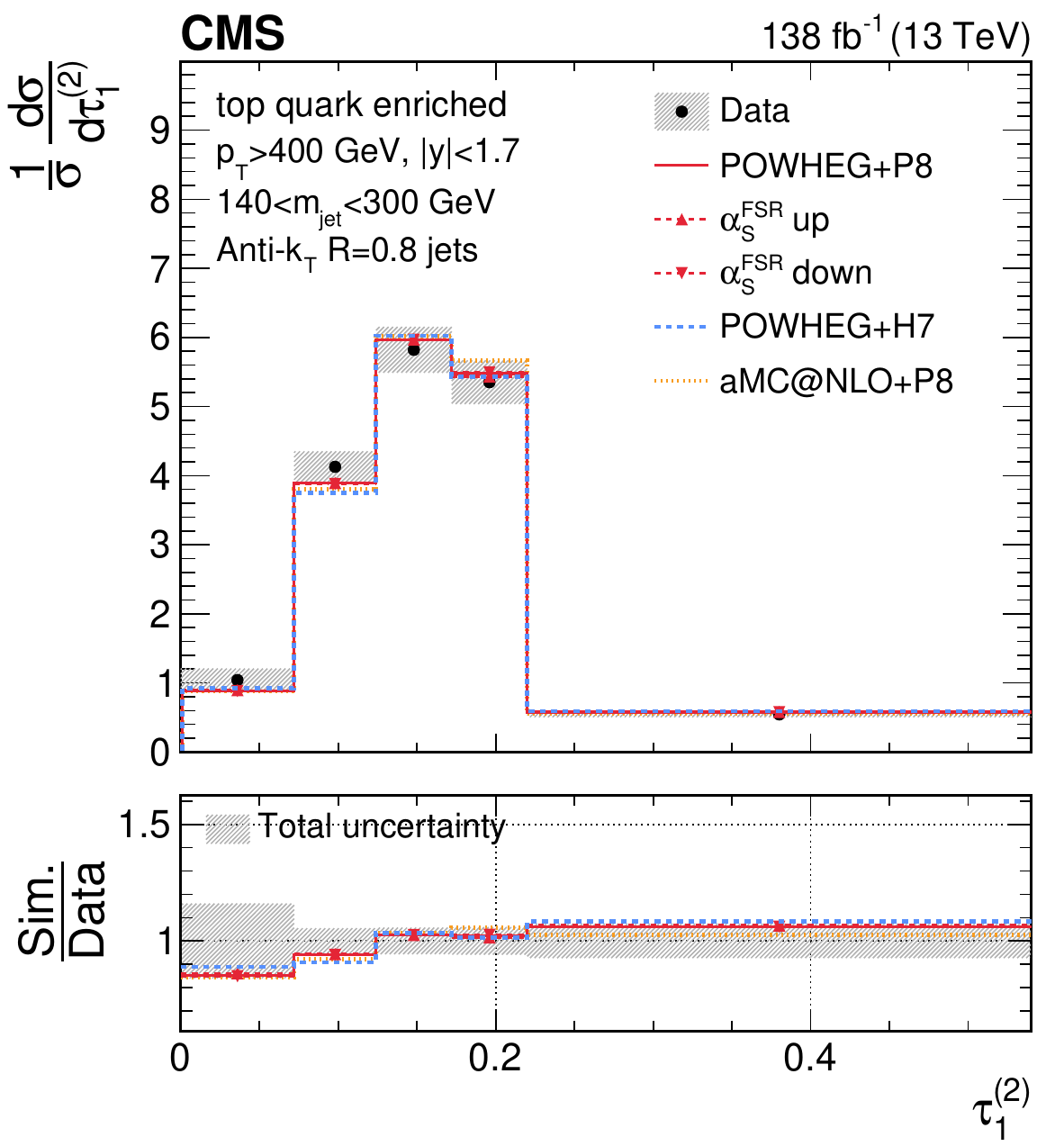} 
	\caption{Unfolded distributions of 1-subjettiness observables, \Nsub{1}{0.25}, \Nsub{1}{0.5}, \Nsub{1}{1}, \Nsub{1}{1.5}, and \Nsub{1}{2}, 
		measured for AK8 jets in the boosted top quark-enriched region, extracted from the normalized, combined distribution after unfolding; the bin contents and the error bars are scaled by the bin widths for the distributions of the individual observables.  
		For comparisons with particle-level predictions, the error bars in data correspond to the total unfolding uncertainties, 
		and the lower panels present the ratio of particle-level predictions to the unfolded data. 
		The dark grey hashed region illustrates the total uncertainties per bin in the unfolded result.}
	\label{fig:addlresults2bodytop}
\end{figure}

The measurements of \Nsub{1}{0.25}, \Nsub{1}{0.5}, \Nsub{1}{1}, \Nsub{1}{1.5}, and \Nsub{1}{2} are presented. 
The unfolded results for the dijet selection are shown in Fig.~\ref{fig:addlresults2bodyDijet}, and results for the boosted \PW boson- and top quark-enriched regions are shown in Figs.~\ref{fig:addlresults2bodyW} and \ref{fig:addlresults2bodytop}, respectively.

\clearpage
\newpage

\subsection{3-body phase space}
\label{sec:addlmeasurements3body}
\begin{figure}[!htb]
	\centering
	\includegraphics[width=.395\textwidth]{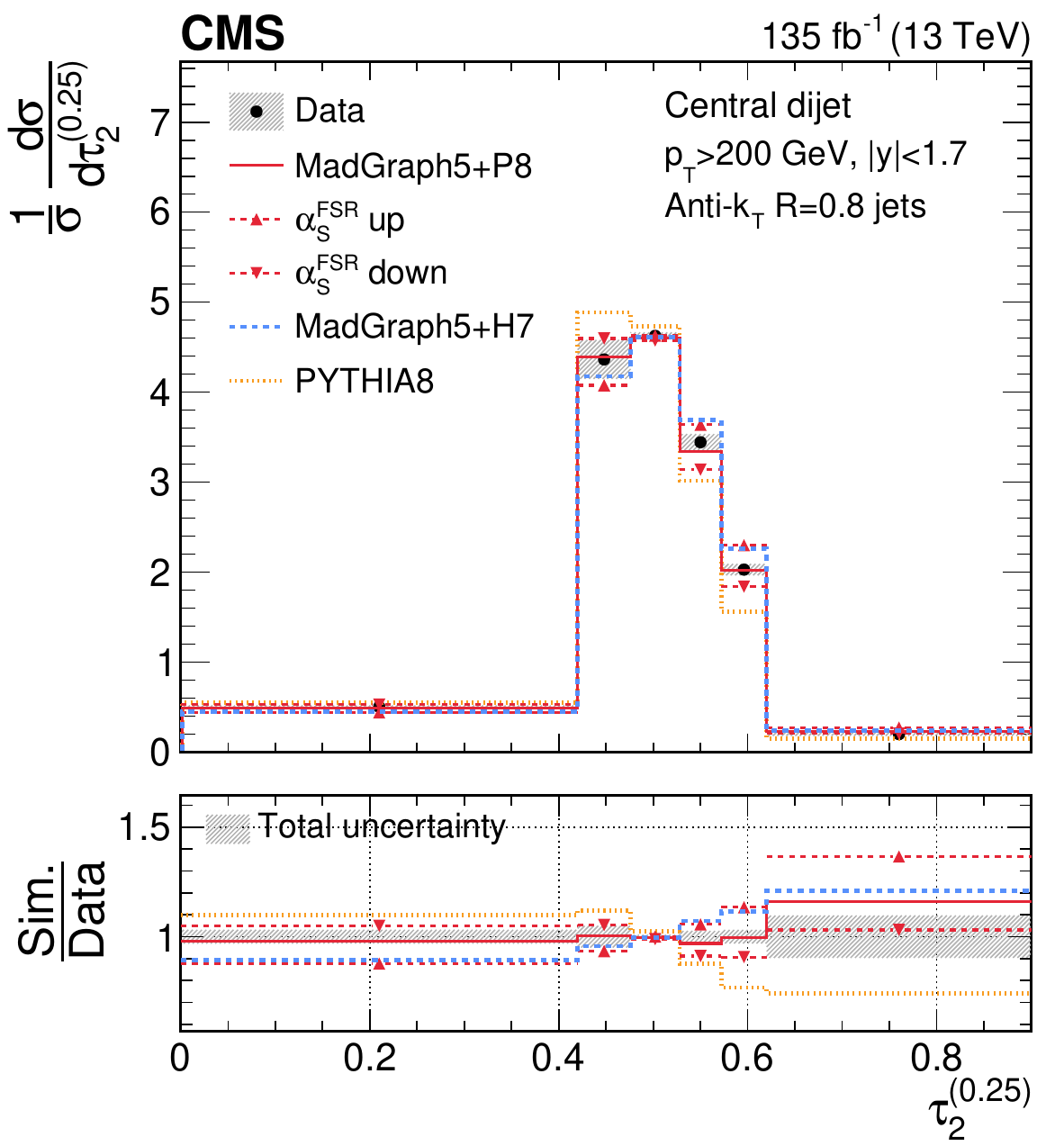} 		
	\includegraphics[width=.395\textwidth]{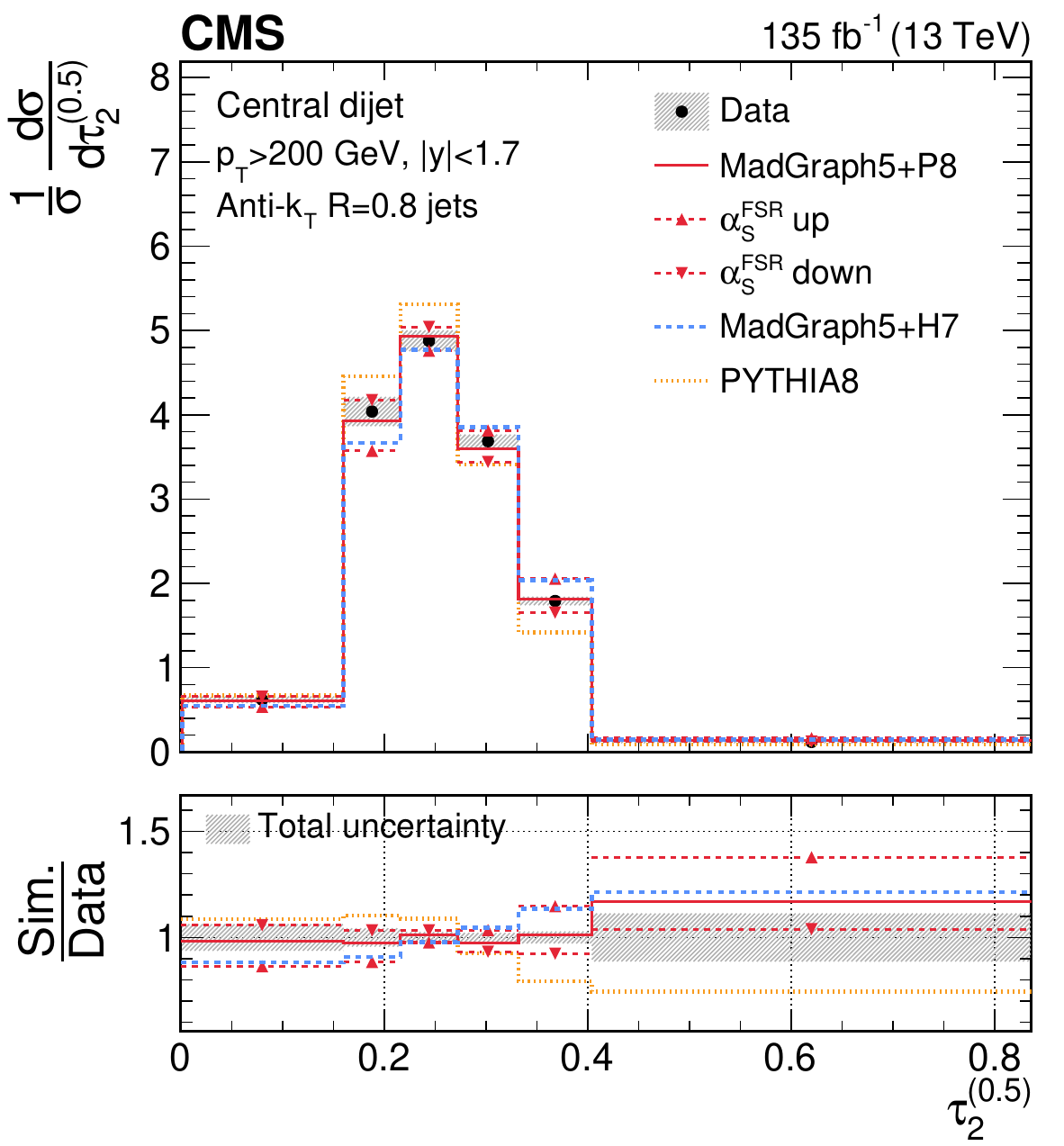} 
	\includegraphics[width=.395\textwidth]{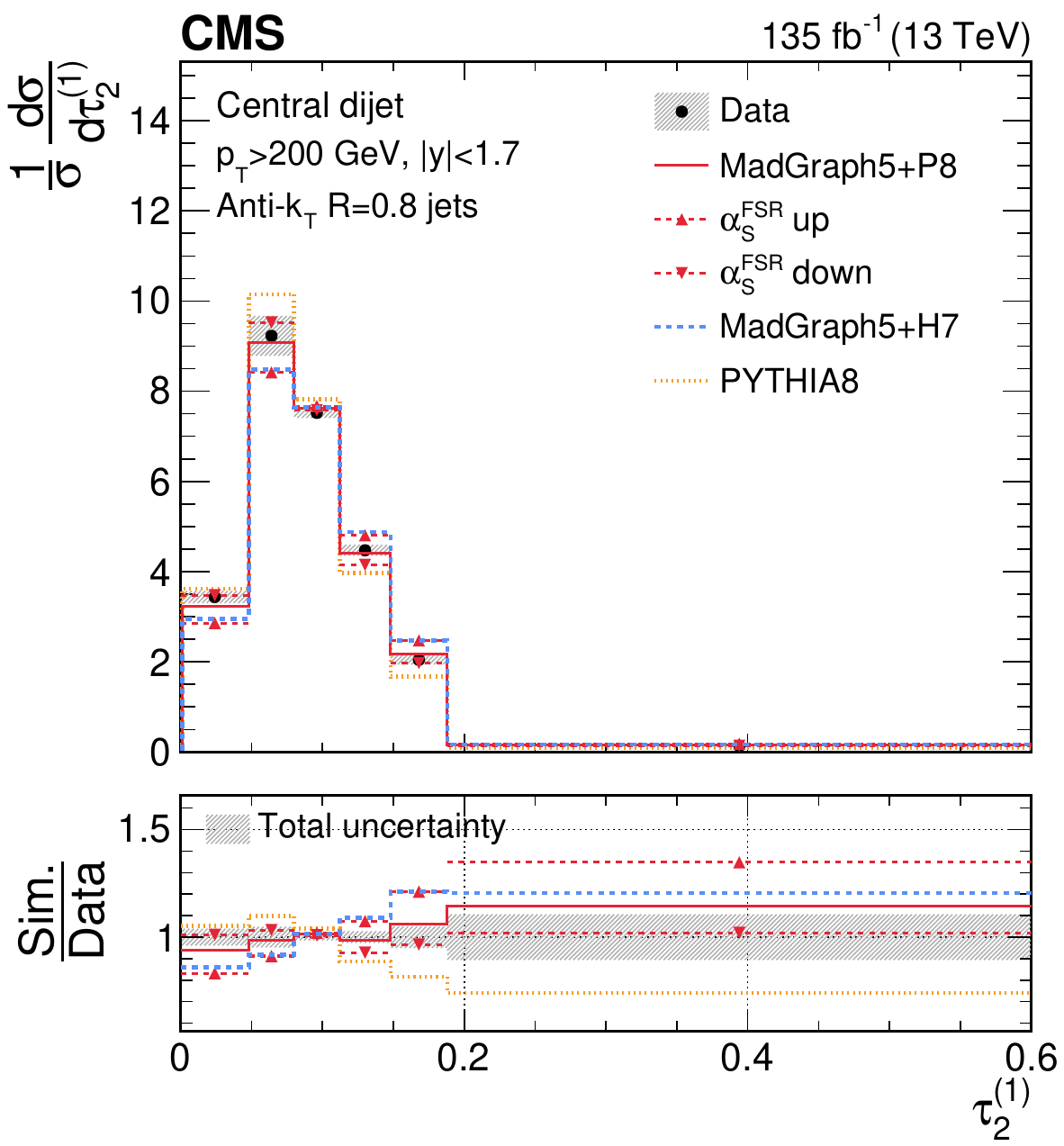} 
	\includegraphics[width=.395\textwidth]{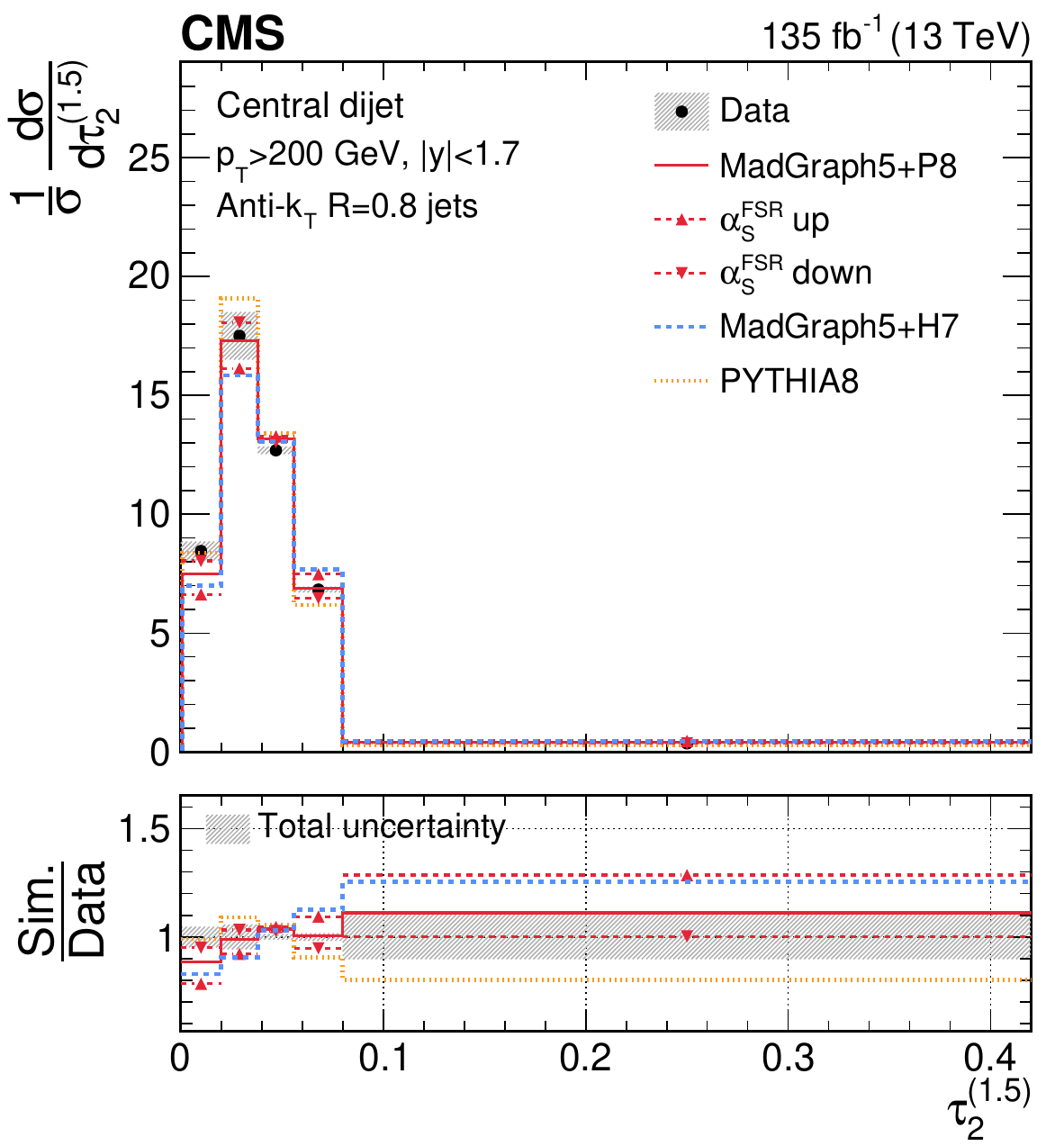} 
	\includegraphics[width=.395\textwidth]{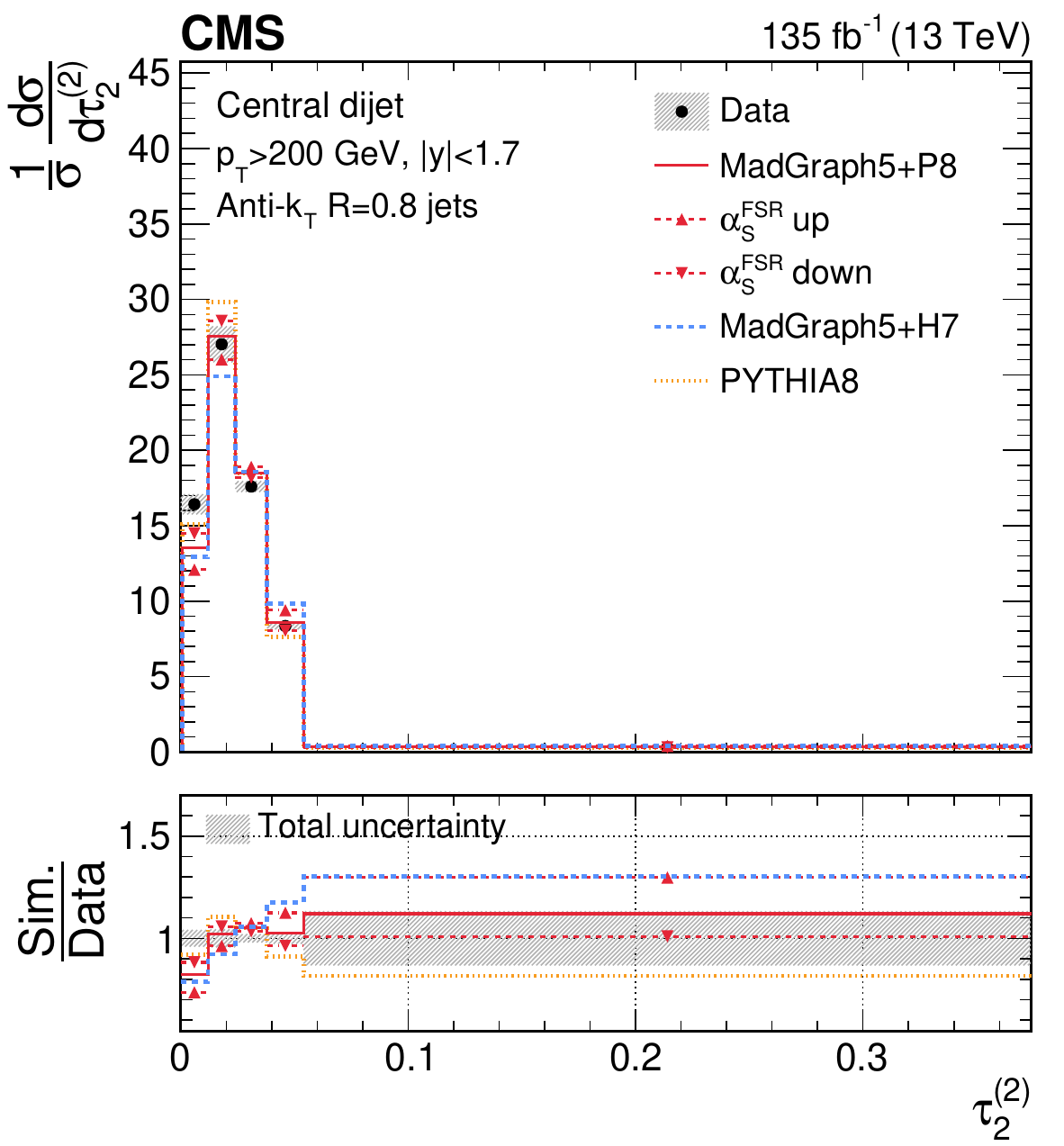} 
	\caption{Unfolded distributions of 2-subjettiness observables, \Nsub{2}{0.25}, \Nsub{2}{0.5}, \Nsub{2}{1}, \Nsub{2}{1.5}, and \Nsub{2}{2}, 
		measured for AK8 jets in the QCD dijet event selection, extracted from the normalized, combined distribution after unfolding; the bin contents and the error bars are scaled by the bin widths for the distributions of the individual observables.  
		For comparisons with particle-level predictions, the error bars in data correspond to the total unfolding uncertainties, 
		and the lower panels present the ratio of particle-level predictions to the unfolded data. 
		The dark grey hashed region illustrates the total uncertainties per bin in the unfolded result.}
	\label{fig:addlresults3bodyDijet}
\end{figure}

\begin{figure}[!htb]
	\centering
	\includegraphics[width=.395\textwidth]{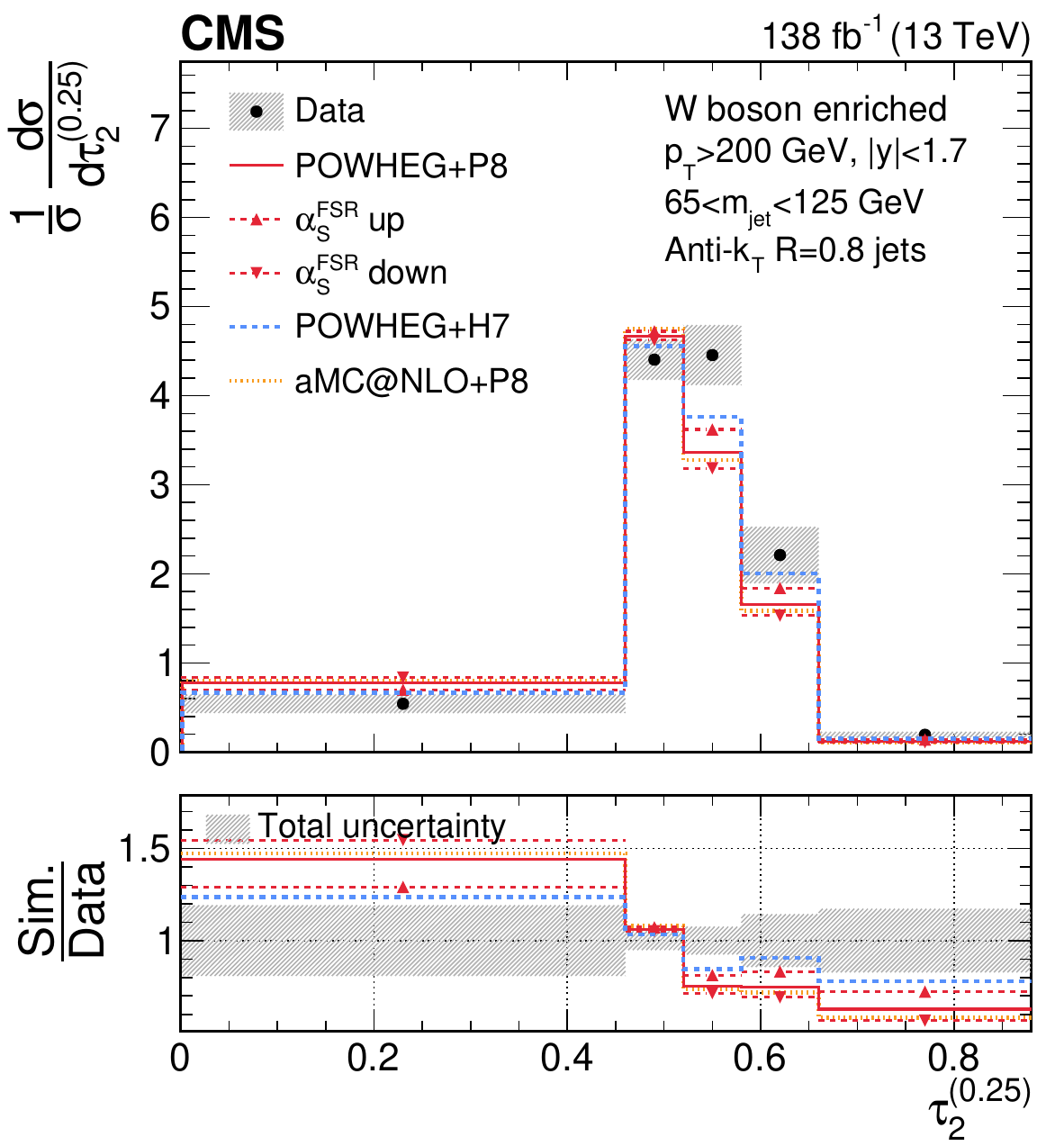} 		
	\includegraphics[width=.395\textwidth]{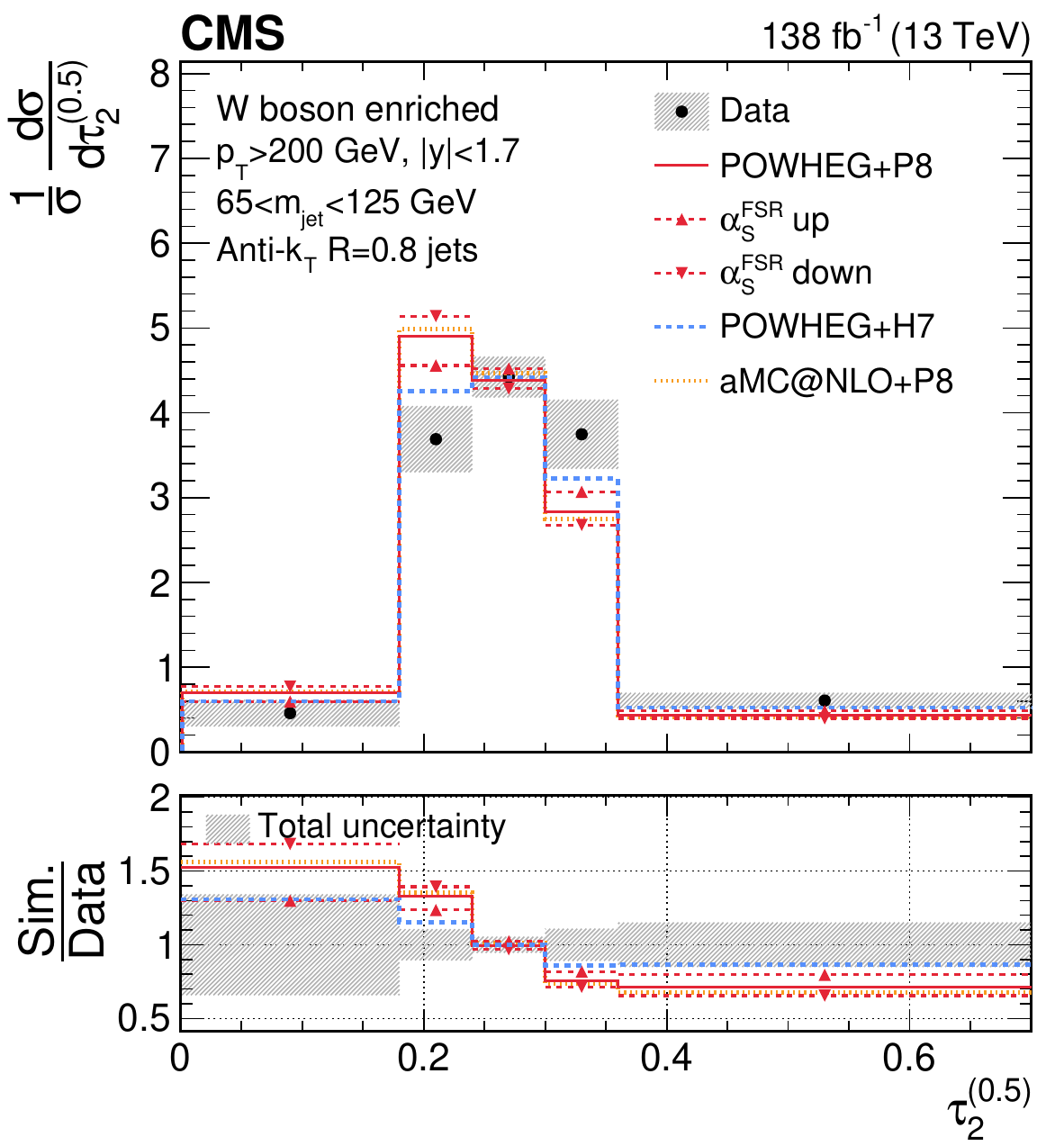} 
	\includegraphics[width=.395\textwidth]{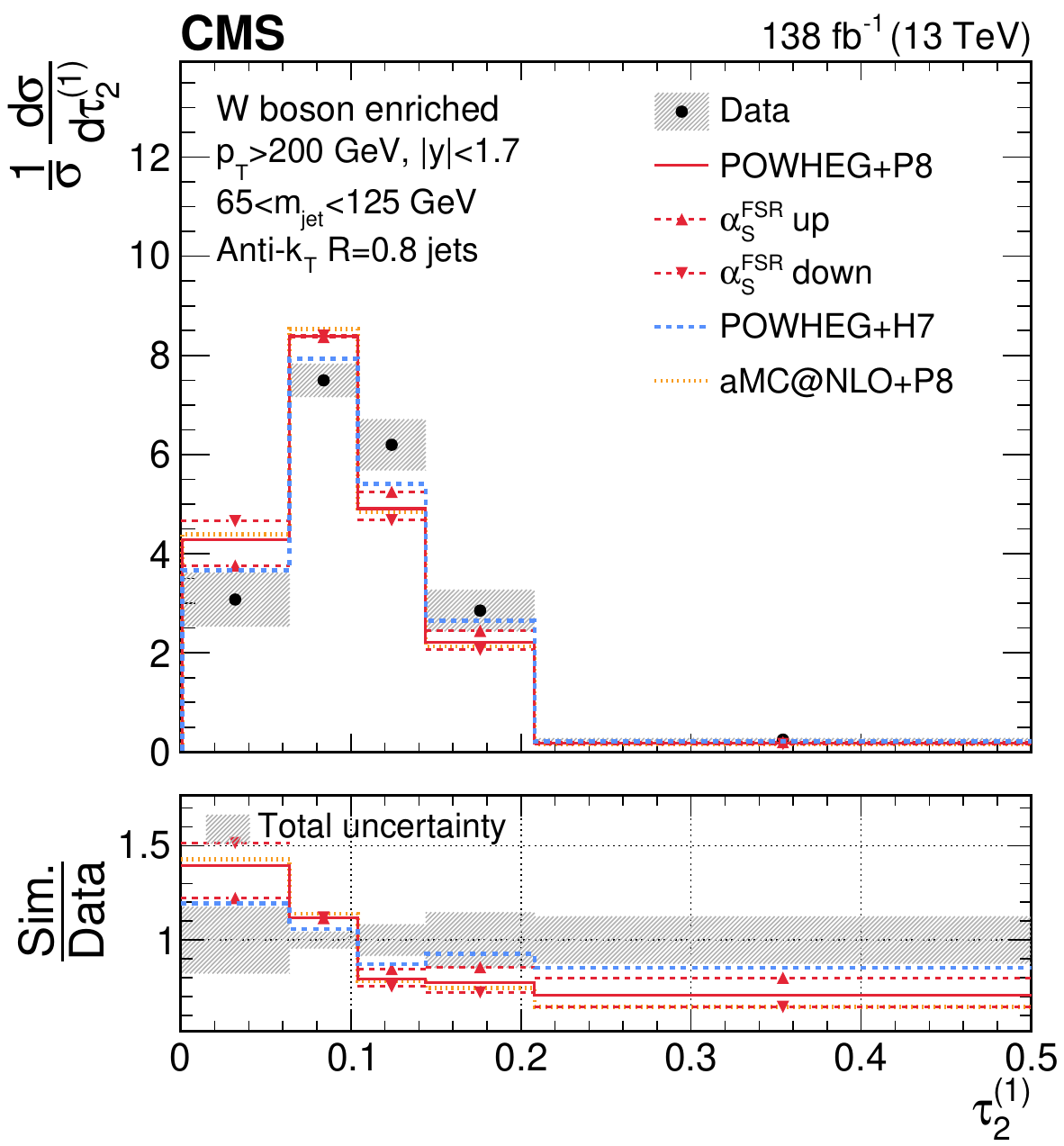} 
	\includegraphics[width=.395\textwidth]{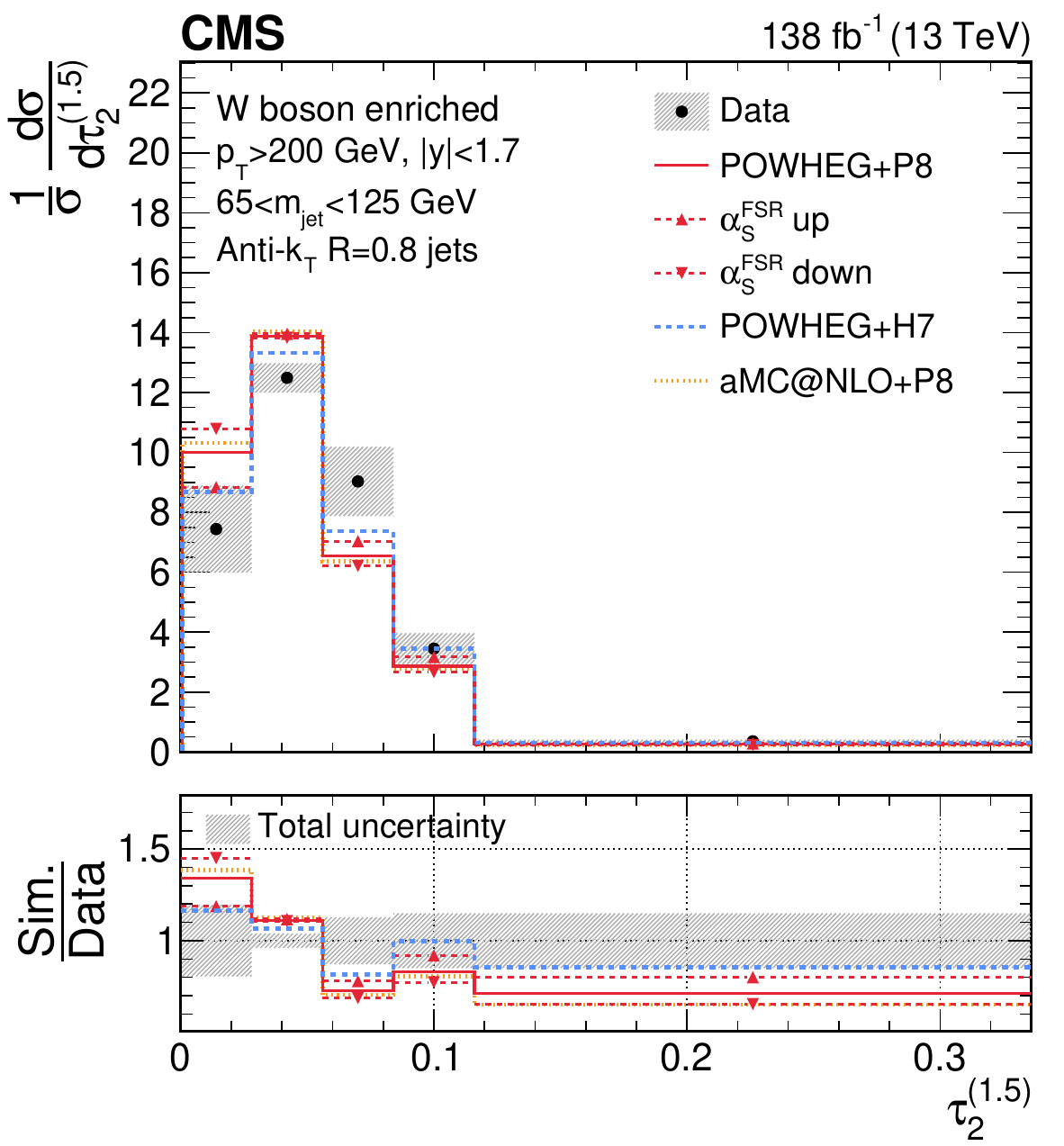} 
	\includegraphics[width=.395\textwidth]{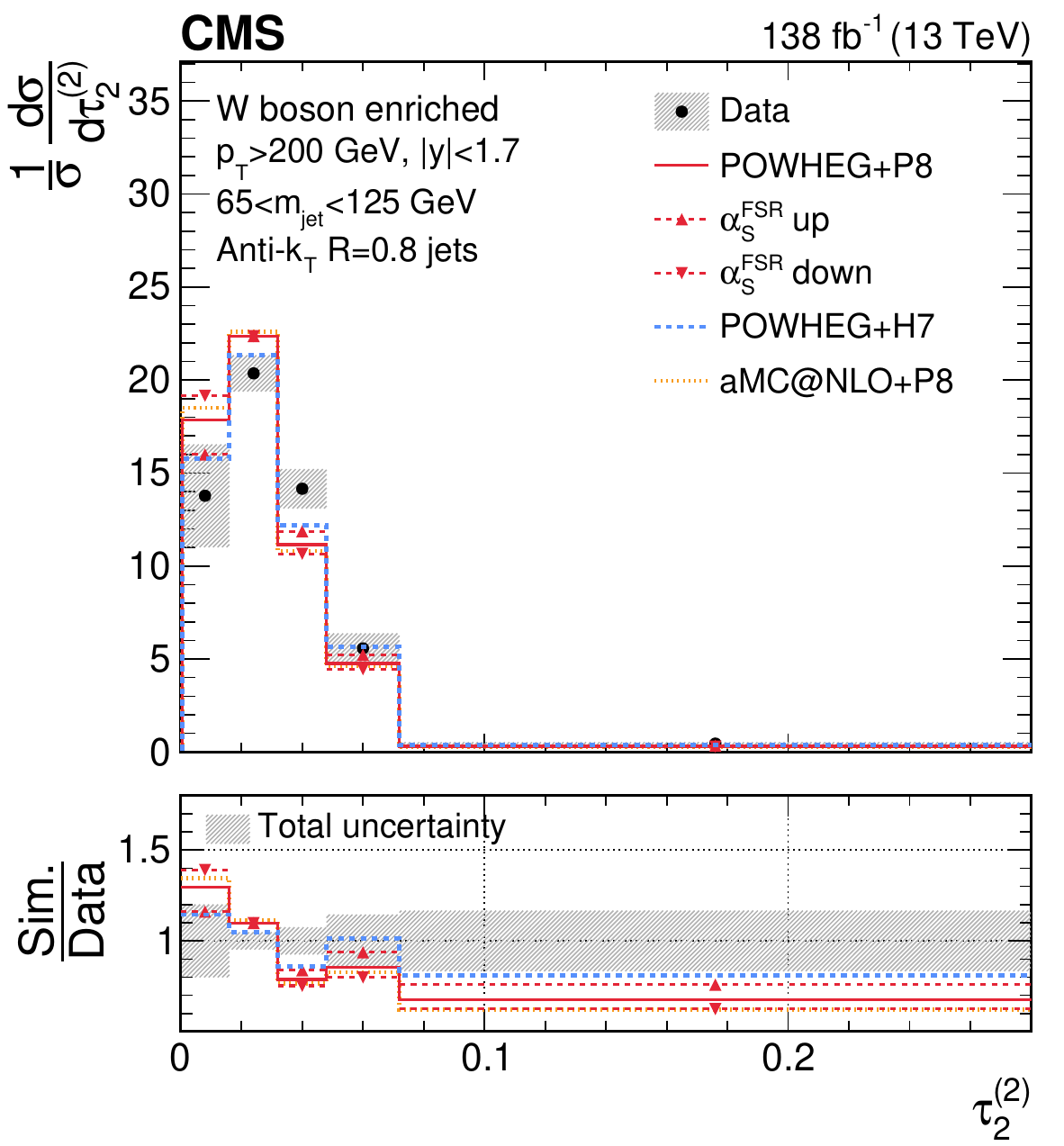} 
	\caption{Unfolded distributions of 2-subjettiness observables, \Nsub{2}{0.25}, \Nsub{2}{0.5}, \Nsub{2}{1}, \Nsub{2}{1.5}, and \Nsub{2}{2}, 
		measured for AK8 jets in boosted \PW boson-enriched events, extracted from the normalized, combined distribution after unfolding; the bin contents and the error bars are scaled by the bin widths for the distributions of the individual observables.  
		For comparisons with particle-level predictions, the error bars in data correspond to the total unfolding uncertainties, 
		and the lower panels present the ratio of particle-level predictions to the unfolded data. 
		The dark grey hashed region illustrates the total uncertainties per bin in the unfolded result.}
	\label{fig:addlresults3bodyW}
\end{figure}

\begin{figure}[!htb]
	\centering
	\includegraphics[width=.395\textwidth]{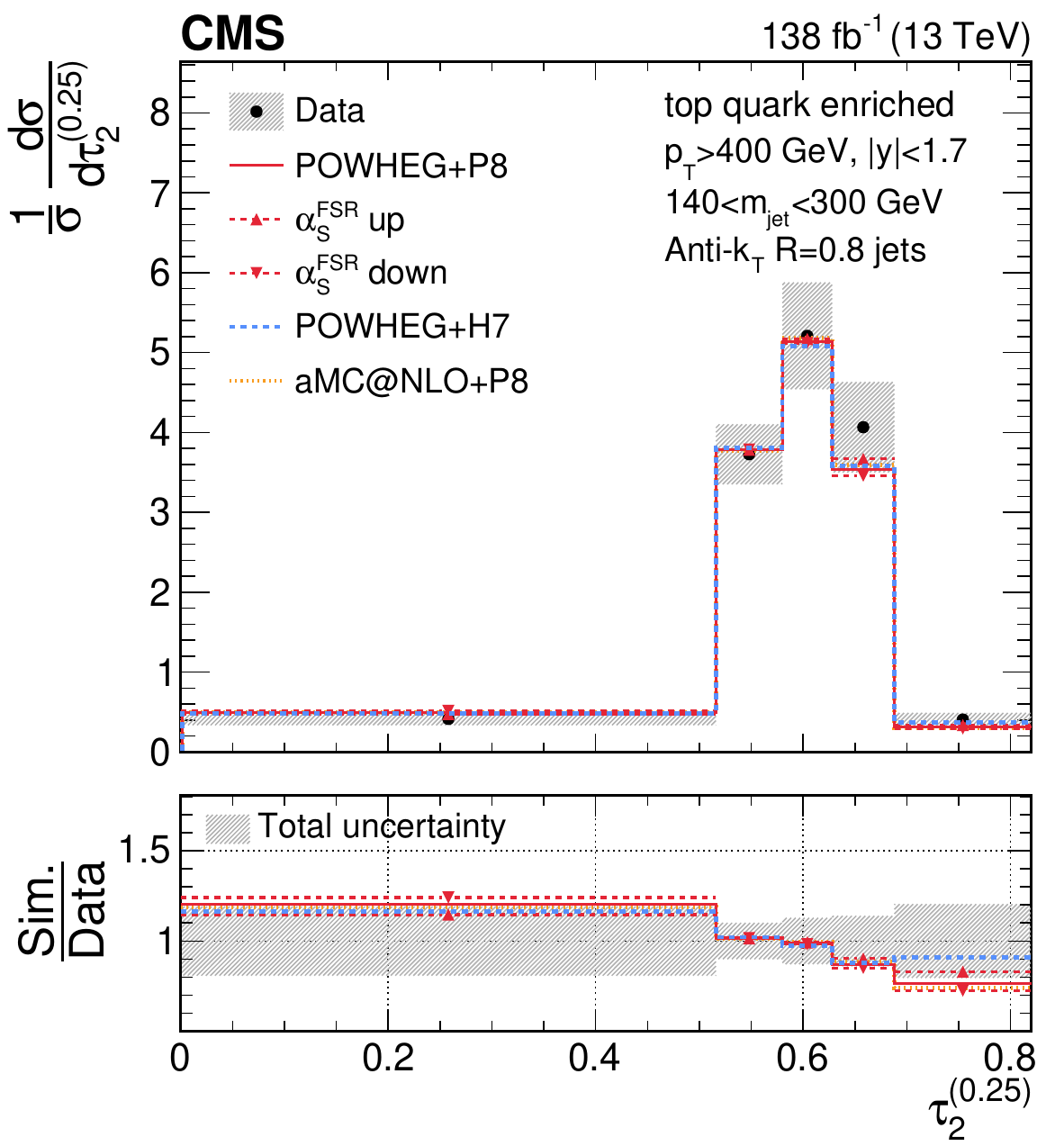} 		
	\includegraphics[width=.395\textwidth]{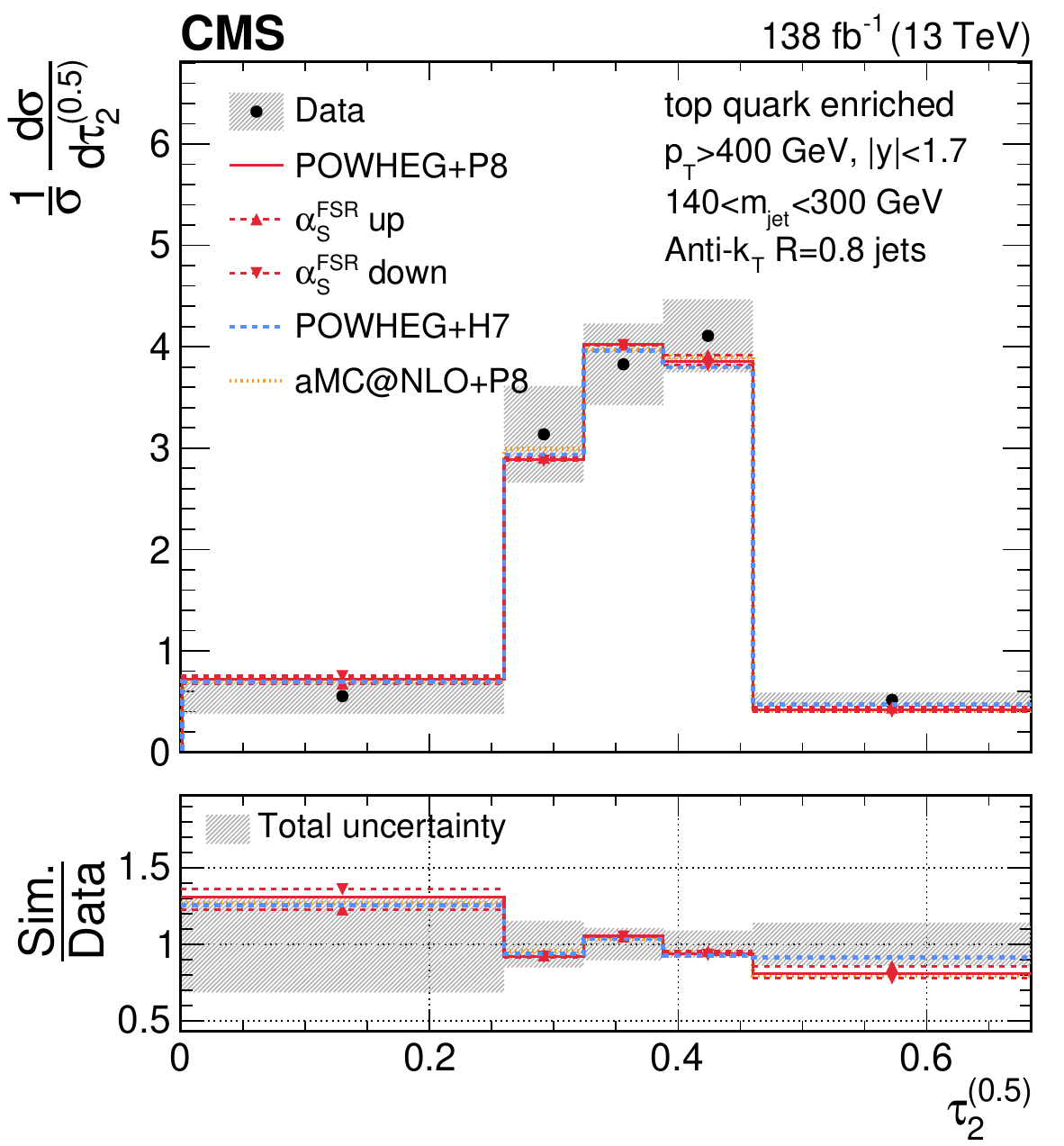} 
	\includegraphics[width=.395\textwidth]{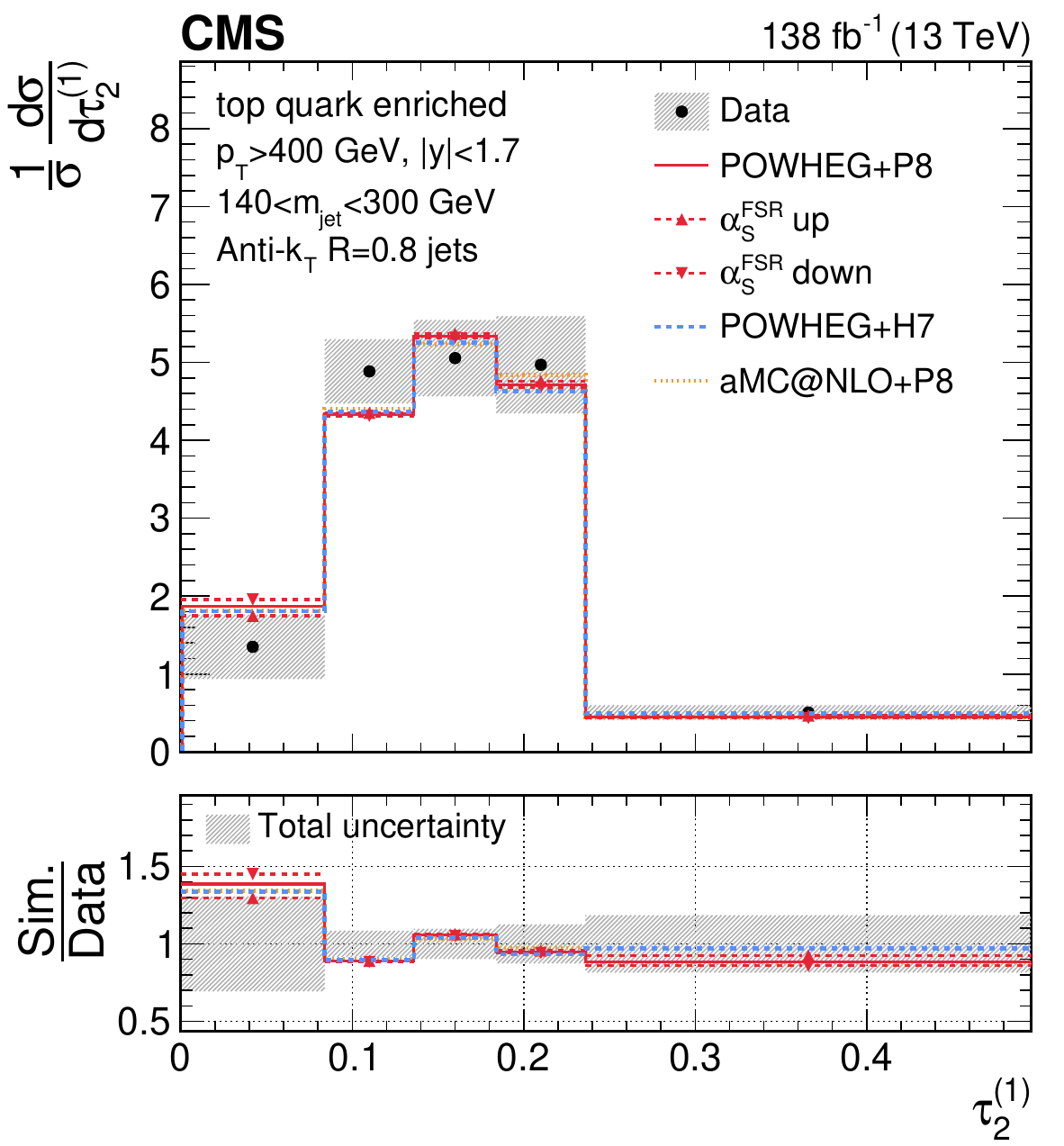} 
	\includegraphics[width=.395\textwidth]{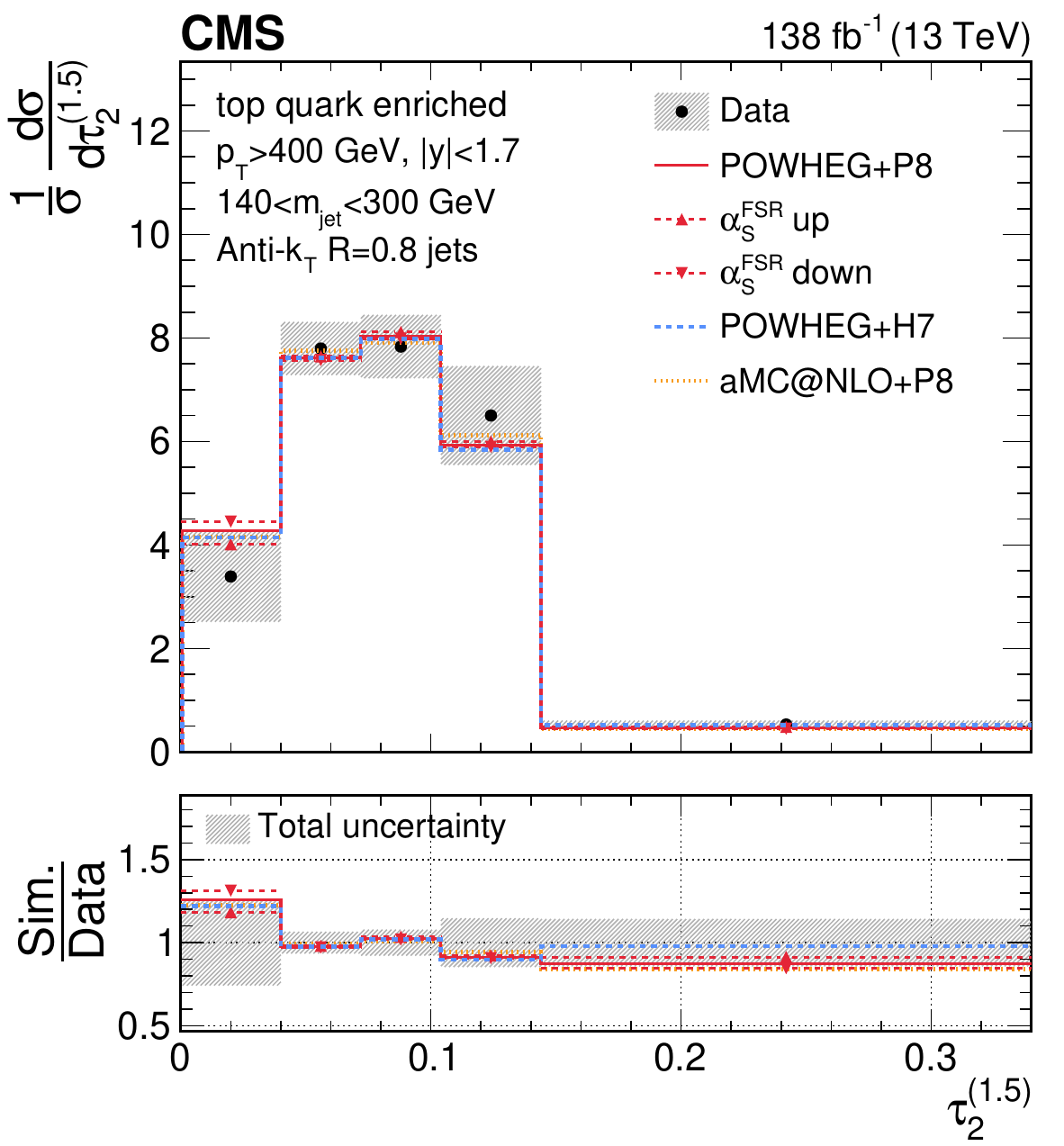} 
	\includegraphics[width=.395\textwidth]{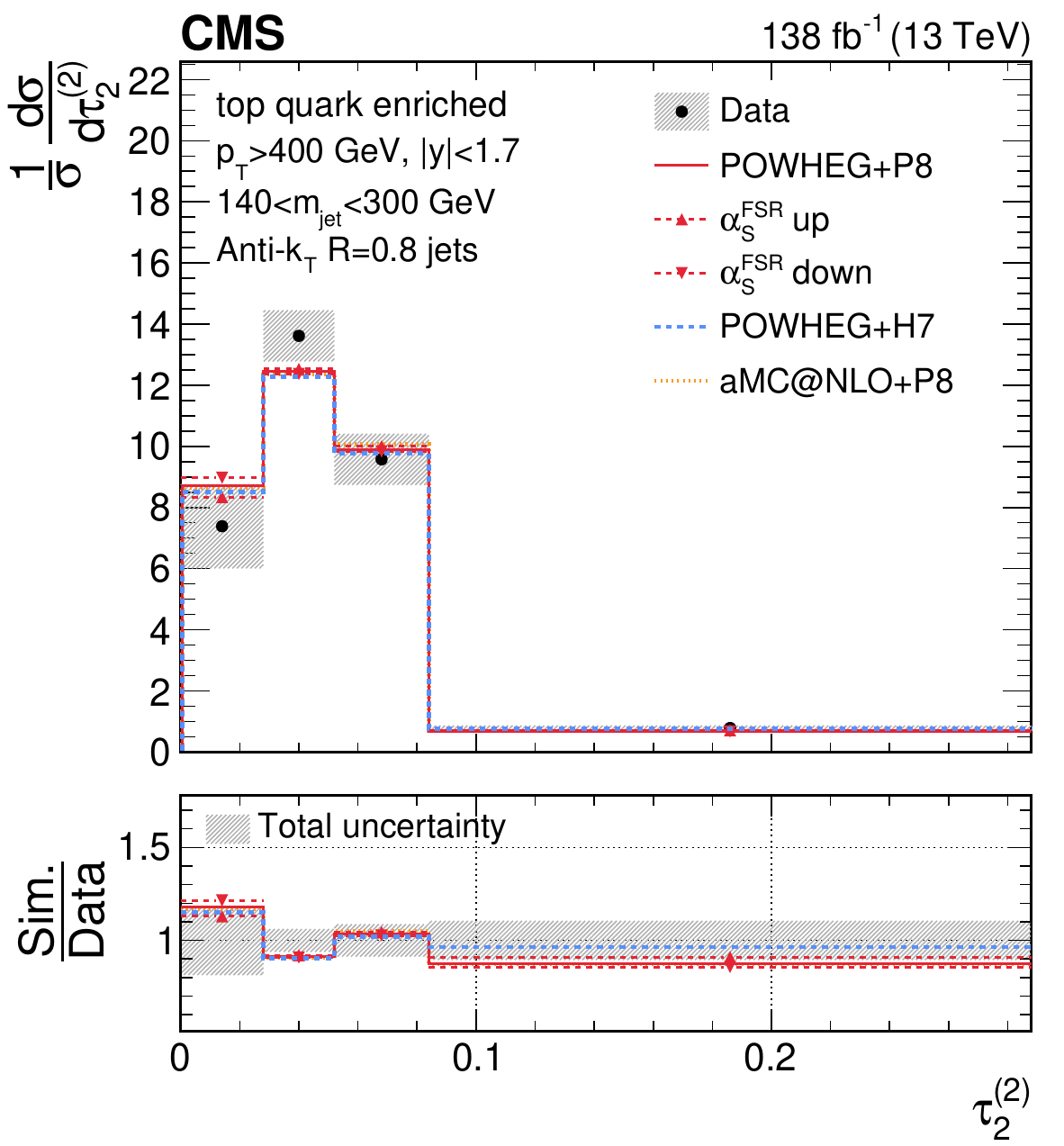} 
	\caption{Unfolded distributions of 2-subjettiness observables, \Nsub{2}{0.25}, \Nsub{2}{0.5}, \Nsub{2}{1}, \Nsub{2}{1.5}, and \Nsub{2}{2}, 
		measured for AK8 jets in the boosted top quark-enriched region, extracted from the normalized, combined distribution after unfolding; the bin contents and the error bars are scaled by the bin widths for the distributions of the individual observables.  
		For comparisons with particle-level predictions, the error bars in data correspond to the total unfolding uncertainties, 
		and the lower panels present the ratio of particle-level predictions to the unfolded data. 
		The dark grey hashed region illustrates the total uncertainties per bin in the unfolded result.}
	\label{fig:addlresults3bodytop}
\end{figure}

The measurements of \Nsub{2}{0.25}, \Nsub{2}{0.5}, \Nsub{2}{1}, \Nsub{2}{1.5}, and \Nsub{2}{2} are presented. 
The unfolded results for the dijet selection are shown in Fig.~\ref{fig:addlresults3bodyDijet}, and results for the boosted \PW boson- and top quark-enriched regions are shown in Figs.~\ref{fig:addlresults3bodyW} and \ref{fig:addlresults3bodytop}, respectively.

\clearpage
\newpage

\subsection{4-body phase space}
\label{sec:addlmeasurements4body}
\begin{figure}[!htb]
	\centering
	\includegraphics[width=.395\textwidth]{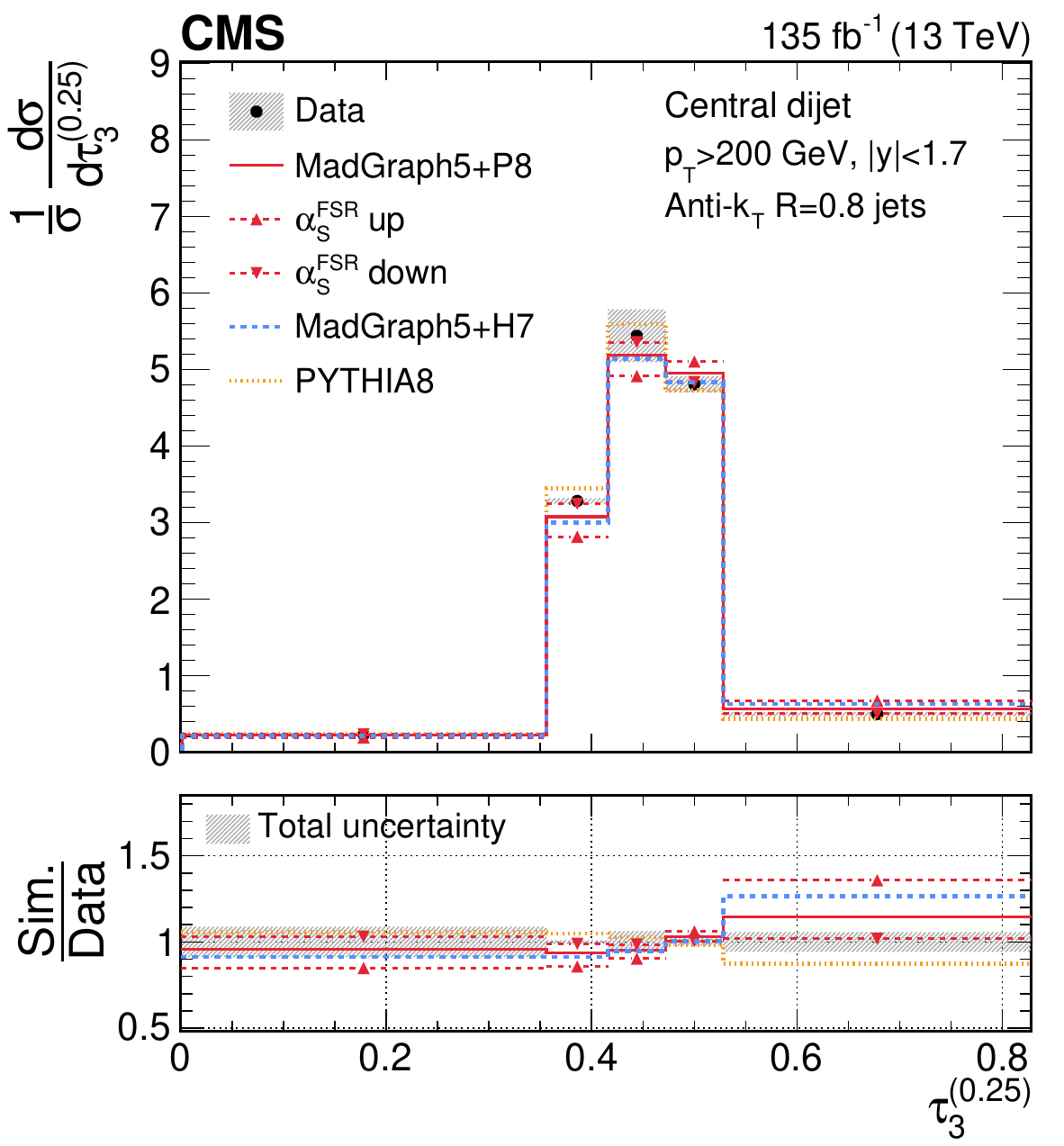} 		
	\includegraphics[width=.395\textwidth]{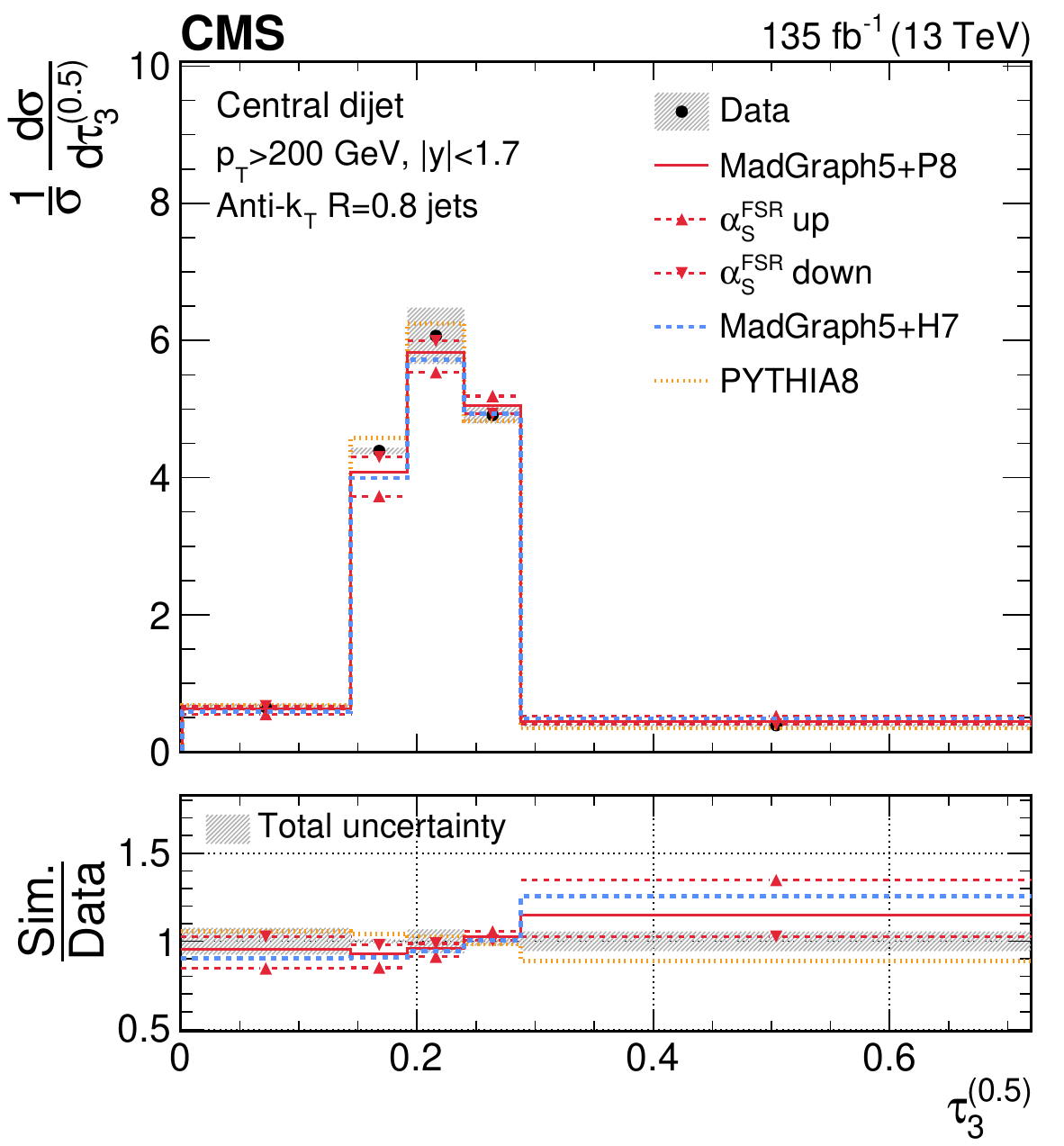} 
	\includegraphics[width=.395\textwidth]{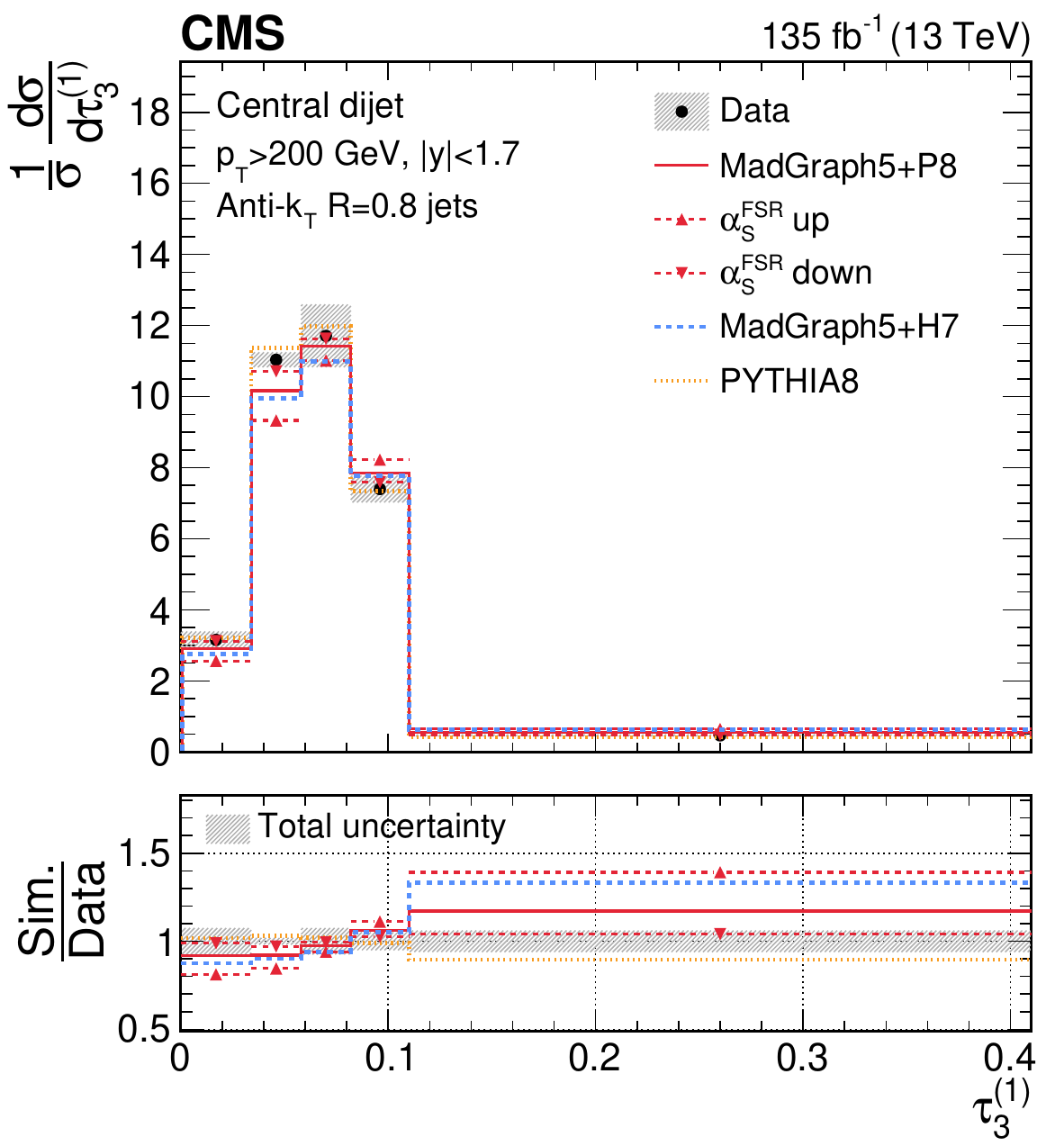} 
	\includegraphics[width=.395\textwidth]{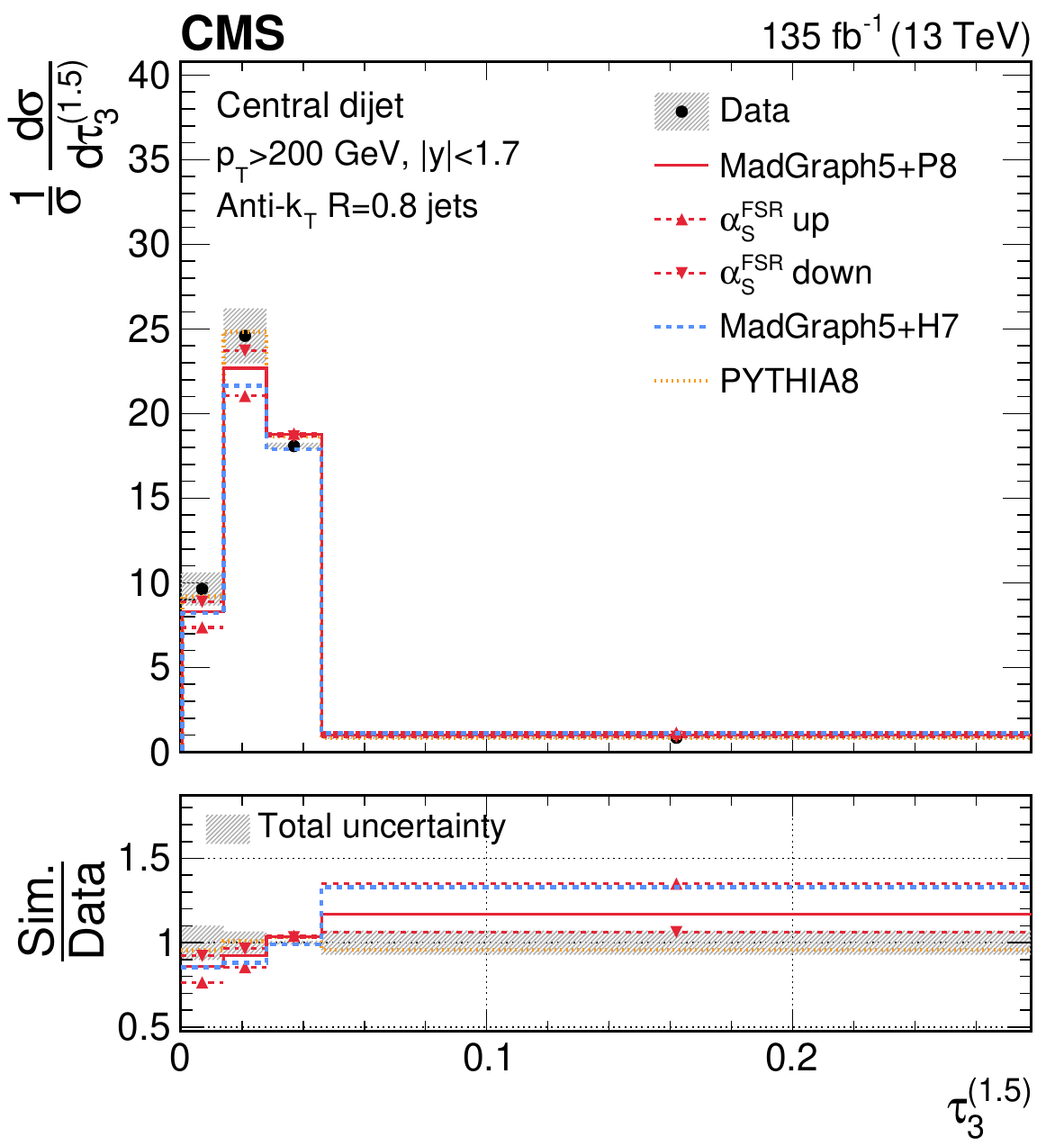} 
	\includegraphics[width=.395\textwidth]{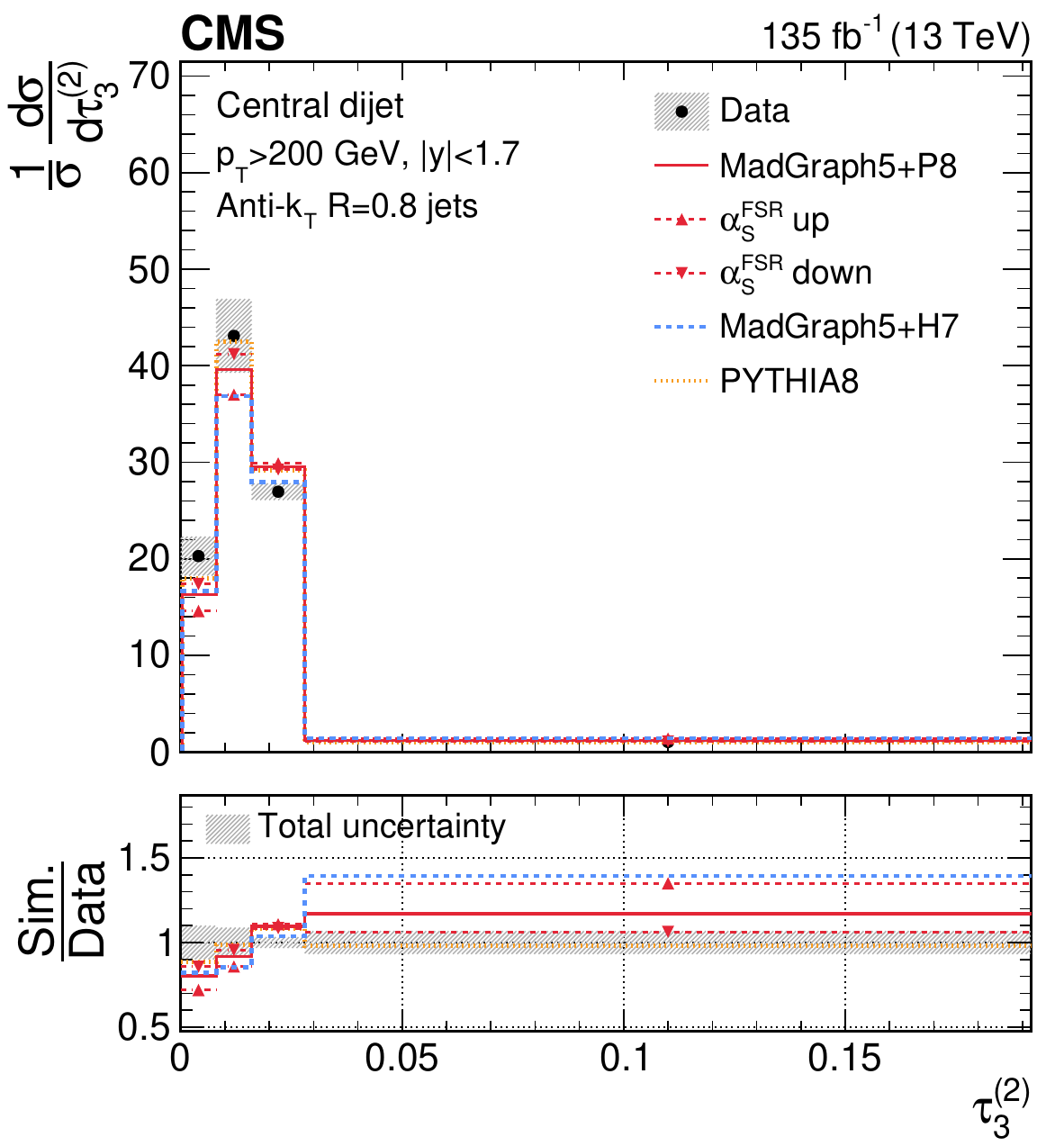} 
	\caption{Unfolded distributions of 3-subjettiness observables, \Nsub{3}{0.25}, \Nsub{3}{0.5}, \Nsub{3}{1}, \Nsub{3}{1.5}, and \Nsub{3}{2}, 
		measured for AK8 jets in the QCD dijet event selection, extracted from the normalized, combined distribution after unfolding; the bin contents and the error bars are scaled by the bin widths for the distributions of the individual observables.  
		For comparisons with particle-level predictions, the error bars in data correspond to the total unfolding uncertainties, 
		and the lower panels present the ratio of particle-level predictions to the unfolded data. 
		The dark grey hashed region illustrates the total uncertainties per bin in the unfolded result.}
	\label{fig:addlresults4bodyDijet}
\end{figure}

\begin{figure}[!htb]
	\centering
	\includegraphics[width=.395\textwidth]{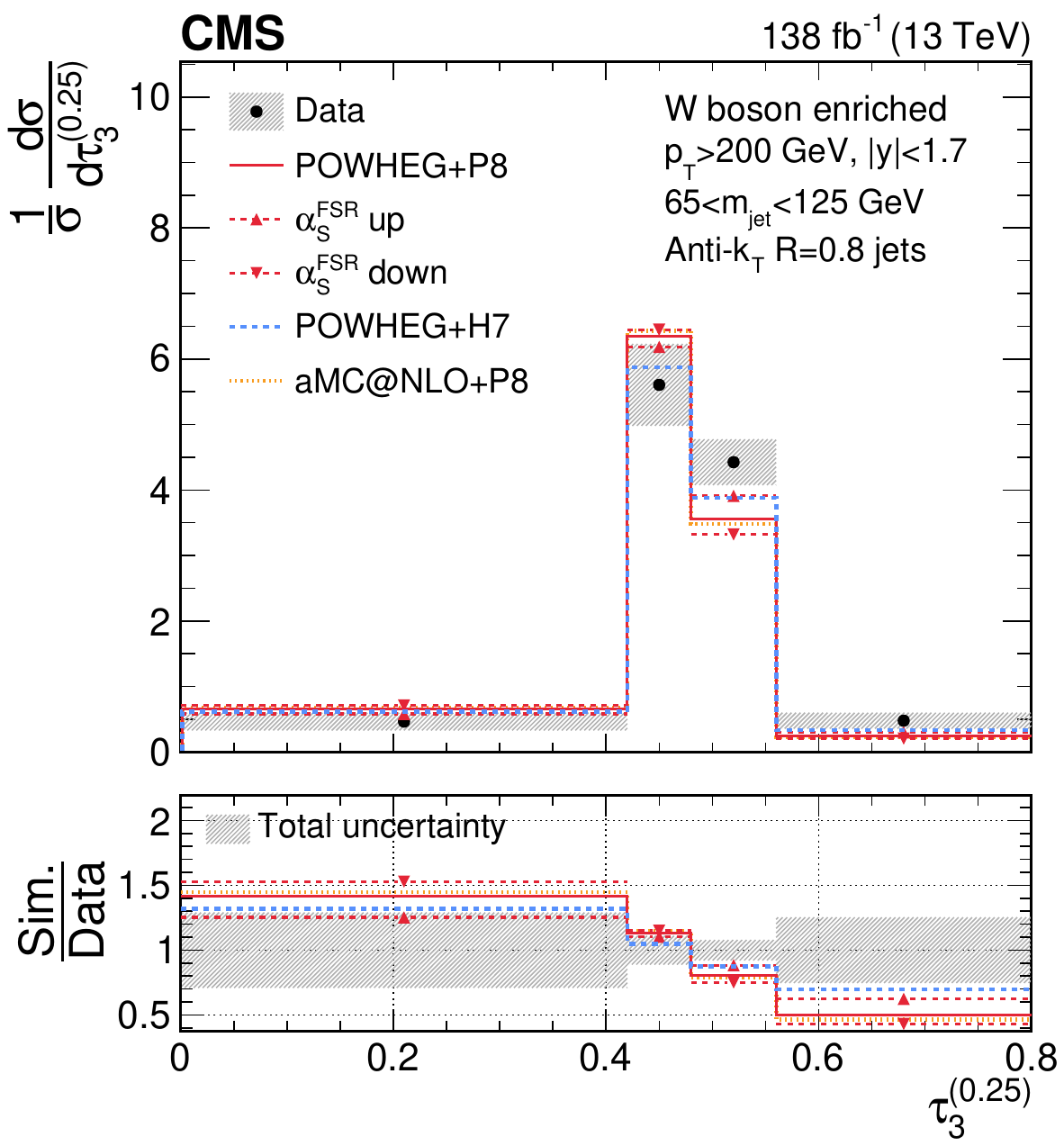} 		
	\includegraphics[width=.395\textwidth]{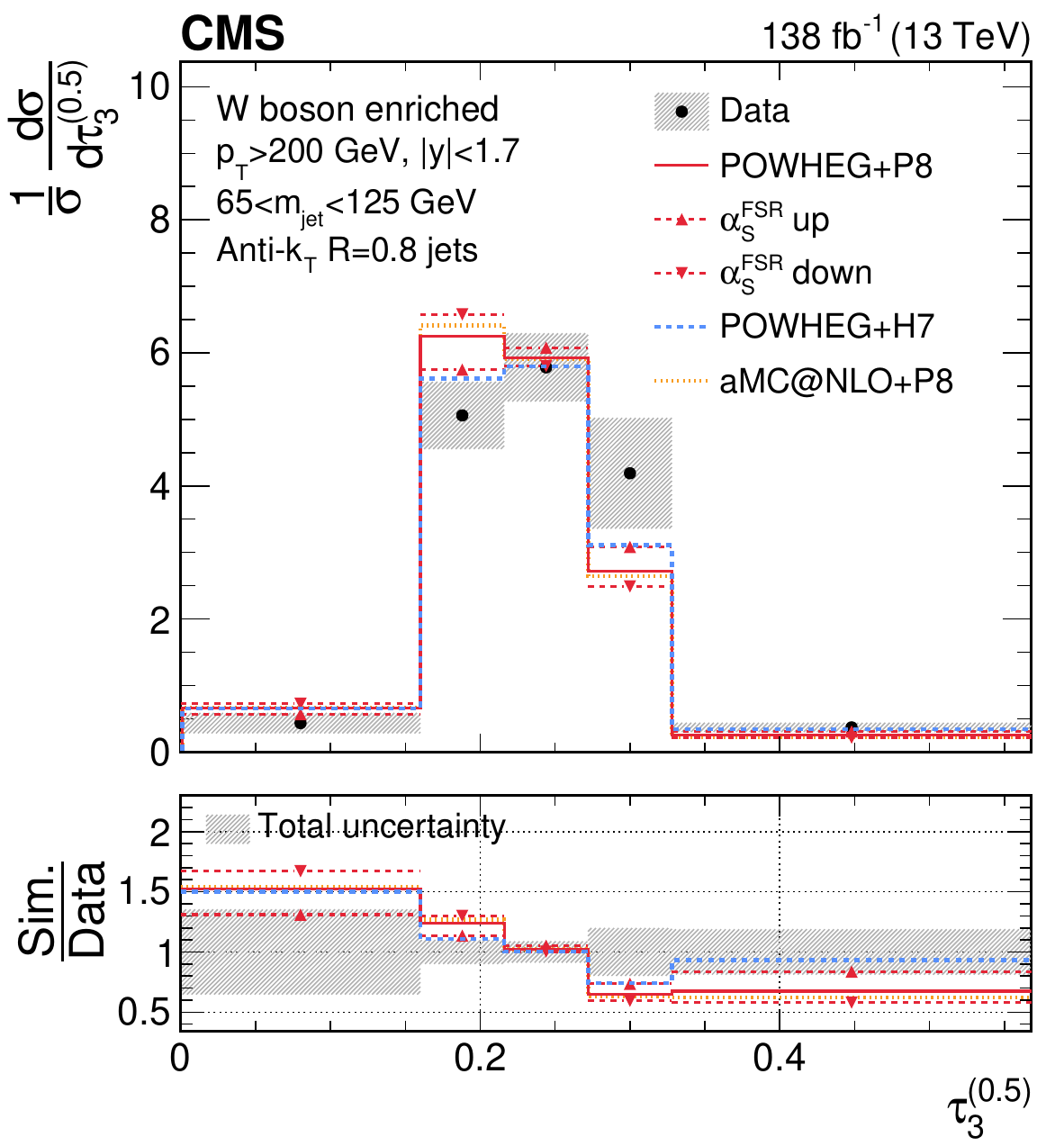} 
	\includegraphics[width=.395\textwidth]{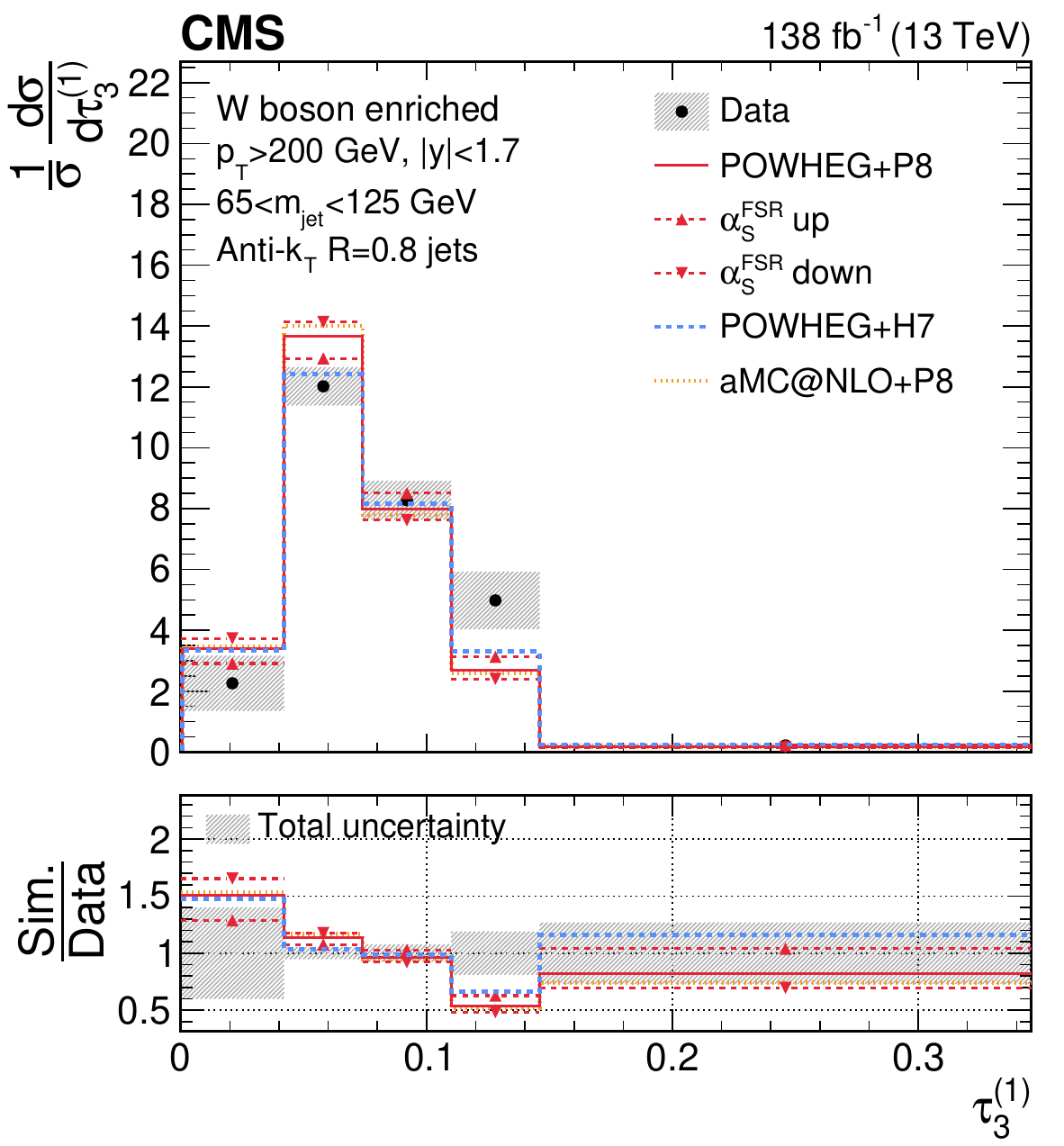} 
	\includegraphics[width=.395\textwidth]{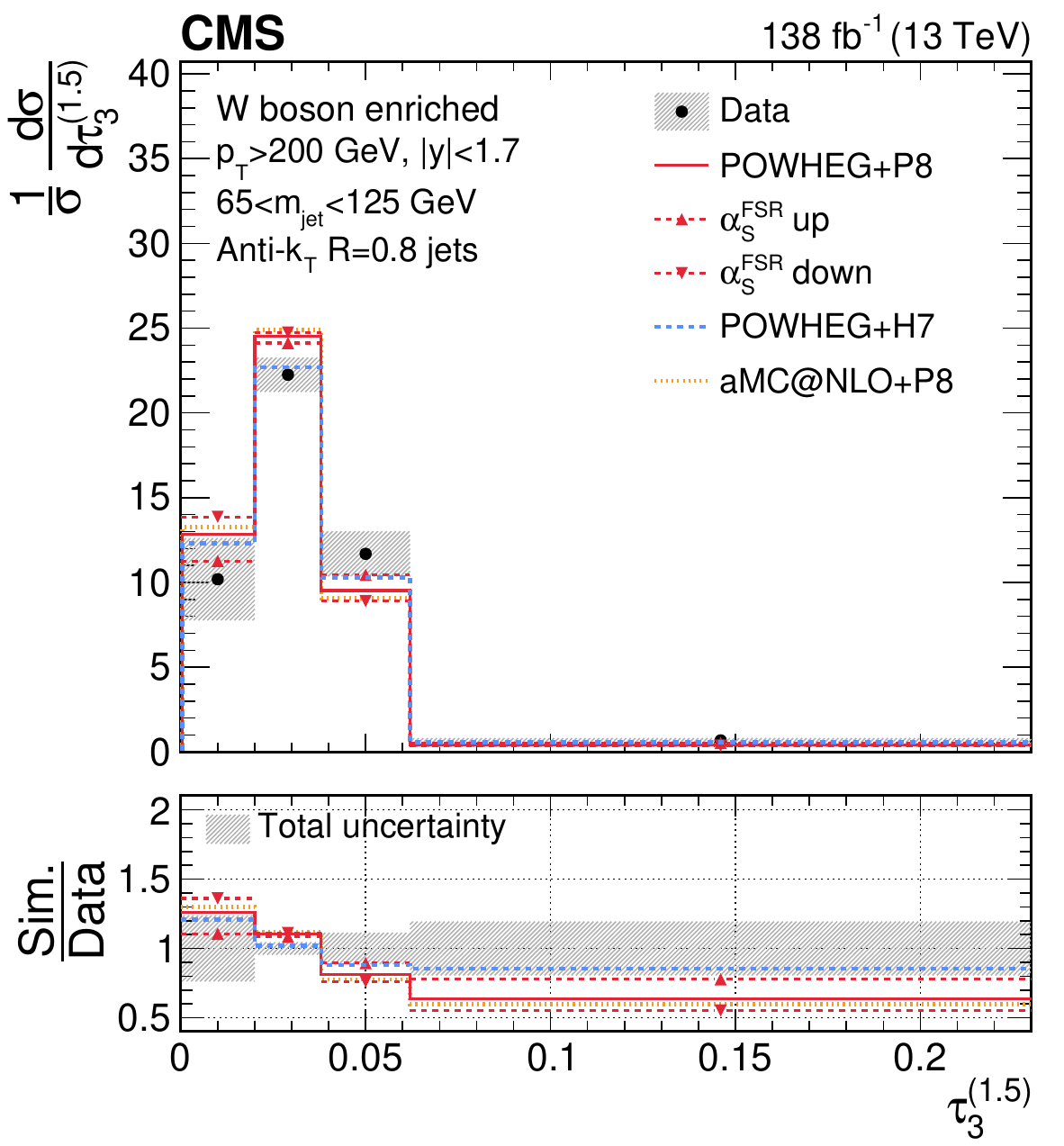} 
	\includegraphics[width=.395\textwidth]{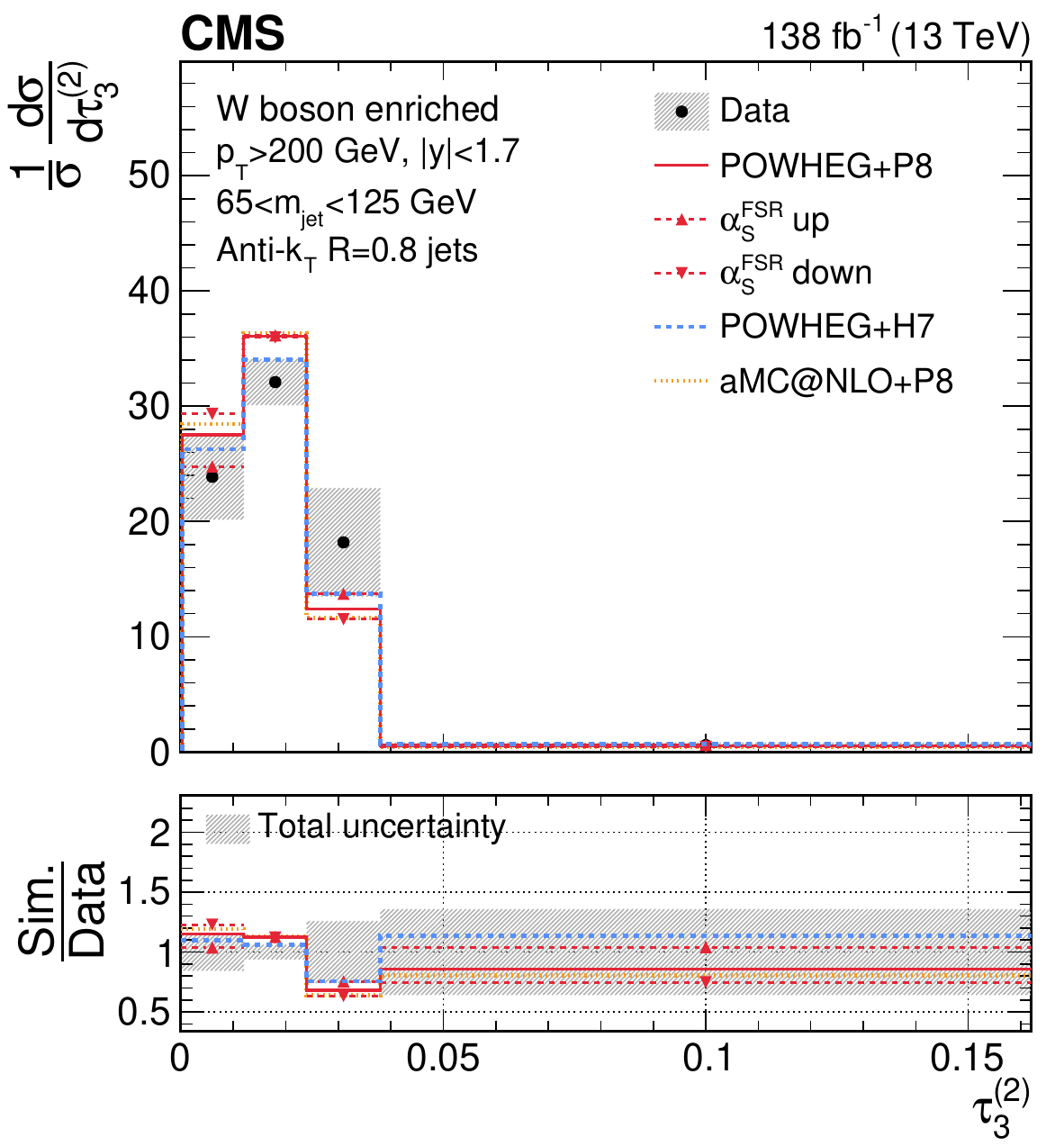} 
	\caption{Unfolded distributions of 3-subjettiness observables, \Nsub{3}{0.25}, \Nsub{3}{0.5}, \Nsub{3}{1}, \Nsub{3}{1.5}, and \Nsub{3}{2}, 
		measured for AK8 jets in boosted \PW boson-enriched events, extracted from the normalized, combined distribution after unfolding; the bin contents and the error bars are scaled by the bin widths for the distributions of the individual observables.  
		For comparisons with particle-level predictions, the error bars in data correspond to the total unfolding uncertainties, 
		and the lower panels present the ratio of particle-level predictions to the unfolded data. 
		The dark grey hashed region illustrates the total uncertainties per bin in the unfolded result.}
	\label{fig:addlresults4bodyW}
\end{figure}

\begin{figure}[!htb]
	\centering
	\includegraphics[width=.395\textwidth]{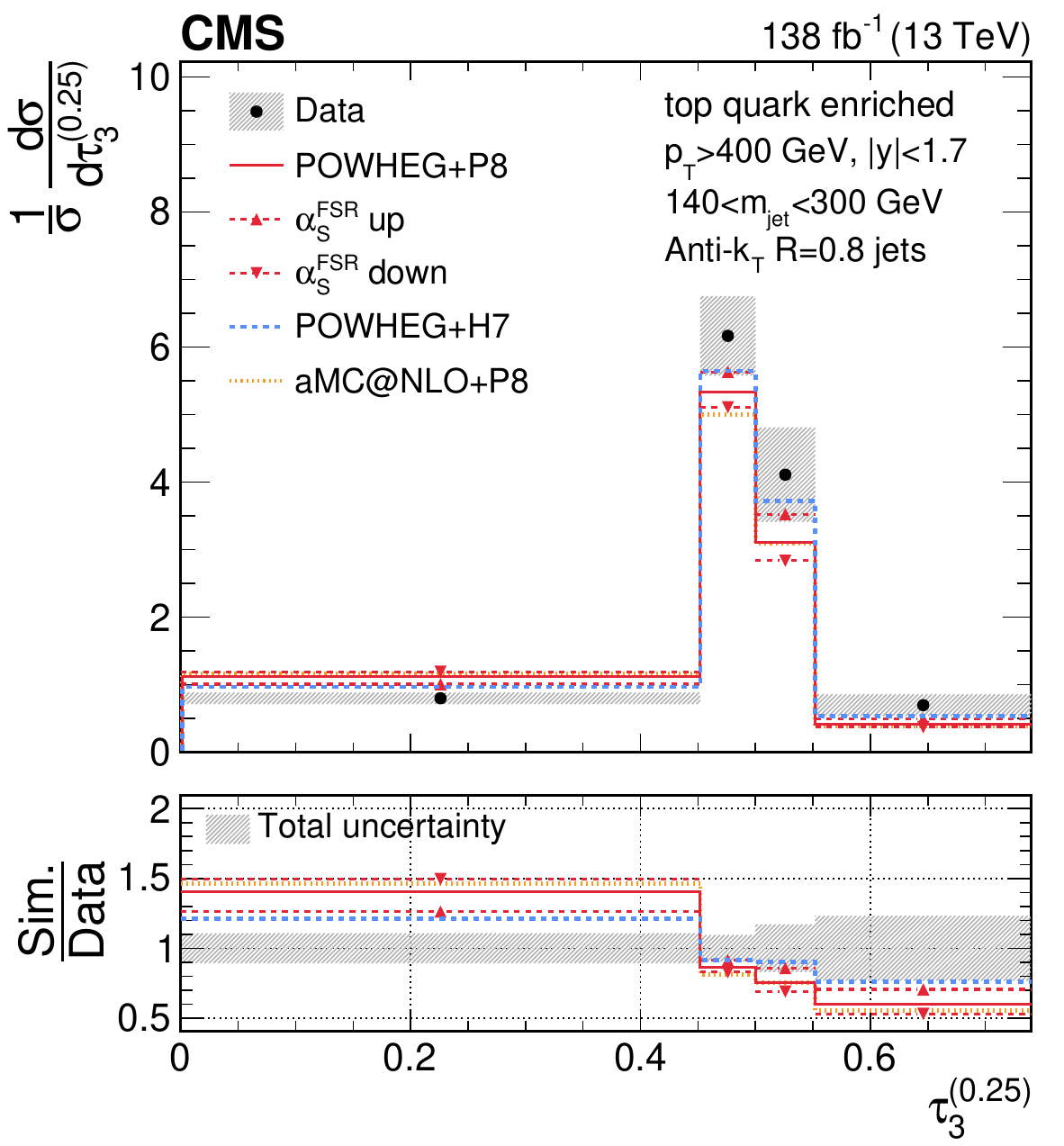} 		
	\includegraphics[width=.395\textwidth]{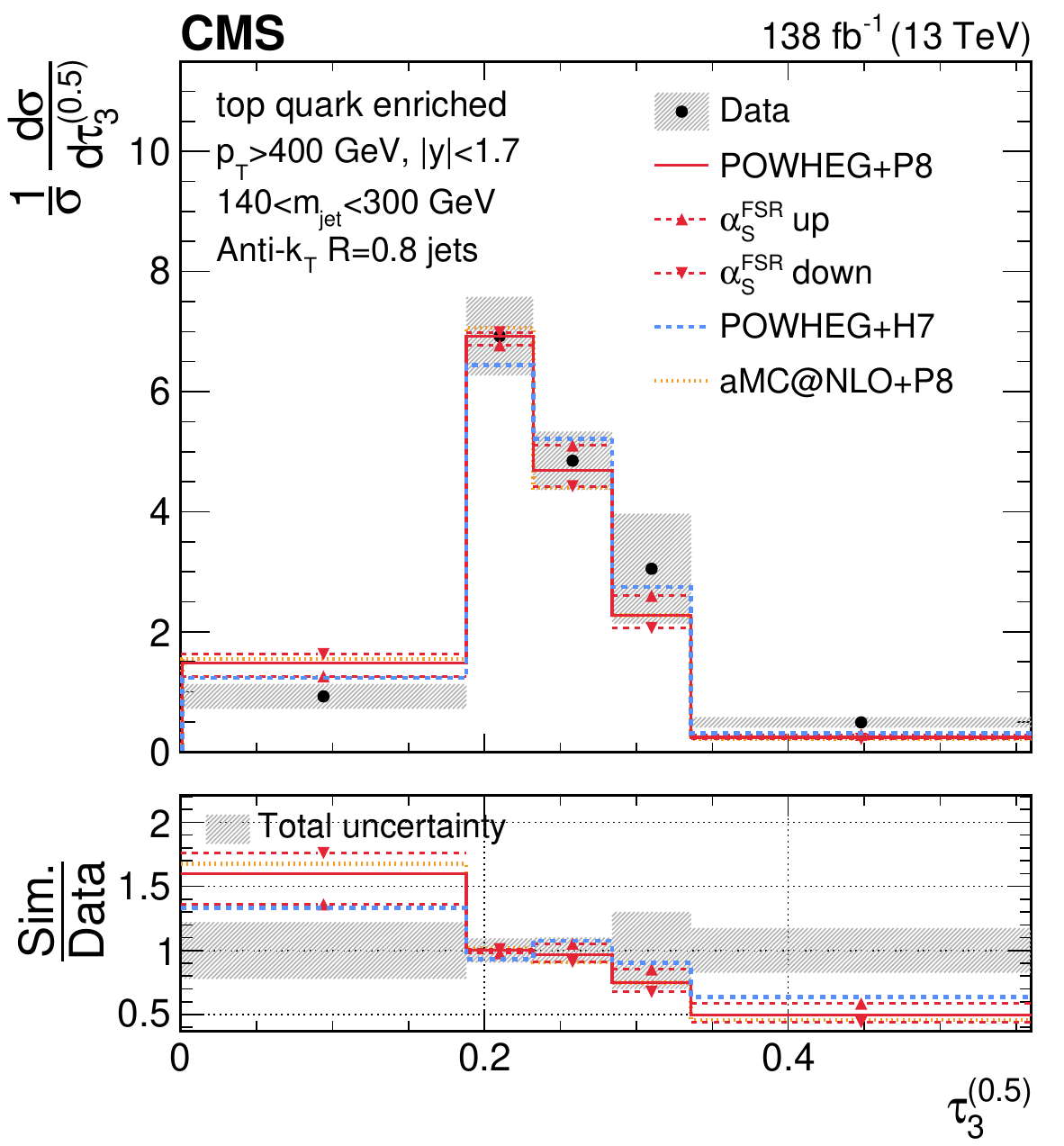} 
	\includegraphics[width=.395\textwidth]{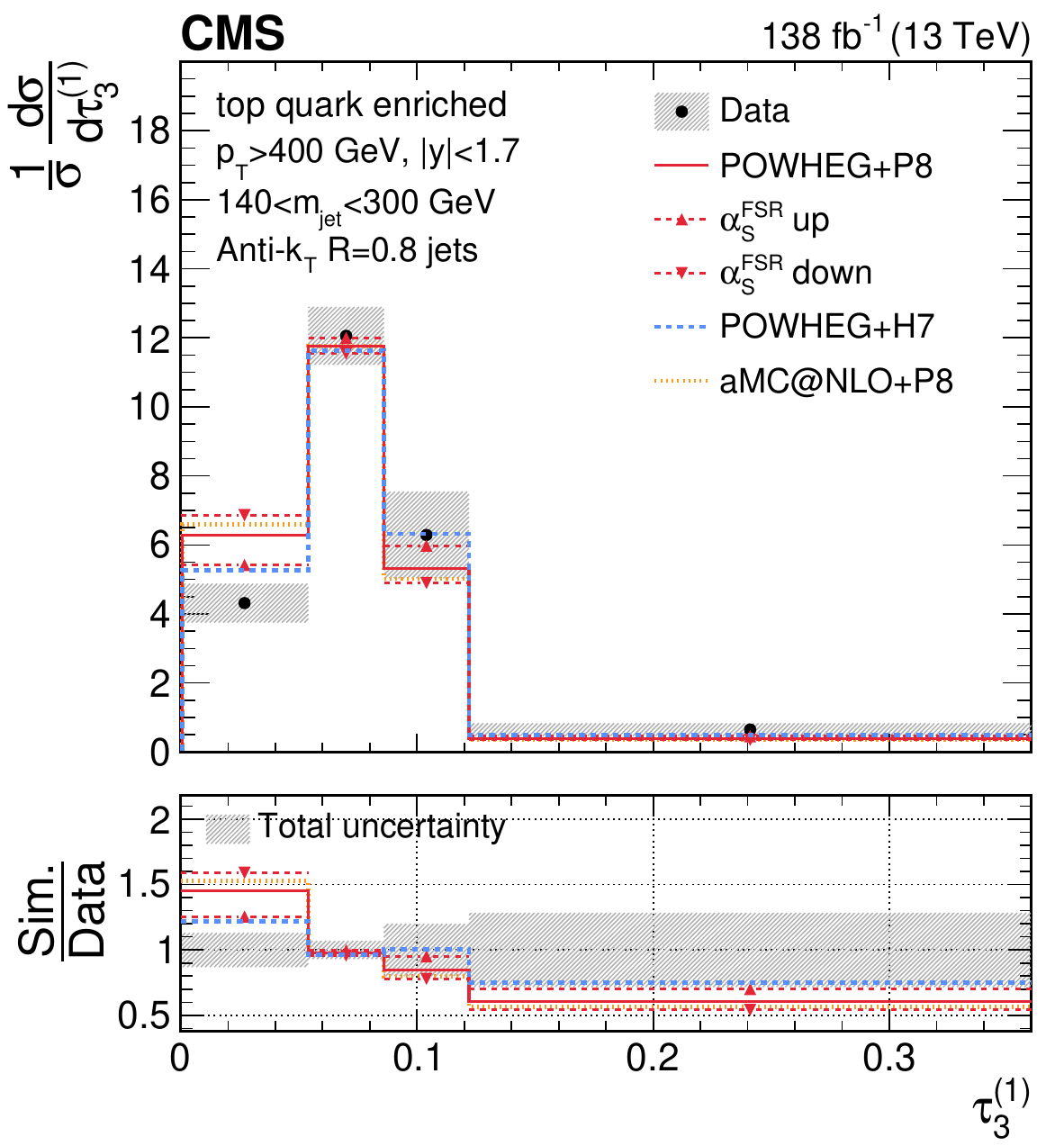} 
	\includegraphics[width=.395\textwidth]{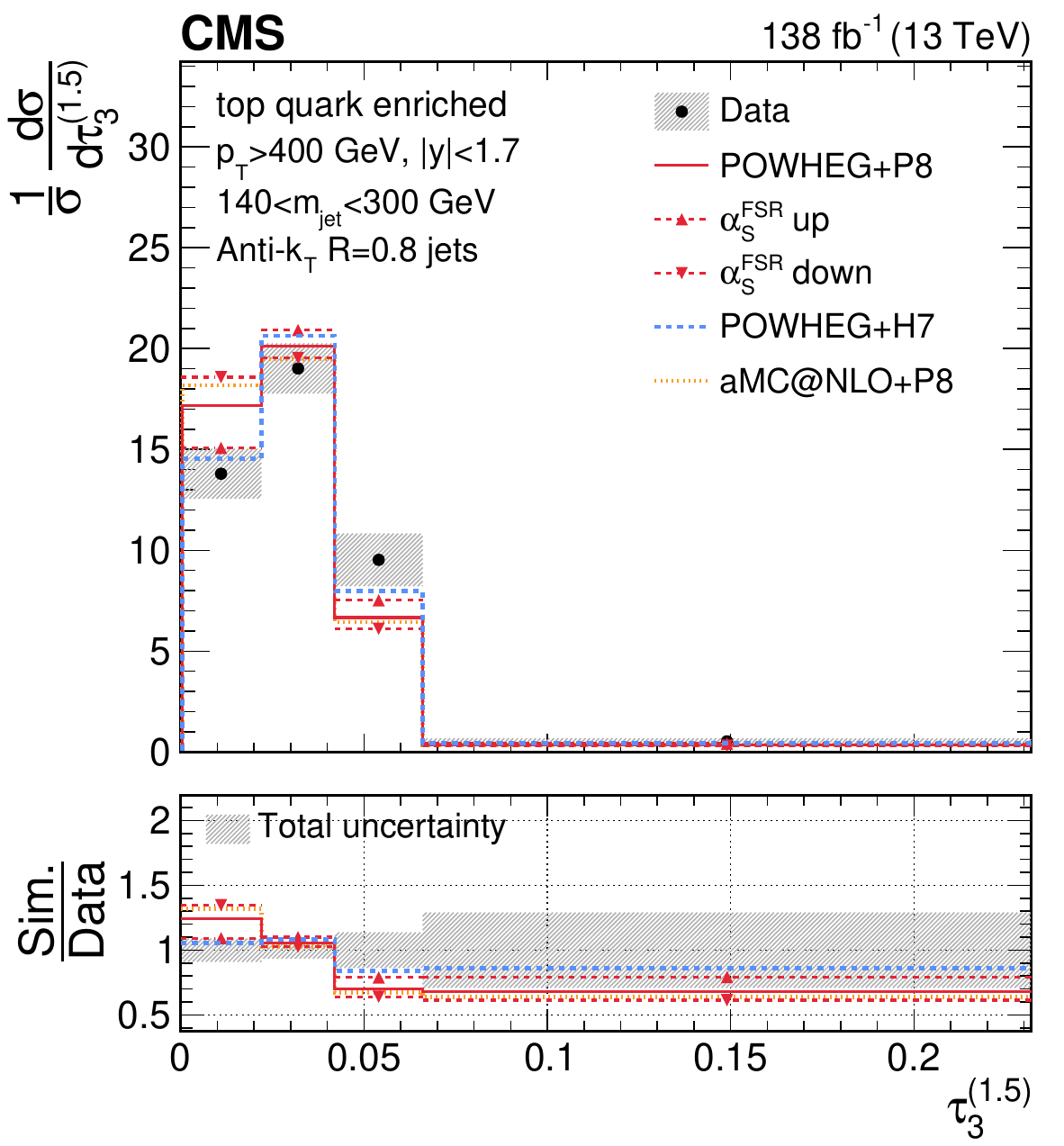} 
	\includegraphics[width=.395\textwidth]{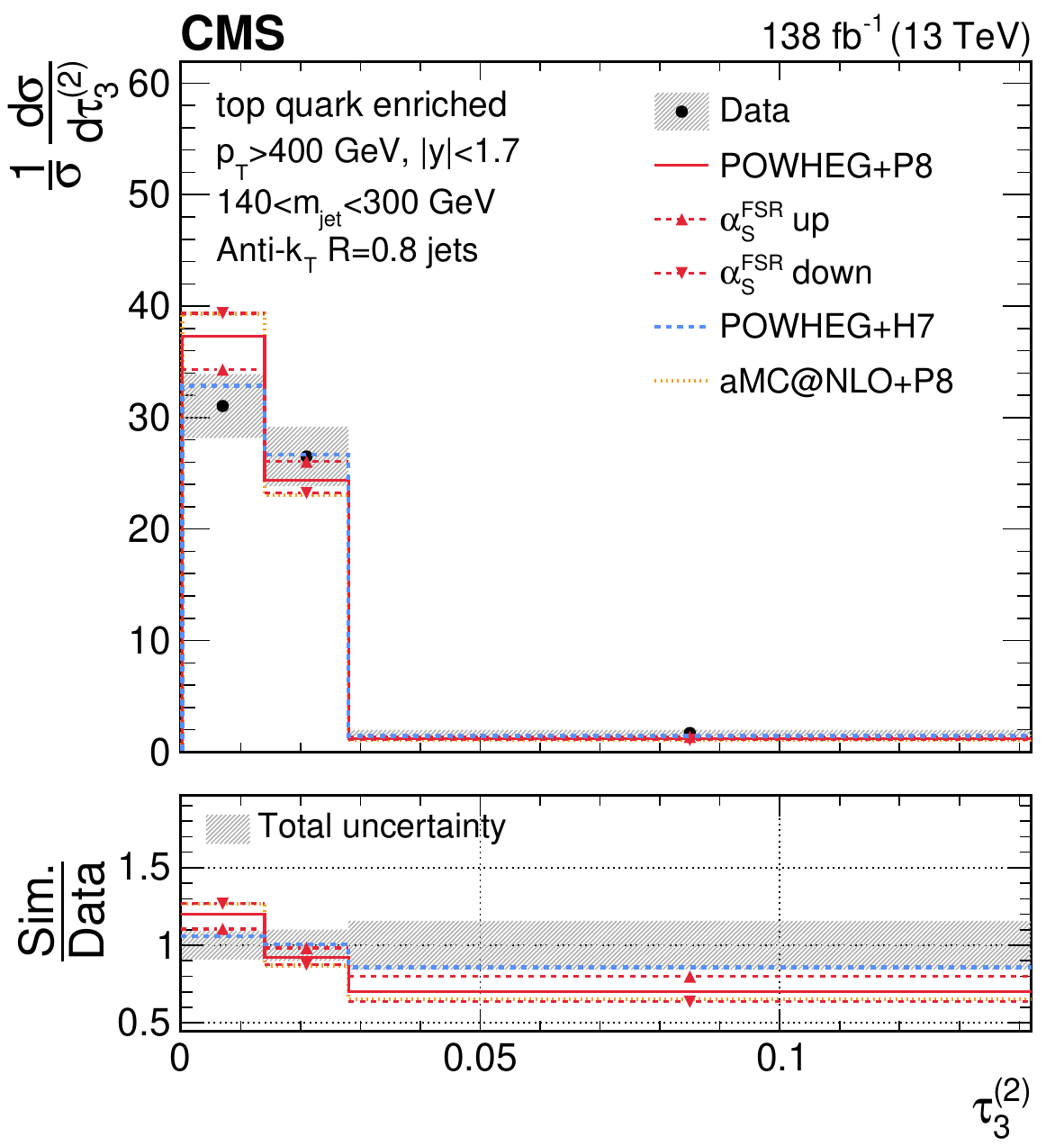} 
	\caption{Unfolded distributions of 3-subjettiness observables, \Nsub{3}{0.25}, \Nsub{3}{0.5}, \Nsub{3}{1}, \Nsub{3}{1.5}, and \Nsub{3}{2}, 
		measured for AK8 jets in the boosted top quark-enriched region, extracted from the normalized, combined distribution after unfolding; the bin contents and the error bars are scaled by the bin widths for the distributions of the individual observables.  
		For comparisons with particle-level predictions, the error bars in data correspond to the total unfolding uncertainties, 
		and the lower panels present the ratio of particle-level predictions to the unfolded data. 
		The dark grey hashed region illustrates the total uncertainties per bin in the unfolded result.}
	\label{fig:addlresults4bodytop}
\end{figure}

The measurements of \Nsub{3}{0.25}, \Nsub{3}{0.5}, \Nsub{3}{1}, \Nsub{3}{1.5}, and \Nsub{3}{2} are presented. 
The unfolded results for the dijet selection are shown in Fig.~\ref{fig:addlresults4bodyDijet}, and results for the boosted \PW boson- and top quark-enriched regions are shown in Figs.~\ref{fig:addlresults4bodyW} and \ref{fig:addlresults4bodytop}, respectively.

\clearpage
\newpage

\subsection{5-body phase space}
\label{sec:addlmeasurements5body}
\begin{figure}[!htb]
	\centering
	\includegraphics[width=.395\textwidth]{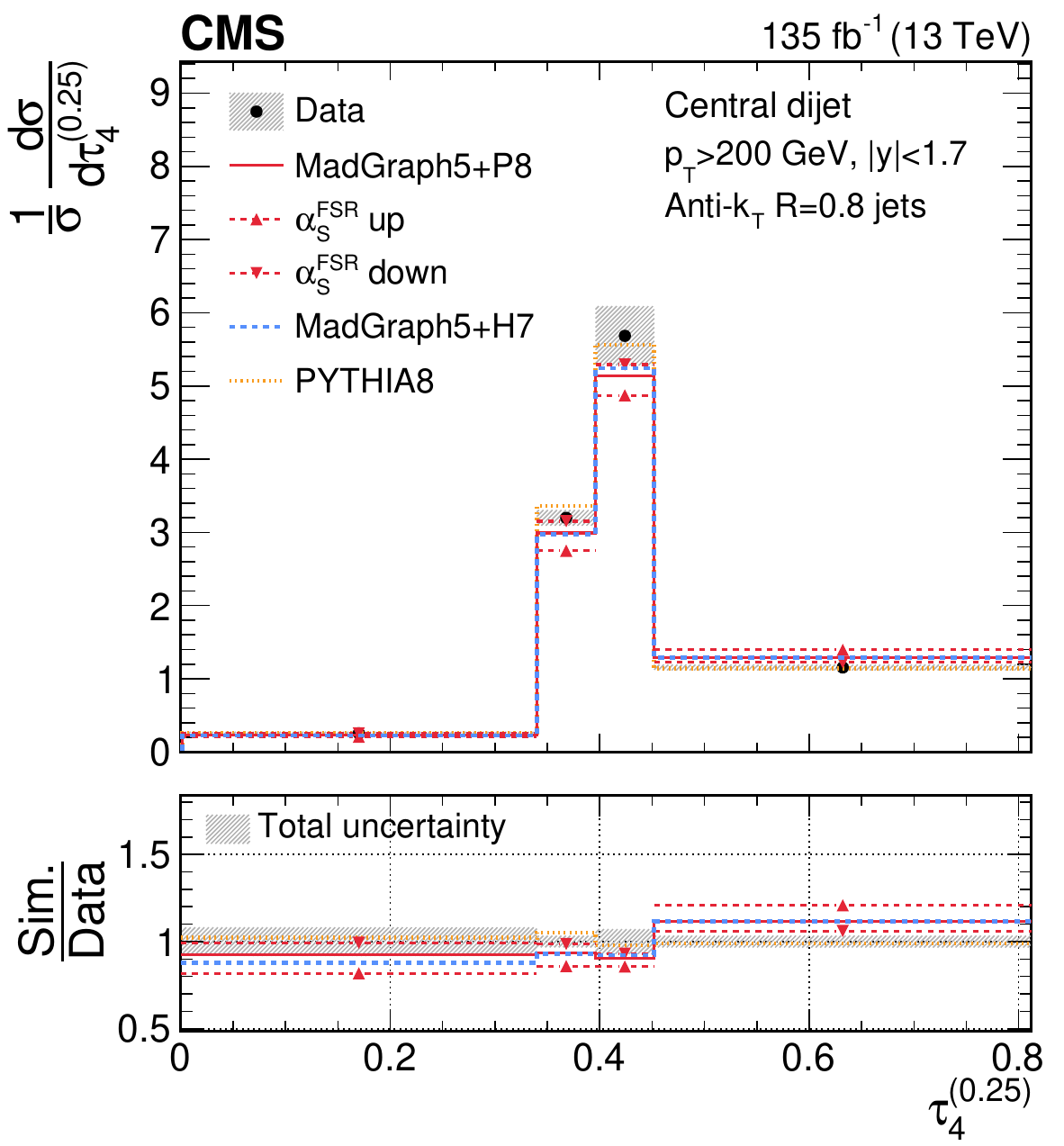} 		
	\includegraphics[width=.395\textwidth]{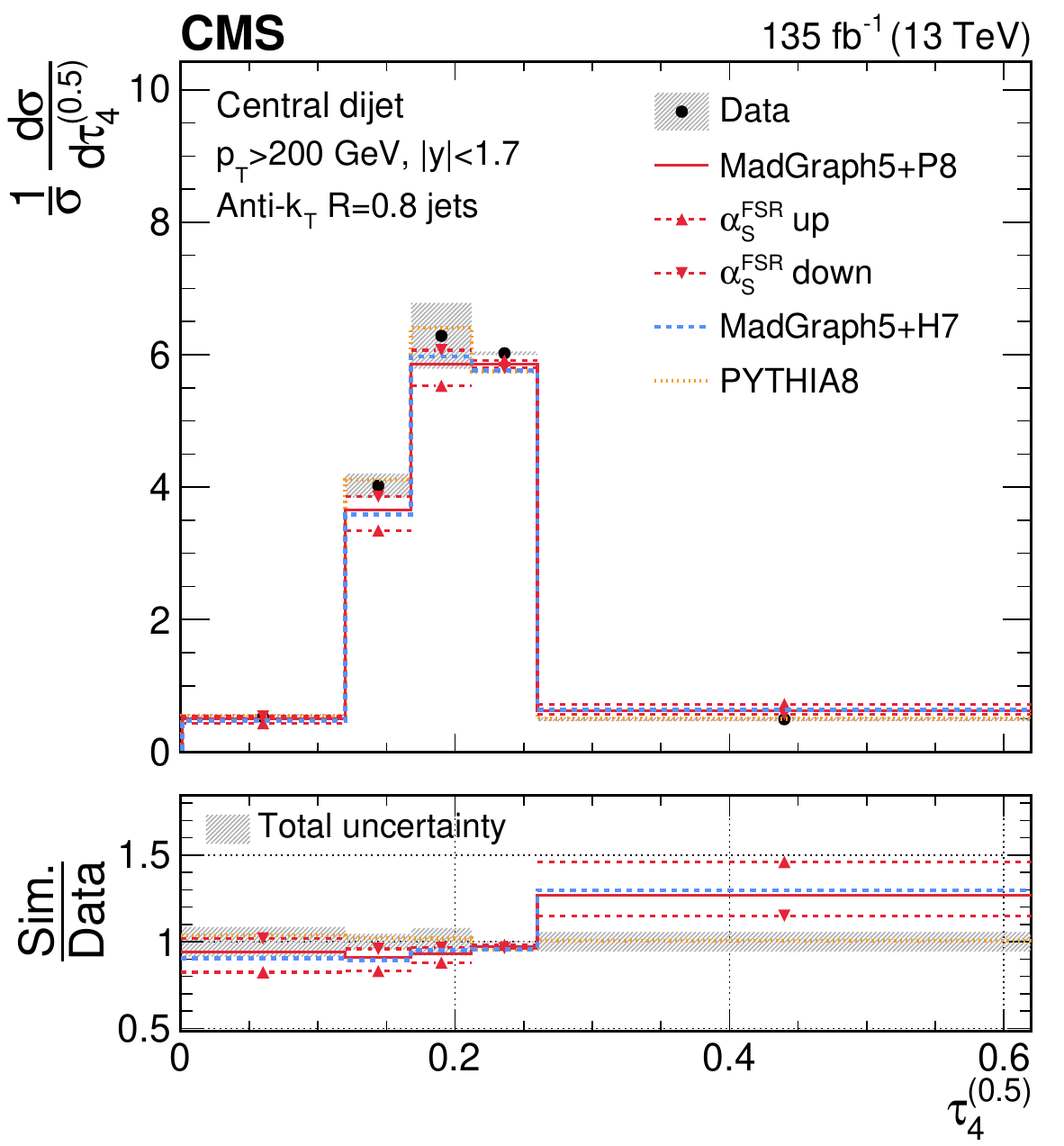} 
	\includegraphics[width=.395\textwidth]{Figure_013-b.pdf} 
	\includegraphics[width=.395\textwidth]{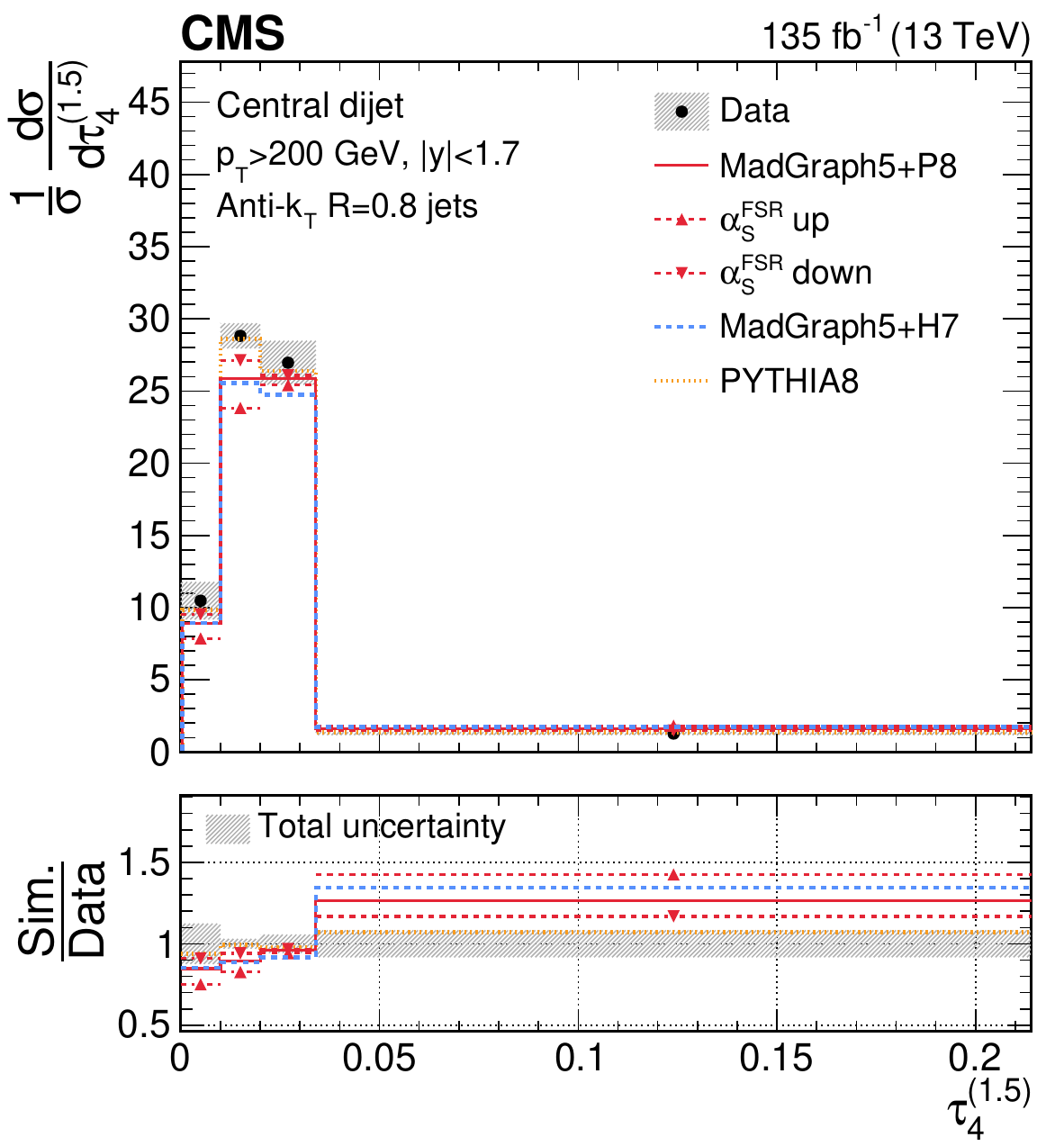} 
	\includegraphics[width=.395\textwidth]{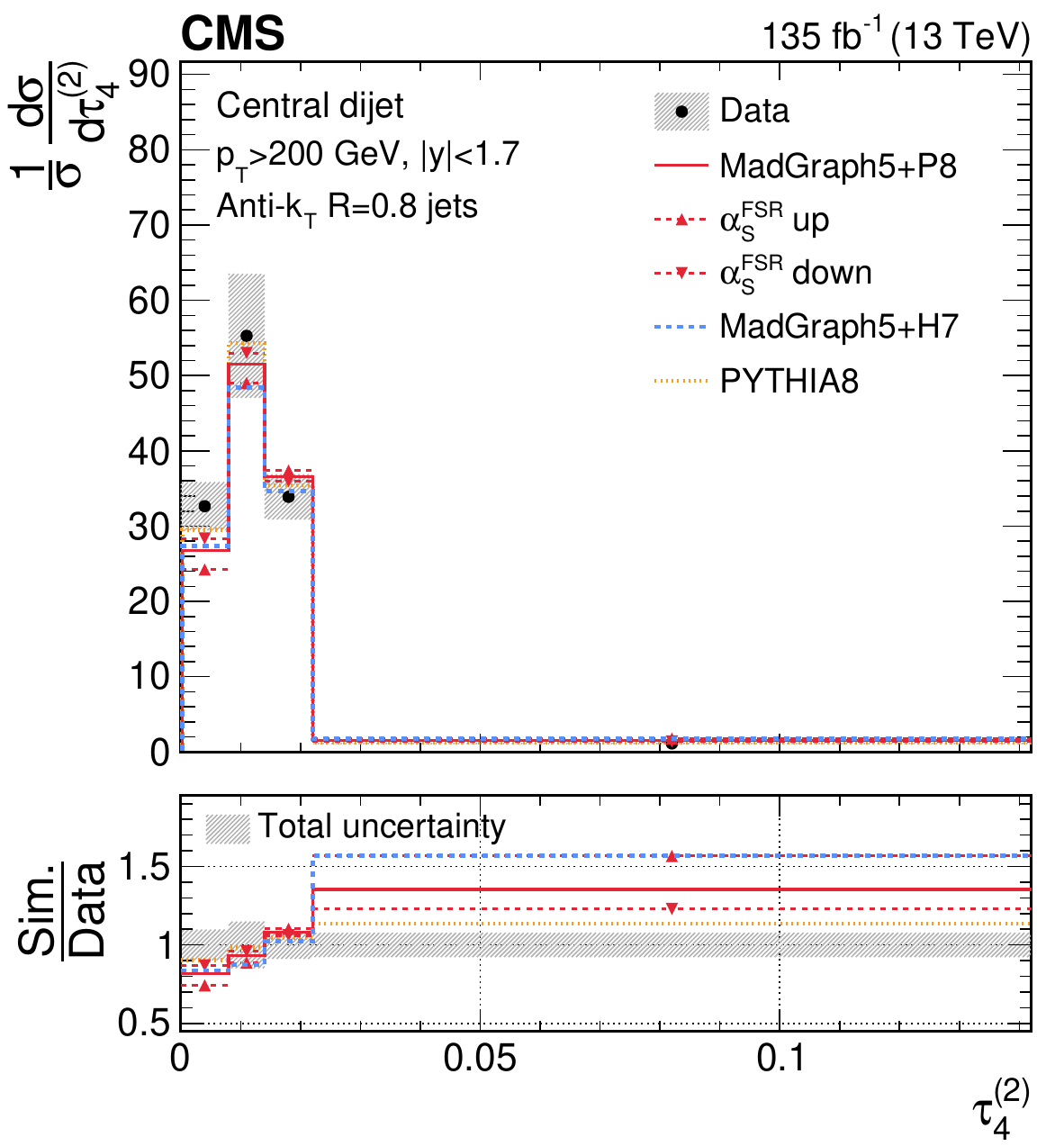} 
	\caption{Unfolded distributions of 4-subjettiness observables, \Nsub{4}{0.25}, \Nsub{4}{0.5}, \Nsub{4}{1}, \Nsub{4}{1.5}, and \Nsub{4}{2}, 
		measured for AK8 jets in the QCD dijet event selection, extracted from the normalized, combined distribution after unfolding; the bin contents and the error bars are scaled by the bin widths for the distributions of the individual observables.  
		For comparisons with particle-level predictions, the error bars in data correspond to the total unfolding uncertainties, 
		and the lower panels present the ratio of particle-level predictions to the unfolded data. 
		The dark grey hashed region illustrates the total uncertainties per bin in the unfolded result.}
	\label{fig:addlresults5bodyDijet}
\end{figure}

\begin{figure}[!htb]
	\centering
	\includegraphics[width=.395\textwidth]{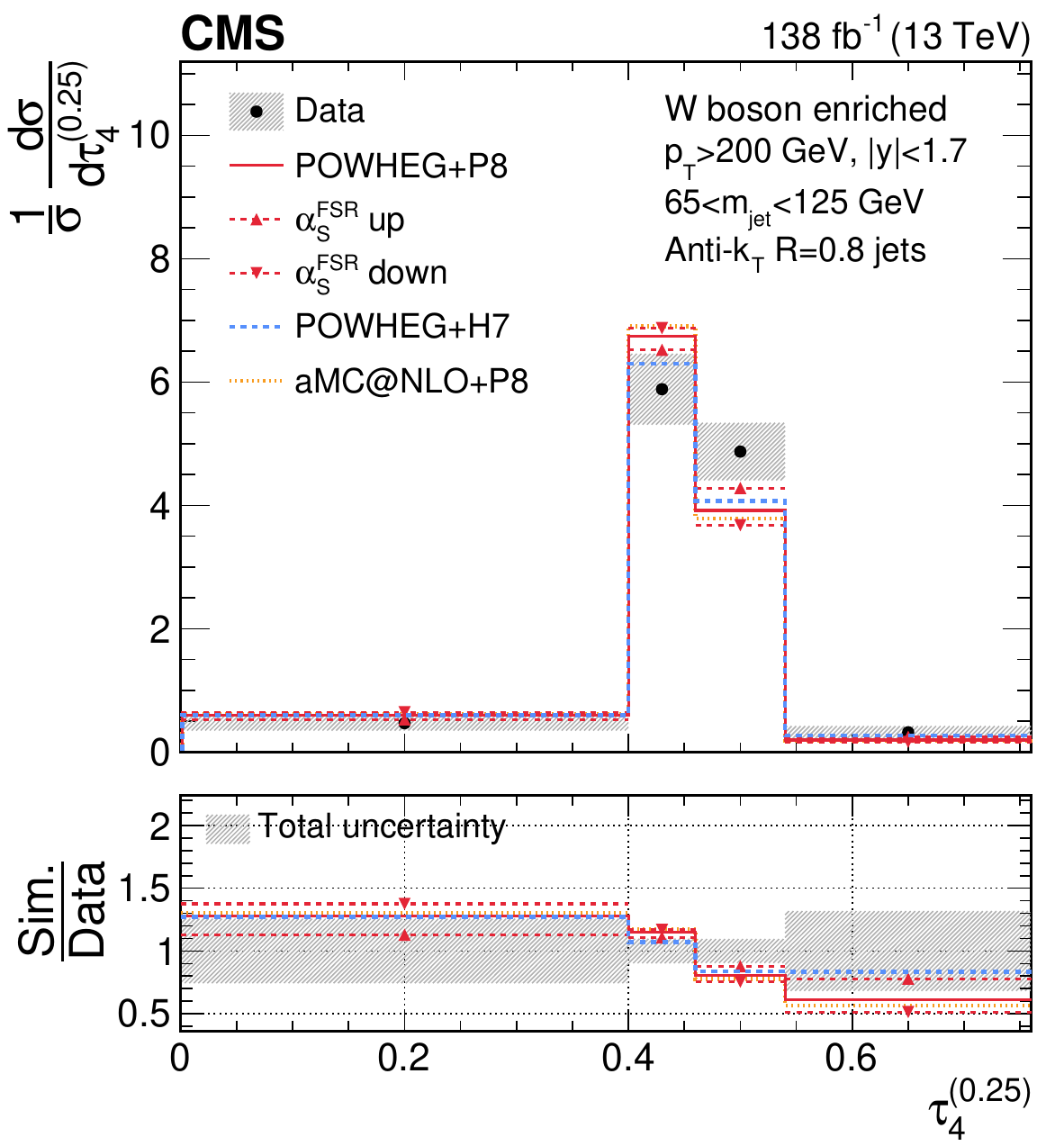} 		
	\includegraphics[width=.395\textwidth]{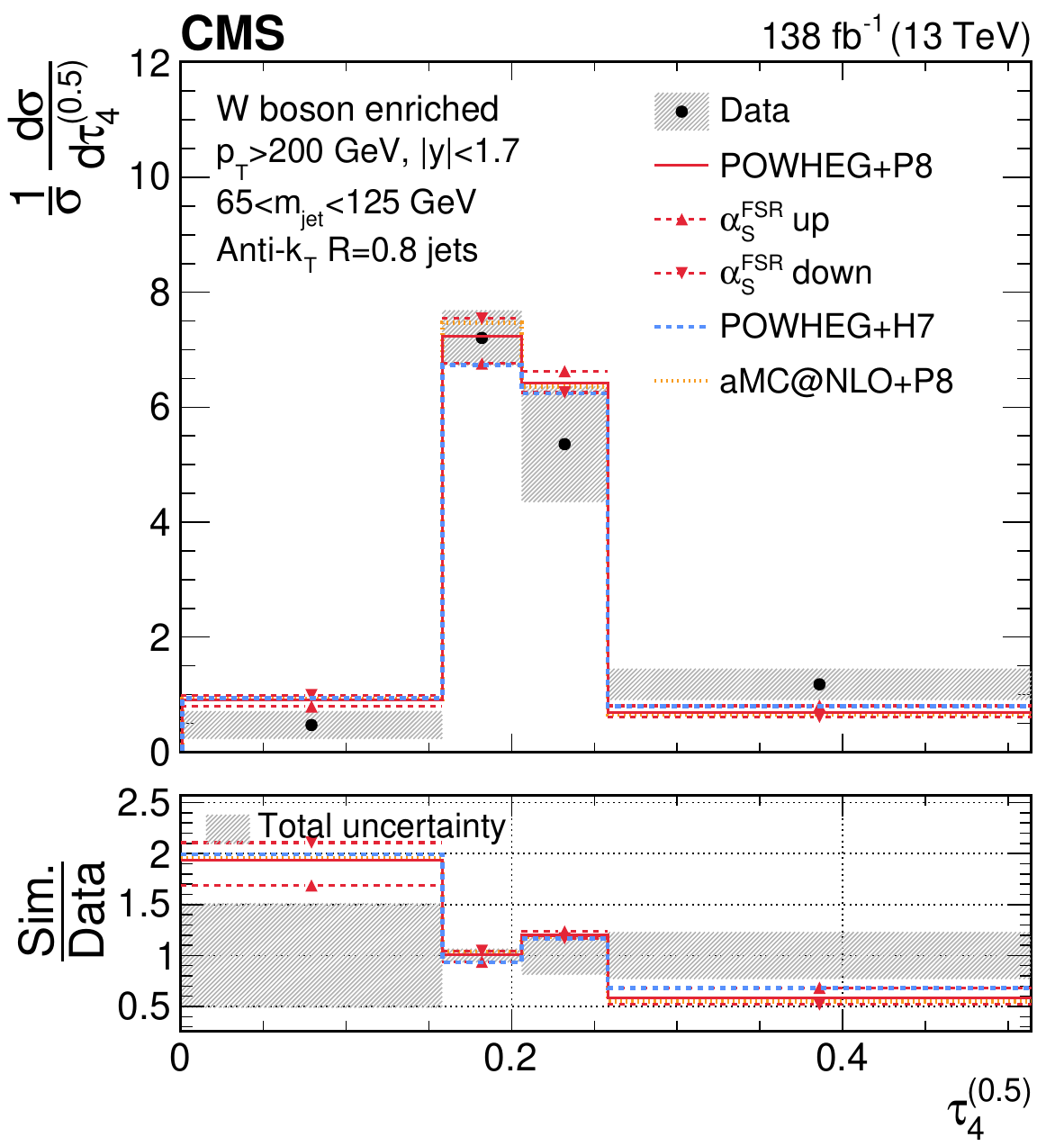} 
	\includegraphics[width=.395\textwidth]{Figure_014-b.pdf} 
	\includegraphics[width=.395\textwidth]{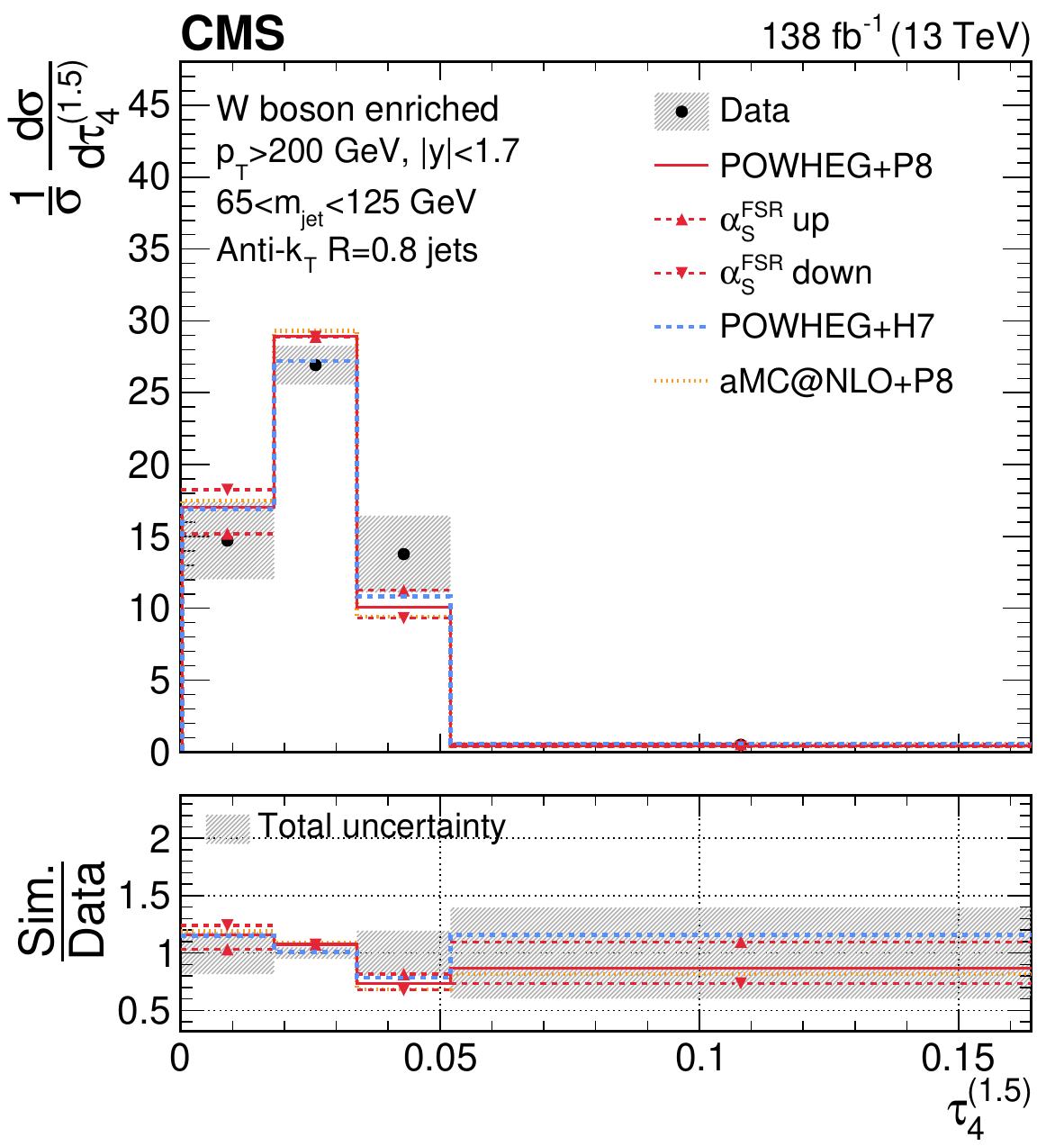} 
	\includegraphics[width=.395\textwidth]{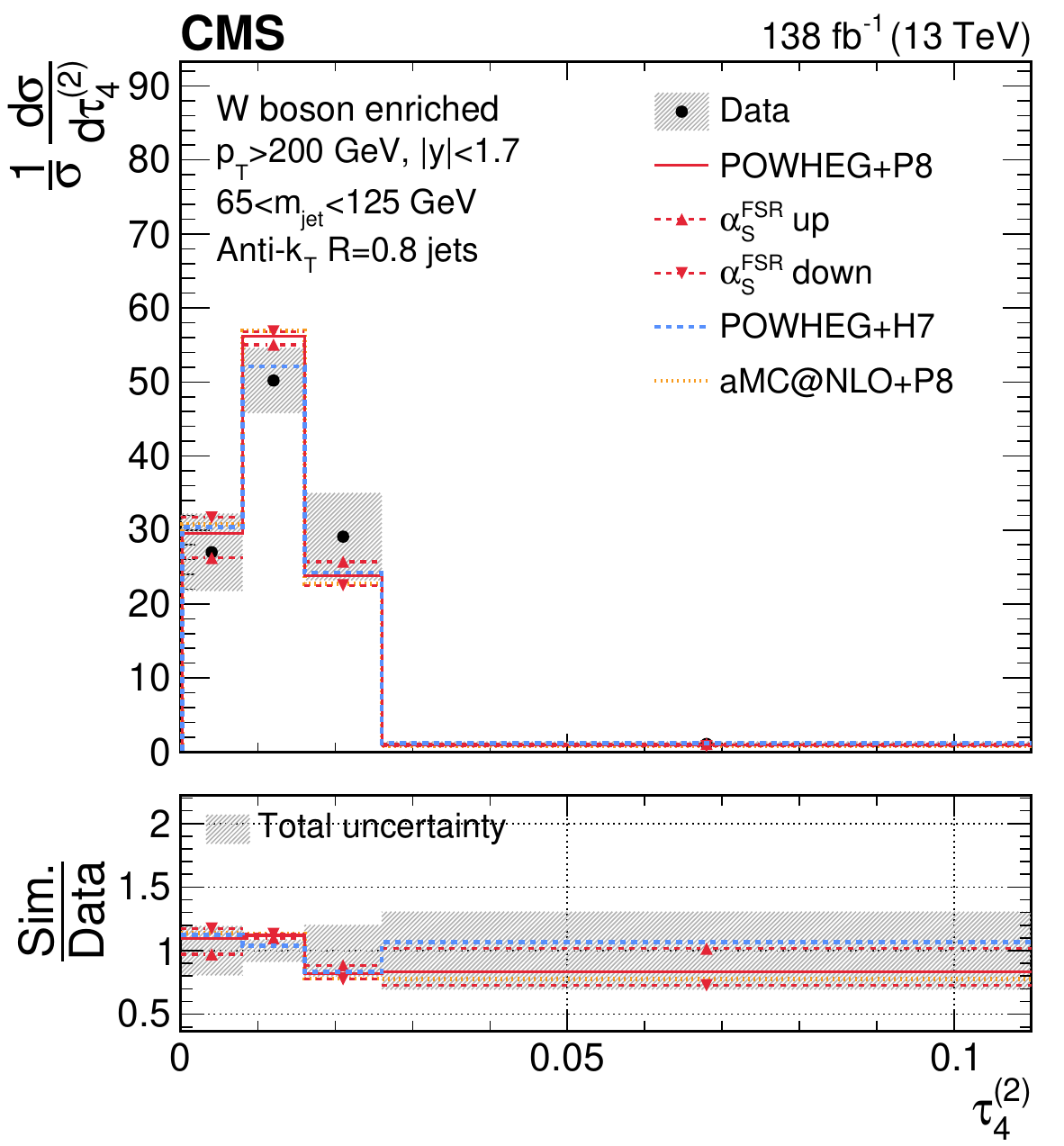} 
	\caption{Unfolded distributions of 4-subjettiness observables, \Nsub{4}{0.25}, \Nsub{4}{0.5}, \Nsub{4}{1}, \Nsub{4}{1.5}, and \Nsub{4}{2}, 
		measured for AK8 jets in boosted \PW boson-enriched events, extracted from the normalized, combined distribution after unfolding; the bin contents and the error bars are scaled by the bin widths for the distributions of the individual observables.  
		For comparisons with particle-level predictions, the error bars in data correspond to the total unfolding uncertainties, 
		and the lower panels present the ratio of particle-level predictions to the unfolded data. 
		The dark grey hashed region illustrates the total uncertainties per bin in the unfolded result.}
	\label{fig:addlresults5bodyW}
\end{figure}

\begin{figure}[!htb]
	\centering
	\includegraphics[width=.395\textwidth]{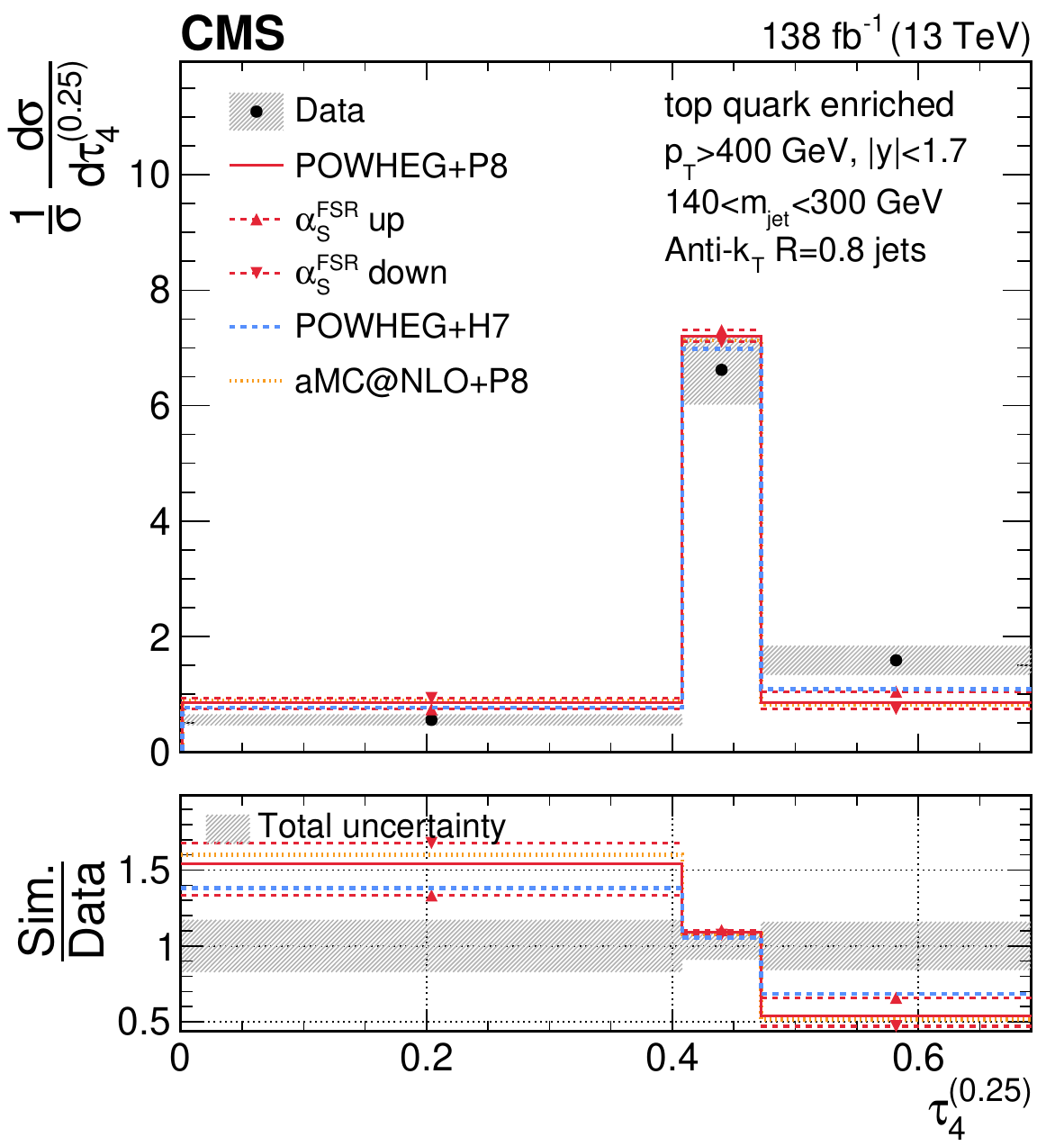} 		
	\includegraphics[width=.395\textwidth]{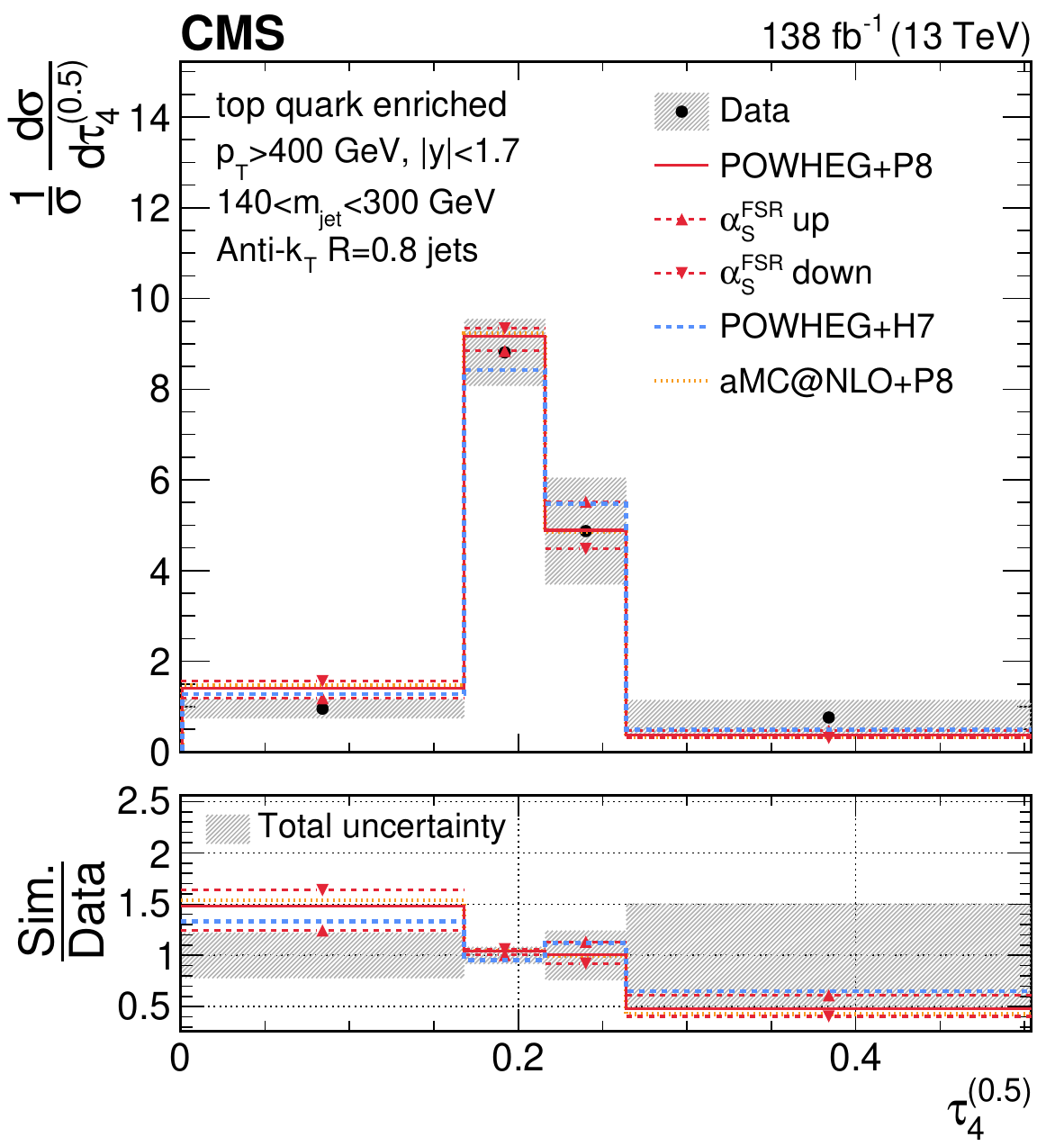} 
	\includegraphics[width=.395\textwidth]{Figure_015-b.pdf} 
	\includegraphics[width=.395\textwidth]{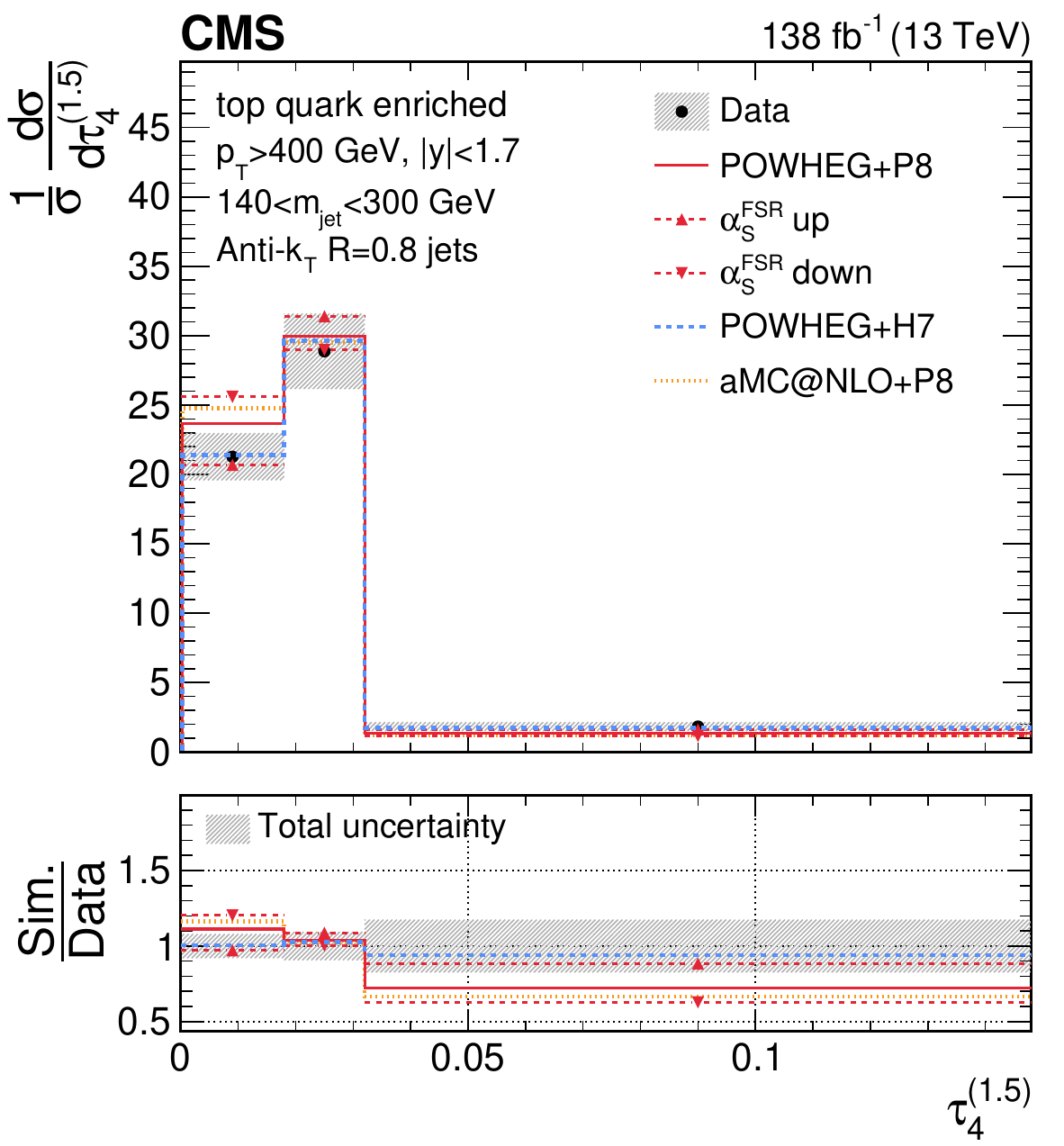} 
	\includegraphics[width=.395\textwidth]{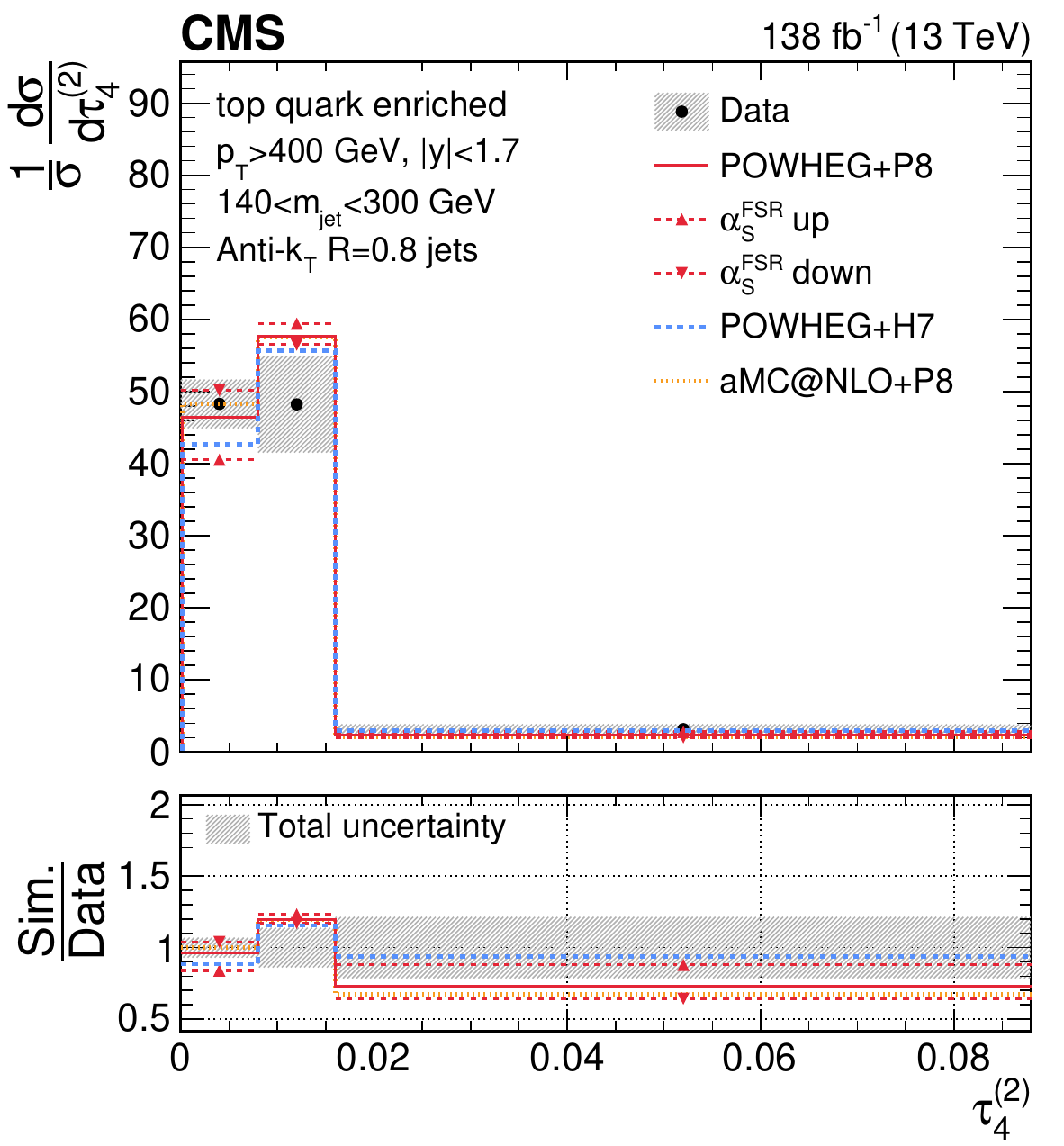} 
	\caption{Unfolded distributions of 4-subjettiness observables, \Nsub{4}{0.25}, \Nsub{4}{0.5}, \Nsub{4}{1}, \Nsub{4}{1.5}, and \Nsub{4}{2}, 
		measured for AK8 jets in the boosted top quark-enriched region, extracted from the normalized, combined distribution after unfolding; the bin contents and the error bars are scaled by the bin widths for the distributions of the individual observables.  
		For comparisons with particle-level predictions, the error bars in data correspond to the total unfolding uncertainties, 
		and the lower panels present the ratio of particle-level predictions to the unfolded data. 
		The dark grey hashed region illustrates the total uncertainties per bin in the unfolded result.}
	\label{fig:addlresults5bodytop}
\end{figure}

The measurements of \Nsub{4}{0.25}, \Nsub{4}{0.5}, \Nsub{4}{1}, \Nsub{4}{1.5}, and \Nsub{4}{2} are presented. 
The unfolded results for the dijet selection are shown in Fig.~\ref{fig:addlresults5bodyDijet}, and results for the boosted \PW boson- and top quark-enriched regions are shown in Figs.~\ref{fig:addlresults5bodyW} and \ref{fig:addlresults5bodytop}, respectively.

\clearpage
\newpage

\subsection{6-body phase space}
\label{sec:addlmeasurements6body}
\begin{figure}[!htb]
	\centering
	\includegraphics[width=.395\textwidth]{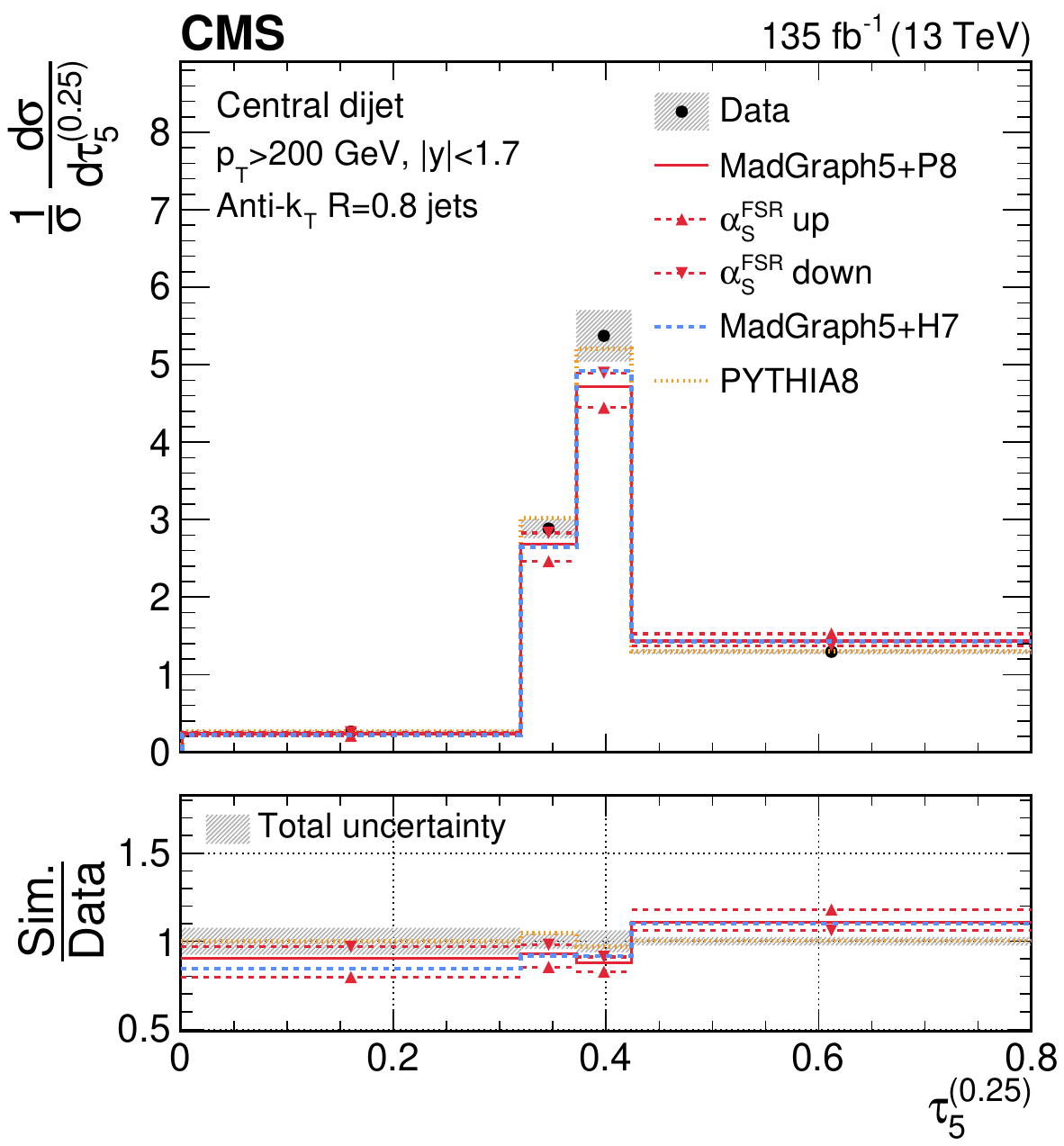} 		
	\includegraphics[width=.395\textwidth]{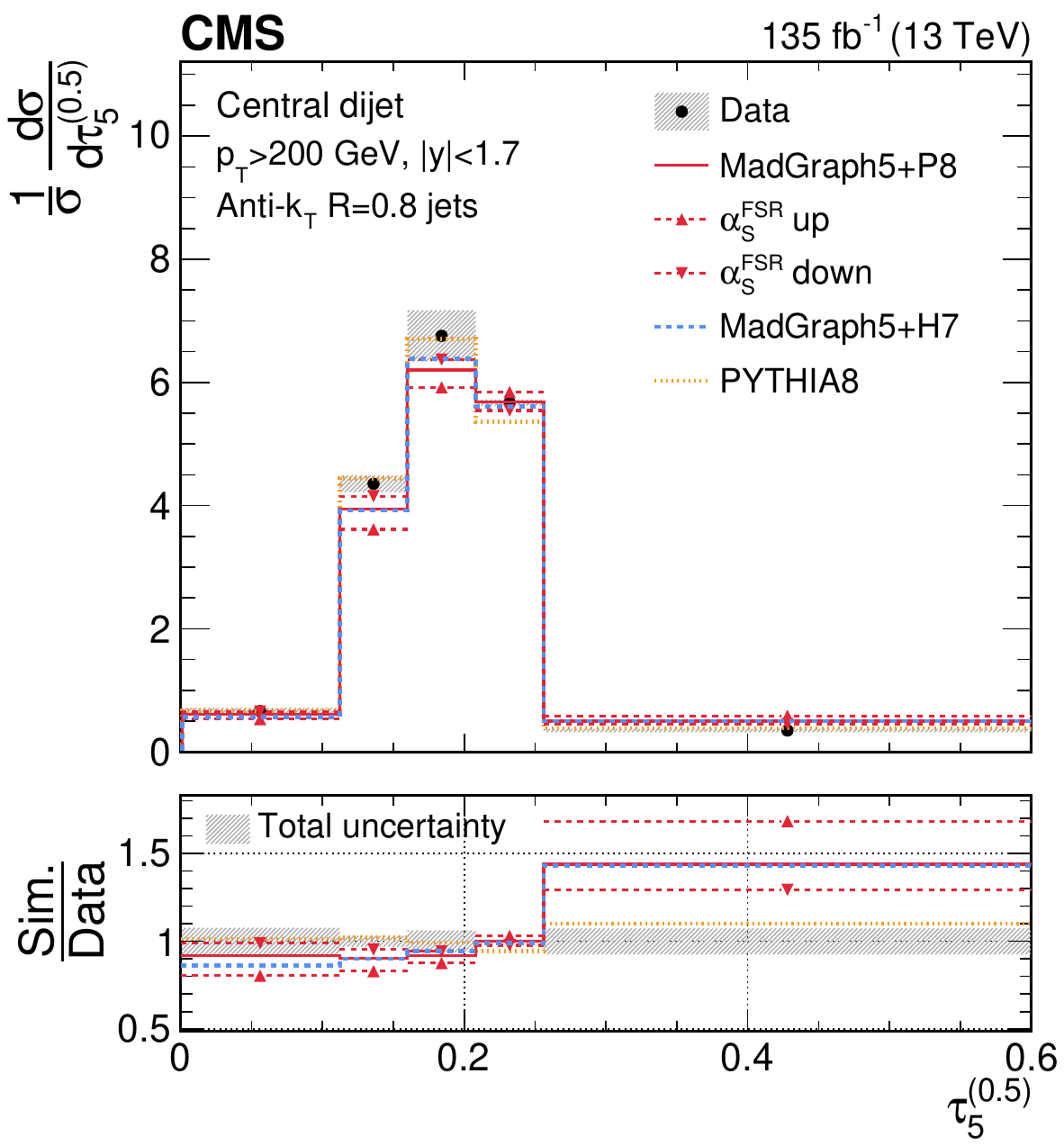} 
	\includegraphics[width=.395\textwidth]{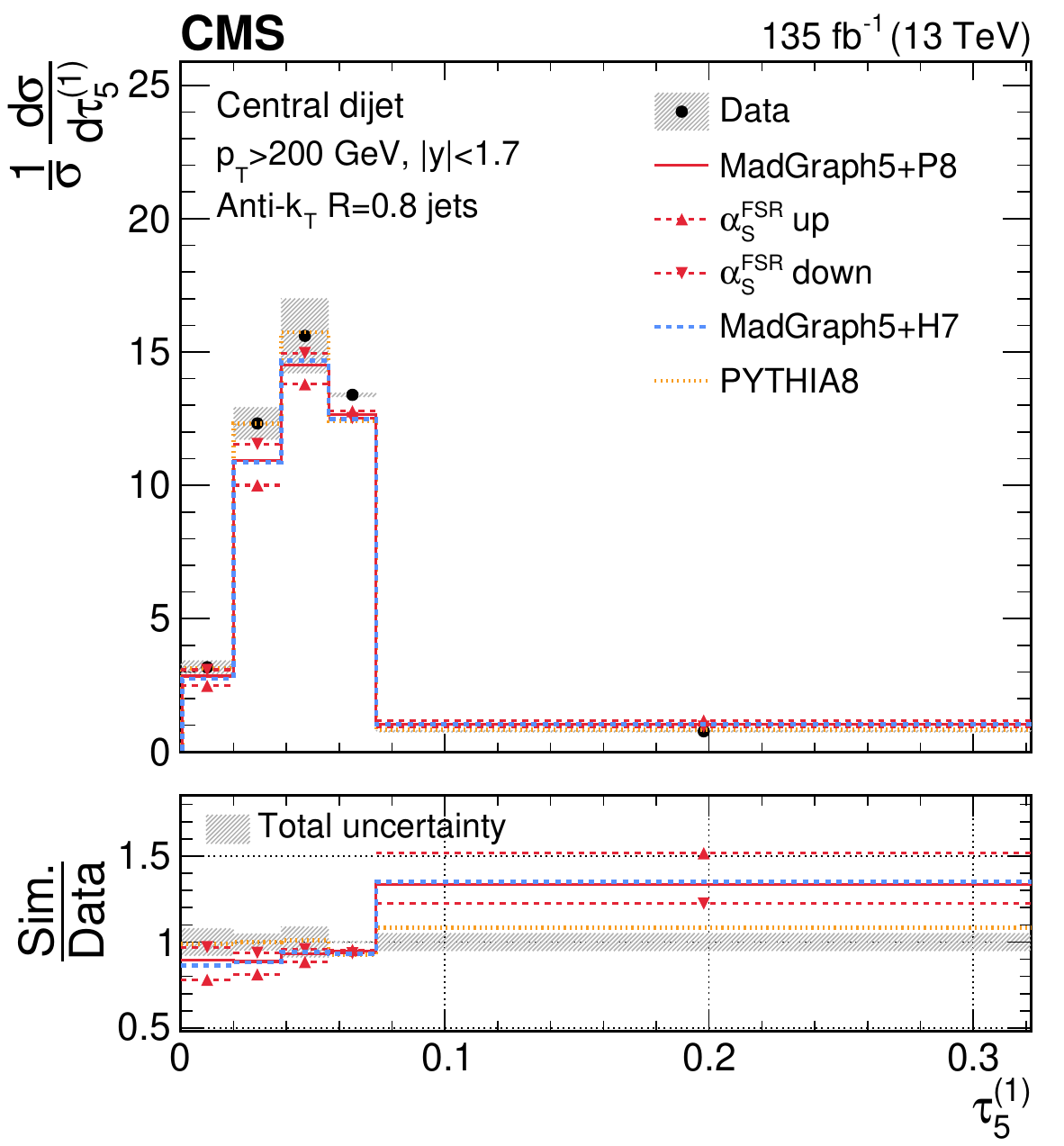} 
	\includegraphics[width=.395\textwidth]{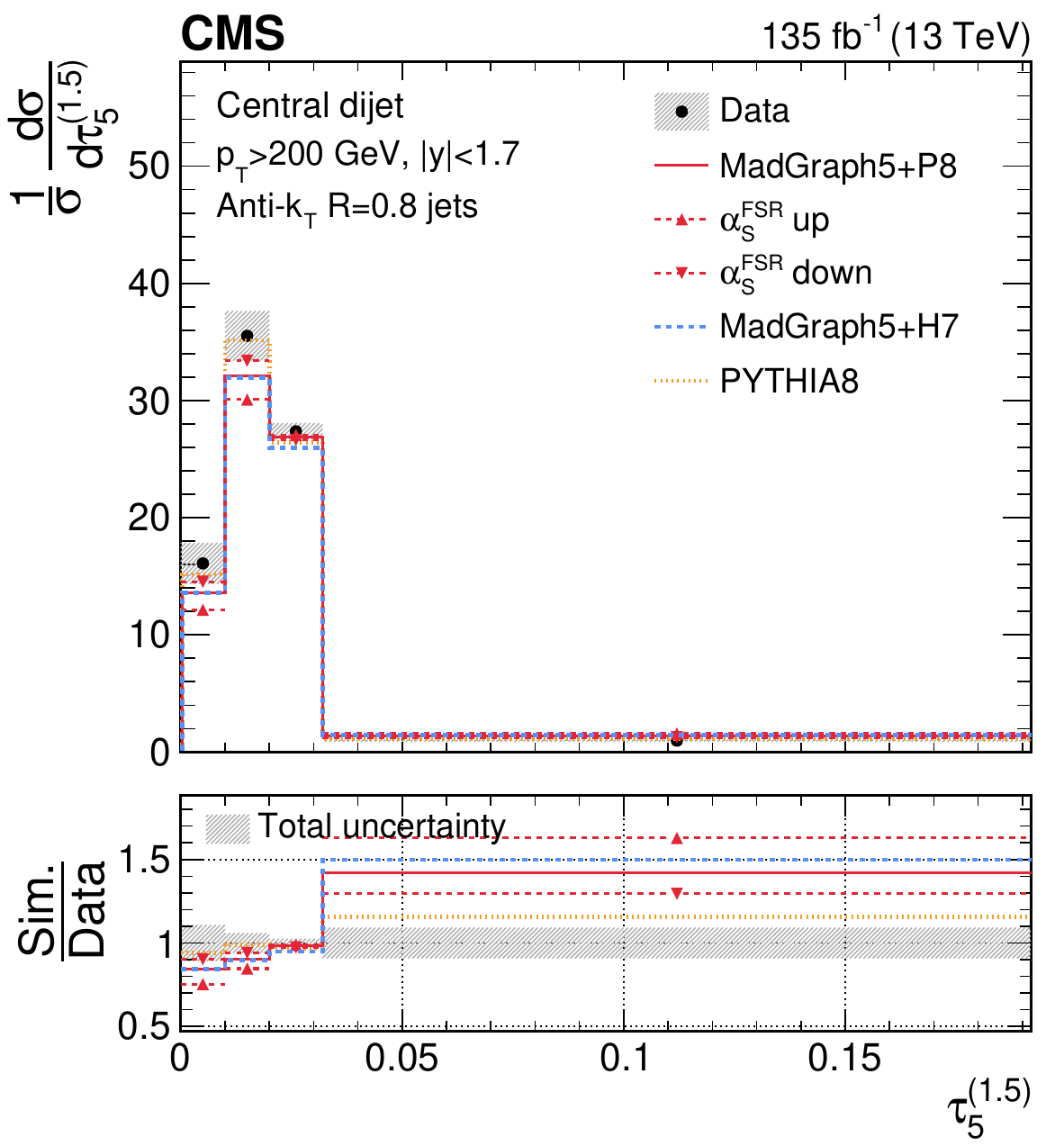} 
	\includegraphics[width=.395\textwidth]{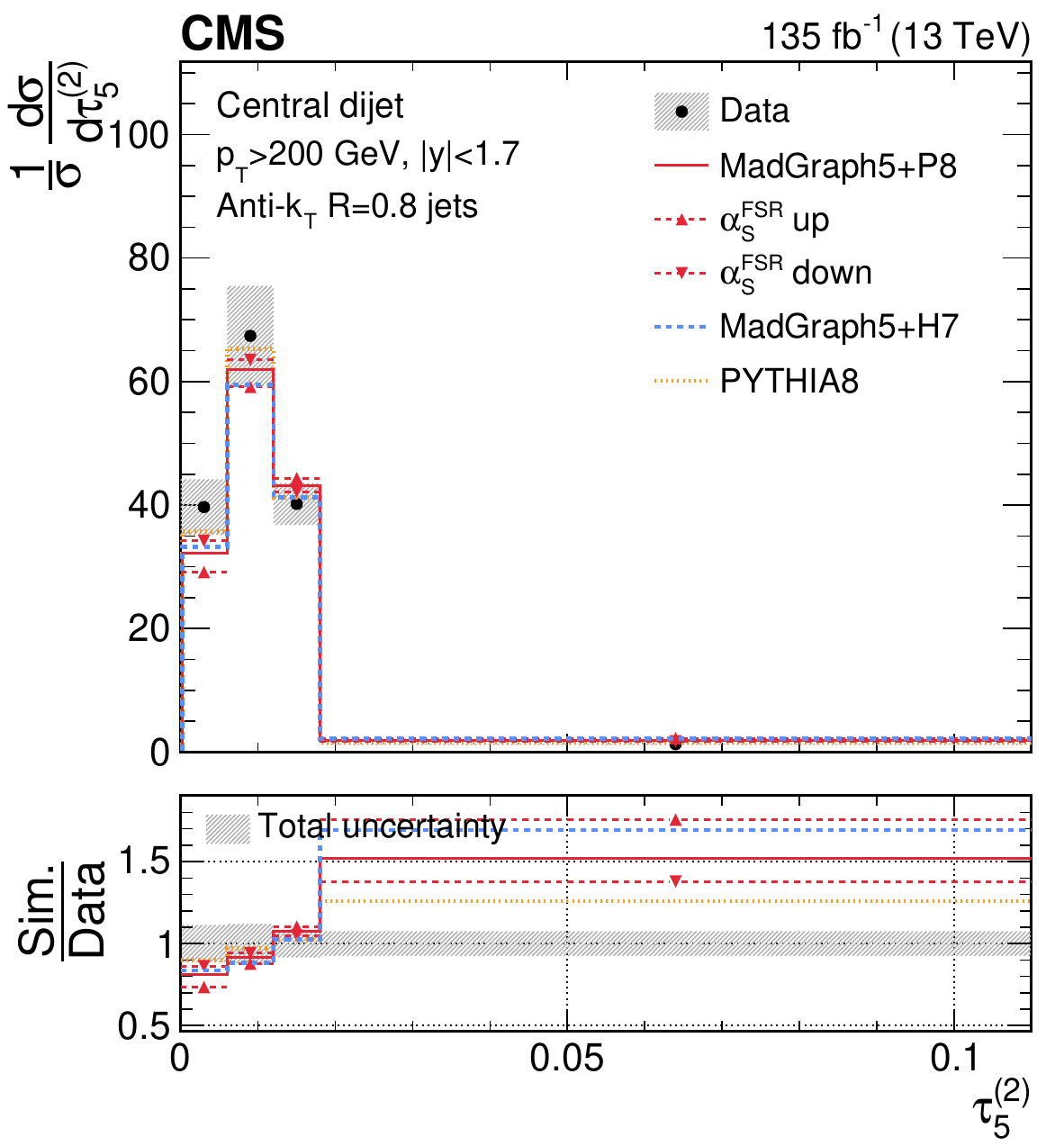} 
	\caption{Unfolded distributions of 5-subjettiness observables, \Nsub{5}{0.25}, \Nsub{5}{0.5}, \Nsub{5}{1}, \Nsub{5}{1.5}, and \Nsub{5}{2}, 
		measured for AK8 jets in the QCD dijet event selection, extracted from the normalized, combined distribution after unfolding; the bin contents and the error bars are scaled by the bin widths for the distributions of the individual observables.  
		For comparisons with particle-level predictions, the error bars in data correspond to the total unfolding uncertainties, 
		and the lower panels present the ratio of particle-level predictions to the unfolded data. 
		The dark grey hashed region illustrates the total uncertainties per bin in the unfolded result.}
	\label{fig:addlresults6bodyDijet}
\end{figure}

\begin{figure}[!htb]
	\centering
	\includegraphics[width=.395\textwidth]{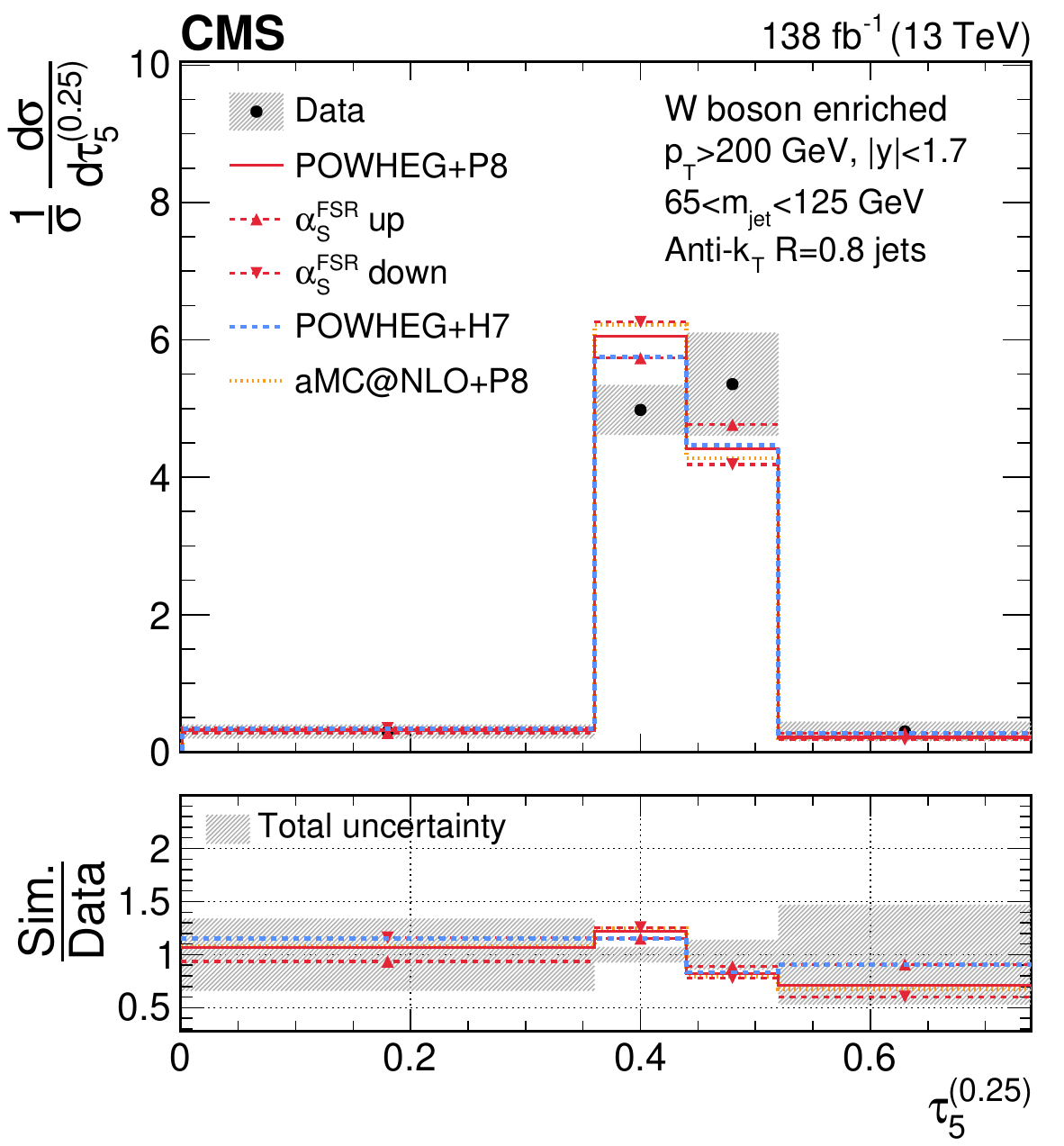} 		
	\includegraphics[width=.395\textwidth]{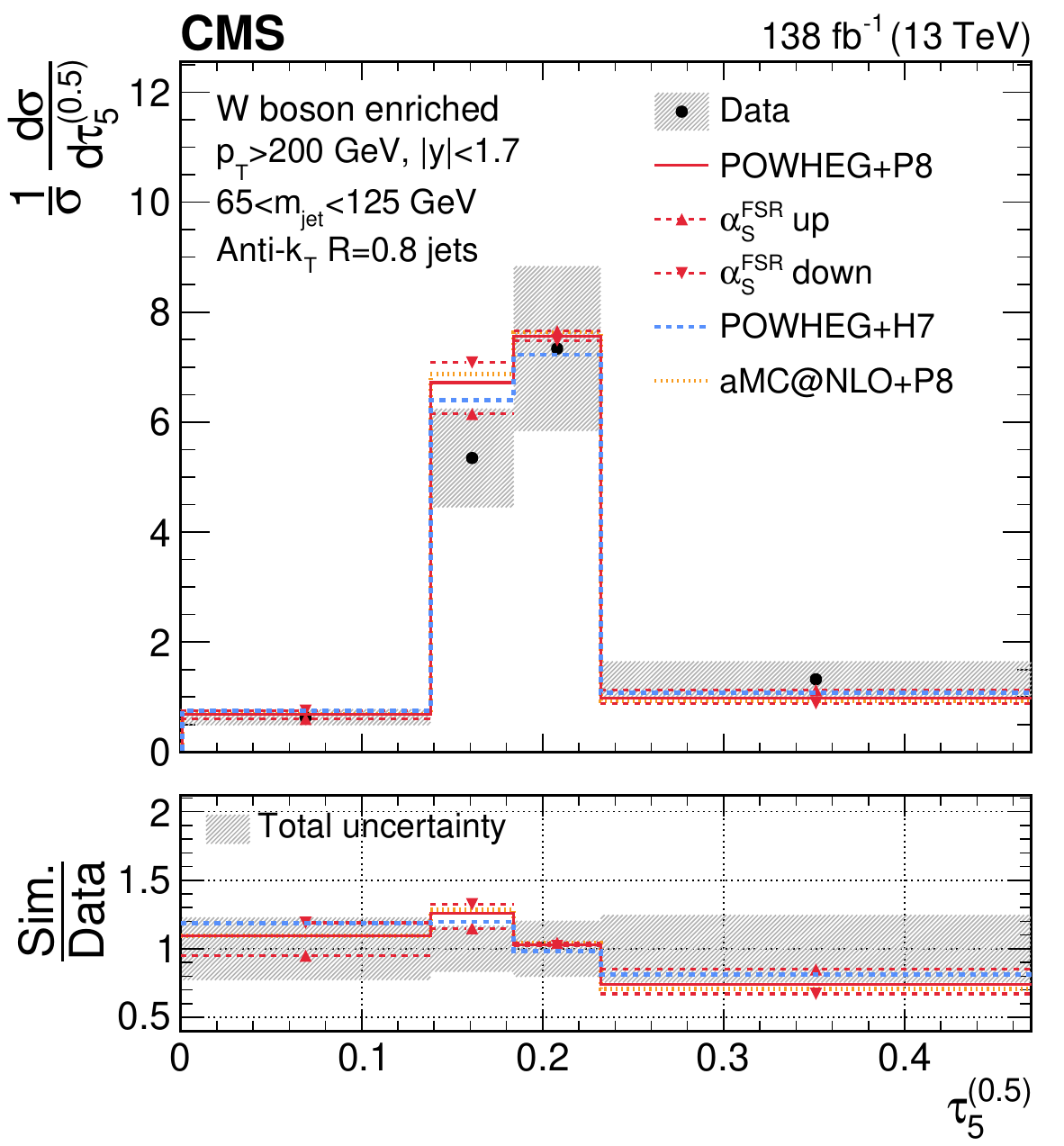} 
	\includegraphics[width=.395\textwidth]{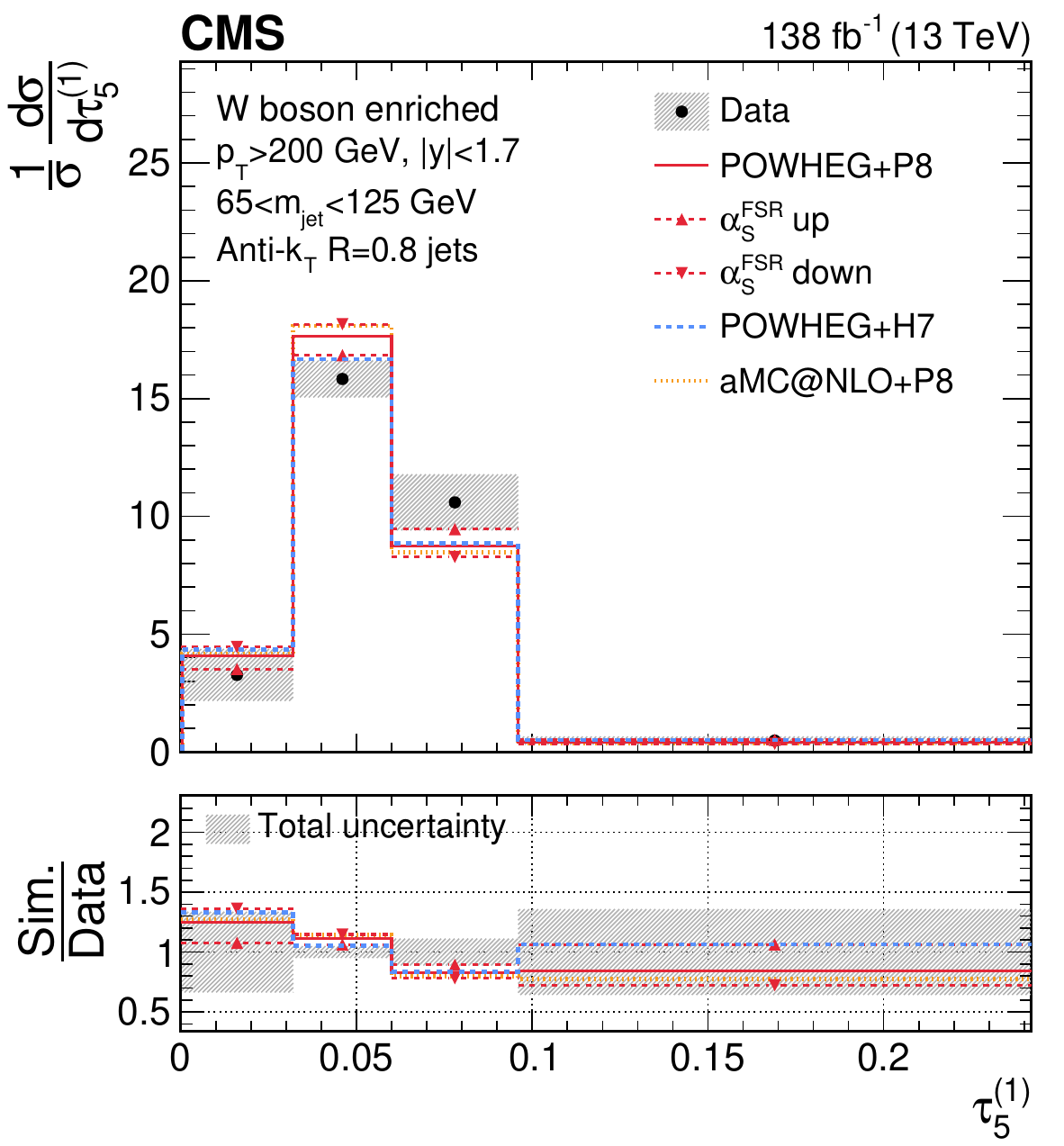} 
	\includegraphics[width=.395\textwidth]{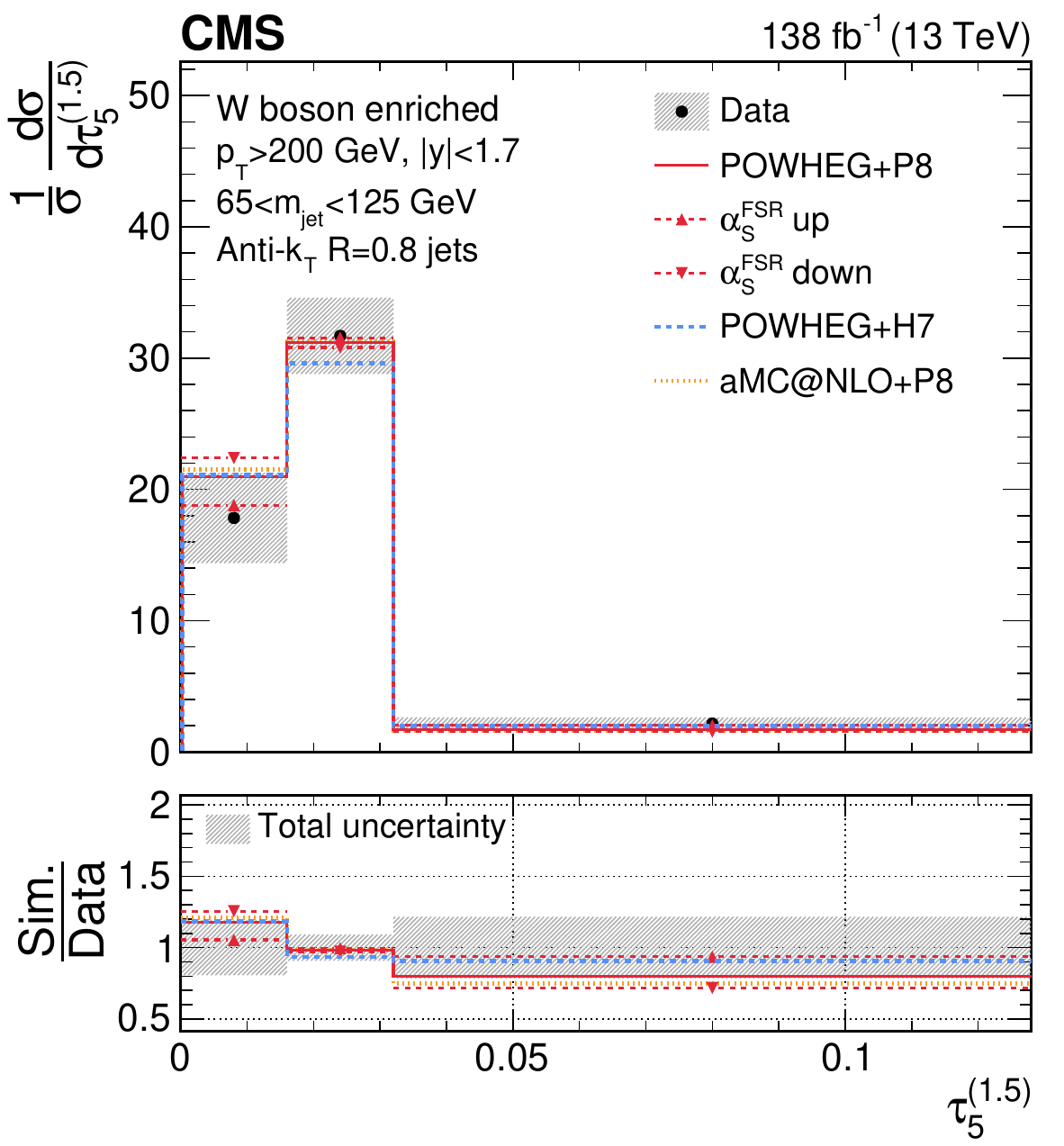} 
	\includegraphics[width=.395\textwidth]{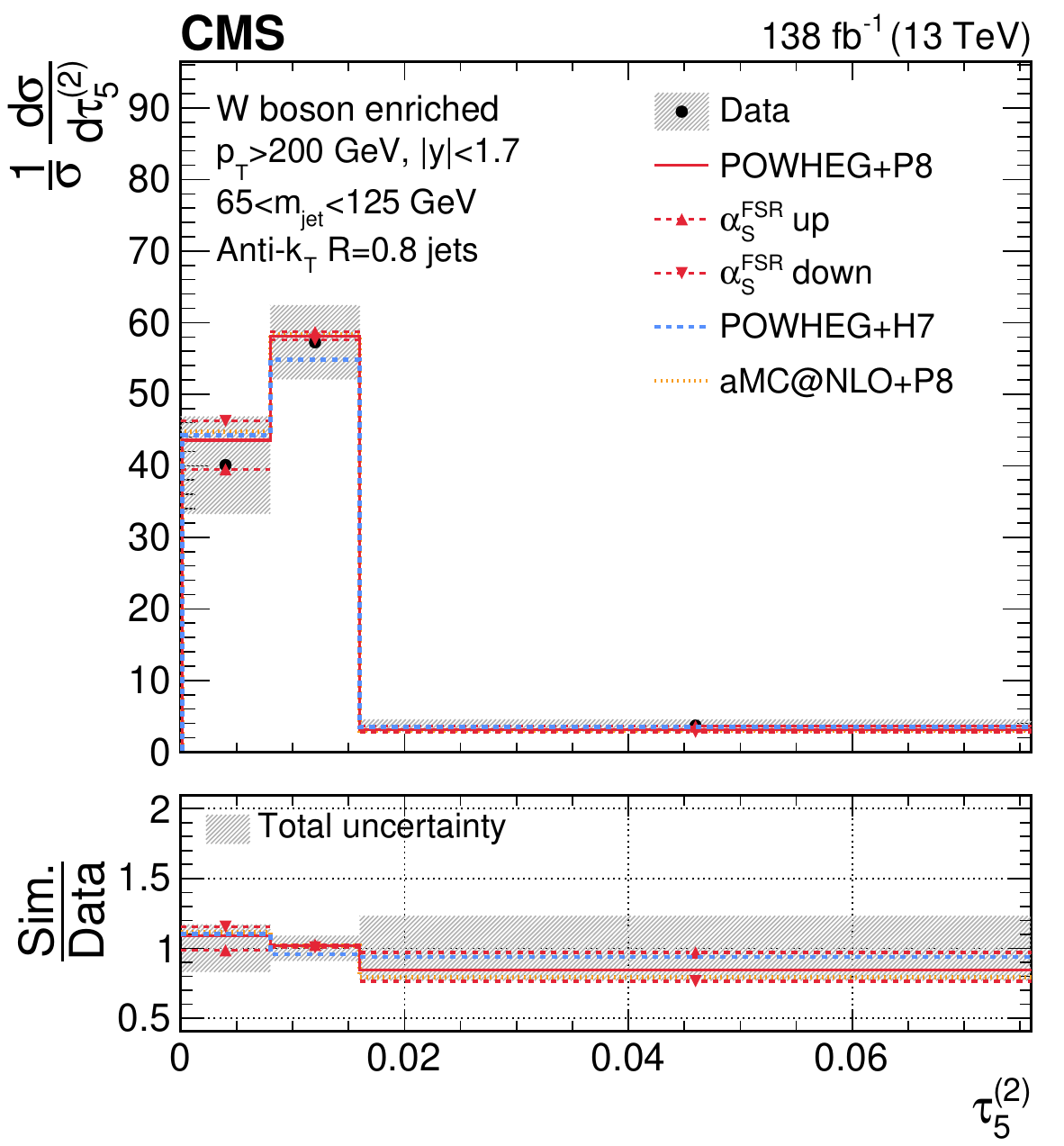} 
	\caption{Unfolded distributions of 5-subjettiness observables, \Nsub{5}{0.25}, \Nsub{5}{0.5}, \Nsub{5}{1}, \Nsub{5}{1.5}, and \Nsub{5}{2}, 
		measured for AK8 jets in boosted \PW boson-enriched events, extracted from the normalized, combined distribution after unfolding; the bin contents and the error bars are scaled by the bin widths for the distributions of the individual observables.  
		For comparisons with particle-level predictions, the error bars in data correspond to the total unfolding uncertainties, 
		and the lower panels present the ratio of particle-level predictions to the unfolded data. 
		The dark grey hashed region illustrates the total uncertainties per bin in the unfolded result.}
	\label{fig:addlresults6bodyW}
\end{figure}

\begin{figure}[!htb]
	\centering
	\includegraphics[width=.395\textwidth]{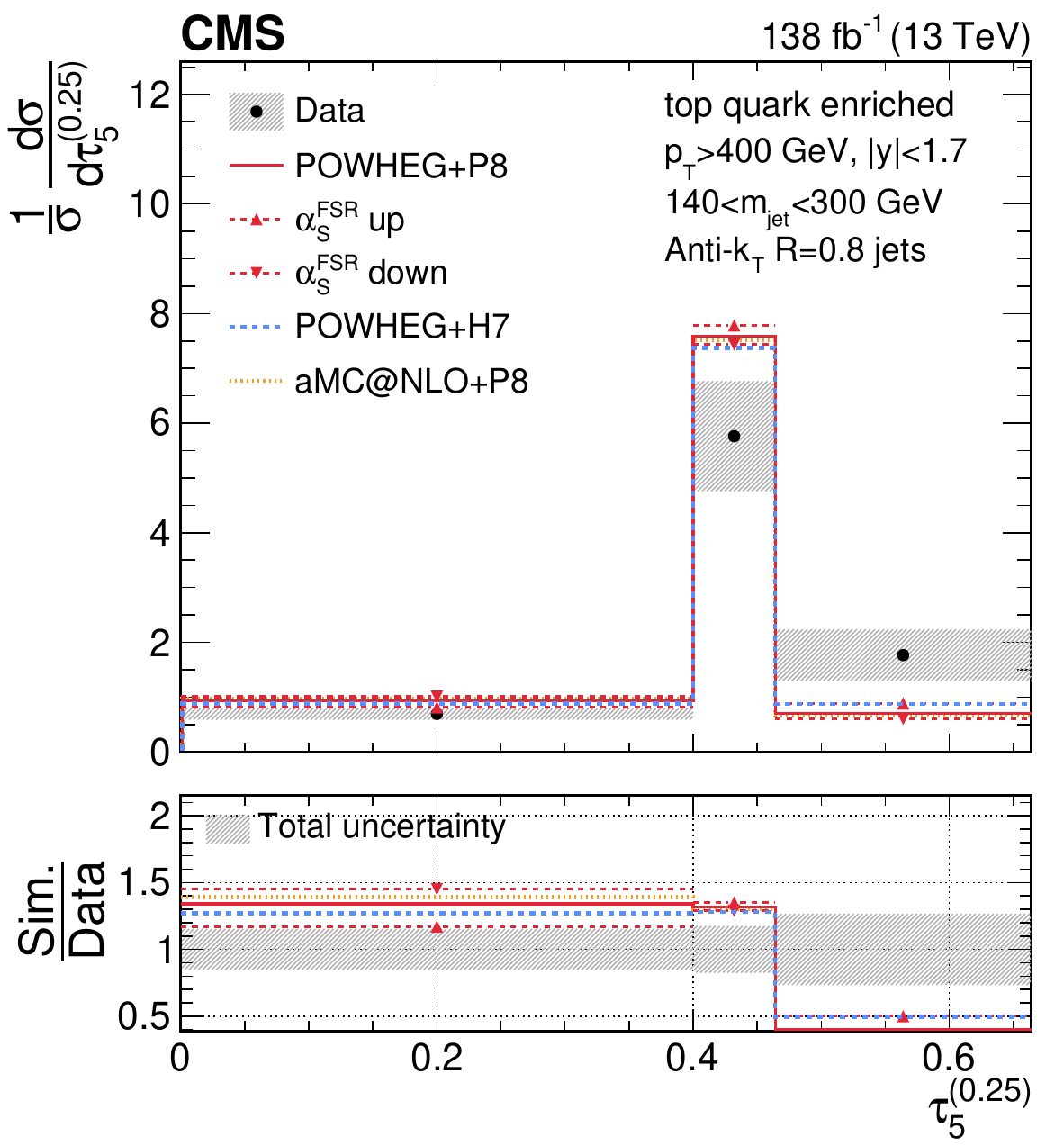} 		
	\includegraphics[width=.395\textwidth]{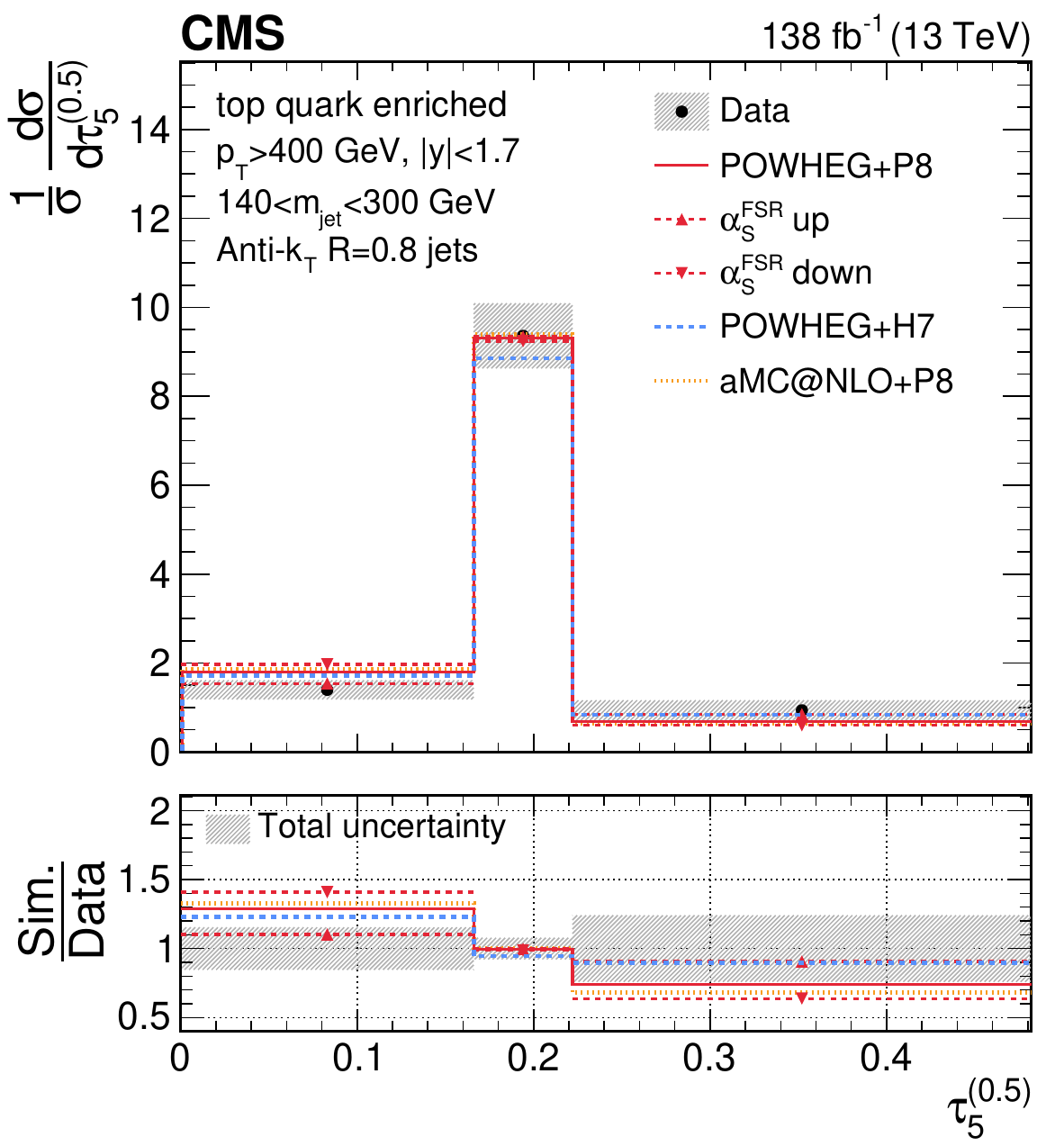} 
	\includegraphics[width=.395\textwidth]{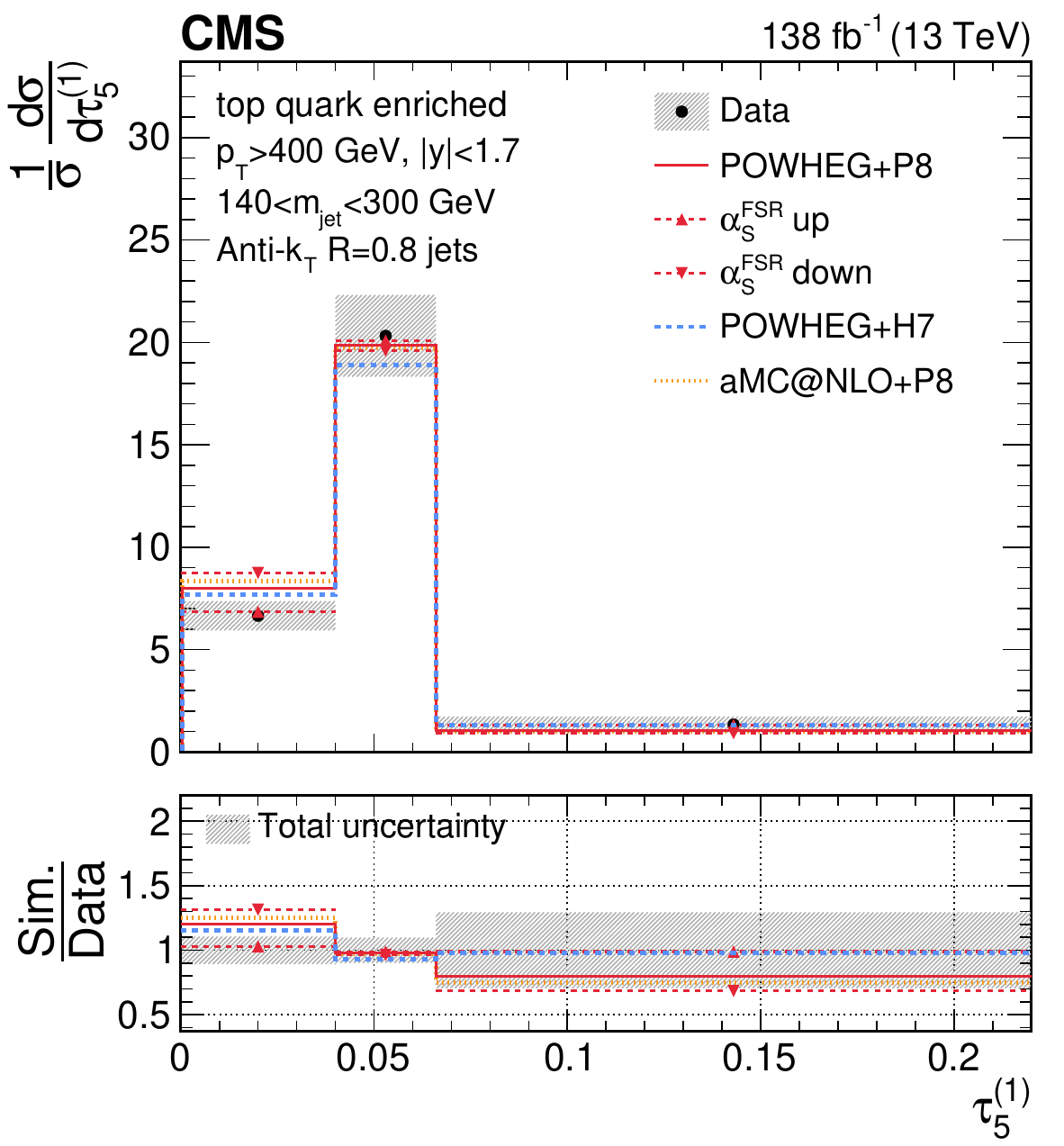} 
	\includegraphics[width=.395\textwidth]{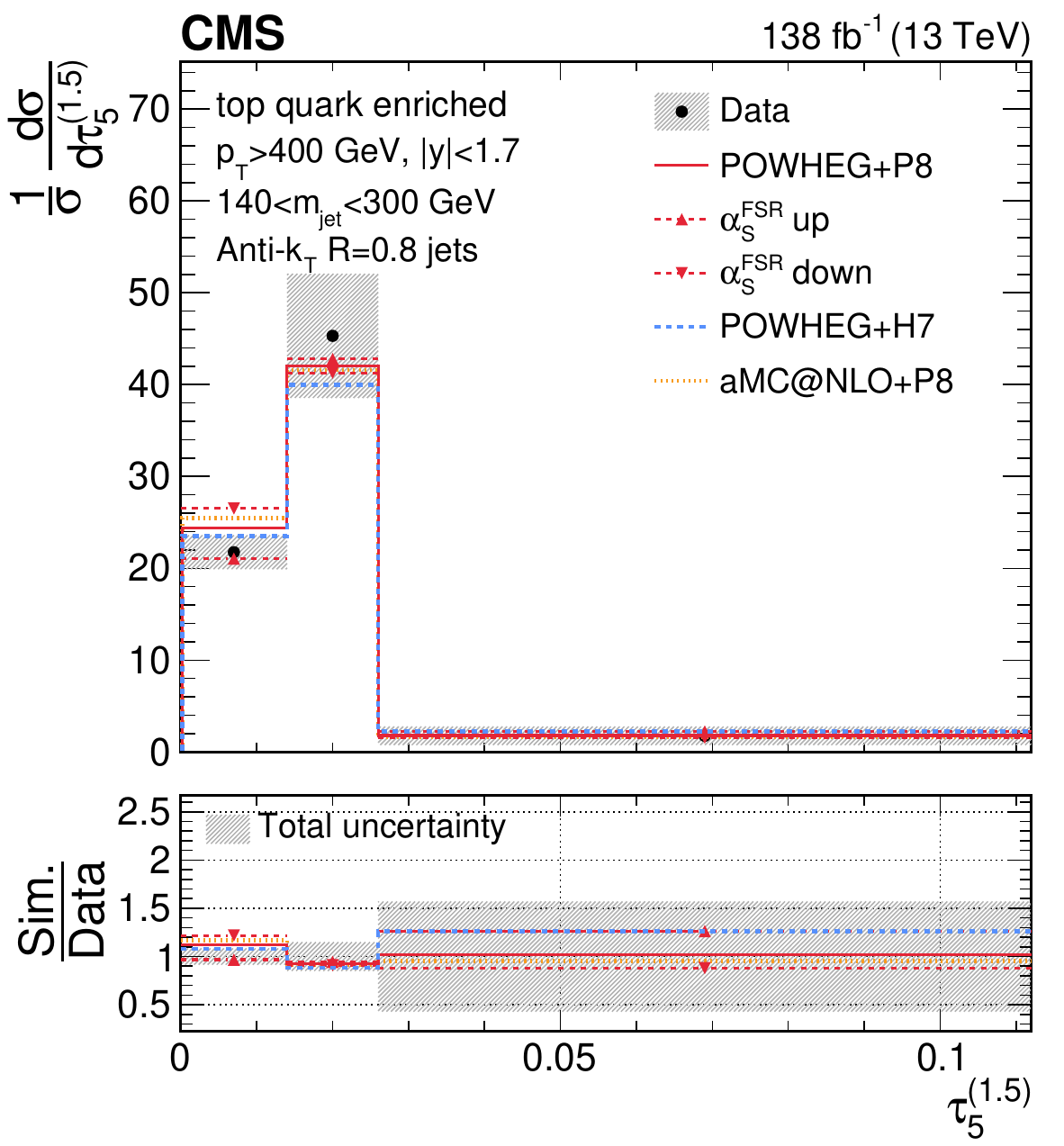} 
	\includegraphics[width=.395\textwidth]{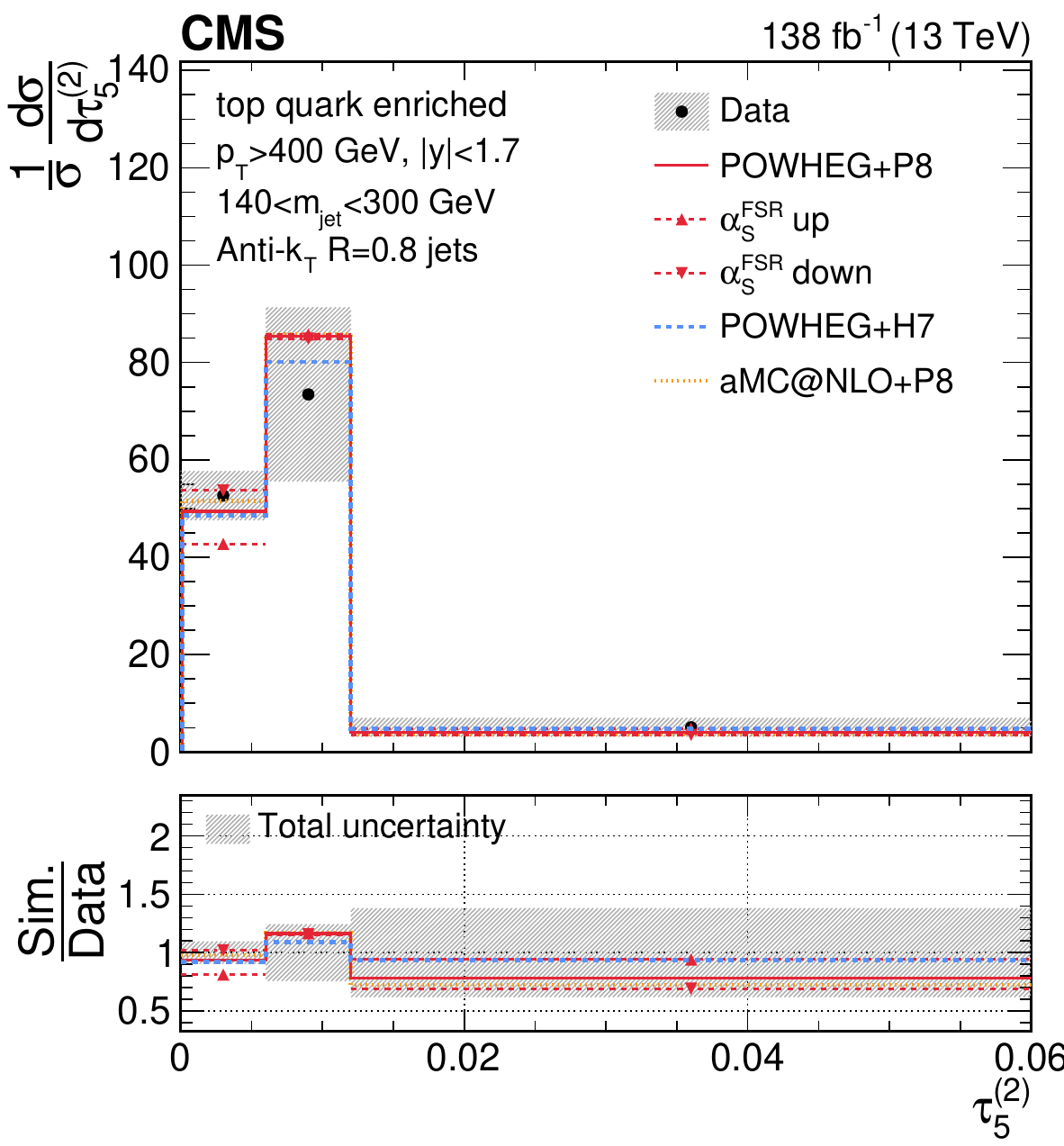} 
	\caption{Unfolded distributions of 5-subjettiness observables, \Nsub{5}{0.25}, \Nsub{5}{0.5}, \Nsub{5}{1}, \Nsub{5}{1.5}, and \Nsub{5}{2}, 
		measured for AK8 jets in the boosted top quark-enriched region, extracted from the normalized, combined distribution after unfolding; the bin contents and the error bars are scaled by the bin widths for the distributions of the individual observables.  
		For comparisons with particle-level predictions, the error bars in data correspond to the total unfolding uncertainties, 
		and the lower panels present the ratio of particle-level predictions to the unfolded data. 
		The dark grey hashed region illustrates the total uncertainties per bin in the unfolded result.}
	\label{fig:addlresults6bodytop}
\end{figure}

The measurements of \Nsub{5}{0.25}, \Nsub{5}{0.5}, \Nsub{5}{1}, \Nsub{5}{1.5}, and \Nsub{5}{2} are presented. 
The unfolded results for the dijet selection are shown in Fig.~\ref{fig:addlresults6bodyDijet}, and results for the boosted \PW boson- and top quark-enriched regions are shown in Figs.~\ref{fig:addlresults6bodyW} and \ref{fig:addlresults6bodytop}, respectively.

\clearpage
\newpage

\subsection{Unfolding uncertainties: gluon and light-flavour quark jets}
\label{sec:dijetUnfUncs}
\begin{figure}[htpb]
	\centering
	\includegraphics[width=.42\textwidth]{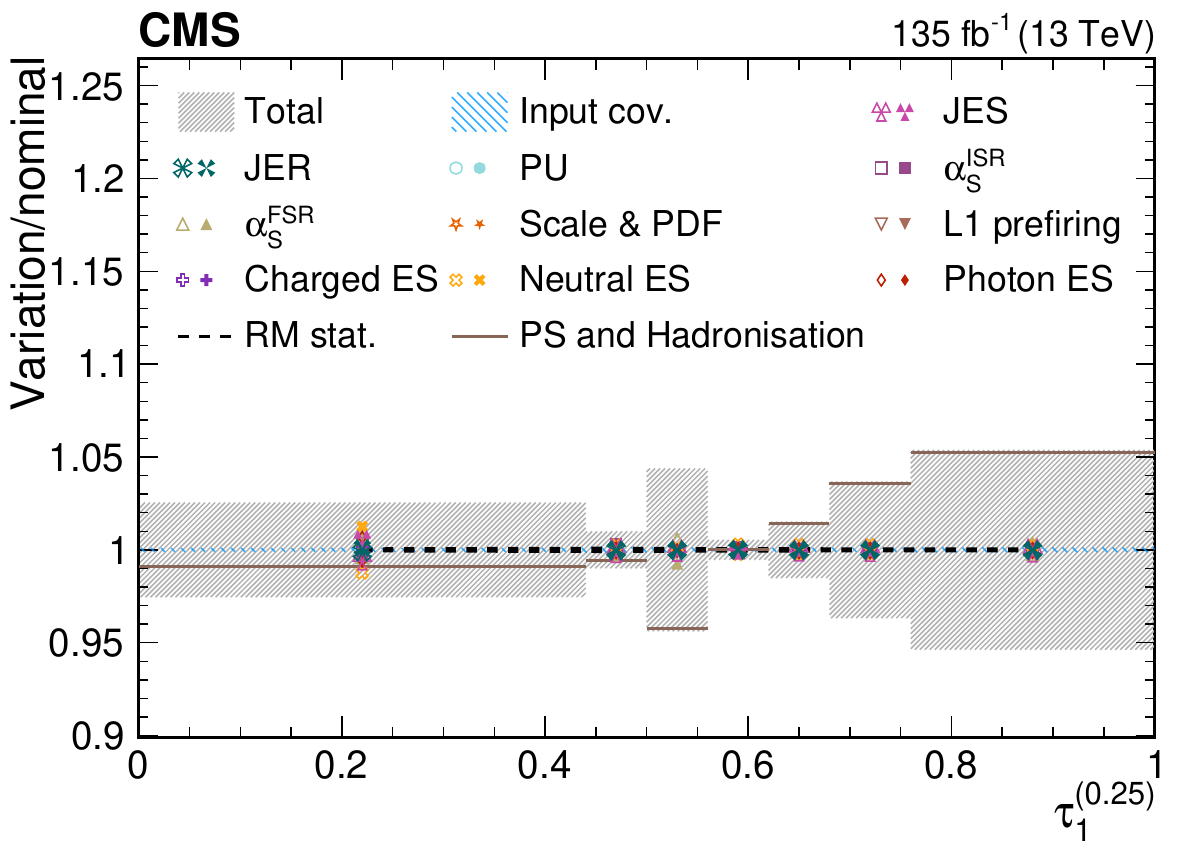}
	\includegraphics[width=.42\textwidth]{Figure_016-a.pdf}
	\includegraphics[width=.42\textwidth]{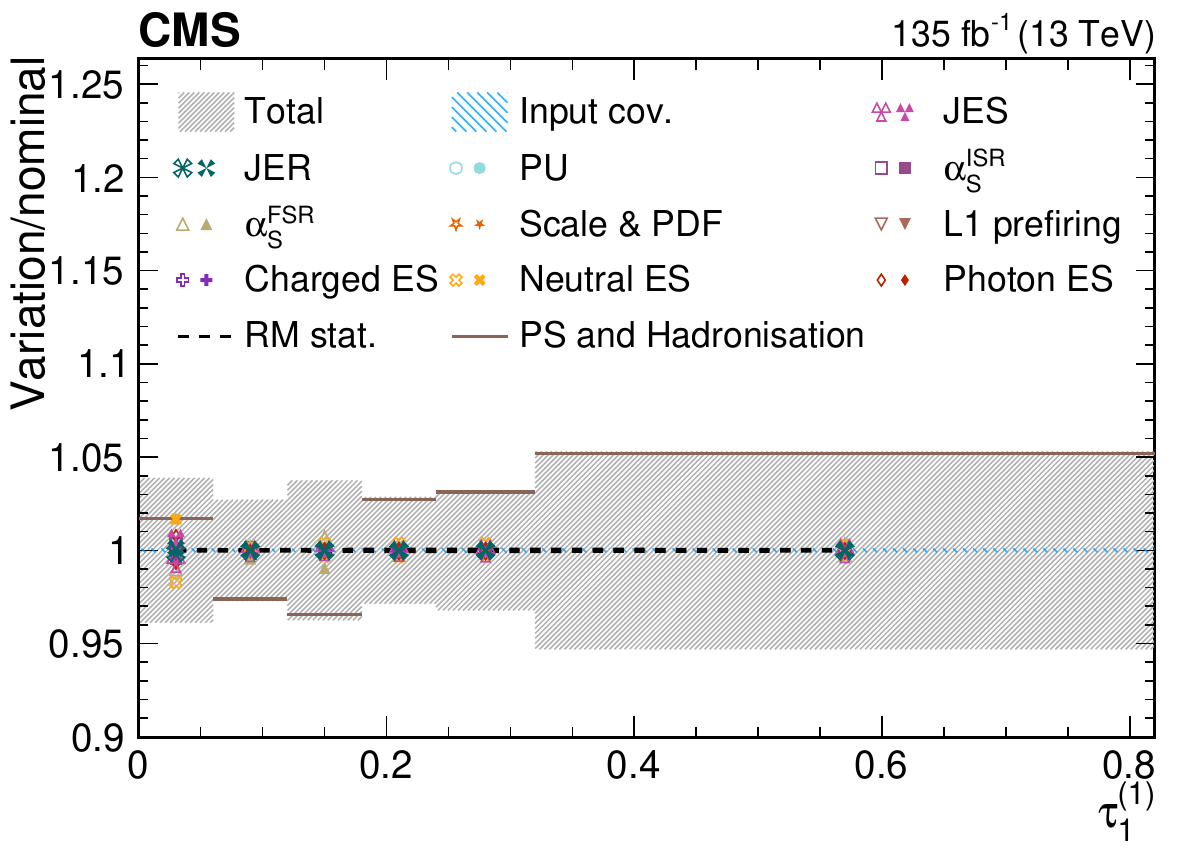}
	\includegraphics[width=.42\textwidth]{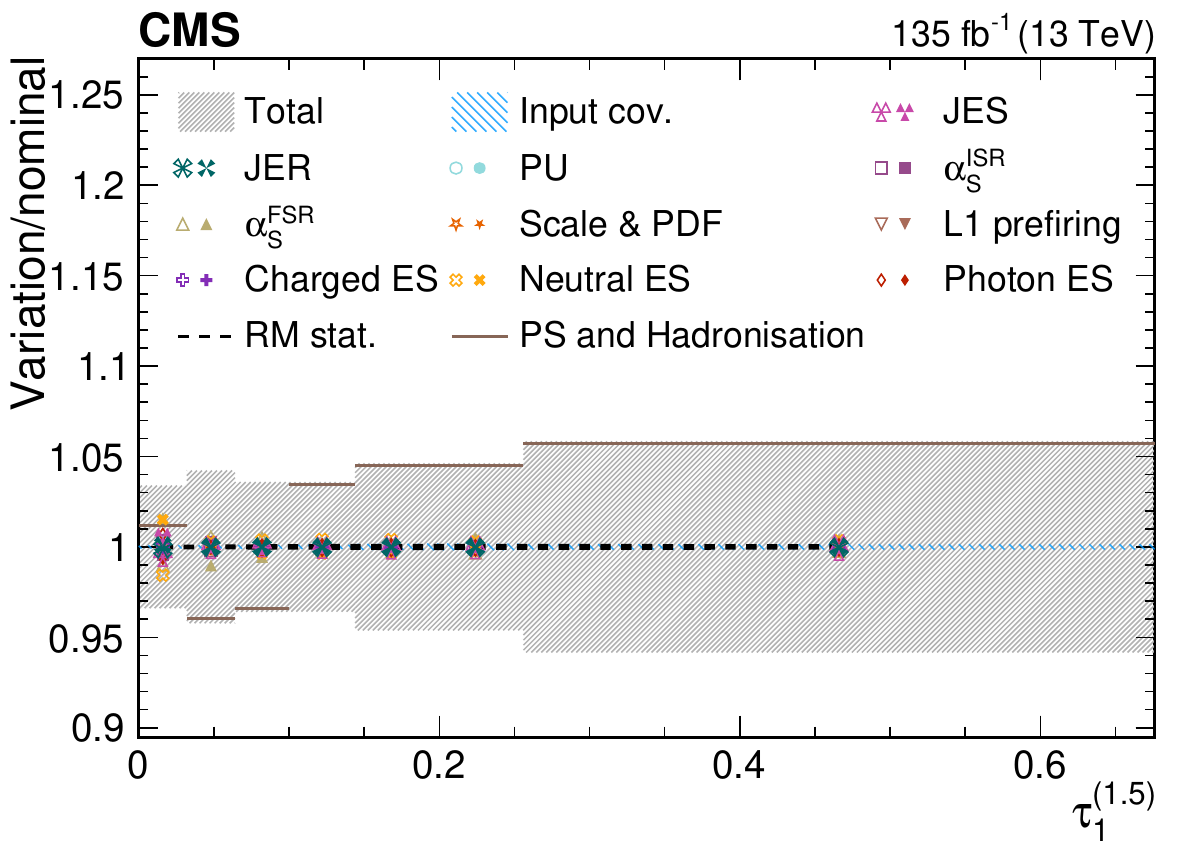}
	\includegraphics[width=.42\textwidth]{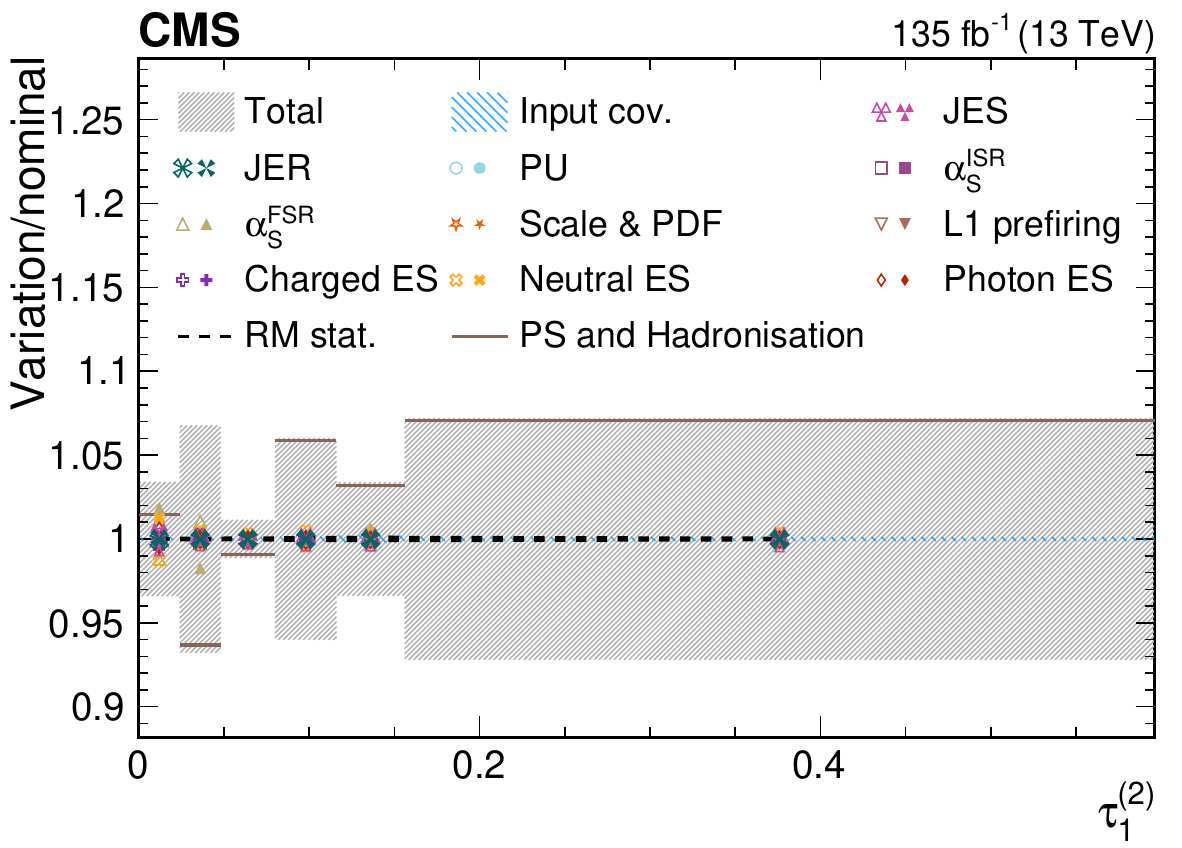}
	\caption{Contributions from various systematic variations to the normalized, unfolded distribution for $\tau_1^{(\beta)}$ observables measured for AK8 jets in the QCD dijet selection. 
		The total unfolding uncertainty is indicated with the dark grey, hashed region, while the blue hashed region indicates the contributions from the input covariance matrix, which includes the propagated effects of the statistical uncertainties of the input data after background subtraction. Contributions from statistical uncertainties of the simulated sample used to construct the nominal response matrix are indicated with the dashed black line. The physics model uncertainty is computed as a one-sided shift compared to the nominal unfolding, and up (down) contributions from other sources are indicated with filled (open) markers of the same type and colour.}
	\label{fig:unfUncsDijet_tau1}
\end{figure}

\begin{figure}[htpb]
	\centering
	\includegraphics[width=.42\textwidth]{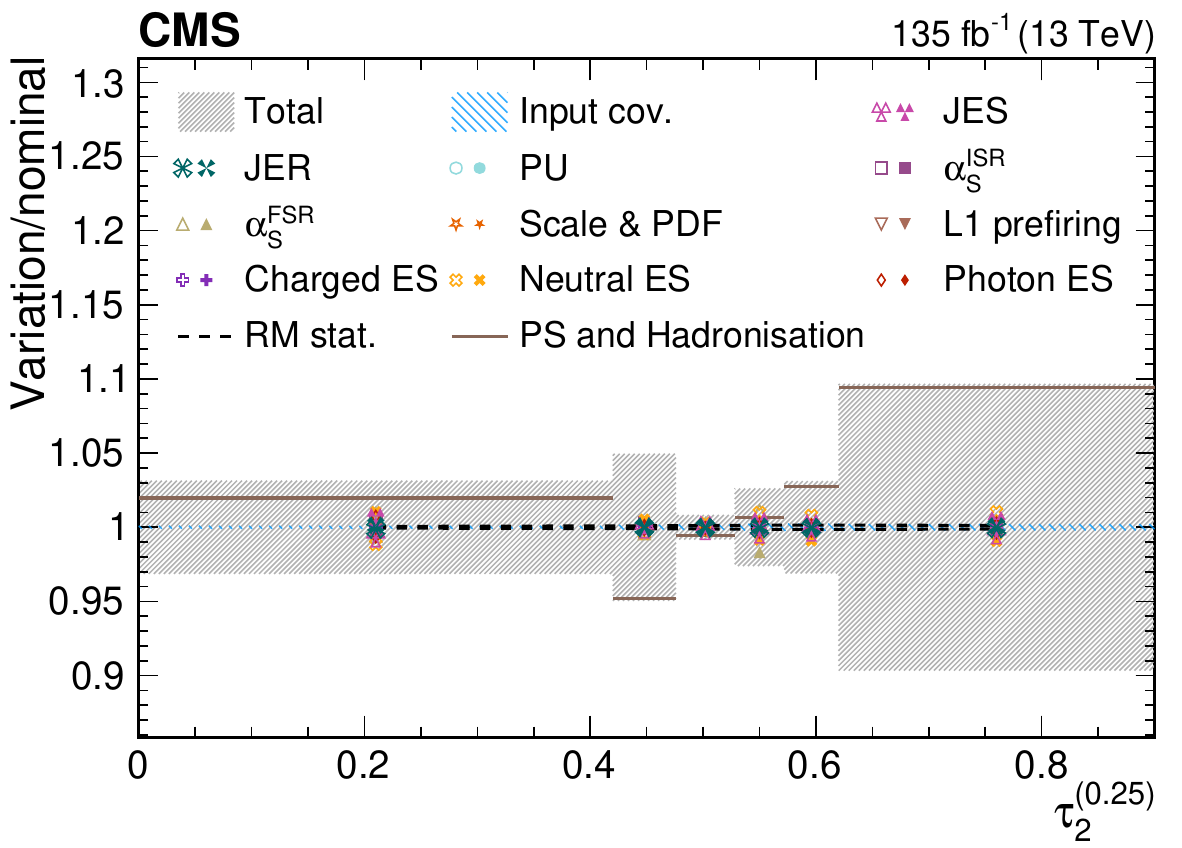}
	\includegraphics[width=.42\textwidth]{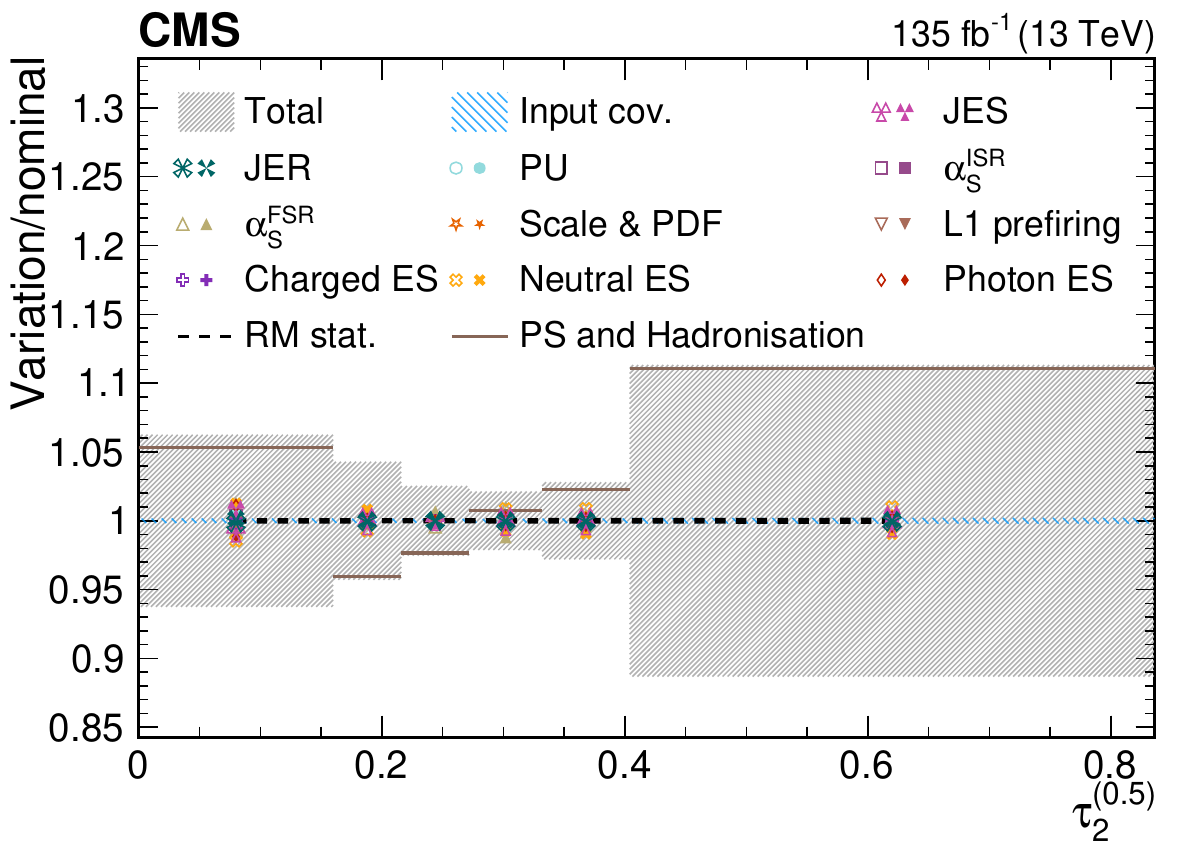}
	\includegraphics[width=.42\textwidth]{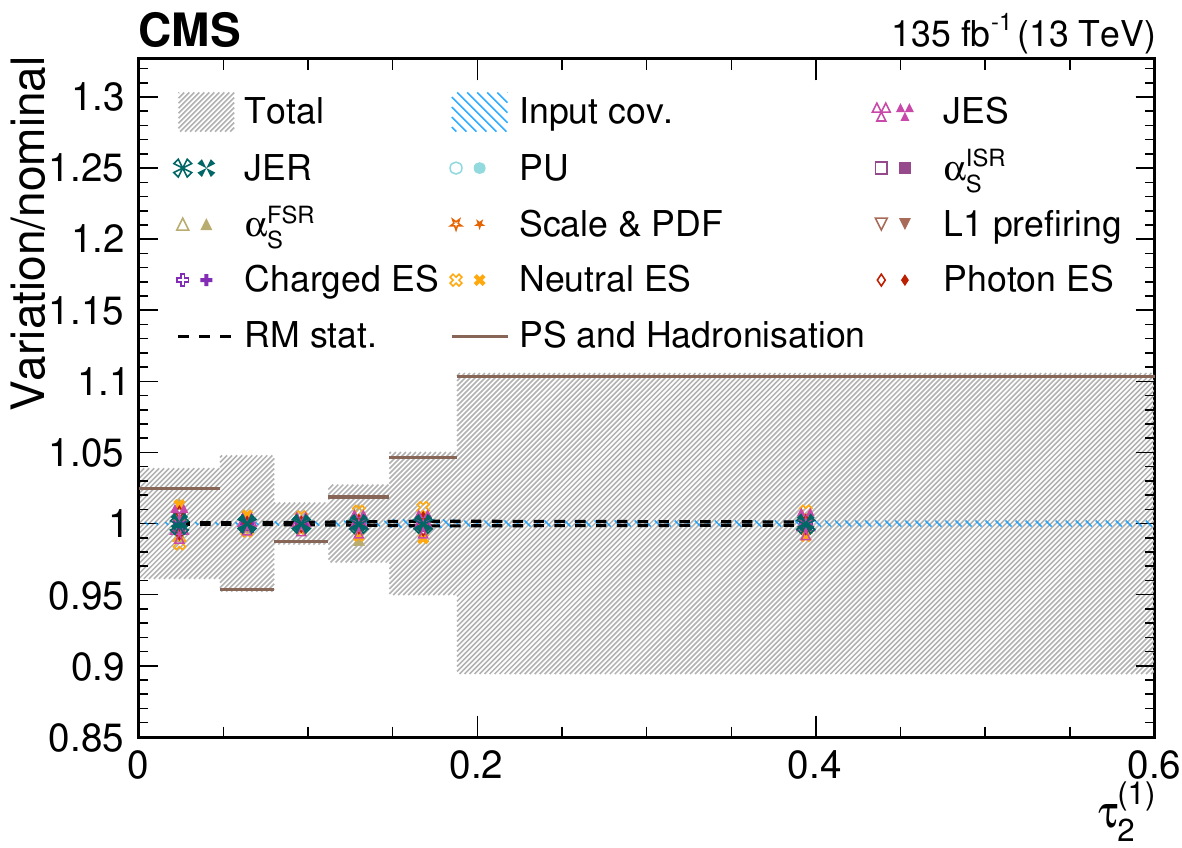}
	\includegraphics[width=.42\textwidth]{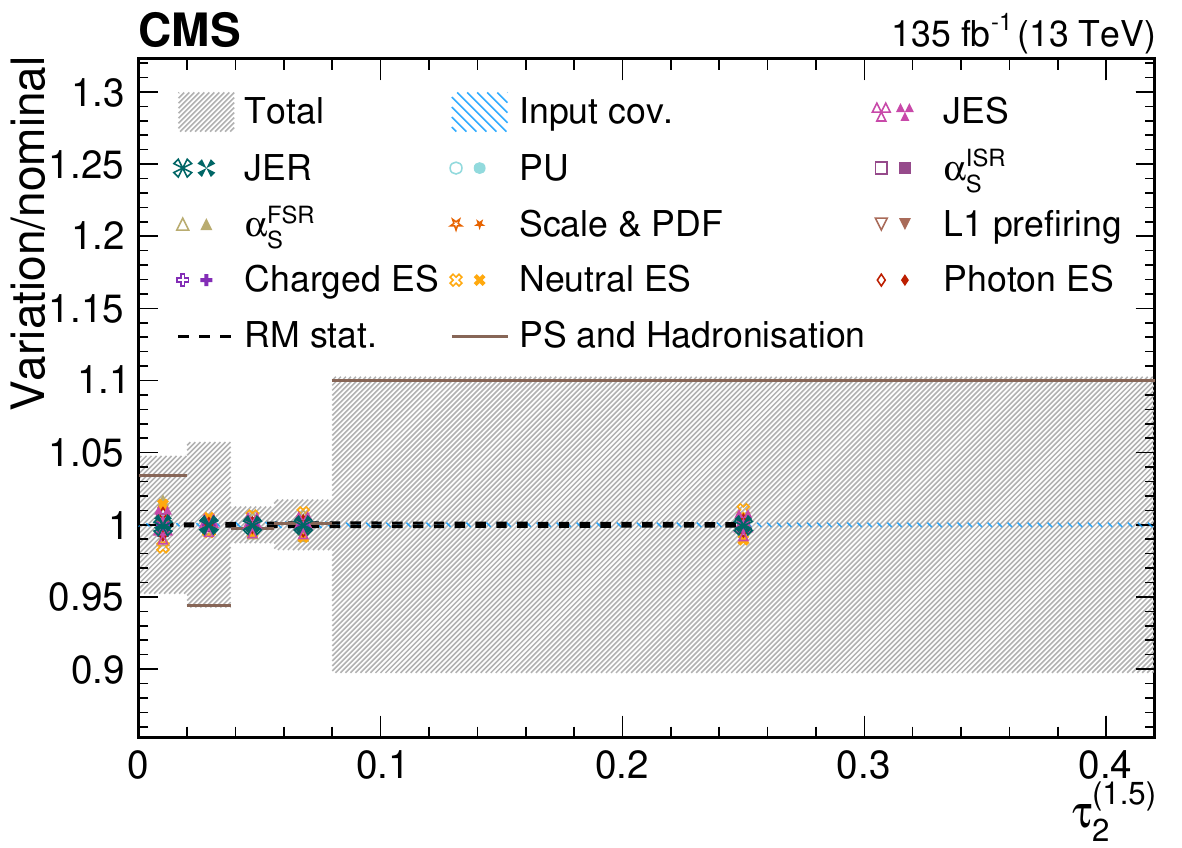}
	\includegraphics[width=.42\textwidth]{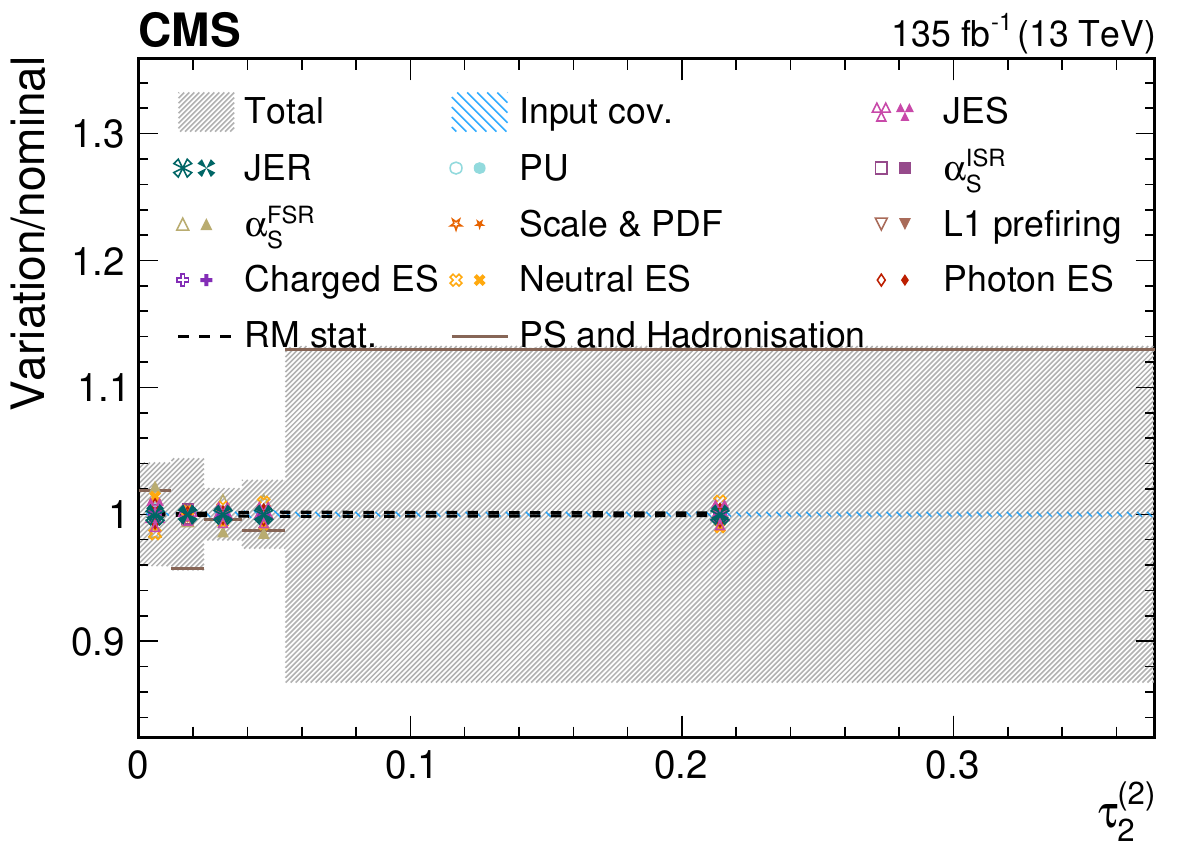}
	\caption{Contributions from various systematic variations to the normalized, unfolded distribution for $\tau_2^{(\beta)}$ observables measured for AK8 jets in the QCD dijet selection. 
		The total unfolding uncertainty is indicated with the dark grey, hashed region, while the blue hashed region indicates the contributions from the input covariance matrix, which includes the propagated effects of the statistical uncertainties of the input data after background subtraction. Contributions from statistical uncertainties of the simulated sample used to construct the nominal response matrix are indicated with the dashed black line. The physics model uncertainty is computed as a one-sided shift compared to the nominal unfolding, and up (down) contributions from other sources are indicated with filled (open) markers of the same type and colour.}
	\label{fig:unfUncsDijet_tau2}
\end{figure}

\begin{figure}[htpb]
	\centering
	\includegraphics[width=.42\textwidth]{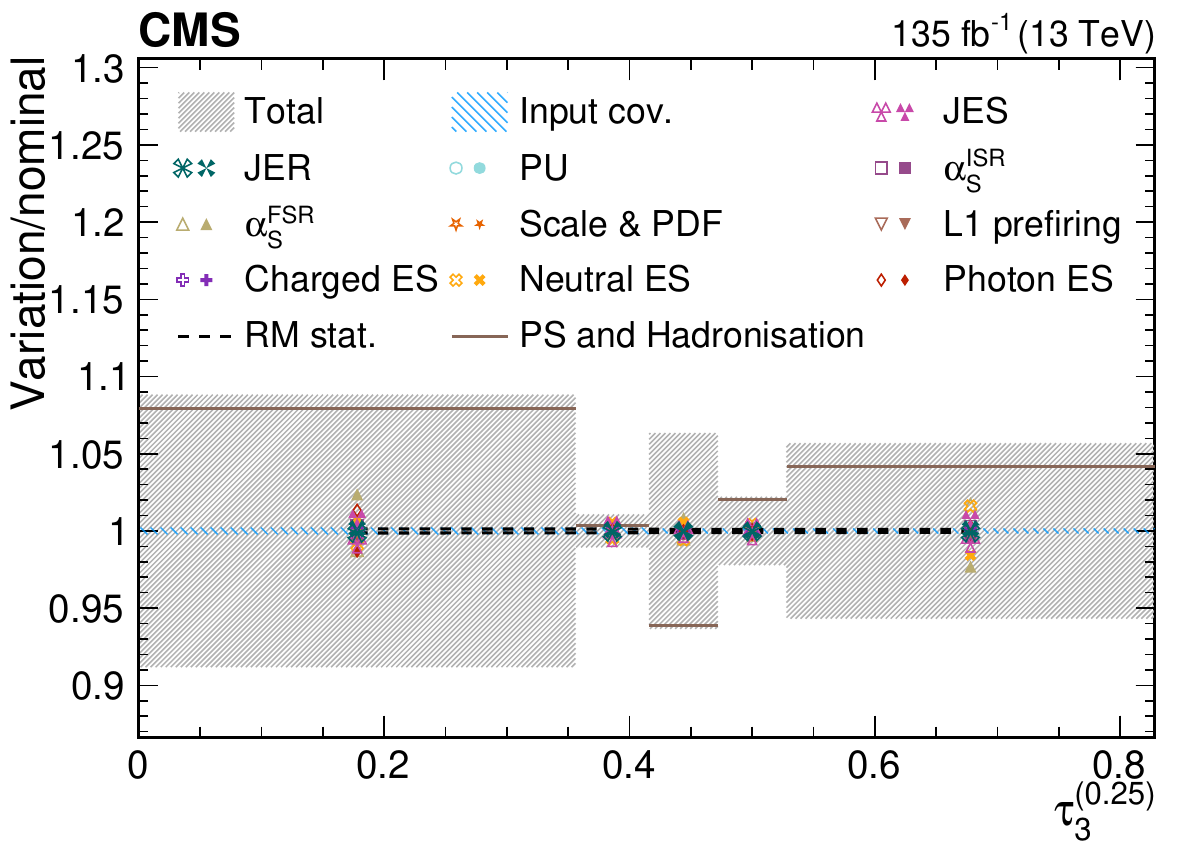}
	\includegraphics[width=.42\textwidth]{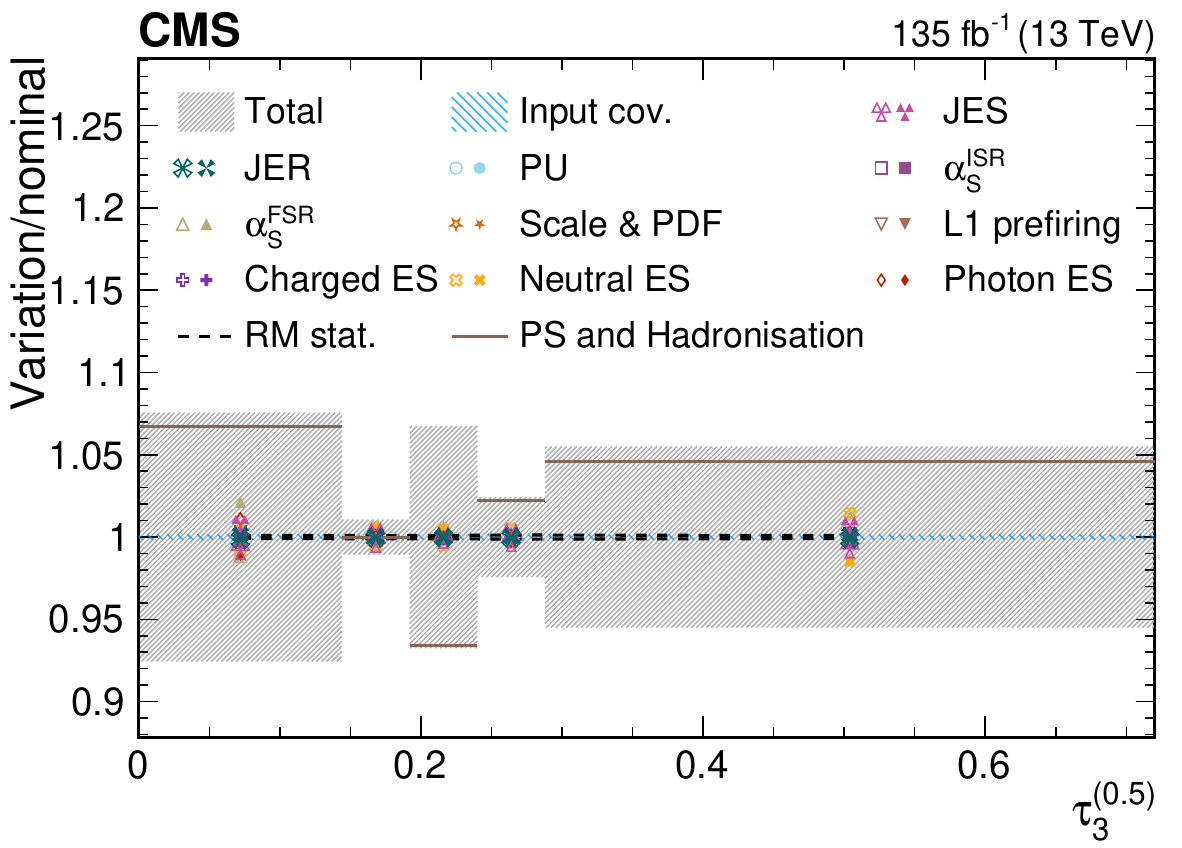}
	\includegraphics[width=.42\textwidth]{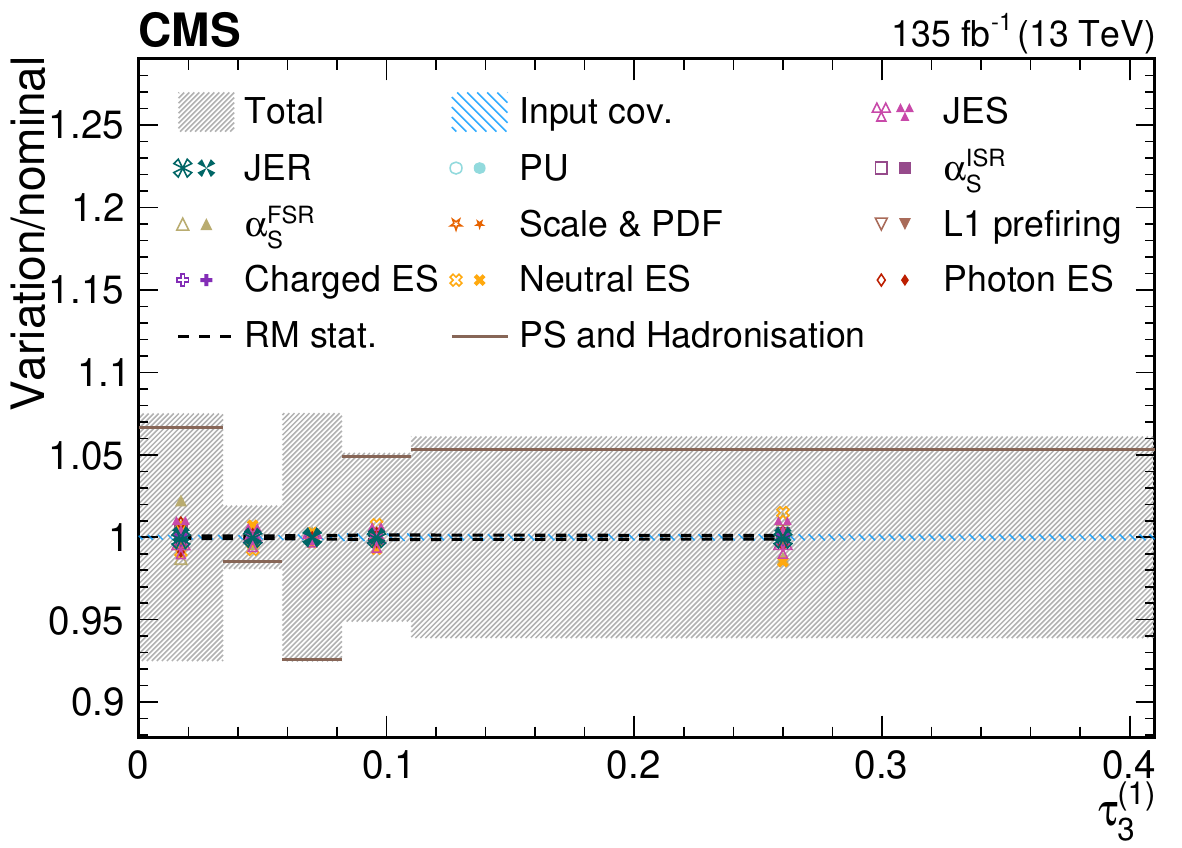}
	\includegraphics[width=.42\textwidth]{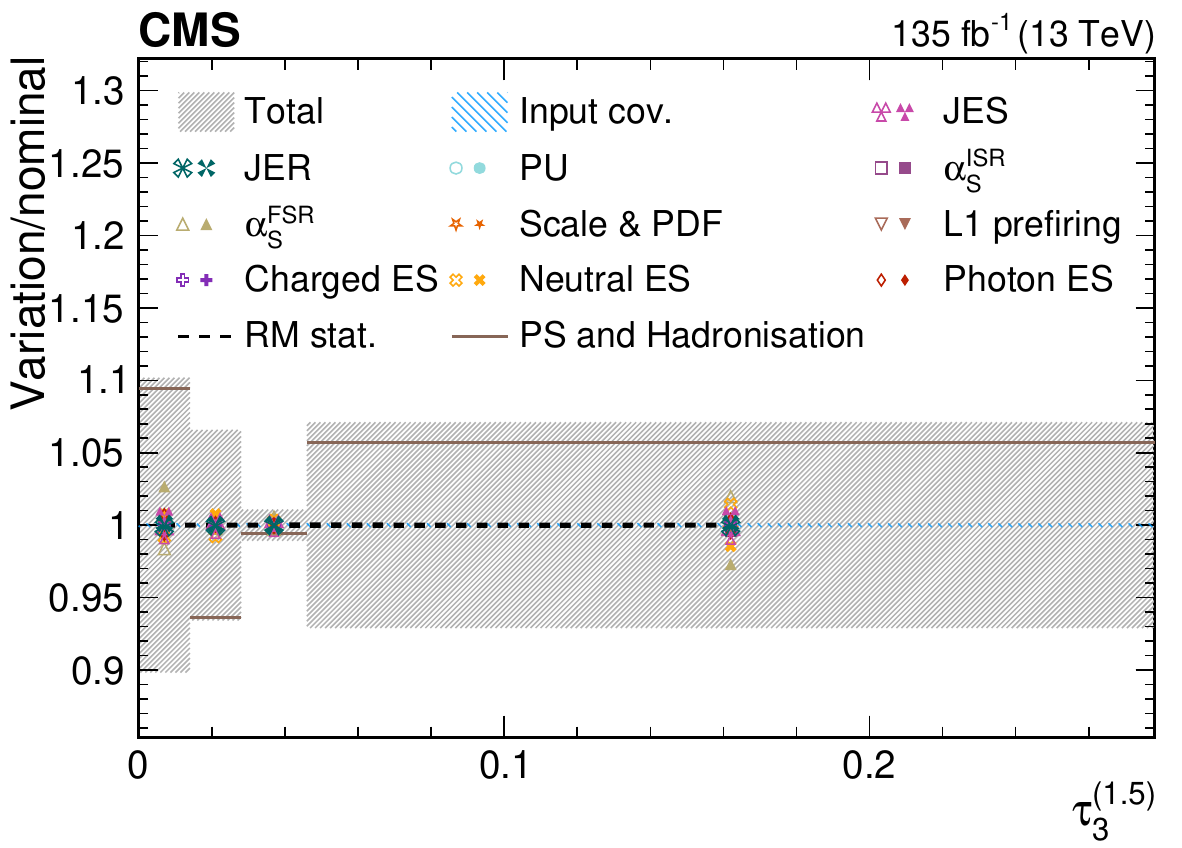}
	\includegraphics[width=.42\textwidth]{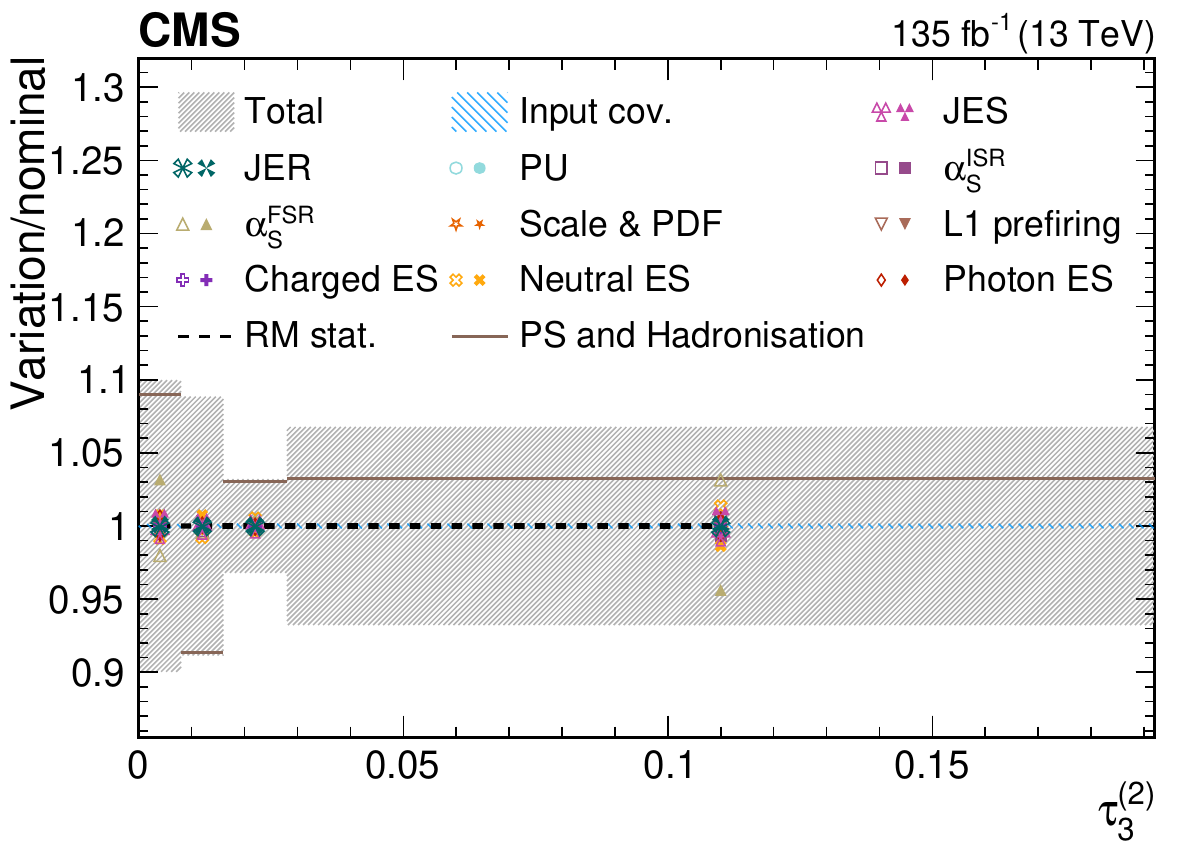}
	\caption{Contributions from various systematic variations to the normalized, unfolded distribution for $\tau_3^{(\beta)}$ observables measured for AK8 jets in the QCD dijet selection. 
		The total unfolding uncertainty is indicated with the dark grey, hashed region, while the blue hashed region indicates the contributions from the input covariance matrix, which includes the propagated effects of the statistical uncertainties of the input data after background subtraction. Contributions from statistical uncertainties of the simulated sample used to construct the nominal response matrix are indicated with the dashed black line. The physics model uncertainty is computed as a one-sided shift compared to the nominal unfolding, and up (down) contributions from other sources are indicated with filled (open) markers of the same type and colour.}
	\label{fig:unfUncsDijet_tau3}
\end{figure}

\begin{figure}[htpb]
	\centering
	\includegraphics[width=.42\textwidth]{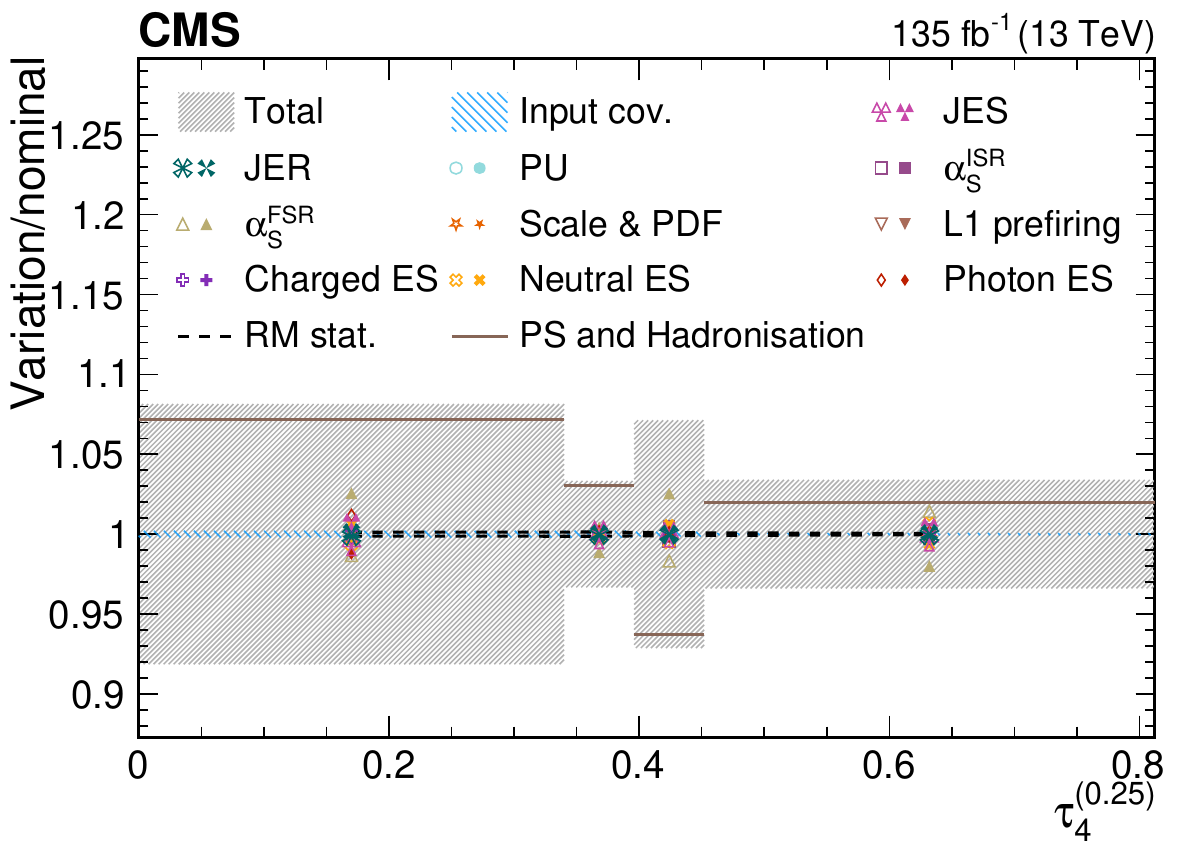}
	\includegraphics[width=.42\textwidth]{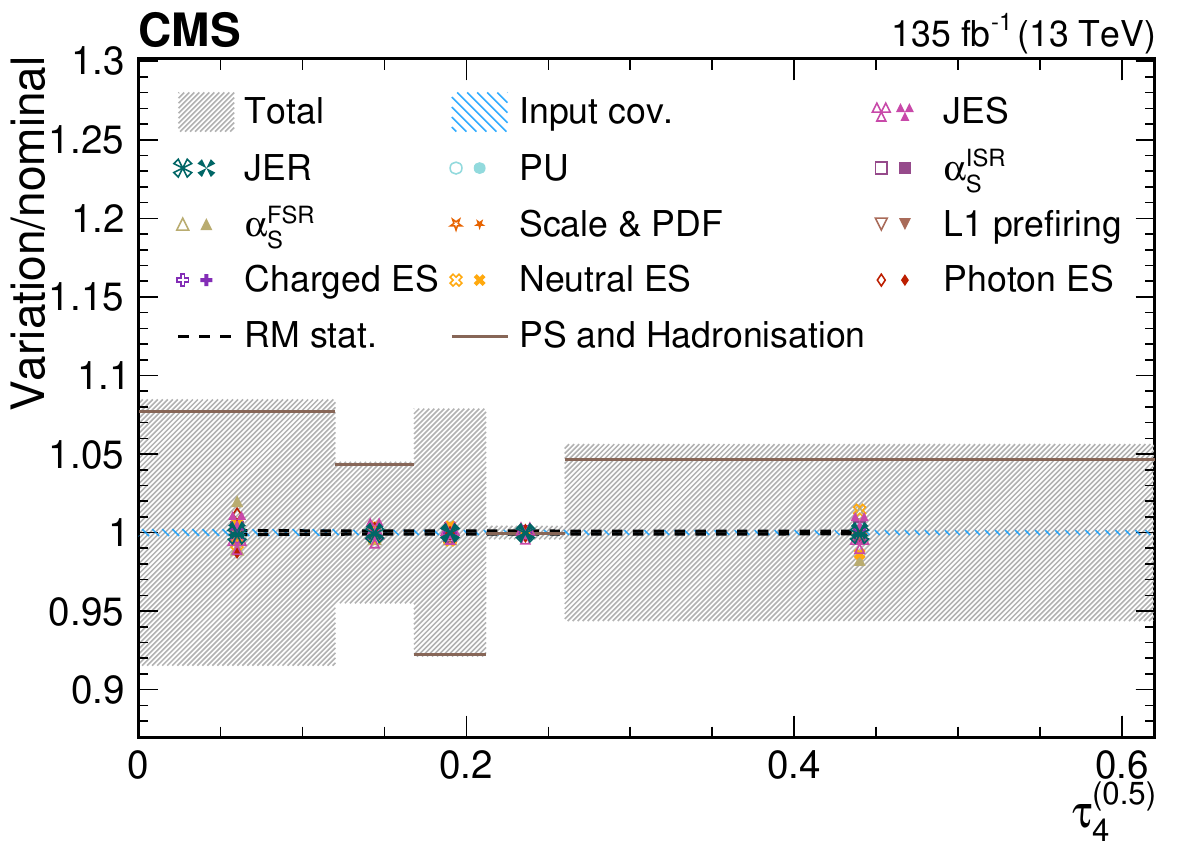}
	\includegraphics[width=.42\textwidth]{Figure_016-b.pdf}
	\includegraphics[width=.42\textwidth]{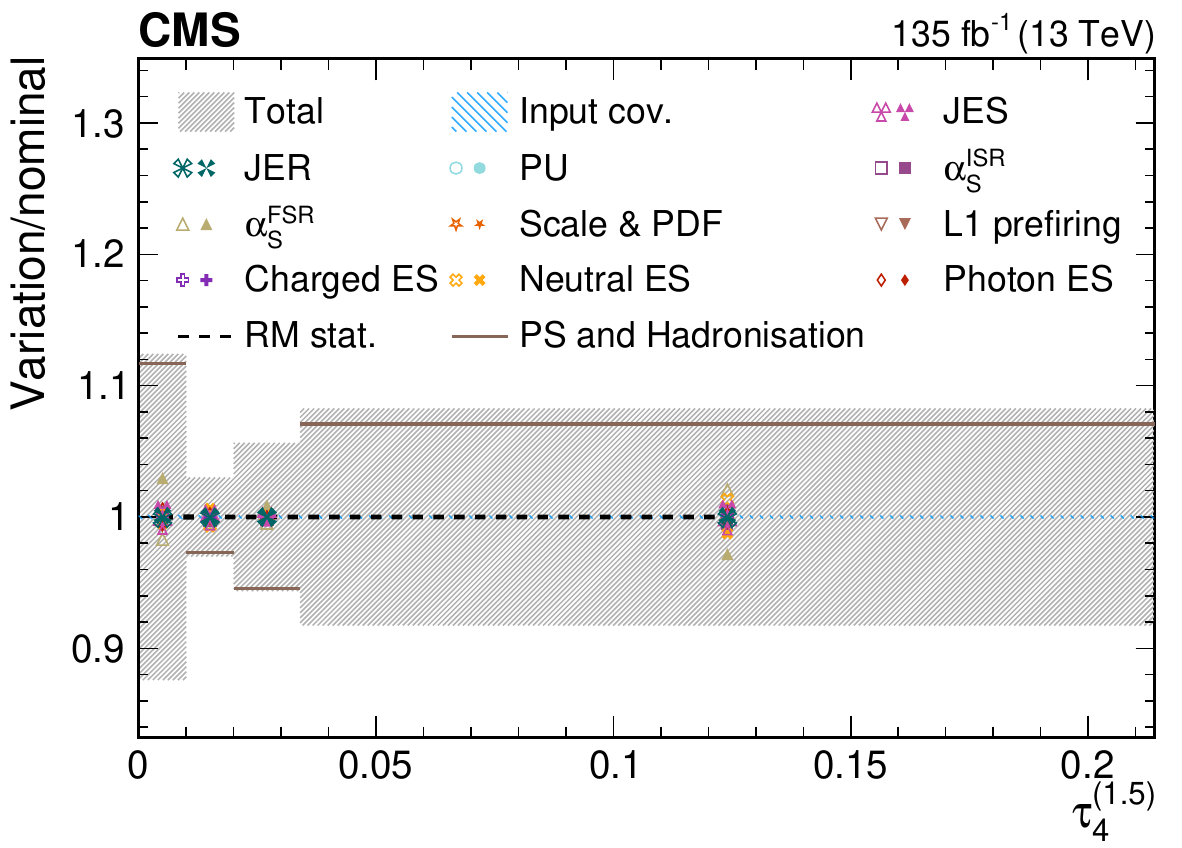}
	\includegraphics[width=.42\textwidth]{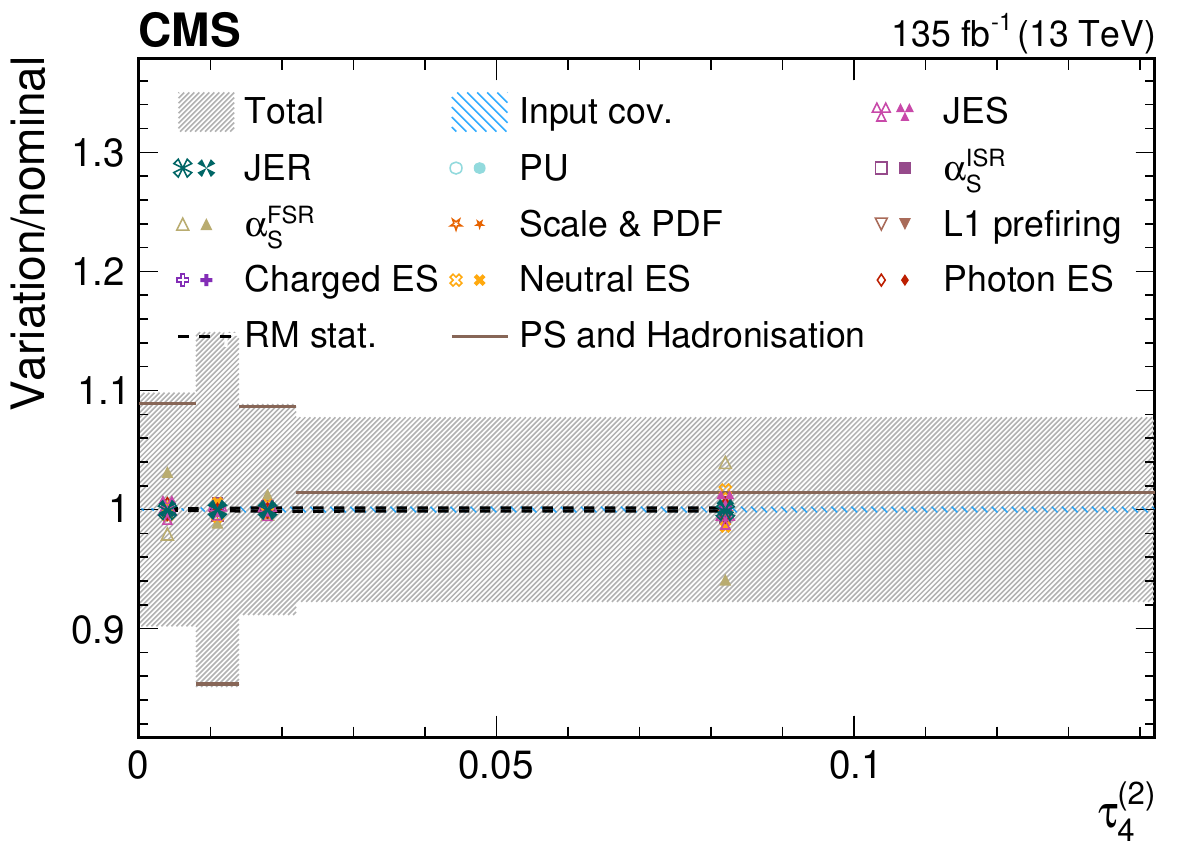}
	\caption{Contributions from various systematic variations to the normalized, unfolded distribution for $\tau_4^{(\beta)}$ observables measured for AK8 jets in the QCD dijet selection. 
		The total unfolding uncertainty is indicated with the dark grey, hashed region, while the blue hashed region indicates the contributions from the input covariance matrix, which includes the propagated effects of the statistical uncertainties of the input data after background subtraction. Contributions from statistical uncertainties of the simulated sample used to construct the nominal response matrix are indicated with the dashed black line. The physics model uncertainty is computed as a one-sided shift compared to the nominal unfolding, and up (down) contributions from other sources are indicated with filled (open) markers of the same type and colour.}
	\label{fig:unfUncsDijet_tau4}
\end{figure}

\begin{figure}[htpb]
	\centering
	\includegraphics[width=.42\textwidth]{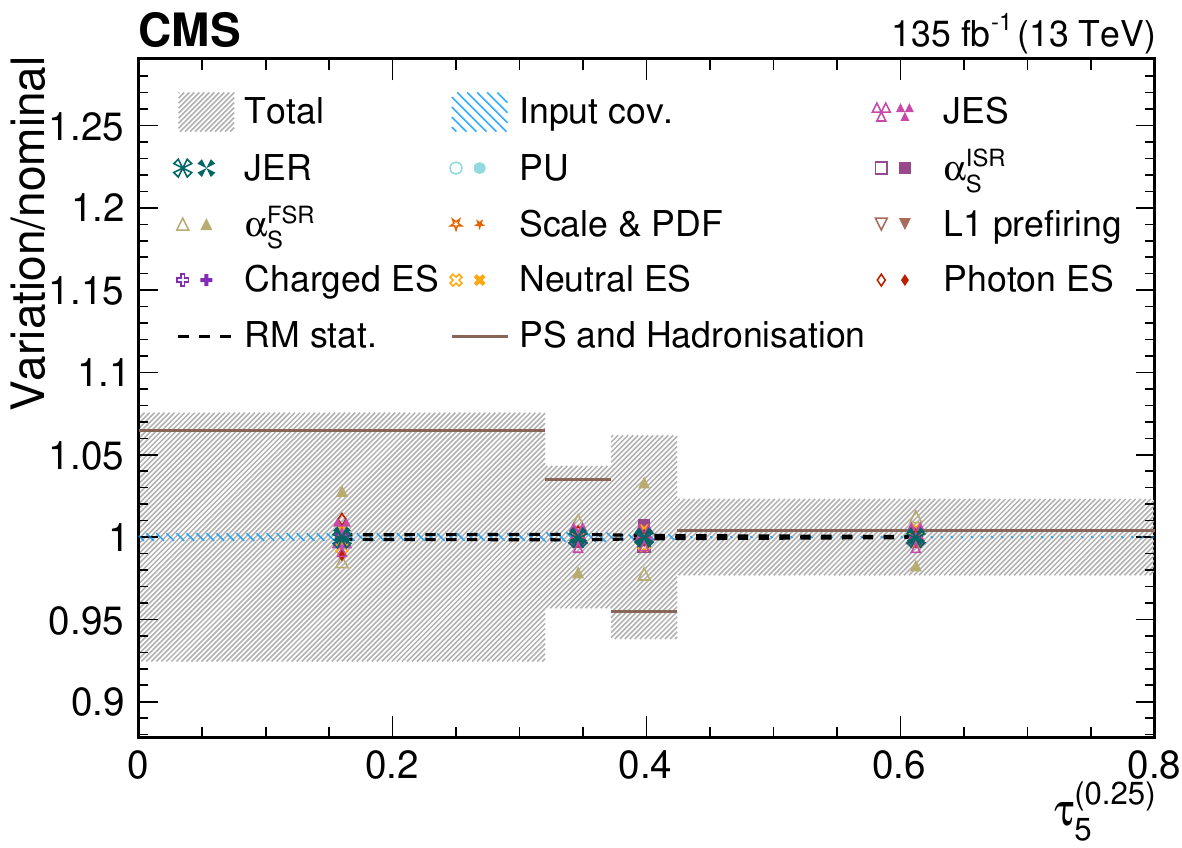}
	\includegraphics[width=.42\textwidth]{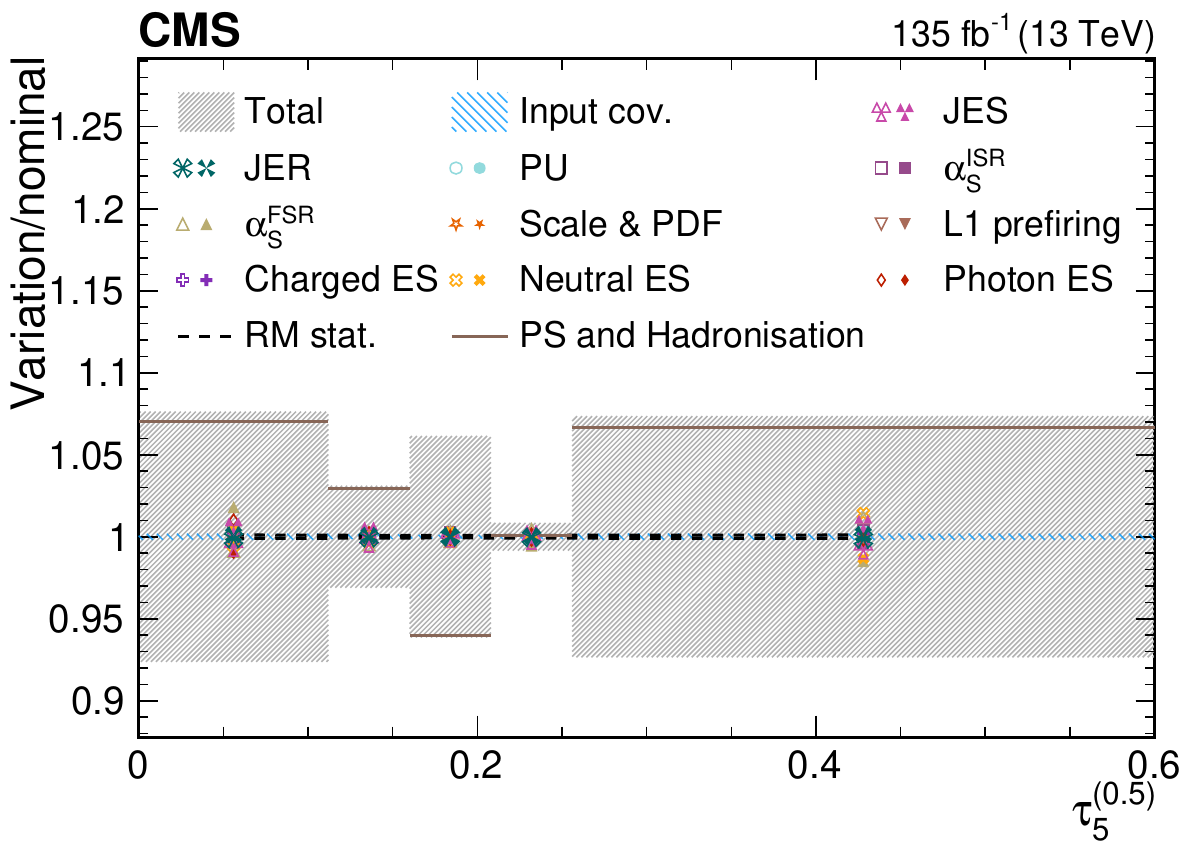}
	\includegraphics[width=.42\textwidth]{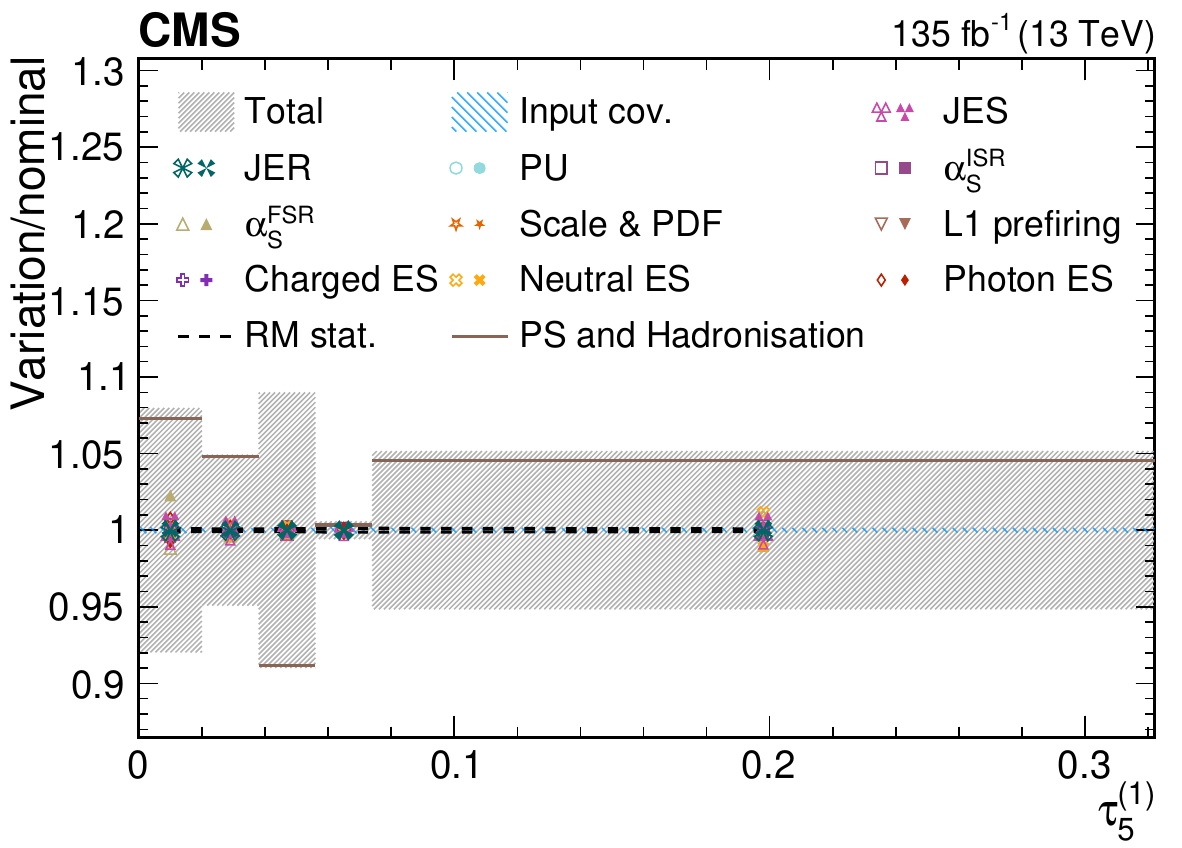}
	\includegraphics[width=.42\textwidth]{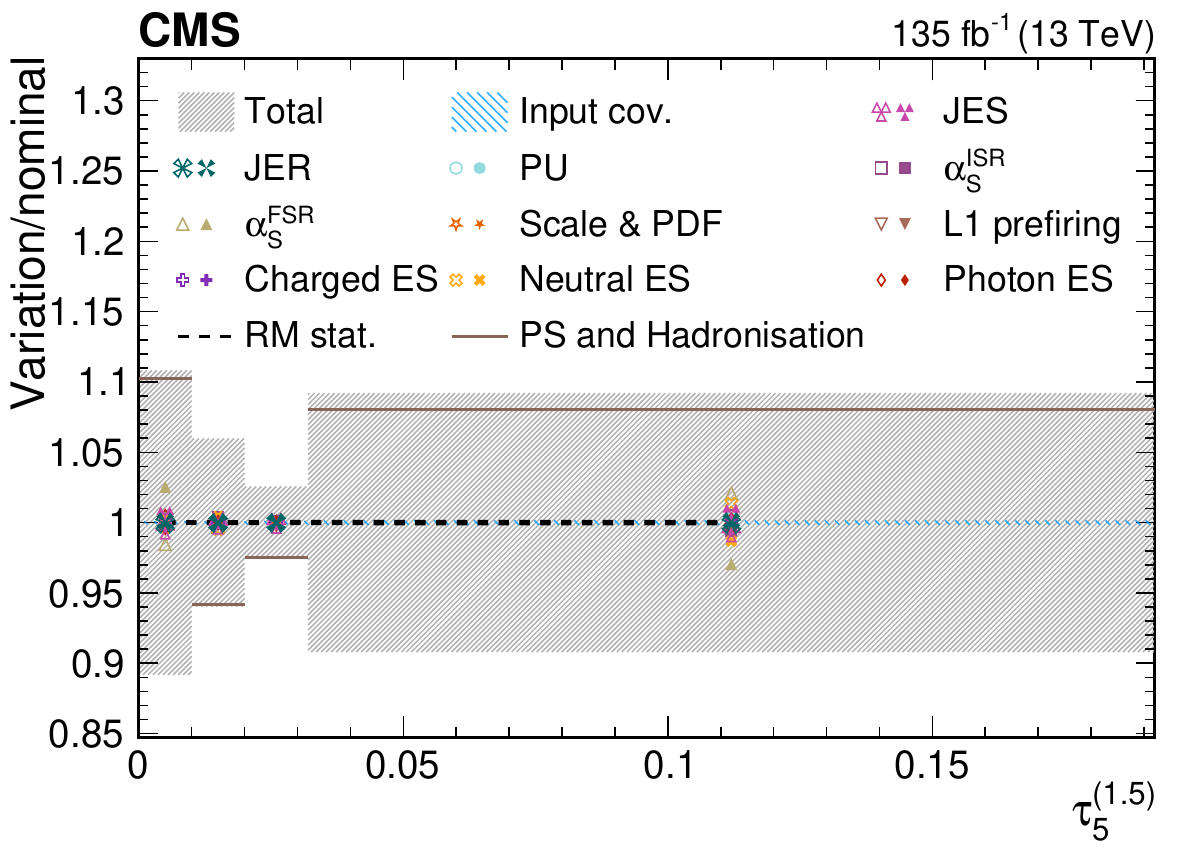}
	\includegraphics[width=.42\textwidth]{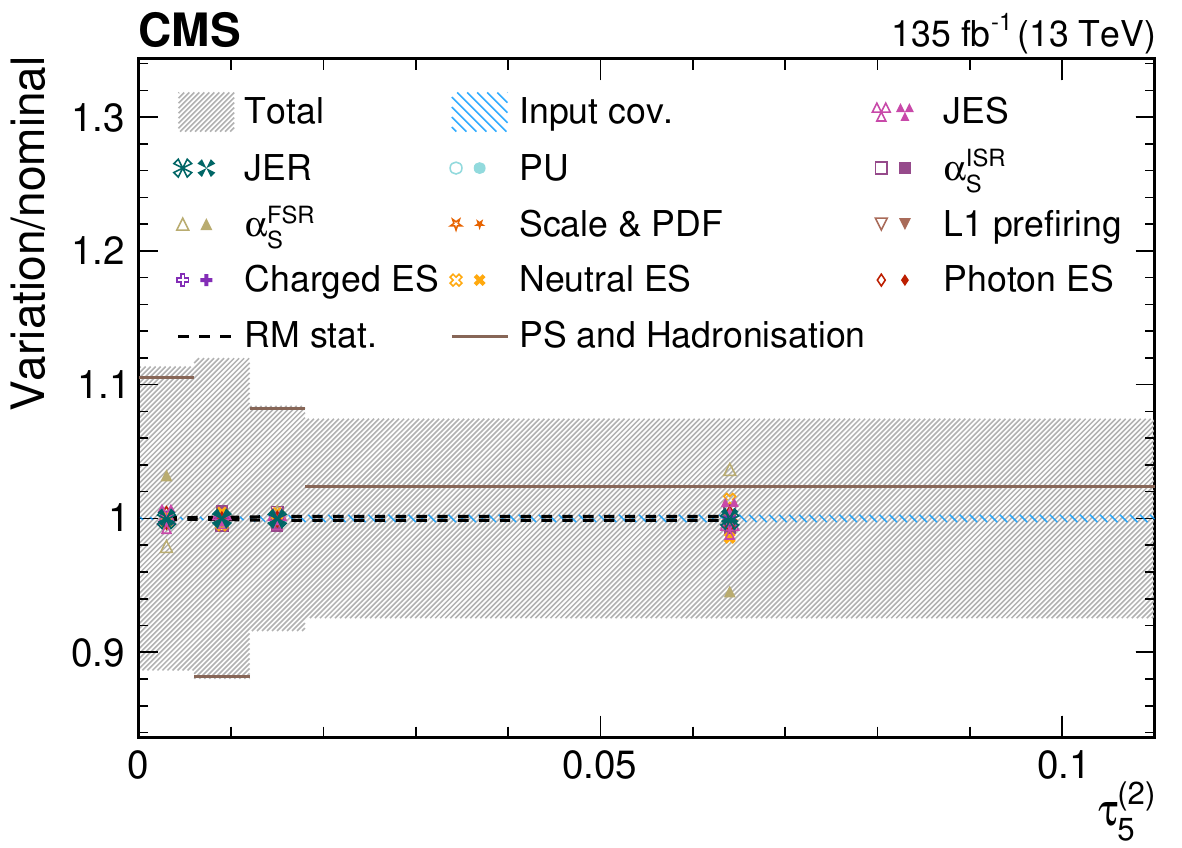}
	\caption{Contributions from various systematic variations to the normalized, unfolded distribution for $\tau_5^{(\beta)}$ observables measured for AK8 jets in the QCD dijet selection. 
		The total unfolding uncertainty is indicated with the dark grey, hashed region, while the blue hashed region indicates the contributions from the input covariance matrix, which includes the propagated effects of the statistical uncertainties of the input data after background subtraction. Contributions from statistical uncertainties of the simulated sample used to construct the nominal response matrix are indicated with the dashed black line. The physics model uncertainty is computed as a one-sided shift compared to the nominal unfolding, and up (down) contributions from other sources are indicated with filled (open) markers of the same type and colour.}
	\label{fig:unfUncsDijet_tau5}
\end{figure}

Estimated contributions of various sources of experimental and modelling uncertainty are presented for the measurement of $1$- through $5$-subjettiness in the dijet selection in Figs.~\ref{fig:unfUncsDijet_tau1}--\ref{fig:unfUncsDijet_tau5}.

\clearpage
\newpage

\subsection{Unfolding uncertainties: boosted \texorpdfstring{\PW}{W} boson jets}
\label{sec:WUnfUncs}
\begin{figure}[htpb]
	\centering
	\includegraphics[width=.42\textwidth]{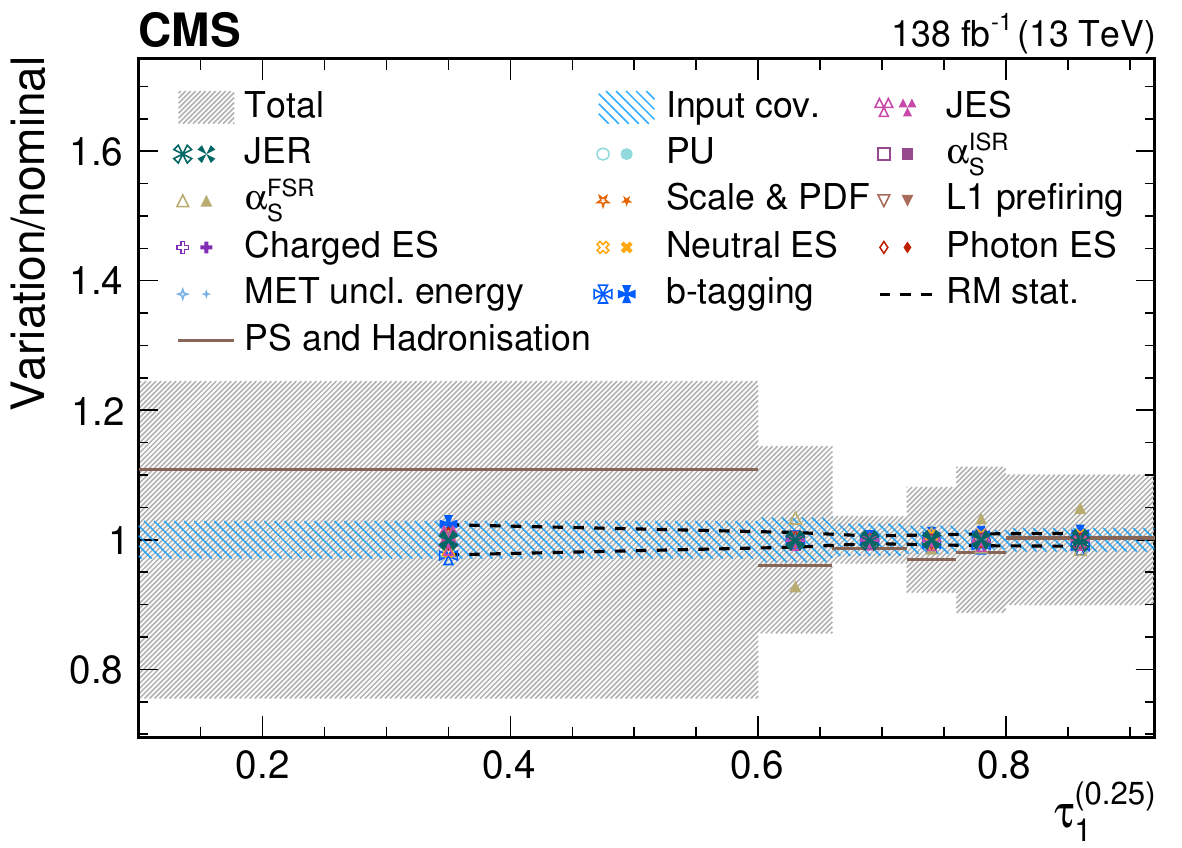}
	\includegraphics[width=.42\textwidth]{Figure_017-a.pdf}
	\includegraphics[width=.42\textwidth]{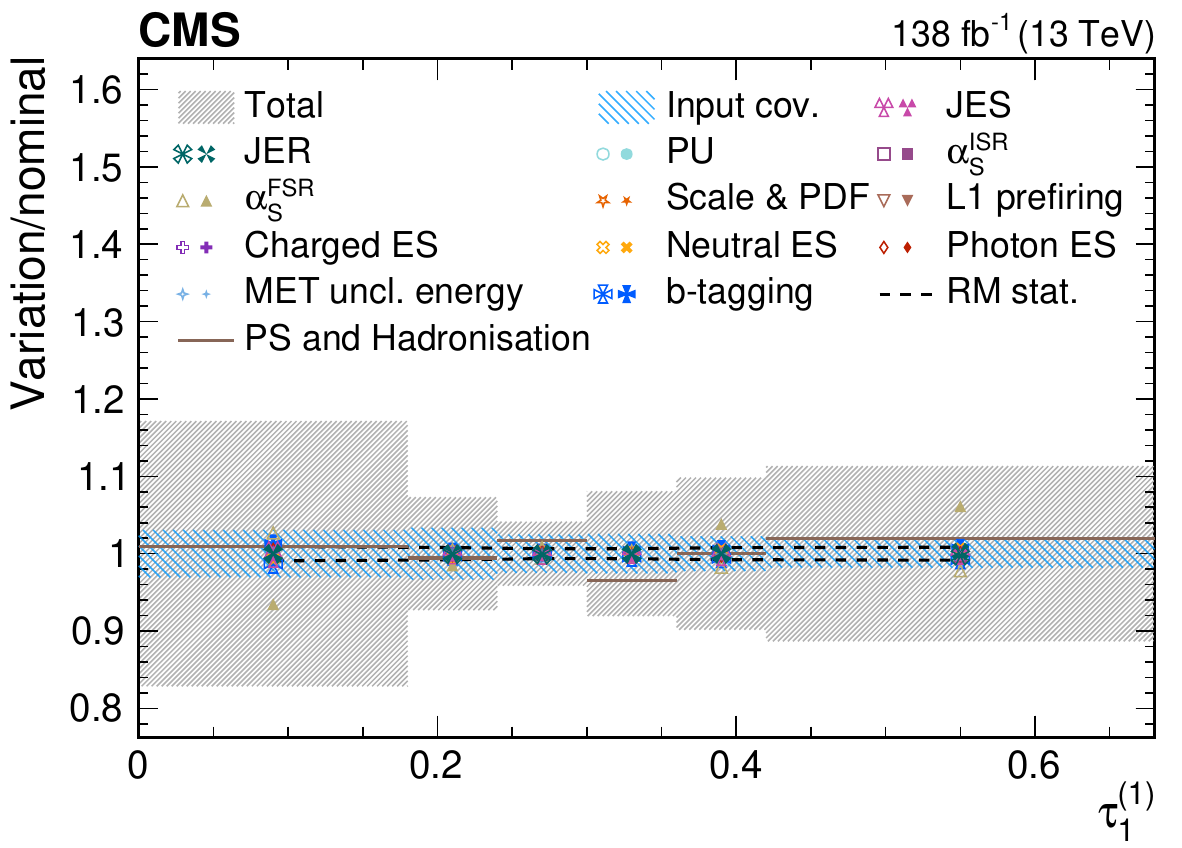}
	\includegraphics[width=.42\textwidth]{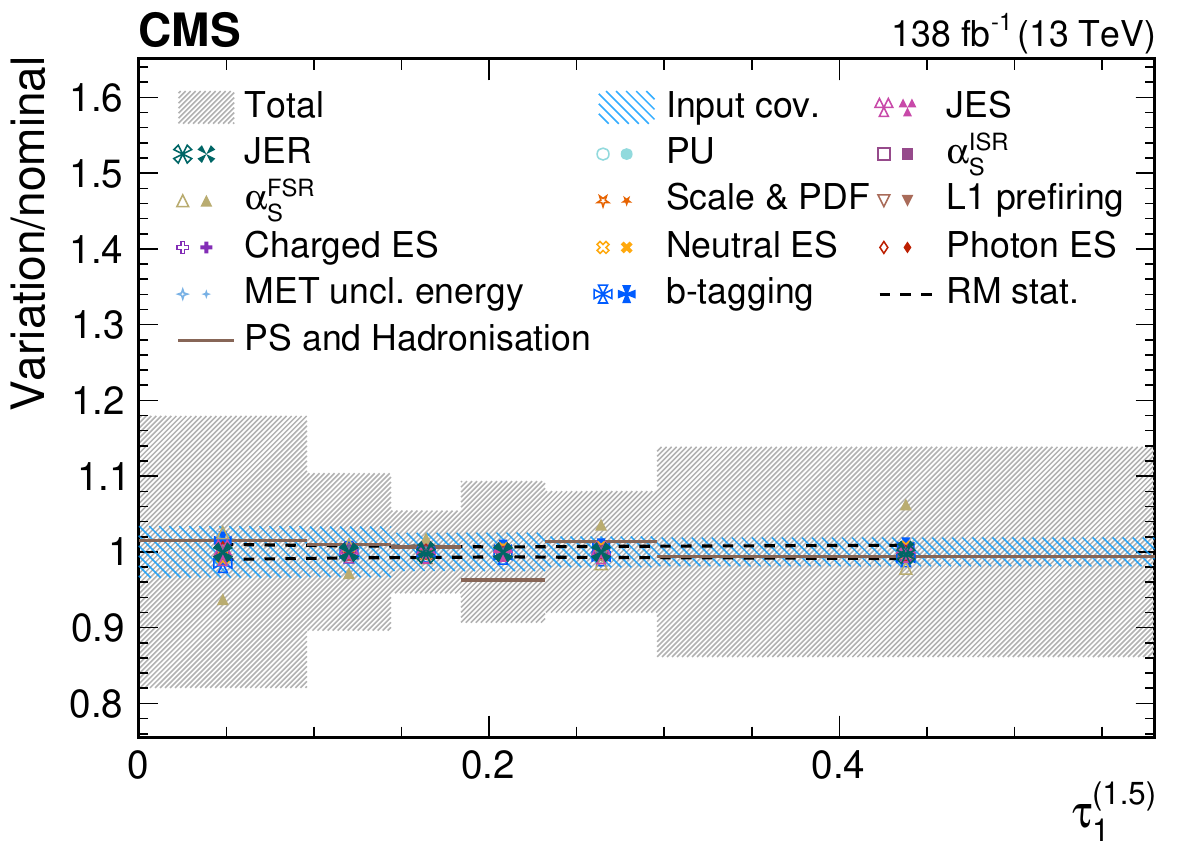}
	\includegraphics[width=.42\textwidth]{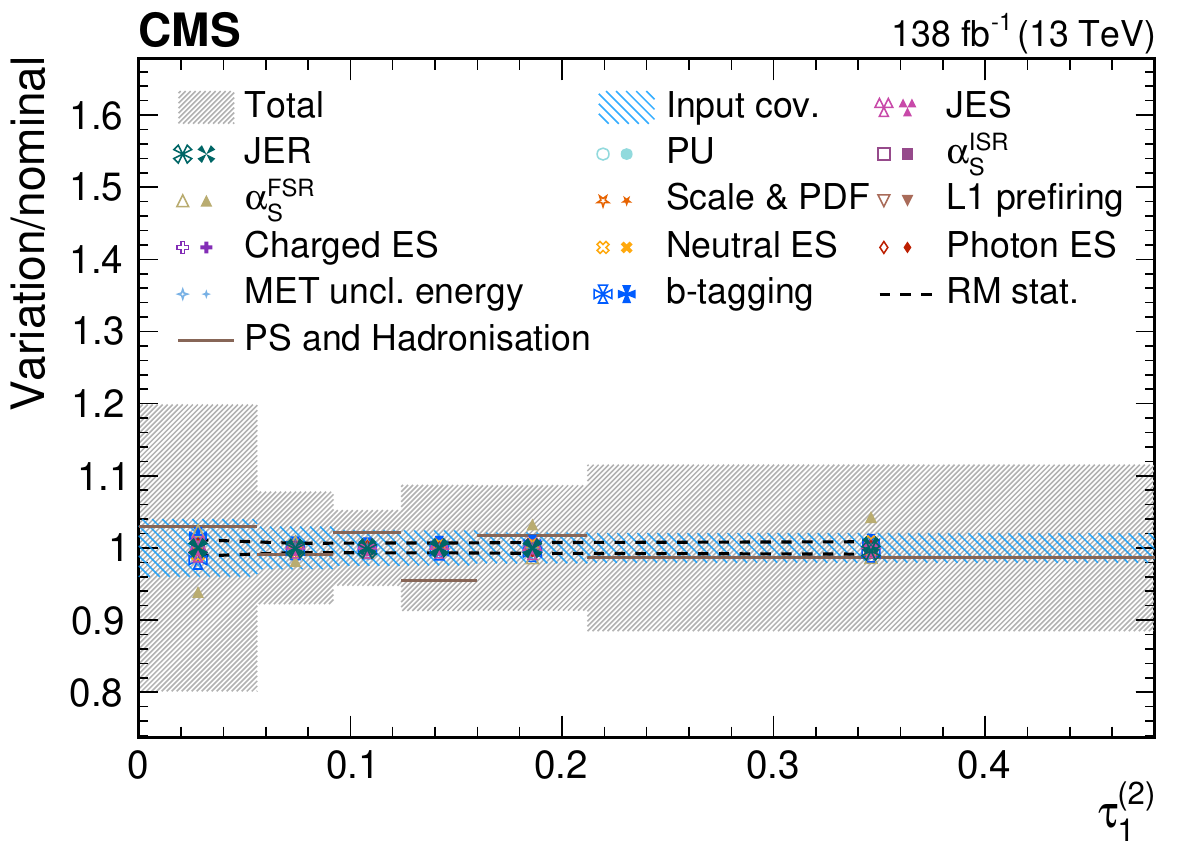}
	\caption{Contributions from various systematic variations to the normalized, unfolded distribution for $\tau_1^{(\beta)}$ observables measured for AK8 jets passing the boosted \PW boson-enriched selection in $\PGm$+jets \ttbar events. 
		The total unfolding uncertainty is indicated with the dark grey, hashed region, while the blue hashed region indicates the contributions from the input covariance matrix, which includes the propagated effects of the statistical uncertainties of the input data after background subtraction. Contributions from statistical uncertainties of the simulated sample used to construct the nominal response matrix are indicated with the dashed black line. The physics model uncertainty is computed as a one-sided shift compared to the nominal unfolding, and up (down) contributions from other sources are indicated with filled (open) markers of the same type and colour.}
	\label{fig:unfUncsW_tau1}
\end{figure}

\begin{figure}[htpb]
	\centering
	\includegraphics[width=.42\textwidth]{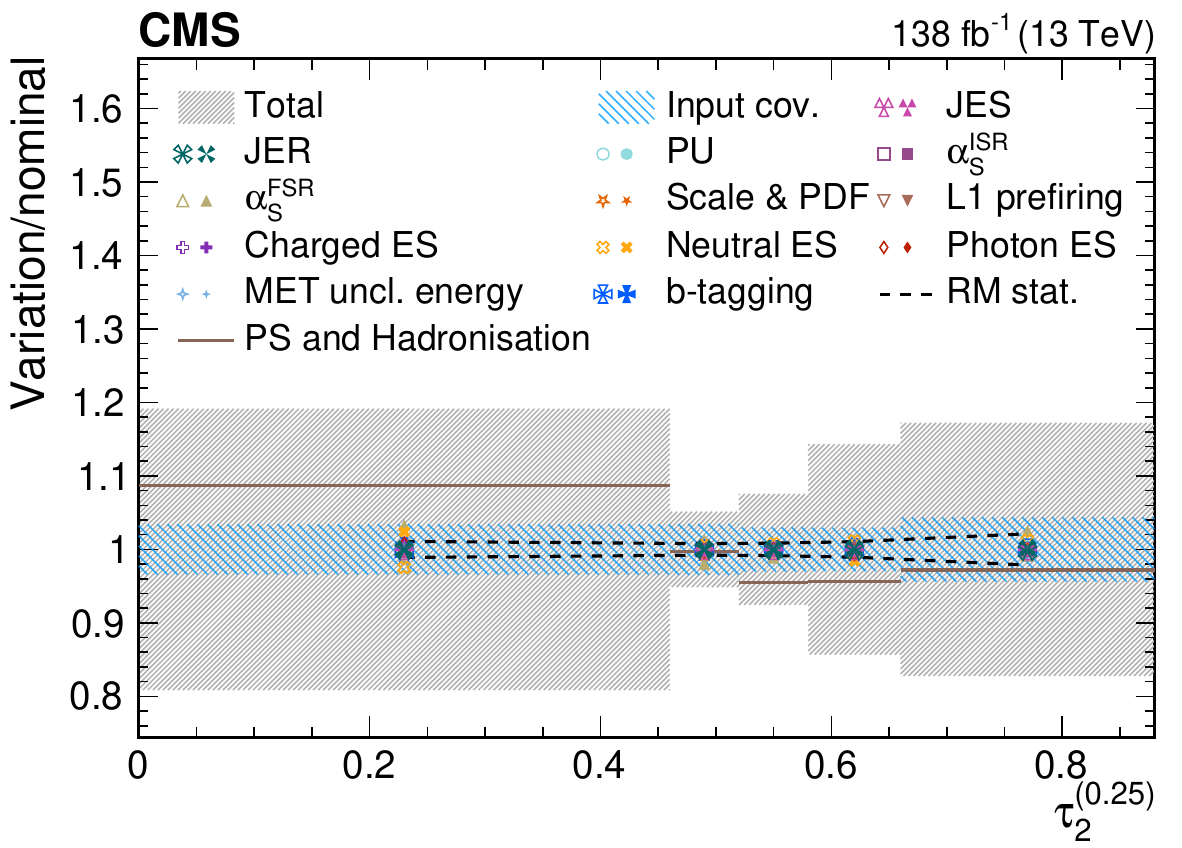}
	\includegraphics[width=.42\textwidth]{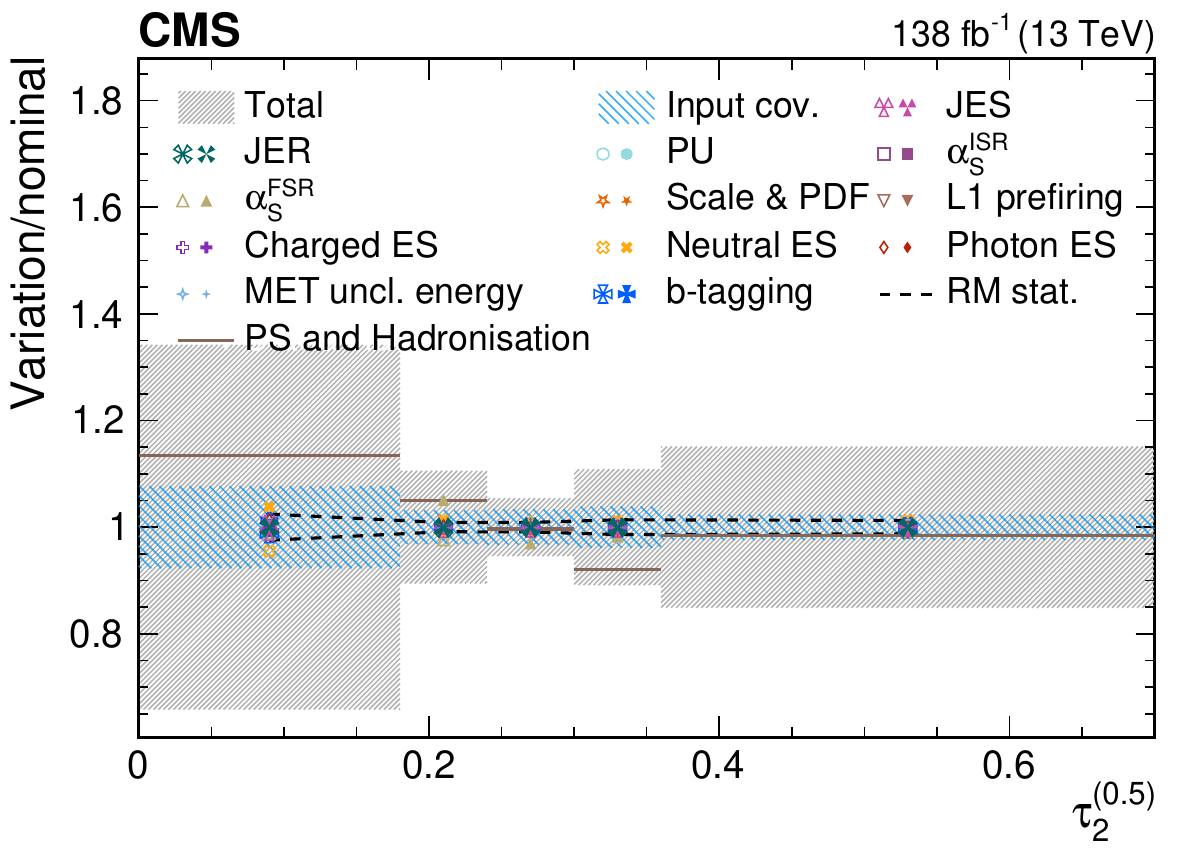}
	\includegraphics[width=.42\textwidth]{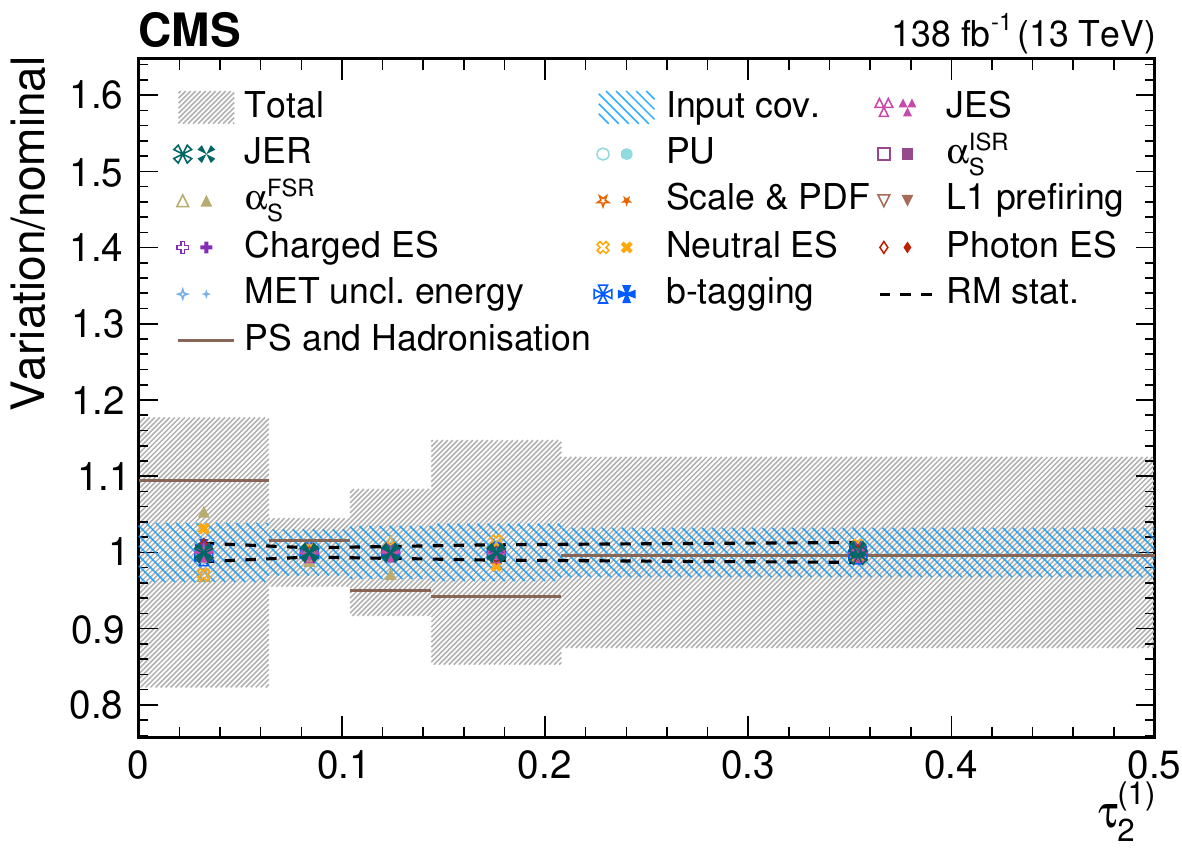}
	\includegraphics[width=.42\textwidth]{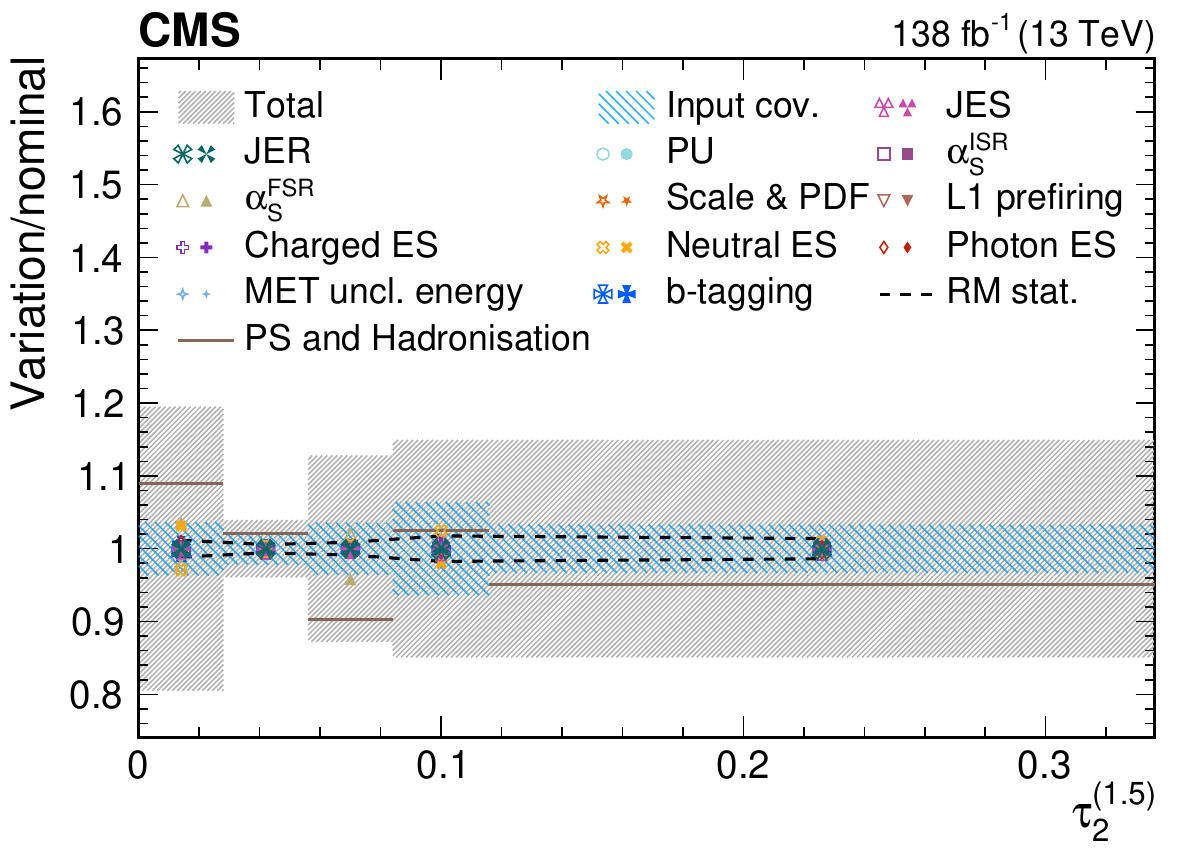}
	\includegraphics[width=.42\textwidth]{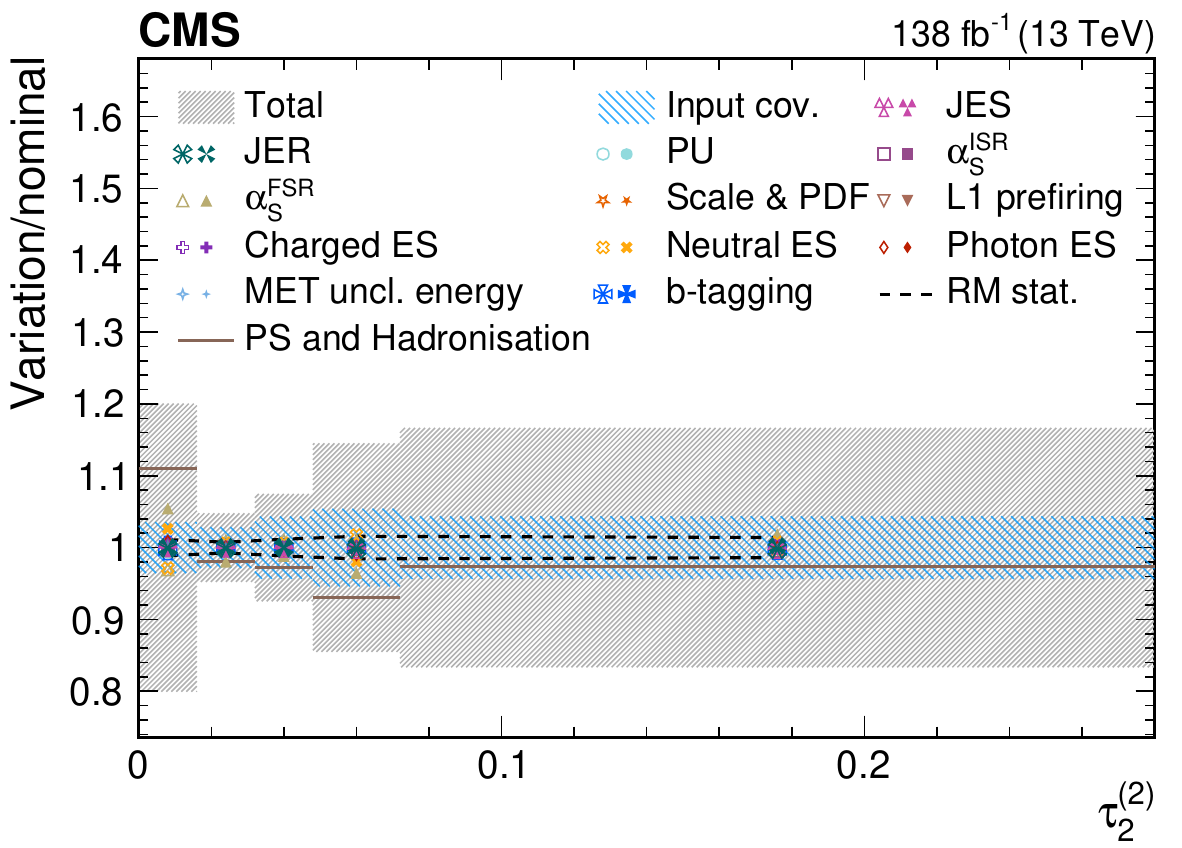}
	\caption{Contributions from various systematic variations to the normalized, unfolded distribution for $\tau_2^{(\beta)}$ observables measured for AK8 jets passing the boosted \PW boson-enriched selection in $\PGm$+jets \ttbar events. 
		The total unfolding uncertainty is indicated with the dark grey, hashed region, while the blue hashed region indicates the contributions from the input covariance matrix, which includes the propagated effects of the statistical uncertainties of the input data after background subtraction. Contributions from statistical uncertainties of the simulated sample used to construct the nominal response matrix are indicated with the dashed black line. The physics model uncertainty is computed as a one-sided shift compared to the nominal unfolding, and up (down) contributions from other sources are indicated with filled (open) markers of the same type and colour.}
	\label{fig:unfUncsW_tau2}
\end{figure}

\begin{figure}[htpb]
	\centering
	\includegraphics[width=.42\textwidth]{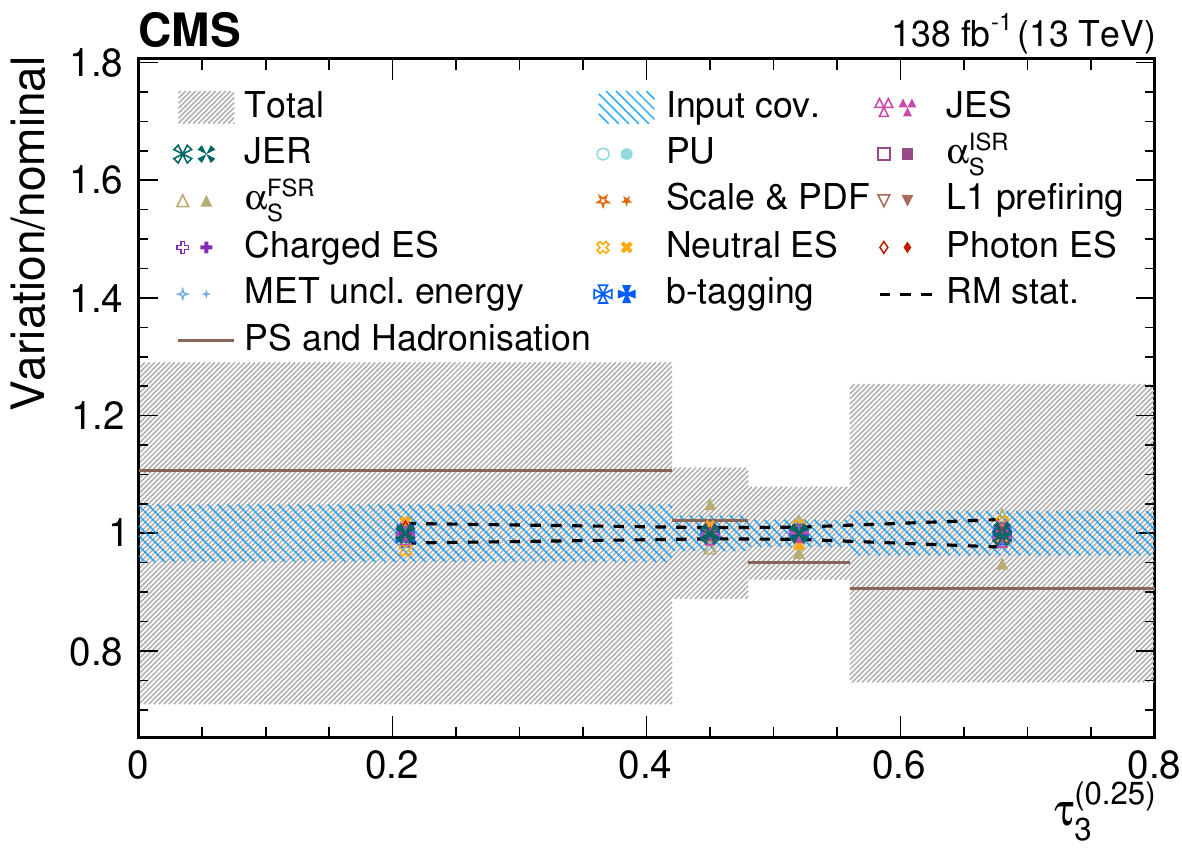}
	\includegraphics[width=.42\textwidth]{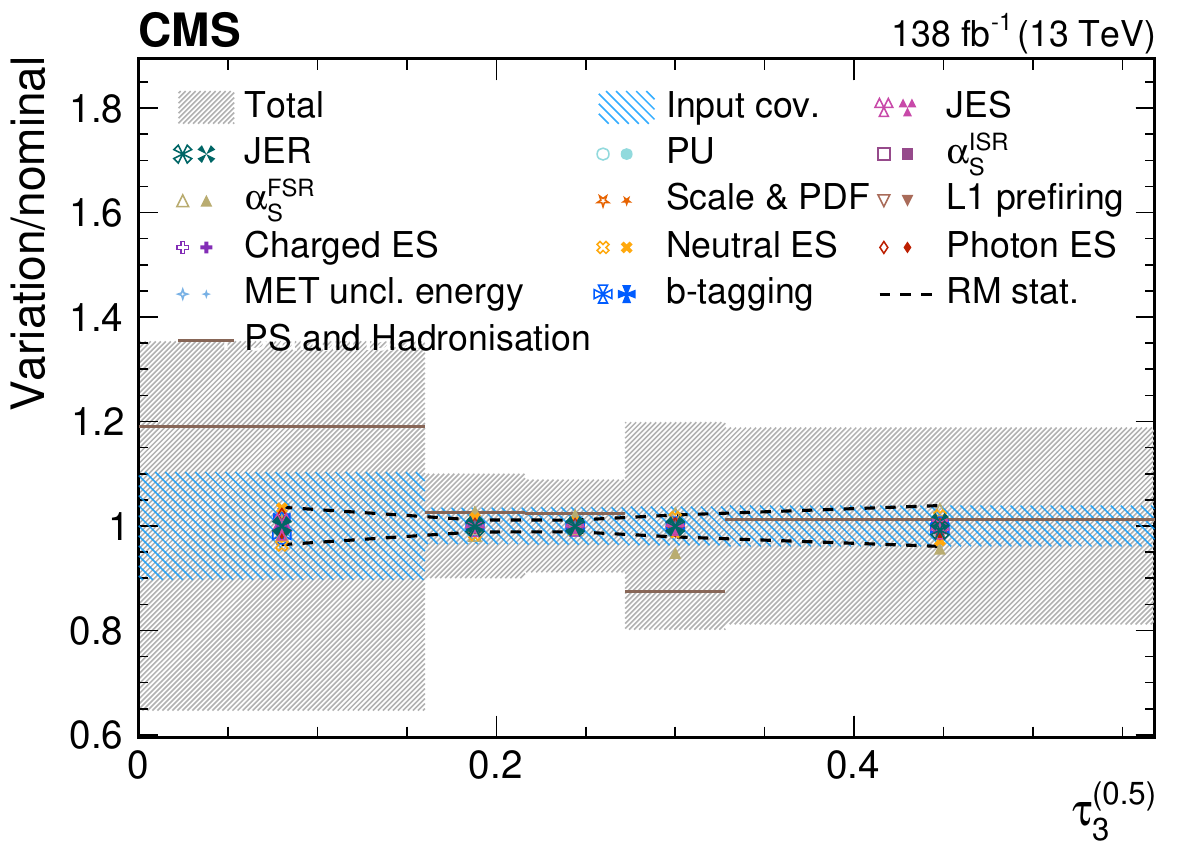}
	\includegraphics[width=.42\textwidth]{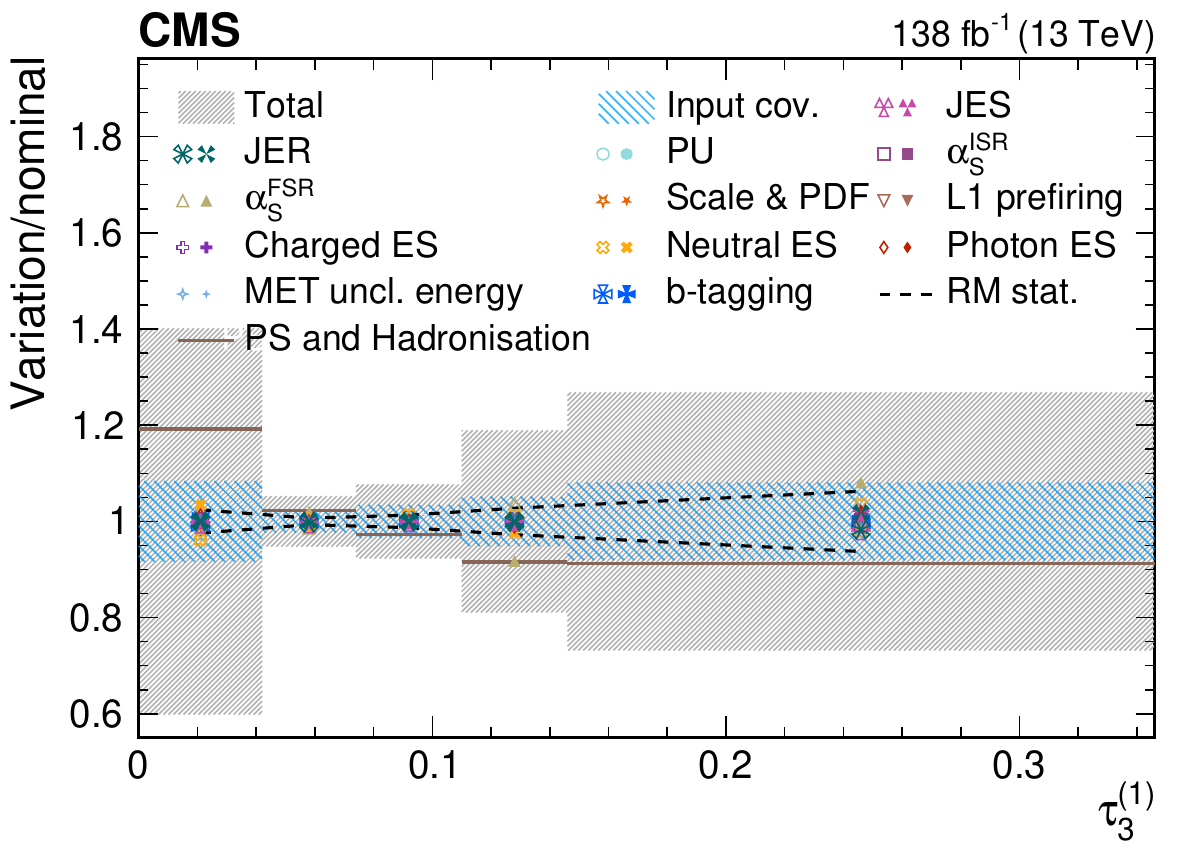}
	\includegraphics[width=.42\textwidth]{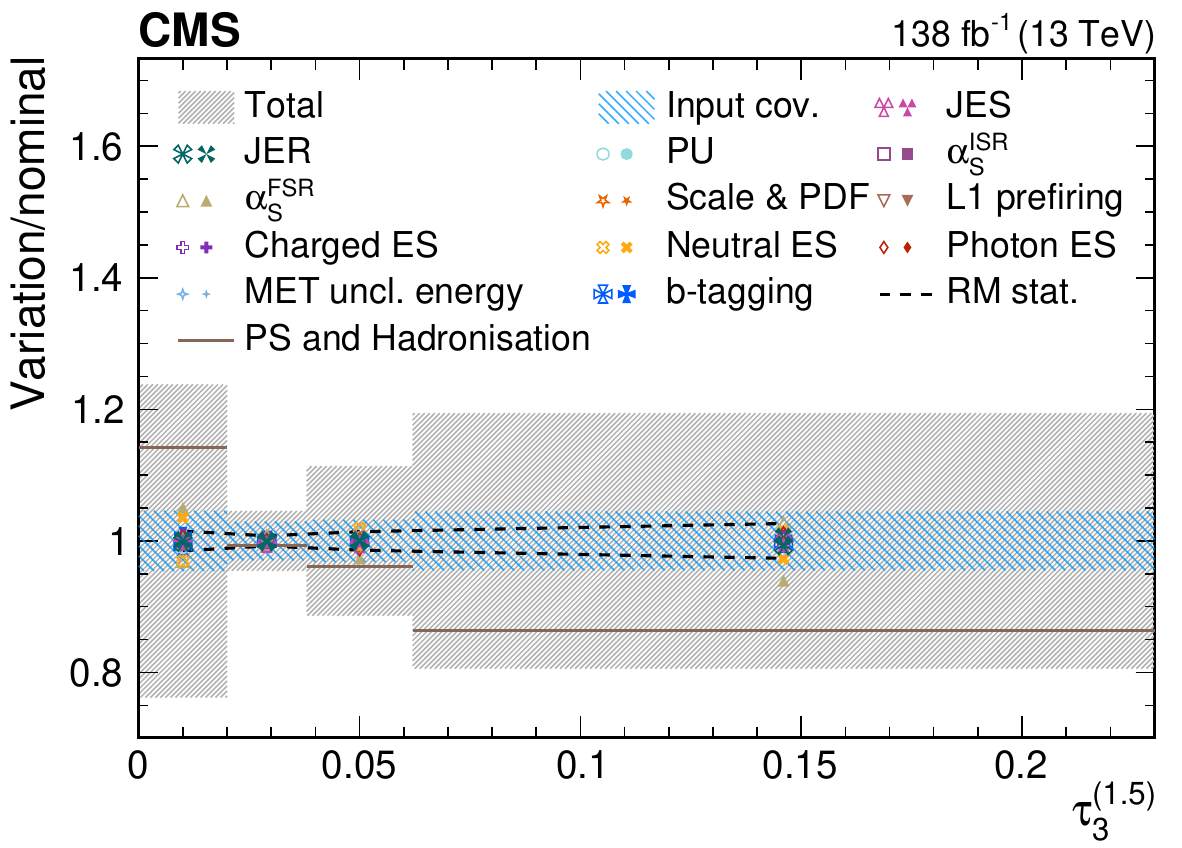}
	\includegraphics[width=.42\textwidth]{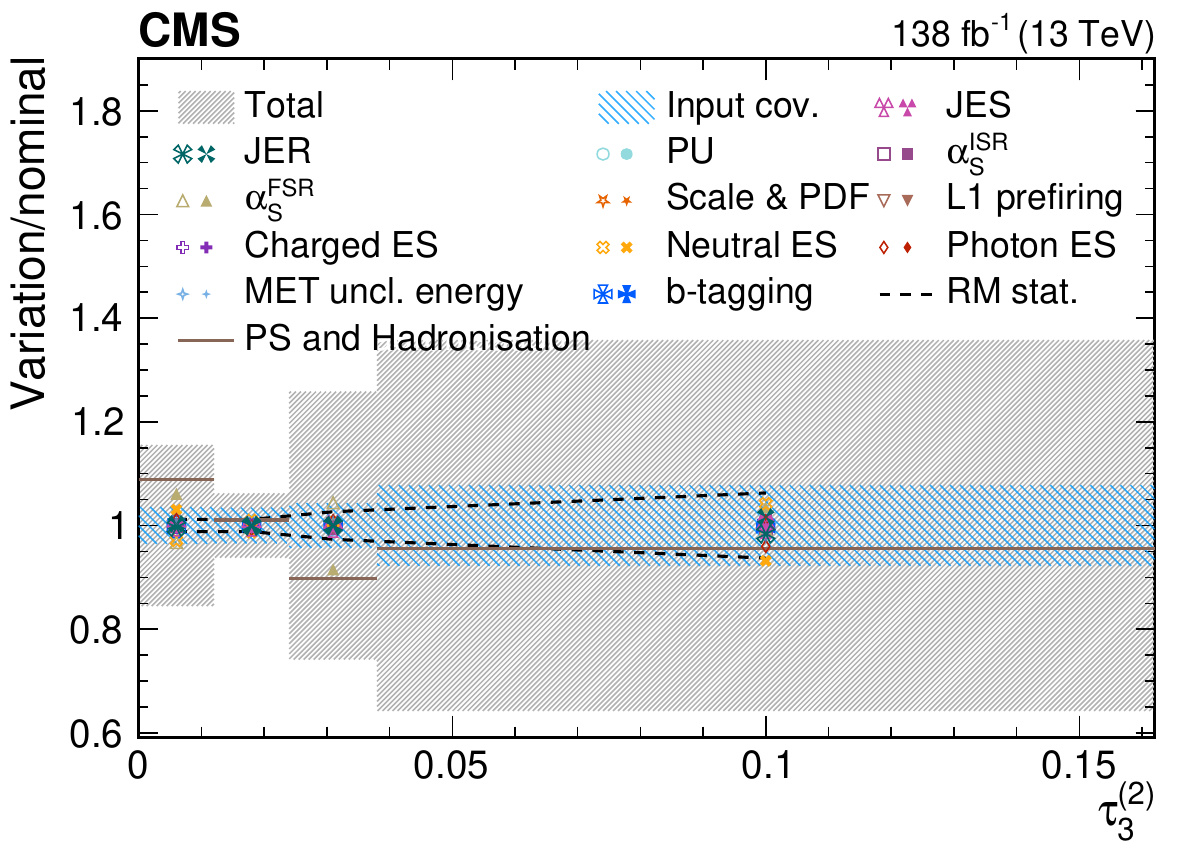}
	\caption{Contributions from various systematic variations to the normalized, unfolded distribution for $\tau_3^{(\beta)}$ observables measured for AK8 jets passing the boosted \PW boson-enriched selection in $\PGm$+jets \ttbar events. 
		The total unfolding uncertainty is indicated with the dark grey, hashed region, while the blue hashed region indicates the contributions from the input covariance matrix, which includes the propagated effects of the statistical uncertainties of the input data after background subtraction. Contributions from statistical uncertainties of the simulated sample used to construct the nominal response matrix are indicated with the dashed black line. The physics model uncertainty is computed as a one-sided shift compared to the nominal unfolding, and up (down) contributions from other sources are indicated with filled (open) markers of the same type and colour.}
	\label{fig:unfUncsW_tau3}
\end{figure}

\begin{figure}[htpb]
	\centering
	\includegraphics[width=.42\textwidth]{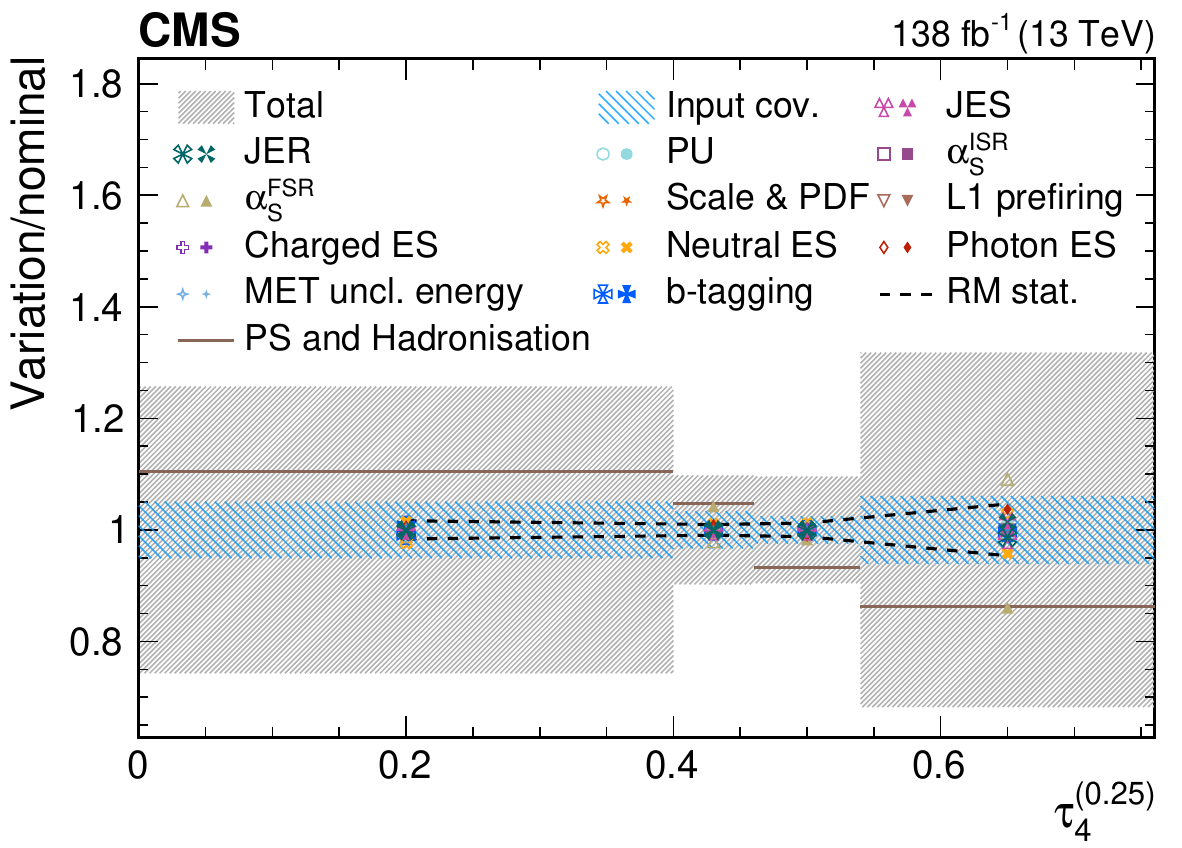}
	\includegraphics[width=.42\textwidth]{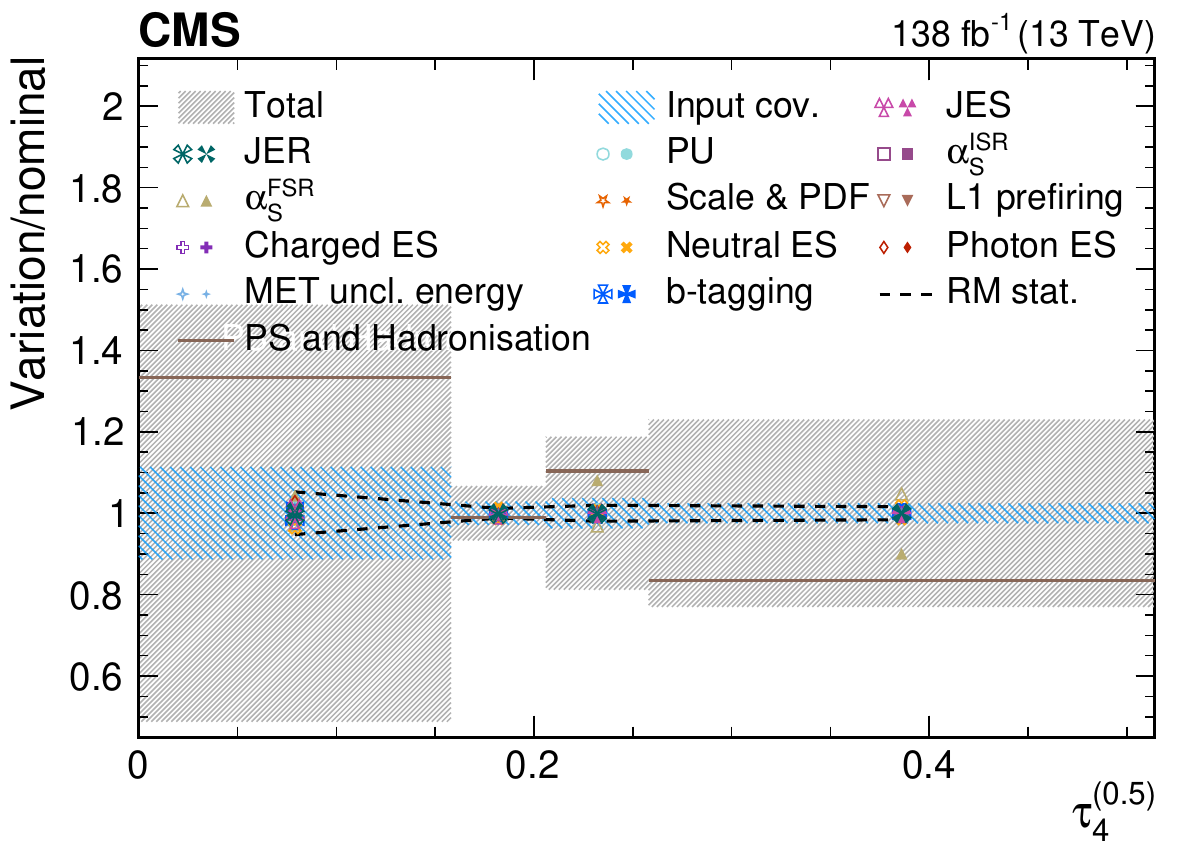}
	\includegraphics[width=.42\textwidth]{Figure_017-c.pdf}
	\includegraphics[width=.42\textwidth]{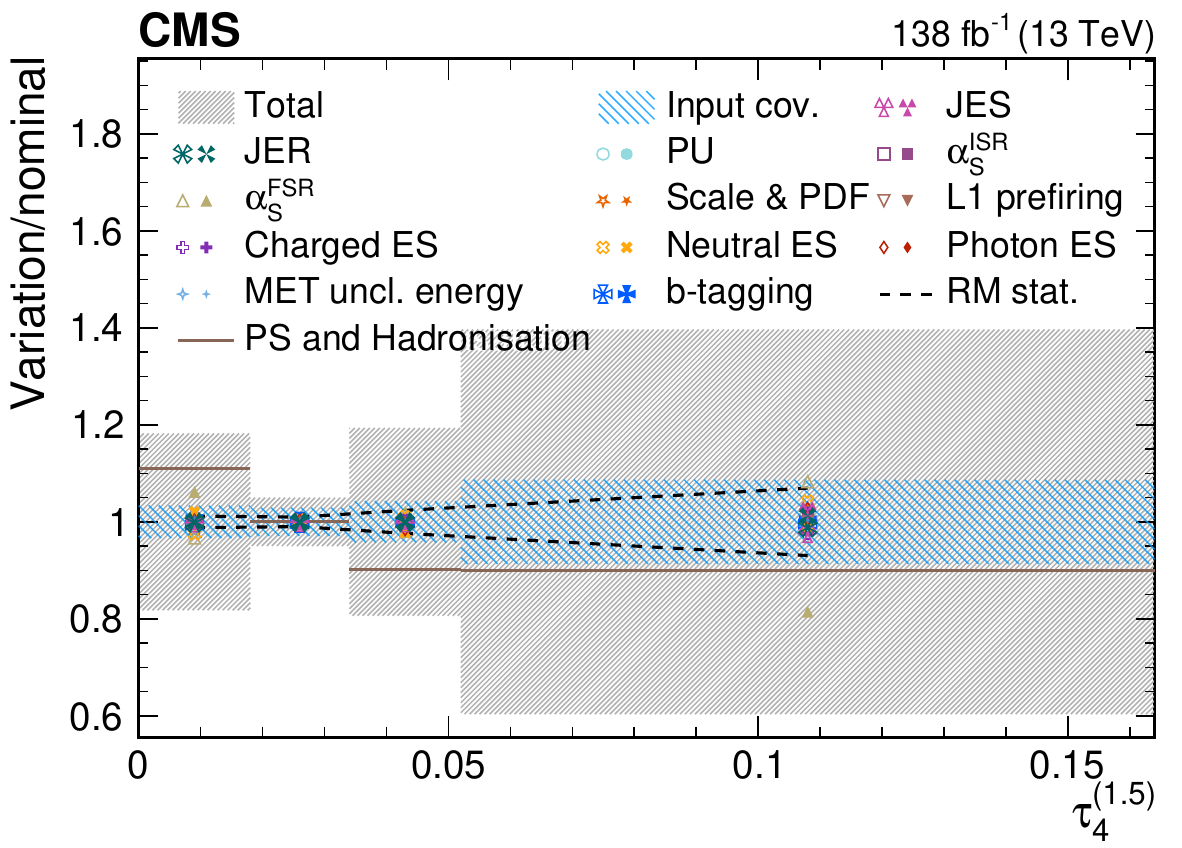}
	\includegraphics[width=.42\textwidth]{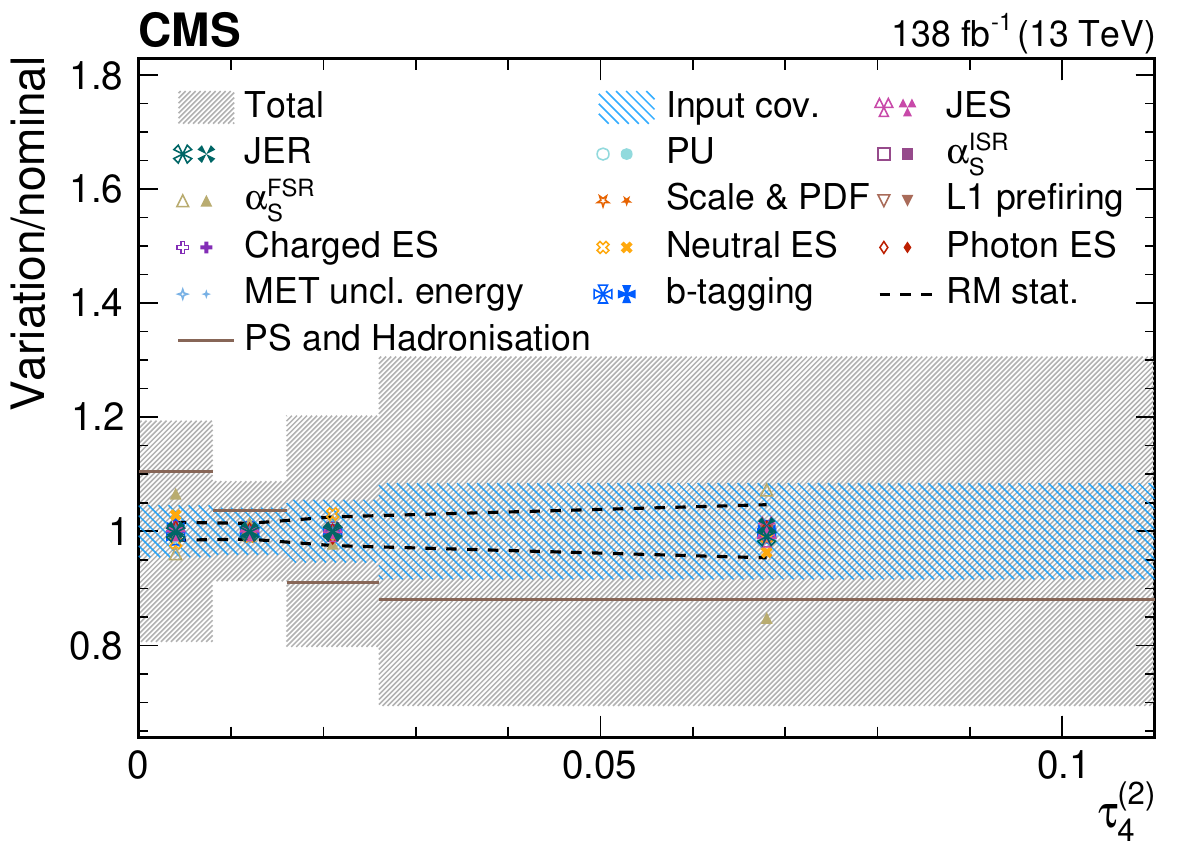}
	\caption{Contributions from various systematic variations to the normalized, unfolded distribution for $\tau_4^{(\beta)}$ observables measured for AK8 jets passing the boosted \PW boson-enriched selection in $\PGm$+jets \ttbar events. 
		The total unfolding uncertainty is indicated with the dark grey, hashed region, while the blue hashed region indicates the contributions from the input covariance matrix, which includes the propagated effects of the statistical uncertainties of the input data after background subtraction. Contributions from statistical uncertainties of the simulated sample used to construct the nominal response matrix are indicated with the dashed black line. The physics model uncertainty is computed as a one-sided shift compared to the nominal unfolding, and up (down) contributions from other sources are indicated with filled (open) markers of the same type and colour.}
	\label{fig:unfUncsW_tau4}
\end{figure}

\begin{figure}[htpb]
	\centering
	\includegraphics[width=.42\textwidth]{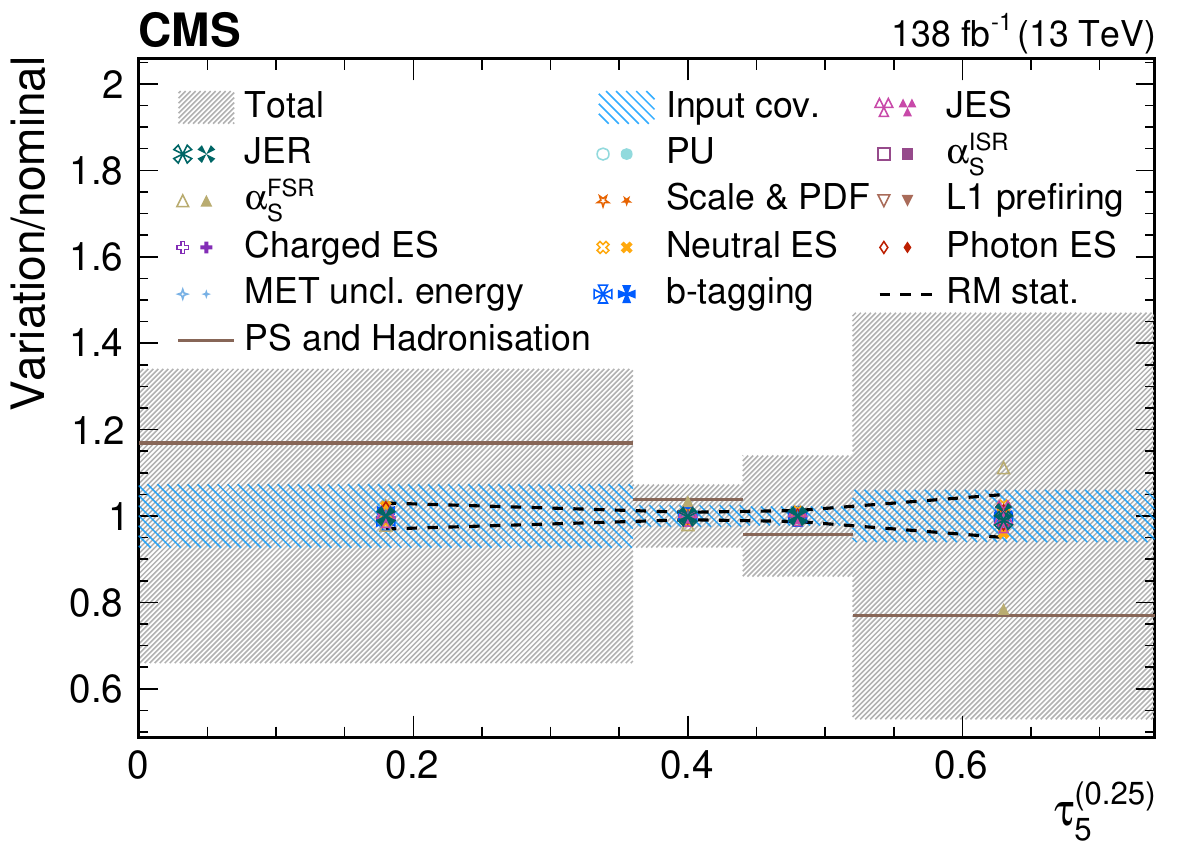}
	\includegraphics[width=.42\textwidth]{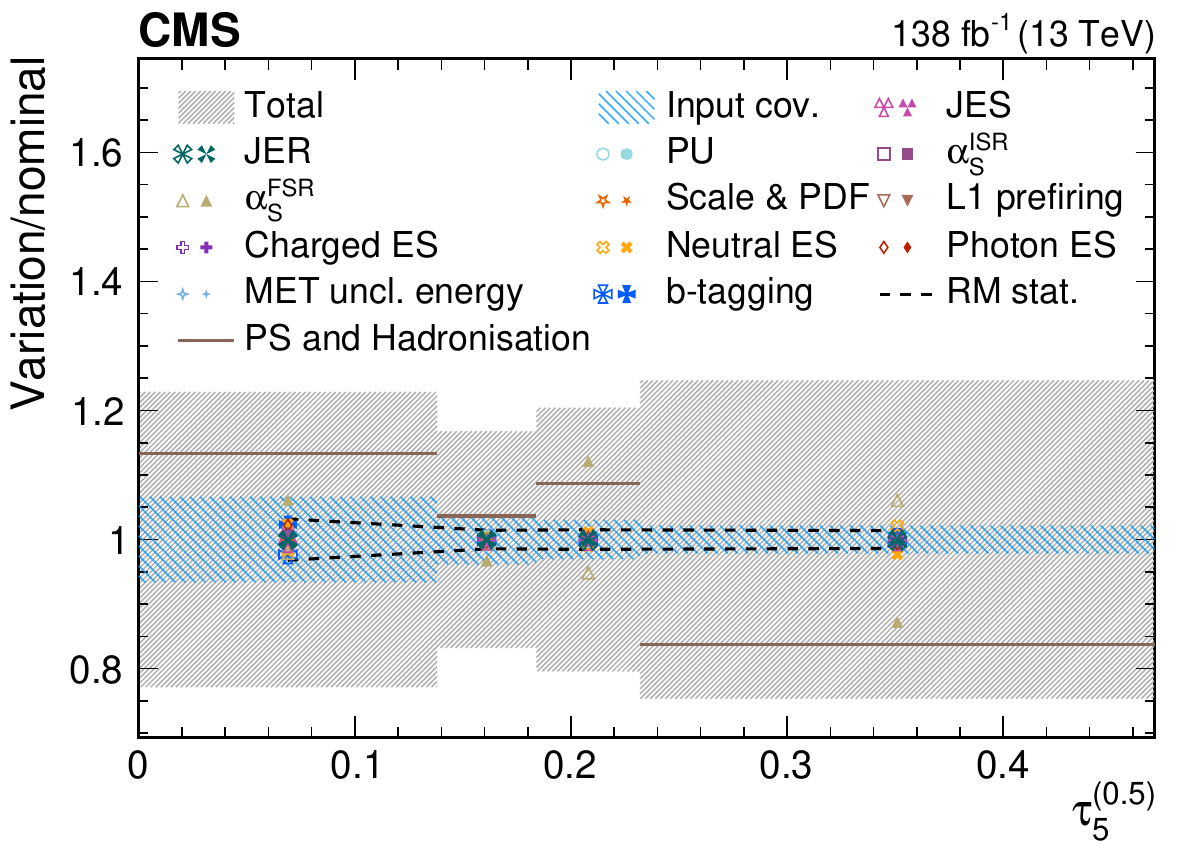}
	\includegraphics[width=.42\textwidth]{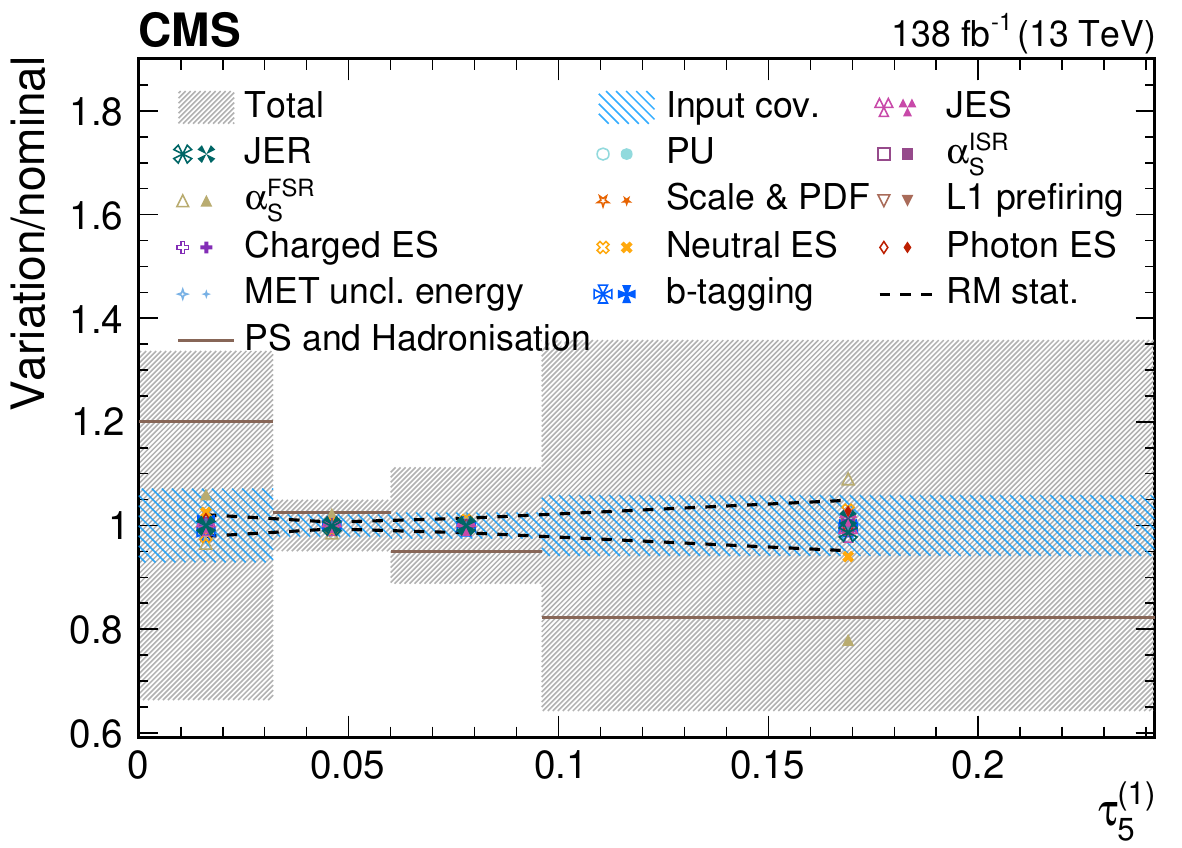}
	\includegraphics[width=.42\textwidth]{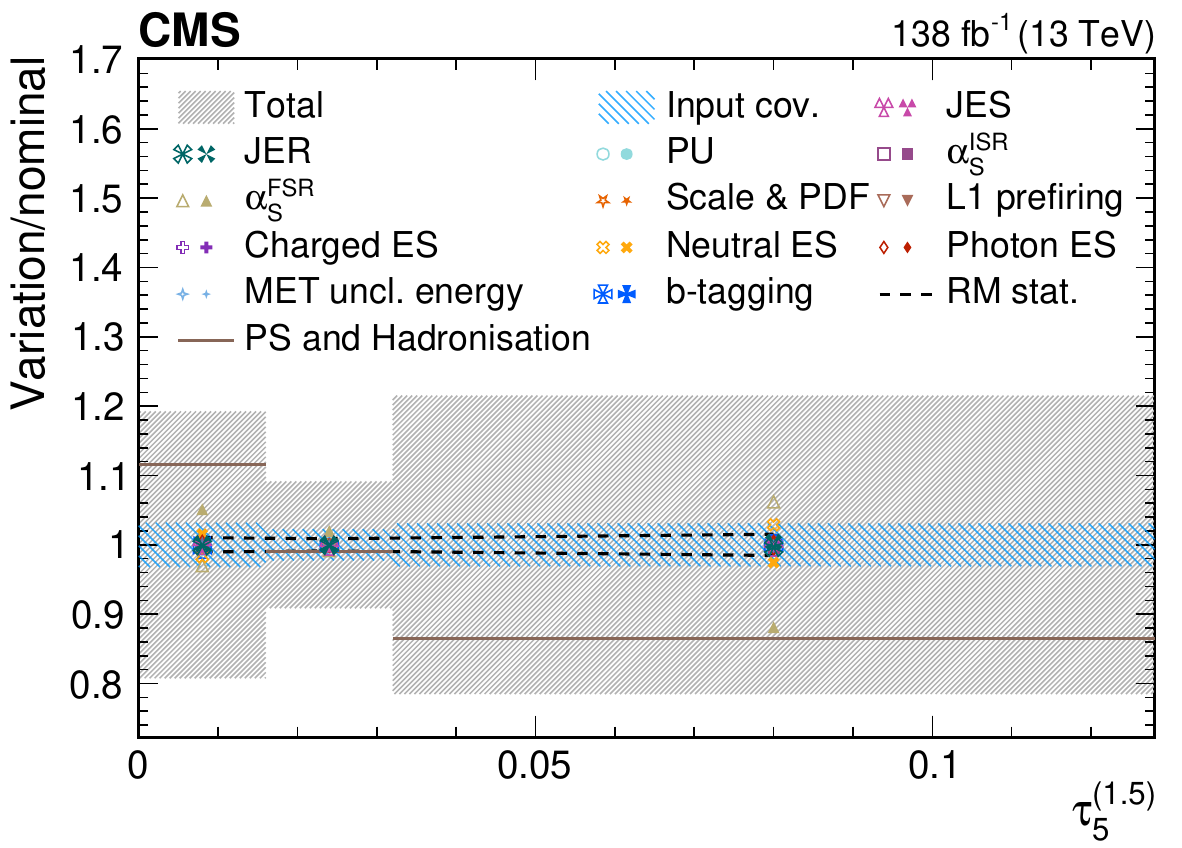}
	\includegraphics[width=.42\textwidth]{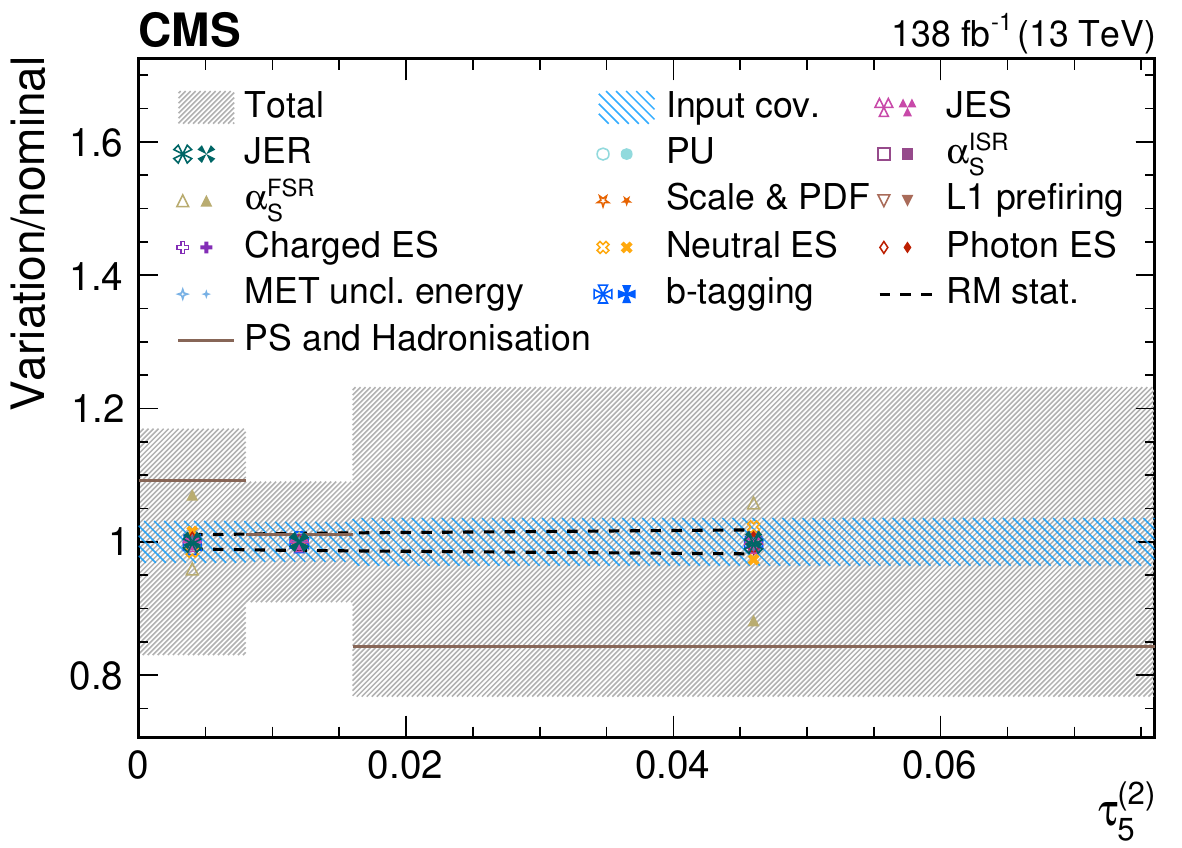}
	\caption{Contributions from various systematic variations to the normalized, unfolded distribution for $\tau_5^{(\beta)}$ observables measured for AK8 jets passing the boosted \PW boson-enriched selection in $\PGm$+jets \ttbar events. 
		The total unfolding uncertainty is indicated with the dark grey, hashed region, while the blue hashed region indicates the contributions from the input covariance matrix, which includes the propagated effects of the statistical uncertainties of the input data after background subtraction. Contributions from statistical uncertainties of the simulated sample used to construct the nominal response matrix are indicated with the dashed black line. The physics model uncertainty is computed as a one-sided shift compared to the nominal unfolding, and up (down) contributions from other sources are indicated with filled (open) markers of the same type and colour.}
	\label{fig:unfUncsW_tau5}
\end{figure}

\begin{figure}[htpb]
	\centering
	\includegraphics[width=.42\textwidth]{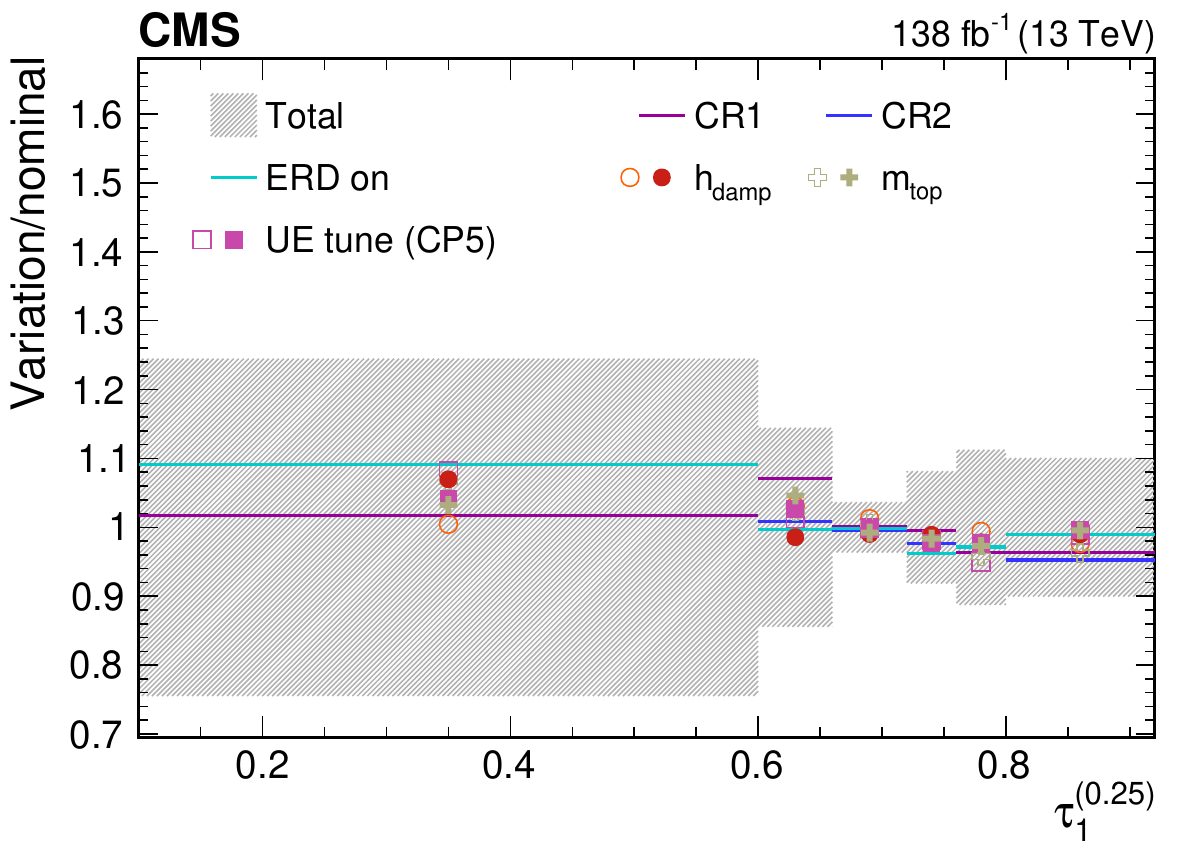}
	\includegraphics[width=.42\textwidth]{Figure_017-b.pdf}
	\includegraphics[width=.42\textwidth]{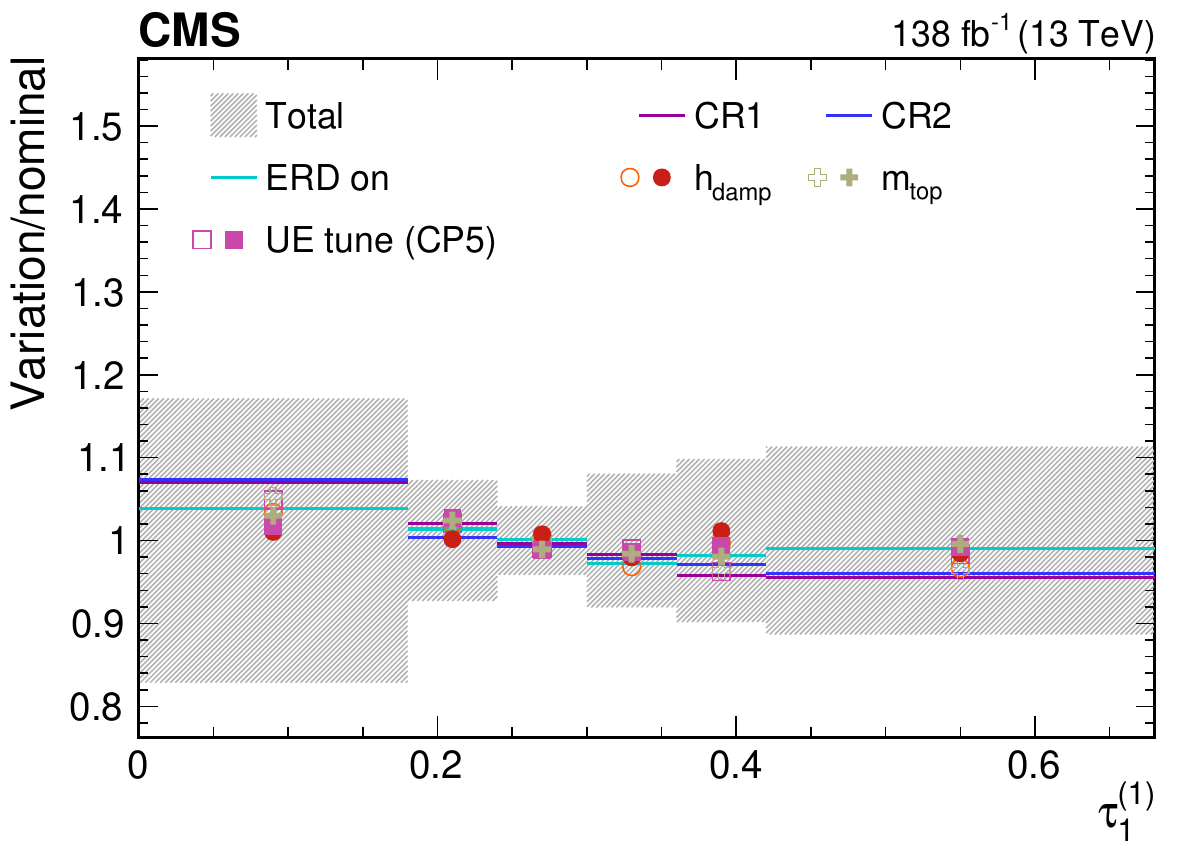}
	\includegraphics[width=.42\textwidth]{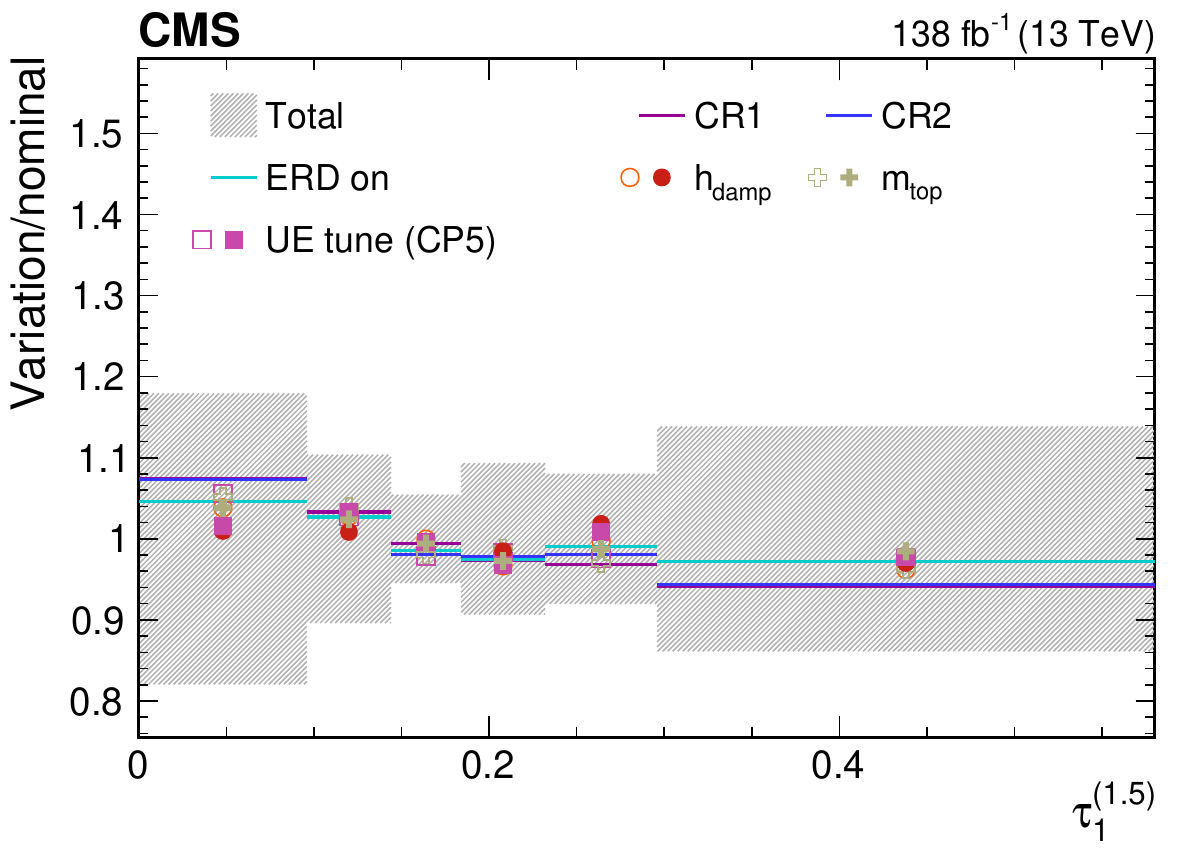}
	\includegraphics[width=.42\textwidth]{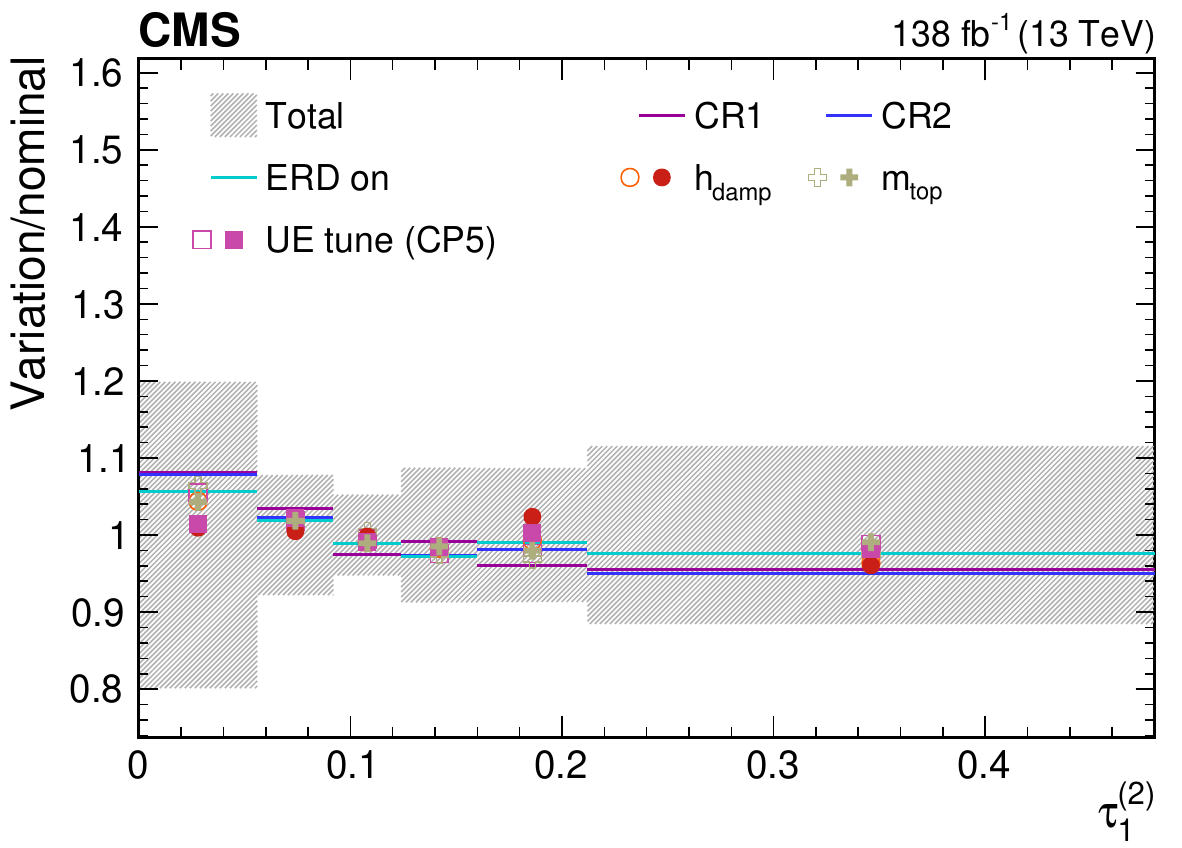}
	\caption{Contributions from various theory model systematic variations to the normalized, unfolded distribution for $\tau_1^{(\beta)}$ observables measured for AK8 jets passing the boosted \PW boson-enriched selection in $\PGm$+jets \ttbar events. 
		The total unfolding uncertainty is indicated with the dark grey, hashed region, while the blue hashed region indicates the contributions from the input covariance matrix, which includes the propagated effects of the statistical uncertainties of the input data after background subtraction. Contributions from statistical uncertainties of the simulated sample used to construct the nominal response matrix are indicated with the dashed black line. The uncertainty contributions for different choices of colour reconnection models are illustrated as one-sided shifts compared to the nominal unfolding, and up (down) contributions from other sources are indicated with filled (open) markers of the same type and colour.}
	\label{fig:unfUncsTheoryW_tau1}
\end{figure}

\begin{figure}[htpb]
	\centering
	\includegraphics[width=.42\textwidth]{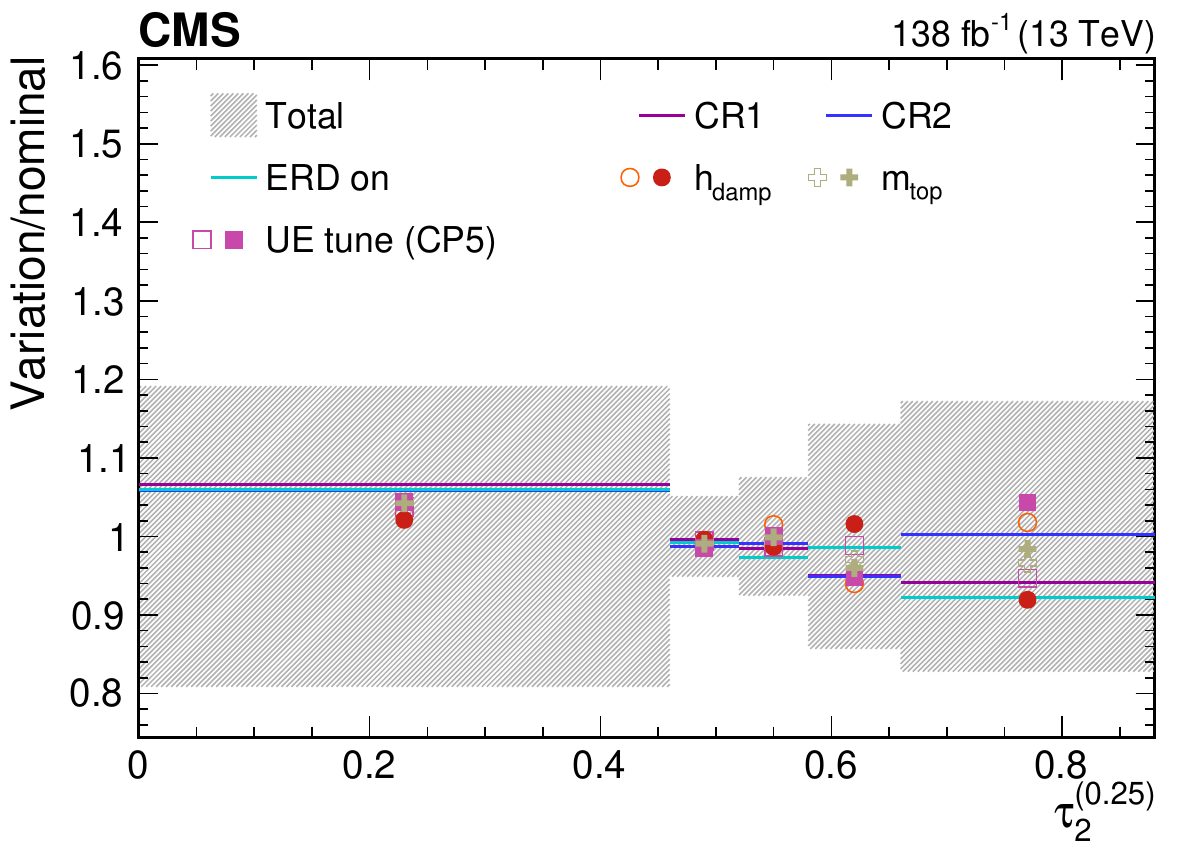}
	\includegraphics[width=.42\textwidth]{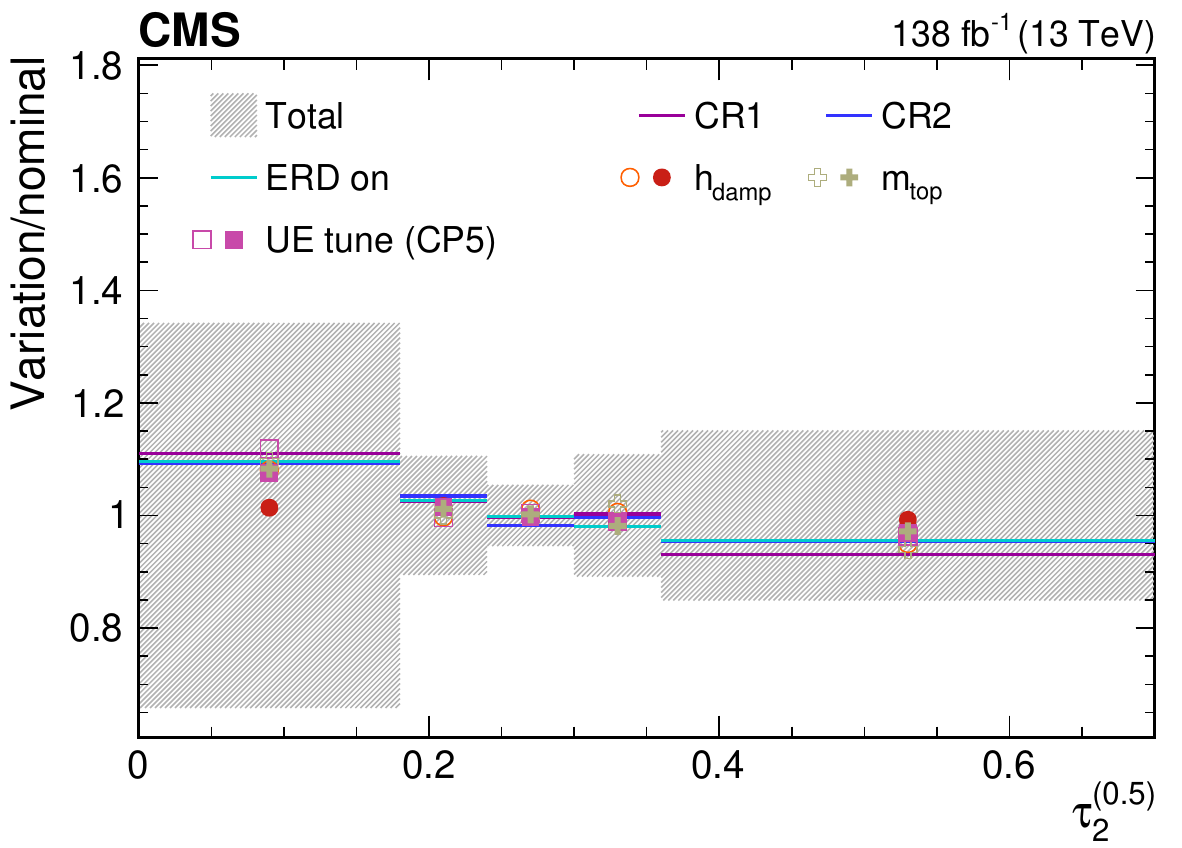}
	\includegraphics[width=.42\textwidth]{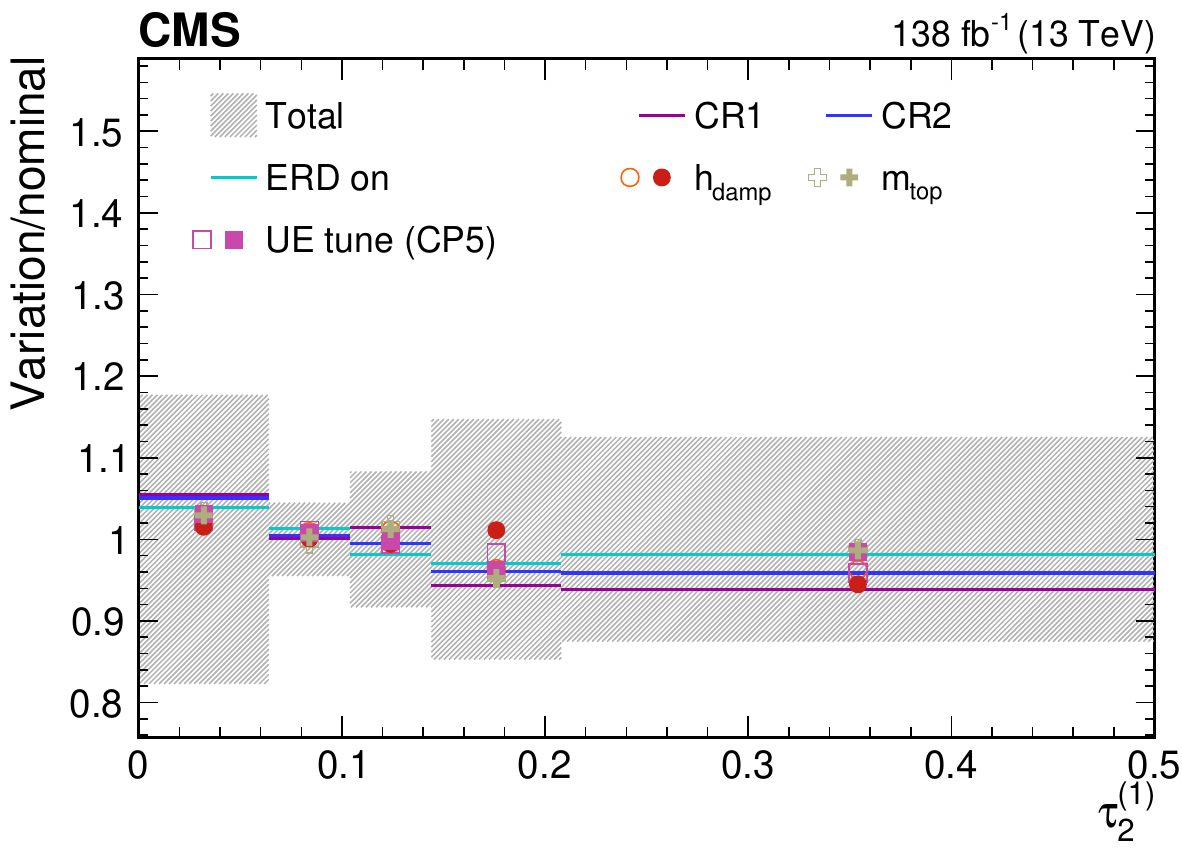}
	\includegraphics[width=.42\textwidth]{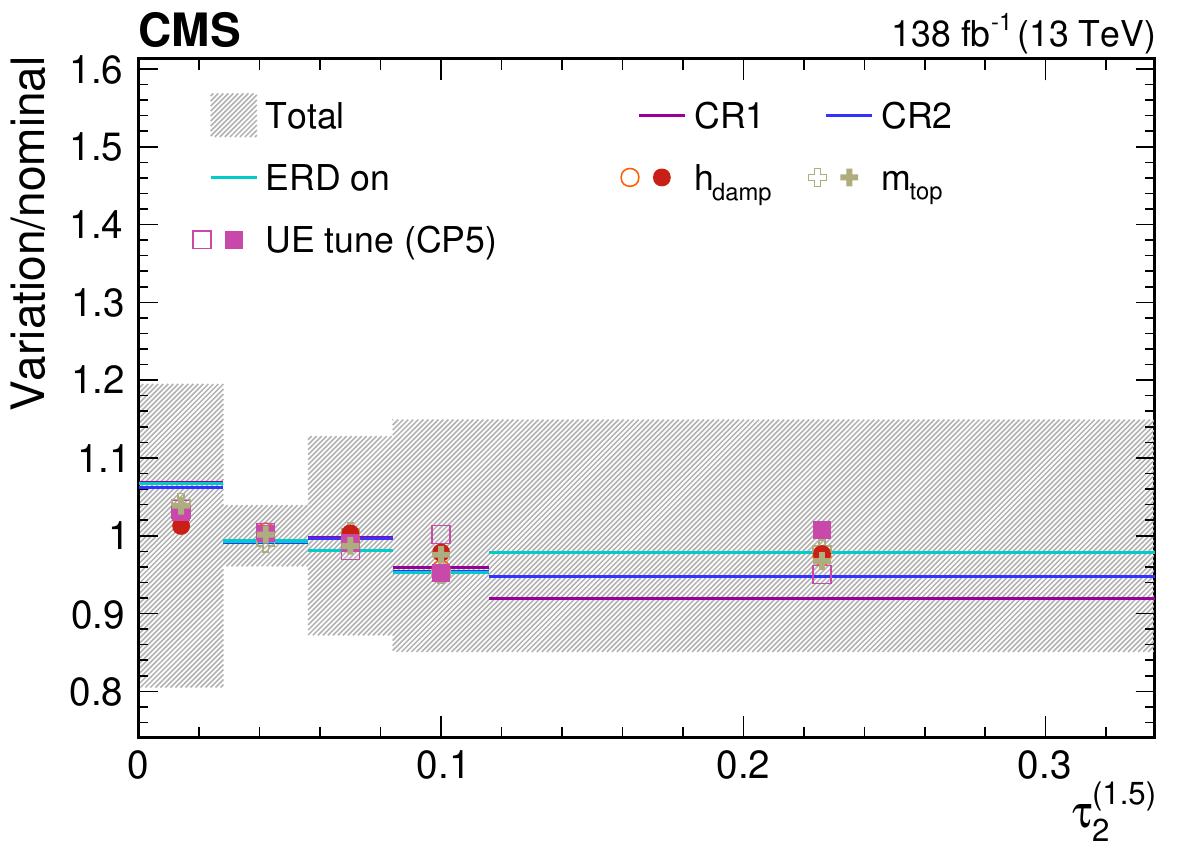}
	\includegraphics[width=.42\textwidth]{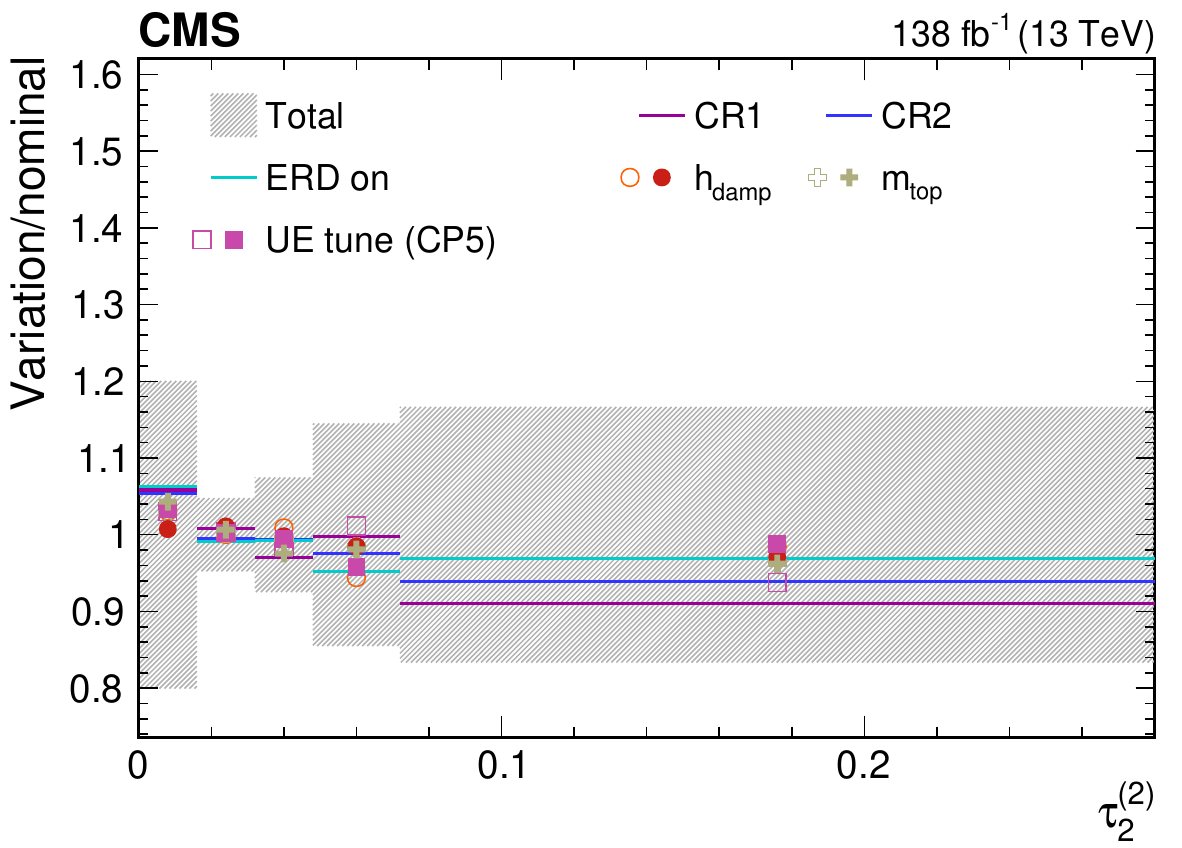}
	\caption{Contributions from various theory model systematic variations to the normalized, unfolded distribution for $\tau_2^{(\beta)}$ observables measured for AK8 jets passing the boosted \PW boson-enriched selection in $\PGm$+jets \ttbar events. 
		The total unfolding uncertainty is indicated with the dark grey, hashed region, while the blue hashed region indicates the contributions from the input covariance matrix, which includes the propagated effects of the statistical uncertainties of the input data after background subtraction. Contributions from statistical uncertainties of the simulated sample used to construct the nominal response matrix are indicated with the dashed black line. The uncertainty contributions for different choices of colour reconnection models are illustrated as one-sided shifts compared to the nominal unfolding, and up (down) contributions from other sources are indicated with filled (open) markers of the same type and colour.}
	\label{fig:unfUncsTheoryW_tau2}
\end{figure}

\begin{figure}[htpb]
	\centering
	\includegraphics[width=.42\textwidth]{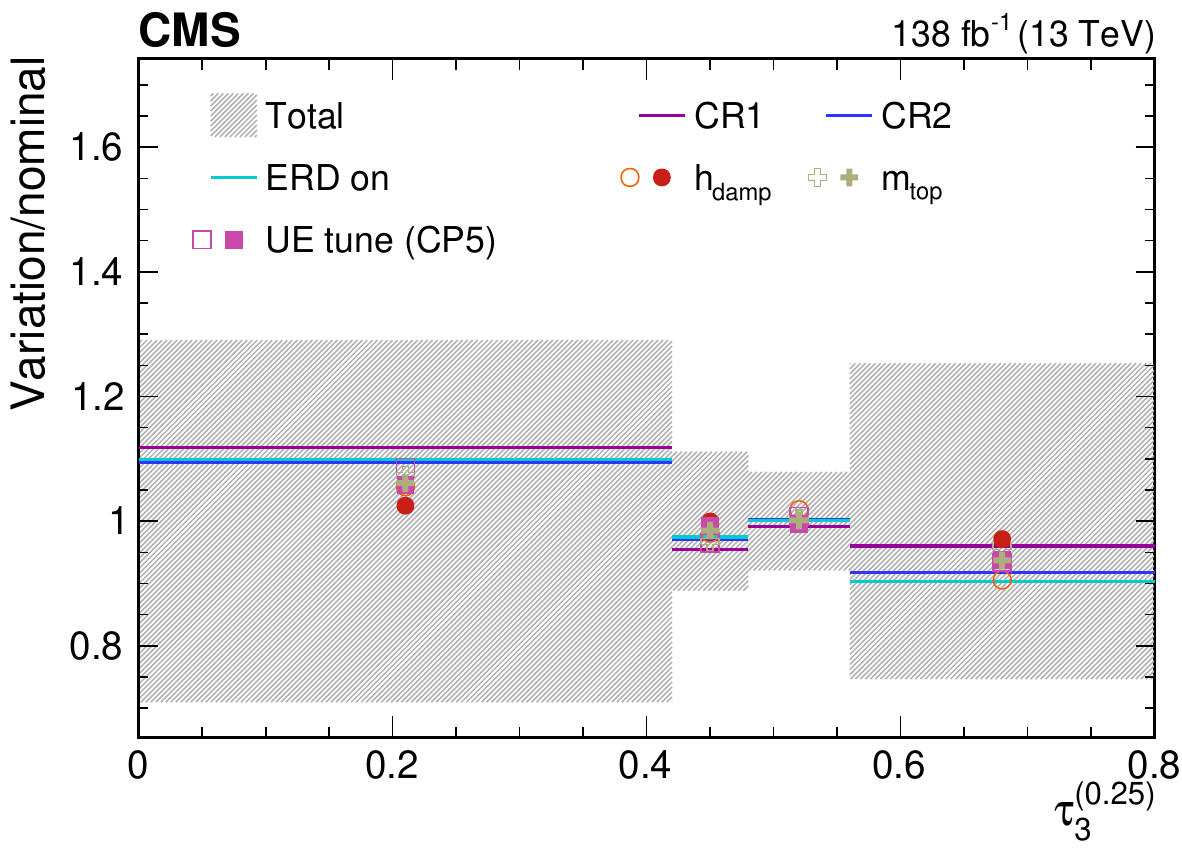}
	\includegraphics[width=.42\textwidth]{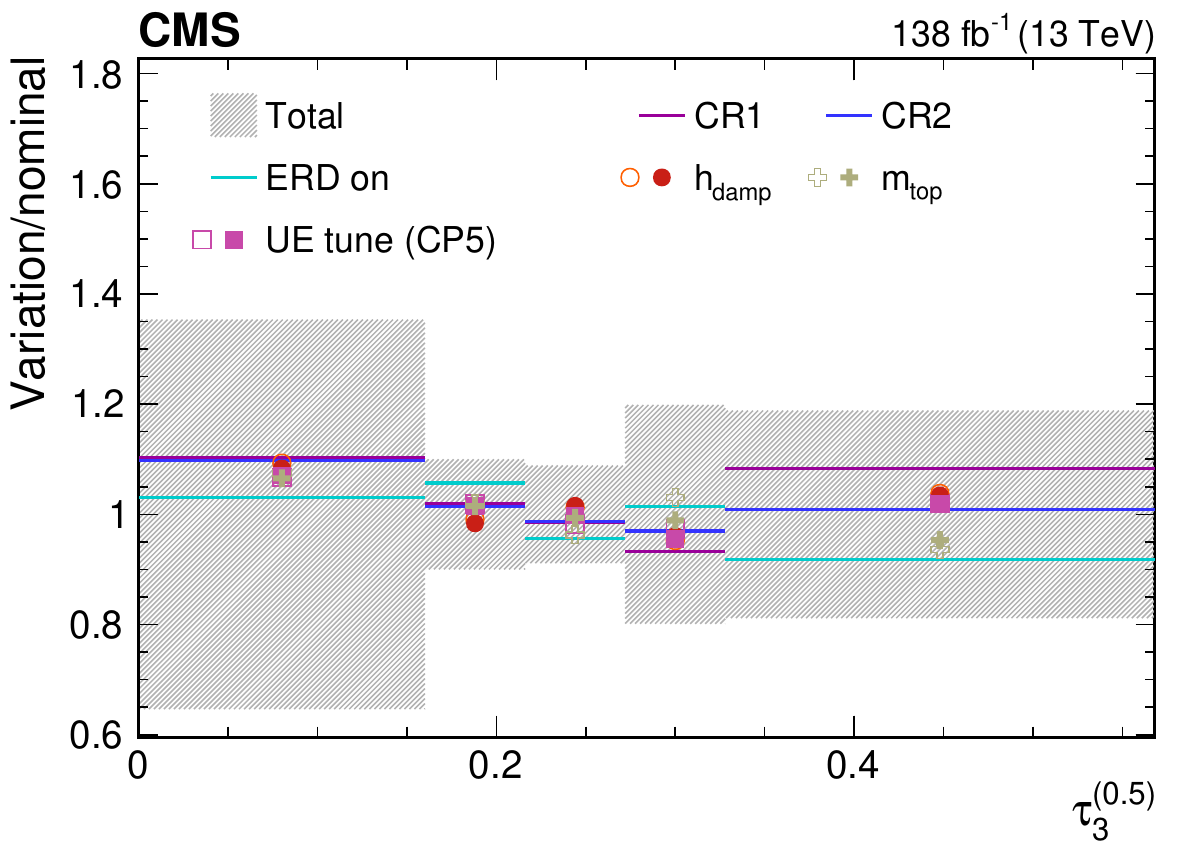}
	\includegraphics[width=.42\textwidth]{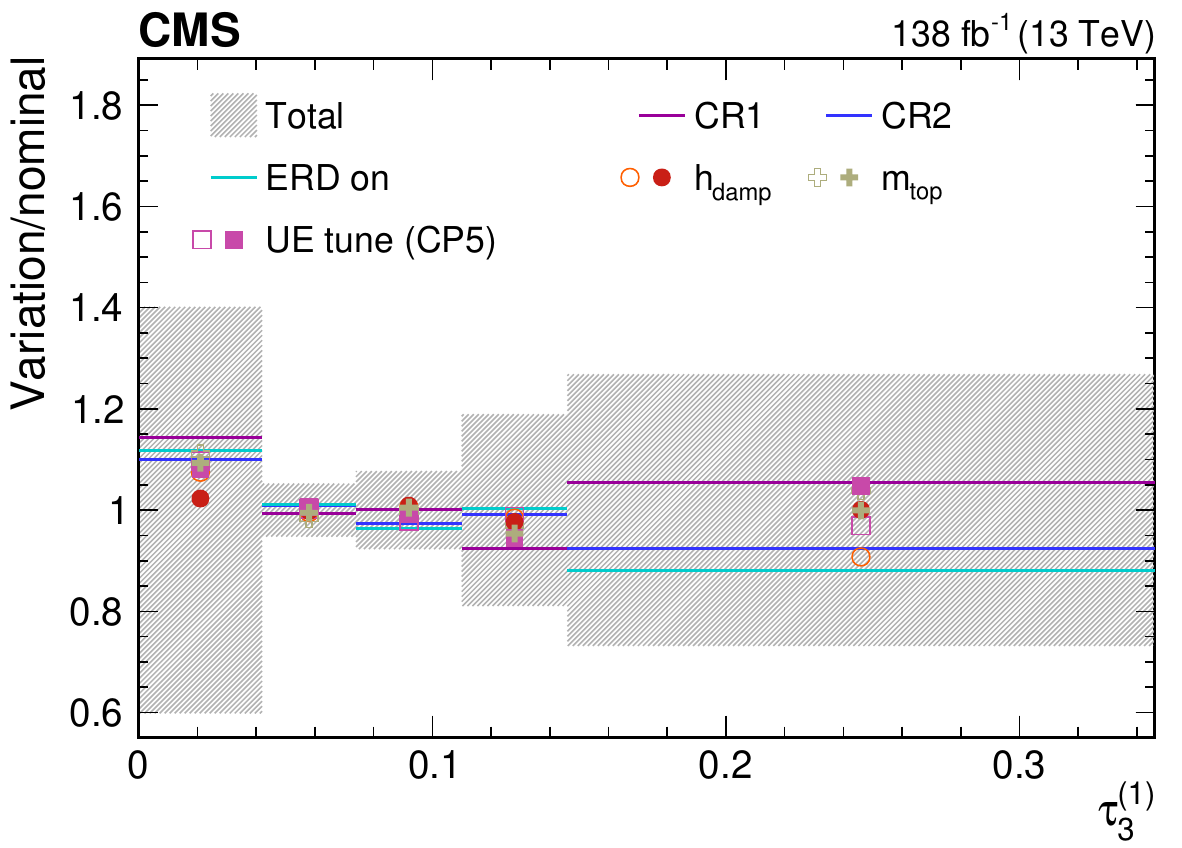}
	\includegraphics[width=.42\textwidth]{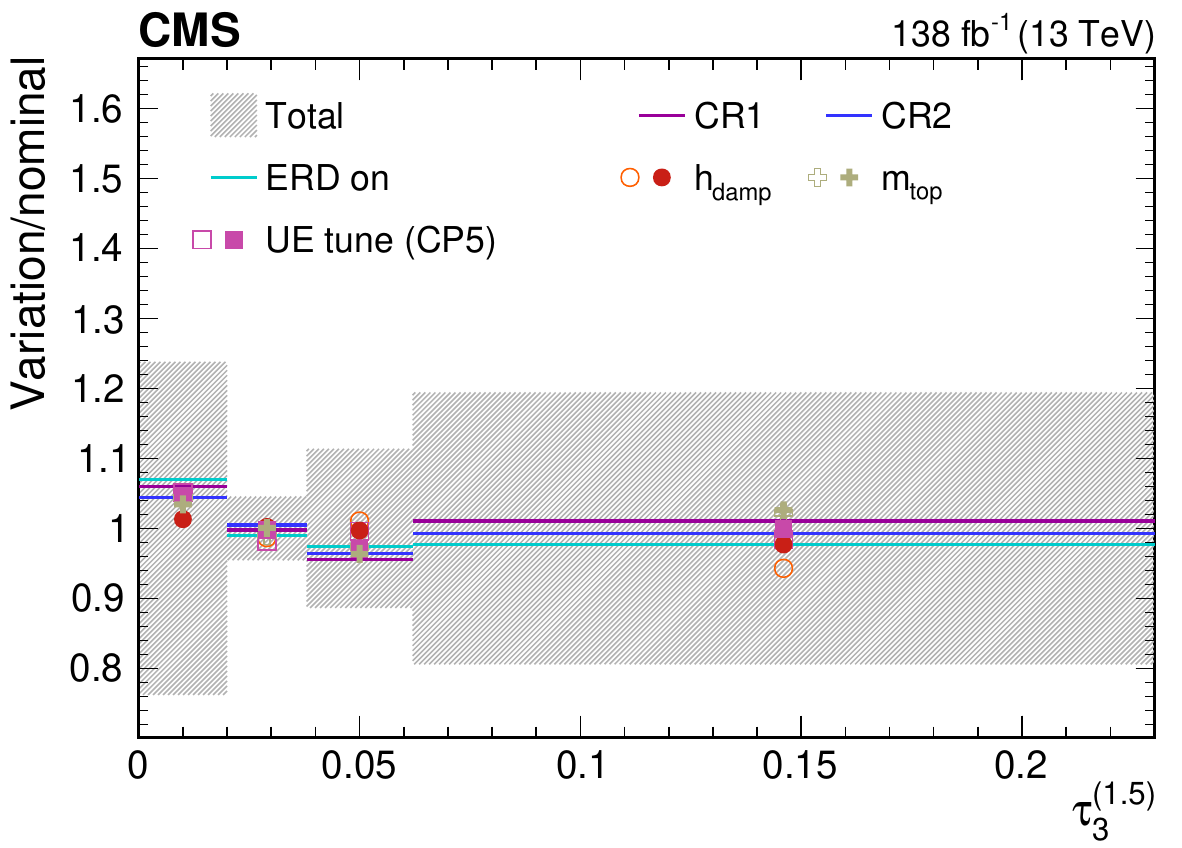}
	\includegraphics[width=.42\textwidth]{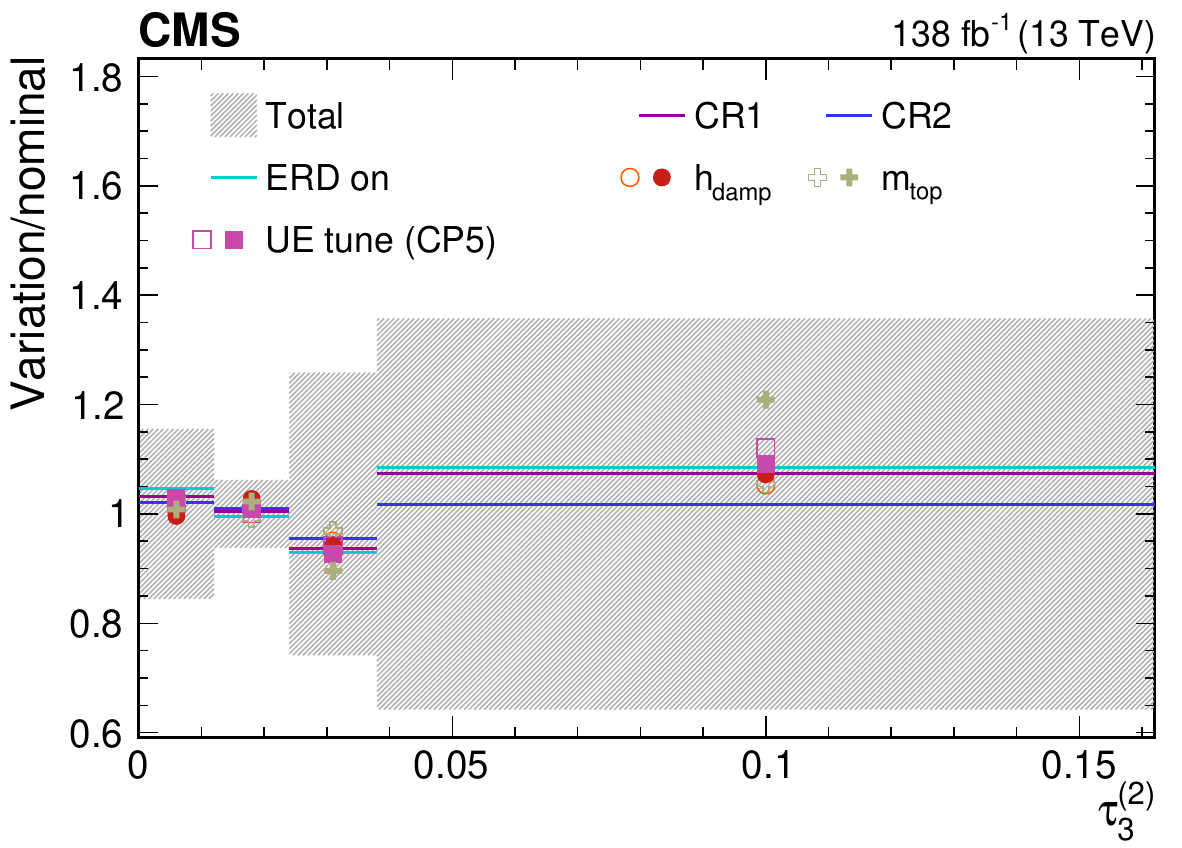}
	\caption{Contributions from various theory model systematic variations to the normalized, unfolded distribution for $\tau_3^{(\beta)}$ observables measured for AK8 jets passing the boosted \PW boson-enriched selection in $\PGm$+jets \ttbar events. 
		The total unfolding uncertainty is indicated with the dark grey, hashed region, while the blue hashed region indicates the contributions from the input covariance matrix, which includes the propagated effects of the statistical uncertainties of the input data after background subtraction. Contributions from statistical uncertainties of the simulated sample used to construct the nominal response matrix are indicated with the dashed black line. The uncertainty contributions for different choices of colour reconnection models are illustrated as one-sided shifts compared to the nominal unfolding, and up (down) contributions from other sources are indicated with filled (open) markers of the same type and colour.}
	\label{fig:unfUncsTheoryW_tau3}
\end{figure}

\begin{figure}[htpb]
	\centering
	\includegraphics[width=.42\textwidth]{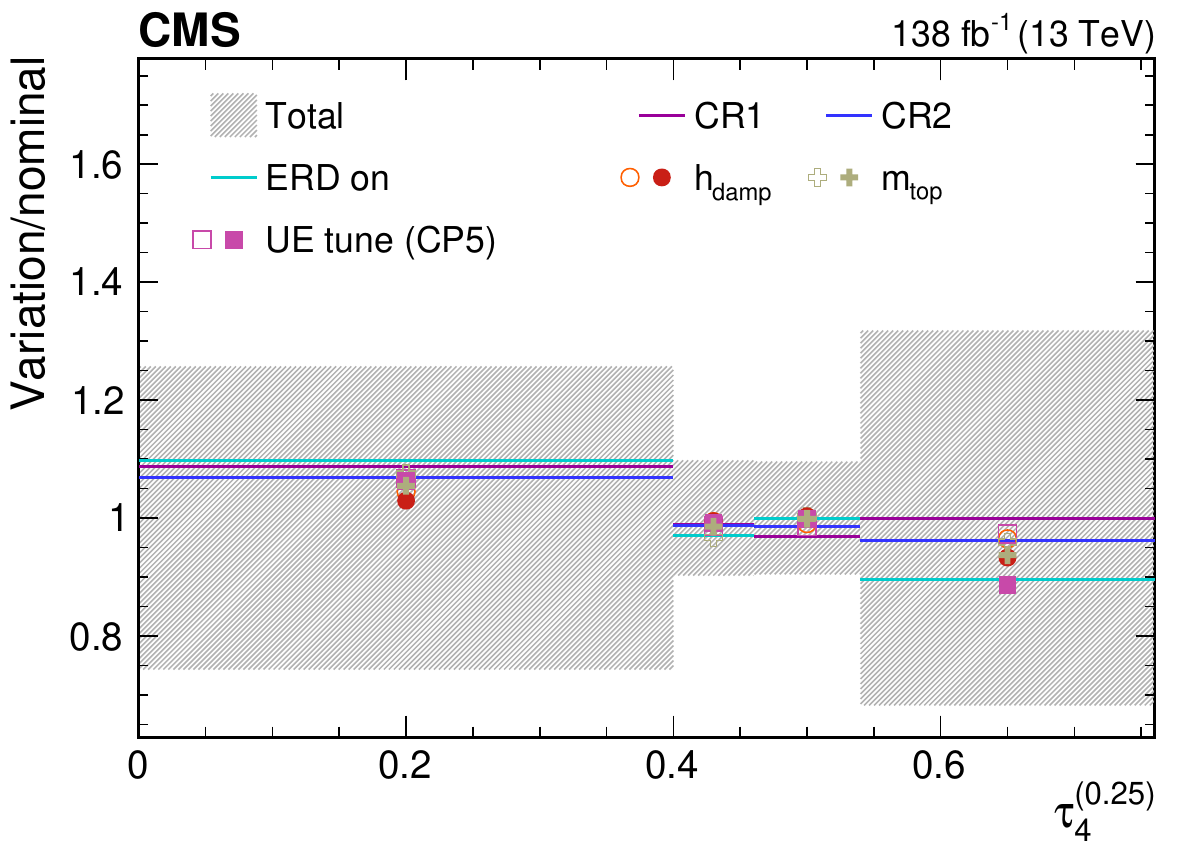}
	\includegraphics[width=.42\textwidth]{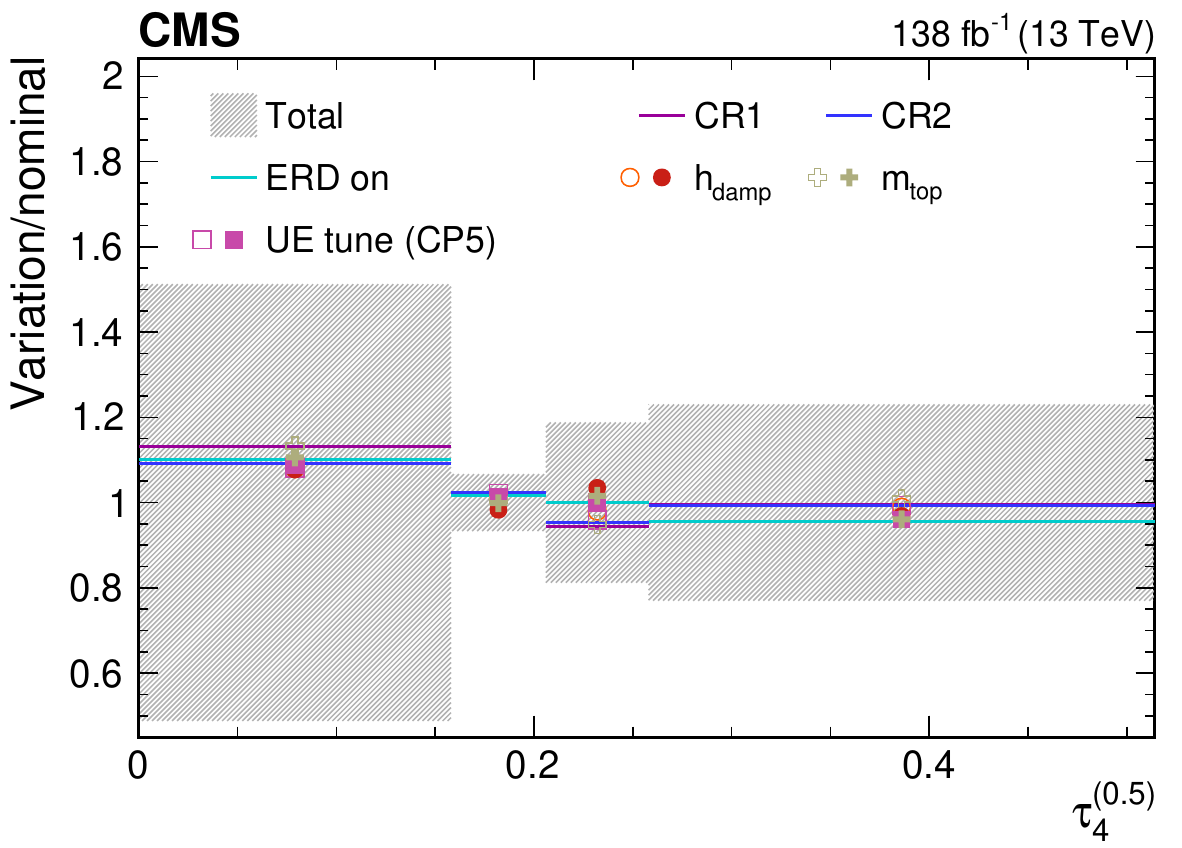}
	\includegraphics[width=.42\textwidth]{Figure_017-d.pdf}
	\includegraphics[width=.42\textwidth]{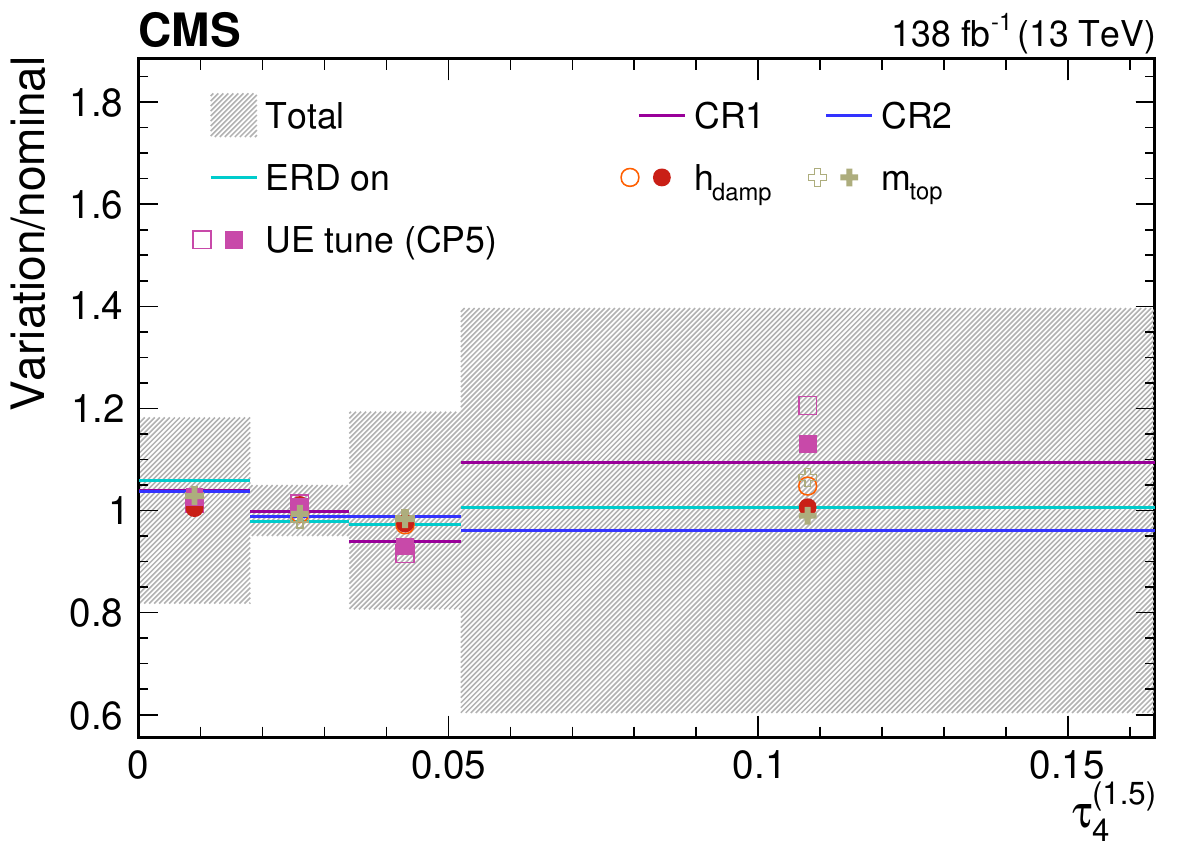}
	\includegraphics[width=.42\textwidth]{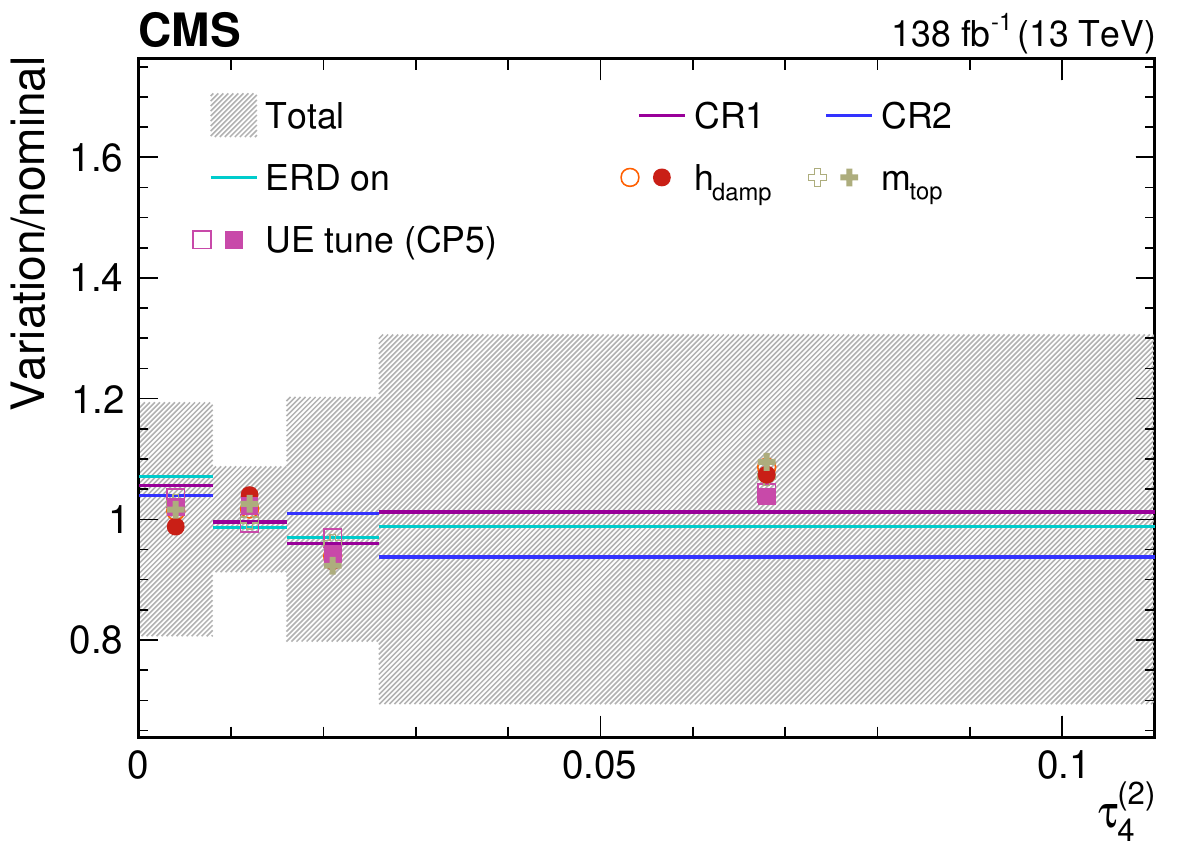}
	\caption{Contributions from various theory model systematic variations to the normalized, unfolded distribution for $\tau_4^{(\beta)}$ observables measured for AK8 jets passing the boosted \PW boson-enriched selection in $\PGm$+jets \ttbar events. 
		The total unfolding uncertainty is indicated with the dark grey, hashed region, while the blue hashed region indicates the contributions from the input covariance matrix, which includes the propagated effects of the statistical uncertainties of the input data after background subtraction. Contributions from statistical uncertainties of the simulated sample used to construct the nominal response matrix are indicated with the dashed black line. The uncertainty contributions for different choices of colour reconnection models are illustrated as one-sided shifts compared to the nominal unfolding, and up (down) contributions from other sources are indicated with filled (open) markers of the same type and colour.}
	\label{fig:unfUncsTheoryW_tau4}
\end{figure}

\begin{figure}[htpb]
	\centering
	\includegraphics[width=.42\textwidth]{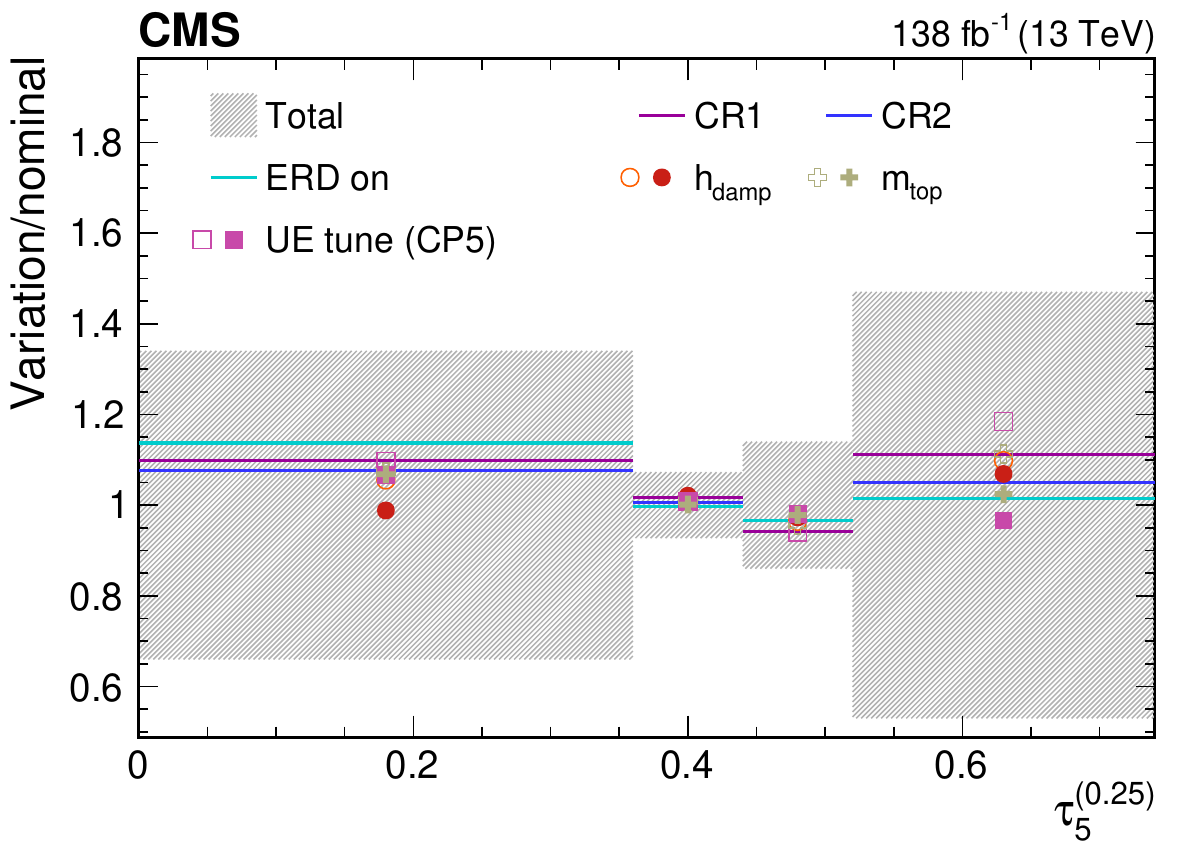}
	\includegraphics[width=.42\textwidth]{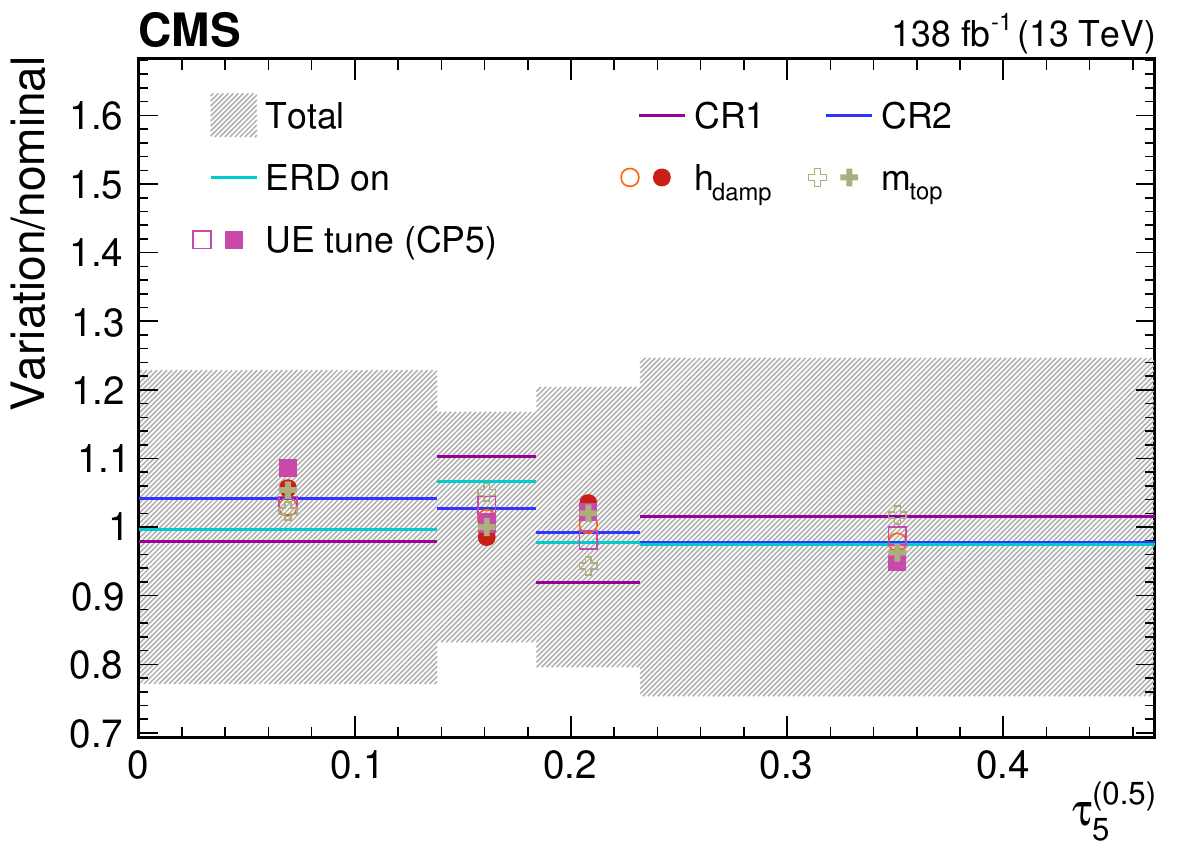}
	\includegraphics[width=.42\textwidth]{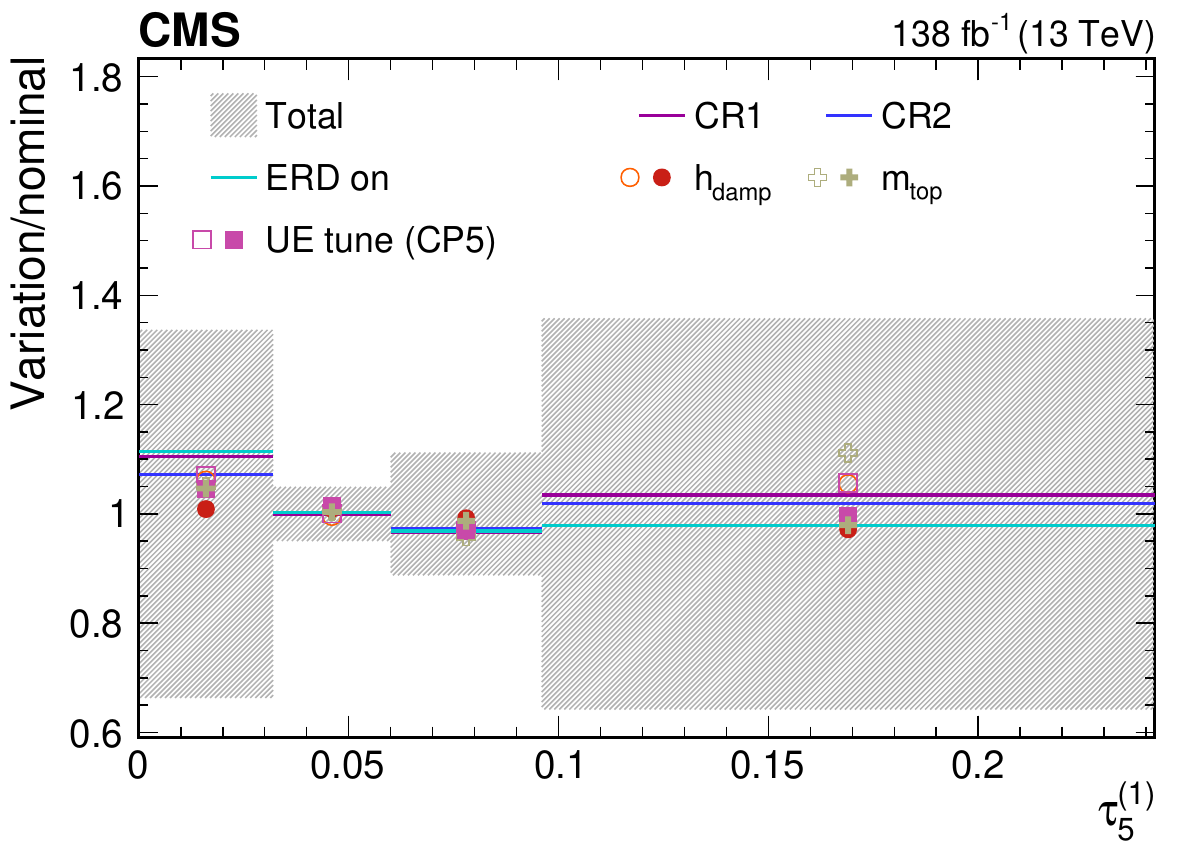}
	\includegraphics[width=.42\textwidth]{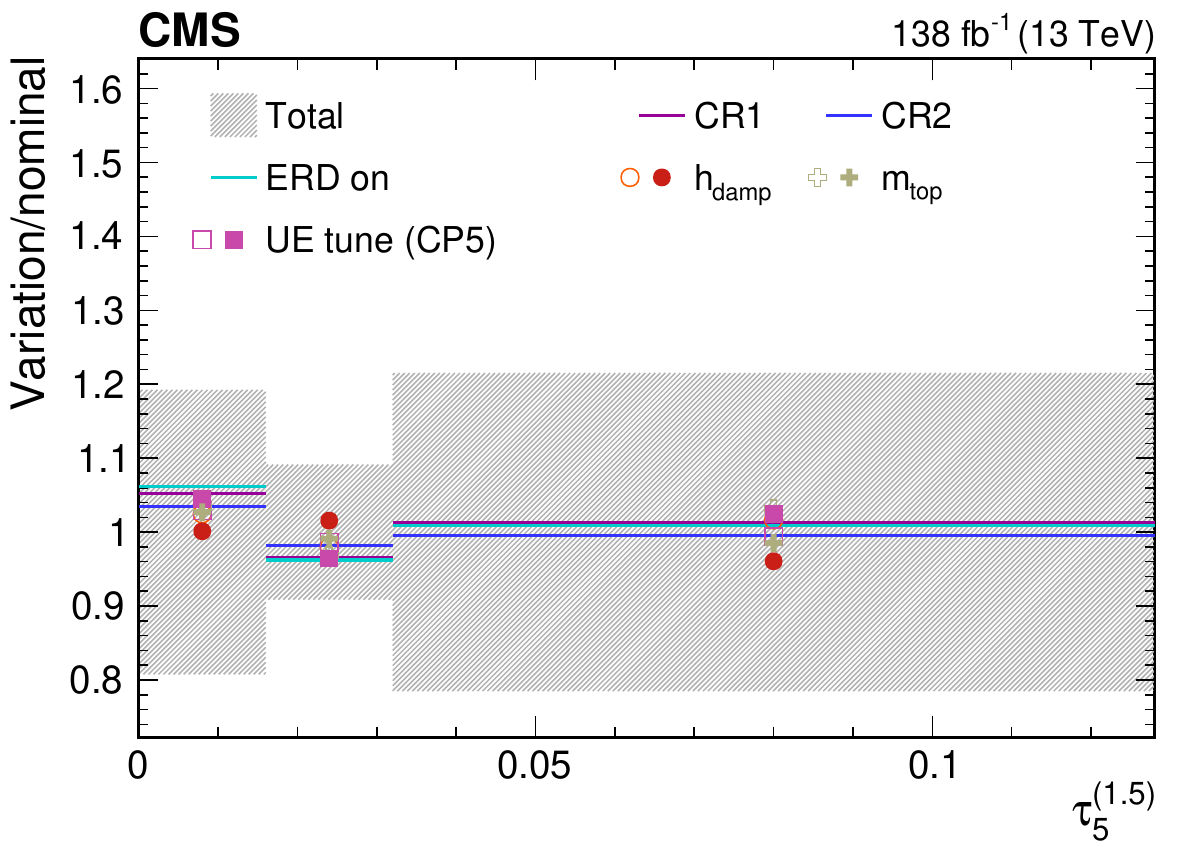}
	\includegraphics[width=.42\textwidth]{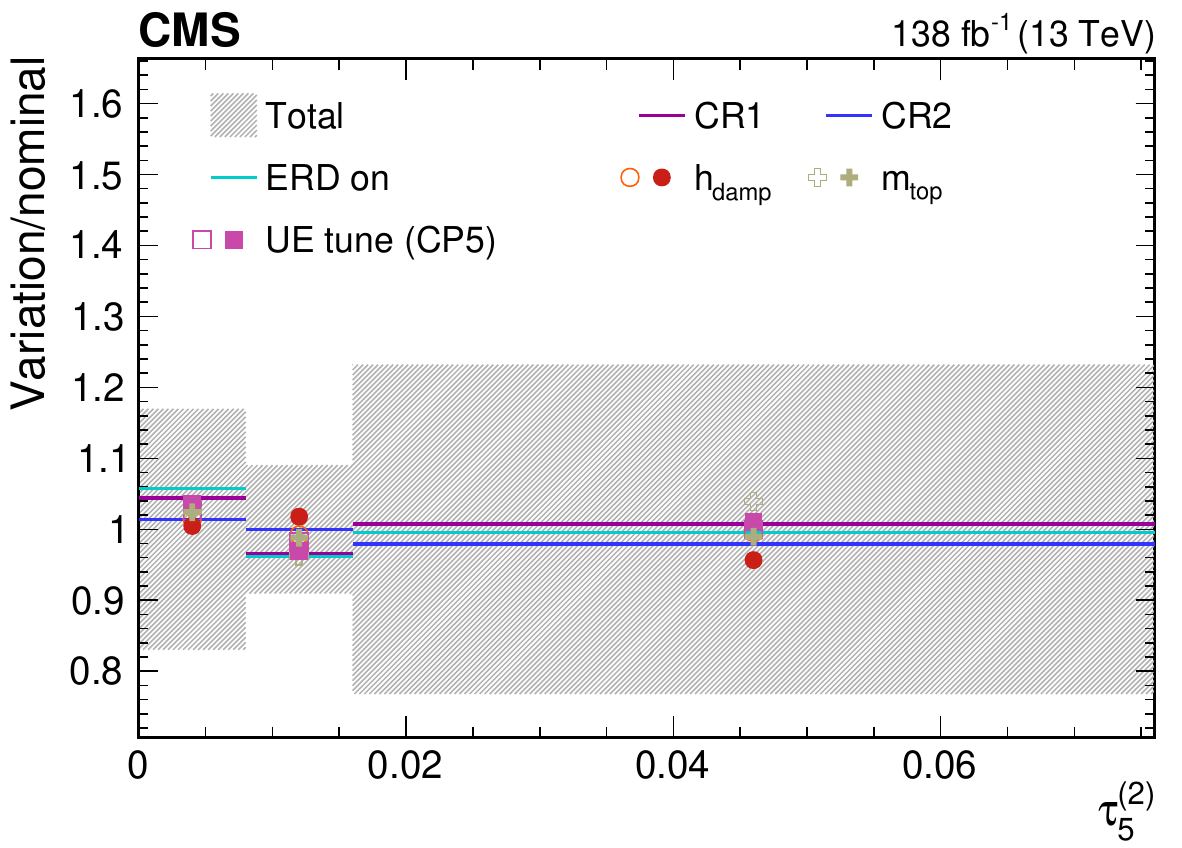}
	\caption{Contributions from various theory model systematic variations to the normalized, unfolded distribution for $\tau_5^{(\beta)}$ observables measured for AK8 jets passing the boosted \PW boson-enriched selection in $\PGm$+jets \ttbar events. 
		The total unfolding uncertainty is indicated with the dark grey, hashed region, while the blue hashed region indicates the contributions from the input covariance matrix, which includes the propagated effects of the statistical uncertainties of the input data after background subtraction. Contributions from statistical uncertainties of the simulated sample used to construct the nominal response matrix are indicated with the dashed black line. The uncertainty contributions for different choices of colour reconnection models are illustrated as one-sided shifts compared to the nominal unfolding, and up (down) contributions from other sources are indicated with filled (open) markers of the same type and colour.}
	\label{fig:unfUncsTheoryW_tau5}
\end{figure}

Estimated contributions of various sources of experimental and modelling uncertainty are presented for the measurement of $1$- through $5$-subjettiness in boosted \PW boson-enriched events.
Uncertainty sources that are common also to the dijet selection, as well as sources of experimental uncertainty considered only for the boosted \PW boson-/top quark-enriched regions, are presented in one set of figures: Figs.~\ref{fig:unfUncsW_tau1}--\ref{fig:unfUncsW_tau5}, while those arising from model variations considered only in the boosted \PW boson-/top quark-enriched regions are presented in a separate set of figures: Figs.~\ref{fig:unfUncsTheoryW_tau1}--\ref{fig:unfUncsTheoryW_tau5}. 

\clearpage
\newpage

\subsection{Unfolding uncertainties: boosted top quark jets}
\label{sec:topUnfUncs}
\begin{figure}[htpb]
	\centering
	\includegraphics[width=.42\textwidth]{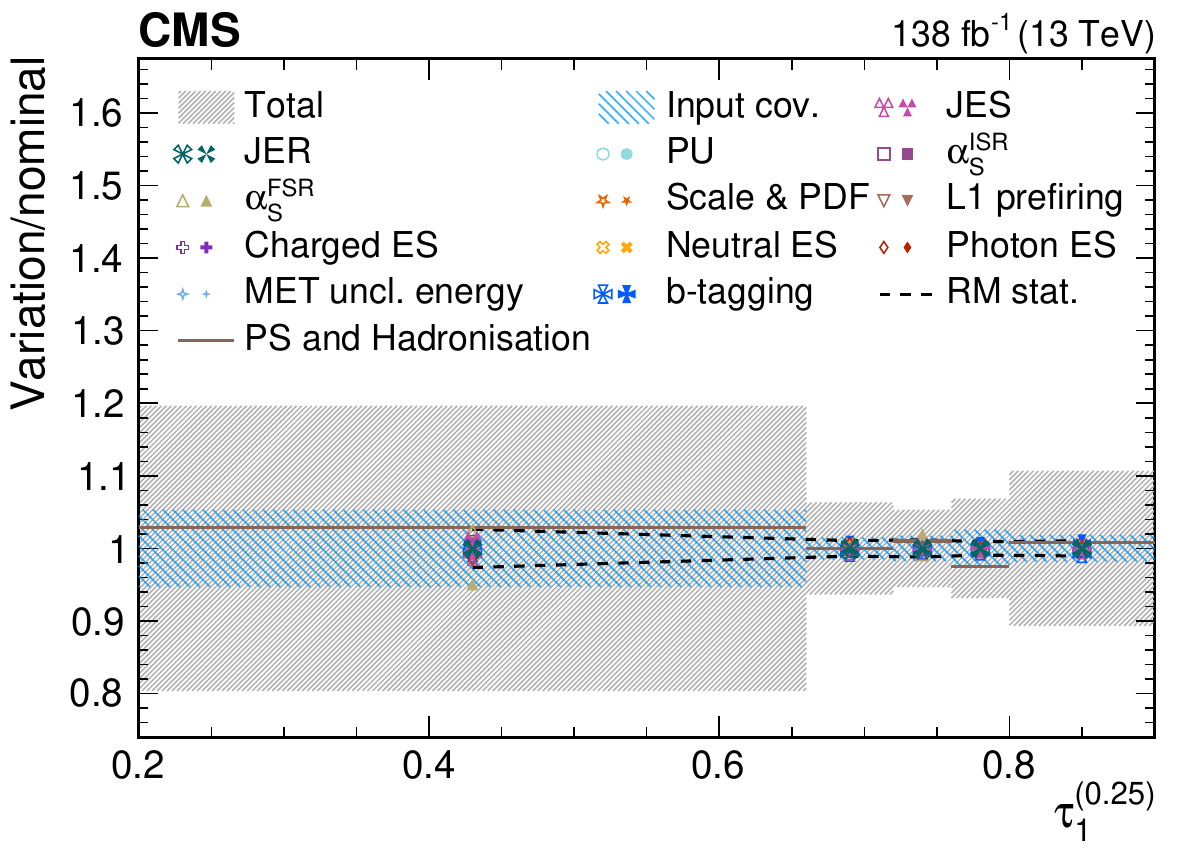}
	\includegraphics[width=.42\textwidth]{Figure_018-a.pdf}
	\includegraphics[width=.42\textwidth]{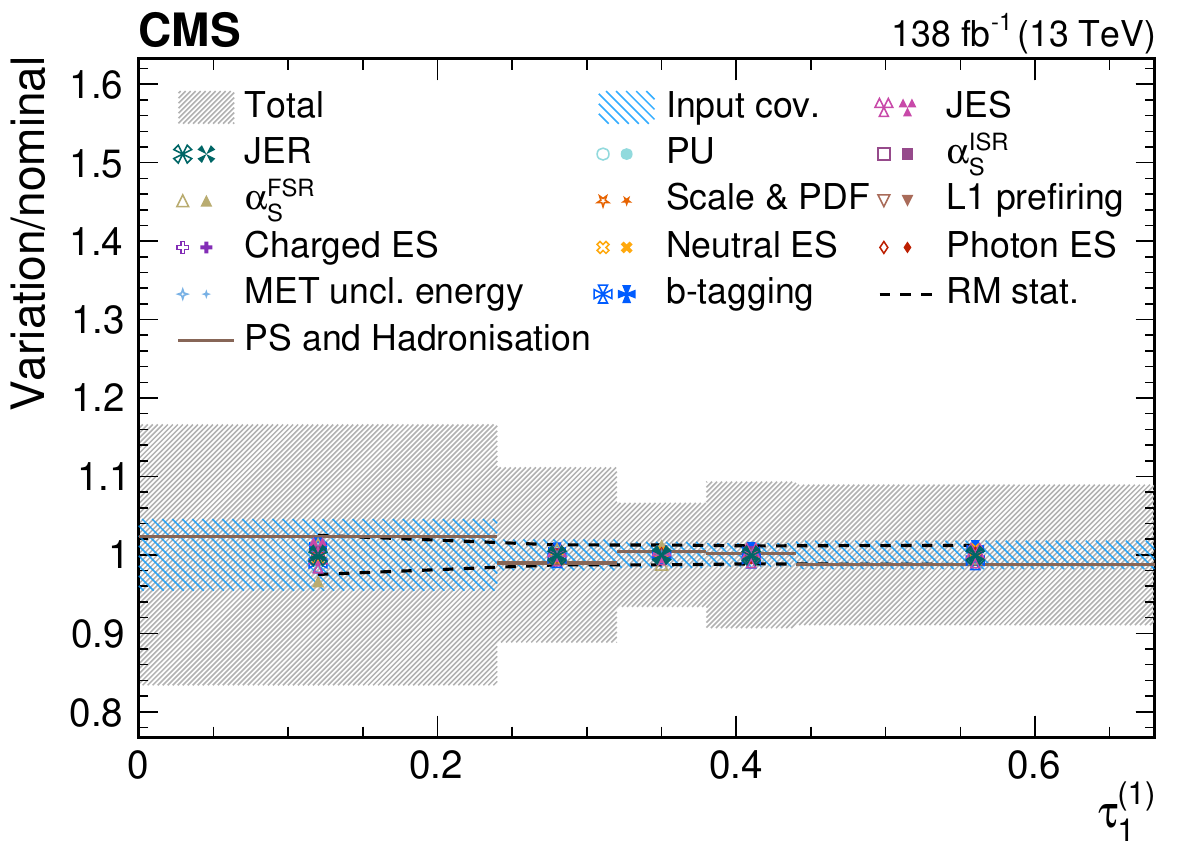}
	\includegraphics[width=.42\textwidth]{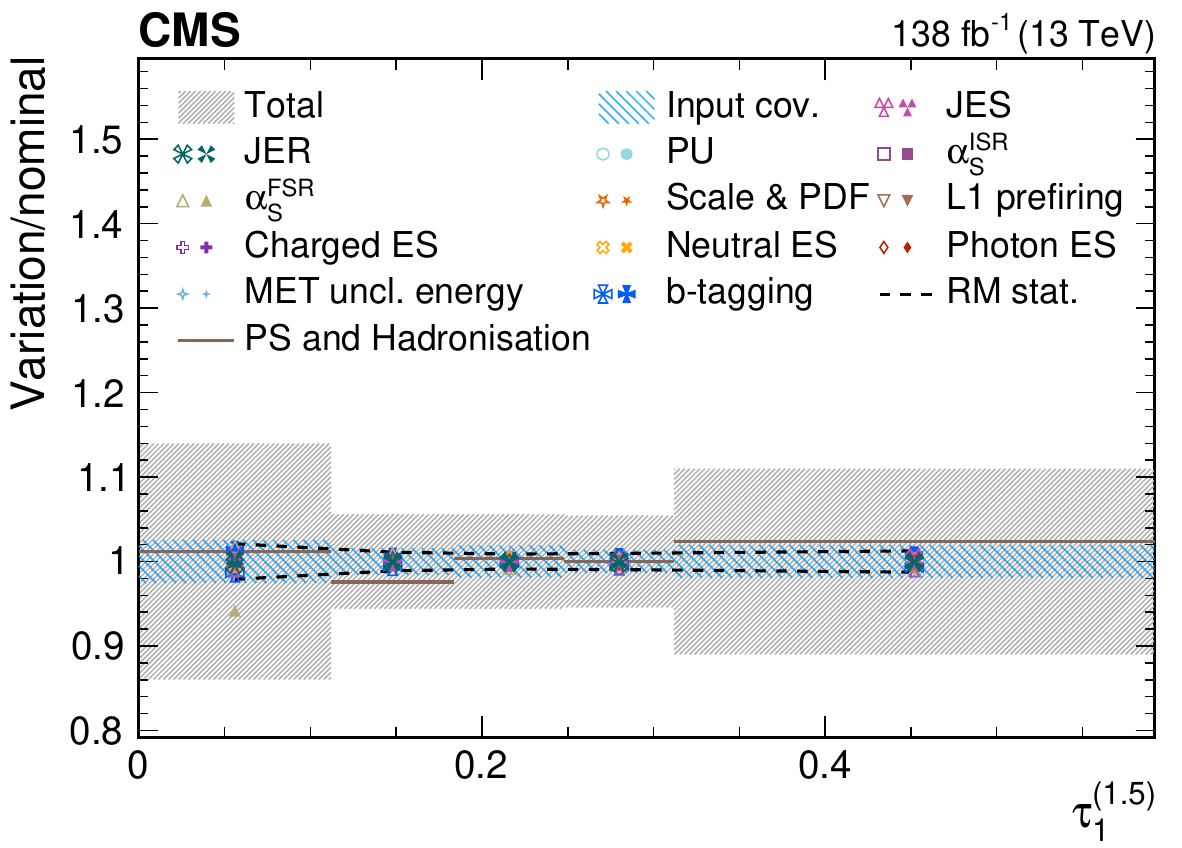}
	\includegraphics[width=.42\textwidth]{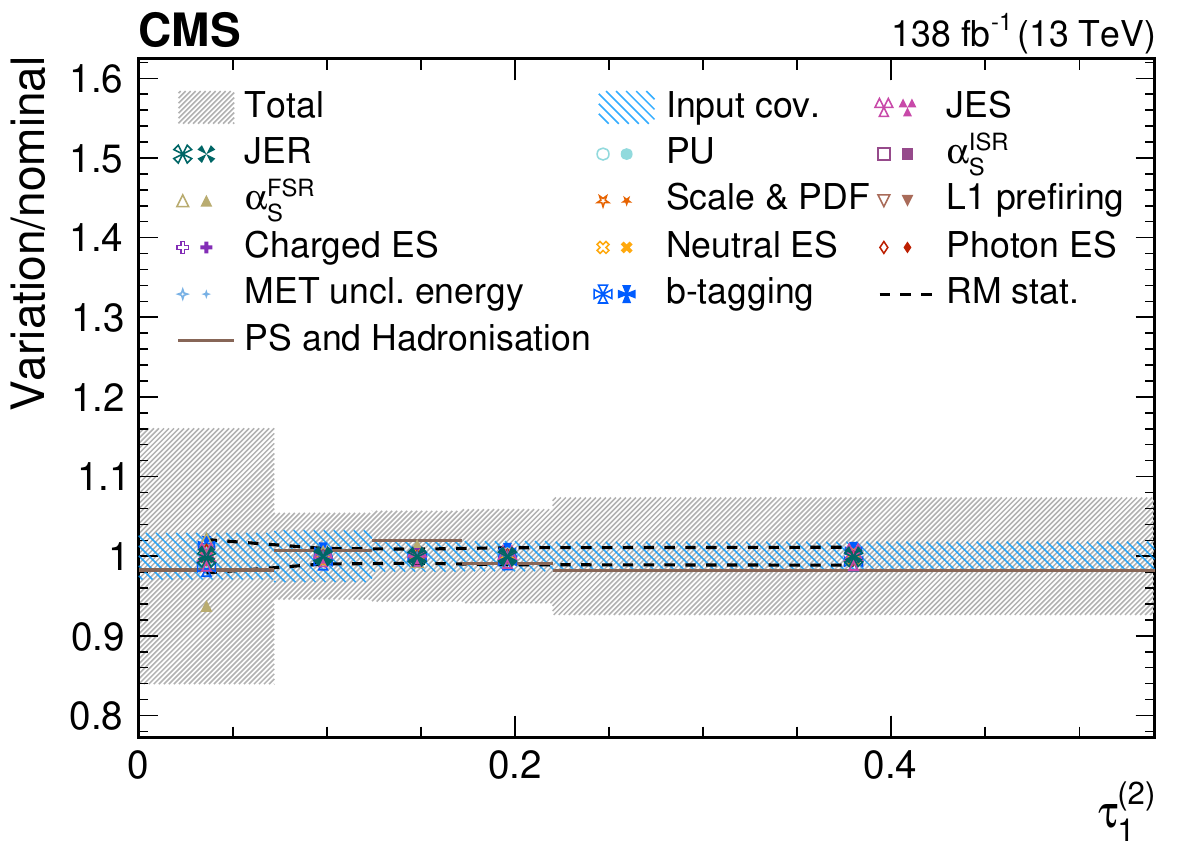}
	\caption{Contributions from various systematic variations to the normalized, unfolded distribution for $\tau_1^{(\beta)}$ observables measured for AK8 jets passing the boosted top quark-enriched selection in $\PGm$+jets \ttbar events. 
		The total unfolding uncertainty is indicated with the dark grey, hashed region, while the blue hashed region indicates the contributions from the input covariance matrix, which includes the propagated effects of the statistical uncertainties of the input data after background subtraction. Contributions from statistical uncertainties of the simulated sample used to construct the nominal response matrix are indicated with the dashed black line. The physics model uncertainty is computed as a one-sided shift compared to the nominal unfolding, and up (down) contributions from other sources are indicated with filled (open) markers of the same type and colour.}
	\label{fig:unfUncstop_tau1}
\end{figure}

\begin{figure}[htpb]
	\centering
	\includegraphics[width=.42\textwidth]{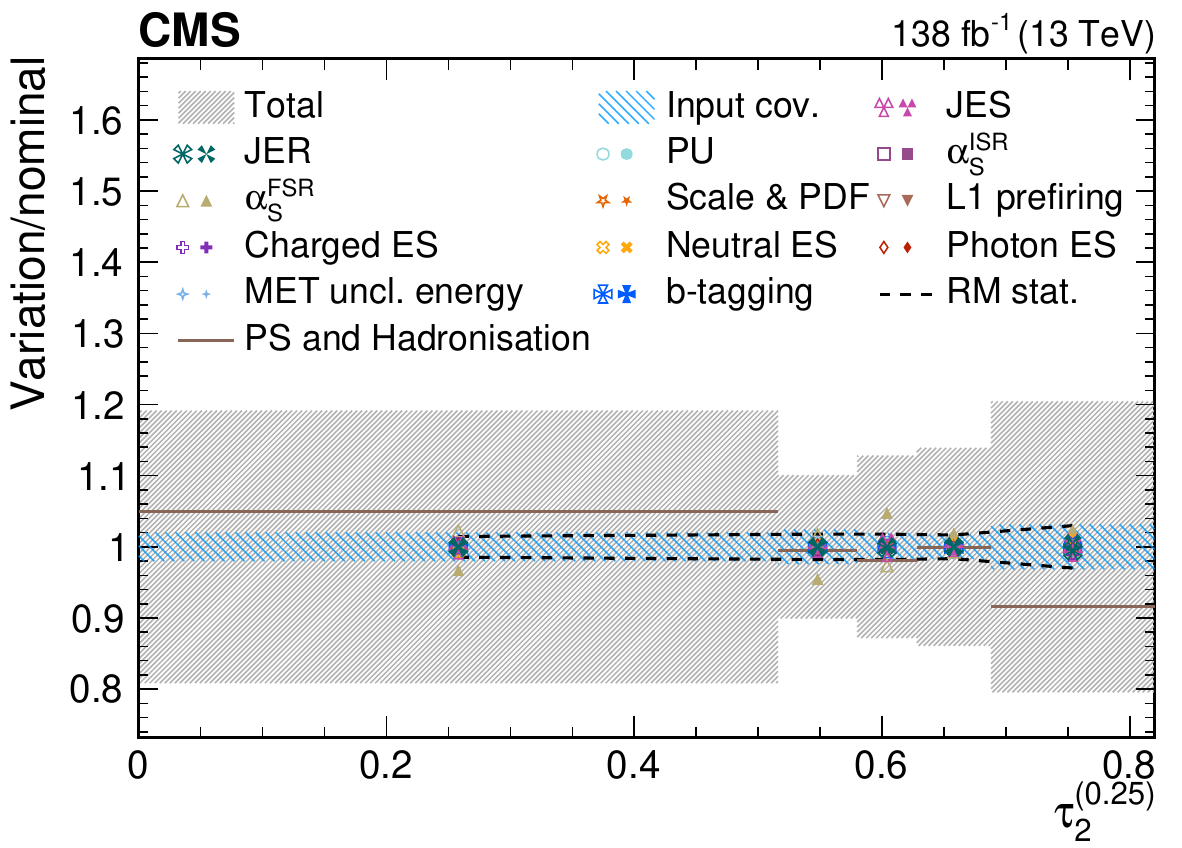}
	\includegraphics[width=.42\textwidth]{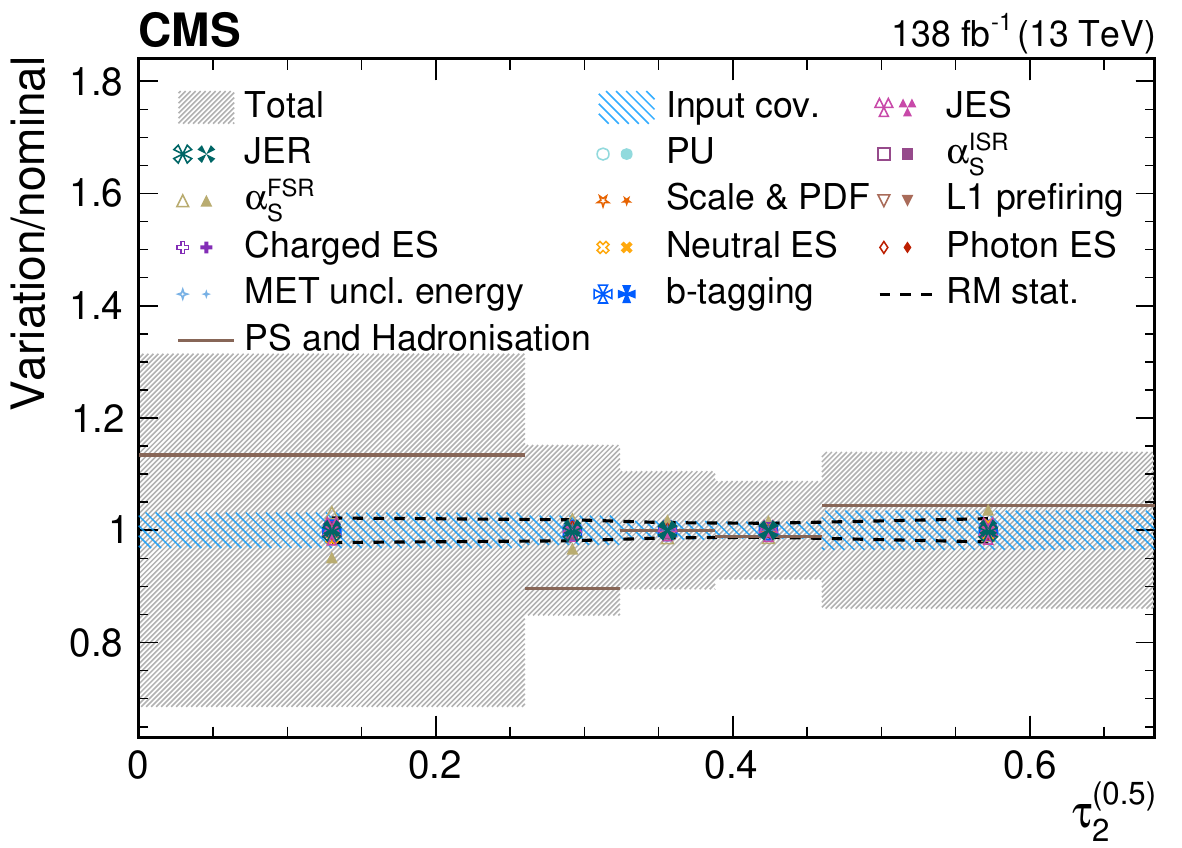}
	\includegraphics[width=.42\textwidth]{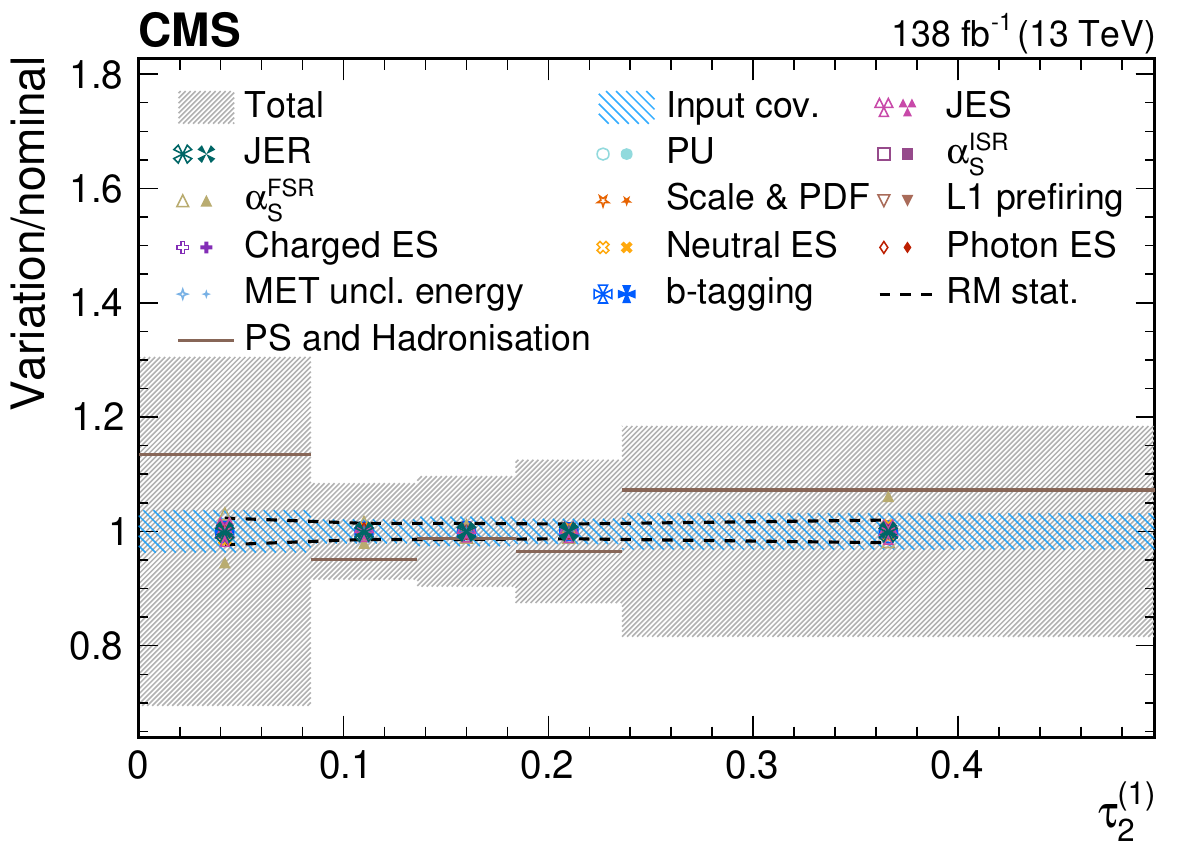}
	\includegraphics[width=.42\textwidth]{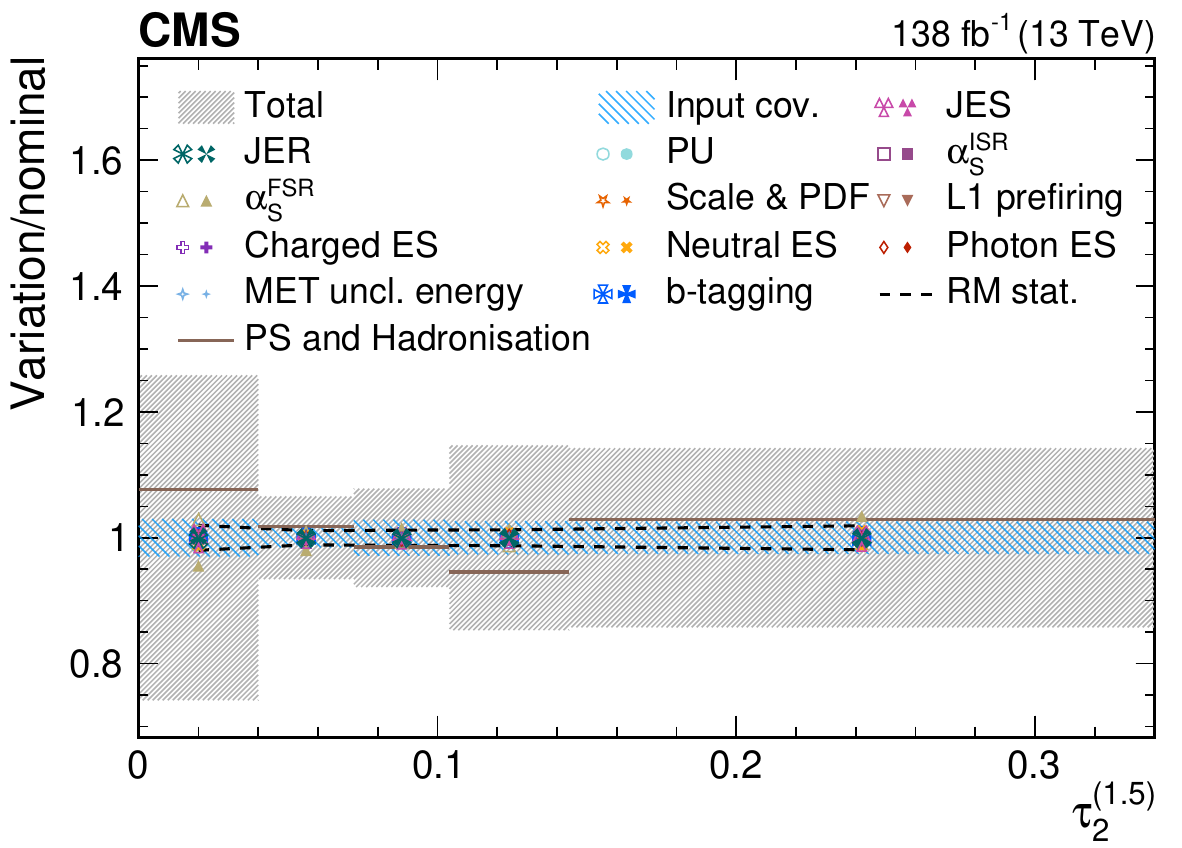}
	\includegraphics[width=.42\textwidth]{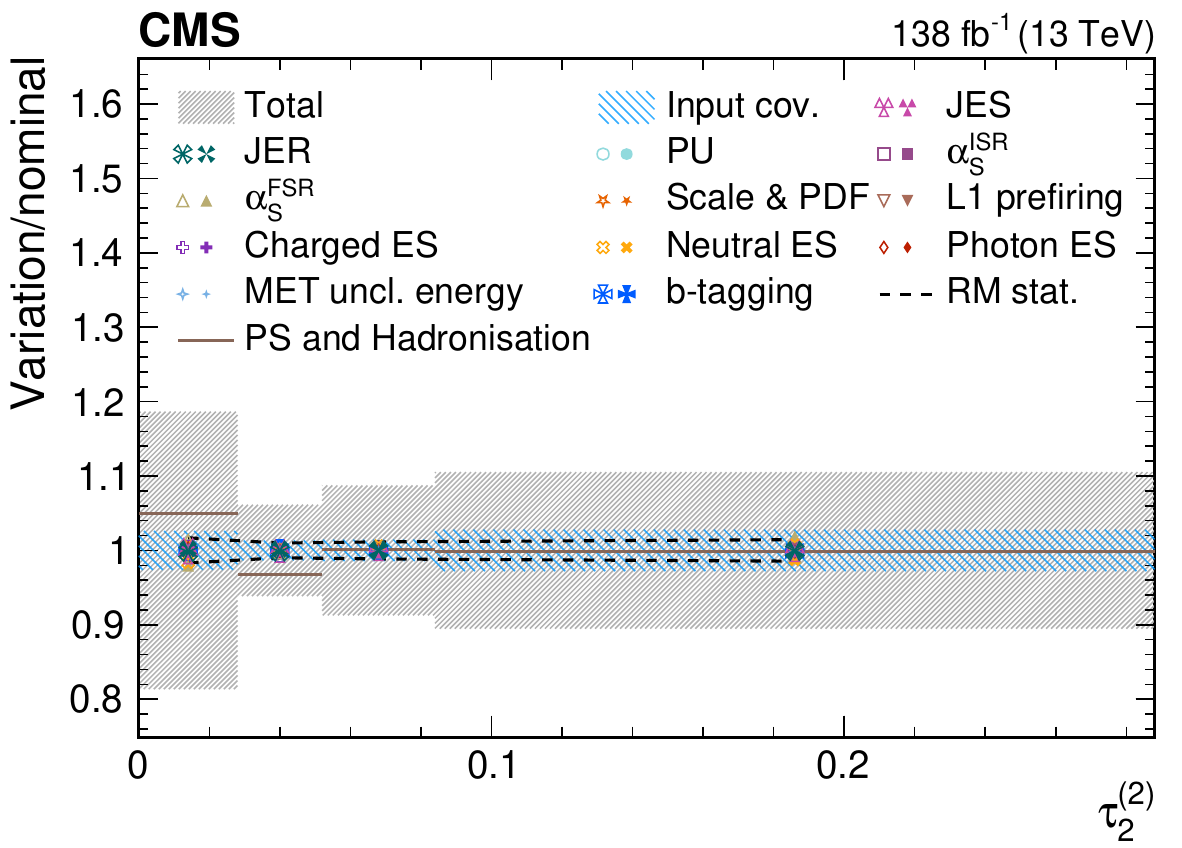}
	\caption{Contributions from various systematic variations to the normalized, unfolded distribution for $\tau_2^{(\beta)}$ observables measured for AK8 jets passing the boosted top quark-enriched selection in $\PGm$+jets \ttbar events. 
		The total unfolding uncertainty is indicated with the dark grey, hashed region, while the blue hashed region indicates the contributions from the input covariance matrix, which includes the propagated effects of the statistical uncertainties of the input data after background subtraction. Contributions from statistical uncertainties of the simulated sample used to construct the nominal response matrix are indicated with the dashed black line. The physics model uncertainty is computed as a one-sided shift compared to the nominal unfolding, and up (down) contributions from other sources are indicated with filled (open) markers of the same type and colour.}
	\label{fig:unfUncstop_tau2}
\end{figure}

\begin{figure}[htpb]
	\centering
	\includegraphics[width=.42\textwidth]{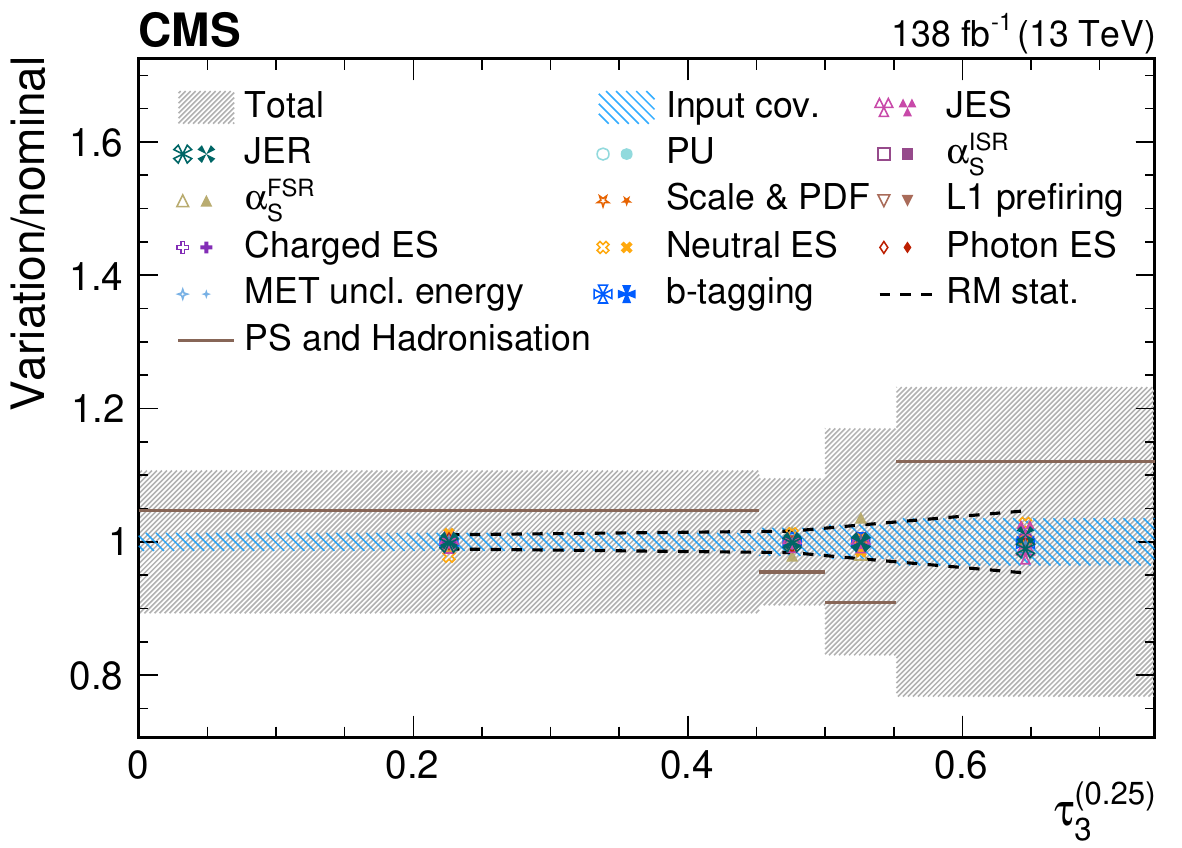}
	\includegraphics[width=.42\textwidth]{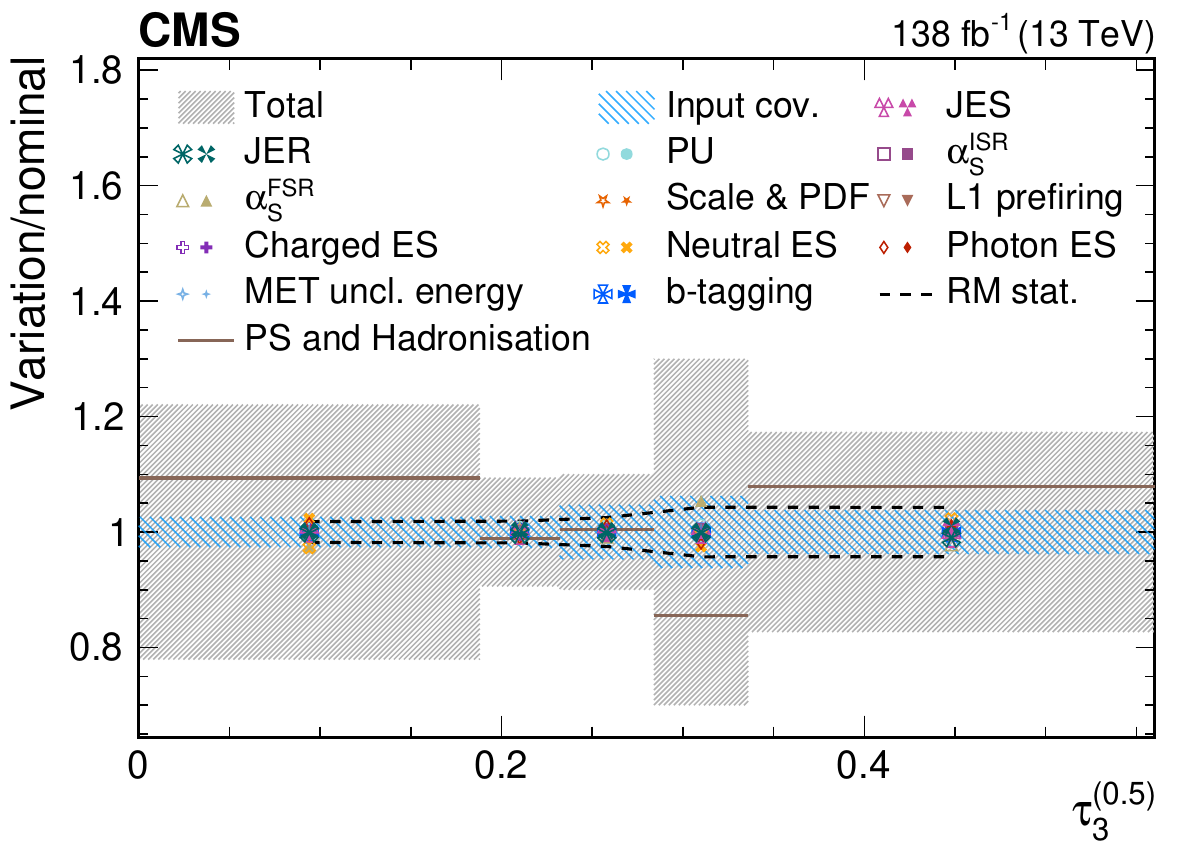}
	\includegraphics[width=.42\textwidth]{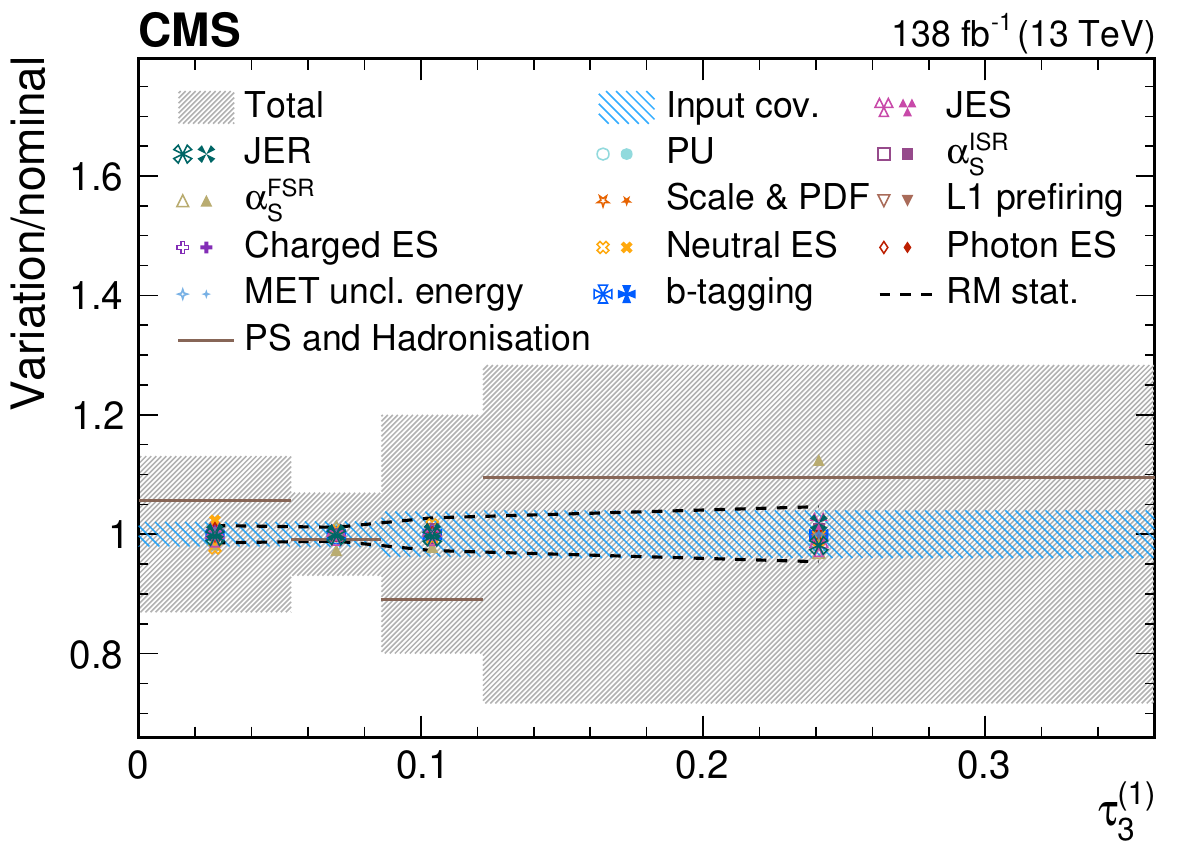}
	\includegraphics[width=.42\textwidth]{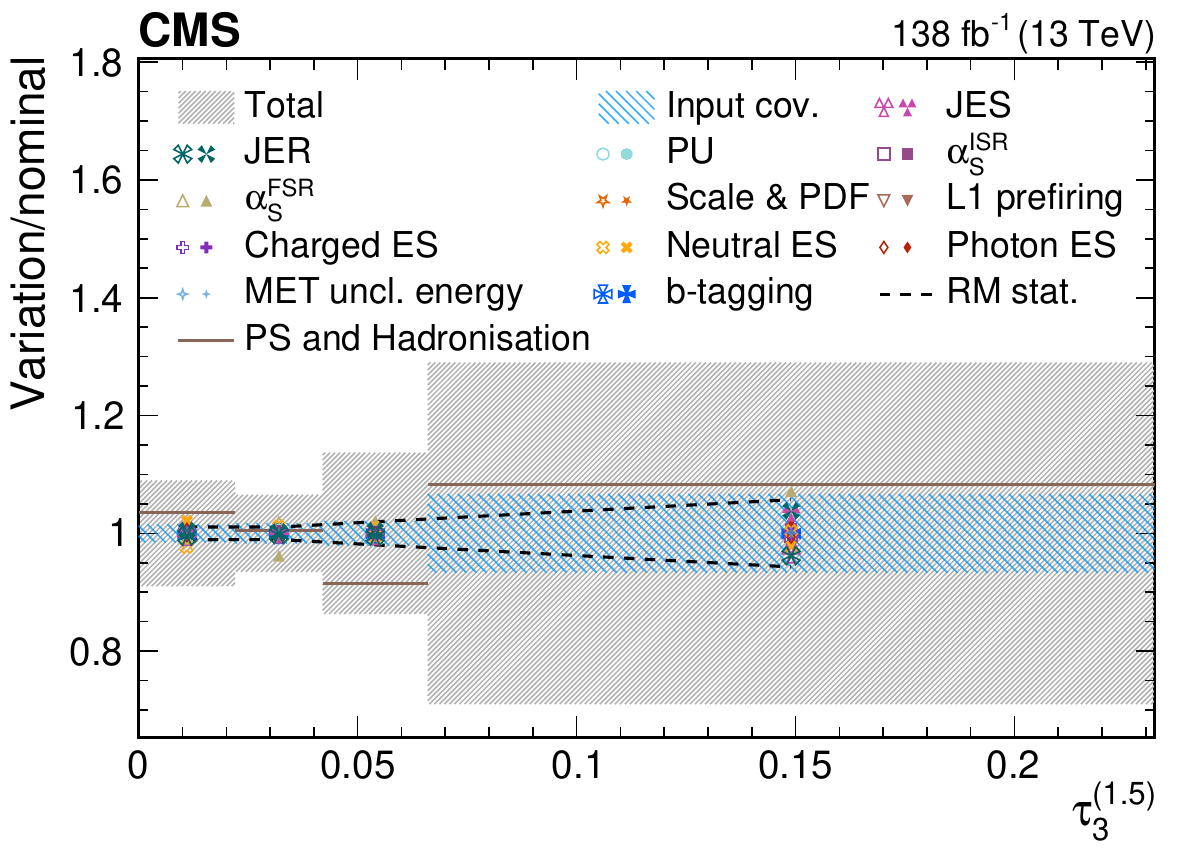}
	\includegraphics[width=.42\textwidth]{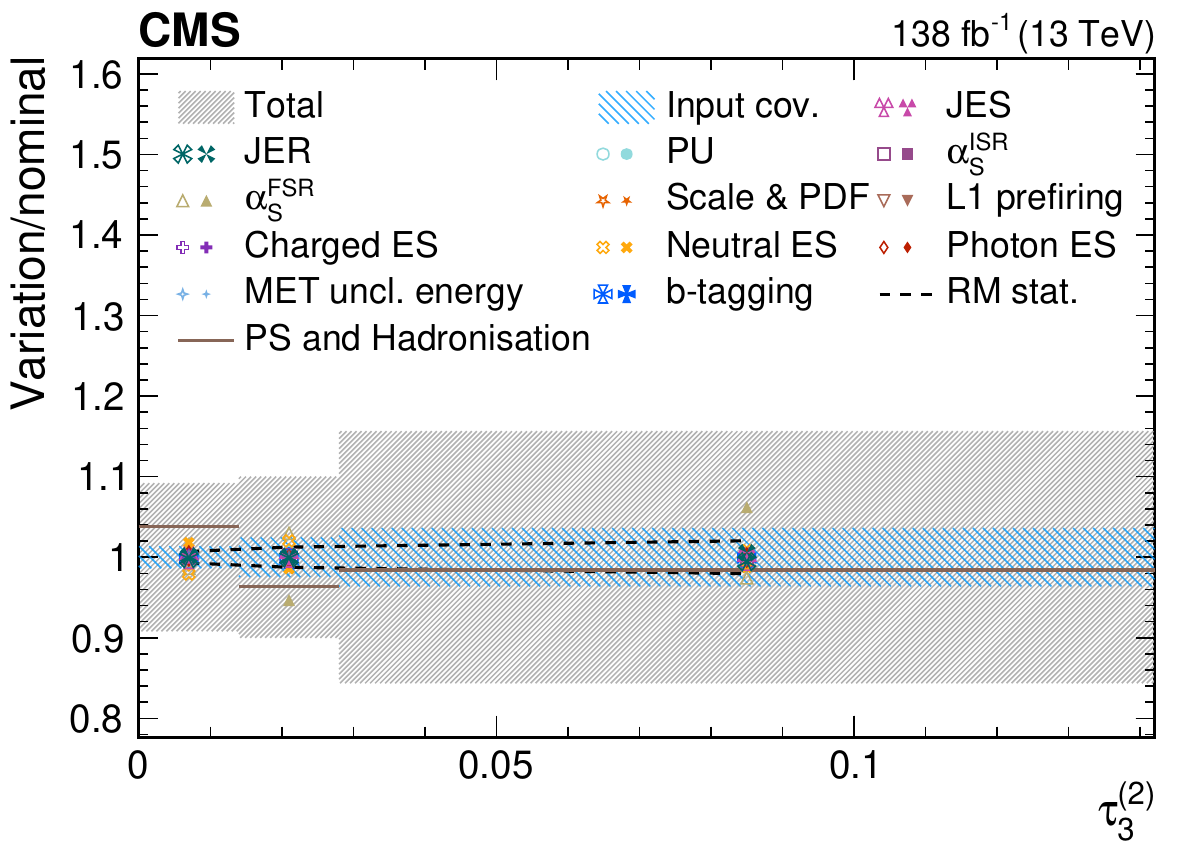}
	\caption{Contributions from various systematic variations to the normalized, unfolded distribution for $\tau_3^{(\beta)}$ observables measured for AK8 jets passing the boosted top quark-enriched selection in $\PGm$+jets \ttbar events. 
		The total unfolding uncertainty is indicated with the dark grey, hashed region, while the blue hashed region indicates the contributions from the input covariance matrix, which includes the propagated effects of the statistical uncertainties of the input data after background subtraction. Contributions from statistical uncertainties of the simulated sample used to construct the nominal response matrix are indicated with the dashed black line. The physics model uncertainty is computed as a one-sided shift compared to the nominal unfolding, and up (down) contributions from other sources are indicated with filled (open) markers of the same type and colour.}
	\label{fig:unfUncstop_tau3}
\end{figure}

\begin{figure}[htpb]
	\centering
	\includegraphics[width=.42\textwidth]{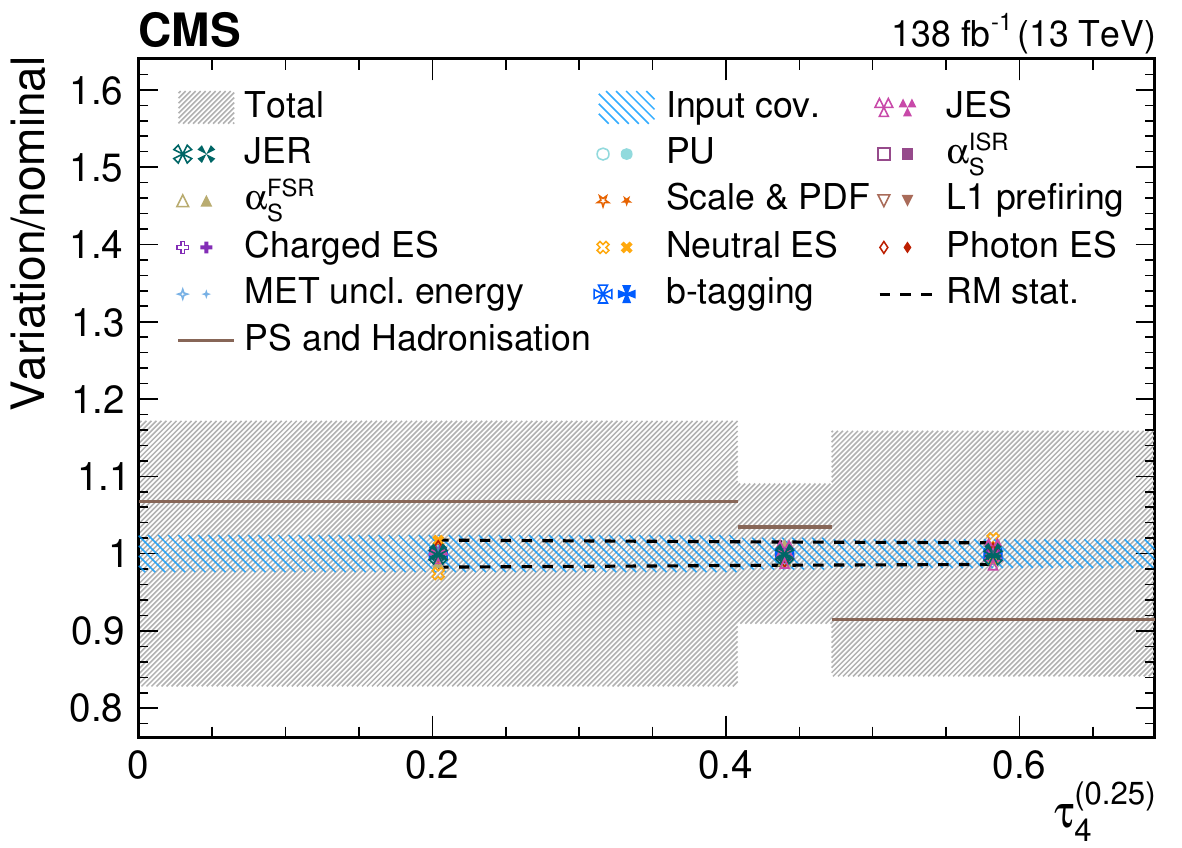}
	\includegraphics[width=.42\textwidth]{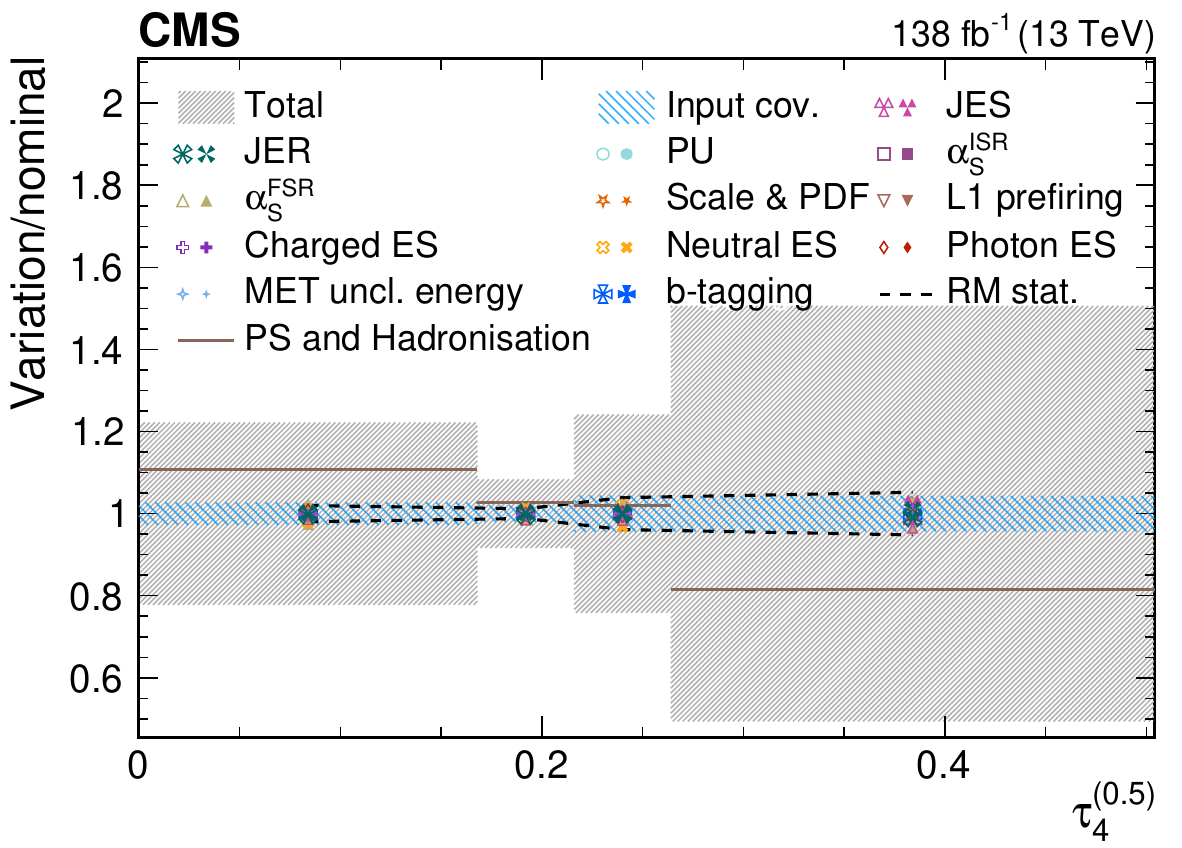}
	\includegraphics[width=.42\textwidth]{Figure_018-c.pdf}
	\includegraphics[width=.42\textwidth]{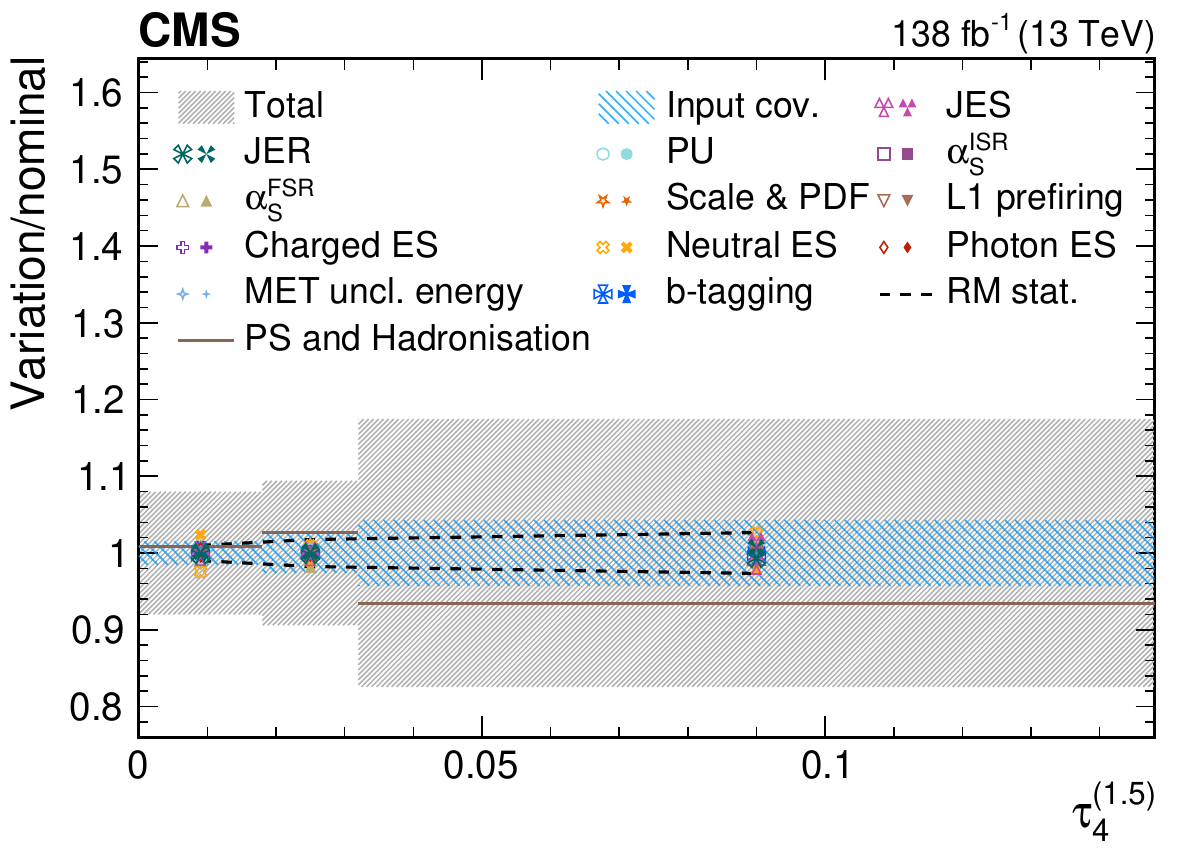}
	\includegraphics[width=.42\textwidth]{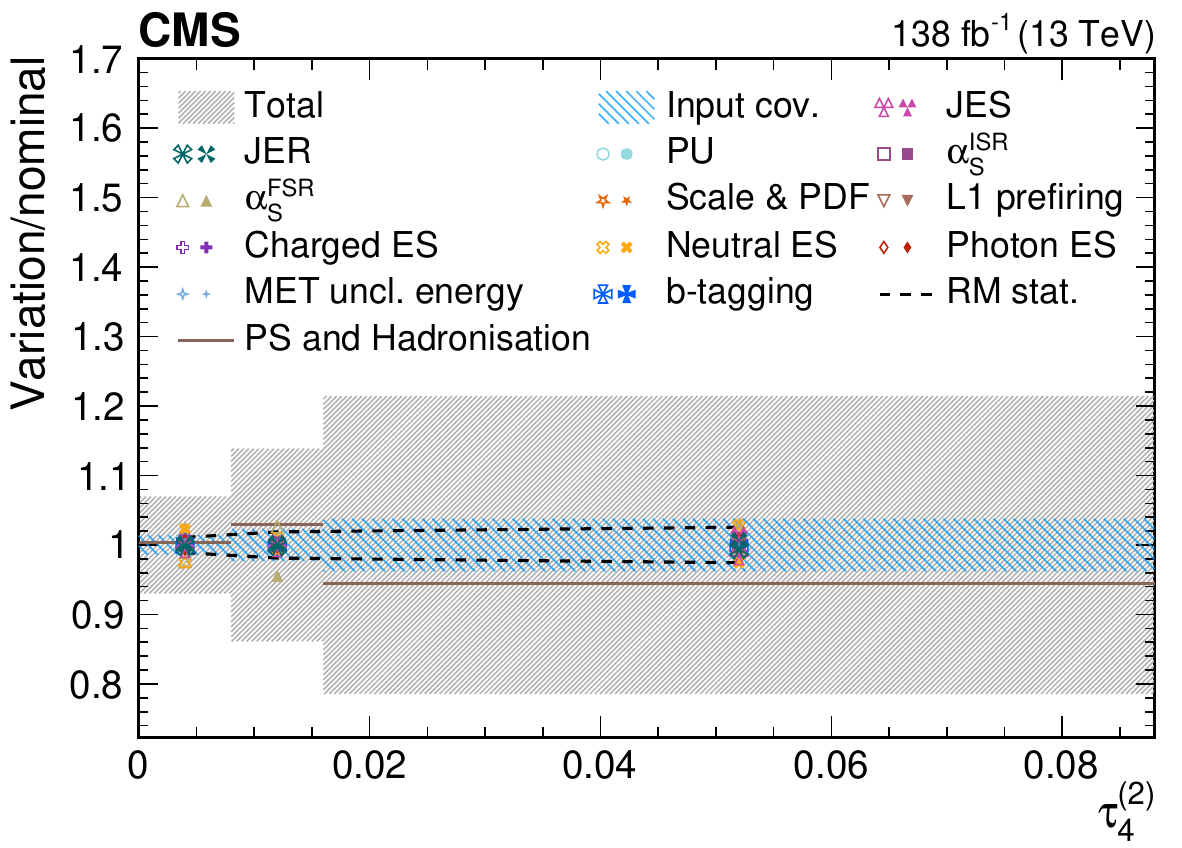}
	\caption{Contributions from various systematic variations to the normalized, unfolded distribution for $\tau_4^{(\beta)}$ observables measured for AK8 jets passing the boosted top quark-enriched selection in $\PGm$+jets \ttbar events. 
		The total unfolding uncertainty is indicated with the dark grey, hashed region, while the blue hashed region indicates the contributions from the input covariance matrix, which includes the propagated effects of the statistical uncertainties of the input data after background subtraction. Contributions from statistical uncertainties of the simulated sample used to construct the nominal response matrix are indicated with the dashed black line. The physics model uncertainty is computed as a one-sided shift compared to the nominal unfolding, and up (down) contributions from other sources are indicated with filled (open) markers of the same type and colour.}
	\label{fig:unfUncstop_tau4}
\end{figure}

\begin{figure}[htpb]
	\centering
	\includegraphics[width=.42\textwidth]{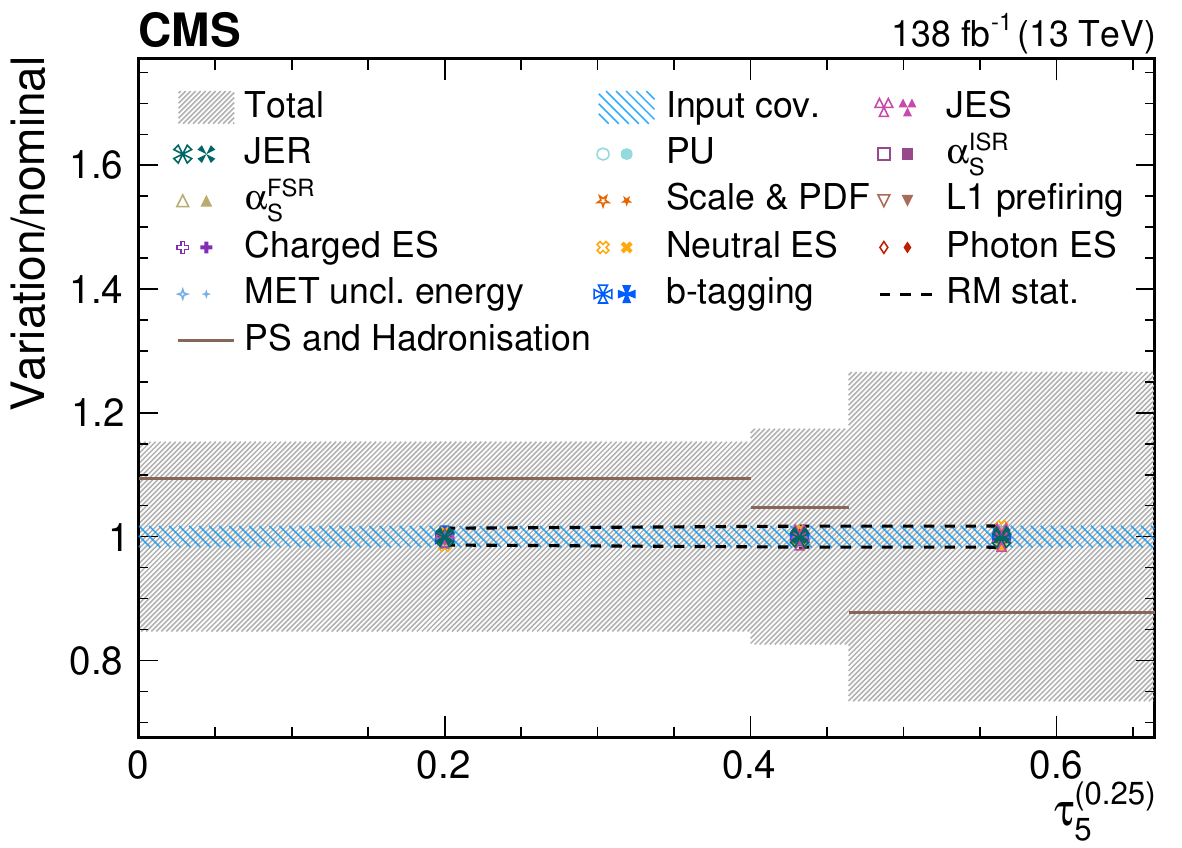}
	\includegraphics[width=.42\textwidth]{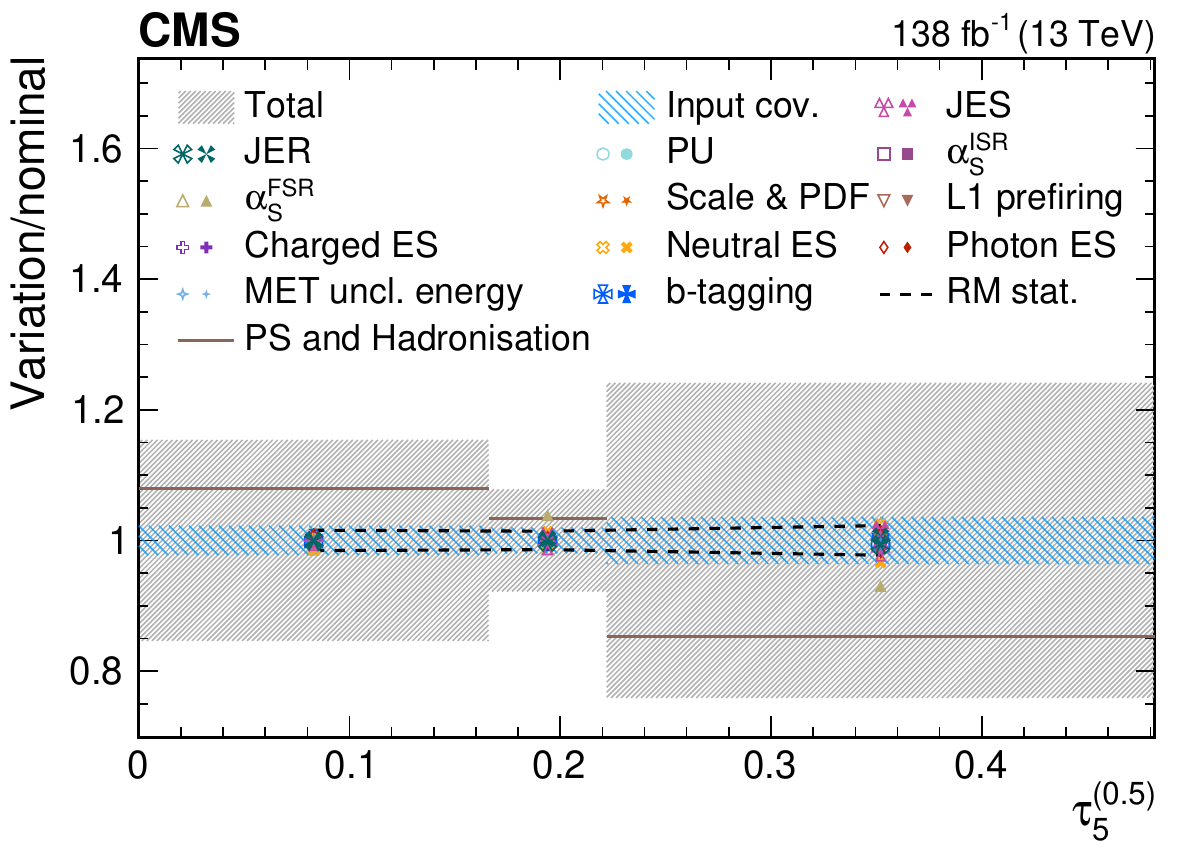}
	\includegraphics[width=.42\textwidth]{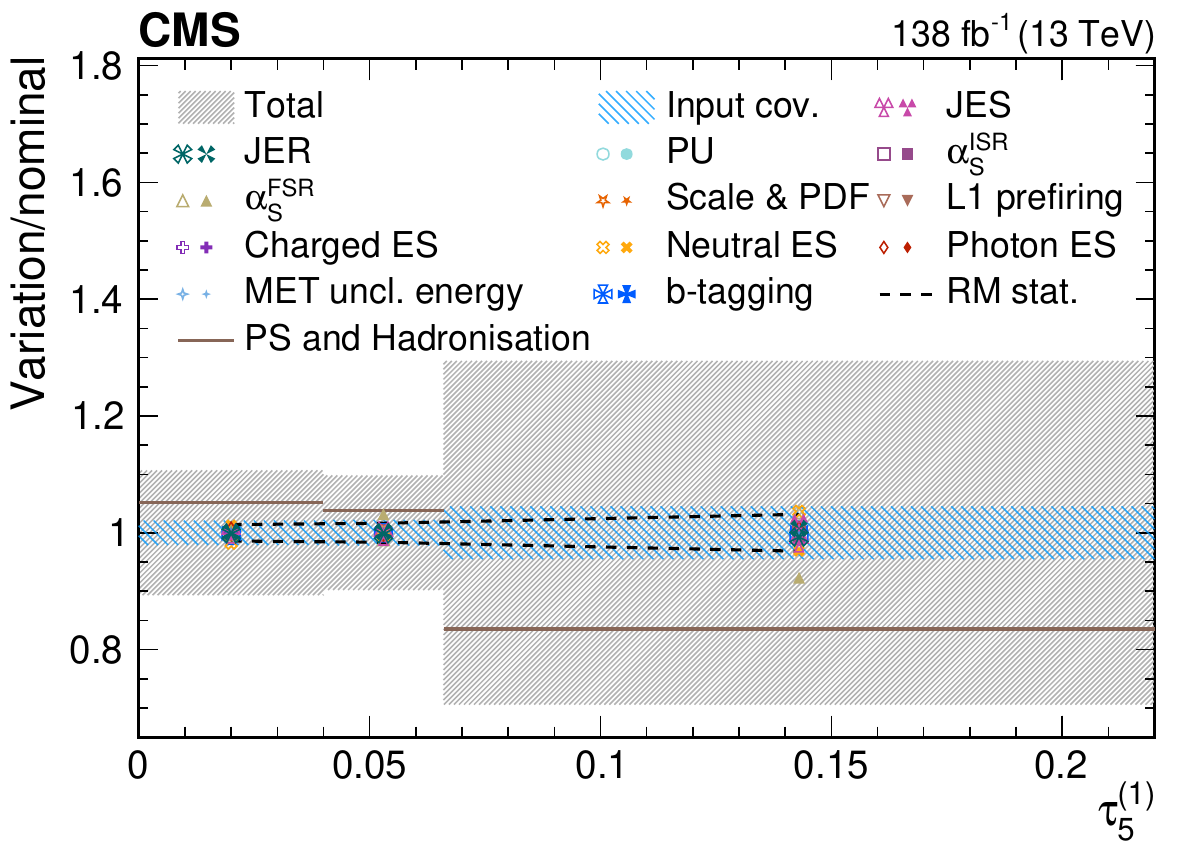}
	\includegraphics[width=.42\textwidth]{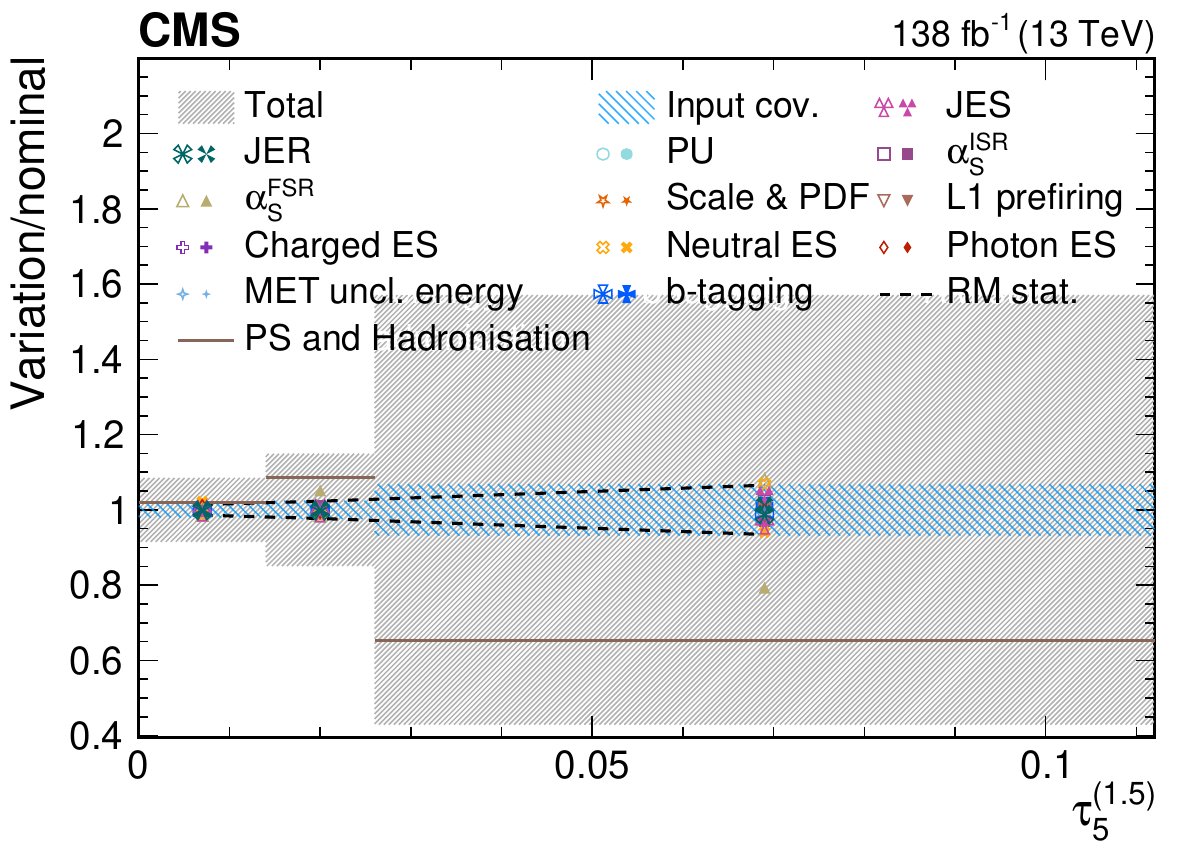}
	\includegraphics[width=.42\textwidth]{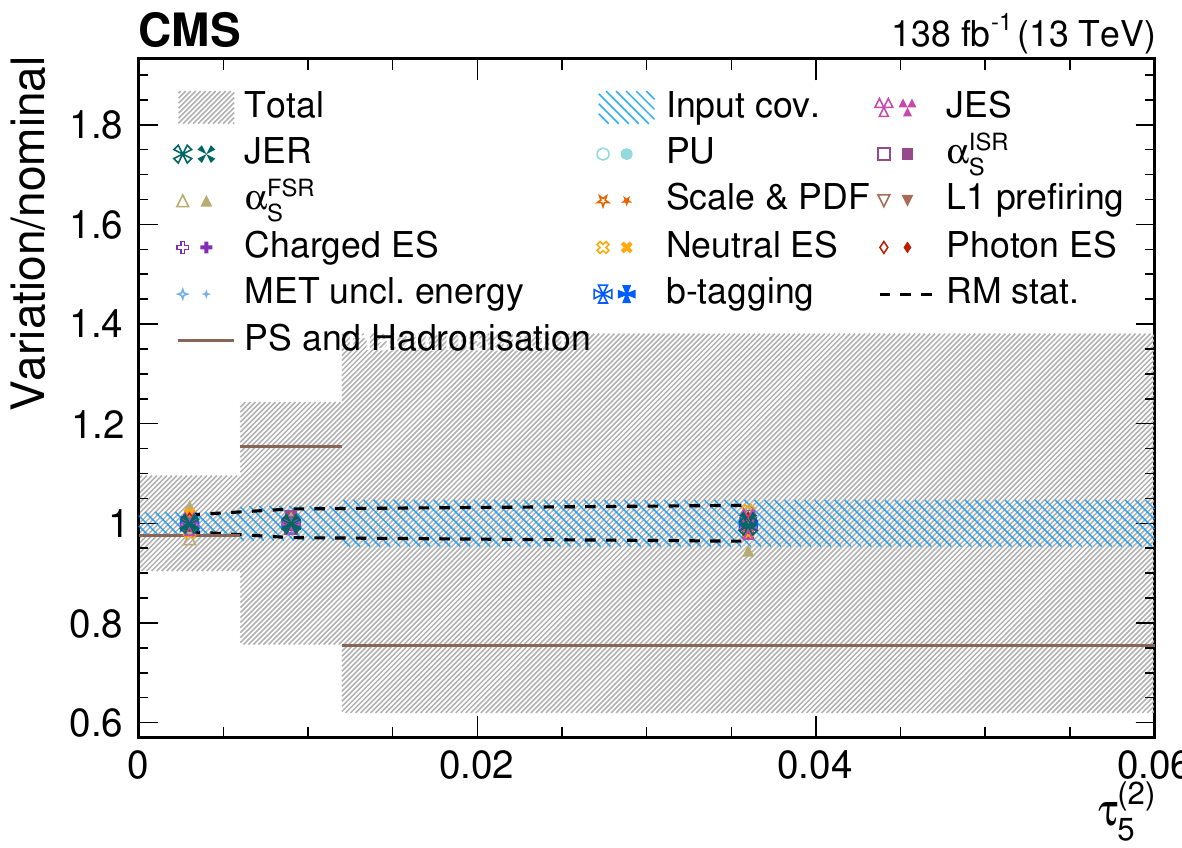}
	\caption{Contributions from various systematic variations to the normalized, unfolded distribution for $\tau_5^{(\beta)}$ observables measured for AK8 jets passing the boosted top quark-enriched selection in $\PGm$+jets \ttbar events. 
		The total unfolding uncertainty is indicated with the dark grey, hashed region, while the blue hashed region indicates the contributions from the input covariance matrix, which includes the propagated effects of the statistical uncertainties of the input data after background subtraction. Contributions from statistical uncertainties of the simulated sample used to construct the nominal response matrix are indicated with the dashed black line. The physics model uncertainty is computed as a one-sided shift compared to the nominal unfolding, and up (down) contributions from other sources are indicated with filled (open) markers of the same type and colour.}
	\label{fig:unfUncstop_tau5}
\end{figure}

\begin{figure}[htpb]
	\centering
	\includegraphics[width=.42\textwidth]{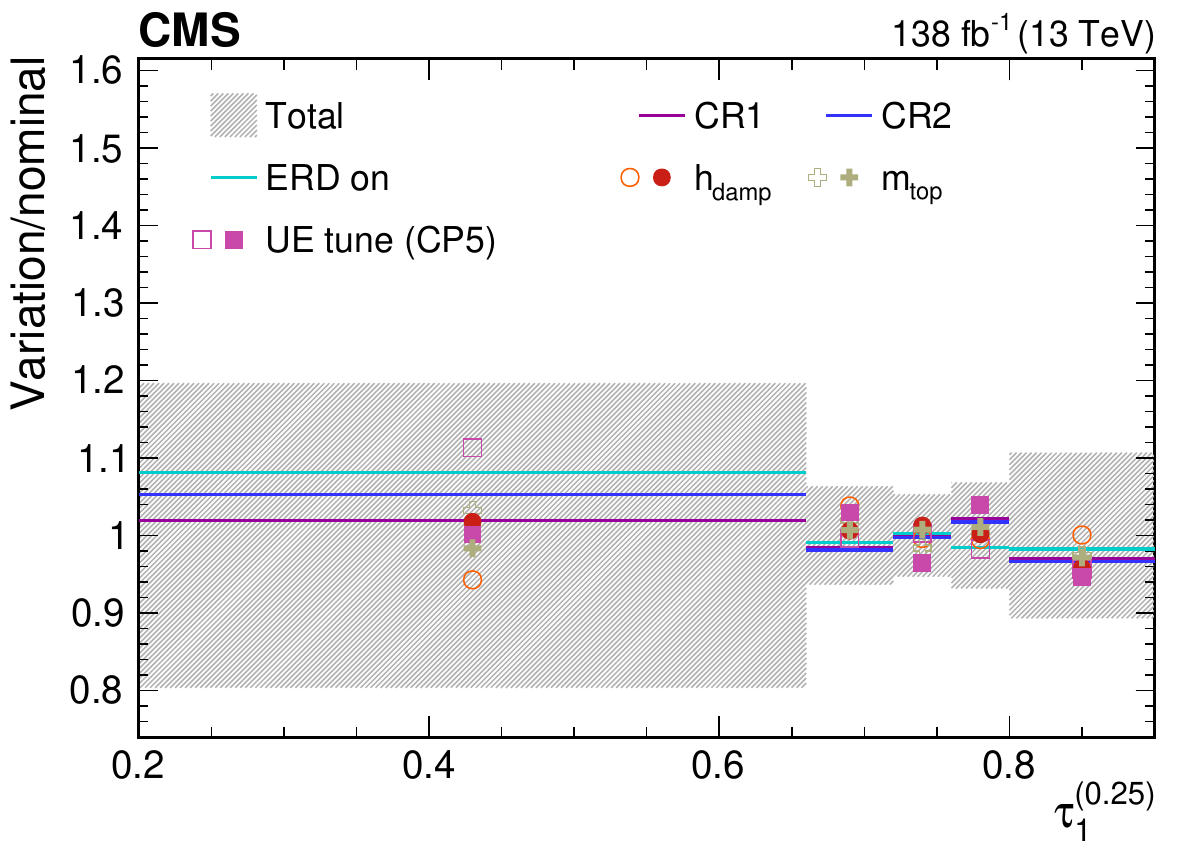}
	\includegraphics[width=.42\textwidth]{Figure_018-b.pdf}
	\includegraphics[width=.42\textwidth]{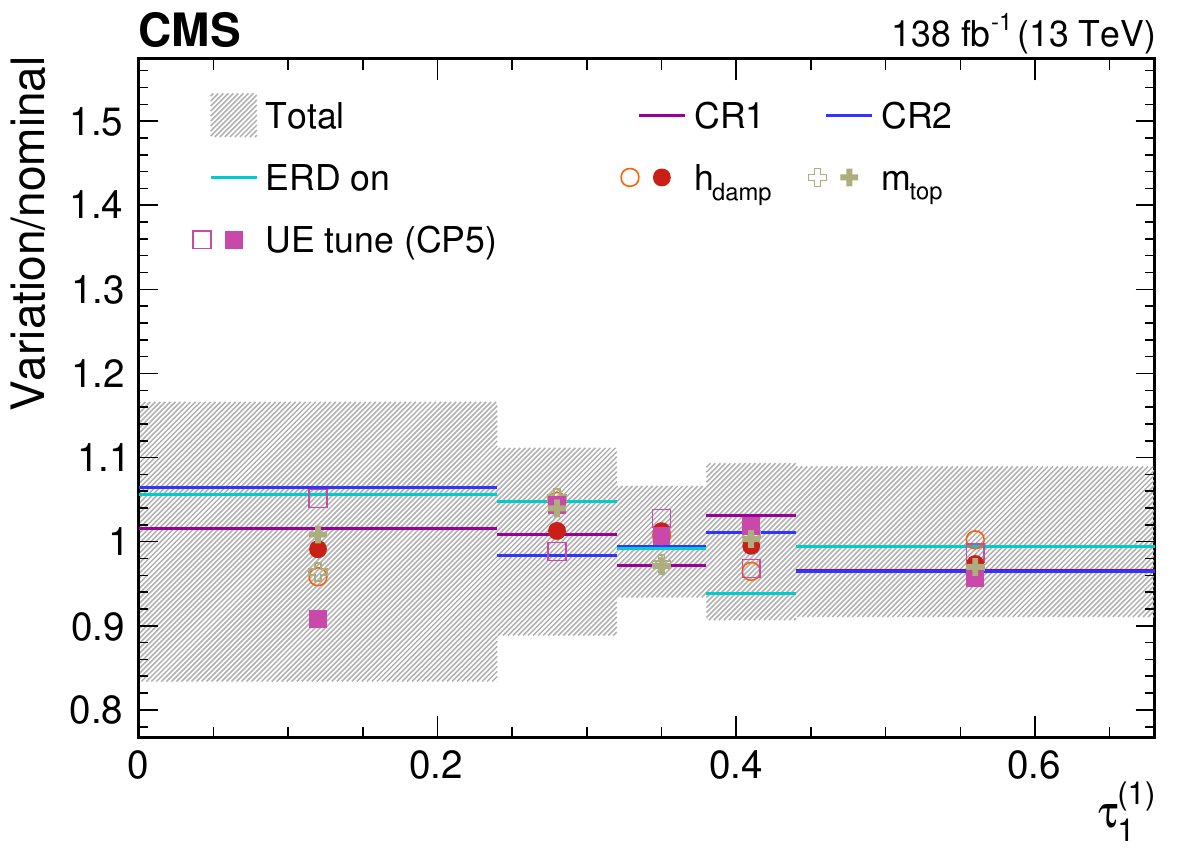}
	\includegraphics[width=.42\textwidth]{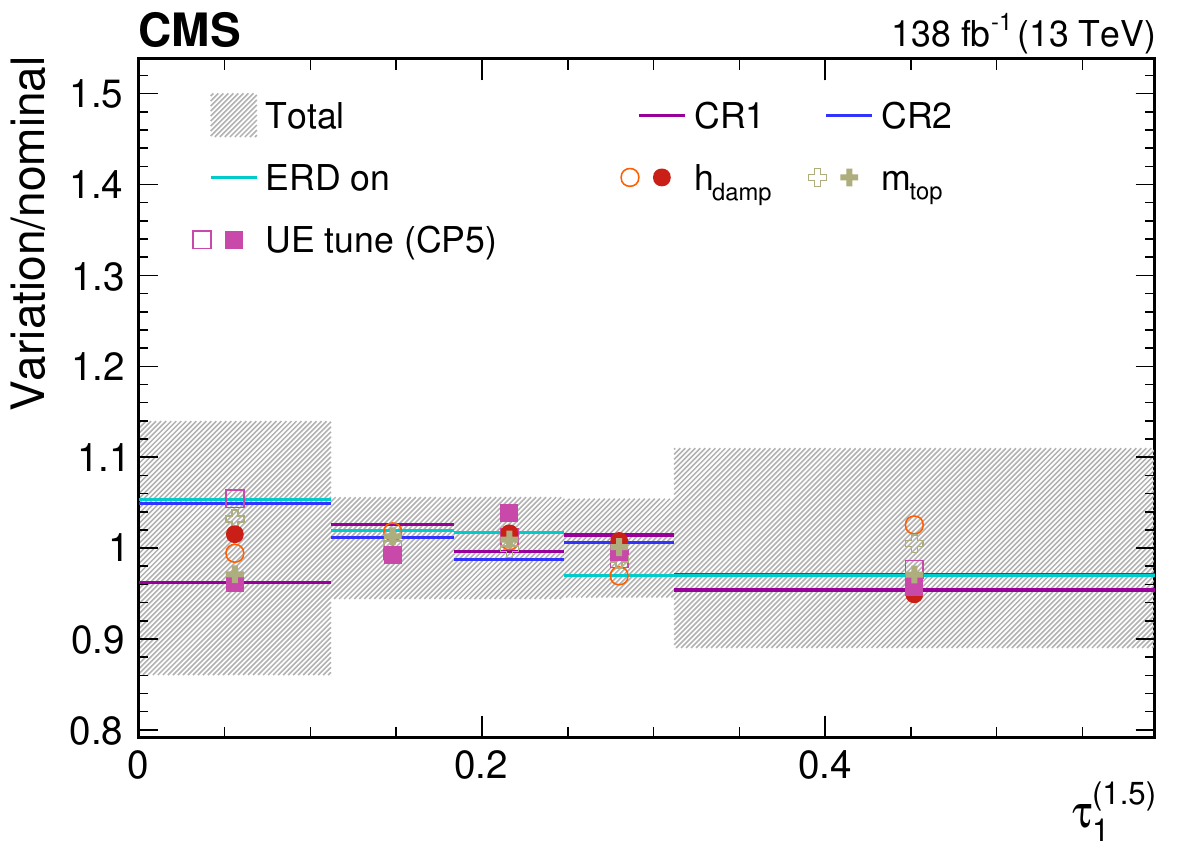}
	\includegraphics[width=.42\textwidth]{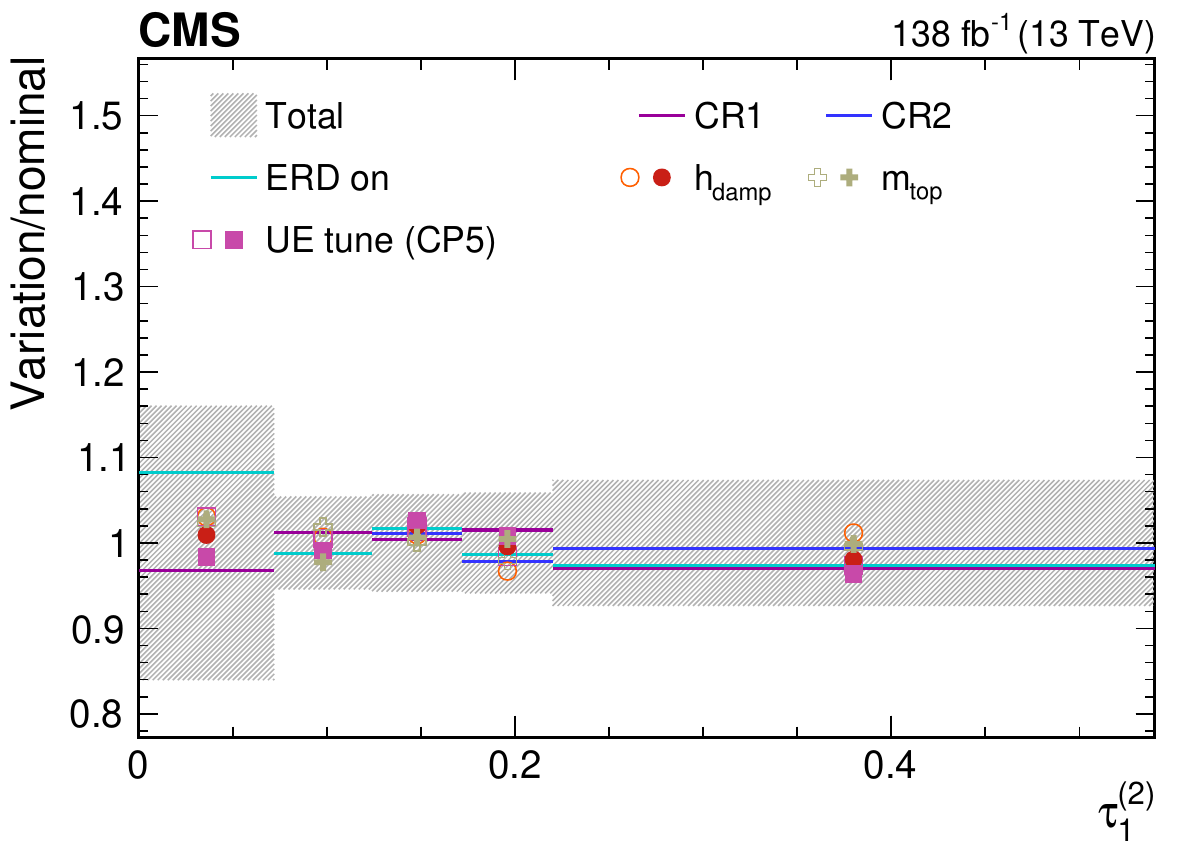}
	\caption{Contributions from various theory model systematic variations to the normalized, unfolded distribution for $\tau_1^{(\beta)}$ observables measured for AK8 jets passing the boosted top quark-enriched selection in $\PGm$+jets \ttbar events. 
		The total unfolding uncertainty is indicated with the dark grey, hashed region, while the blue hashed region indicates the contributions from the input covariance matrix, which includes the propagated effects of the statistical uncertainties of the input data after background subtraction. Contributions from statistical uncertainties of the simulated sample used to construct the nominal response matrix are indicated with the dashed black line. The uncertainty contributions for different choices of colour reconnection models are illustrated as one-sided shifts compared to the nominal unfolding, and up (down) contributions from other sources are indicated with filled (open) markers of the same type and colour.}
	\label{fig:unfUncsTheorytop_tau1}
\end{figure}

\begin{figure}[htpb]
	\centering
	\includegraphics[width=.42\textwidth]{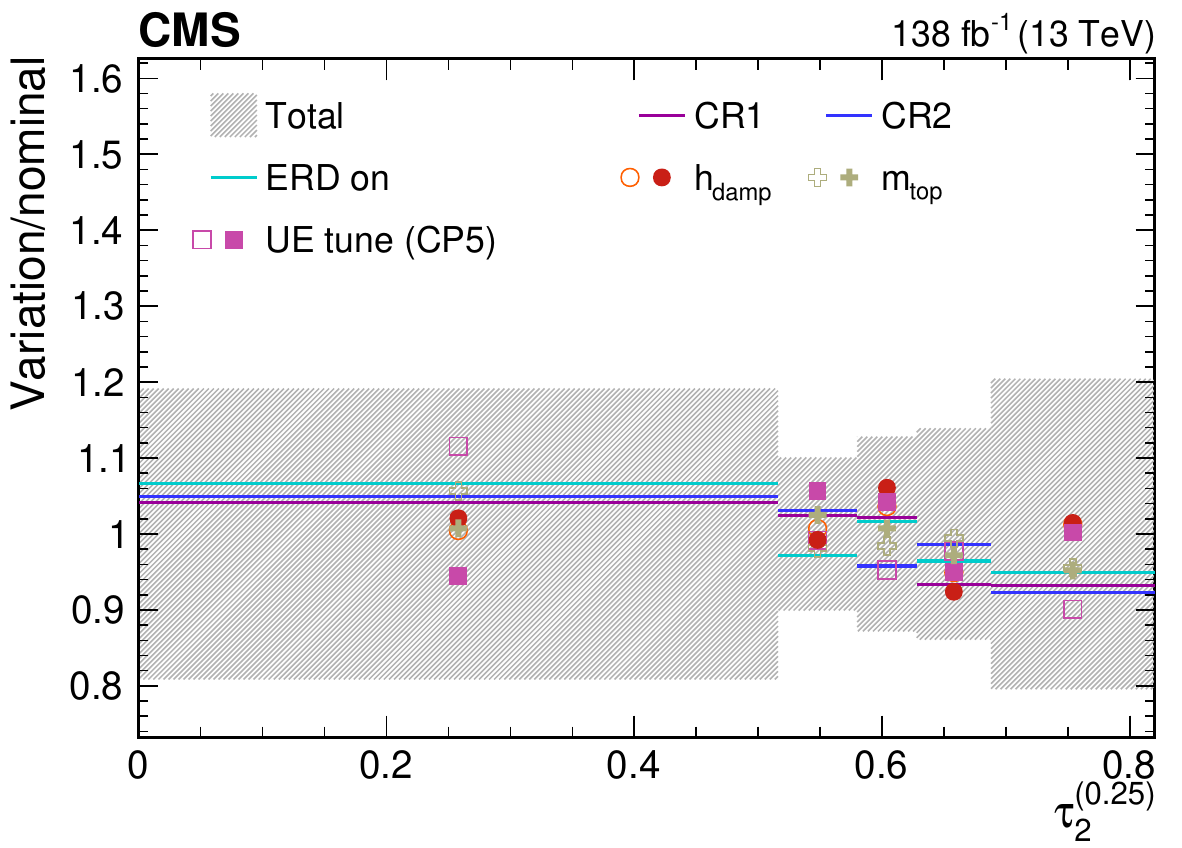}
	\includegraphics[width=.42\textwidth]{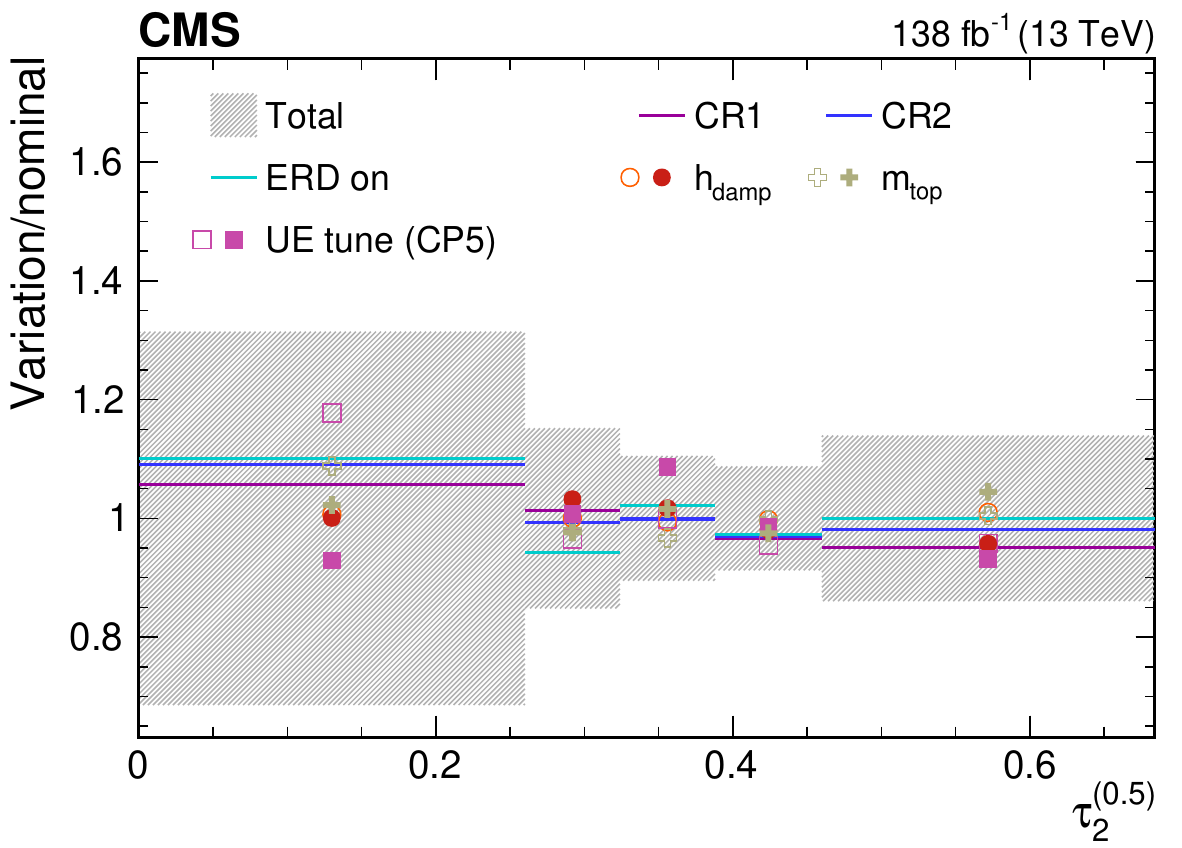}
	\includegraphics[width=.42\textwidth]{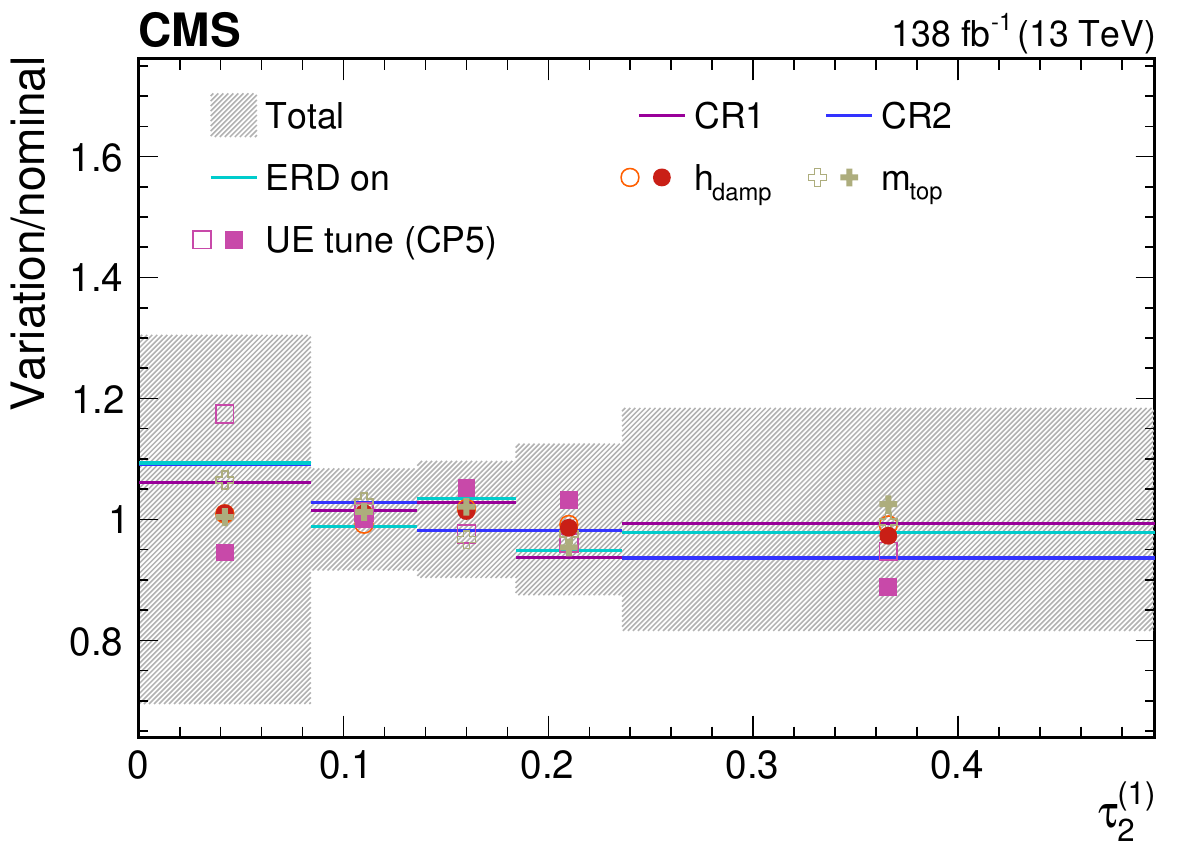}
	\includegraphics[width=.42\textwidth]{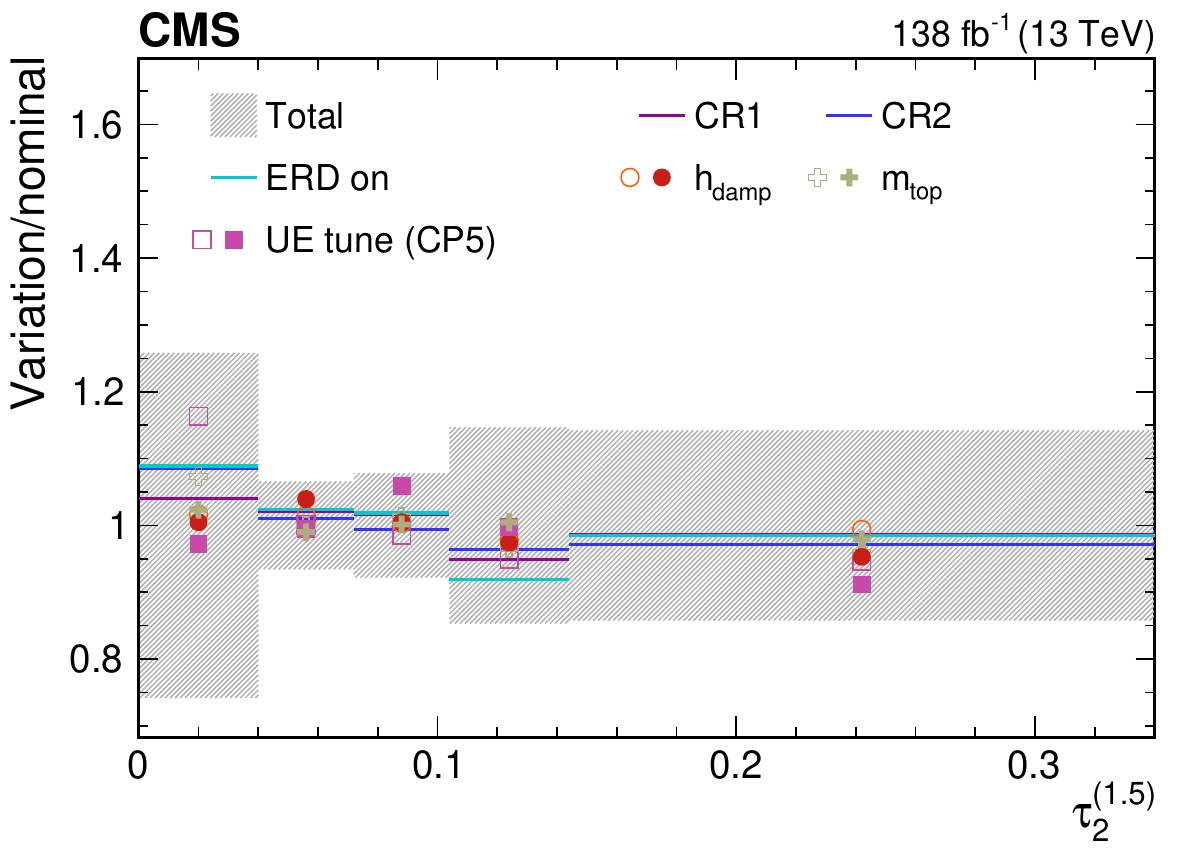}
	\includegraphics[width=.42\textwidth]{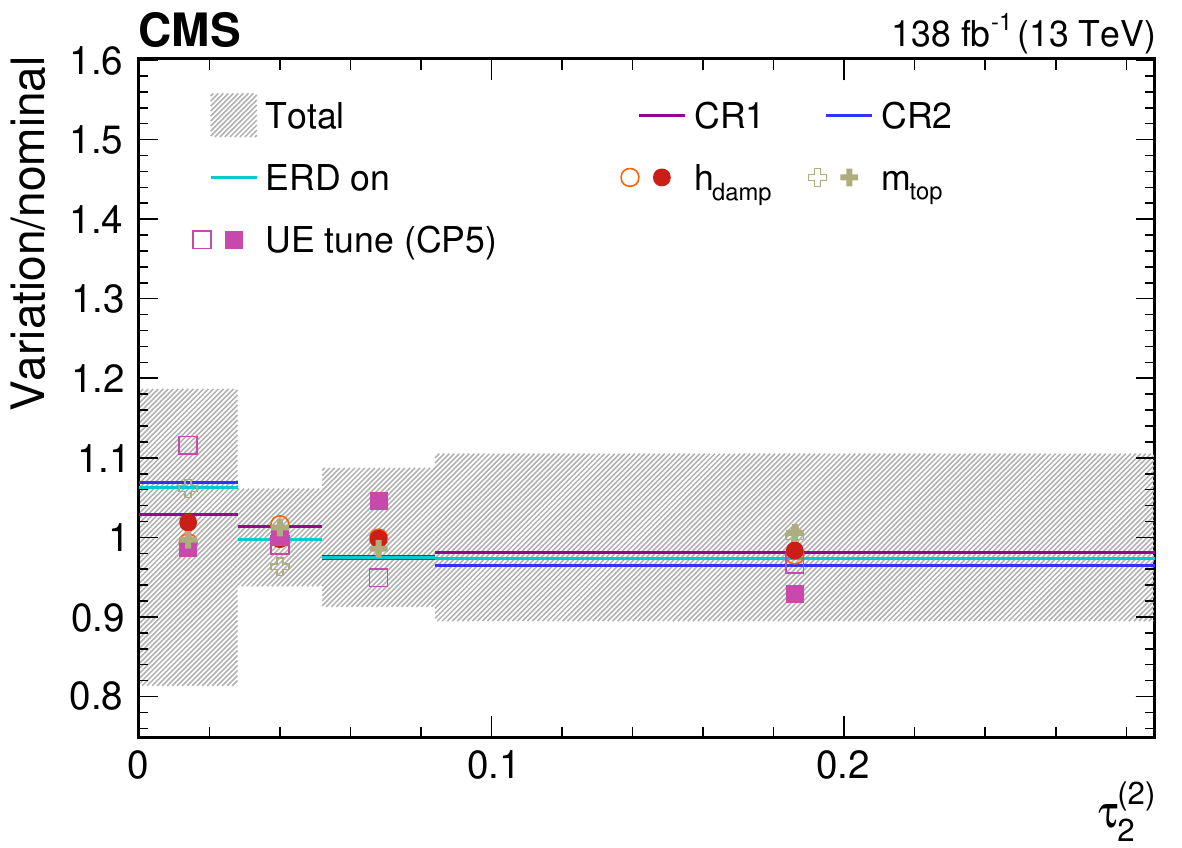}
	\caption{Contributions from various theory model systematic variations to the normalized, unfolded distribution for $\tau_2^{(\beta)}$ observables measured for AK8 jets passing the boosted top quark-enriched selection in $\PGm$+jets \ttbar events. 
		The total unfolding uncertainty is indicated with the dark grey, hashed region, while the blue hashed region indicates the contributions from the input covariance matrix, which includes the propagated effects of the statistical uncertainties of the input data after background subtraction. Contributions from statistical uncertainties of the simulated sample used to construct the nominal response matrix are indicated with the dashed black line. The uncertainty contributions for different choices of colour reconnection models are illustrated as one-sided shifts compared to the nominal unfolding, and up (down) contributions from other sources are indicated with filled (open) markers of the same type and colour.}
	\label{fig:unfUncsTheorytop_tau2}
\end{figure}

\begin{figure}[htpb]
	\centering
	\includegraphics[width=.42\textwidth]{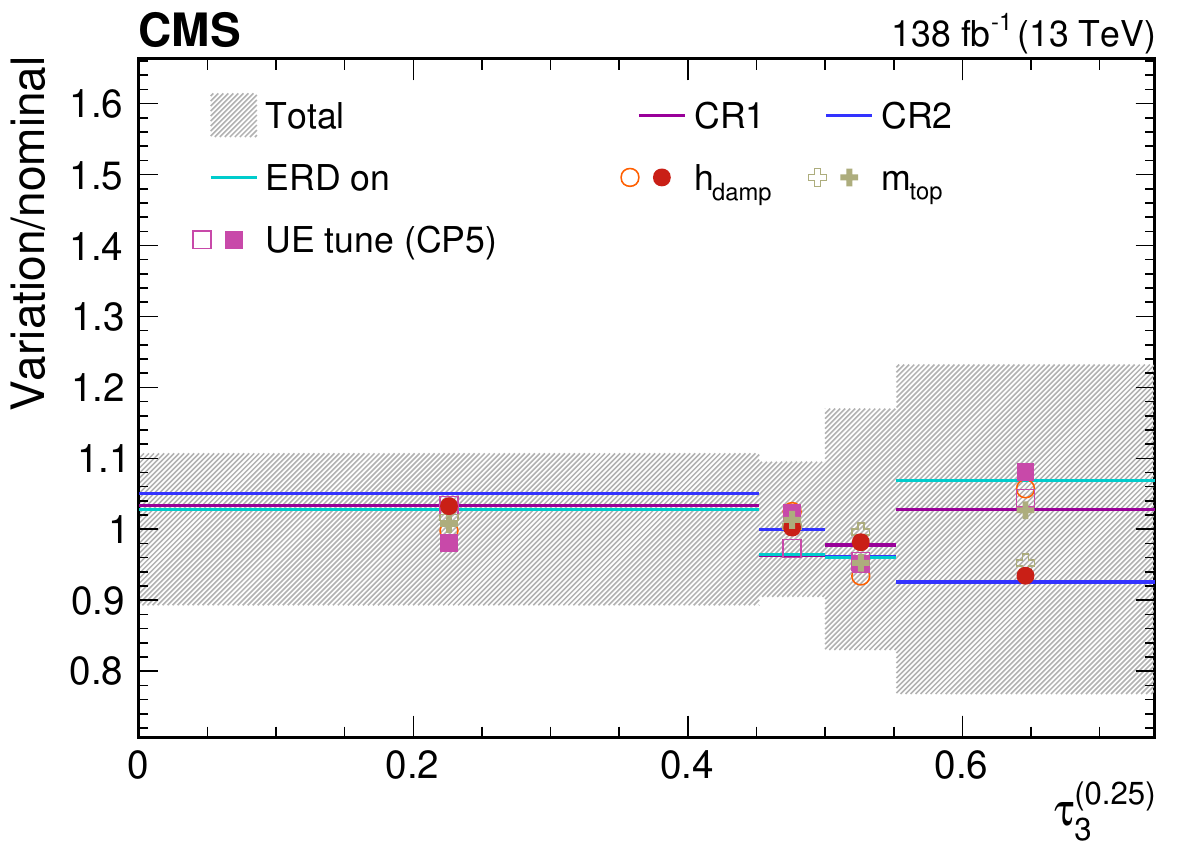}
	\includegraphics[width=.42\textwidth]{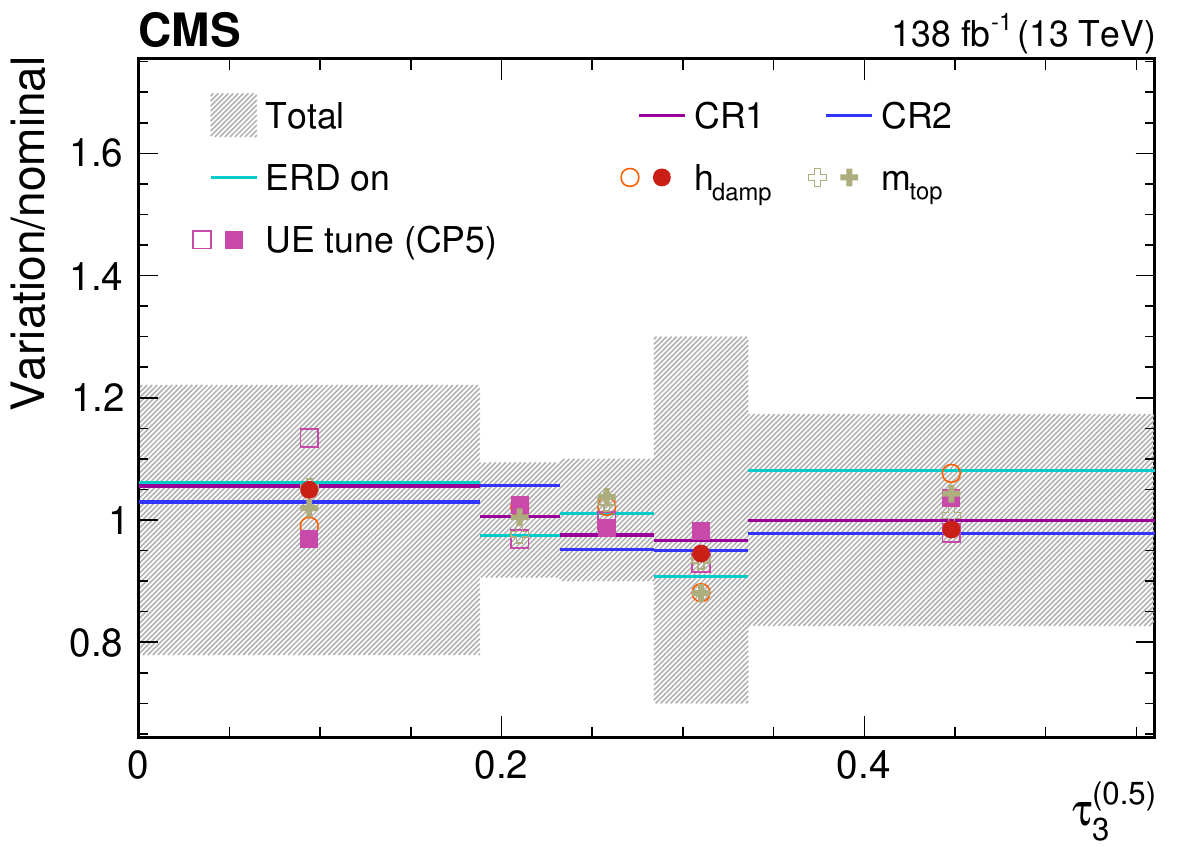}
	\includegraphics[width=.42\textwidth]{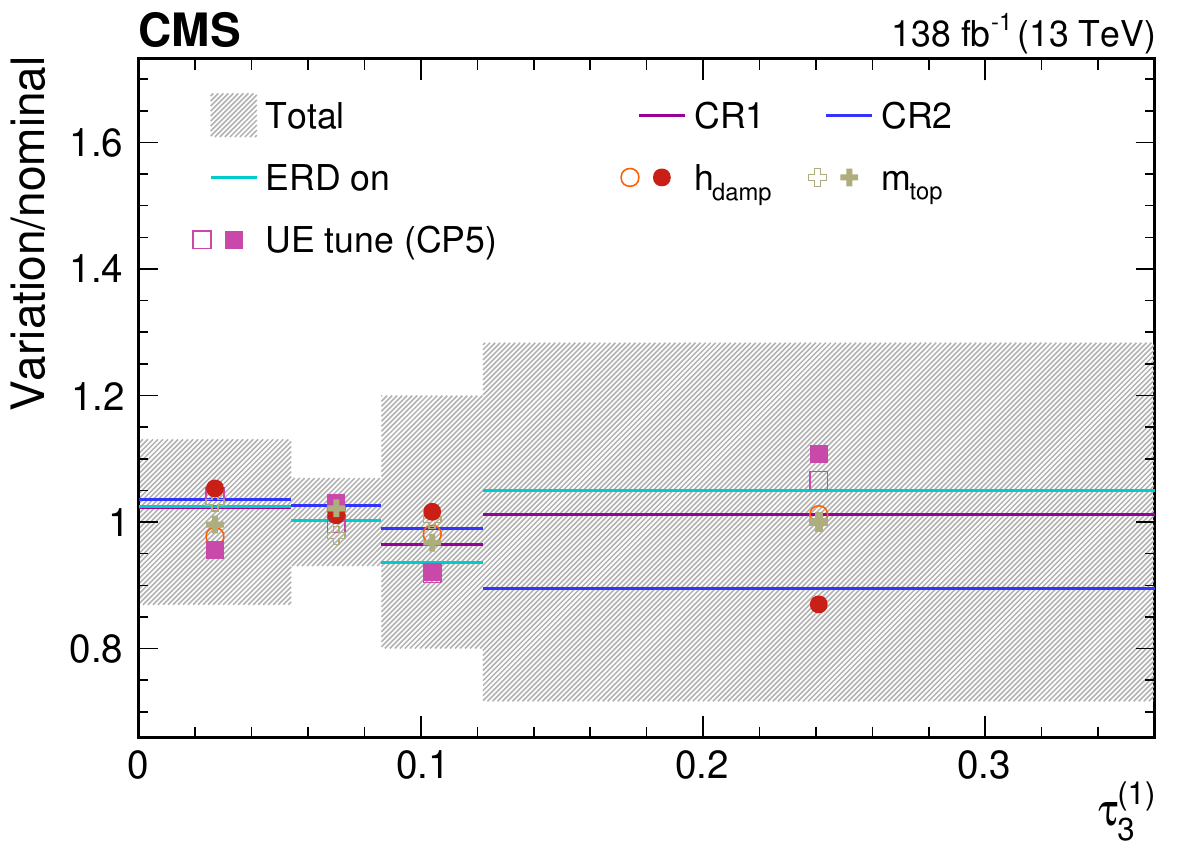}
	\includegraphics[width=.42\textwidth]{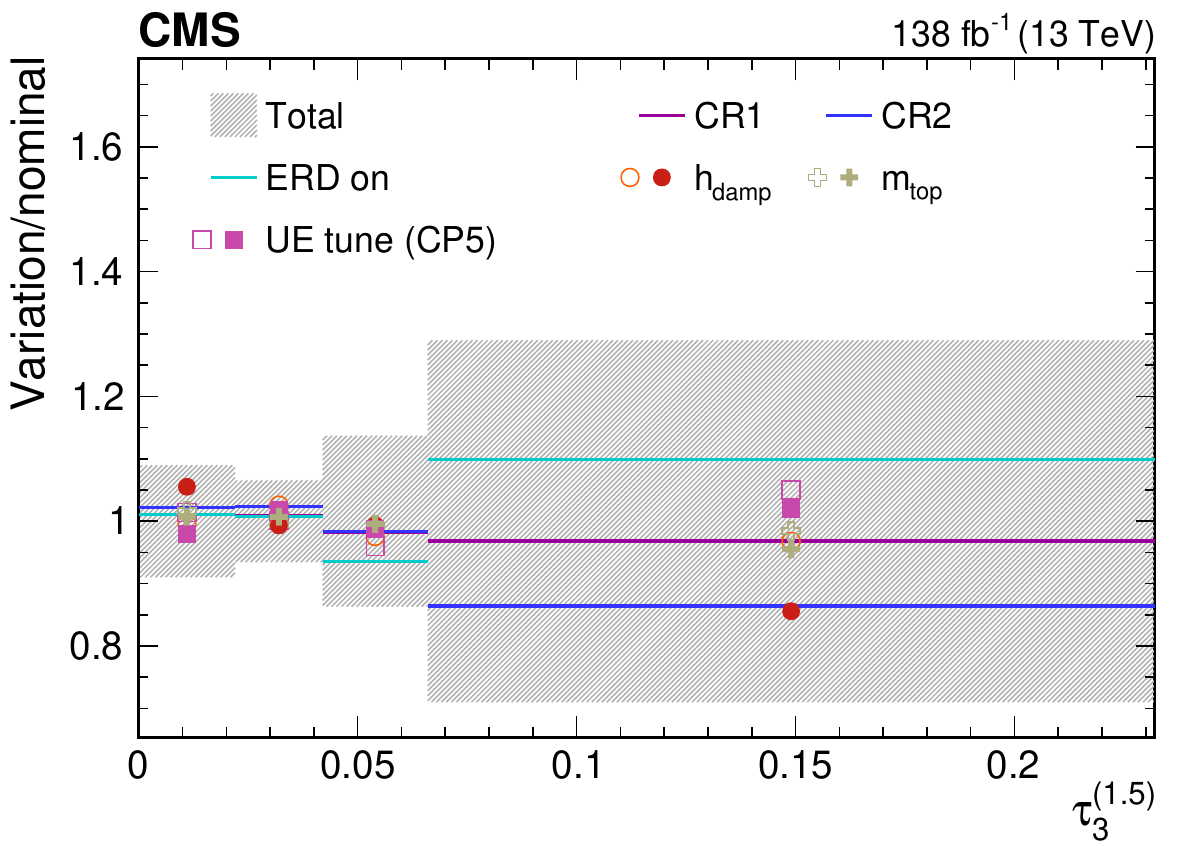}
	\includegraphics[width=.42\textwidth]{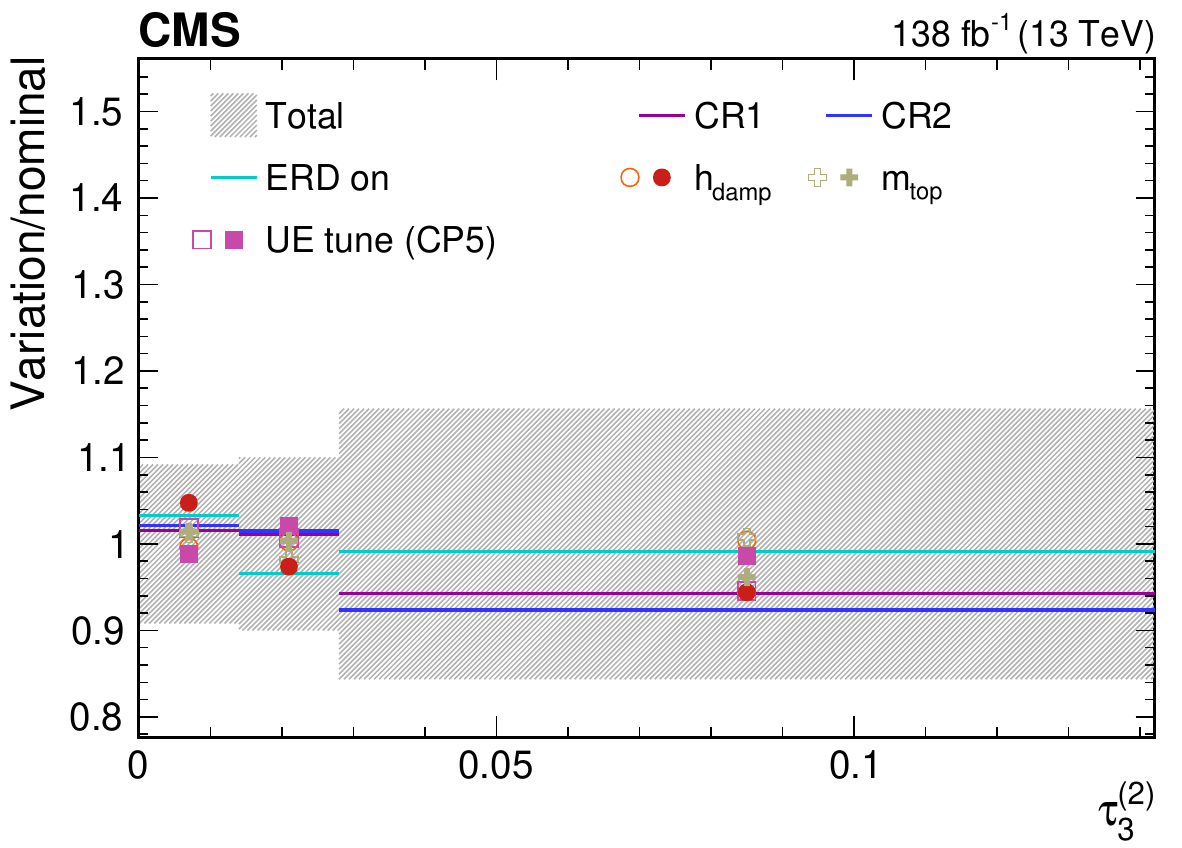}
	\caption{Contributions from various theory model systematic variations to the normalized, unfolded distribution for $\tau_3^{(\beta)}$ observables measured for AK8 jets passing the boosted top quark-enriched selection in $\PGm$+jets \ttbar events. 
		The total unfolding uncertainty is indicated with the dark grey, hashed region, while the blue hashed region indicates the contributions from the input covariance matrix, which includes the propagated effects of the statistical uncertainties of the input data after background subtraction. Contributions from statistical uncertainties of the simulated sample used to construct the nominal response matrix are indicated with the dashed black line. The uncertainty contributions for different choices of colour reconnection models are illustrated as one-sided shifts compared to the nominal unfolding, and up (down) contributions from other sources are indicated with filled (open) markers of the same type and colour.}
	\label{fig:unfUncsTheorytop_tau3}
\end{figure}

\begin{figure}[htpb]
	\centering
	\includegraphics[width=.42\textwidth]{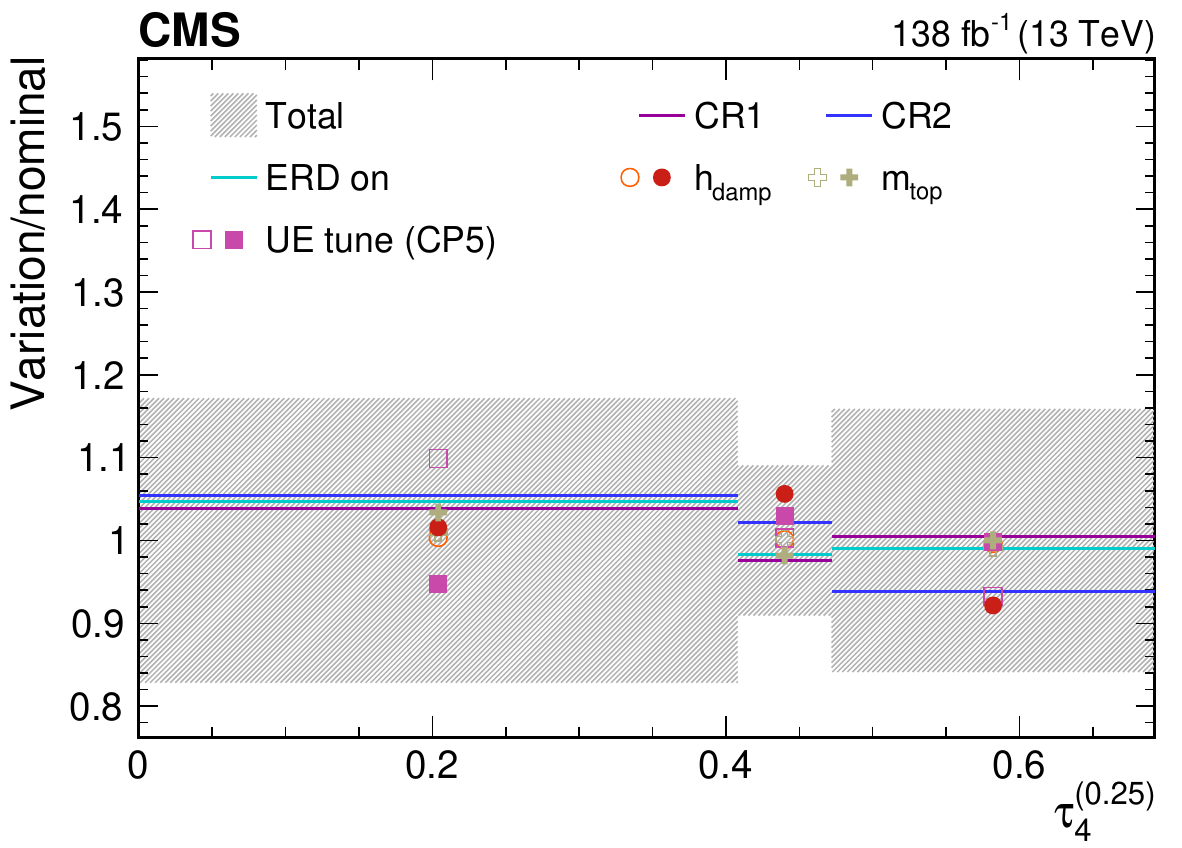}
	\includegraphics[width=.42\textwidth]{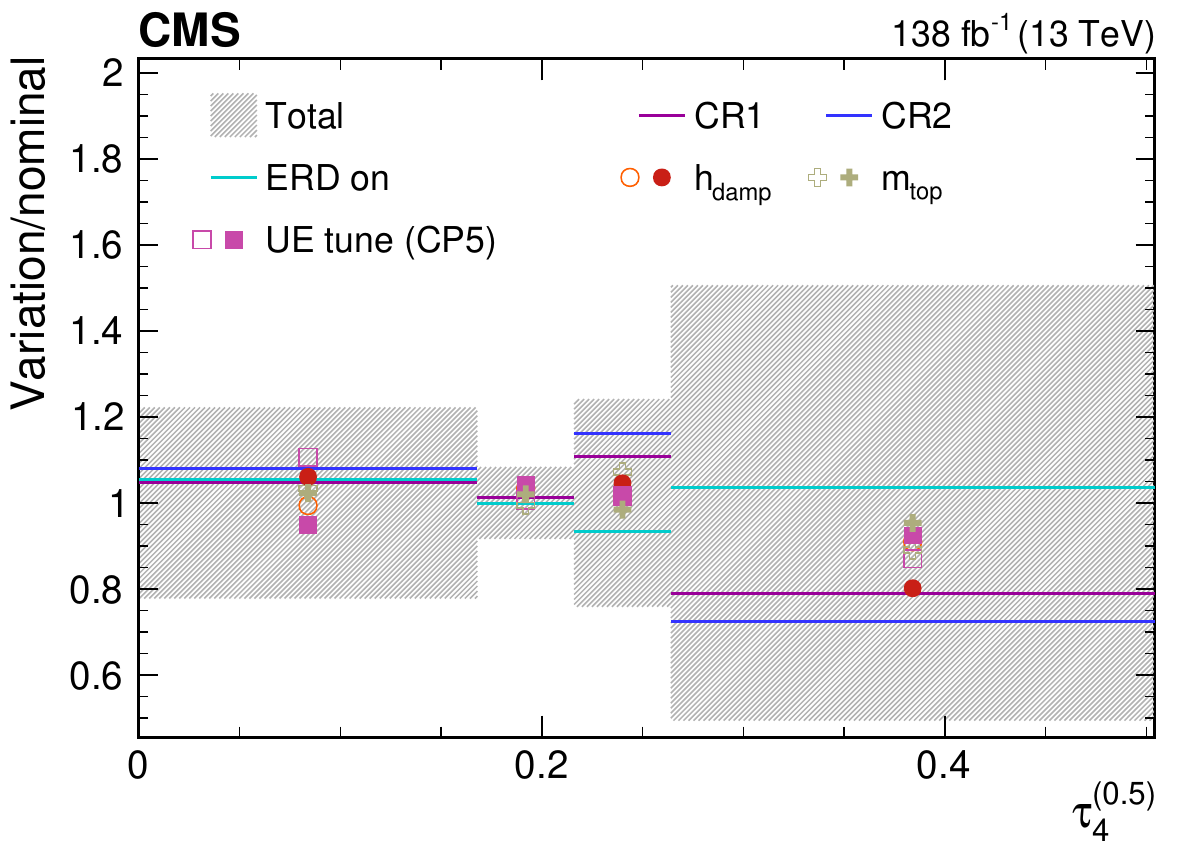}
	\includegraphics[width=.42\textwidth]{Figure_018-d.pdf}
	\includegraphics[width=.42\textwidth]{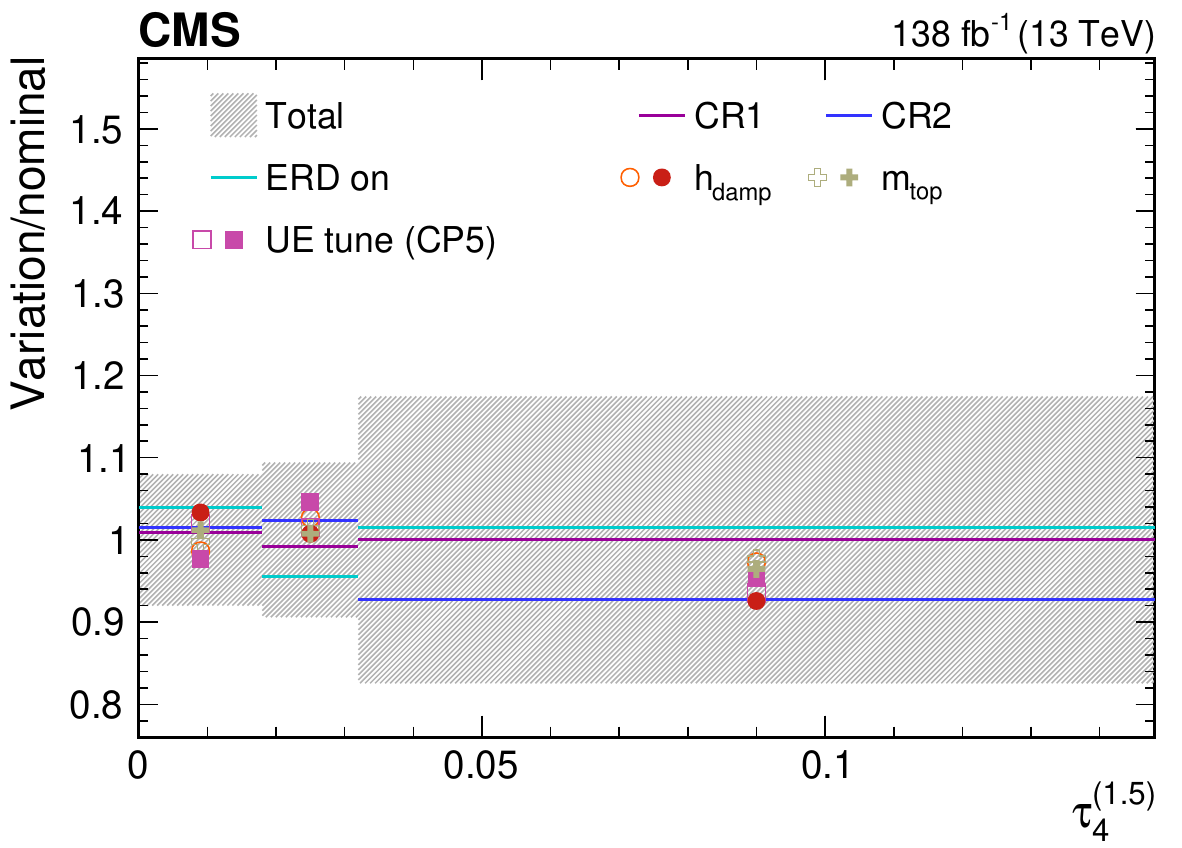}
	\includegraphics[width=.42\textwidth]{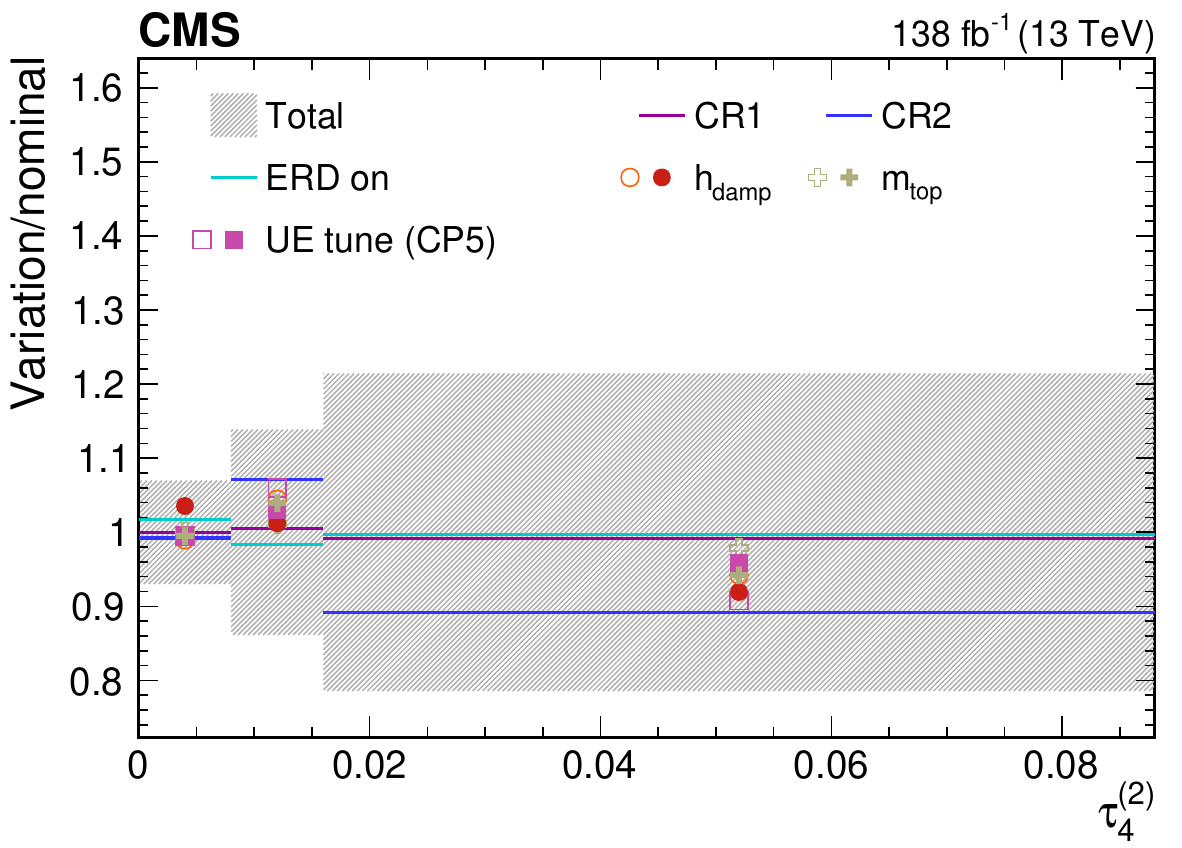}
	\caption{Contributions from various theory model systematic variations to the normalized, unfolded distribution for $\tau_4^{(\beta)}$ observables measured for AK8 jets passing the boosted top quark-enriched selection in $\PGm$+jets \ttbar events. 
		The total unfolding uncertainty is indicated with the dark grey, hashed region, while the blue hashed region indicates the contributions from the input covariance matrix, which includes the propagated effects of the statistical uncertainties of the input data after background subtraction. Contributions from statistical uncertainties of the simulated sample used to construct the nominal response matrix are indicated with the dashed black line. The uncertainty contributions for different choices of colour reconnection models are illustrated as one-sided shifts compared to the nominal unfolding, and up (down) contributions from other sources are indicated with filled (open) markers of the same type and colour.}
	\label{fig:unfUncsTheorytop_tau4}
\end{figure}

\begin{figure}[htpb]
	\centering
	\includegraphics[width=.42\textwidth]{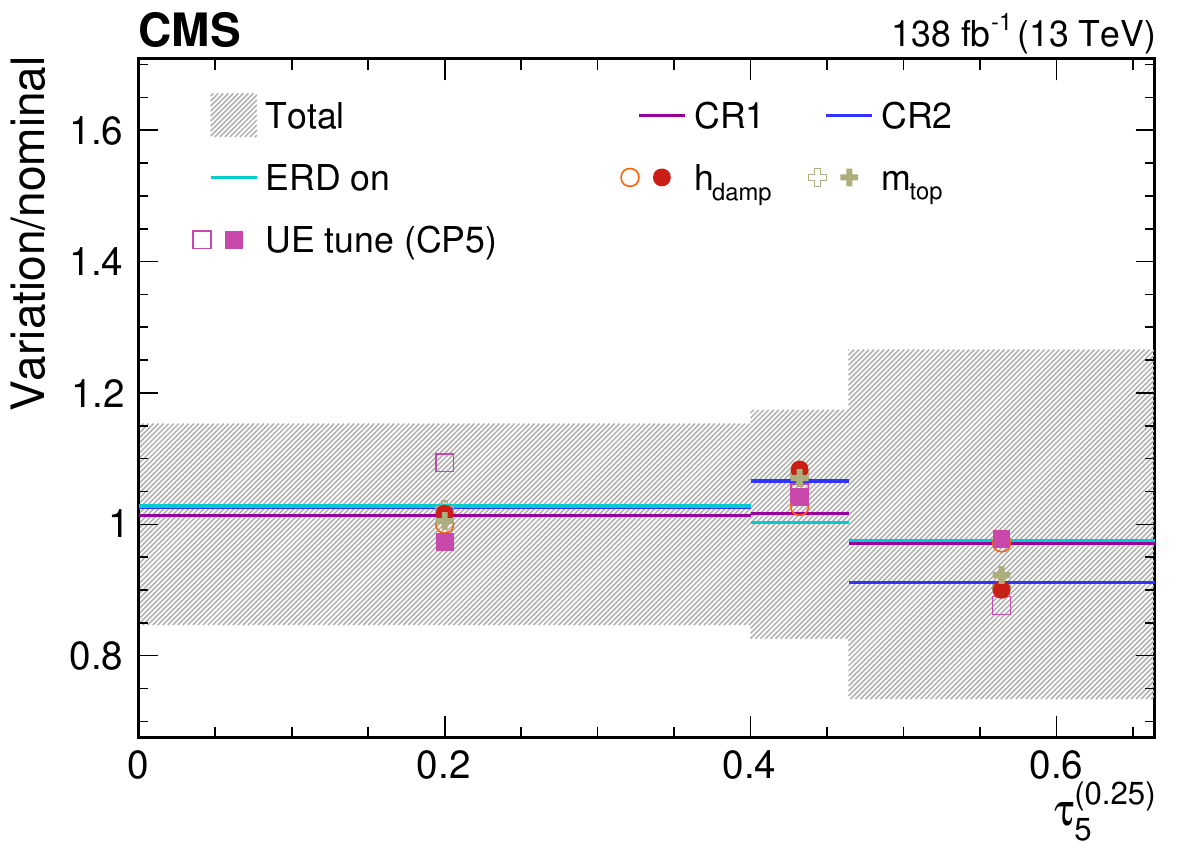}
	\includegraphics[width=.42\textwidth]{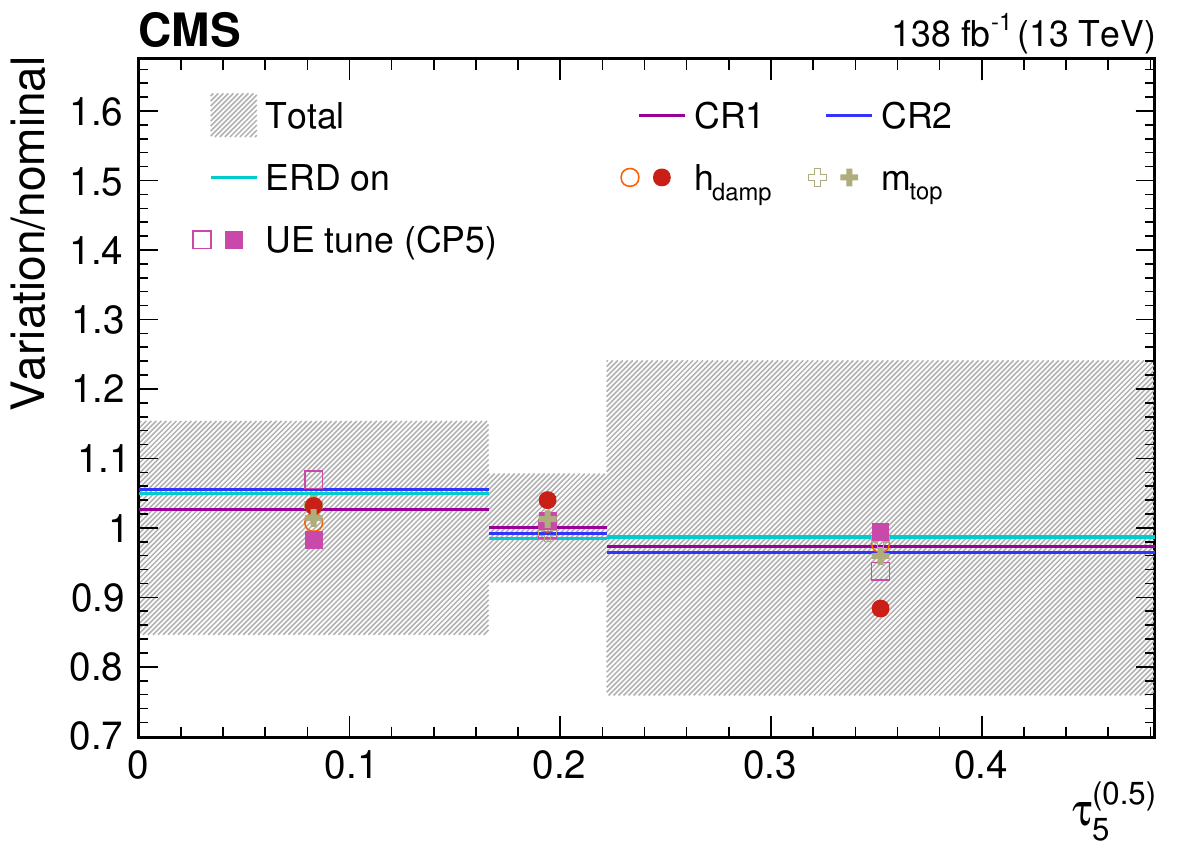}
	\includegraphics[width=.42\textwidth]{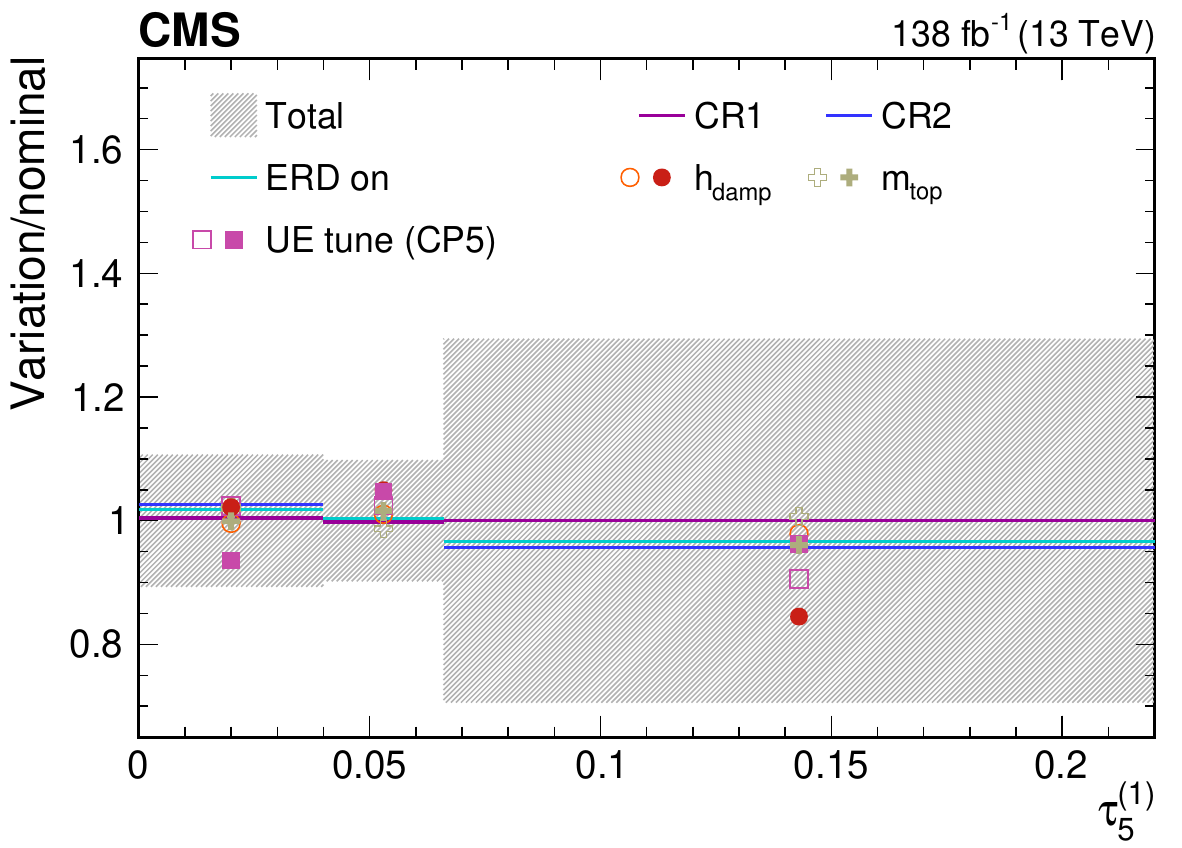}
	\includegraphics[width=.42\textwidth]{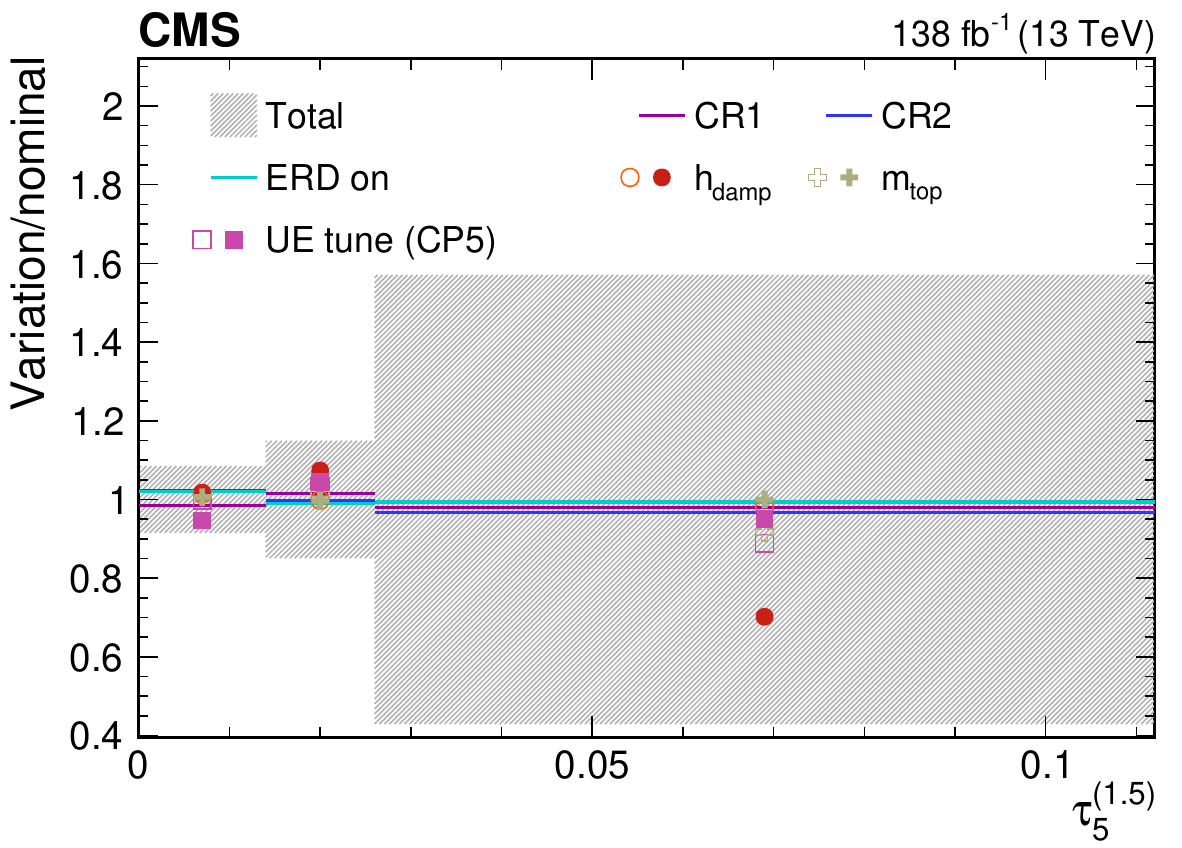}
	\includegraphics[width=.42\textwidth]{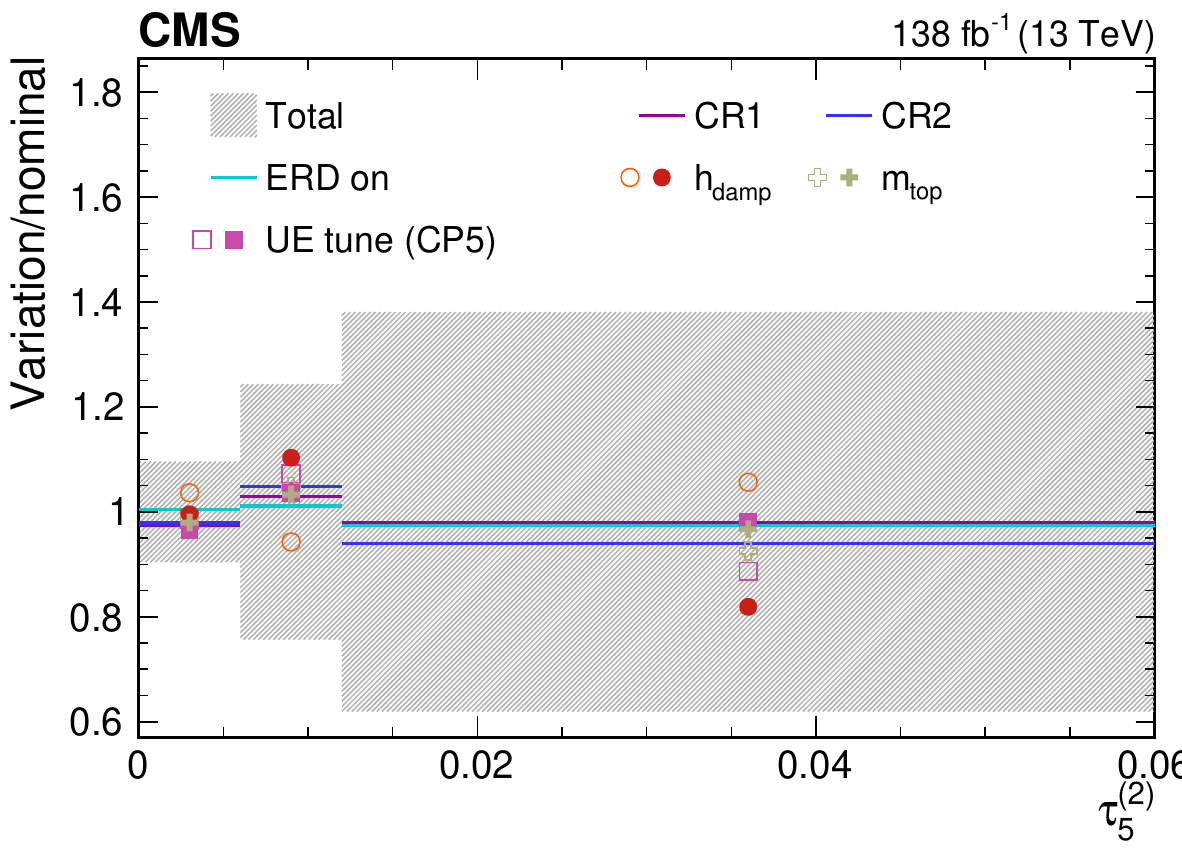}
	\caption{Contributions from various theory model systematic variations to the normalized, unfolded distribution for $\tau_5^{(\beta)}$ observables measured for AK8 jets passing the boosted top quark-enriched selection in $\PGm$+jets \ttbar events. 
		The total unfolding uncertainty is indicated with the dark grey, hashed region, while the blue hashed region indicates the contributions from the input covariance matrix, which includes the propagated effects of the statistical uncertainties of the input data after background subtraction. Contributions from statistical uncertainties of the simulated sample used to construct the nominal response matrix are indicated with the dashed black line. The uncertainty contributions for different choices of colour reconnection models are illustrated as one-sided shifts compared to the nominal unfolding, and up (down) contributions from other sources are indicated with filled (open) markers of the same type and colour.}
	\label{fig:unfUncsTheorytop_tau5}
\end{figure}

Estimated contributions of various sources of experimental and modelling uncertainty are presented for the measurement of 1- through 5-subjettiness in boosted top quark jets.
Uncertainty sources that are common also to the dijet selection, as well as sources of experimental uncertainty considered only for the boosted \PW boson-/top quark-enriched regions, are presented in one set of figures: Figs.~\ref{fig:unfUncstop_tau1}--\ref{fig:unfUncstop_tau5}, while those arising from model variations considered only in the boosted \PW boson-/top quark-enriched regions are presented in a separate set of figures: Figs.~\ref{fig:unfUncsTheorytop_tau1}--\ref{fig:unfUncsTheorytop_tau5}. 

\cleardoublepage \section{The CMS Collaboration \label{app:collab}}\begin{sloppypar}\hyphenpenalty=5000\widowpenalty=500\clubpenalty=5000\cmsinstitute{Yerevan Physics Institute, Yerevan, Armenia}
{\tolerance=6000
A.~Hayrapetyan, V.~Makarenko\cmsorcid{0000-0002-8406-8605}, A.~Tumasyan\cmsAuthorMark{1}\cmsorcid{0009-0000-0684-6742}
\par}
\cmsinstitute{Institut f\"{u}r Hochenergiephysik, Vienna, Austria}
{\tolerance=6000
W.~Adam\cmsorcid{0000-0001-9099-4341}, L.~Benato\cmsorcid{0000-0001-5135-7489}, T.~Bergauer\cmsorcid{0000-0002-5786-0293}, M.~Dragicevic\cmsorcid{0000-0003-1967-6783}, C.~Giordano\cmsorcid{0000-0001-6317-2481}, P.S.~Hussain\cmsorcid{0000-0002-4825-5278}, M.~Jeitler\cmsAuthorMark{2}\cmsorcid{0000-0002-5141-9560}, N.~Krammer\cmsorcid{0000-0002-0548-0985}, A.~Li\cmsorcid{0000-0002-4547-116X}, D.~Liko\cmsorcid{0000-0002-3380-473X}, M.~Matthewman, J.~Schieck\cmsAuthorMark{2}\cmsorcid{0000-0002-1058-8093}, D.~Schwarz\cmsorcid{0000-0002-3821-7331}, R.~Sch\"{o}fbeck\cmsAuthorMark{2}\cmsorcid{0000-0002-2332-8784}, M.~Shooshtari\cmsorcid{0009-0004-8882-4887}, M.~Sonawane\cmsorcid{0000-0003-0510-7010}, W.~Waltenberger\cmsorcid{0000-0002-6215-7228}, C.E.~Wulz\cmsAuthorMark{2}\cmsorcid{0000-0001-9226-5812}
% ...
\par}
$^{1}$Also at Yerevan State University, Yerevan, Armenia\\
$^{2}$Also at Technische Universit\"{a}t Wien, Vienna, Austria\\
% ...
\end{sloppypar}
%%% END EDITABLE REGION %%%
% skeleton_end
\end{document}